\documentclass[fleqn,usenatbib]{mnras}

\usepackage{newtxtext,newtxmath}

\usepackage[T1]{fontenc}
\usepackage{ae,aecompl}

\usepackage{graphicx}	
\usepackage{amsmath}	
\usepackage{amssymb}	
\usepackage{pdflscape}

\title[The VISCACHA survey - II. Structure of MC clusters]{The VISCACHA survey - II. Structure of star clusters in the Magellanic Clouds periphery}

\author[J. F. C. Santos Jr. et al.]{Jo\~ao F. C. Santos Jr.,$^{1,2}$\thanks{E-mail: jsantos@fisica.ufmg.br}
Francisco F. S. Maia,$^{3}$
Bruno Dias,$^{4}$
Leandro de O. Kerber,$^{5}$
\newauthor
Andr\'es E. Piatti,$^{6,7}$
Eduardo Bica,$^{8}$
Mateus S. Angelo,$^{9}$
Dante Minniti,$^{10,11,12}$
\newauthor
Angeles P\'erez-Villegas,$^{13}$
Alexandre Roman-Lopes,$^{2}$
Pieter Westera,$^{14}$
Luciano Fraga,$^{15}$
\newauthor
Bruno Quint$^{16}$ and
David Sanmartim$^{17}$
\\
$^{1}$Departamento de F\'isica, ICEx - UFMG, Av. Ant\^onio Carlos 6627, Belo Horizonte 31270-901, Brazil \\ 
$^{2}$Departamento de Astronom\'ia, Universidad de La Serena, Av. Juan Cisternas 1200 North, La Serena, Chile \\
$^{3}$Instituto de F\'isica, Universidade Federal do Rio de Janeiro, 21941-972, Rio de Janeiro, RJ, Brazil\\
$^{4}$Instituto de Alta Investigaci\'on, Universidad de Tarapac\'a, Casilla 7D, Arica, Chile\\
$^{5}$Departamento de Ci\^encias Exatas e Tecnol\'ogicas, UESC, Rodovia Jorge Amado km 16, 45662-900, Brazil \\
$^{6}$Instituto Interdisciplinario de Ciencias B\'asicas (ICB), CONICET-UNCUYO, Padre J. Contreras 1300, M5502JMA, Mendoza, Argentina\\
$^{7}$Consejo Nacional de Investigaciones Cient\'{\i}ficas y T\'ecnicas (CONICET), Godoy Cruz 2290, C1425FQB,  Buenos Aires, Argentina\\
$^{8}$Departamento de Astronomia, IF - UFRGS, Av. Bento Gon\c calves 9500, 91501-970, Brazil \\ 
$^{9}$Centro Federal de Educa\c c\~ao Tecnol\'ogica de Minas Gerais, Av. Monsenhor Luiz de Gonzaga, 103, 37250-000 Nepomuceno, MG, Brazil\\
$^{10}$Departamento de Ciencias F\'isicas, Universidad Andres Bello, Fernandez Concha 700, Las Condes, Santiago, Chile\\
$^{11}$Millennium Institute of Astrophysics, Av. Vicuna Mackenna 4860, 782-0436, Santiago, Chile\\
$^{12}$Vatican Observatory, V00120 Vatican City State, Italy\\
$^{13}$Universidade de S\~ao Paulo, IAG, Rua do Mat\~ao 1226, Cidade Universit\'aria, S\~ao Paulo 05508-900, Brazil\\
$^{14}$Universidade Federal do ABC, Centro de Ci\^encias Naturais e Humanas, Avenida dos Estados, 5001, 09210-580, Brazil \\
$^{15}$Laborat\'orio Nacional de Astrof\'isica, Rua Estados Unidos 154, Itajub\'a 37504-364, Brazil\\
$^{16}$NSF's OIR Lab - Gemini Observatory, c/o AURA - Casilla 603, La Serena, Chile\\
$^{17}$Las Campanas Observatory, Carnegie Institution of Washington, Colina el Pino, 601 Casilla, La Serena, Chile }

\date{Accepted XXX. Received YYY; in original form ZZZ}

\pubyear{2020}

\begin{document}
\label{firstpage}
\pagerange{\pageref{firstpage}--\pageref{lastpage}}
\maketitle

\begin{abstract}
  We provide a homogeneous set of structural parameters of 83  star clusters located at the periphery of the Small Magellanic Cloud (SMC) and the Large Magellanic Cloud (LMC). The clusters' stellar density and surface brightness profiles were built from deep, AO assisted optical images, and uniform analysis techniques. The structural parameters were obtained from King and Elson et al. model fittings. Integrated magnitudes and masses (for a subsample) are also provided.  The sample contains mostly low surface brightness clusters with distances between 4.5 and 6.5\,kpc and between 1 and 6.5\,kpc from the LMC and SMC centres, respectively.  We analysed their spatial distribution and structural properties,  comparing them with those of inner clusters. Half-light and Jacobi radii were estimated, allowing an evaluation of the Roche volume tidal filling. We found that: (i) for  our sample of LMC clusters, the tidal radii are, on average, larger than those of inner clusters from previous studies; (ii) the core radii dispersion tends to be greater for LMC clusters located towards the southwest, with position angles of $\sim$200 degrees and about $\sim$5 degrees from the  LMC centre, i.e., those LMC clusters nearer to the SMC; (iii) the core radius evolution for clusters with known age is similar to that of inner clusters; (iv) SMC clusters with galactocentric distances closer than 4\,kpc are overfilling; (v) the recent Clouds collision did not leave marks on the LMC clusters' structure that our analysis could reveal.
\end{abstract}


\section{Introduction}

 Star clusters located at the outskirts of the Small Magellanic Cloud (SMC) and the Large Magellanic Cloud (LMC) are  witnesses of the disturbed environment generated  by the galaxies' interaction with each other  and the Milky Way (MW). The changing tidal field produced when galaxies interact gravitationally induces star formation by compression of the gas in certain regions \citep[e.g.][]{rbk14}. Recent star formation (in cluster complexes) was detected in tidal tails of merging galaxies \citep[e.g.][]{wzl99}. Studies linking epochs of enhanced star formation with the MCs' approach are numerous \citep[e.g.][]{bgd98,s04,ggk10,ldk13,rgk15,shz19}.

 The Magellanic Clouds (MCs) are $\sim 20$\,kpc apart and the distance between them is increasing \citep{zkm18}. Convincing evidence that the MCs had a recent and close encounter are: (i) The bridge of gas and stars between them seems to be a tidal feature \citep[e.g.][]{yn03,bkh12}; (ii) The relative orientation of their three-dimensional velocity vectors, obtained from their proper motions and Doppler redshifts, implies at least one collision within the last 500\,Myr \citep{kmb13}.

\cite{zkm18} modeled the MCs past mutual interactions based on proper motions obtained from Hubble Space Telescope (HST) images, resulting in a relative velocity between them of $103\pm 26$\,km\,s$^{-1}$. They found that in 97 per cent of the simulated cases the MCs had a direct collision $147\pm33$\,Myr ago, with a mean impact parameter of $7.5\pm 2.5$\,kpc. Considering the escape velocity of the LMC of $90$\,km\,s$^{-1}$ (assuming that its mass is $1.7\times 10^{10}$\,M$_\odot$ according to \cite{df16} and the present distance between the MCs is $\sim 20$\,kpc), it would be unlikely that they existed as a binary system for a long time, unless the LMC was much more massive than current observations indicate \citep{b15}.

Given these uncertainties, the process of interaction has been debated in several studies. The classical scenario in which the MCs are orbiting the Galaxy \citep{gn96,db12} has an alternative one where these two dwarf irregular galaxies are approaching the MW for the first time \citep{bkh07,bkh10}. In the classical scenario, simulations show that the HST proper motions and models with a high mass for the Milky Way ($1-2 \times 10^{12}$\,M$_\odot$) imply excentric orbits of the MCs with periods of 3-9\,Gyr, leading to the conclusion that if it is the correct perspective, the Clouds have performed no more than two to three revolutions around the Galaxy \citep{kmb13,df16}. In both cases, gravitational forces generate tidal effects on the gas and stellar content of the galaxies, making their structures complex, which challenges interpretation \citep{mkm01,mmm05,tbp19}. The high accuracy of the HST MCs proper motions \citep{kmb13} favoured the scenario of first approach, raising doubts about classical orbital models. The first accepted models on a first encounter \citep{bkh12} predicts a direct collision between the MCs in the last 500\,Myr, triggering a ring-like structure in the periphery of the LMC, disaggregating gas and stellar content from the SMC and producing tidal effects like the Magellanic Bridge. 

 \citet[][hereafter WZ11]{wz11} found that the LMC lacks star clusters that are as large as those in the SMC, and suggested that this could be a signature of stronger tidal forces in the LMC. However, since they only covered the central part of the galaxy, they could not explore such effects in the LMC outer disc. By using a sizable sample of clusters in the LMC periphery we  aim to study the clusters' structures in order to probe WZ11's results.

 Under the influence of a steady tidal field, a cluster evolves dynamically by evaporation of stars through two-body relaxation. The tidal field contributes to lower the escape energy of stars that then may leave the cluster \citep{sp87,hh03}. By losing mass via stellar evolution, evaporation, tidal stripping and shocks \citep[e.g.][]{lbg10,wrk19}, a cluster changes its internal energy, which flows from the inner core to the  outer region \citep{sp87,hh03}. The energy flow leads to the collapse of the cluster's core, increasing its binding energy and causing an overall expansion as the outer regions heat. The core eventually stops shrinking and expands due to the injection of energy from newly formed binaries \citep{gh89}.  As a consequence, the cluster's structure is altered, with the sizes of the core and outer regions varying non homologously \citep{pmg10}. 

 For clusters  on eccentric orbits, the slowly varying tidal field contributes to the tidal heating which increases mass loss. In addition, since the Jacobi radius (a gravitational limit for stars bound to the cluster) shrinks when a cluster passes by the perigalacticon, energetic stars in the cluster outskirts may change status from bound to unbound, contributing to mass loss \citep[e.g.][]{wls14}.  If the cluster stars' orbital periods exceed the time of the effective interaction, then a tidal shock ensues. Besides the perigalacticon passage, a tidal shock may also occur when a cluster moves closeby a molecular cloud \citep{gr16}  or when a cluster crosses the Milky Way disc \citep{osc72}, or in a changing environment as settled by collisions of galaxies that eventually merge \citep{kpl12,rg13}. In all cases, the tidal shock would also heat preferentially the outer regions of the cluster, where the tides are more effective.   The strength of the tidal field and the duration of the cluster interaction with an enhanced density matters to define the cluster mass loss rate, when it can survive longer or dissolve faster. Strong tidal forces can dominate the evolution of star clusters in merging galaxies and determine their mass loss rates and lifetimes \citep{mmv17}.

 All these dynamical mechanisms ultimately lead the cluster to dissolution on different time-scales,  altering the cluster structure. Detailed reviews on cluster evolution in a broad context can be found in \cite{v10,pmg10,r18,kmb19}.

According to \cite{bbds08}'s catalogue, the LMC outer disc ($r>5\degr$) contains about 260 star clusters.  The vast majority of them has only their positions and visual sizes catalogued. The picture is similar for the SMC outer clusters \citep{bwk20}. Therefore, an increase of the number of MC clusters with well-known properties is desirable for a comprehensive knowledge of the formation and chemodynamical evolution of the Clouds. Besides our observational campaign \citep[][hereafter paper I]{mds19}, other research groups are working in the field to fulfill this gap, e.g. the DES collaboration \citep{psb16}, the SMASH \citep{now17}, the OGLE-IV \citep{sss16,ssu17} and the VMC \citep{ccg11} surveys.

In this work, we present structural parameters for 51 LMC and 32 SMC clusters located in the galaxies' outskirts. The 4.1-m SOAR Telescope Adaptive Module \citep[SAM;][]{tct16} was used to carry out optical observations in the context of the VIsible Soar photometry of star Clusters in tApii and Coxi HuguA (VISCACHA\footnote{\url{http://www.astro.iag.usp.br/~viscacha/}} - see paper~I for details) Survey. A summary of the observations and data reduction is presented in Sect~\ref{sec:obs}. Our sample consists of an homogeneous data set of clusters observed towards low reddening (Galactic and extragalactic) sightlines. At the distance of the MCs, the instrument field-of-view ($3\arcmin\times3\arcmin$) and spatial resolution ($0.09\arcsec$/pixel, binned array) is particularly suitable to investigate star clusters inner and outer structures, whose  angular sizes are typically 1\arcmin. All these aspects make the VISCACHA Survey a qualified database to explore the properties of the MCs via star clusters, whose distribution in spatially representative groups is provided in Sect.~\ref{sec:distrib}.   Homogeneous results on structural and photometric parameters  based on empirical model fittings to cluster radial surface brightness and stellar density are presented in Sect.~\ref{sec:king}. The general methodology used in the present study is detailed in paper~I. Structural parameters are investigated in connection with clusters' distances to the parent galaxy centre and age in Sect.~\ref{sec:res}. In Sect.~\ref{sec:rh_rj}, we determined the half-light and the Jacobi radii of the clusters.  A discussion of the results is provided in Sect.~\ref{sec:disc} and the concluding remarks are given in Sect~\ref{sec:conc}.

\section{Observations and data reduction}
\label{sec:obs}

A description of the VISCACHA Survey and the related observations is given in paper~I, where the instrument setup, observational strategy, and the methodology for the data reduction and calibration are fully explained. Here we provide a brief summary.

SAM is a ground-layer 
adaptive optics (GLAO) instrument using
a Rayleigh laser guide star (LGS) at $\sim$7 km from the telescope \citep{t13}. 
SAM was employed with its 
internal CCD detector, SAMI (4K$\times$ 4K CCD). It was set to a gain of 
2.1\, e$^-$/ADU and a readout noise of 4.7\,e$^-$. The CCD
binning (2 $\times$ 2) provides a plate-scale of 0.091\,arcsec/pixel with
the detector covering 3.1$\times$3.1\,arcmin on the sky.

Photometric image data with $BVI$ filters were obtained for the cluster sample during semesters 2015A, 2015B and 2016B. 
Total integration times were 1350s ($B$), 1125s ($V$) and 1680s ($I$) for LMC clusters, 1200s ($V$), 1800s ($I$) for old SMC clusters and 300s ($V$) and 400s ($I$) for young SMC clusters. Short exposures were also acquired (2x30s in all bands for all clusters) to replace stars saturated in the longer exposures. 
The data were processed for bias subtraction and division by skyflats in a standard way. Astrometric calibration was performed with IRAF
tasks using The Guide Star Catalog \citep{llm08} as astrometric reference. The three images obtained per filter were
combined (removing cosmic rays) unless their seeing differed significantly, with the poorer seeing images being discarded. The combined images were then calibrated photometrically using \cite{s00} standards or stars in fields in common with the Magellanic Clouds Photometric Survey \citep[MCPS][]{zht02,zht04}.
See \cite{fkt13} for additional reduction and astrometric calibration details.

\section{Location of the observed star clusters}
\label{sec:distrib}

\begin{figure}
\includegraphics[width=0.99\linewidth]{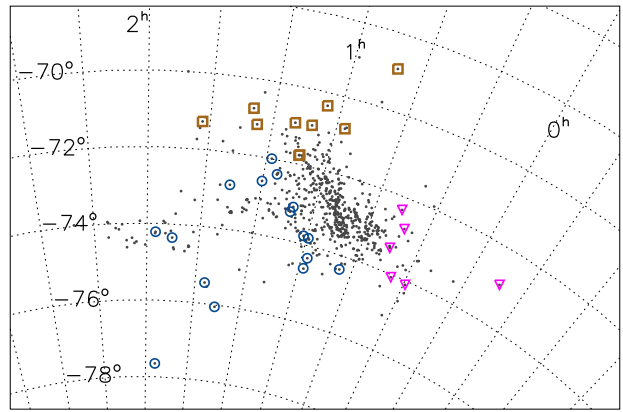}
\caption{On-sky projected spatial  distribution of SMC star clusters 
\citep[][small dots]{bbds08}. 
The observed clusters are represented by coloured symbols according to their locations following \citet{dkb14,dkb16}: 
west halo (inverted triangles), wing/bridge (circles), counter-bridge (squares).}
\label{fig:chart}
\end{figure}

The selected MCs peripheral star clusters for this study are represented by the coloured symbols in Figs.~\ref{fig:chart} and ~\ref{fig:map_deproj}. The distribution of all clusters in the catalogue by \cite{bbds08}, recently updated for the SMC and Bridge by \cite{bwk20}, is indicated by small, grey dots, helping to reveal the galaxies main structures. The different symbols and colours separate clusters in different external regions of the galaxies. The SMC clusters were discriminated according to the regions defined by \citet{dkb14,dkb16}, i.e., west halo, wing/bridge and counter-bridge (Fig.~\ref{fig:chart}). The LMC clusters were divided in four regions along the outer ring (Fig.~\ref{fig:map_deproj}). 

\subsection{Deprojected distance to the LMC centre}

The complex morphology of the MCs originated by the MW-LMC-SMC gravitational interaction can be traced back from simulations and observations \citep[e.g.][]{bkh12}. Although several analyses provide evidence of the dynamical complexity of the LMC \citep[e.g.][]{mke16,cno18a,mkd18,pac19,be19}, the bulk motion of the stellar and gas components as revealed by early radial velocities measurements \citep{dv54} and recent proper motion observations \citep{vk14,v18} assure that the LMC conforms to clockwise rotating disc dynamics. \cite{sso92} analysed radial velocities for $\sim 40$ LMC clusters located beyond $5\degr$ from the LMC centre, also implying disc-like kinematics.

Therefore, taking into account the kinematical observations and the available models, we employed LMC disc parameters to deproject our LMC cluster sample. Our objective is to search for any possible connection between the deprojected clusters' distances from the dynamical LMC centre and the clusters' structural parameters. 

\begin{figure}
\hskip -0.5cm \includegraphics[width=1.1\linewidth]{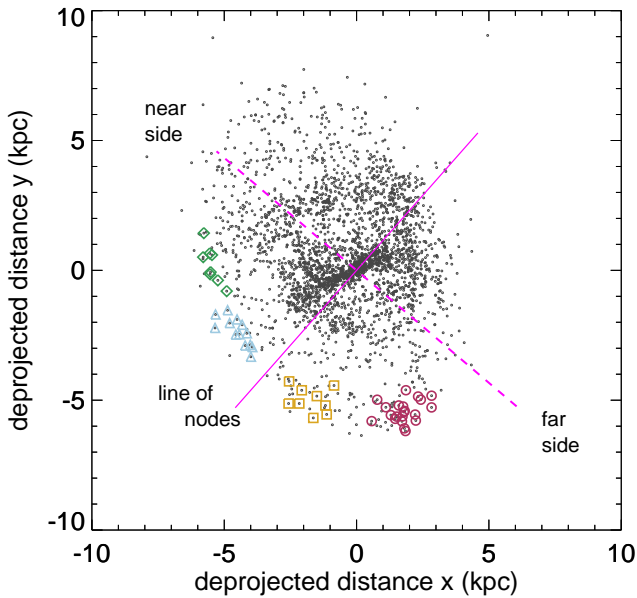}
\caption{Deprojected distribution of LMC star clusters. Clusters from the \citet{bbds08} catalogue are represented by dots. Coloured symbols identify azimuthally distinct star cluster groups in our sample. The line of nodes (continuous magenta line) separating the closer and far away sides of the LMC is also indicated. East is towards the left and north is up.}
\label{fig:map_deproj}
\end{figure}

 Star clusters deprojected distances from the LMC dynamical centre were computed according to the equation:

\begin{equation}
d = s\left[1 + \sin^2(PA - \theta) \tan^2i\right]^{1/2},
\end{equation}

 \noindent where $s$ is the projected distance, $PA$ is the position angle of the cluster, $\theta$ is the position angle of the line of nodes and $i$ the inclination angle of the LMC plane. This deprojection is shown in Fig.~\ref{fig:map_deproj}. The adopted values of $\theta =139.1^\circ$ and $i=34^\circ$ were based on a model built from proper motion and radial velocities of the disc old population \citep{vk14}. The coordinates of the LMC dynamical centre are $\alpha$=5$^{\rm h}$ 19$^{\rm m}$ 31$^{\rm s}$ and $\delta$= -69$^\circ$ 35$^\prime$ 24$^{\prime\prime}$ and its adopted distance modulus $(m-M)_\circ=18.5\pm0.1$ \citep{vk14,gwb14}.

 The overall LMC cluster deprojected distribution peaks at $\sim 2.5$\,kpc from the galaxy centre, away from the mean deprojected distance of our LMC cluster sample (between 4.5 and 6.5\,kpc). Therefore, our sample is located in less crowded areas of the LMC, where the contrast between field and cluster stars is enhanced.

\subsection{Clusters distance to the SMC centre}

The interaction with the LMC and the MW disturbs the SMC, transforming the galaxy into a more complex structure that resembles an ellipsoid elongated along the line of sight according to different tracers \citep{mfv86,csp01,ss12,jsm17}. As a first guess, we used the cluster projected distances from the SMC centre as our reference to search for spatial variation of the clusters' structural parameters. A database with individual MC cluster distances (and deprojected distances) is one of the goals of the VISCACHA survey and  will be presented in  Kerber et al. (in preparation). In the present study, we adopted for the SMC $(m-M)_\circ=18.96\pm0.20$ \citep{gb15}, where the error accounts for the SMC depth. The adopted centre is $\alpha=00^{\rm h}52^{\rm m}45^{\rm s}$, $\delta=-72^\circ 49^\prime 43^{\prime\prime}$, estimated by \cite{csp01} on the basis of the positions of 12 SMC clusters. A similar centre -- about 10\,arcmin north and 1\,arcmin west from \cite{csp01} centre -- was obtained by \cite{gb15} who compiled distances for 25 clusters, including the clusters from \cite{csp01}.

\section{Cluster structures from empirical model fittings}
\label{sec:king}

 To provide a database of structural parameters obtained from different models and access how well they fit the profiles we used the empirical models by \cite{k62} and \citet[][hereafter EFF]{eff87}.  Although dynamical models \citep[e.g.][]{k66,w75} would be preferred, they are mainly applied to globular clusters, where the large number of stars allows a detailed analysis yielding a robust inference of the clusters' parameters. Dynamical models also need the velocity distribution to better constrain the clusters' outer structure \citep[e.g.][]{gz15}, which we do not have available. Also, \cite{k66} pointed out that his 1962 empirical model closely follow the dynamical one for $W_\circ\leq 7$, which corresponds to concentration parameters $\log(r_t/r_c)\leq 1.53$ that are compatible with our sample. Therefore, to evaluate how well the clusters' structures are fitted by a non-tidally truncated model, we fitted the EFF model, which was shown to reproduce well the profiles of young LMC clusters.

The first step for the characterization of the clusters' structural parameters is to determine their centres. To do this, stellar positions and fluxes were extracted from the reduced images using the program {\sc DAOPHOT} \citep{s87} as implemmented in {\sc IDL}. Only sources brighter than 3\,$\sigma$ above the sky level were considered. The centre of each cluster was determined iteratively starting with an average of the stars' coordinates within a circle defined by the cluster's visual radius (i.e. the radius where the stellar density falls to the field density as judged visually), centered at the catalogued cluster coordinates \citep{bbds08}. By allowing this circle to shift and recentering it accordingly to the previous centre estimate, the process is iterated  until the new centre position repeats. $V-$ and $I-$band images were analysed independently to control the method's stability using the most populous clusters, i.e., their centres (and the structural parameters determined as follows) should converge for the same values within uncertainties. Given the predominantly older nature of the field population, clusters' radial profiles are contaminated by field stars more strongly in the $I-$band images than in the $V-$band images. As most of our sample contains poorly populated clusters, more affected by field contamination, $V-$band images were preferentially analysed. For three clusters, we employed $I-$band images given the poor quality of their $V-$band images.

\citet[][hereafter HZ06]{hz06} investigated structural properties of 204 SMC star clusters using $V-$band data from MCPS.  They found  that the King model provides slightly better fits to SMC clusters than the EFF model. We followed the general strategy undertaken by HZ06, with a few differences: (i) our  centre estimate is based on the average  of stellar positions (accurate to $\sim 0.1$\,arcsec; see paper~I) within a circular area, and then determined iteractively (see above); (ii) we fit King  and EFF models (weighted by the datapoint uncertainties) to both, surface brightness profiles (SBPs) and radial number density profiles (RDPs) using the Levenberg-Marquardt least-squares minimization algorithm; (iii) the background/foreground was taken into account by averaging the flux (counts) in circular areas in the frame borders for the SBP (RDP). Whenever the fit did not converge or return reasonable parameter values, we did not report them. Also, for $\sim 10$ per cent of the clusters we obtained tidal radii ($r_t$\footnote{The tidal radius is defined in this work as the truncation radius parameter of the King model according to equation \ref{eq:k62}}) values beyond the FoV limits ($r_t>100$\,arcsec), leading to uncertainties that are probably underestimated in these cases.

The methods employed to fit the King model to RDPs and SBPs are fully described and justified in paper~I. Here, we briefly summarize the procedures (also applied to fit the EFF model) and argue further for a combined use of both profiles as an efficient and accurate way to extract structural parameters of star clusters.  

The central parts of the RDP require an evaluation of the cluster completeness fraction. For SBPs, however, the brighter stars dominate in the central profile, so that photometric incompleteness of faint stars is not an issue. Instead, we followed an approach that combines the information obtained from RDPs and SBPs, recognizing that RDPs are superior measurements of the clusters' outer structures and the SBPs are better gauges of their inner stellar distributions.  

Another issue affecting the determination of structural parameters via RDP (particularly $r_t$), is the photometric depth of the data set. Limiting data analyses to a given magnitude (or mass), even if corrected for completeness,  may introduce a bias in the tidal radius estimate, since outer regions are expected to be dominated by low-mass stars. Nevertheless, clusters with a negligible degree of mass segregation, having a spatially uniform mass function, are less affected by the photometric depth. The RDP is also influenced by the magnitude limit since field stars not detected do not contribute to the background level.

\cite{bb08} studied how the photometric depth of simulated data affects the cluster radial profiles and the derived radii (core, half-number or half-mass, and tidal radii) as modeled by the 3-parameter King function. They found that the RDP and the stellar mass density profiles are more sensitive to the photometric depth than the SBP, if the stellar mass function is spatially variable. On the other hand, for a uniform mass function throughout the cluster, there is no such difference. The SBP does not change with photometric depth, except for in very young clusters, because of the presence of a few bright stars, causing the cluster radii to be underestimated. For profiles derived from photometry deeper than the turnoff, regardless of the cluster age, RDPs lead to radii (especially the core radius) systematically larger than SBPs. In general, SBPs produce more uniform structural parameters, since they are almost insensitive to photometric depth. However, at large radii the high level of stellar background/foreground contamination may difficult an accurate analysis.

\subsection{Model fittings to radial density profiles}
\label{sec:rdp}

 The RDPs were built from the stellar densities computed in annular bins of various sizes. The sample was limited to stars brighter than the magnitude corresponding to the peak value of the cluster plus field luminosity function ($V_{lim}$). 

The background/foreground stellar density ($\sigma_{bg}$) was evaluated from the frame corners, and supposed to be nearly constant throughout the cluster region, since the SAMI FoV is relatively small, covering $\sim 44$\,pc on a side for the LMC and $\sim 52$\,pc for the SMC.  $\sigma_{bg}$ was then kept fixed in the $\chi^2$ minimization procedure employed to fit the King and EFF models to the RDP. Since many studied MC clusters are near each other in projection, we could check for background consistency from their similar surrounding fields.

\subsubsection{King}

The structural parameters, namely, central surface stellar density ($\sigma_\circ$), core radius ($r_c$) and tidal radius ($r_t$) were estimated by fitting the King model to the clusters' RDPs according to the expression:

\begin{equation}
\label{eq:k62}
\sigma(r)=\sigma_\circ\left[\frac{1}{\sqrt{1+(r/r_c)^2}}-\frac{1}{\sqrt{1+(r_t/r_c)^2}} \right]^2 +\sigma_{bg}
\end{equation}

\subsubsection{EFF}

 We also obtained the structural parameters according to the empirical function by EFF, i.e.:

\begin{equation}
\sigma(r)=\sigma_\circ\left(1+\frac{r^2}{a^2}\right)^{-\gamma/2} +\sigma_{bg}
\end{equation}

\noindent where $a$ is the scale length related to the core radius  and $\gamma$ probes the outermost cluster structure. Since $r_c$ represents the radius at which the stellar density falls by half of its central value, $r_c=a\sqrt{2^{2/\gamma}-1}$.

\subsubsection{Results}
 
\begin{figure*}
\includegraphics[width=0.33\linewidth]{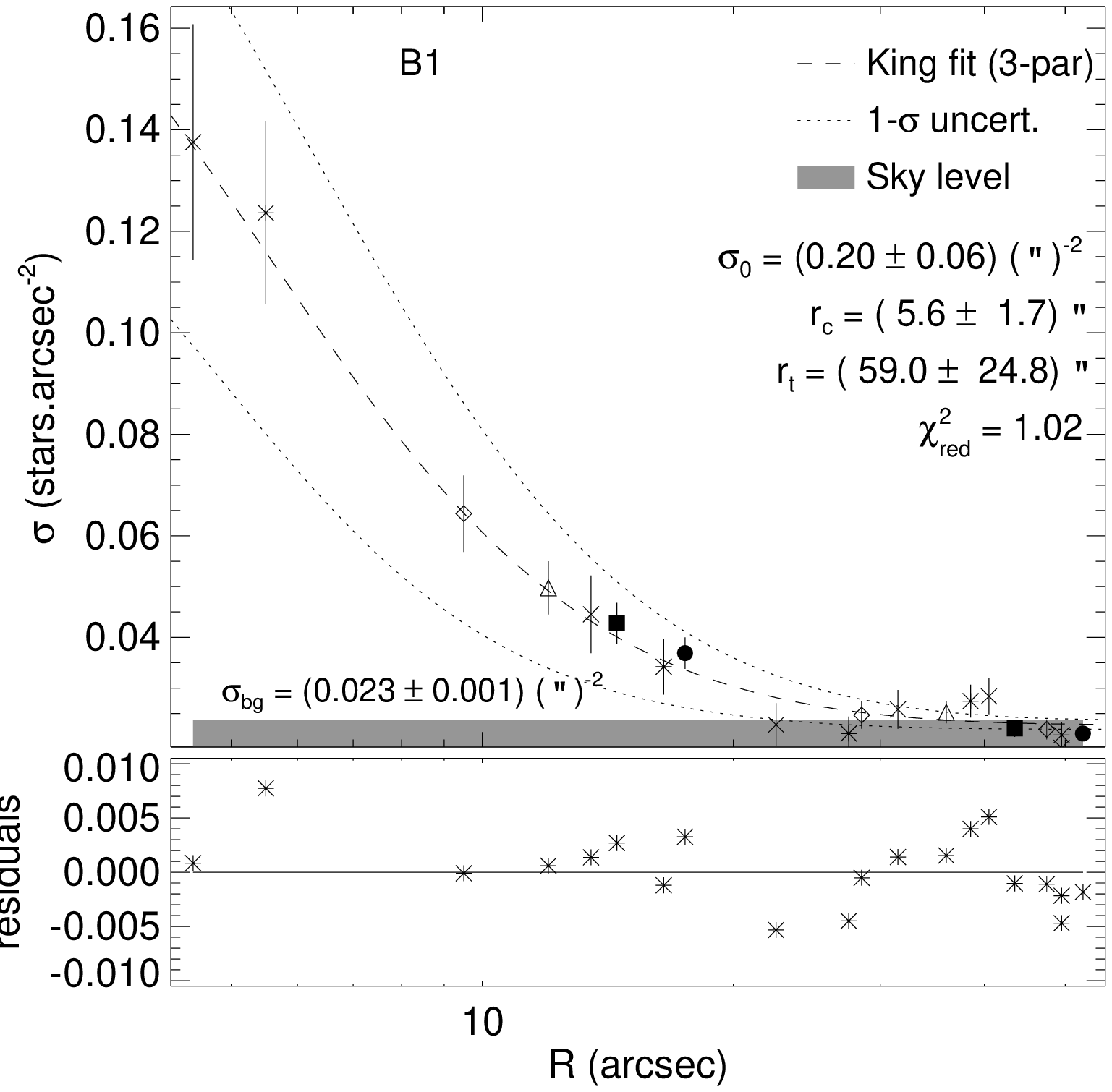}\includegraphics[width=0.33\linewidth]{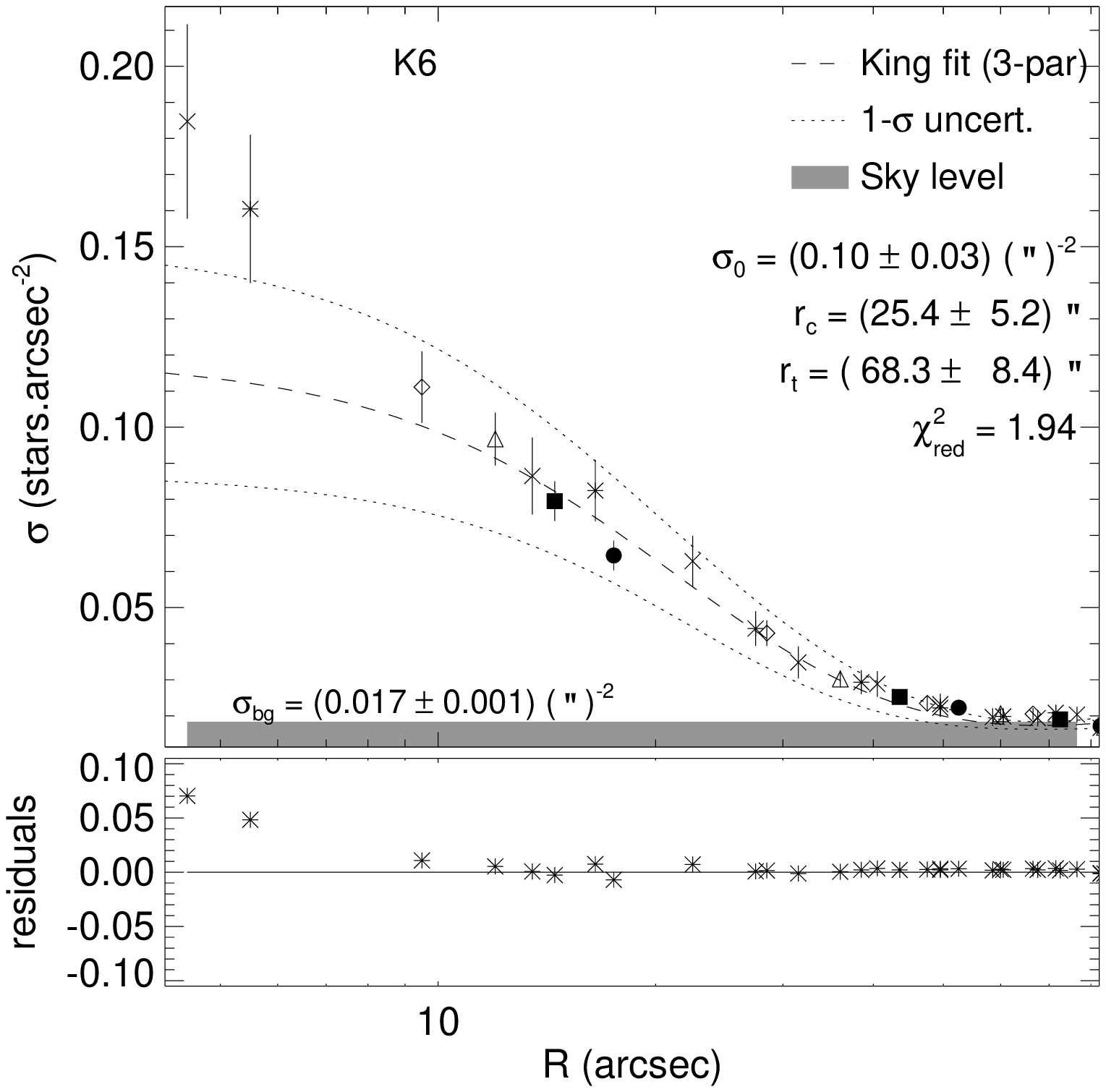}\includegraphics[width=0.33\linewidth]{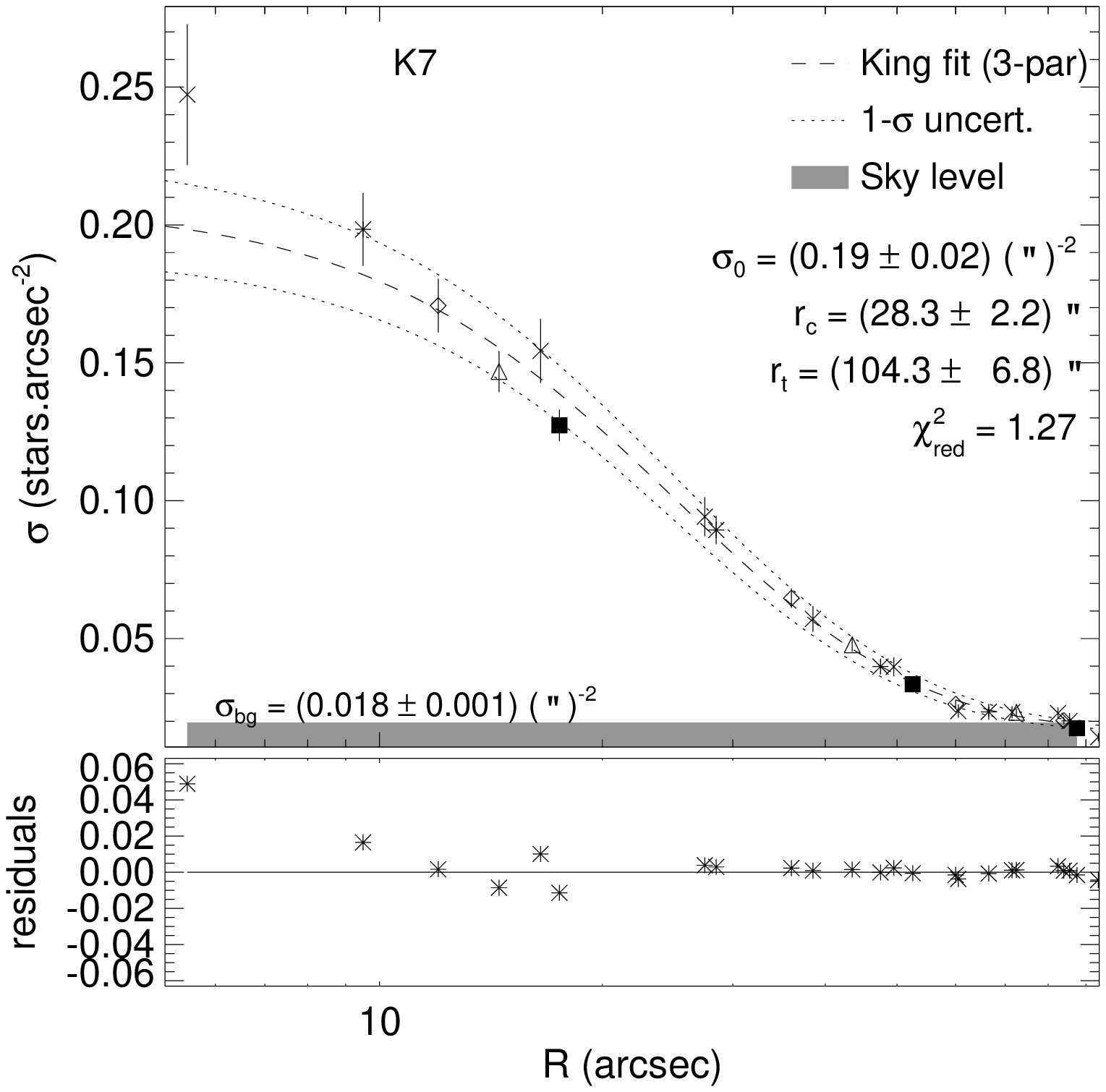}

\includegraphics[width=0.33\linewidth]{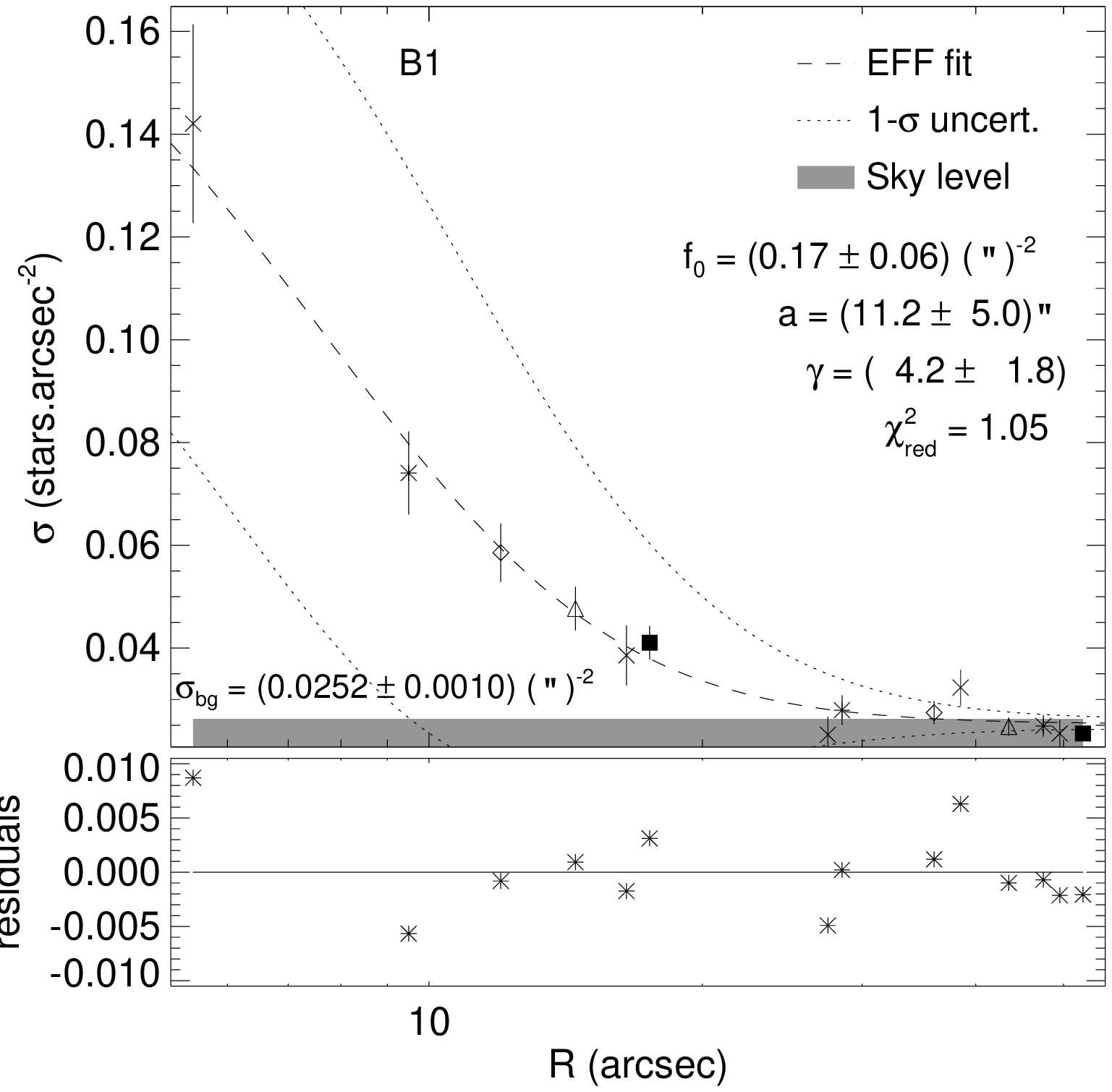}\includegraphics[width=0.33\linewidth]{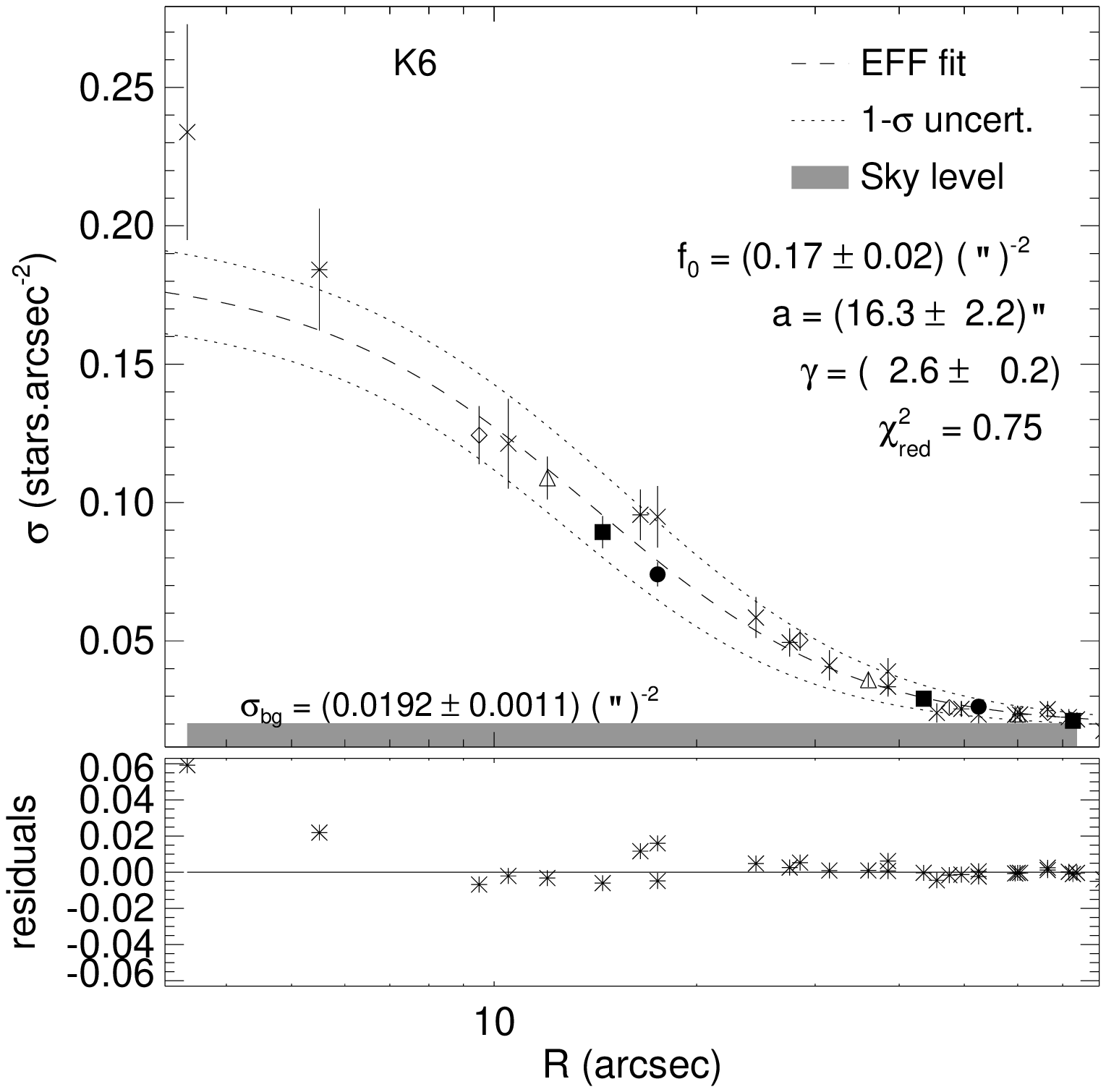}\includegraphics[width=0.33\linewidth]{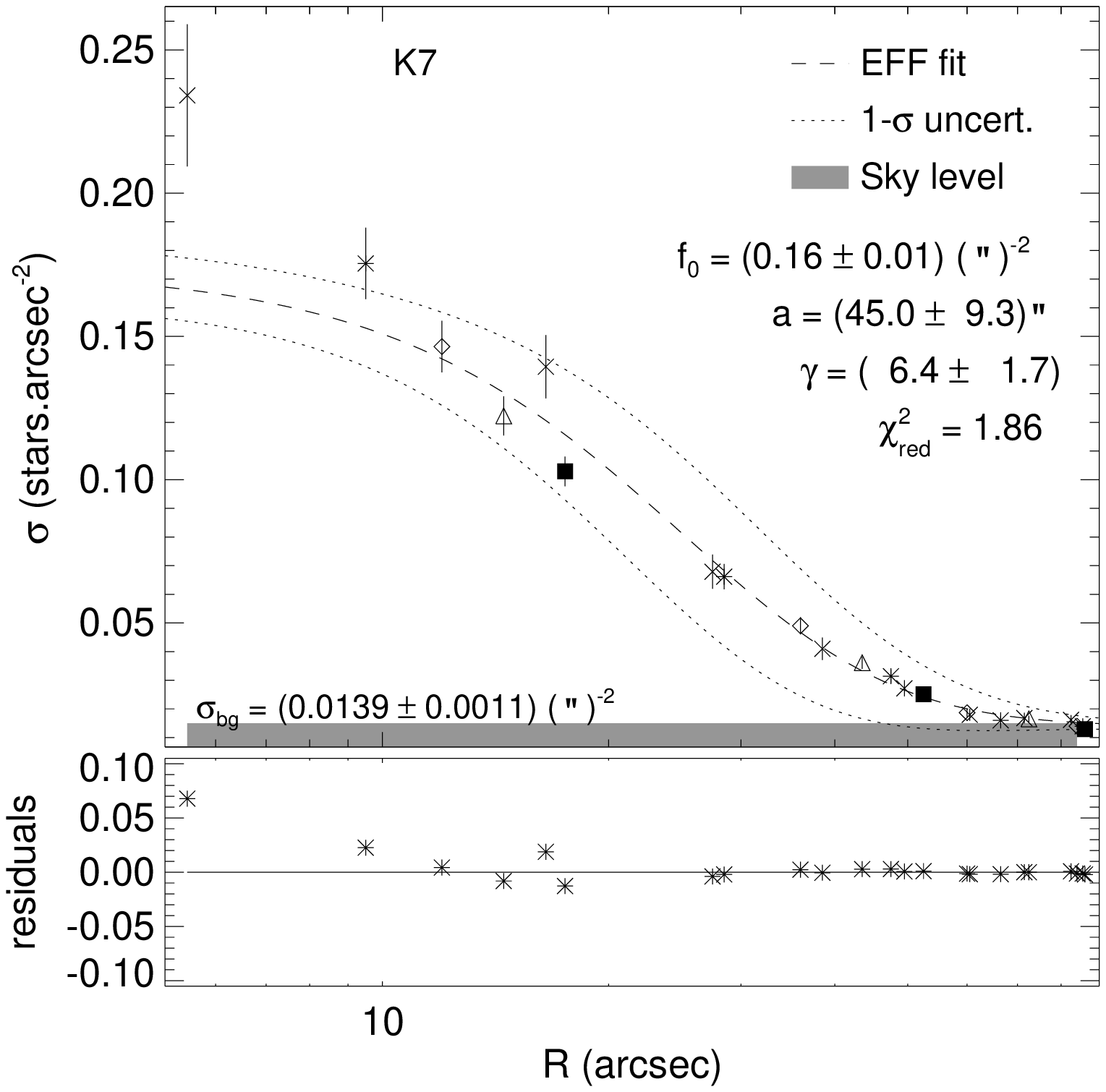}

\includegraphics[width=0.33\linewidth]{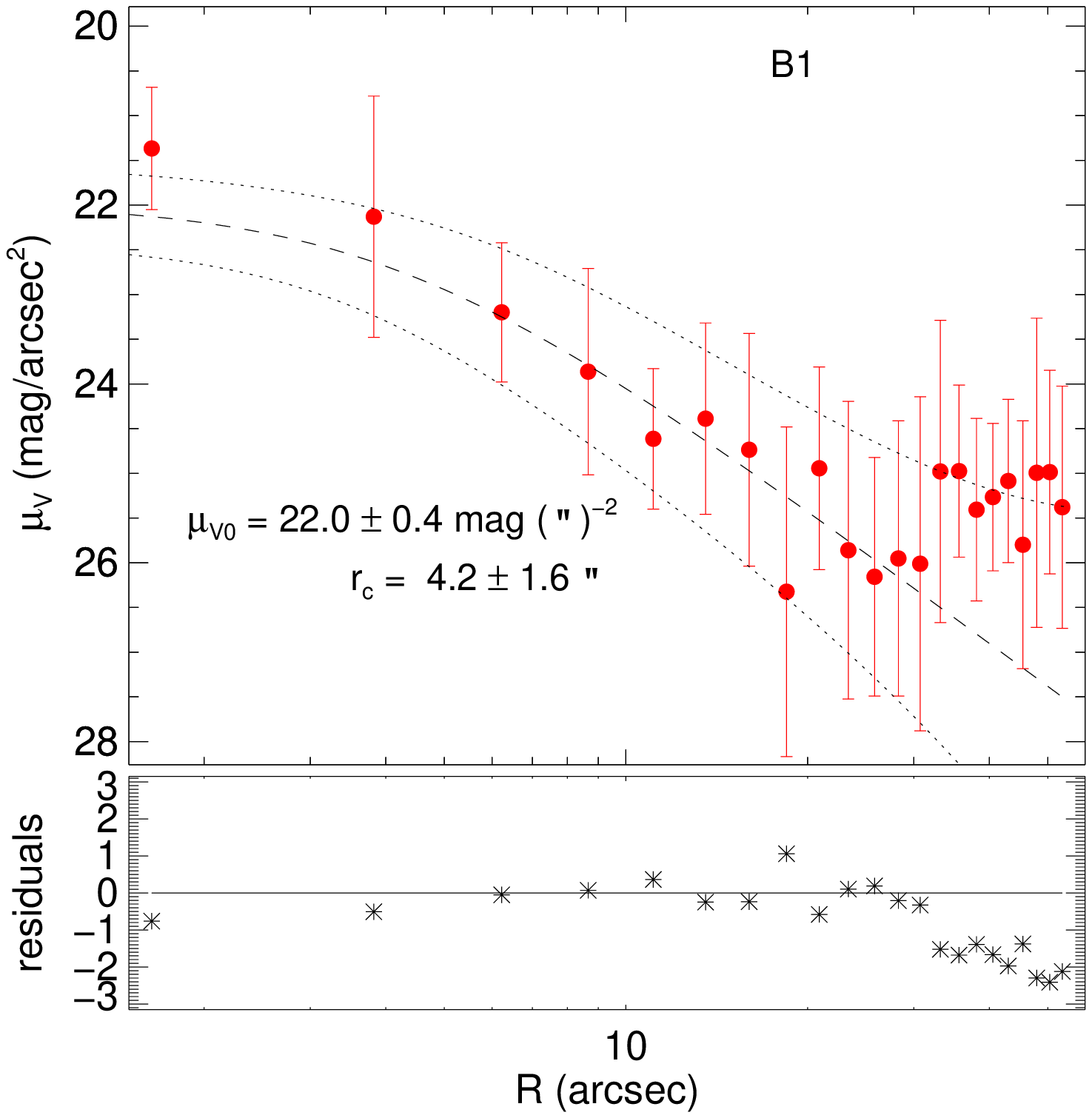}\includegraphics[width=0.33\linewidth]{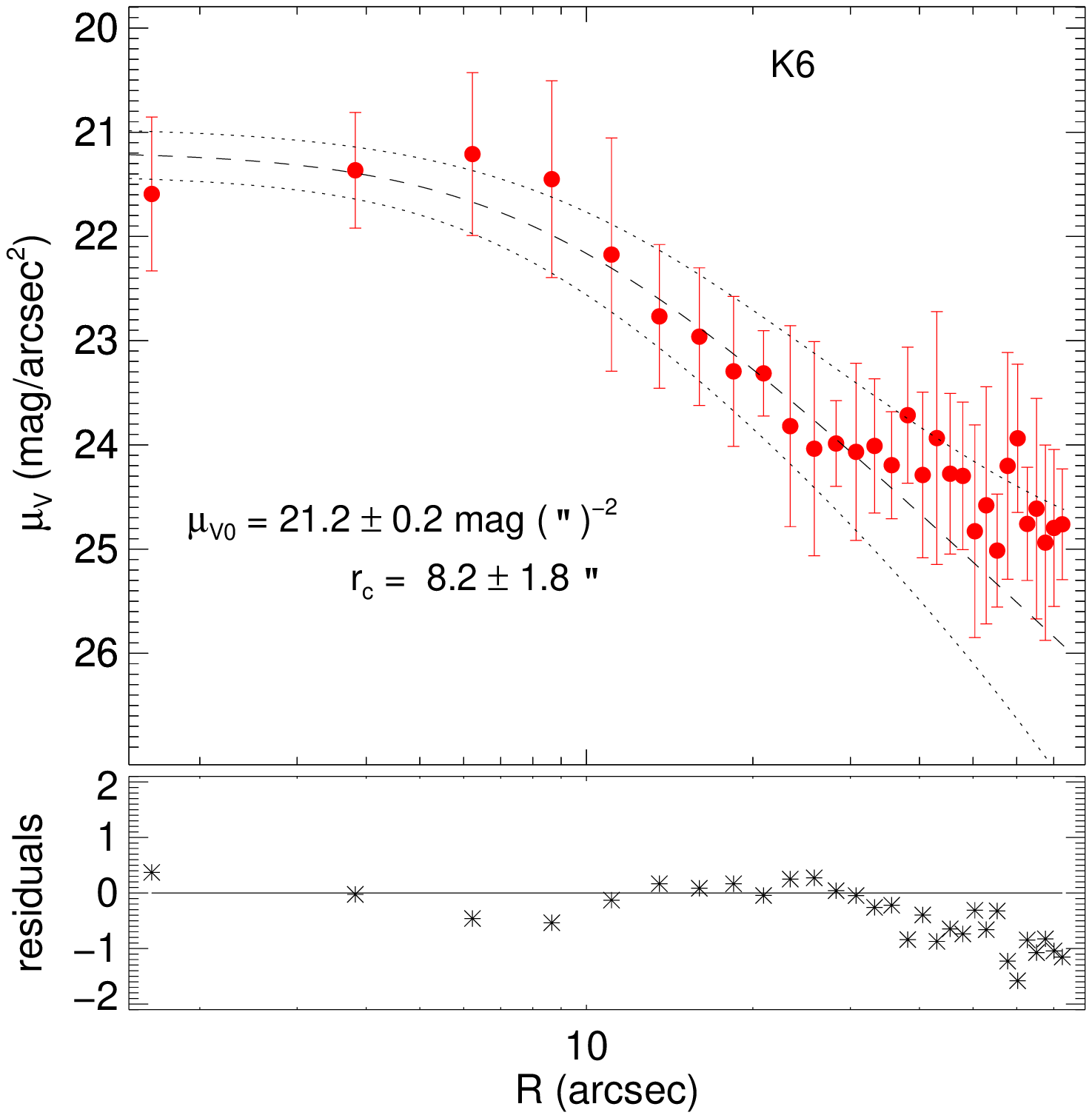}\includegraphics[width=0.33\linewidth]{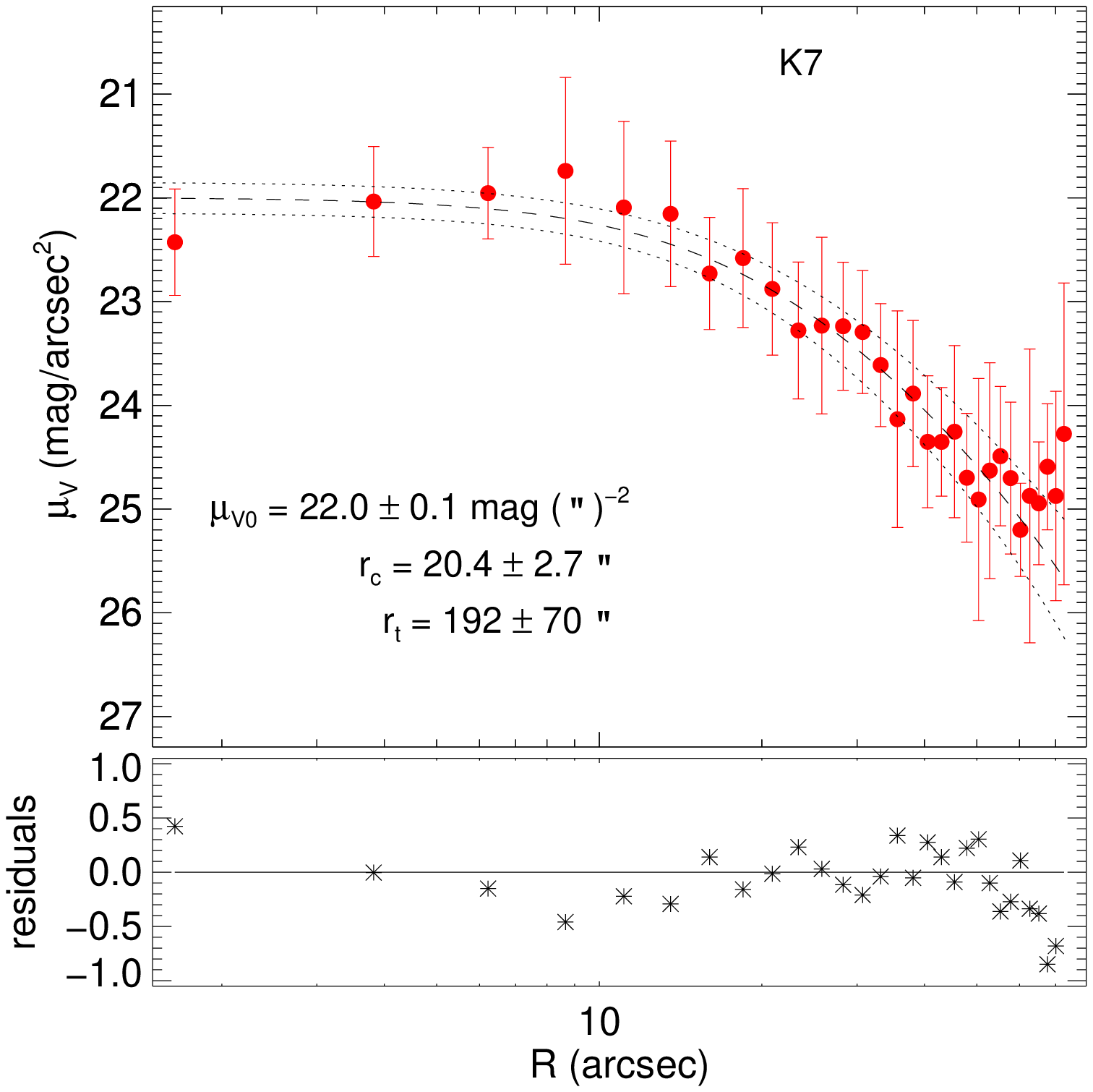}

\includegraphics[width=0.33\linewidth]{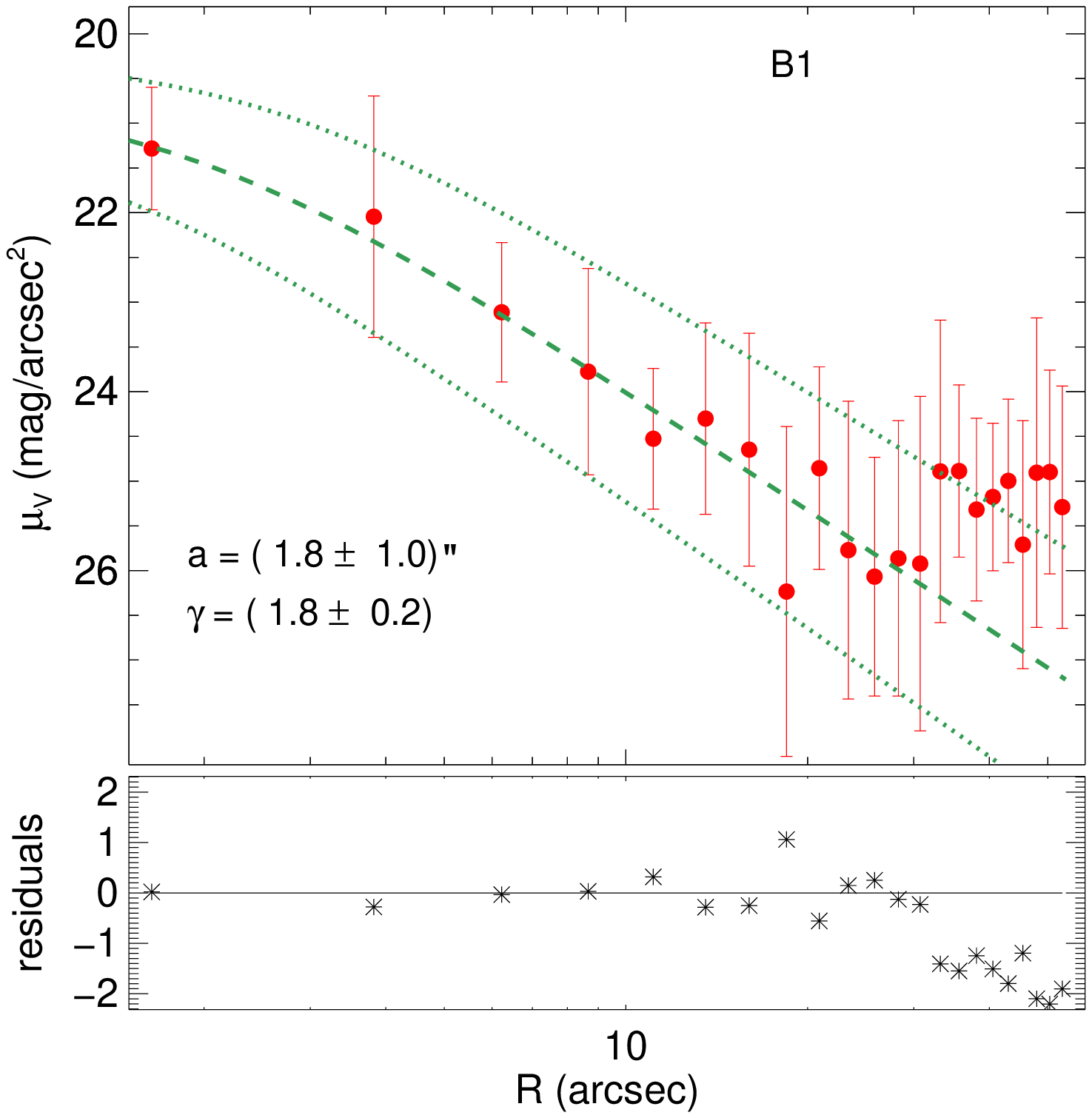}\includegraphics[width=0.33\linewidth]{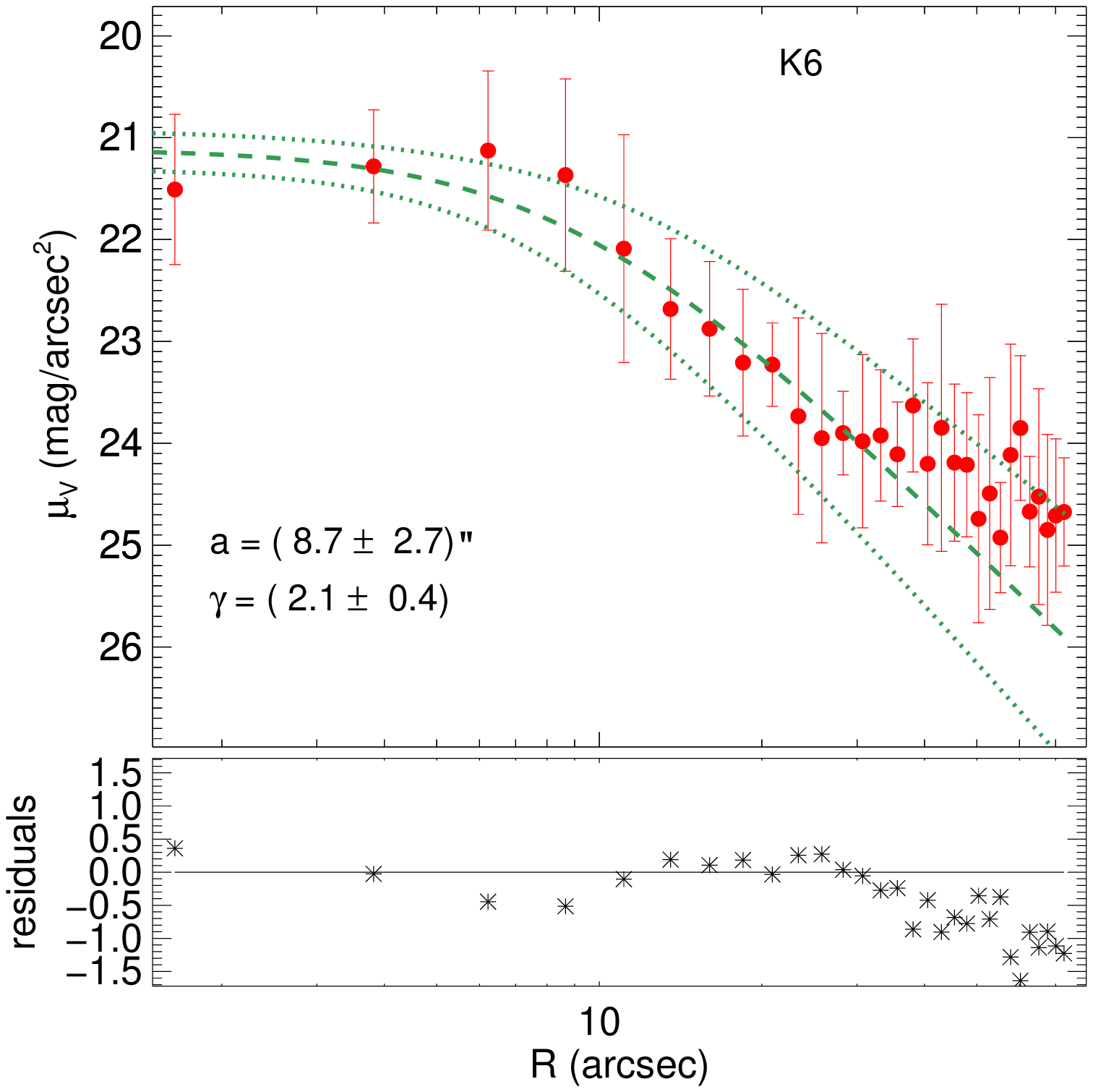}\includegraphics[width=0.33\linewidth]{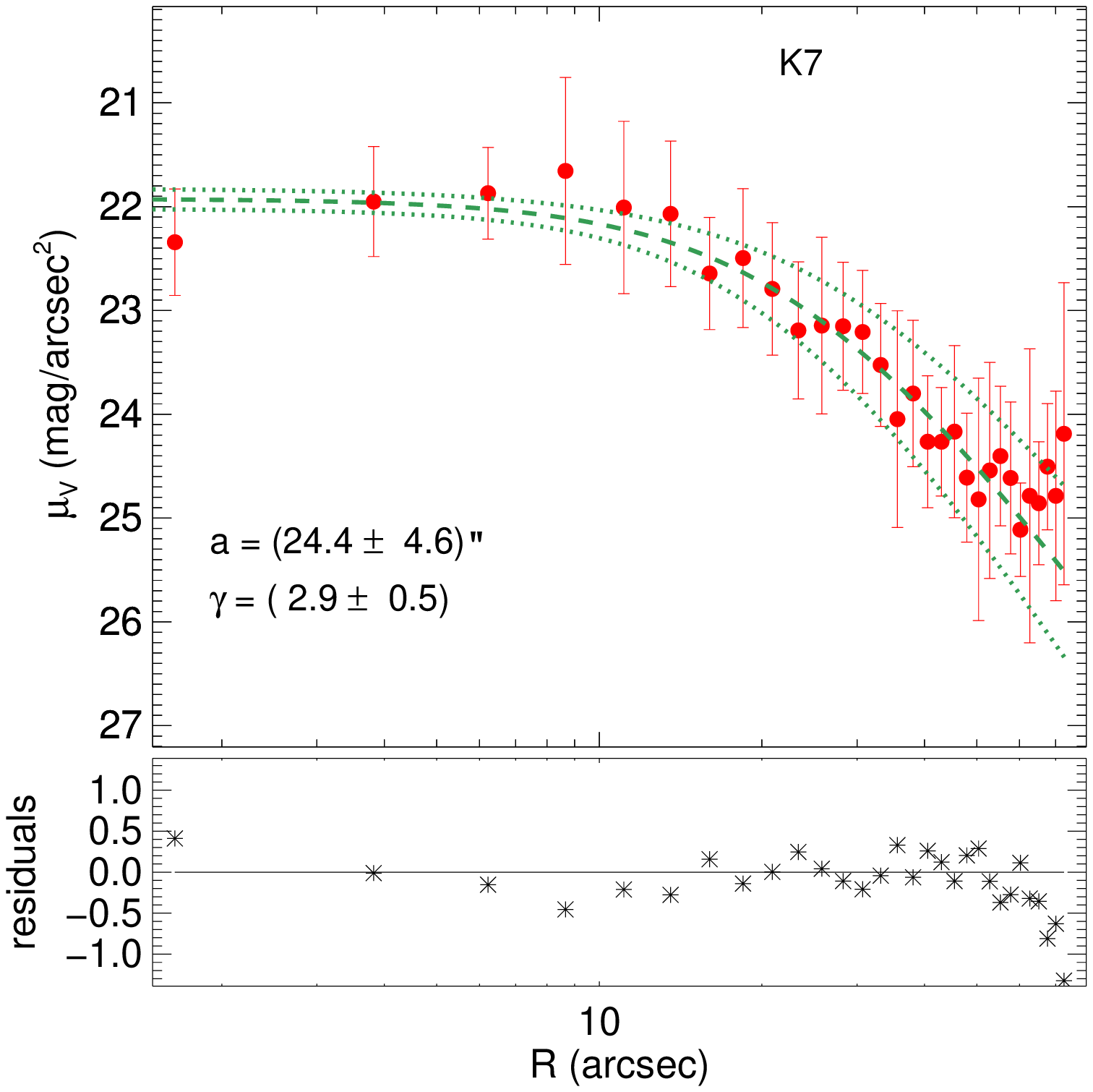}

\caption{ Radial density profiles for 3 SMC clusters and their respective King (first row) and EFF (second row) model fits (dashed line) with envelopes of 1\,$\sigma$ uncertainty (dotted lines). Different symbols correspond to the various widths of the annular bins employed.   Surface brightness profiles for the same 3 SMC clusters and their respective King (third row) and EFF (fourth row) model fits (dashed line) and uncertainty (dotted lines). The best-fitting  parameters are indicated and the fit residuals are plotted in the lower panels. The full set for the whole sample is published online.}
\label{fig:rdp_sbp}
\end{figure*}

Fig.~\ref{fig:rdp_sbp} presents King and EFF model fittings to the RDPs of three clusters. The full set of plots  are available on the online version of the paper. Tables~\ref{tab:sample_SMC_RDP} and \ref{tab:sample_LMC_RDP} provide the best-fitting parameters extracted from the RDPs. Core radii  and scale length are provided for completeness purposes, as the subsequent analysis will be based on their values derived from the SBP (Sect.~\ref{sec:sbp}). This is justified because the SBP and the RDP are measurements of cluster structure that complement each other. For the cluster central regions, while SBPs are based on integrated flux without the need to resolve stars, RDPs rely on counting the number of stars, which is affected by crowding. For the cluster  outskirts, stochasticity and heterogeneity of field stars  make the fluctuations on the SBP much higher than those of the RDP. As the outer regions are normally not crowded, RDPs provide an accurate tool there, even without completeness correction.

\subsection{Model fittings to surface brightness profiles}
\label{sec:sbp}

 The clusters' SBPs were built from the calibrated $V-$ and $I-$band images considering annular bins  divided in eight sectors, for which the median flux was calculated. The sky level, obtained from the whole image, was subtracted before the fitting procedure. Although the image quality is better in the $I-$band images, their relatively more numerous detected field stars makes the profiles noisier. For this reason we adopt in this work the parameters derived from the $V-$band images, except whenever  they were not useful, in which cases the $I-$band images were analysed (see Tables~\ref{tab:sample_SMC_SBP} and \ref{tab:sample_LMC_SBP}).

 The fits were performed from the cluster centre to the limiting radius, which is defined here as the radius where the cluster density profile merges with the background taking into account its fluctuation. From the limiting radius outward, the flux density provided the stellar background level, which was in turn subtracted from the cluster profile. We did not estimate the tidal radius for some clusters because the background level dominates their outer profiles.

\subsubsection{King}

 The structural parameters central surface brightness ($\mu_\circ$), $r_c$ and $r_t$ were estimated by fitting the following expression to the clusters' surface brightness\footnote{The equation was mistakenly written with a plus sign in paper~I}: 

\begin{equation}
\mu(r)=\mu'_\circ-5\log\left[\frac{1}{\sqrt{1+(r/r_c)^2}}-\frac{1}{\sqrt{1+(r_t/r_c)^2}} \right]  
\label{eq:king_sbp}
\end{equation}
\noindent
where
\begin{equation}
\mu'_\circ=\mu_\circ+5\log\left[1-\frac{1}{\sqrt{1+(r_t/r_c)^2}} \right]  .
\end{equation}
\noindent

\subsubsection{EFF}

 The clusters' SBP was also fitted by the EFF model using the following expression:

\begin{equation}
\mu(r)=\mu'_\circ +1.25\gamma\log\left( 1 + \frac{r^2}{a^2}\right) 
\label{eq:elson_sbp}
\end{equation}

\subsubsection{Results}

Fig.~\ref{fig:rdp_sbp} presents a sample of King and EFF model fits to the SBP of the same three clusters for which the RDP is shown. The full set of plots is available in the online version of the journal. Tables~\ref{tab:sample_SMC_SBP} and \ref{tab:sample_LMC_SBP} provide the best-fitting parameters extracted from the SBPs. Tidal radii and $\gamma$ derived from SBPs are listed for completeness purposes; we used in the subsequent analysis those derived from the RDPs.

\subsection{Integrated absolute {\it V} magnitude}
\label{sec:mv}
  
The integrated apparent magnitudes ($V_{int}$), given in Tables~\ref{tab:sample_SMC_SBP} and \ref{tab:sample_LMC_SBP}, were determined from the clusters' SBP by integrating the flux from the centre out to the limiting radius after subtracting the stellar foreground/background flux.  The $V_{int}$ uncertainties were obtained by propagating the errors from the measured fluxes in the rings. We then converted $V_{int}$ to the absolute one ($M_V$) by using the average distance modulus for the galaxies, namely $(m-M)_\circ=18.5\pm0.1$ for the LMC \citep{vk14,gwb14} and $(m-M)_\circ=18.96\pm0.20$ for the SMC \citep{gb15}, and the individual extinction towards each cluster according to the COBE/DIRBE and IRAS dust maps \citep{sfd98} recalibrated by \cite{sf11} with SDSS spectra (see Table~\ref{tab:red_age}). The uncertainties  in $V_{int}$, extinction and $(m-M)_\circ$ were propagated to the final $M_V$ value. We verified the effect of adopting the overall $(m-M)_\circ$ on the final $M_V$ magnitudes by evaluating their differences against those calculated from individual distances, derived in paper~I for the 9 clusters in common with the present sample. We confirmed that all of them are within the $M_V$ uncertainties. Nevertheless, the final magnitudes will be fine tuned when we derive the individual distances and reddening for each cluster with VISCACHA data (Kerber et al., in preparation).

\begin{landscape}
\begin{table}
\tiny
\caption{SMC clusters' structural parameters from RDPs. The full table is available online.} 
\label{tab:sample_SMC_RDP}
\begin{tabular}{lccccccccccc}
\hline
         &              &               &\multicolumn{5}{c}{\underline{\hspace{2.8cm}King$^{\ \dagger}$\hspace{2.8cm}}}&\multicolumn{3}{c}{\underline{\hspace{1.2cm}EFF$^{\ \dagger}$\hspace{1.2cm}}}\\
 Cluster  &   $\alpha$(J2000) &   $\delta$(J2000) & $\sigma_\circ$ & $r_c$ & $r_t$ & $\sigma_{bg}$ & $\chi^2$ & $a$ & $\gamma$ & $\chi^2$ &$V_{lim}$ \\
        &    (h:m:s)    &   (\ $^\circ$\ :\ $^\prime$\ :\ $^{\prime\prime}$\ )& (arcsec$^{-2}$) & (arcsec) & (arcsec) &  (arcsec$^{-2}$)& &  (arcsec) & & &(mag)\\
\hline
B1         & 00:19:20 &  -74:06:24  &$0.24\pm 0.07$&$  \ 6\pm  2$&$\   59\pm    25$& $0.023 \pm 0.001$&$1.02$& $11\pm\ 5$ & $ 4.2\pm 1.8$ & $1.05$&  $22.75$\\
K6         & 00:25:27 &  -74:04:30  &$0.24\pm 0.05$&$   25\pm  5$&$\   68\pm\    8$& $0.017 \pm 0.001$&$1.94$& $16\pm\ 2$ & $ 2.6\pm 0.2$ & $0.75$&  $21.75$\\
K7         & 00:27:45 &  -72:46:53  &$0.35\pm 0.02$&$   28\pm  2$&$   104\pm \   7$& $0.018 \pm 0.001$&$1.27$& $45\pm\ 9$ & $ 6.4\pm 1.7$ & $1.86$&  $22.25$\\
K9         & 00:30:00 &  -73:22:40  &$0.14\pm 0.04$&$   23\pm  8$&$\   92\pm    35$& $0.062 \pm 0.002$&$2.86$& $21\pm\ 5$ & $ 3.0\pm 0.6$ & $0.73$&  $22.75$\\
HW5        & 00:31:01 &  -72:20:30  &$0.46\pm 0.04$&$  \ 8\pm  1$&$   192\pm    35$& $0.020 \pm 0.001$&$0.70$& $10\pm\ 1$ & $ 2.9\pm 0.2$ & $0.96$&  $22.75$\\
\hline
\end{tabular}

Notes: $^\ast$ SBP $I-$band filter measurements. $^{\ \dagger}$ $r_c$ and $a$ were adopted from the SBP fit; $r_t$ and $\gamma$ from the RDP fit (see discussion in Sects.~\ref{sec:rdp} and \ref{sec:sbp}). 

\end{table}

\begin{table}
\tiny
\caption{LMC clusters' structural parameters from RDPs. The full table is available online.} 
\label{tab:sample_LMC_RDP}
\begin{tabular}{lccccccccccc}
\hline
         &              &               &\multicolumn{5}{c}{\underline{\hspace{2.8cm}King$^{\ \dagger}$\hspace{2.8cm}}}&\multicolumn{3}{c}{\underline{\hspace{1.2cm}EFF$^{\ \dagger}$\hspace{1.2cm}}}\\
 Cluster  &   $\alpha$(J2000) &   $\delta$(J2000) & $\sigma_\circ$ & $r_c$ & $r_t$ & $\sigma_{bg}$ & $\chi^2$ &  $a$ & $\gamma$ & $\chi^2$ &$V_{lim}$ \\
        &    (h:m:s)    &   (\ $^\circ$\ :\ $^\prime$\ :\ $^{\prime\prime}$\ )& (arcsec$^{-2}$) & (arcsec) & (arcsec) &  (arcsec$^{-2}$) &  & (arcsec) & &&(mag)\\
\hline
LW15      & 04:38:26&  -74:27:48&$ 0.22\pm 0.03$&$   10\pm 2$&$    153\pm   39$&$ 0.015 \pm 0.001 $ &1.56& $21\pm 6$ & $ 3.7\pm 0.9$ & $1.94$& 23.25 \\
SL13      & 04:39:42&  -74:01:00&$ 0.14\pm 0.02$&$   22\pm 3$&$   \ 61\pm \  5$&$ 0.008 \pm 0.001 $ &0.83& $14\pm 3$ & $ 2.5\pm 0.3$ & $0.69$& 22.25 \\
SL28      & 04:44:40&  -74:15:36&$ 0.59\pm 0.05$&$   25\pm 2$&$    132\pm   12$&$ 0.031 \pm 0.002 $ &2.02& $37\pm 5$ & $ 4.3\pm 0.6$ & $2.35$& 23.25 \\
SL29      & 04:45:13&  -75:07:00&$ 0.09\pm 0.01$&$   17\pm 2$&$  \  64\pm \  6$&$ 0.005 \pm 0.001 $ &0.39& $20\pm 4$ & $ 4.6\pm 0.8$ & $0.45$& 21.75 \\
SL36      & 04:46:09&  -74:53:18&$ 0.68\pm 0.06$&$ \  9\pm 1$&$  \  79\pm \  9$&$ 0.025 \pm 0.001 $ &1.64& $15\pm 2$ & $ 3.9\pm 0.4$ & $1.18$& 23.75 \\
\hline
\end{tabular}

Notes: $^\ast$ SBP $I-$band filter measurements. $^{\ \dagger}$ $r_c$ and $a$ were adopted from the SBP fit; $r_t$ and $\gamma$ from the RDP fit (see discussion in Sects.~\ref{sec:rdp} and \ref{sec:sbp}).

\end{table}

\begin{table}
\tiny
\caption{SMC clusters' structural parameters from SBPs. The full table is available online.} 
\label{tab:sample_SMC_SBP}
\begin{tabular}{lccccccccccc}
\hline
         &              &               &\multicolumn{5}{c}{\underline{\hspace{3.2cm}King$^{\ \dagger}$\hspace{3.2cm}}}&\multicolumn{3}{c}{\underline{\hspace{1.1cm}EFF$^{\ \dagger}$\hspace{1.3cm}}}\\
 Cluster  &   $\alpha$(J2000) &   $\delta$(J2000) & $\mu_{{\rm v},\circ}$ & $r_c$ & $r_t$ & $\mu_{{\rm v},bg}$ & $\chi^2$ & $a$ & $\gamma$ & $\chi^2$ &$V_{int}$\\
        &    (h:m:s)    &   (\ $^\circ$\ :\ $^\prime$\ :\ $^{\prime\prime}$\ )&(mag.arcsec$^{-2}$) & (arcsec) & (arcsec)&(mag.arcsec$^{-2}$)&& (arcsec)&&&(mag)\\
\hline
B1         & 00:19:20 &  -74:06:24  &$   21.97\pm0.41$ &  $\ 4.2\pm1.6$ & $-$                          &  $26.64\pm0.06$ &    $0.29$   & $ 2\pm 1$ & $ 1.8\pm 0.2$ & $0.10$ &$15.51\pm     0.16$\\
K6         & 00:25:27 &  -74:04:30  &$   21.18\pm     0.18$&$ \    8.2\pm     1.8$&$         -       $&$  26.81\pm     0.08$&$    0.14$& $ 9\pm 3$ & $ 2.1\pm 0.4$ & $0.12$ &$13.90\pm     0.06$\\
K7         & 00:27:45 &  -72:46:53  &$   22.00\pm     0.11$&$    20.4\pm     2.7$&$   192\pm    70$&$  26.81\pm     0.08$&$    0.13$   & $24\pm 5$ & $ 2.9\pm 0.5$ & $0.12$ &$13.20\pm     0.06$\\
K9         & 00:30:00 &  -73:22:40  &$   22.86\pm     0.19$&$    23.3\pm     3.6$&$         -       $&$  26.70\pm     0.07$&$    0.19$ & $10\pm 7$ & $ 1.0\pm 0.4$ & $0.12$ &$13.09\pm     0.12$\\
HW5        & 00:31:01 &  -72:20:30  &$   19.95\pm     0.21$&$   \  2.3\pm     0.4$&$         -       $&$  26.52\pm     0.07$&$    0.90$& $ 2\pm 0$ & $ 1.7\pm 0.1$ & $0.16$ &$14.35\pm     0.07$\\
\hline
\end{tabular}

Notes: $^\ast$ SBP $I-$band filter measurements. $^{\ \dagger}$ $r_c$ was adopted from the SBP fit and $r_t$ from the RDP fit (see discussion in Sects.~\ref{sec:rdp} and \ref{sec:sbp}). 

\end{table}

\begin{table}
\tiny
\caption{LMC clusters' structural parameters from SBPs. The full table is available online.} 
\label{tab:sample_LMC_SBP}
\begin{tabular}{lccccccccccc}
\hline
         &              &               &\multicolumn{5}{c}{\underline{\hspace{3.2cm}King$^{\ \dagger}$\hspace{3.2cm}}}&\multicolumn{3}{c}{\underline{\hspace{1.1cm}EFF$^{\ \dagger}$\hspace{1.3cm}}}\\
 Cluster  &   $\alpha$(J2000) &   $\delta$(J2000) & $\mu_{{\rm v},\circ}$ & $r_c$ & $r_t$ & $\mu_{{\rm v},bg}$ & $\chi^2$ & $a$ & $\gamma$ & $\chi^2$ &$V_{int}$\\
        &    (h:m:s)    &   (\ $^\circ$\ :\ $^\prime$\ :\ $^{\prime\prime}$\ )&(mag.arcsec$^{-2}$) & (arcsec) & (arcsec)&(mag.arcsec$^{-2}$)&& (arcsec)&&&(mag)\\
\hline
LW15      & 04:38:26&  -74:27:48&$  22.78\pm 0.56$&$   17.6\pm     7.8$&$  \  44\pm    11$&$   26.91\pm     0.06$&     0.09 &               $ 9\pm 6$ & $ 2.3\pm 1.0$ & $0.15$&$   14.79\pm     0.16$ \\      
SL13      & 04:39:42&  -74:01:00&$  23.50\pm 0.30$&$   16.0\pm     5.4$&$     -$&$   26.88\pm     0.07$&     0.14 &                         $19\pm10$ & $ 2.6\pm 1.3$ & $0.10$&$   14.84\pm     0.08$ \\      
SL28      & 04:44:40&  -74:15:36&$    21.35\pm     0.59$&$   21.0\pm     5.8$&$  \  68\pm    32$&$   26.92\pm     0.08$&     0.07 &         $28\pm15$ & $ 4.9\pm 3.9$ & $0.06$&$   13.62\pm     0.05$ \\      
SL29      & 04:45:13&  -75:07:00&$\ 23\pm\ 2$&$\ 28\pm 22$&$ \ 53\pm 33$&$ 26.80\pm 0.10$&    $0.13$ &                                      $-$ & $ -$ & $ -$                 &$   15.28\pm     0.18$ \\  
SL36      & 04:46:09&  -74:53:18&$    20.18\pm 0.10$&$   \   3.8\pm     0.4$&$  \  51\pm  \   7$&$   26.94\pm     0.07$&     0.05 &         $ 5\pm 1$ & $ 3.0\pm 0.2$ & $0.05$&$   14.80\pm     0.55$ \\      
\hline
\end{tabular}

Notes: $^\ast$ SBP $I-$band filter measurements. $^{\ \dagger}$ $r_c$ and $a$ were adopted from the SBP fit; $r_t$ and $\gamma$ from the RDP fit (see discussion in Sects.~\ref{sec:rdp} and \ref{sec:sbp}). 

\end{table}
\end{landscape}

\begin{table*}
\tiny
\caption{MC clusters astrophysical parameters} 
\label{tab:red_age}
\begin{tabular}{lcccccccc}
\hline
 Cluster  &  $d$\,(kpc)$^{\ \rm{a}}$ & $E$($B-V$)$^{\ \rm{b}}$  & $M_V$\,$^{\ \rm{c}}$ & $\log{M/M_\odot}^{\ \rm{d}}$ & $r_h$\,(pc)$^{\ \rm{e}}$ & $r_J$\,(pc)$^{\ \rm{f}}$  & $\log[t({\rm yr})]^{\ \rm{g}}$ & ref. \\
\hline
B1         &  $ 2.91\pm 0.27$ &  $ 0.033\pm 0.001$ & $ -3.55\pm    0.26 $    & $ -$ &  $ 3.3\pm 1.6 $& $ -$     & $-$& \\ 
K6         &  $2.49\pm0.23$ &  $ 0.040\pm 0.001$ & $ -5.19\pm    0.21  $   & $3.73\pm0.15$ &  $ 5.0\pm 1.3 $& $19.2\pm 2.8$     & $9.20\pm  0.12$& 1\\ 
K7         &  $2.00\pm0.18$ &  $ 0.028\pm 0.001$ & $ -5.85\pm    0.21 $    & $4.20\pm0.16$ &  $ 9.5\pm 1.5 $& $23.8\pm 3.5$     & $9.54\pm  0.14$& 2\\ 
K9         &  $1.88\pm0.17$ &  $ 0.033\pm 0.004$ & $ -5.97\pm  0.24 $      & $3.70\pm0.20$ &  $ 9.5\pm 2.6 $& $15.6\pm 2.8$     & $8.64\pm  0.25$& 3,4,5\\ 
HW5        &  $1.83\pm0.17$ &  $ 0.026\pm 0.001$ & $ -4.69\pm    0.21 $    & $3.79\pm0.15$ &  $ 4.4\pm 0.9 $& $16.4\pm 2.3$     & $9.63\pm  0.10$& 6\\ 
HW20       &  $1.76\pm0.16$ &  $ 0.053\pm 0.002$ & $ -4.96\pm    0.21 $    & $3.54\pm0.13$ &  $ 5.1\pm 1.3 $& $13.2\pm 1.8$     & $9.04\pm  0.04$& 7\\ 
L32        &  $4.25\pm0.39$ &  $ 0.020\pm 0.001$ & $ -4.08\pm    0.21  $   & $3.57\pm0.13$ &  $ 6.2\pm 1.5 $& $24.2\pm 3.2$     & $9.66\pm  0.02$& 8 \\ 
HW33       &  $2.22\pm0.20$ &  $ 0.025\pm 0.001$ & $  -3.82\pm    0.26 $   & $2.51\pm0.15$ &  $ 4.2\pm 1.5 $& $\ 7.0\pm 1.0$     & $8.10\pm  0.10$& 9\\ 
K37        &  $1.66\pm0.15$ &  $ 0.039\pm 0.001$ & $  -5.35\pm    0.21$    & $3.83\pm0.14$ &  $ 7.0\pm 2.0 $& $15.9\pm 2.2$     & $9.26\pm  0.06$& 7,8\\ 
B94        &  $ 1.97\pm 0.18$ &  $ 0.043\pm 0.001$ & $ -3.74\pm    0.24  $   & $ -$ &  $ 4.9\pm 2.3 $& $ -$     & $-$& \\ 
HW38       &  $1.19\pm0.11$ &  $ 0.054\pm 0.003$ & $  -6.28\pm    0.21 $   & $3.92\pm0.22$ &  $ 9.3\pm 2.6 $& $13.6\pm 2.6$     & $8.80\pm  0.30$& 3\\ 
HW44       &  $ 1.23\pm 0.11$ &  $ 0.047\pm 0.002$ & $ -4.55\pm    0.48  $   & $ -$ &  $ - $& $ -$     & $-$& \\ 
L73        &  $ 2.86\pm 0.26$ &  $ 0.023\pm 0.001$ & $ -4.50\pm    0.23  $   & $ -$ &  $ 6.6\pm 2.3 $& $ -$     & $-$& \\ 
K55        &  $1.21\pm0.11$ &  $ 0.096\pm 0.008$ & $ -6.52\pm    0.20 $    & $3.85\pm0.16$ &  $ 6.2\pm 2.4 $& $13.0\pm 1.9$    & $8.52\pm  0.15$& 3,10,11\\ 
HW56       &  $ 2.40\pm 0.22$ &  $ 0.026\pm 0.001$ & $ -3.09\pm    0.30 $    & $ -$ &  $ 3.6\pm 0.9 $& $ -$     & $-$& \\ 
K57        &  $1.30\pm0.12$ &  $ 0.091\pm 0.022$ & $ -5.99\pm    0.22  $   & $3.72\pm0.14$ &  $ 6.7\pm 2.6 $& $12.3\pm 1.7$     & $8.65\pm  0.05$& 3,11\\ 
NGC422     &  $1.79\pm0.16$ &  $ 0.067\pm 0.012$ & $ -5.81\pm  0.22  $     & $3.31\pm0.15$ &  $ 3.4\pm 1.2 $& $11.1\pm 1.6$     & $8.10\pm  0.12$& 3,12\\ 
IC1641     &  $1.80\pm0.17$ &  $ 0.064\pm 0.010$ & $ -4.68\pm  0.28 $      & $3.22\pm0.15$ &  $ 4.0\pm 1.9 $& $10.5\pm 1.5$     & $8.70\pm  0.03$& 3,12\\ 
HW67       &  $2.64\pm0.24$ &  $ 0.030\pm 0.001$ & $ -4.35\pm    0.40 $    & $3.49\pm0.20$ &  $ 5.4\pm 2.1 $& $16.6\pm 2.9$    & $9.36\pm  0.09$& 8,13,14\\ 
HW71NW     &  $ 1.90\pm 0.17$ &  $ 0.051\pm 0.004$ & $ -5.21\pm    0.62 $    & $ -$ &  $ 5.0\pm 2.2 $& $ -$     & $-$& \\ 
L100       &  $2.26\pm0.21$ &  $ 0.039\pm 0.001$ & $ -5.12\pm    0.22 $    & $3.78\pm0.15$ &  $ 4.9\pm 1.2 $& $18.7\pm 2.7$     & $9.32\pm  0.10$& 8,14\\ 
HW77       &  $2.21\pm0.20$ &  $ 0.035\pm 0.002$ & $ -4.67\pm    0.21 $    & $3.49\pm0.14$ &  $ 8.2\pm 1.5 $& $14.8\pm 2.1$     & $9.15\pm  0.10$& 14\\ 
IC1708     &  $3.22\pm0.30$ &  $ 0.038\pm 0.002$ & $ -4.93\pm    0.21 $    & $3.57\pm0.14$ &  $ 5.1\pm 0.7 $& $20.1\pm 2.8$     & $9.10\pm  0.10$& 14\\ 
B168       &  $3.61\pm0.33$ &  $ 0.026\pm 0.001$ & $ -4.11\pm    0.31 $    & $3.54\pm0.17$ &  $ 4.9\pm 1.6 $& $21.3\pm 3.4$     & $9.60\pm  0.10$& 14\\ 
L106       &  $4.43\pm0.41$ &  $ 0.045\pm 0.002$ & $ -5.29\pm    0.22 $    & $3.80\pm0.16$ &  $ 6.9\pm 1.1 $& $29.7\pm 4.5$     & $9.25\pm  0.15$& 8,14\\ 
BS95-187   &  $3.05\pm0.28$ &  $ 0.035\pm 0.001$ & $  -2.84\pm    0.37 $   & $2.85\pm0.19$ &  $ 5.1\pm 3.3 $& $11.2\pm 1.9$     & $9.30\pm  0.10$& 14\\ 
L112       &  $4.26\pm0.39$ &  $ 0.061\pm 0.001$ &  $-$                    & $-$ &  $ 3.7\pm 1.0 $& $-$     & $9.73\pm  0.10$& 8,14\\ 
HW85       &  $4.45\pm0.41$ &  $ 0.028\pm 0.001$ & $ -4.37\pm    0.29 $    & $3.47\pm0.17$ &  $ 3.4\pm 1.1 $& $23.1\pm 3.6$     & $9.31\pm  0.10$& 8,13\\ 
L114       &  $4.68\pm0.43$ &  $ 0.038\pm 0.001$ & $   -7.14\pm    0.24 $  & $3.87\pm0.16$ &  $ 3.2\pm 1.6 $& $32.6\pm5.0$     & $8.15\pm  0.15$& 14\\ 
L116       &  $6.73\pm0.62$ &  $ 0.045\pm 0.002$ & $ -  $                  & $-$ &  $ 5.3\pm 2.3 $& $-$     & $9.43\pm  0.08$& 8\\ 
NGC796     &  $5.11\pm0.47$ &  $ 0.040\pm 0.001$ & $ -6.87\pm    0.22 $    & $3.34\pm0.16$ &  $ 3.6\pm 0.7 $& $23.0\pm 3.4$     & $7.45\pm  0.15$& 7,15\\ 
AM3        &  $5.05\pm0.47$ &  $ 0.029\pm 0.001$ & $ -2.30\pm    0.68 $    & $2.90\pm0.30$ &  $ 3.7\pm 1.5 $& $16.3\pm 4.0$     & $9.74\pm  0.06$& 7,8,16\\
\hline
LW15       & $5.99\pm 0.30$ &  $ 0.075\pm 0.002$ & $ -3.94\pm    0.19 $ &  $ -$ &$ 8.9\pm 4.6$ & $ -$ & - & \\
SL13       & $5.60\pm0.26$ &  $ 0.077\pm 0.002$ & $ -3.48\pm    0.16 $ &  $3.24\pm0.14$ &$ 5.0\pm 4.3$ & $16.1\pm 2.8$ &$9.40\pm  0.07$& 19\\ 
SL28       & $5.54\pm0.25$ &  $ 0.086\pm 0.002$ & $ -5.15\pm    0.11 $ &  $3.76\pm0.14$ &$ 8.3\pm 2.7$ & $23.9\pm 4.2$ &$9.18\pm  0.10$& 17\\ 
SL29       & $6.20\pm0.28$ &  $ 0.100\pm 0.004$ & $ -3.53\pm    0.21 $ &  $3.19\pm0.16$ &$ 6.8\pm 5.9$ & $16.6\pm 3.1$ &$9.30\pm  0.10$& 20\\ 
SL36       & $5.98\pm0.27$ &  $ 0.098\pm 0.004$ & $ -4.00\pm    0.56 $ &  $3.38\pm0.26$ &$ 3.0\pm 0.4$ & $18.7\pm 4.6$ &$9.30\pm  0.10$& 20\\ 
LW62       & $5.39\pm 0.27$ &  $ 0.080\pm 0.002$ & $ -2.78\pm    0.19 $ &  $ -$ &$ 2.8\pm 1.9$ & $ -$ &$-$& \\   
SL53       & $ 6.46\pm 0.32$ &  $ 0.096\pm 0.005$ & $ -3.92\pm    0.13 $ &  $ -$ &$ 5.3\pm 1.9$ & $ -$ &$-$& \\   
SL61       & $6.35\pm0.29$ &  $ 0.108\pm 0.007$ & $ -5.90\pm    0.12 $ &  $4.11\pm0.14$ &$ 9.2\pm 2.1$ & $34.2\pm 6.0$ &$9.26\pm  0.10$& 7,17\\ 
SL74       & $ 5.71\pm 0.29$ &  $ 0.096\pm 0.004$ & $ -5.62\pm    0.19 $ &  $ -$ &$ 5.1\pm 1.2$ & $ -$ &$-$& \\   
SL80       & $ 5.74\pm 0.29$ &  $ 0.098\pm 0.005$ & $ -4.11\pm    1.59 $ &  $ -$ &$ 4.9\pm 2.1$ & $ -$ &$-$& \\   
OHSC1      & $ 6.06\pm 0.30$ &  $ 0.116\pm 0.002$ & $ -2.95\pm    0.13 $ &  $ -$ &$ 5.5\pm 2.7$ & $ -$ &$-$& \\   
SL84       & $ 5.88\pm 0.29$ &  $ 0.106\pm 0.007$ & $ -5.58\pm    0.14 $ &  $ -$ &$ 5.6\pm 1.5$ & $ -$ &$-$& \\   
KMHK228    & $ 4.98\pm 0.25$ &  $ 0.082\pm 0.001$ & $ -2.64\pm    1.00 $ &  $ -$ &$ 5.6\pm 4.0$ & $ -$ &$-$& \\
OHSC2      & $ 5.53\pm 0.28$ &  $ 0.087\pm 0.001$ & $ -2.86\pm    0.19 $ &  $ -$ &$ 3.2\pm 0.9$ & $ -$ &$-$& \\
SL118      & $ 5.44\pm 0.27$ &  $ 0.082\pm 0.003$ & $ -4.07\pm    0.13 $ &  $ -$ &$ 4.7\pm 1.2$ & $ -$ &$-$& \\
KMHK343    & $ 5.83\pm 0.29$ &  $ 0.088\pm 0.005$ & $ -3.56\pm    0.15 $ &  $ -$ &$ 4.5\pm 1.9$ & $ -$ &$-$& \\
OHSC3      & $5.90\pm0.27$ &  $ 0.096\pm 0.007$ & $ -3.77\pm    1.16 $ &  $3.26\pm0.48$ &$ 2.0\pm 1.1$ & $16.8\pm 6.7$ &$9.25\pm  0.05$& 7\\ 
OHSC4      & $ 5.72\pm 0.29$ &  $ 0.080\pm 0.002$ & $ -5.50\pm    0.25 $ &  $ -$ &$ 4.0\pm 2.4$ & $ -$ &$-$& \\
SL192      & $ 5.39\pm 0.27$ &  $ 0.078\pm 0.001$ & $ -4.26\pm    0.53 $ &  $ -$ &$ 7.0\pm 2.3$ & $ -$ &$-$& \\
LW141      & $ 5.04\pm 0.25$ &  $ 0.073\pm 0.002$ & $ -3.65\pm    0.17 $ &  $ -$ &$ 4.2\pm 1.4$ & $ -$ &$-$& \\
SL295      & $ 5.83\pm 0.29$ &  $ 0.074\pm 0.001$ & $ -4.35\pm    0.21 $ &  $ -$ &$ 5.5\pm 1.9$ & $ -$ &$-$& \\
SL576      & $4.52\pm0.21$ &  $ 0.078\pm 0.004$ & $ -6.86\pm    0.27 $ &  $4.33\pm0.16$ &$ 6.6\pm 1.3$ & $32.0\pm 6.0$ &$8.99\pm  0.05$& 7\\ 
IC2148     & $ 5.66\pm 0.28$ &  $ 0.075\pm 0.001$ & $ -4.38\pm    0.15 $ &  $ -$ &$ 4.0\pm 1.1$ & $ -$ &$-$& \\
SL647      & $ 5.34\pm 0.27$ &  $ 0.082\pm 0.002$ & $ -4.20\pm    0.13 $ &  $ -$ &$ 4.7\pm 1.6$ & $ -$ &$-$& \\
SL703      & $ 5.07\pm 0.25$ &  $ 0.092\pm 0.002$ & $ -4.69\pm    0.47 $ &  $ -$ &$ 7.5\pm 2.4$ & $ -$ &$-$& \\
SL737      & $ 5.91\pm 0.30$ &  $ 0.070\pm 0.001$ & $ -4.38\pm    0.41 $ &  $ -$ &$ 4.6\pm 1.6$ & $ -$ &$-$& \\
SL783      & $ 5.06\pm 0.25$ &  $ 0.118\pm 0.001$ & $ -5.16\pm    0.11 $ &  $ -$ &$ 5.2\pm 0.9$ & $ -$ &$-$& \\
IC2161     & $ 5.57\pm 0.28$ &  $ 0.089\pm 0.006$ & $ -4.61\pm    0.25 $ &  $ -$ &$ 7.0\pm 2.0$ & $ -$ &$-$& \\
SL828      & $ 4.98\pm 0.25$ &  $ 0.114\pm 0.006$ & $ -5.20\pm    0.11 $ &  $ -$ &$ 6.5\pm 1.0$ & $ -$ &$-$& \\
SL835      & $5.74\pm0.26$ &  $ 0.090\pm 0.001$ & $ - $ &$ -$ &  $3.4\pm0.2$ & $-$ &$9.30\pm  0.10$& 20\\ 
SL882      & $ 4.93\pm 0.25$ &  $ 0.070\pm 0.002$ & $ -4.35\pm    0.13 $ &  $ -$ &$ 4.0\pm 2.1$ & $ -$ &$-$& \\
LW458      & $ 5.95\pm 0.30$ &  $ 0.051\pm 0.001$ & $-$                  &  $ -$ &$- $ & $ -$ &$-$& \\
LW460      & $ 4.80\pm 0.24$ &  $ 0.080\pm 0.003$ & $ -3.03\pm    0.33 $ &  $ -$ &$ 4.3\pm 3.3$ & $ -$ &$-$& \\
LW459      & $ 5.50\pm 0.27$ &  $ 0.049\pm 0.001$ & $ -2.96\pm    0.18 $ &  $ -$ &$ 3.8\pm 2.2$ & $ -$ &$-$& \\
LW462      & $ 4.93\pm 0.25$ &  $ 0.080\pm 0.003$ & $-$                  &  $ -$ &$ -$ & $ -$ &$-$& \\
LW463      & $ 4.80\pm 0.24$ &  $ 0.076\pm 0.002$ & $ - $ &  $ -$ &$ -$ & $ -$ &$-$& \\
KMHK1732   & $ 4.97\pm 0.25$ &  $ 0.083\pm 0.001$ & $ -4.31\pm    0.18 $ &  $ -$ &$ 4.6\pm 2.6$ & $ -$ &$-$& \\
SL883      & $ 5.58\pm 0.28$ &  $ 0.050\pm 0.002$ & $ -4.00\pm    0.49 $ &  $ -$ &$ 3.5\pm 2.2$ & $ -$ &$-$& \\
KMHK1739   & $ 4.90\pm 0.24$ &  $ 0.072\pm 0.002$ & $ -5.01\pm    0.29 $ &  $ -$ &$ 4.2\pm 1.9$ & $ -$ &$-$& \\
SL886      & $ 5.26\pm 0.26$ &  $ 0.078\pm 0.001$ & $ -4.09\pm    0.39 $ &  $ -$ &$ 5.1\pm 2.2$ & $ -$ &$-$& \\
LW469      & $5.20\pm0.24$ &  $ 0.068\pm 0.001$ & $ -4.79\pm    0.14 $ &  $3.36\pm0.13$ &$ 2.4\pm 0.6$ & $16.8\pm 2.9$ &$8.78\pm  0.07$& 18\\ 
LW470      & $ 5.11\pm 0.25$ &  $ 0.076\pm 0.002$ & $ -4.21\pm    0.48 $ &  $ -$ &$ 4.9\pm 1.6$ & $ -$ &$-$& \\
NGC2241    & $5.52\pm0.25$ &  $ 0.070\pm 0.002$ & $ -5.54\pm    0.12 $ &  $3.98\pm0.14$ &$ 4.8\pm 1.0$ & $28.1\pm 4.9$ &$9.28\pm  0.08$& 17\\ 
SL890      & $ 5.07\pm 0.25$ &  $ 0.074\pm 0.001$ & $ -3.50\pm    0.59 $ &  $ -$ &$ 3.3\pm 1.4$ & $ -$ &$-$& \\
LW472      & $ 5.82\pm 0.29$ &  $ 0.062\pm 0.001$ & $ -3.56\pm    0.15 $ &  $ -$ &$ 2.6\pm 0.8$ & $ -$ &$-$& \\
LW475      & $ 5.11\pm 0.25$ &  $ 0.085\pm 0.002$ & $ -3.73\pm    0.30 $ &  $ -$ &$ 4.8\pm 1.4$ & $ -$ &$-$& \\
SL889      & $ 5.54\pm 0.28$ &  $ 0.076\pm 0.003$ & $ -3.01\pm    0.79 $ &  $ -$ &$ 3.5\pm 0.8$ & $ -$ &$-$& \\
SL891      & $ 5.19\pm 0.26$ &  $ 0.073\pm 0.002$ & $ -4.59\pm    0.22 $ &  $ -$ &$ 4.6\pm 0.9$ & $ -$ &$-$& \\
SL892      & $ 5.21\pm 0.26$ &  $ 0.076\pm 0.002$ & $ -4.20\pm    0.24 $ &  $ -$ &$ 2.4\pm 0.6$ & $ -$ &$-$& \\
OHSC36     & $ 5.59\pm 0.28$ &  $ 0.071\pm 0.002$ & $ -3.48\pm    0.16 $ &  $ -$ &$ 4.0\pm 1.7$ & $ -$ &$-$& \\
SL897      & $ 5.79\pm 0.29$ &  $ 0.072\pm 0.001$ & $ -4.91\pm    0.30 $ &  $ -$ &$ 5.8\pm 1.8$ & $ -$ &$-$& \\
\hline
\end{tabular}

Notes: $^{\rm a}$~ galactocentric distances; $^{\rm b}$~ obtained from \cite{sf11} maps; $^{\rm c}$~ integrated from the surface brightness profiles (see Sect.~\ref{sec:mv}; $^{\rm d}$~ derived from $M_V$ and $\log(t)$ (see eqs.~\ref{eq:mass} and ~\ref{eq:emass}); $^{\rm e}$ $r_h$ is the 3D half-light radius calculated from $r_c$ (SBP) and $r_t$ (RDP) (see Sect.~\ref{sec:rh_rj}); $^{\rm f}$ $r_J$ is the Jacobi radius estimated in Sect.~\ref{sec:rh_rj}; $^{\rm g}$ taken from (1) \cite{psc05}, (2) \cite{mjd92}, (3) \cite{ggk10}, (4) \cite{mrb02}, (5) \cite{nsc18}, (6) \cite{dkb16}, (7) \cite{mds19}, (8) \cite{pgc14}, (9) \cite{p14}, (10) \cite{cvh06}, (11) \cite{mps14}, (12) \cite{ssb10}, (13) \cite{p11b}, (14) \cite{pgr15}, (15) \cite{kce18}, (16) \cite{dkb14}, (17) \cite{gbd97}, (18) \cite{p12}, (19) \cite{p11}, (20) \cite{ldk13}

\end{table*}

\subsection{Comparison with previous studies}
\label{sec:comp}

 In paper~I we studied 9 clusters in common with the present sample. In general, the structural parameters are in agreement within $\sim 1 \sigma$.  However, there is one case where the discrepancy is significant: for SL\,576 we obtained a larger $r_t$ than the one given in paper~I, in which we employed a single average magnitude limit (supposedly where we found the photometry is statistically complete).  In the present study, we improved this criterion and performed an individual evaluation of this limit for each image independently, resulting in a different $r_t$ for SL\,576.  Also, given that $\sim 10$ per cent of the present sample has $r_t$ beyond the FoV limits ($r_t>100$\,arcsec), the uncertainties may be underestimated for these cases (including SL\,576).

\begin{figure}
\includegraphics[width=1.02\linewidth]{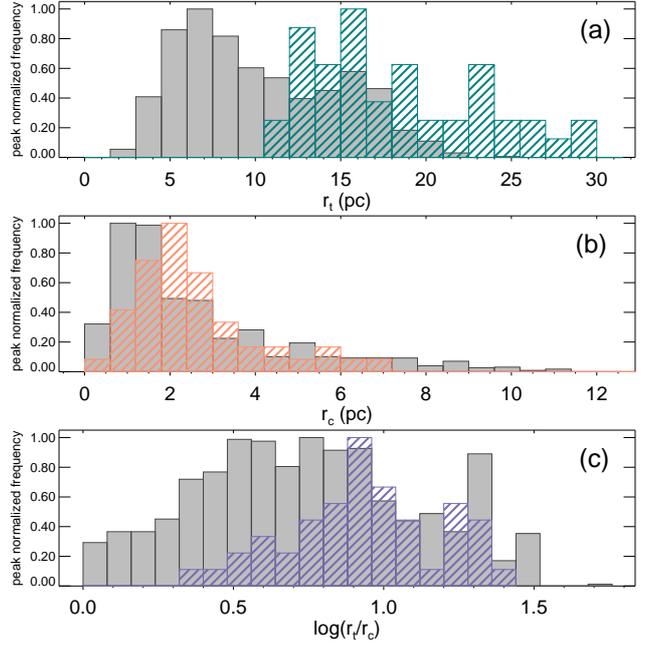}
\caption{LMC cluster distribution of tidal radius (a), core radius (b) and concentration parameter (c) comparing our sample (coloured, hatched histograms) with the WZ11 sample (grey bars).}
\label{fig:param_hist_LMC}
\end{figure}

\begin{figure}
\includegraphics[width=1.02\linewidth]{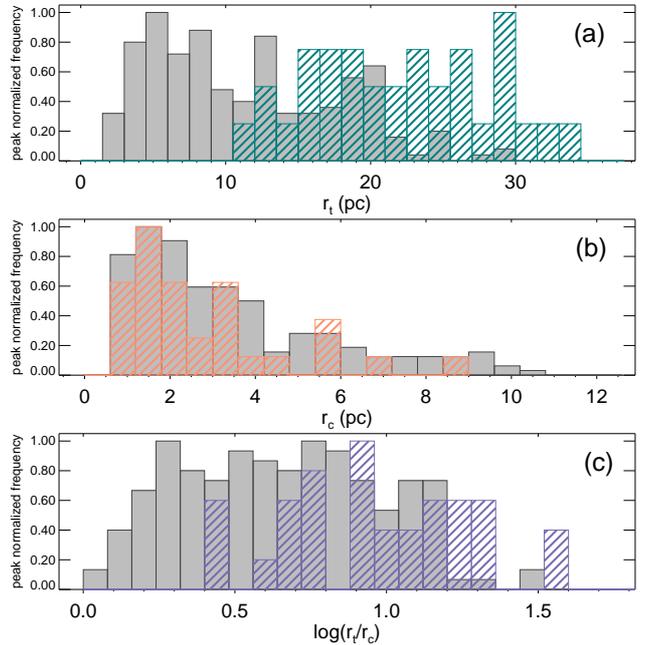}
\caption{SMC cluster distribution of tidal radius (a), core radius (b) and concentration parameter (c) comparing our sample (coloured, hatched histograms) with the HZ06 sample (grey bars).}
\label{fig:param_hist_SMC}
\end{figure}

The distributions of the structural parameters $r_t$, $r_c$ and the concentration parameter ($\log(r_t/r_c)$) for our LMC cluster sample and that of WZ11 for inner LMC clusters are plotted in Fig.~\ref{fig:param_hist_LMC}. The histograms are normalized to the peak of each distribution. Similar histograms for the SMC are shown in Fig.~\ref{fig:param_hist_SMC}, where our sample is compared with the HZ06  sample, although in this case there is not a clear separation of outer and inner clusters, with some clusters present in both samples. Since WZ11 and HZ06 did not provide the tidal radii but the 90 per cent light radii, we converted their values to the full radii enclosing 100 per cent of the clusters (model) light.  A relevant question is that the magnitude limit of the VISCACHA survey is deeper than that of MCPS, and thus lower mass stars are reached by the former. This difference is taken into account when the radius at 90 per cent of the cluster light was converted to the full light profile.

 According to Fig.~\ref{fig:param_hist_LMC}, there is a tendency for larger clusters to be located in the LMC outskirts, which we cannot assert for the SMC given the mixed samples. Indeed, it is known that  outer clusters are able to expand to larger sizes because the gravitational field is weaker \citep[e.g.][]{v94,brg15,bh18,asc20}.  In the inner galaxy regions, clusters are more compact due to the stronger gravitational field that tidally strips stars as the cluster expands. On the other hand, inner cluster regions are less affected by tidal effects \citep[e.g.][]{pm18}, like shocks from passing satellites.

\begin{figure}
\includegraphics[width=0.99\linewidth]{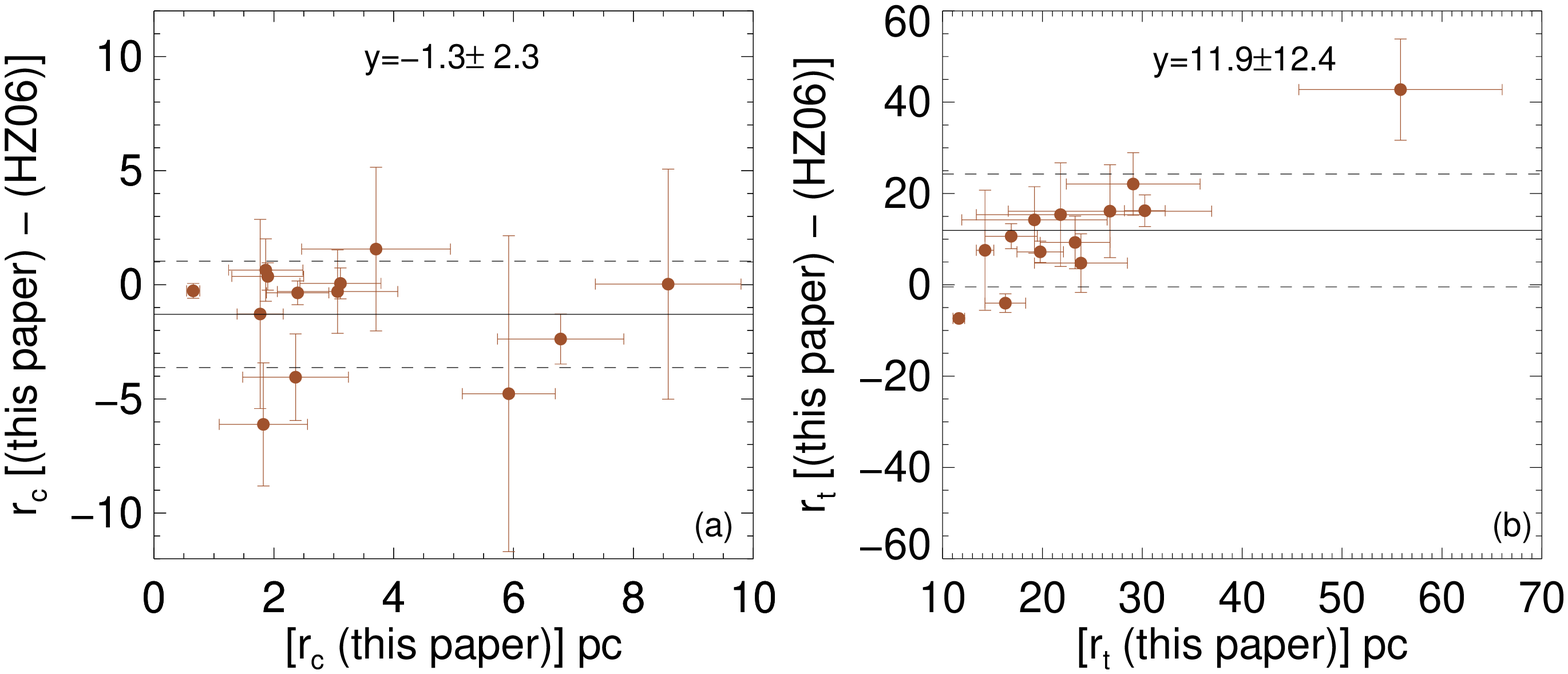}
\caption{Comparison of (a) $r_c$ and (b) $r_t$ for 13 SMC clusters in our sample in common with the HZ06 sample. The continuous line is the average of the difference between our structural parameter values and those by HZ06, while the dashed line is the corresponding $1\ \sigma$ deviation. }
\label{fig:rhz}
\end{figure}

In Fig.~\ref{fig:rhz}, we compare directly $r_c$ and $r_t$ for our sample clusters in common with the HZ06 SMC sample. There are no entries in common with the WZ11 LMC sample. Both SMC and LMC clusters' tidal radii from HZ06 and WZ11, respectively, are in general smaller than our $r_t$, which is possibly a residual systematic difference caused by the limiting magnitudes of the VISCACHA  and MCPS surveys.  Therefore, when comparing our sample $r_t$  with the  literature ones (Figs.~\ref{fig:param_hist_LMC} and \ref{fig:param_hist_SMC}), we should take into account this effect. The core size $r_c$, that we evaluated from the SBP fit, should be less affected by photometric depth, since the core is the brightest cluster region.  Fig.~\ref{fig:rhz} gives the difference between our structural parameter values and those by HZ06 as a function of our values. The average and standard deviation ($1\ \sigma$) of this difference are represented by the continuous and the dashed lines, respectively. We note that although there is a good agreement of our $r_c$ values with the ones of HZ06, the $r_t$ values have a systematic difference with a large standard deviation.

\begin{figure}
\includegraphics[width=0.99\linewidth]{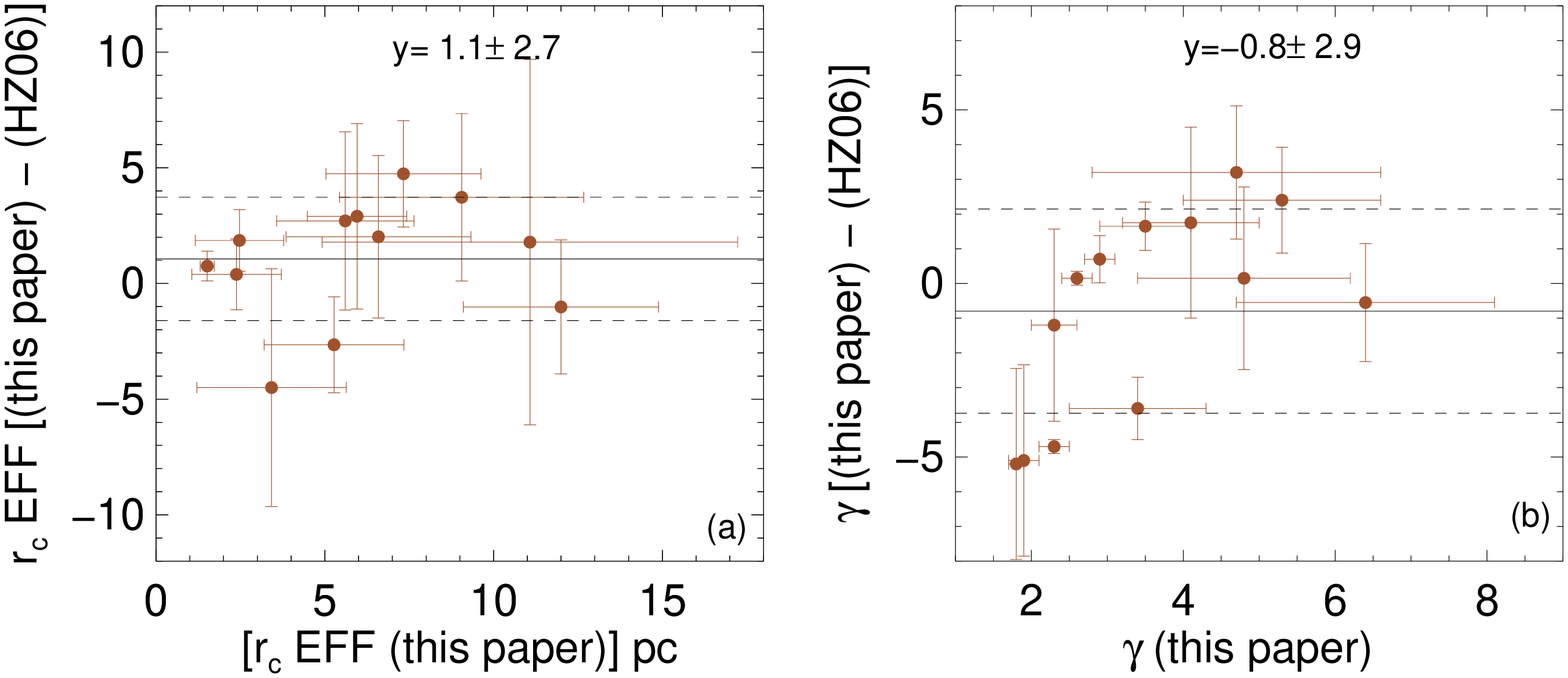}
\caption{ Comparison of (a) $r_c=a\sqrt{2^{2/\gamma}-1}$ and (b) $\gamma$ for 13 SMC clusters in our sample in common with the HZ06 sample. The continuous line is the average of the difference between our structural parameter values and those by HZ06, while the dashed line is the corresponding $1\ \sigma$ deviation. }
\label{fig:eff_ag}
\end{figure}

 We also compared structural parameters obtained from the EFF fits with those by HZ06 for the clusters in common, showing a good agreement (Fig.~\ref{fig:eff_ag}).

\subsection{Assessing the fit quality}

\begin{figure}
\includegraphics[width=1.0\linewidth]{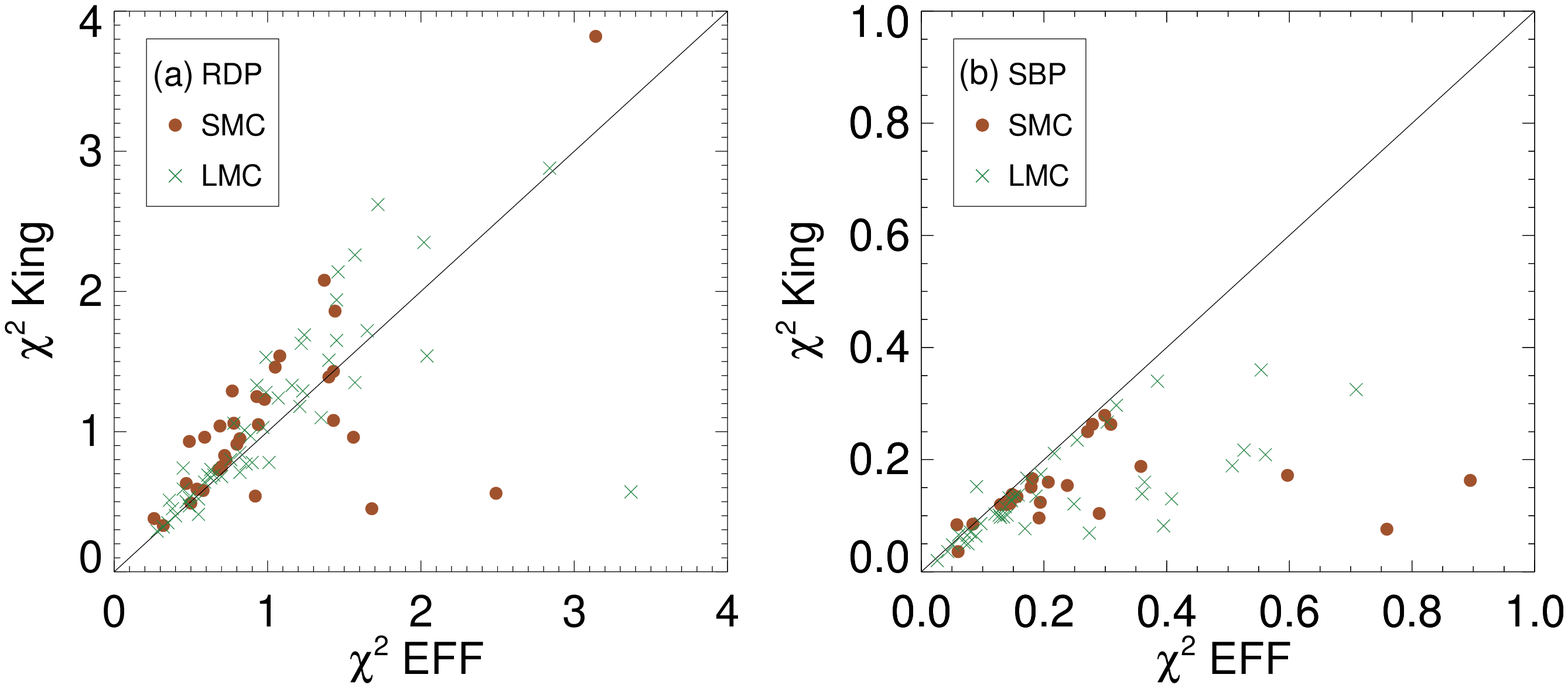}
\caption{ Comparison between $\chi^2$ resulting from the King and EFF fitting to the RDP (a) and SBP (b) of SMC (filled circles) and LMC (crosses) clusters. The straight line indicates the 1:1 relation.}
\label{fig:chi2}
\end{figure}

 Fig.~\ref{fig:chi2} shows the comparison between the $\chi^2$ obtained from the King and EFF fittings to the RDP (panel a) and to the SBP (panel b). In agreement with HZ06 and WZ11, our study reveals that most of the clusters are well fitted by both models, although lower values of $\chi^2$ are achieved, on average, for King models fitted to the SBPs. EFF and King models yield fittings of similar quality on the RDPs, as suggested by the $\chi^2$ distribution in panel (a).

\section{Structural parameters and their spatio-temporal distribution}
\label{sec:res}

\subsection{Spatial properties}

\begin{figure}
\includegraphics[width=0.99\linewidth]{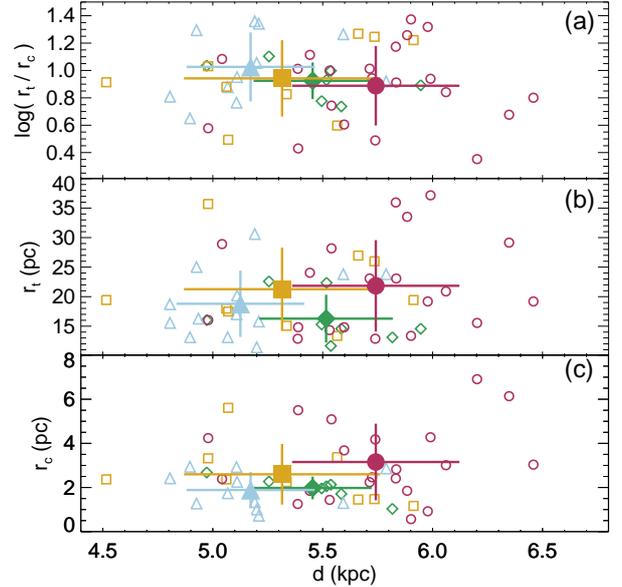}
\caption{Structural parameters as a function of deprojected distance from the LMC centre for clusters in the four groups identified by empty symbols as in Fig.~\ref{fig:map_deproj}. (a) Concentration parameter, (b) tidal radius, (c) core radius. The  mean values are represented by filled symbols with error bars determined by the standard deviations.}
\label{fig:rctdist}
\end{figure}

\begin{figure}
\includegraphics[width=0.99\linewidth]{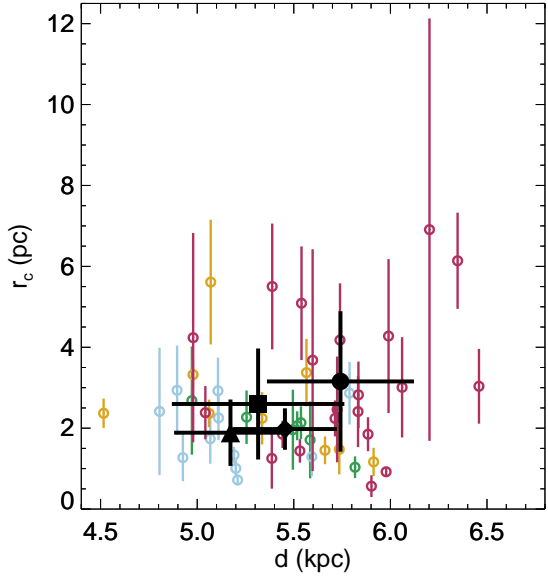}
\caption{Core radii and their uncertainties as a function of deprojected distance from the LMC centre. Clusters discriminated by groups as in Fig.~\ref{fig:map_deproj}.  The mean values of the groups are represented by filled symbols with error bars determined by the standard deviations.}
\label{fig:rcdist}
\end{figure}

The relationship between the structural parameters and deprojected distance from the LMC centre are presented in Fig.~\ref{fig:rctdist}. Empty symbols represent individual clusters grouped with different colours as in Fig.~\ref{fig:map_deproj}, while filled symbols indicate the groups' mean and the error bars account for the  standard deviation. The three panels of Fig.~\ref{fig:rctdist} correspond to (a) the concentration parameter, (b) the tidal radius and (c) the core radius, against the deprojected distance from the LMC centre. The mean values of the concentration parameter for the four groups, between $0.8<\log{(r_t/r_c)}<1.0$, is similar to that of open clusters in our Galaxy \citep{bm98}. The two westernmost cluster groups (red circles and yellow squares), the closest ones to the SMC, have $r_t$  dispersion above those for the easternmost groups (blue triangles and green diamonds),  although this trend is very weak. The same occurs for $r_c$.  For better visualisation of the larger $r_c$ spread of the westernmost groups compared with the easternmost ones, we reproduce Fig.~\ref{fig:rctdist}(c) with error bars in Fig.~\ref{fig:rcdist}.

\begin{figure}
\includegraphics[width=0.99\linewidth]{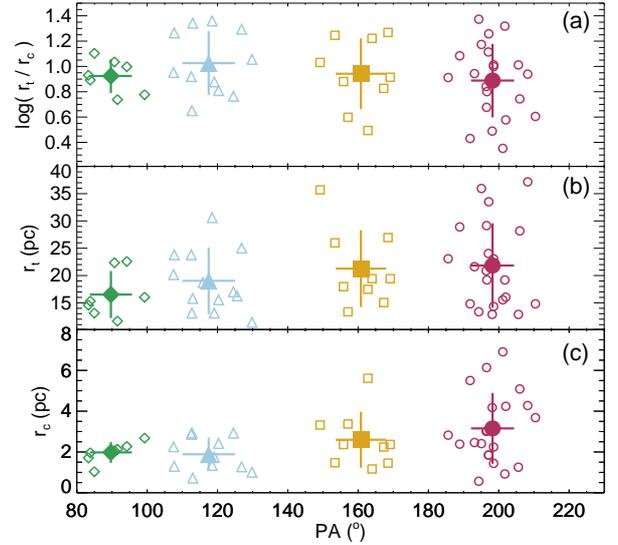}
\caption{Structural parameters as a function of the position angle over the LMC centre for clusters identified by empty symbols as in Fig.~\ref{fig:map_deproj}. (a) Concentration parameter, (b) tidal radius, (c) core radius. The group parameters means are represented by filled symbols with error bars determined by the  standard deviations.}
\label{fig:pa}
\end{figure}

 There is a marginal tendency for an increase of the clusters' core and tidal radii dispersions towards the region where the LMC warp (Choi et al. 2018a) starts, suposedly triggered by the interaction with the SMC. If this interaction characterizes a tidal shock strong enough to disturb the clusters' structure is difficult to assess. Nevertheless, to confirm this tendency statistically, more data with better accuracy is needed. Fig.~\ref{fig:pa} shows a clearer separation of the groups where the clusters are characterized by their position angles over the LMC centre. From northeast to southwest the number of clusters per group with $r_c>4$\,pc amounts to 0, 0, 11, and 33 per cent, respectively, and for $r_t>25$\,pc the fractions are 0, 15, 33, and 29 per cent.

\begin{figure}
\includegraphics[width=0.99\linewidth]{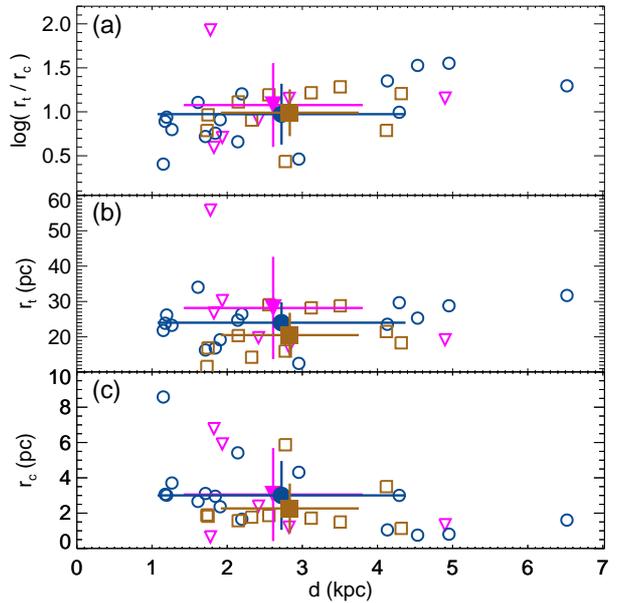}
\caption{Structural parameters as a function of the projected distance from the SMC centre for clusters in the three groups identified by empty symbols as in Fig.~\ref{fig:chart}. (a) Concentration parameter, (b) tidal radius and (c) core radius. The mean values are represented by filled symbols with error bars determined by the standard deviations}.
\label{fig:rctdist_SMC}
\end{figure}

Because of the lack of precise individual cluster distances (to be derived with VISCACHA data in the near future) and the more complicated geometry of the SMC, the projected distance was employed to investigate the distribution of SMC cluster structural parameters. The relations between structural parameters and the projected distance to the SMC centre are shown for SMC clusters in Fig.~\ref{fig:rctdist_SMC}.  Some trends are found, especially in concentration and core radius, but we cannot draw any firm conclusions given the uncertainties involved at this time. A clear scenario will only emerge when the clusters' individual distances are determined (Kerber et al., in prep.) and  a full statistical exploration of the database, including new observations, is presented in a forthcoming study.

\subsection{Structure versus age}
\label{sec:core}

\begin{figure}
\includegraphics[width=1.0\linewidth]{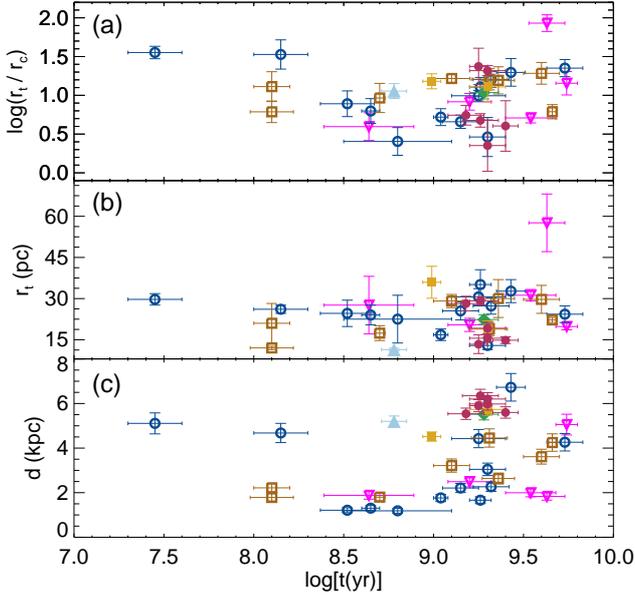}
\caption{Concentration parameter (a), tidal radius (b) and galactocentric distance (c) as a function of age for SMC (open symbols) and LMC (filled symbols) clusters with available age. Symbol colours are the same as in Figs.~\ref{fig:chart} and \ref{fig:map_deproj}.}
\label{fig:par_age}
\end{figure}

We compiled age information from the literature for 10 LMC clusters (5 of them from paper I) and 26 SMC clusters (4 from paper I) in our sample (see Table~\ref{tab:red_age}). The difference between the number of LMC and SMC clusters with age information reflects the usual selection of  high surface brightness targets for observation, which are primarily found in our SMC sample. Most of the outer ring LMC clusters are low surface brightness objects. Whenever two or more different age sources were available, we simply averaged the provided ages and propagated the uncertainties, unless one method was judged clearly more accurate than another, in which case we took its age and uncertainty. For instance, ages determinated from isochrone fitting to the cluster CMD were considered superior to those obtained from integrated photometry. Distributions of tidal radius and concentration parameter with age are presented in Fig.~\ref{fig:par_age}. 

\begin{figure}
\includegraphics[width=1.0\linewidth]{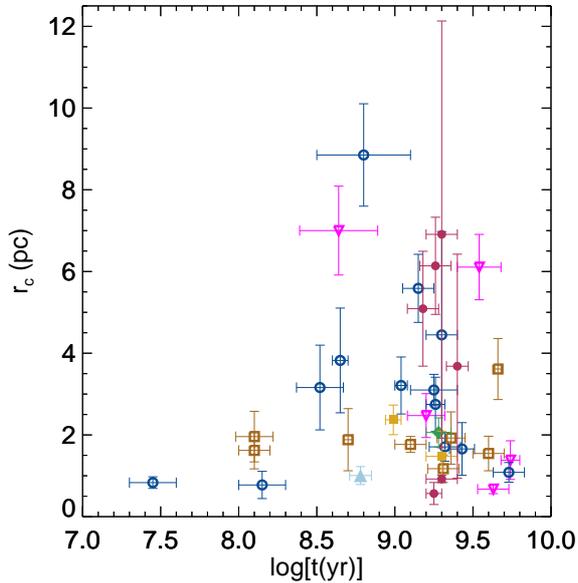}
\caption{Core radius as a function of age for SMC (open symbols) and LMC (filled symbols) clusters with available age. Symbol colours are the same as in Figs.~\ref{fig:chart} and \ref{fig:map_deproj}.}
\label{fig:rc_evol1}
\end{figure}

Fig.~\ref{fig:rc_evol1} shows the core radius evolution, with clusters identified by their locations as in Figs.~\ref{fig:chart} and \ref{fig:map_deproj}. It reproduces the \cite{mg03b} results (see their fig. 2), namely, there is a spread of core radius for older clusters that seems to start at $\log(t)\sim 8.5$. 

\subsection{Mass distribution}
\label{sec:mass}

For clusters with age information, we determined their masses using simple stellar population models and derived relations according to \cite{mps14}. Specifically, these models are based on Padova isochrones \citep{mgb08} for ages between $\log[t(yr)]=6.6$ and 10.1, metallicities $Z=$0.019, 0.008 and 0.004, with star masses distributed in the range $0.08<m/$M$_\odot<120$  as a \cite{kwp13} initial mass function. Linear relations for $\log[t(yr)]>7.3$ were obtained from these models to provide mass as a function of integrated absolute magnitude (in several bands) and age \citep{mps14}. We reproduce below those relations for $Z=$0.008, as representative of the LMC overall metallicity, and $Z=$0.004, as representative of the SMC overall metallicity \citep{w97}. 

\begin{equation}
\label{eq:mass}
\log (M/{\rm M}_\odot)=a +b \log[t(yr)] -0.4(M_V-M_{V\odot})
\end{equation}

\noindent where $a=-6.14\pm 0.08$ and $b=0.644\pm 0.009$ for $Z=0.008$; and $a=-5.87\pm 0.07$, $b=0.608\pm 0.008$ for $Z=0.004$;  $M_{V\odot}=4.83$.
 
The propagated uncertainty is then:

\begin{equation}
\label{eq:emass}
\sigma_{\log(M/{\rm M}_\odot)}=\sqrt{\sigma_a^2+\sigma_b^2 \log^2(t)+0.4^2 \sigma^2_{M_V}}
\end{equation}

\begin{figure}
\includegraphics[width=1.0\linewidth]{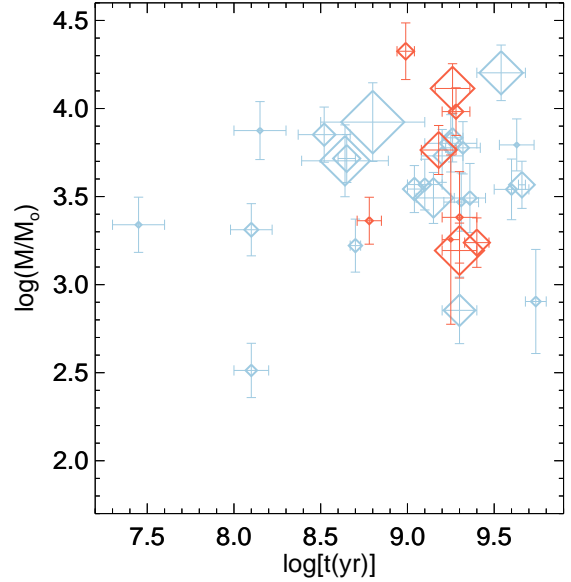}
\caption{SMC (blue) and LMC (red) clusters' mass as a function of age. Symbol sizes indicate the core radius sizes.}
\label{fig:rc_evol3}
\end{figure}

It is worth noticing that the assumption of a single average metallicity for all clusters within a galaxy does not significantly affect their calculated masses since the metallicity difference yields mass values that are within the uncertainties. The results are shown in Fig.~\ref{fig:rc_evol3}, where the SMC (blue) and LMC (red) cluster masses are plotted versus their ages with symbol sizes representing the core sizes ($r_c$). Most cluster masses are in the range $10^3<M/{\rm M}_\odot<10^4$. 
 
It calls the attention that groups of clusters with similar age ($\log(t)\sim 9.0$) and mass ($\log M/{\rm M}_\odot \sim 3.7$) span a range of core radius values, indicating that the dynamical evolution of the clusters strongly depends on their initial conditions, such as, different binary fractions, small variations of the initial mass function leading to different fractions of BHs and blue stragglers \citep{mwd08,fld19},  although we do not expect the retention of BHs by the relatively low mass clusters in our sample.

\section{Half-light and Jacobi radii}
\label{sec:rh_rj}

 The half-light radius ($r_h$) has been used as a reference radius in many studies \citep[e.g.][]{h09,lbg10,agl14} because it stays nearly constant for a significant timespan of a cluster existence, whenever two-body relaxation is dominant \citep{bpg10}. We estimated $r_{hp}$, the projected half-light radius, from the fitted King model parameters using the expression (see Appendix~\ref{sec:app}):

\begin{equation}
\label{eq:final}
\log\left(\frac{r_{hp}}{r_c}\right)=-(0.339\pm0.009)+(0.602\pm0.015)c-(0.037\pm0.005)c^2
\end{equation}

\noindent where $c=\log{(r_t/r_c)}$ is the concentration parameter. 
 
 We also calculated the Jacobi radius for all clusters with mass determined in Sect.~\ref{sec:mass} by means of \citep{ihw83,a08}:

\begin{equation}
\label{eq:jac}
r_J=\frac{2^{2/3}}{3} \left(\frac{G M_{cl}}{v_c^2}\right)^{1/3} d^{2/3} 
\end{equation}

\noindent where $M_{cl}$ is our estimate for the cluster mass, $d$ is the cluster galactocentric distance (deprojected in the case of the LMC clusters) and $v_c$ is the circular velocity for the MCs with flat rotation curves. We used $v_{c,SMC}= 55\pm5$ km\,s$^{-1}$ \citep{dmj19} and $v_{c,LMC} = 91.7\pm 18.8$ km\,s$^{-1}$ \citep{vk14}. 

The ratio between the deprojected half-light radius ($r_h$) and $r_J$, the Roche volume filling factor, determines how tidally filling a cluster is \citep[e.g.][]{agl14}. Even more directly, the ratio between the tidal radius and $r_J$ would be close to 1 for clusters filling their Roche volume \citep{ej13}. This information is relevant to infer on the clusters' dynamical state, since those systems that fill their Roche volumes are more susceptible to tidal effects leading to mass loss \citep[e.g.][]{hh03,ebj15}. To estimate this ratio, we considered the three-dimensional value $r_h=1.33 r_{hp}$ \citep[e.g.][]{bpg10}. The $r_h$ and $r_J$ values are presented in Table~\ref{tab:red_age}.

\begin{figure}
\includegraphics[width=1.0\linewidth]{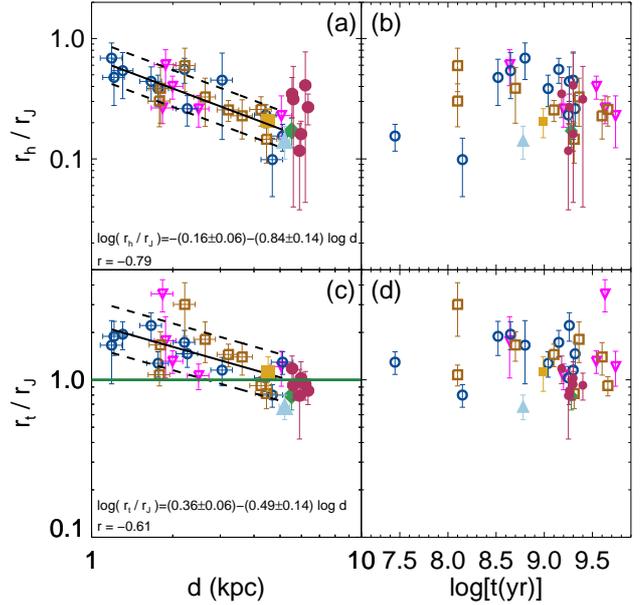}
\caption{Roche volume filling factor versus galactocentric distance and age (in log scale) for SMC (open symbols)  and LMC (filled symbols) clusters. Symbol colours are the same as in Figs.~\ref{fig:chart} and \ref{fig:map_deproj}. The ratio between the half-light and Jacobi radii are displayed in panels (a) and (b) while the ratio between the tidal and the Jacobi radius are shown in panels (c) and (d). In panels (a) and (c), linear fits with their 1\,$\sigma$ uncertainties (continuous and dashed lines, respectively) were performed for the SMC cluster data. The fit coefficients are indicated. The green horizontal line in panel (c) separates clusters that are tidally underfilling from those that are tidally overfilling.}
\label{fig:rhrj}
\end{figure}

Fig.~\ref{fig:rhrj} shows the distribution of the Roche volume filling factor as a function of the galactocentric distance and age. Considering only SMC clusters, which are distributed throughout a wide range of distances from the SMC centre, it can be seen a significant negative correlation between the galactocentric distance and $r_h/r_J$ or $r_t/r_J$. Even with projected distances that we are using, the relations stand out. Straight lines, shown in panels (a) and (c), were fitted to this data (only SMC clusters) yielding  correlation coefficients of $r=-0.79$ for $r_h/r_J$ versus $d$ and $r=-0.61$ for $r_t/r_J$ versus $d$. The fitted functions and their associated 1\,$\sigma$ uncertainties are also displayed in Fig.~\ref{fig:rhrj}. We did not find a correlation between the Roche volume filling factor, either using $r_h/r_J$ or $r_t/r_J$, and the cluster ages (Fig.~\ref{fig:rhrj}(b,d)).

 Under influence of a steady tidal field, as $r_h$ remains approximately constant during cluster evolution and $r_J$ shrinks due to cluster mass loss, the ratio $r_h/r_J$ rises, at least for single stellar mass clusters \citep[e.g.][]{kkb08}. At a given time, equal mass clusters at larger galactocentric distances are expected to have larger $r_J$ than that for clusters at smaller galactocentric distances and, consequently, they also have smaller $r_h/r_J$. This simplified scenario agrees with what is seen in Fig.~\ref{fig:rhrj}, although the clusters' phase of dynamical evolution and their different masses contribute to the scatter around the mean relation.  In the case of a more complex varying tidal field, as that originated by the recent past MCs encounter, the behaviour is not expected to be very different, as clusters rapidly adjust their structure to the tidal field in which they are located. Indeed, \cite{mws16} performed simulations showing that a cluster belonging to a galaxy accreted by the MW should have its half-mass radius quickly adjusted to the new tidal field, erasing any structural feature that would reveal its origin.

If an outer LMC cluster is close to filling or overfilling, which is expected because they are under a weak tidal field, the passage of the SMC would strip stars located beyond the Roche volume (that shrunk because of the stronger tidal field during the encounter). After the SMC passage, the cluster is again  predominantly subjected to the steady LMC tidal field, increasing its $r_J$. The cluster half-mass radius would readjust after 1 or 2 relaxation times \citep{mws16}, until then it would not fill its Roche volume. Therefore, we would expect to find any trace of the interaction on the LMC clusters structure only if they were underfilling, which cannot be seen in Fig.~\ref{fig:rhrj}. We shall further explore  this analysis in a forthcoming study.

\section{Discussion}
\label{sec:disc}

On average, the outer LMC clusters of our sample (located between 4.5 and 6.5\,kpc from the LMC centre) have larger $r_t$ than most of the inner clusters in the sample studied by WZ11. If the LMC tidal field strength is sufficient to affect significantly the clusters structure, then our sample clusters should be under weaker tidal forces than inner clusters, generating the surplus of larger clusters in our outer sample (see Fig.~\ref{fig:param_hist_LMC}). Nevertheless, this result is to be taken with caution due to  a selection effect: low surface brightness clusters when immersed in rich fields have their limiting radius difficult to estimate because of the low contrast between field and member stars, especially at the cluster periphery. Therefore, WZ11 may be missing larger clusters located in the denser fields of the LMC.  Our sample is not immersed in dense stellar fields and was subjected to analyses with homogeneous techniques, from which we obtained accurate results.

 \cite{m87} studied the structure of populous LMC clusters located at different galactocentric distances and found that clusters closer to the LMC centre have smaller core radii than those farther away. The tidal field effect on the inner structure of LMC clusters was investigated by \cite{whm03} through N-body simulations. They studied clusters in elliptical orbits spanning 2 to 8\,kpc (peri and apogalacticon, respectively) from the LMC centre showing that the observed spread of core radii cannot be explained by a time varying tidal field or binary fraction. Since their analysis involves cluster orbits at distances from the LMC centre that are similar to our observed sample, it seems that the effect is not tidal in origin.  Therefore, according to \cite{whm03} results, the $r_c$ variations found in the present work would be due to internal dynamical effects, not external tidal effects, and the cluster to cluster differences in $r_c$ would come from distinct dynamical ages between  them \citep[see also][]{mwd08}. \cite{fld19} suggested that the existence of populous, old clusters ($\sim 13$\, Gyr) with a range of core radii is a consequence of the initial conditions at formation and dynamical evolution. Indeed, \citet{efl89} already presented the core expansion as a result of high-mass stars mass-loss, linking initial conditions, more specifically the initial mass function, with the observed core size spread among intermediate-age clusters. 

 Nevertheless, we found a possible increase on $r_c$ dispersion for LMC clusters closer to the SMC (Fig.~\ref{fig:pa}), but we shall need to include more clusters to investigate this tendency. We also cannot rule out the possibility that the clusters physical sizes might be biased due to the adoption of a single distance modulus. However, should these clusters lie on the LMC disc, then those on the near side would have their sizes reduced while those on the far side would have their sizes increased, further reinforcing the trend found.  The determination of individual distances for these (and additional) clusters would help to settle this question.

In this context, it is interesting to note that \cite{cno18a} used field red clump stars from the Survey of the MAgellanic Stellar History (SMASH), to constrain the 3D structure of the LMC, showing a warp towards the southwest of the outer disc,  possibly associated to the past close encounter with the SMC. Also, \cite{cno18b} recovered a density enhancement of red clump stars tracing an extended arc or ring about $6\degr$ from the LMC centre, attributed to the tidal interaction with the SMC. Simulations involving a galaxy with a bar and a smaller galaxy targeting perpendicular to its disc yield a ring (or spiral arm) around the bar, mimicking the outer LMC clusters spatial distribution \citep{a96}. A simulation by \cite{bkh12} involving three encounters between the Clouds including a direct collision,  yields this warp and indicates that the MCs, not the MW, are responsible for the distorted features observed in the LMC. Particularly, \cite{mke16} highlight a warp, a protruding off-disc, or tilted bar and low density stellar arcs at $\sim 15$\,kpc from the LMC centre towards north.

 The SMC cluster sample covers a larger distance range from the SMC centre than the LMC sample does from the LMC centre. Thereafter, the SMC regions sampled are more heterogeneous \citep{dkb16} as compared to the LMC ring clusters. There appears to be a trend between the concentration parameter and the projected galactocentric distance for SMC clusters (Fig.~\ref{fig:rctdist_SMC}(a)); i.e. clusters closer to the centre present less concentrated structures. To elucidate this feature, a larger sample of clusters is needed. Since we have adopted an overall distance modulus for the SMC clusters and given the large depth derived for this galaxy \citep{csp01,ss12}, we cannot draw any strong conclusions regarding the structural parameters distributions. Only when individual cluster distances are obtained from isochronal fittings (Kerber et al., in prep.)  will it be possible to build a more precise picture of such distributions around the SMC. The concentration parameter, however, should be unaffected by such distances constraints. 

 In order to search for any signature of the past recent MCs collision on the clusters structure, we first discuss general aspects on the clusters' filling factor in tidal fields.

N-body simulations by \cite{ebj15} show that the Roche volume filling factor is important to define the cluster dissolution mechanism, either by mass loss driven by the varying cluster potential (overfilling) or by two-body relaxation from internal dynamical processes (underfilling). 

\citet{bpg10} studied the Roche volume filling factor for Galactic globular clusters, noting a lack of clusters with $r_h/r_J>0.5$, explained by the fast dissolution time-scale induced by the strong tidal field \citep[see also][]{pwc19}. The biggest difference between our sample and that of Galactic globular clusters are the cluster masses and the tidal field in which they are immersed. However, the distribution of $r_h/r_J$  for Galactic globular clusters with galactocentric distances smaller than 8\,kpc, as can be seen in \cite{bpg10} fig. 2 and \cite{pwc19} fig. 7, follows qualitatively our SMC sample distribution. Nevertheless, Galactic open clusters provide a better comparison with our sample, given the mass similarity. 

\cite{ej13} studied this ratio and $r_t/r_J$ for 236 Galactic open and 38 globular clusters using data from \cite{psk07} and \cite{dgv99}, respectively. The Jacobi radius was calculated according to realistic approximations for the clusters' orbits. They concluded that the median of the $r_h/r_J$ distributions are 3 to 5 times larger for (solar neighbourhood) open clusters than for globular clusters, suggesting that most globular clusters formed underfilling their Roche volumes, while open clusters may fill their Roche volumes after the initial gas expulsion.  How their sample of open clusters compares to ours in terms of the median of $r_h/r_J$ and $r_t/r_J$ values? 

To perform such a comparison we should match clusters under similar tidal fields in the three galaxies. As the tidal field changes with the galaxy mass interior to the galactocentric distance and inversely with this distance to the third power, an open cluster located at 8\,kpc from the Milky Way centre is subject to the same tidal field as a LMC cluster placed at $\sim$4.4\,kpc and a SMC cluster placed at $\sim$2.3\,kpc from the respective galaxy centre. For this estimate, we used the masses of $1.7\times 10^{10}$\,M$_\odot$ \citep[LMC;][]{vk14}, $2.4\times 10^{9}$\,M$_\odot$ \citep[SMC;][]{dmj19} and $1.0\times 10^{11}$\,M$_\odot$ \citep[MW;][]{kbi20}. Therefore, since all LMC clusters in our sample are farther than 4.4\,kpc from the LMC centre, they are under the influence of a smaller tidal field than open clusters at 8\,kpc from the Milky Way  centre. This is compatible with the LMC clusters being overfilling or close to it, as Fig.~\ref{fig:rhrj}(c) indicates. 

The sample of open clusters analysed by \cite{ej13} has the median values $\langle r_h/r_J\rangle=0.38$ and $\langle r_t/r_J\rangle=0.81$, while our sample of SMC clusters within galactocentric distances between 2.1 and 2.5\,kpc (equivalent to the open clusters Galacticentric distances for the same  tidal field), has the  median values $\langle r_h/r_J\rangle=0.55$ and $\langle r_t/r_J\rangle=1.73$. In consequence, although the spread of our sample is large in this distance range, a larger fraction of the SMC clusters, compared to open clusters, may have filled their Roche volumes. A bigger sample would be needed to verify this issue on a statistical basis.  The comparison between our sample with Galactic clusters illustrates the effect of different steady tidal fields on the clusters' structural properties. 

 In addition, the Roche volume filling analysis is inconclusive regarding the influence of the MCs recent collision on the clusters' structure. This is because all outer clusters are overfilling or close to it, precluding us to distinguish clusters evolving in a steady tidal field and clusters that were perturbed by the MCs encounter, since the latter would quickly {\it disguise} as the former ones.  Particularly, the passage of the SMC by the outer LMC could possibly have affected the structure of the LMC outer clusters, but because they are closer to filling prevents us of drawing any conclusions.

In summary, a varying tidal field as produced by the past passage of the SMC does not seem to leave detectable marks on the clusters' structure, due to their quick response to the local tidal field.

\section{Concluding remarks}
\label{sec:conc}

In this work, we provide a homogeneous set of structural parameters of 83  star clusters located at the periphery of the MCs, based on the clusters stellar density and surface brightness profiles determined from uniform observations and analysis techniques. The structural parameters were analysed, aided by available ages and derived photometric properties (integrated magnitude and mass), allowing us to investigate relations between them and the clusters' distances to the galaxies' centres, which are related to the tidal field strength.

 The outer LMC clusters have deprojected distances to the LMC centre that do not differ by more than 2 kpc (assuming that the clusters lie in a disc), but are distributed azimuthally from northeast to southwest throughout $\approx 130^\circ$, leading us to expect different dynamical effects. Our results indicated  that the outer clusters' $r_t$ are on average larger than the ones of inner clusters from WZ11. Furthermore, the analysis of structural parameters along the clusters position angle revealed that LMC clusters closer to the SMC (towards southwest) have increasingly larger $r_c$ dispersion. Although this preliminary result suggests a connection with the beginning of the warp at $\sim 6$\,kpc towards the southwest of the outer LMC disc \citep{cno18a,cno18b}, more clusters need to be analysed to shed light on this issue.  

 The SMC clusters in our sample closer to the SMC centre have a tendency to present less concentrated structures than those farther out, although this issue should be better investigated with a larger sample with individual distances to the clusters well determined.

 The distribution of $r_c$ with age for outer SMC and LMC clusters appears to mimic the one of  inner populous clusters \citep{mg03b}, suggesting that tidal forces are less significant in shaping cluster inner structure than their internal dynamical processes as found, e.g., by \cite{pwc19} for Galactic globular clusters.

 The Roche volume filling factor was determined for clusters with age information from the literature. Its analysis shows that the great majority of the SMC clusters  closer than $\sim$\,4\,kpc from the SMC centre overfills this volume. This suggests that these clusters are dissolving by mass loss as their gravitational potential weakens, while a few clusters beyond $\sim$\,4\,kpc evolve mainly via two-body relaxation from internal dynamical processes. The LMC sample, confined to a narrow range of galactocentric distances, presents clusters closer to overfilling their Roche volumes. 

The MCs peripheral clusters investigated in this study are located in an agitated environment with a variable tidal field produced by the MCs encounter. Therefore, the sample clusters are conditioned to such surroundings and possibly most of them doomed to unbind themselves from the MCs, which may alter their structure and internal dynamical evolution during the MCs closest approach.  As the clusters in our sample are all overfilling or close to it, their structure may either reflect the effect of a steady weak tidal field or the quick adjustment after the shock generated by the MCs collision. Properties differentiating outer and inner clusters in both galaxies that would betray this variable tidal field have not been as yet identified.  

Together with astrophysical parameters and derived information on additional clusters from recent observations, the VISCACHA database will be fully explored in the near future with analyses that shall contribute to the knowledge of the Clouds dynamical and chemical evolution.

\section*{Acknowledgements}

 We thank the referee for the very detailed, constructive and helpful review. Based on observations obtained at the Southern Astrophysical Research (SOAR) telescope, which is a joint project of the Minist\'erio da Ci\^encia, Tecnologia, e Inova\c c\~ao (MCTI) da Rep\'ublica Federativa do Brasil, the U.S. National Optical Astronomy Observatory (NOAO), the University of North Carolina at Chapel Hill (UNC), and Michigan State University (MSU). This study was financed in part by the Coordena\c c\~ao de Aperfei\c coamento de Pessoal de N\'ivel Superior - Brasil (CAPES) - Finance Code 001. We also acknowledge support from CNPq. ARL thanks financial support provided by the FOUNDECYT regular project 1170476. DM is supported by the BASAL Center for Astrophysics and Associated Technologies (CATA) through grant AFB 170002, by the Programa Iniciativa Cient\'ifica Milenio grant IC120009, awarded to the Millennium Institute of Astrophysics (MAS), and by Proyecto FONDECYT Regular No. 1170121. APV acknowledges the FAPESP postdoctoral fellowship no. 2017/15893-1.

\section*{data availability}

The data underlying this article are available in the NOIRLab Astro Data Archive (https://astroarchive.noao.edu/).

\bibliographystyle{mnras}
\bibliography{struct_supmat} 

\appendix
\section{Half-light radius estimate from King model parameters} 
\label{sec:app}

By developing the square power of eq.~\ref{eq:k62}, one gets:

\begin{equation}
\begin{split}
f & & = f_\circ \left[\frac{1}{1+(r/r_c)^2}-\frac{2}{\sqrt{(1+(r/r_c)^2)(1+(r_t/r_c)^2)}} \right. \\
& & \left. + \frac{1}{1+(r_t/r_c)^2}\right]
\end{split}
\end{equation}

\noindent where we replaced $\sigma(r)$ by $f$, and note that the single-mass King models may describe interchangeably the radial distributions of stellar number density, mass density or surface brightness.

Rearranging the terms:

\begin{equation}
f=f_\circ \left[\frac{1}{1+x^2}-\frac{2}{\sqrt{a(1+x^2)}}+\frac{1}{a}\right]
\end{equation}

\noindent where $x\equiv r/r_c$ and\ \  $a\equiv 1+(r_t/r_c)^2=1+x_t^2$.

By integrating $f$ over the radial profile with area element $2\pi r dr$ (ring), one obtains the cluster total flux (or total number of stars):
   
\begin{equation}
\label{eq:tot}
f_{tot}=r_c^2\int^{x_t}_0 2\pi x f(x)dx
\end{equation}

\noindent where the integral upper limit corresponds to the cluster limiting radius, and a change of variable of $r$ into $x$ was made so that $dx=dr/r_c$.

Rewriting,

\begin{equation}
f_{tot}=2\pi r_c^2 f_\circ \left[ \int_0^{x_t} \frac{xdx}{1+x^2} - \frac{2}{\sqrt{a}} \int_0^{x_t} \frac{xdx}{\sqrt{1+x^2}} +\frac{1}{a} \int_0^{x_t} xdx \right]
\end{equation}

By definition of $r_{h}$, the integral limits may be split into two intervals with equal flux:

\begin{equation}
\begin{split}
f_{tot} & &=2\pi r_c^2 f_\circ \left[ \int_0^{x_h} \frac{xdx}{1+x^2} - \frac{2}{\sqrt{a}} \int_0^{x_h} \frac{xdx}{\sqrt{1+x^2}} +\frac{1}{a} \int_0^{x_h} xdx \right]\\
&& +2\pi r_c^2 f_\circ \left[ \int_{x_h}^{x_t} \frac{xdx}{1+x^2} - \frac{2}{\sqrt{a}} \int_{x_h}^{x_t} \frac{xdx}{\sqrt{1+x^2}} +\frac{1}{a} \int_{x_h}^{x_t} xdx \right] 
\end{split}
\end{equation}

\noindent where $x_h=r_{h}/r_c$.

Since the two quantities between brackets are equal, 

\begin{equation}
f_{tot} =4\pi r_c^2 f_\circ \left[ \int_0^{x_h} \frac{xdx}{1+x^2} - \frac{2}{\sqrt{a}} \int_0^{x_h} \frac{xdx}{\sqrt{1+x^2}} +\frac{1}{a} \int_0^{x_h} xdx \right]
\end{equation}

And solving the integrals, one gets:

\begin{equation}
f_{tot} =4\pi r_c^2 f_\circ \left[ \frac{\ln{(x_h^2+1)}}{2}-\frac{2}{\sqrt{a}}(\sqrt{x_h^2+1}-1)+\frac{x_h^2}{2a}  \right]
\end{equation}

Or (compare to equation 18 in \cite{k62}, which gives the total number of stars within a radius $r$):

\begin{equation}
\label{eq:ltot}
f_{tot} =2\pi r_c^2 f_\circ \left[ \ln{(x_h^2+1)}-\frac{4}{\sqrt{a}}(\sqrt{x_h^2+1}-1)+\frac{x_h^2}{a}  \right]
\end{equation}

The quantity within brackets depends on $r_{h}$ and is called $\beta$ here:

\begin{equation}
\beta \equiv  \ln{(x_h^2+1)}-\frac{4}{\sqrt{a}}(\sqrt{x_h^2+1}-1)+\frac{x_h^2}{a}
\end{equation}

To obtain the half-light radius (or half-number radius) one can consider the left side of eq.~\ref{eq:tot} as the total flux. We know then that the radius containing half of this total quantity corresponds to the half-light (or half-number radius). The resulting integral is eq.~\ref{eq:ltot} with $f_{tot}$ replaced by $n$ for simplicity. We can calculate the ratio of half the total flux (from $r=0$ to $r=r_{h}$) to the total flux (from $r=0$ to $r=r_t$):

\begin{equation}
\frac{n(x_{h})}{n(x_t)}=\frac{1}{2}=\frac{\beta(x_{h})}{\beta(x_t)}
\end{equation}

Then, to find $x_{h}$ we need to obtain the roots of the equation:

\begin{equation}
\label{eq:root}
2\beta(x_{h})-\beta(x_t)=0
\end{equation}

As $\beta$ depends on the concentration parameter, the numerical solution should be obtained for each value of $c$. The algorithm used to obtain the root is the Muller's method as implemented in the IDL function \textsc{FX\_ROOT}. Table~\ref{tab:rhc} shows the results for several values of $c$. There is no convergence for $c>2.5$. 

\begin{table}
\caption{Roots of eq.~\ref{eq:root} according to the concentration parameter} 
\label{tab:rhc}
\begin{tabular}{cc}
\hline
  $c=\log{x_t}=\log(r_t/r_c)$ & $x_{h}=r_{h}/r_c$  \\
\hline
    0.150   & 0.535    \\
    0.238   & 0.620    \\
    0.477   & 0.889    \\
    0.602   & 1.052    \\
    0.699   & 1.191    \\
    0.778   & 1.313    \\
    0.845   & 1.422    \\
    0.903   & 1.523    \\
     1.000  & 1.703    \\
     1.349  & 2.511    \\
     1.422  & 2.718    \\
     1.477  & 2.884    \\
     1.849  & 4.317    \\
     1.922  & 4.673    \\
     2.000  & 5.083    \\
     2.150  & 5.990    \\
     2.301  & 7.064    \\
     2.389  & 7.782    \\
     2.477  & 8.576    \\
     2.500  & 8.795    \\
\hline
\end{tabular}
\end{table}

Within the range of $0.15<c<2.5$, representative of the structure of most star clusters, $\log(r_{h}/r_c)$ is a monotonically increasing function of $c$.

To provide with a simple means for obtaining $r_{h}$ from King model parameters, we fit a polynomial function to the data of Table~\ref{tab:rhc}, resulting in eq.~\ref{eq:final}.

\section{Full tables 1 to 4}
\label{sec:app2}

Tables B1 to B4 are the complete version of tables 1 to 4 and are available online as supplementary material.

\section{Additional figures}
\label{sec:app3}

RDPs and SBPs with model fittings for the whole sample are presented in Figs.~C1 to C8 (available online as supplementary material), except for those clusters presented in the main manuscript (Fig.~\ref{fig:rdp_sbp}). 

\vfill\eject
\noindent {\bf SUPPLEMENTARY MATERIAL}\\ 

\noindent This document contains the full tables 1 to 4, and the complete set of radial density profiles and surface density profiles for all clusters not shown in Fig.~\ref{fig:rdp_sbp}.

\appendix

\setcounter{section}{1}
\section{Full tables 1-4}

 The full tables of structural parameters are presented below.
 
\begin{landscape}
\begin{table}
\tiny
\caption{SMC clusters' structural parameters from RDPs.} 
\label{tab:sample_SMC_RDPa}
\begin{tabular}{lccccccccccc}
\hline
         &              &               &\multicolumn{5}{c}{\underline{\hspace{2.8cm}King$^{\ \dagger}$\hspace{2.8cm}}}&\multicolumn{3}{c}{\underline{\hspace{1.2cm}EFF$^{\ \dagger}$\hspace{1.2cm}}}\\
 Cluster  &   $\alpha$(J2000) &   $\delta$(J2000) & $\sigma_\circ$ & $r_c$ & $r_t$ & $\sigma_{bg}$ & $\chi^2$ & $a$ & $\gamma$ & $\chi^2$ &$V_{lim}$ \\
        &    (h:m:s)    &   (\ $^\circ$\ :\ $^\prime$\ :\ $^{\prime\prime}$\ )& (arcsec$^{-2}$) & (arcsec) & (arcsec) &  (arcsec$^{-2}$)& &  (arcsec) & & &(mag)\\
\hline
B1         & 00:19:20 &  -74:06:24  &$0.24\pm 0.07$&$  \ 6\pm  2$&$\   59\pm    25$& $0.023 \pm 0.001$&$1.02$& $11\pm\ 5$ & $ 4.2\pm 1.8$ & $1.05$&  $22.75$\\
K6         & 00:25:27 &  -74:04:30  &$0.24\pm 0.05$&$   25\pm  5$&$\   68\pm\    8$& $0.017 \pm 0.001$&$1.94$& $16\pm\ 2$ & $ 2.6\pm 0.2$ & $0.75$&  $21.75$\\
K7         & 00:27:45 &  -72:46:53  &$0.35\pm 0.02$&$   28\pm  2$&$   104\pm \   7$& $0.018 \pm 0.001$&$1.27$& $45\pm\ 9$ & $ 6.4\pm 1.7$ & $1.86$&  $22.25$\\
K9         & 00:30:00 &  -73:22:40  &$0.14\pm 0.04$&$   23\pm  8$&$\   92\pm    35$& $0.062 \pm 0.002$&$2.86$& $21\pm\ 5$ & $ 3.0\pm 0.6$ & $0.73$&  $22.75$\\
HW5        & 00:31:01 &  -72:20:30  &$0.46\pm 0.04$&$  \ 8\pm  1$&$   192\pm    35$& $0.020 \pm 0.001$&$0.70$& $10\pm\ 1$ & $ 2.9\pm 0.2$ & $0.96$&  $22.75$\\
HW20       & 00:44:48 &  -74:21:47  &$0.19\pm 0.03$&$   18\pm  4$&$ \  56\pm  \  7$& $0.070 \pm 0.002$&$0.46$& $13\pm\ 4$ & $ 3.4\pm 0.9$ & $0.91$&  $23.25$\\
L32        & 00:47:24 &  -68:55:12  &$0.30\pm 0.01$&$   22\pm  1$&$ \  74\pm\    3$& $0.005 \pm 0.001$&$0.74$& $44\pm10$ & $ 8.5\pm 2.8$ & $1.54$&  $22.75$\\
HW33       & 00:57:23 &  -70:48:36  &$0.06\pm 0.01$&$   12\pm  4$&$\   70\pm    24$& $0.007 \pm 0.001$&$0.90$& $19\pm\ 5$ & $ 4.2\pm 1.0$ & $0.33$&  $21.75$\\
K37        & 00:57:49 &  -74:19:32  &$0.54\pm 0.04$&$   22\pm  3$&$   117\pm    18$& $0.102 \pm 0.003$&$3.62$& $30\pm\ 6$ & $ 4.1\pm 0.9$ & $3.82$&  $23.75$\\
B94        & 00:58:17 &  -74:36:28  &$0.17\pm 0.04$&$   10\pm  3$&$ \  66\pm    25$& $0.041 \pm 0.002$&$1.08$& $19\pm\ 7$ & $ 4.7\pm 1.9$ & $0.95$&  $22.75$\\
HW38       & 00:59:25 &  -73:49:01  &$0.21\pm 0.04$&$   14\pm  4$&$ \  75\pm    29$& $0.094 \pm 0.003$&$1.65$& $\ 6\pm\ 1$ & $ 1.8\pm 0.1$ & $0.45$&  $22.75$\\
HW44       & 01:01:22 &  -73:47:12  &$0.15\pm 0.03$&$   16\pm  5$&$ \  90\pm    38$& $0.091 \pm 0.003$&$1.35$& $\ 9\pm\ 5$ & $ 1.6\pm 0.3$ & $1.39$&  $22.75$\\
L73        & 01:04:25 &  -70:20:42  &$0.18\pm 0.02$&$   18\pm  3$&$ \  55\pm \   5$& $0.010 \pm 0.001$&$1.07$& $30\pm12$ & $ 8.3\pm 4.6$ & $1.25$&  $22.75$\\
K55        & 01:07:33 &  -73:07:17  &$0.22\pm 0.02$&$   16\pm  3$&$ \  82\pm    16$& $0.051 \pm 0.002$&$1.04$& $22\pm\ 4$ & $ 5.3\pm 1.3$ & $0.58$&  $21.75$\\
HW56       & 01:07:41 &  -70:56:04  &$0.21\pm 0.02$&$   14\pm  2$&$ \  49\pm  \  3$& $0.011 \pm 0.001$&$1.38$& $\ 7\pm\ 2$ & $ 2.3\pm 0.3$ & $0.96$&  $22.25$\\
K57        & 01:08:14 &  -73:15:27  &$0.16\pm 0.02$&$   23\pm  4$&$ \  80\pm    12$& $0.045 \pm 0.002$&$0.55$& $30\pm\ 8$ & $ 4.8\pm 1.4$ & $0.80$&  $21.75$\\
NGC422     & 01:09:25 &  -71:45:59  &$0.31\pm 0.03$&$   17\pm  4$&$ \  40\pm  \  2$& $0.014 \pm 0.001$&$3.09$& $10\pm\ 2$ & $ 2.3\pm 0.2$ & $0.56$&  $20.75$\\
IC1641     & 01:09:40 &  -71:46:03  &$0.17\pm 0.02$&$   11\pm  2$&$ \  58\pm  \  9$& $0.012 \pm 0.001$&$0.55$& $\ 5\pm\ 2$ & $ 1.9\pm 0.2$ & $0.54$&  $20.25$\\
HW67       & 01:13:02 &  -70:57:47  &$0.28\pm 0.04$&$  \ 7\pm  1$&$   100\pm    23$& $0.015 \pm 0.001$&$1.17$& $12\pm\ 3$ & $ 3.5\pm 0.6$ & $1.06$&  $22.75$\\
HW71NW     & 01:15:30 &  -72:22:36  &$0.12\pm 0.03$&$  \ 9\pm  2$&$ \  58\pm    16$& $0.015 \pm 0.001$&$1.65$& $\ 6\pm\ 4$ & $ 2.4\pm 0.6$ & $1.43$&  $21.25$\\
L100       & 01:18:16 &  -72:00:06  &$0.42\pm 0.03$&$   11\pm  1$&$ \  91\pm    10$& $0.018 \pm 0.001$&$1.04$& $19\pm\ 3$ & $ 4.4\pm 0.6$ & $1.23$&  $22.25$\\
HW77       & 01:20:10 &  -72:37:12  &$0.15\pm 0.03$&$   30\pm  5$&$ \  85\pm    11$& $0.041 \pm 0.002$&$0.50$& $35\pm10$ & $ 4.5\pm 1.4$ & $0.83$&  $23.25$\\
IC1708     & 01:24:56 &  -71:11:00  &$0.29\pm 0.02$&$   14\pm  1$&$ \  97\pm  \  8$& $0.007 \pm 0.001$&$0.45$& $15\pm\ 2$ & $ 3.6\pm 0.4$ & $1.46$&  $22.25$\\
B168       & 01:26:43 &  -70:46:48  &$0.20\pm 0.03$&$  \ 8\pm  1$&$ \  99\pm    17$& $0.009 \pm 0.001$&$0.45$& $13\pm\ 3$ & $ 3.6\pm 0.5$ & $0.63$&  $22.75$\\
L106       & 01:30:38 &  -76:03:18  &$0.16\pm 0.01$&$   21\pm  3$&$   102\pm    10$& $0.003 \pm 0.001$&$1.30$& $40\pm\ 7$ & $ 6.3\pm 1.3$ & $1.29$&  $22.25$\\
BS95-187   & 01:31:01 &  -72:50:48  &$0.08\pm 0.02$&$  \ 6\pm  1$&$ \  43\pm  \  5$& $0.002 \pm 0.000$&$0.55$& $\ 6\pm\ 3$ & $ 2.6\pm 0.6$ & $0.59$&  $21.25$\\
L112$^\ast$   & 01:36:01 &  -75:27:30  &$0.14\pm 0.01$&$   17\pm  3$&$ \  81\pm    10$& $0.005 \pm 0.001$&$0.71$& $25\pm\ 5$ & $ 4.6\pm 1.0$ & $1.04$&  $22.75$\\
HW85       & 01:42:28 &  -71:16:48  &$0.21\pm 0.02$&$   12\pm  2$&$ \  63\pm  \  8$& $0.008 \pm 0.001$&$1.01$& $20\pm\ 4$ & $ 5.9\pm 1.5$ & $0.93$&  $23.25$\\
L114       & 01:50:19 &  -74:21:24  &$0.40\pm 0.03$&$  \ 9\pm  1$&$ \  87\pm  \  5$& $0.004 \pm 0.001$&$0.56$& $14\pm\ 1$ & $ 3.6\pm 0.2$ & $0.38$&  $21.75$\\
L116$^\ast$   & 01:55:33 &  -77:39:18  &$0.16\pm 0.01$&$   16\pm  2$&$   109\pm    14$& $0.002 \pm 0.001$&$1.53$& $18\pm\ 4$ & $ 3.3\pm 0.5$ & $2.08$&  $22.75$\\
NGC796     & 01:56:44 &  -74:13:12  &$0.36\pm 0.05$&$  \ 8\pm  1$&$ \  99\pm  \  7$& $0.002 \pm 0.001$&$0.36$& $13\pm\ 1$ & $ 3.5\pm 0.2$ & $0.49$&  $21.75$\\
AM3        & 23:48:59 &  -72:56:42  &$0.11\pm 0.01$&$   11\pm  2$&$ \  66\pm  \  4$& $0.002 \pm 0.001$&$0.23$& $11\pm\ 2$ & $ 3.1\pm 0.3$ & $1.08$&  $22.75$\\
\hline
\end{tabular}

Notes: $^\ast$ SBP $I-$band filter measurements. $^{\ \dagger}$ $r_c$ and $a$ were adopted from the SBP fit; $r_t$ and $\gamma$ from the RDP fit (see discussion in Sects.~4.1 and 4.2). 

\end{table}
\end{landscape}

\begin{landscape}
\begin{table}
\tiny
\caption{LMC clusters' structural parameters from RDPs.} 
\label{tab:sample_LMC_RDPa}
\begin{tabular}{lccccccccccc}
\hline
         &              &               &\multicolumn{5}{c}{\underline{\hspace{2.8cm}King$^{\ \dagger}$\hspace{2.8cm}}}&\multicolumn{3}{c}{\underline{\hspace{1.2cm}EFF$^{\ \dagger}$\hspace{1.2cm}}}\\
 Cluster  &   $\alpha$(J2000) &   $\delta$(J2000) & $\sigma_\circ$ & $r_c$ & $r_t$ & $\sigma_{bg}$ & $\chi^2$ &  $a$ & $\gamma$ & $\chi^2$ &$V_{lim}$ \\
        &    (h:m:s)    &   (\ $^\circ$\ :\ $^\prime$\ :\ $^{\prime\prime}$\ )& (arcsec$^{-2}$) & (arcsec) & (arcsec) &  (arcsec$^{-2}$) &  & (arcsec) & &&(mag)\\
\hline
LW15      & 04:38:26&  -74:27:48&$ 0.22\pm 0.03$&$   10\pm 2$&$    153\pm   39$&$ 0.015 \pm 0.001 $ &1.56& $21\pm 6$ & $ 3.7\pm 0.9$ & $1.94$& 23.25 \\
SL13      & 04:39:42&  -74:01:00&$ 0.14\pm 0.02$&$   22\pm 3$&$   \ 61\pm \  5$&$ 0.008 \pm 0.001 $ &0.83& $14\pm 3$ & $ 2.5\pm 0.3$ & $0.69$& 22.25 \\
SL28      & 04:44:40&  -74:15:36&$ 0.59\pm 0.05$&$   25\pm 2$&$    132\pm   12$&$ 0.031 \pm 0.002 $ &2.02& $37\pm 5$ & $ 4.3\pm 0.6$ & $2.35$& 23.25 \\
SL29      & 04:45:13&  -75:07:00&$ 0.09\pm 0.01$&$   17\pm 2$&$  \  64\pm \  6$&$ 0.005 \pm 0.001 $ &0.39& $20\pm 4$ & $ 4.6\pm 0.8$ & $0.45$& 21.75 \\
SL36      & 04:46:09&  -74:53:18&$ 0.68\pm 0.06$&$ \  9\pm 1$&$  \  79\pm \  9$&$ 0.025 \pm 0.001 $ &1.64& $15\pm 2$ & $ 3.9\pm 0.4$ & $1.18$& 23.75 \\
LW62      & 04:46:18&  -74:09:36&$ 0.21\pm 0.05$&$   20\pm 5$&$ \   53\pm \  7$&$ 0.027 \pm 0.001 $ &0.93& $17\pm 3$ & $ 3.3\pm 0.5$ & $0.50$& 23.25 \\
SL53      & 04:49:54&  -75:37:42&$ 0.15\pm 0.02$&$   24\pm 4$&$   \ 79\pm \  8$&$ 0.011 \pm 0.001 $ &0.99& $43\pm 9$ & $ 7.0\pm 2.1$ & $0.51$& 22.75 \\
SL61      & 04:50:45&  -75:32:00&$ 0.25\pm 0.01$&$   32\pm 2$&$    120\pm\   6$&$ 0.005 \pm 0.001 $ &1.26& $55\pm 9$ & $ 6.4\pm 1.3$ & $2.14$& 21.75 \\
SL74      & 04:52:01&  -74:50:42&$ 0.34\pm 0.02$&$   19\pm 3$&$   \ 95\pm   13$&$ 0.031 \pm 0.002 $ &1.36& $32\pm 9$ & $ 5.5\pm 1.8$ & $2.62$& 23.25 \\
SL80      & 04:52:22&  -74:53:24&$ 0.09\pm 0.02$&$   14\pm 4$&$   \ 53\pm   15$&$ 0.027 \pm 0.001 $ &0.54& $ 6\pm 3$ & $ 2.2\pm 0.4$ & $0.68$& 23.25 \\
OHSC1     & 04:52:41&  -75:16:36&$ 0.14\pm 0.02$&$   14\pm 3$&$  \  86\pm   24$&$ 0.020 \pm 0.001 $ &1.97& $41\pm21$ & $ 8.9\pm 6.8$ & $1.33$& 23.25 \\
SL84      & 04:52:45&  -75:04:30&$ 0.15\pm 0.01$&$   24\pm 2$&$    138\pm   15$&$ 0.010 \pm 0.001 $ &0.96& $25\pm 4$ & $ 3.0\pm 0.4$ & $0.97$& 22.25 \\
KMHK228   & 04:52:56&  -74:00:58&$ 0.06\pm 0.01$&$   15\pm 5$&$  \  66\pm   21$&$ 0.015 \pm 0.001 $ &1.03& $19\pm 8$ & $ 3.8\pm 1.5$ & $0.85$& 22.25 \\
OHSC2     & 04:53:10&  -74:40:54&$ 0.16\pm 0.05$&$  \ 7\pm 2$&$  \  59\pm   20$&$ 0.009 \pm 0.001 $ &1.53& $ 4\pm 1$ & $ 1.8\pm 0.1$ & $0.41$& 21.75 \\
SL118     & 04:55:32&  -74:40:36&$ 0.17\pm 0.03$&$  \ 8\pm 1$&$  \  99\pm   24$&$ 0.009 \pm 0.001 $ &1.04& $13\pm 3$ & $ 4.3\pm 1.0$ & $1.10$& 21.75 \\
KMHK343   & 04:55:55&  -75:08:18&$ 0.26\pm 0.06$&$  \ 6\pm 1$&$   \ 72\pm   14$&$ 0.009 \pm 0.001 $ &0.66& $13\pm 4$ & $ 4.1\pm 1.1$ & $1.65$& 21.25 \\
OHSC3     & 04:56:36&  -75:14:29&$ 0.24\pm 0.08$&$  \ 6\pm 2$&$  \  55\pm   14$&$ 0.011 \pm 0.001 $ &2.51& $12\pm 5$ & $ 4.7\pm 1.8$ & $1.72$& 21.75 \\
OHSC4     & 04:59:14&  -75:07:58&$ 0.10\pm 0.02$&$   15\pm 5$&$   \ 55\pm   10$&$ 0.011 \pm 0.001 $ &2.20& $15\pm 3$ & $ 3.6\pm 0.7$ & $0.64$& 21.25 \\
SL192     & 05:02:27&  -74:51:51&$ 0.23\pm 0.07$&$   36\pm 7$&$   \ 80\pm   10$&$ 0.032 \pm 0.002 $ &1.57& $-      $ & $   -       $ & $-   $& 22.75 \\
LW141     & 05:07:34&  -74:38:06&$ 0.19\pm 0.02$&$   12\pm 2$&$  \  62\pm \  9$&$ 0.021 \pm 0.001 $ &0.54& $15\pm 2$ & $ 3.6\pm 0.4$ & $0.29$& 22.25 \\
SL295     & 05:10:09&  -75:32:36&$ 0.20\pm 0.02$&$   22\pm 3$&$   \ 89\pm   11$&$ 0.012 \pm 0.001 $ &1.05& $25\pm 4$ & $ 4.2\pm 0.7$ & $1.28$& 21.75 \\
SL576     & 05:33:13&  -74:22:08&$ 0.21\pm 0.01$&$   22\pm 2$&$    148\pm   24$&$ 0.017 \pm 0.001 $ &0.91& $29\pm 4$ & $ 3.6\pm 0.5$ & $0.67$& 22.25 \\
IC2148    & 05:39:11&  -75:33:47&$ 0.31\pm 0.03$&$  \ 7\pm 1$&$   \ 87\pm \  7$&$ 0.003 \pm 0.000 $ &0.45& $ 9\pm 1$ & $ 2.8\pm 0.2$ & $0.78$& 20.75 \\
SL647     & 05:39:35&  -75:12:30&$ 0.15\pm 0.05$&$   37\pm 8$&$  \  80\pm   10$&$ 0.010 \pm 0.001 $ &1.33& $40\pm11$ & $ 5.7\pm 1.9$ & $1.06$& 21.75 \\
SL703     & 05:44:54&  -74:50:54&$ 0.18\pm 0.03$&$   31\pm 4$&$  \  89\pm \  8$&$ 0.028 \pm 0.001 $ &0.60& $28\pm 5$ & $ 3.8\pm 0.7$ & $0.69$& 22.75 \\
SL737     & 05:48:44&  -75:44:00&$ 0.22\pm 0.02$&$   13\pm 1$&$    147\pm   24$&$ 0.007 \pm 0.001 $ &1.15& $20\pm 4$ & $ 3.1\pm 0.5$ & $1.69$& 22.25 \\
SL783     & 05:54:39&  -74:36:19&$ 0.27\pm 0.02$&$   20\pm 2$&$  \  95\pm   10$&$ 0.013 \pm 0.001 $ &1.08& $28\pm 5$ & $ 4.4\pm 0.8$ & $1.53$& 21.75 \\
IC2161    & 05:57:25&  -75:08:23&$ 0.37\pm 0.02$&$   18\pm 2$&$    119\pm   12$&$ 0.011 \pm 0.001 $ &1.33& $27\pm 5$ & $ 3.9\pm 0.6$ & $2.26$& 22.25 \\
SL828     & 06:02:13&  -74:11:24&$ 0.27\pm 0.02$&$   15\pm 1$&$    107\pm   11$&$ 0.008 \pm 0.001 $ &1.26& $19\pm 2$ & $ 3.4\pm 0.3$ & $0.80$& 21.25 \\
SL835$^\ast$ & 06:04:48&  -75:06:09&$ 0.24\pm 0.04$&$   10\pm 2$&$  \  78\pm   12$&$ 0.008 \pm 0.001 $ &1.15& $12\pm 2$ & $ 3.1\pm 0.3$ & $0.59$& 22.25 \\
SL882     & 06:19:04&  -72:23:09&$ 0.20\pm 0.02$&$   16\pm 2$&$    103\pm   17$&$ 0.013 \pm 0.001 $ &1.27& $23\pm 5$ & $ 3.9\pm 0.8$ & $1.33$& 22.25 \\
LW458     & 06:19:11&  -67:29:37&$ 0.10\pm 0.03$&$   23\pm 6$&$  \  60\pm \  9$&$ 0.015 \pm 0.001 $ &1.10& $10\pm 2$ & $ 2.2\pm 0.2$ & $0.40$& 22.25 \\
LW460     & 06:19:15&  -71:43:36&$ 0.10\pm 0.04$&$   18\pm 8$&$  \  64\pm   22$&$ 0.028 \pm 0.002 $ &2.33& $ 5\pm 2$ & $ 1.6\pm 0.2$ & $1.35$& 23.25 \\
LW459     & 06:19:17&  -68:19:39&$ 0.25\pm 0.03$&$   12\pm 2$&$  \  63\pm   12$&$ 0.028 \pm 0.001 $ &1.54& $11\pm 1$ & $ 2.6\pm 0.2$ & $0.53$& 22.75 \\
LW462     & 06:19:40&  -72:16:02&$ 0.09\pm 0.02$&$   15\pm 4$&$ \   67\pm   19$&$ 0.012 \pm 0.001 $ &1.18& $ 9\pm 1$ & $ 2.2\pm 0.2$ & $0.35$& 22.25 \\
LW463     & 06:19:46&  -71:18:47&$ 0.10\pm 0.02$&$   15\pm 5$&$  \  77\pm   27$&$ 0.015 \pm 0.001 $ &1.83& $11\pm 2$ & $ 2.2\pm 0.2$ & $0.52$& 22.25 \\
KMHK1732  & 06:19:47&  -69:47:28&$ 0.26\pm 0.07$&$   24\pm 6$&$ \   66\pm   11$&$ 0.028 \pm 0.001 $ &2.56& $12\pm 2$ & $ 2.2\pm 0.2$ & $0.57$& 23.25 \\
SL883     & 06:19:55&  -68:15:09&$ 0.22\pm 0.06$&$   23\pm 6$&$   \ 60\pm \  9$&$ 0.034 \pm 0.001 $ &2.37& $11\pm 3$ & $ 2.1\pm 0.2$ & $0.71$& 23.25 \\
KMHK1739  & 06:21:03&  -71:02:01&$ 0.18\pm 0.03$&$   14\pm 3$&$ \   54\pm \  8$&$ 0.022 \pm 0.001 $ &1.67& $18\pm 5$ & $ 4.5\pm 1.5$ & $1.24$& 22.75 \\
SL886     & 06:21:25&  -69:17:56&$ 0.14\pm 0.04$&$   10\pm 3$&$  \  93\pm   51$&$ 0.027 \pm 0.001 $ &2.73& $-$ & $ -$ & $-$& 22.75 \\
LW469     & 06:21:34&  -72:47:24&$ 0.14\pm 0.01$&$   18\pm 3$&$ \   47\pm \  2$&$ 0.003 \pm 0.001 $ &2.03& $ 8\pm 1$ & $ 2.5\pm 0.2$ & $0.77$& 20.75 \\
LW470     & 06:22:24&  -72:14:14&$ 0.09\pm 0.01$&$   10\pm 2$&$   \ 70\pm   12$&$ 0.005 \pm 0.001 $ &0.65& $27\pm 9$ & $ 7.9\pm 3.4$ & $0.47$& 21.25 \\
NGC2241   & 06:22:53&  -68:55:30&$ 0.33\pm 0.02$&$   16\pm 1$&$ \   92\pm\   7$&$ 0.010 \pm 0.001 $ &1.23& $21\pm 3$ & $ 3.7\pm 0.4$ & $1.51$& 21.75 \\
SL890     & 06:23:03&  -71:41:11&$ 0.09\pm 0.01$&$   17\pm 4$&$   \ 54\pm \  5$&$ 0.016 \pm 0.001 $ &0.99& $17\pm 6$ & $ 4.7\pm 2.0$ & $0.74$& 22.75 \\
LW472     & 06:23:11&  -68:19:08&$ 0.14\pm 0.04$&$ \  8\pm 2$&$  \  54\pm   13$&$ 0.016 \pm 0.001 $ &2.16& $ 3\pm 2$ & $ 1.7\pm 0.2$ & $0.78$& 21.25 \\
LW475     & 06:23:23&  -70:33:14&$ 0.12\pm 0.02$&$  \ 9\pm 1$&$   \ 83\pm   14$&$ 0.005 \pm 0.001 $ &0.63& $14\pm 2$ & $ 3.7\pm 0.4$ & $0.32$& 21.25 \\
SL889     & 06:23:29&  -68:59:50&$ 0.10\pm 0.03$&$ \  7\pm 2$&$  \  48\pm  \ 9$&$ 0.005 \pm 0.001 $ &0.97& $-$ & $ -$ & $-$& 21.25 \\
SL891     & 06:24:49&  -71:39:32&$ 0.27\pm 0.02$&$   12\pm 1$&$    126\pm   18$&$ 0.012 \pm 0.001 $ &0.94& $15\pm 3$ & $ 3.0\pm 0.4$ & $1.63$& 22.25 \\
SL892     & 06:25:15&  -71:06:08&$ 0.17\pm 0.03$&$   13\pm 3$&$   \ 65\pm   16$&$ 0.011 \pm 0.001 $ &2.19& $ 9\pm 2$ & $ 2.5\pm 0.3$ & $1.03$& 22.25 \\
OHSC36    & 06:29:41&  -70:35:24&$ 0.09\pm 0.01$&$   19\pm 3$&$  \  98\pm   18$&$ 0.009 \pm 0.001 $ &0.94& $18\pm 5$ & $ 2.9\pm 0.5$ & $1.01$& 22.25 \\
SL897     & 06:33:01&  -71:07:40&$ 0.19\pm 0.01$&$   24\pm 3$&$  \  98\pm  \ 8$&$ 0.007 \pm 0.001 $ &1.19& $45\pm 8$ & $ 6.5\pm 1.4$ & $0.73$& 22.25 \\
\hline
\end{tabular}

Notes: $^\ast$ SBP $I-$band filter measurements. $^{\ \dagger}$ $r_c$ and $a$ were adopted from the SBP fit; $r_t$ and $\gamma$ from the RDP fit (see discussion in Sects.~4.1 and 4.2).

\end{table}
\end{landscape}

\begin{landscape}
\begin{table}
\tiny
\caption{SMC clusters' structural parameters from SBPs.}
\label{tab:sample_SMC_SBPa}
\begin{tabular}{lccccccccccc}
\hline
         &              &               &\multicolumn{5}{c}{\underline{\hspace{3.2cm}King$^{\ \dagger}$\hspace{3.2cm}}}&\multicolumn{3}{c}{\underline{\hspace{1.1cm}EFF$^{\ \dagger}$\hspace{1.3cm}}}\\
 Cluster  &   $\alpha$(J2000) &   $\delta$(J2000) & $\mu_{{\rm v},\circ}$ & $r_c$ & $r_t$ & $\mu_{{\rm v},bg}$ & $\chi^2$ & $a$ & $\gamma$ & $\chi^2$ &$V_{int}$\\
        &    (h:m:s)    &   (\ $^\circ$\ :\ $^\prime$\ :\ $^{\prime\prime}$\ )&(mag.arcsec$^{-2}$) & (arcsec) & (arcsec)&(mag.arcsec$^{-2}$)&& (arcsec)&&&(mag)\\
\hline
B1         & 00:19:20 &  -74:06:24  &$   21.97\pm0.41$ &  $\ 4.2\pm1.6$ & $-$                          &  $26.64\pm0.06$ &    $0.29$   & $ 2\pm 1$ & $ 1.8\pm 0.2$ & $0.10$ &$15.51\pm     0.16$\\
K6         & 00:25:27 &  -74:04:30  &$   21.18\pm     0.18$&$ \    8.2\pm     1.8$&$         -       $&$  26.81\pm     0.08$&$    0.14$& $ 9\pm 3$ & $ 2.1\pm 0.4$ & $0.12$ &$13.90\pm     0.06$\\
K7         & 00:27:45 &  -72:46:53  &$   22.00\pm     0.11$&$    20.4\pm     2.7$&$   192\pm    70$&$  26.81\pm     0.08$&$    0.13$   & $24\pm 5$ & $ 2.9\pm 0.5$ & $0.12$ &$13.20\pm     0.06$\\
K9         & 00:30:00 &  -73:22:40  &$   22.86\pm     0.19$&$    23.3\pm     3.6$&$         -       $&$  26.70\pm     0.07$&$    0.19$ & $10\pm 7$ & $ 1.0\pm 0.4$ & $0.12$ &$13.09\pm     0.12$\\
HW5        & 00:31:01 &  -72:20:30  &$   19.95\pm     0.21$&$   \  2.3\pm     0.4$&$         -       $&$  26.52\pm     0.07$&$    0.90$& $ 2\pm 0$ & $ 1.7\pm 0.1$ & $0.16$ &$14.35\pm     0.07$\\
HW20       & 00:44:48 &  -74:21:47  &$   22.55\pm     0.32$&$    10.7\pm     2.3$&$         -       $&$  26.59\pm     0.07$&$    0.76$ & $ 7\pm 3$ & $ 1.2\pm 0.2$ & $0.08$ &$14.16\pm     0.07$\\
L32        & 00:47:24 &  -68:55:12  &$   22.63\pm     0.17$&$    12.0\pm     2.5$&$ \   96\pm    23$&$  26.79\pm     0.06$&$    0.15$  & $17\pm 5$ & $ 3.5\pm 0.8$ & $0.14$ &$14.94\pm     0.06$\\
HW33       & 00:57:23 &  -70:48:36  &$   22.42\pm     0.47$&$   \  5.4\pm     1.5$&$        -       $&$  26.78\pm     0.06$&$    0.32$ & $-$ & $ -$ & $ -$                  &$15.22\pm     0.17$\\
K37        & 00:57:49 &  -74:19:32  &$   20.80\pm     0.20$&$   \  9.2\pm     2.2$&$         -       $&$  26.68\pm     0.09$&$    0.18$& $ 9\pm 3$ & $ 1.9\pm 0.5$ & $0.15$ &$13.73\pm     0.06$\\
B94        & 00:58:17 &  -74:36:28  &$   22.75\pm     0.24$&$   \  8.1\pm     3.0$&$         -       $&$  26.65\pm     0.07$&$    0.19$& $ 4\pm 2$ & $ 1.3\pm 0.2$ & $0.10$ &$15.36\pm     0.13$\\
HW38       & 00:59:25 &  -73:49:01  &$   22.49\pm     0.16$&$    29.5\pm     4.2$&$         -       $&$  26.37\pm     0.09$&$    0.60$ & $ 3\pm 2$ & $ 0.6\pm 0.1$ & $0.17$ &$12.85\pm     0.05$\\
HW44       & 01:01:22 &  -73:47:12  &$   22.14\pm     0.77$&$    10.4\pm     9.9$&$         -       $&$  26.49\pm     0.08$&$    0.72$ & $-$ & $ -$ & $ -$                  &$14.56\pm     0.43$\\
L73        & 01:04:25 &  -70:20:42  &$   23.56\pm     0.38$&$    20.2\pm     6.1$&$  \  50\pm \    6$&$  26.82\pm     0.05$&$    0.24$ & $-$ & $ -$ & $ -$                  &$14.53\pm     0.11$\\
K55        & 01:07:33 &  -73:07:17  &$   21.20\pm     0.58$&$    10.5\pm     3.5$&$         -       $&$  26.63\pm     0.14$&$    1.99$ & $-$ & $ -$ & $ -$                  &$12.73\pm     0.04$\\
HW56       & 01:07:41 &  -70:56:04  &$   22.24\pm     0.16$&$  \   6.1\pm     1.3$&$         -       $&$  26.22\pm     0.04$&$    0.06$& $ 7\pm 2$ & $ 2.2\pm 0.3$ & $0.04$ &$15.95\pm     0.22$\\
K57        & 01:08:14 &  -73:15:27  &$   21.63\pm     0.27$&$    12.7\pm     4.3$&$         -       $&$  26.59\pm     0.12$&$    0.36$ & $11\pm 4$ & $ 1.7\pm 0.4$ & $0.19$ &$13.25\pm     0.07$\\
NGC422     & 01:09:25 &  -71:45:59  &$   19.80\pm     0.28$&$    \ 6.5\pm     2.1$&$  \  51\pm    15$&$  26.68\pm     0.10$&$    0.31$ & $10\pm 4$ & $ 3.8\pm 1.2$ & $0.26$ &$13.36\pm     0.08$\\
IC1641     & 01:09:40 &  -71:46:03  &$   21.19\pm     0.33$&$    \ 6.3\pm     2.5$&$         -       $&$  26.68\pm     0.10$&$    0.24$& $11\pm 6$ & $ 2.7\pm 1.4$ & $0.15$ &$14.48\pm     0.19$\\
HW67       & 01:13:02 &  -70:57:47  &$   22.27\pm     0.45$&$   \  6.4\pm     2.1$&$         -       $&$  26.78\pm     0.06$&$    0.21$& $ 4\pm 2$ & $ 1.7\pm 0.2$ & $0.16$ &$14.71\pm     0.34$\\
HW71NW     & 01:15:30 &  -72:22:36  &$   21.61\pm     0.26$&$    10.2\pm     3.6$&$  \  55\pm    25$&$  26.84\pm     0.08$&$    0.14$  & $14\pm 7$ & $ 3.9\pm 2.0$ & $0.12$ &$13.91\pm     0.58$\\
L100       & 01:18:16 &  -72:00:06  &$   20.61\pm     0.28$&$  \   5.7\pm     1.2$&$   103\pm    35$&$  26.81\pm     0.06$&$    0.28$  & $ 8\pm 2$ & $ 2.8\pm 0.4$ & $0.26$ &$13.96\pm     0.10$\\
HW77       & 01:20:10 &  -72:37:12  &$   23.32\pm     0.20$&$    18.6\pm     2.8$&$         -       $&$  26.68\pm     0.06$&$    0.28$ & $-$ & $ -$ & $ -$                  &$14.40\pm     0.06$\\
IC1708     & 01:24:56 &  -71:11:00  &$   20.72\pm     0.13$&$   \  5.9\pm     0.7$&$ \   93\pm    17$&$  26.37\pm     0.06$&$    0.06$ & $ 7\pm 1$ & $ 2.5\pm 0.2$ & $0.08$ &$14.14\pm     0.05$\\
B168       & 01:26:43 &  -70:46:48  &$   22.50\pm     0.27$&$    \ 5.2\pm     1.4$&$         -       $&$  26.76\pm     0.05$&$    0.27$& $ 5\pm 2$ & $ 2.1\pm 0.3$ & $0.25$ &$14.94\pm     0.23$\\
L106       & 01:30:38 &  -76:03:18  &$   21.83\pm     0.13$&$    10.3\pm     1.3$&$   270\pm   110$&$  26.52\pm     0.06$&$    0.08$   & $11\pm 2$ & $ 2.3\pm 0.2$ & $0.09$ &$13.81\pm     0.09$\\
BS95-187   & 01:31:01 &  -72:50:48  &$   23.88\pm     0.62$&$    14.8\pm     8.4$&$   \ 42\pm    14$&$  26.96\pm     0.06$&$    0.28$  & $-$ & $ -$ & $ -$                  &$16.23\pm     0.31$\\
L112$^\ast$   & 01:36:01 &  -75:27:30&$   23.04\pm     0.29$&$   \  3.6\pm     0.8$&$         -       $&$  25.49\pm     0.03$&$    0.42$& $-$ & $ -$ & $ -$                  &$18.19\pm     1.04$\\ 
HW85       & 01:42:28 &  -71:16:48  &$   21.74\pm     0.34$&$   \  3.9\pm     1.1$&$         -       $&$  26.57\pm     0.05$&$    0.30$& $ 4\pm 2$ & $ 2.1\pm 0.4$ & $0.28$ &$14.68\pm     0.21$\\
L114       & 01:50:19 &  -74:21:24  &$   18.19\pm     0.79$&$   \  2.6\pm     1.1$&$  \  83\pm    23$&$  26.14\pm     0.09$&$    0.14$ & $ 7\pm 2$ & $ 2.9\pm 0.4$ & $0.12$ &$11.93\pm     0.14$\\
L116$^\ast$   & 01:55:33 &  -77:39:18&$   21.87\pm     0.40$&$    \ 5.5\pm     2.2$&$         -       $&$  25.23\pm     0.02$&$    0.34$& $-$ & $ -$ & $ -$                  &$15.06\pm     0.71$\\ 
NGC796     & 01:56:44 &  -74:13:12  &$   18.44\pm     0.31$&$   \  2.8\pm     0.5$&$   128\pm    18$&$  26.31\pm     0.08$&$    0.16$  & $ 5\pm 1$ & $ 2.7\pm 0.1$ & $0.13$ &$12.21\pm     0.09$\\
AM3        & 23:48:59 &  -72:56:42  &$   22.75\pm     0.29$&$  \   4.6\pm     1.6$&$   \ 54\pm    44$&$  26.36\pm     0.04$&$    0.18$ & $ 5\pm 2$ & $ 2.7\pm 0.9$ & $0.17$ &$16.75\pm     0.66$\\
\hline
\end{tabular}

Notes: $^\ast$ SBP $I-$band filter measurements. $^{\ \dagger}$ $r_c$ was adopted from the SBP fit and $r_t$ from the RDP fit (see discussion in Sects.~4.1 and 4.2). 

\end{table}
\end{landscape}

\begin{landscape}
\begin{table}
\tiny
\caption{LMC clusters' structural parameters from SBPs.} 
\label{tab:sample_LMC_SBPa}
\begin{tabular}{lccccccccccc}
\hline
         &              &               &\multicolumn{5}{c}{\underline{\hspace{3.2cm}King$^{\ \dagger}$\hspace{3.2cm}}}&\multicolumn{3}{c}{\underline{\hspace{1.1cm}EFF$^{\ \dagger}$\hspace{1.3cm}}}\\
 Cluster  &   $\alpha$(J2000) &   $\delta$(J2000) & $\mu_{{\rm v},\circ}$ & $r_c$ & $r_t$ & $\mu_{{\rm v},bg}$ & $\chi^2$ & $a$ & $\gamma$ & $\chi^2$ &$V_{int}$\\
        &    (h:m:s)    &   (\ $^\circ$\ :\ $^\prime$\ :\ $^{\prime\prime}$\ )&(mag.arcsec$^{-2}$) & (arcsec) & (arcsec)&(mag.arcsec$^{-2}$)&& (arcsec)&&&(mag)\\
\hline
LW15      & 04:38:26&  -74:27:48&$  22.78\pm 0.56$&$   17.6\pm     7.8$&$  \  44\pm    11$&$   26.91\pm     0.06$&     0.09 &               $ 9\pm 6$ & $ 2.3\pm 1.0$ & $0.15$&$   14.79\pm     0.16$ \\      
SL13      & 04:39:42&  -74:01:00&$  23.50\pm 0.30$&$   16.0\pm     5.4$&$     -$&$   26.88\pm     0.07$&     0.14 &                         $19\pm10$ & $ 2.6\pm 1.3$ & $0.10$&$   14.84\pm     0.08$ \\      
SL28      & 04:44:40&  -74:15:36&$    21.35\pm     0.59$&$   21.0\pm     5.8$&$  \  68\pm    32$&$   26.92\pm     0.08$&     0.07 &         $28\pm15$ & $ 4.9\pm 3.9$ & $0.06$&$   13.62\pm     0.05$ \\      
SL29      & 04:45:13&  -75:07:00&$\ 23\pm\ 2$&$\ 28\pm 22$&$ \ 53\pm 33$&$ 26.80\pm 0.10$&    $0.13$ &                                      $-$ & $ -$ & $ -$                 &$   15.28\pm     0.18$ \\  
SL36      & 04:46:09&  -74:53:18&$    20.18\pm 0.10$&$   \   3.8\pm     0.4$&$  \  51\pm  \   7$&$   26.94\pm     0.07$&     0.05 &         $ 5\pm 1$ & $ 3.0\pm 0.2$ & $0.05$&$   14.80\pm     0.55$ \\      
LW62      & 04:46:18&  -74:09:36&$  22.48\pm 0.54$&$  \  5.2\pm     3.1$&$     -$&$   26.84\pm     0.08$&     0.40 &                        $ 3\pm 2$ & $ 1.4\pm 0.3$ & $0.08$&$   15.96\pm     0.16$ \\     
SL53      & 04:49:54&  -75:37:42&$ 23.61\pm 0.49$&$     12.5\pm     3.8$&$     -$&$   26.17\pm     0.06$&     0.41 &                        $ 8\pm 6$ & $ 1.2\pm 0.5$ & $0.13$&$   14.88\pm     0.09$ \\     
SL61      & 04:50:45&  -75:32:00&$    22.09\pm     0.17$&$   25.3\pm     4.9$&$   164\pm    64$&$   26.38\pm     0.05$&     0.22 &          $30\pm 9$ & $ 3.2\pm 0.9$ & $0.21$&$   12.94\pm     0.06$ \\      
SL74      & 04:52:01&  -74:50:42&$    21.70\pm 0.18$&$\  9.2\pm     1.8$&$     -$&$   26.61\pm     0.06$&     0.17 &                        $ 9\pm 2$ & $ 1.8\pm 0.3$ & $0.08$&$   13.18\pm     0.16$ \\     
SL80      & 04:52:22&  -74:53:24&$    24.03\pm 0.54$&$  17.2\pm     5.8$&$     -$&$   26.56\pm     0.06$&     0.70 &                        $-$ & $ -$ & $ -$                 &$   14.69\pm     1.59$ \\     
OHSC1     & 04:52:41&  -75:16:36&$    23.46\pm 0.44$&$ 12.4\pm     5.1$&$     -$&$   26.53\pm     0.05$&     0.27 &                         $10\pm 6$ & $ 2.2\pm 1.3$ & $0.07$&$   15.91\pm     0.08$ \\     
SL84      & 04:52:45&  -75:04:30&$    21.18\pm 0.27$&$ \ 7.6\pm     1.7$&$     -$&$   26.86\pm     0.07$&     0.51 &                        $ 4\pm 1$ & $ 1.4\pm 0.1$ & $0.19$&$   13.24\pm     0.09$ \\      
KMHK228   & 04:52:56&  -74:00:58&$    23.77\pm 0.40$&$ \ 17\pm    11$&$     -$&$   26.86\pm     0.07$&     0.63 &                           $-$ & $ -$ & $ -$                 &$   16.11\pm     0.99$ \\  
OHSC2     & 04:53:10&  -74:40:54&$    21.84\pm 0.14$&$ \ 5.9\pm     1.2$&$     -$&$   26.87\pm     0.07$&     0.03 &                        $ 6\pm 2$ & $ 2.3\pm 0.5$ & $0.02$&$   15.91\pm     0.16$ \\     
SL118     & 04:55:32&  -74:40:36&$    22.00\pm 0.29$&$ \ 7.6\pm     1.4$&$     -$&$   26.66\pm     0.06$&     0.36 &                        $12\pm 4$ & $ 3.0\pm 0.7$ & $0.16$&$   14.69\pm     0.08$ \\      
KMHK343   & 04:55:55&  -75:08:18&$    21.91\pm 0.25$&$ \ 9.9\pm     3.6$&$  \  36\pm   \  8$&$   26.75\pm     0.07$&     0.13 &             $18\pm10$ & $ 6.8\pm 4.8$ & $0.10$&$   15.21\pm     0.11$ \\      
OHSC3     & 04:56:36&  -75:14:29&$    19.43\pm 0.70$&$  \ 2.3\pm     1.1$&$  \  47\pm    18$&$   26.77\pm     0.06$&     0.53 &             $ 3\pm 1$ & $ 2.6\pm 0.4$ & $0.22$&$   15.03\pm     1.16$ \\      
OHSC4     & 04:59:14&  -75:07:58&$    22.97\pm 0.31$&$ 10.1\pm     5.4$&$     -$&$   26.68\pm     0.06$&     0.56 &                         $ 4\pm 3$ & $ 1.0\pm 0.2$ & $0.21$&$   13.25\pm     0.23$ \\     
SL192     & 05:02:27&  -74:51:51&$    23.36\pm 0.29$&$ 22.7\pm     6.4$&$     -$&$   26.46\pm     0.06$&     0.36 &                         $ -$ & $ -$ & $-$&$   14.48\pm     0.52$ \\     
LW141     & 05:07:34&  -74:38:06&$    22.38\pm 0.23$&$ \ 9.8\pm     2.7$&$  \  61\pm    23$&$   26.75\pm     0.07$&     0.30 &              $14\pm 6$ & $ 4.0\pm 1.6$ & $0.27$&$   15.07\pm     0.13$ \\      
SL295     & 05:10:09&  -75:32:36&$    21.58\pm 0.23$&$ 11.6\pm     3.4$&$  \  77\pm    46$&$   26.70\pm     0.06$&     0.14 &               $14\pm 6$ & $ 3.2\pm 1.3$ & $0.12$&$   14.38\pm     0.18$ \\      
SL576     & 05:33:13&  -74:22:08&$    19.97\pm 0.15$&$ \ 9.7\pm     1.5$&$  \  39\pm   \  6$&$   26.49\pm     0.26$&     0.08 &             $15\pm 3$ & $ 5.3\pm 1.4$ & $0.05$&$   11.88\pm     0.25$ \\      
IC2148    & 05:39:11&  -75:33:47&$    20.64\pm 0.17$&$ \ 6.0\pm     1.4$&$  \  42\pm    11$&$   26.47\pm     0.06$&     0.15 &              $ 9\pm 3$ & $ 3.8\pm 1.0$ & $0.13$&$   14.35\pm     0.11$ \\      
SL647     & 05:39:35&  -75:12:30&$    22.42\pm 0.22$&$   \ 9.2\pm     2.7$&$     -$&$   26.62\pm     0.06$&     0.13 &                      $ 5\pm 2$ & $ 1.4\pm 0.2$ & $0.10$&$   14.55\pm     0.09$ \\     
SL703     & 05:44:54&  -74:50:54&$    23.02\pm 0.29$&$ 23.1\pm     6.3$&$   125\pm    64$&$   26.80\pm     0.06$&     0.10 &                $-$ & $ -$ & $ -$                 &$   14.09\pm     0.46$ \\      
SL737     & 05:48:44&  -75:44:00&$    20.96\pm  0.38$&$ \ 4.8\pm     1.4$&$  \  90\pm    48$&$   26.73\pm     0.06$&     0.15 &             $ 6\pm 2$ & $ 2.7\pm 0.5$ & $0.14$&$   14.34\pm     0.40$ \\      
SL783     & 05:54:39&  -74:36:19&$    20.87\pm 0.12$&$  \ 9.7\pm     1.4$&$   119\pm    43$&$   26.72\pm     0.07$&     0.17 &              $12\pm 2$ & $ 2.7\pm 0.4$ & $0.17$&$   13.71\pm     0.04$ \\      
IC2161    & 05:57:25&  -75:08:23&$    21.12\pm 0.33$&$ 13.9\pm     3.4$&$  \  38\pm   \  6$&$   26.73\pm     0.07$&     0.13 &              $13\pm 4$ & $ 4.2\pm 1.3$ & $0.10$&$   14.16\pm     0.23$ \\      
SL828     & 06:02:13&  -74:11:24&$    20.74\pm 0.09$&$ 13.7\pm     1.6$&$  \  44\pm  \   3$&$   26.74\pm     0.08$&     0.07 &              $26\pm 5$ & $ 8.3\pm 2.3$ & $0.05$&$   13.65\pm     0.04$ \\      
SL835$^\ast$ & 06:04:48&  -75:06:09&$    20.50\pm 0.30$&$ \ 6.1\pm     2.5$&$  \  24\pm  \   8$&$   25.00\pm     0.02$&     0.17 &             $-$ & $ -$ & $ -                 $&$   14.83\pm   0.16$  \\    
SL882     & 06:19:04&  -72:23:09&$    21.22\pm     0.39$&$    \ 5.2\pm     2.4$&$     -$&$   26.75\pm     0.06$&     0.71 &                 $ 2\pm 2$ & $ 1.3\pm 0.3$ & $0.32$&$   14.37\pm     0.08$ \\     
LW458     & 06:19:11&  -67:29:37&$     -$&$     -$&$     -$&$    -$&  $-$ &                                                                 $-$ & $ -$ & $ -$                 &$    -               $ \\  
LW460     & 06:19:15&  -71:43:36&$    22.64\pm  0.52$&$  \ 9.9\pm     6.5$&$  \  57\pm    47$&$   26.61\pm     0.06$&     0.35 &            $-$ & $ -$ & $ -$                 &$   15.72\pm     0.32$ \\      
LW459     & 06:19:17&  -68:19:39&$    22.54\pm     0.46$&$     \  8.1\pm     4.1$&$     -$&$   26.62\pm     0.06$&     0.39 &               $ 9\pm 6$ & $ 2.1\pm 1.3$ & $0.34$&$   15.69\pm     0.15$ \\     
LW462     & 06:19:40&  -72:16:02&$     -$&$     -$&$     -$&$    -$&    $-$ &                                                               $-$ & $ -$ & $ -$                 &$     -              $ \\  
LW463     & 06:19:46&  -71:18:47&$  -$&$     -$&$    -$&$    -$&     $-$ &                                $-$ & $ -$ & $ -$                 &$   -$ \\    
KMHK1732  & 06:19:47&  -69:47:28&$    22.09\pm     0.33$&$   11.0\pm    5.5$&$  \  66\pm    43$&$   26.78\pm     0.07$&     0.32 &          $10\pm 6$ & $ 2.6\pm 1.1$ & $0.30$&$   14.45\pm     0.15$ \\      
SL883     & 06:19:55&  -68:15:09&$ 21.78\pm 0.44$&$ \    7.0\pm     3.9$&$     -$&$   26.69\pm     0.07$&     0.55 &                        $ 4\pm 4$ & $ 1.4\pm 0.6$ & $0.36$&$   14.65\pm     0.47$ \\     
KMHK1739  & 06:21:03&  -71:02:01&$  22.23\pm 0.33$&$   12.1\pm     4.5$&$ \   50\pm    18$&$   26.50\pm     0.05$&     0.08 &               $19\pm11$ & $ 5.2\pm 3.8$ & $0.07$&$   13.71\pm     0.28$ \\      
SL886     & 06:21:25&  -69:17:56&$    21.81\pm 0.41$&$ \ 9.3\pm     2.7$&$     -$&$   26.76\pm     0.10$&     0.33 &                        $-$ & $ -$ & $ -$                 &$   14.65\pm     0.37$ \\     
LW469     & 06:21:34&  -72:47:24&$ 20.04\pm     0.20$&$ \    4.1\pm     0.9$&$     -$&$   26.51\pm     0.06$&     0.20 &                    $ 4\pm 1$ & $ 2.0\pm 0.3$ & $0.17$&$   13.92\pm     0.10$ \\     
LW470     & 06:22:24&  -72:14:14&$    22.87\pm 0.24$&$ 12.0\pm     3.4$&$     -$&$   26.80\pm     0.06$&     0.25 &                         $14\pm 6$ & $ 2.4\pm 0.8$ & $0.23$&$   14.52\pm     0.47$ \\     
NGC2241   & 06:22:53&  -68:55:30&$    19.62\pm     0.14$&$    \ 8.5\pm     1.5$&$  \  44\pm\     5$&$   26.36\pm     0.06$&     0.19 &      $14\pm 3$ & $ 5.1\pm 1.0$ & $0.14$&$   13.18\pm     0.06$ \\      
SL890     & 06:23:03&  -71:41:11&$    22.30\pm 0.27$&$ \ 7.1\pm     2.5$&$     -$&$   26.60\pm     0.05$&     0.09 &                        $ 8\pm 3$ & $ 2.5\pm 0.7$ & $0.06$&$   15.22\pm     0.58$ \\     
LW472     & 06:23:11&  -68:19:08&$    21.44\pm 0.18$&$ \ 4.3\pm     1.1$&$     -$&$   26.62\pm     0.05$&     0.14 &                        $ 4\pm 2$ & $ 2.0\pm 0.4$ & $0.10$&$   15.13\pm     0.11$ \\      
LW475     & 06:23:23&  -70:33:14&$    22.28\pm 0.30$&$ \ 9.3\pm     2.3$&$     -$&$   26.59\pm     0.05$&     0.25 &                        $10\pm 5$ & $ 2.6\pm 1.3$ & $0.12$&$   15.04\pm     0.28$ \\      
SL889     & 06:23:29&  -68:59:50&$ 22.41\pm 0.16$&$  \   8.8\pm     1.7$&$  \  42\pm \    9$&$   26.57\pm     0.06$&     0.04 &             $13\pm 4$ & $ 4.5\pm 1.4$ & $0.04$&$   15.72\pm     0.78$ \\      
SL891     & 06:24:49&  -71:39:32&$    20.27\pm     0.14$&$  \  5.5\pm     0.8$&$     -$&$   26.62\pm     0.06$&     0.10 &                  $ 5\pm 1$ & $ 2.1\pm 0.3$ & $0.09$&$   14.14\pm     0.19$ \\     
SL892     & 06:25:15&  -71:06:08&$    19.64\pm 0.25$&$   \ 3.0\pm     0.5$&$  \  59\pm    16$&$   26.48\pm     0.05$&     0.06 &            $ 4\pm 1$ & $ 2.6\pm 0.3$ & $0.07$&$   14.54\pm     0.22$ \\      
OHSC36    & 06:29:41&  -70:35:24&$ 21.96\pm  0.42$&$ \    5.3\pm     2.0$&$     -$&$   26.48\pm     0.05$&     0.16 &                       $ 6\pm 3$ & $ 2.3\pm 0.6$ & $0.14$&$   15.24\pm     0.13$ \\     
SL897     & 06:33:01&  -71:07:40&$  21.19\pm  0.23$&$   11.8\pm     3.2$&$  \  74\pm    40$&$   26.27\pm     0.04$&     0.14 &              $14\pm 5$ & $ 3.1\pm 1.2$ & $0.13$&$   13.81\pm     0.28$ \\      
\hline
\end{tabular}

Notes: $^\ast$ SBP $I-$band filter measurements. $^{\ \dagger}$ $r_c$ and $a$ were adopted from the SBP fit; $r_t$ and $\gamma$ from the RDP fit (see discussion in Sects.~4.1 and 4.2). 

\end{table}
\end{landscape}




\section{Additional figures}


\begin{figure*}

\includegraphics[width=0.325\linewidth]{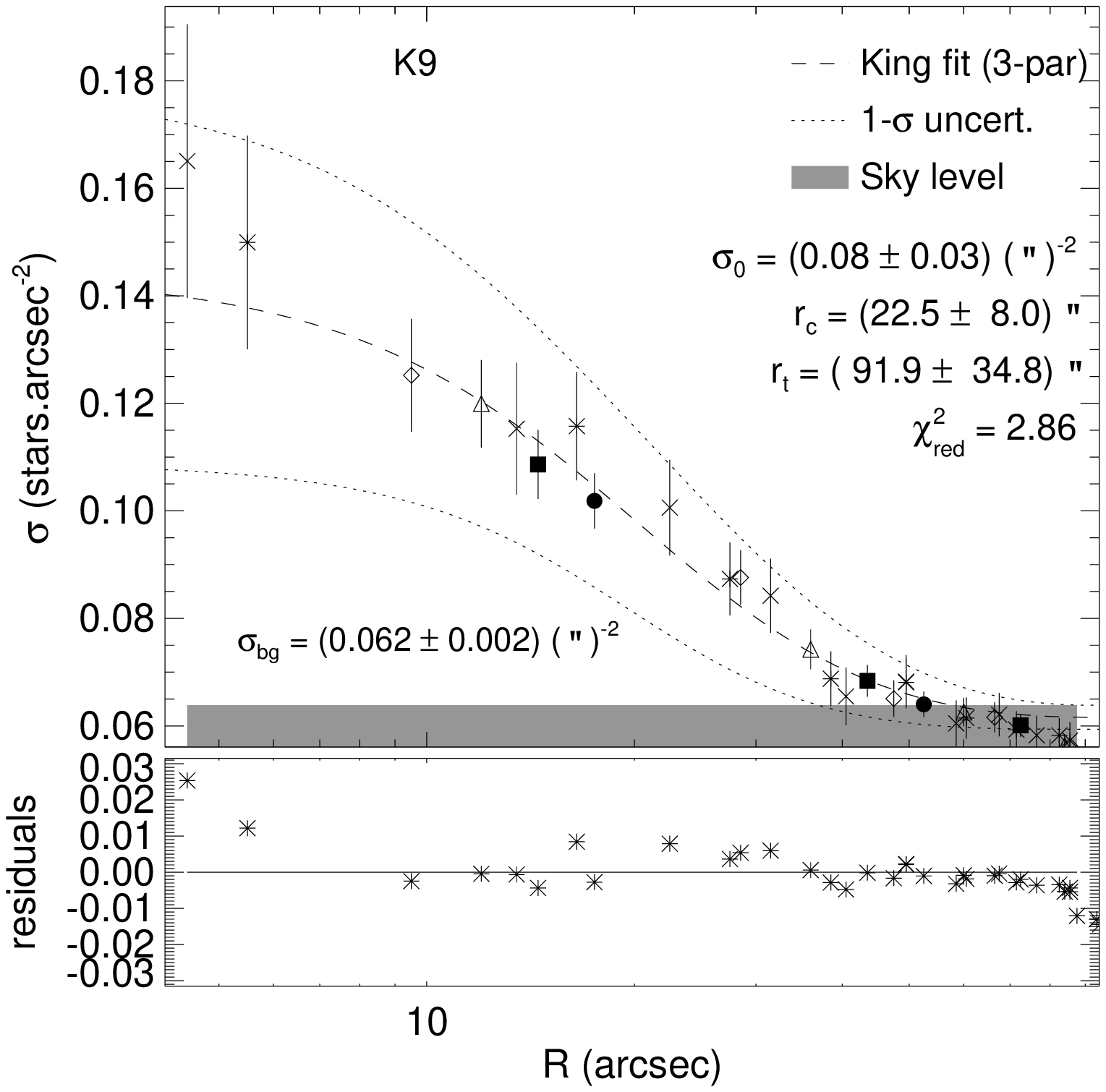}\includegraphics[width=0.325\linewidth]{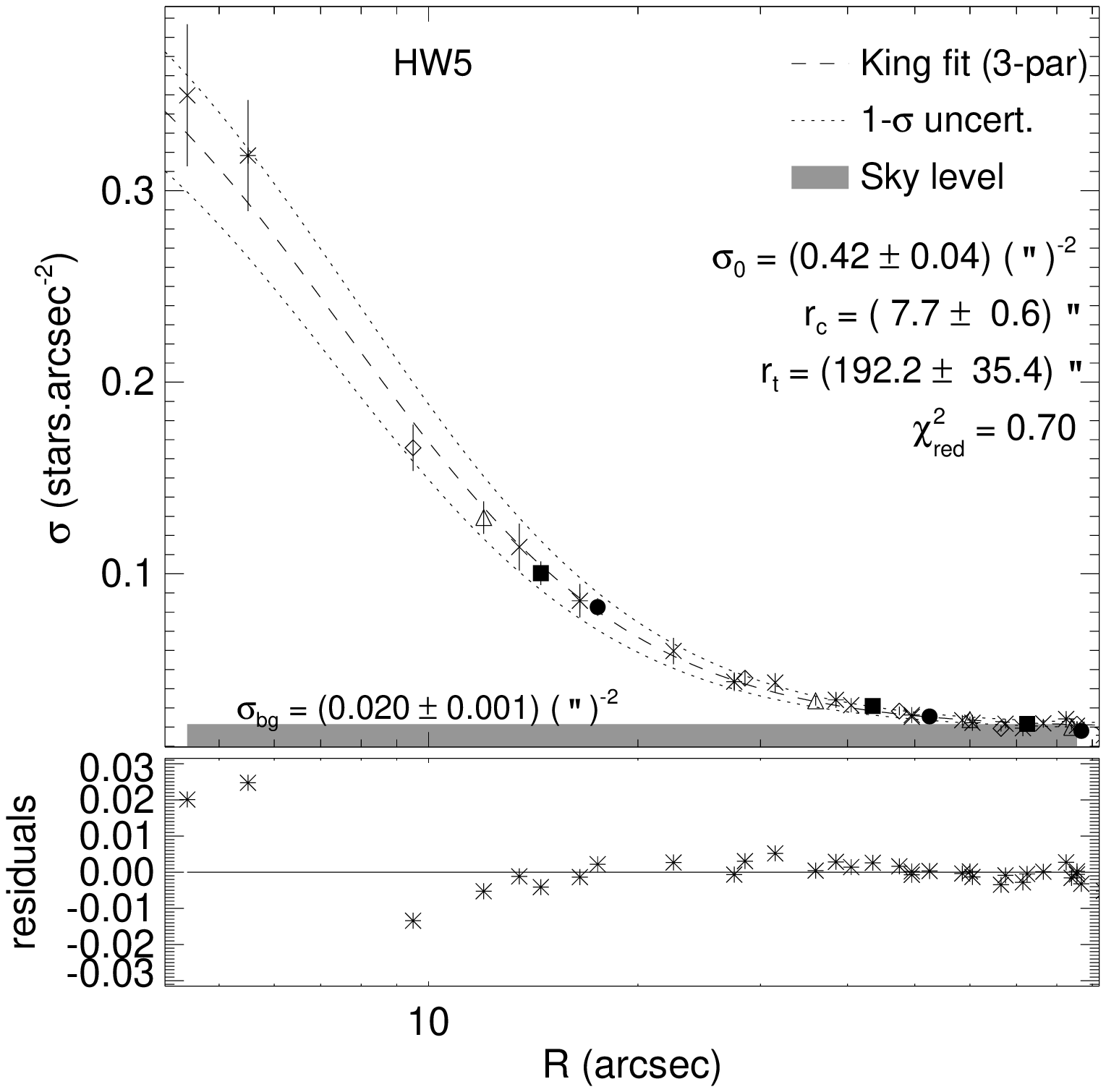}\includegraphics[width=0.325\linewidth]{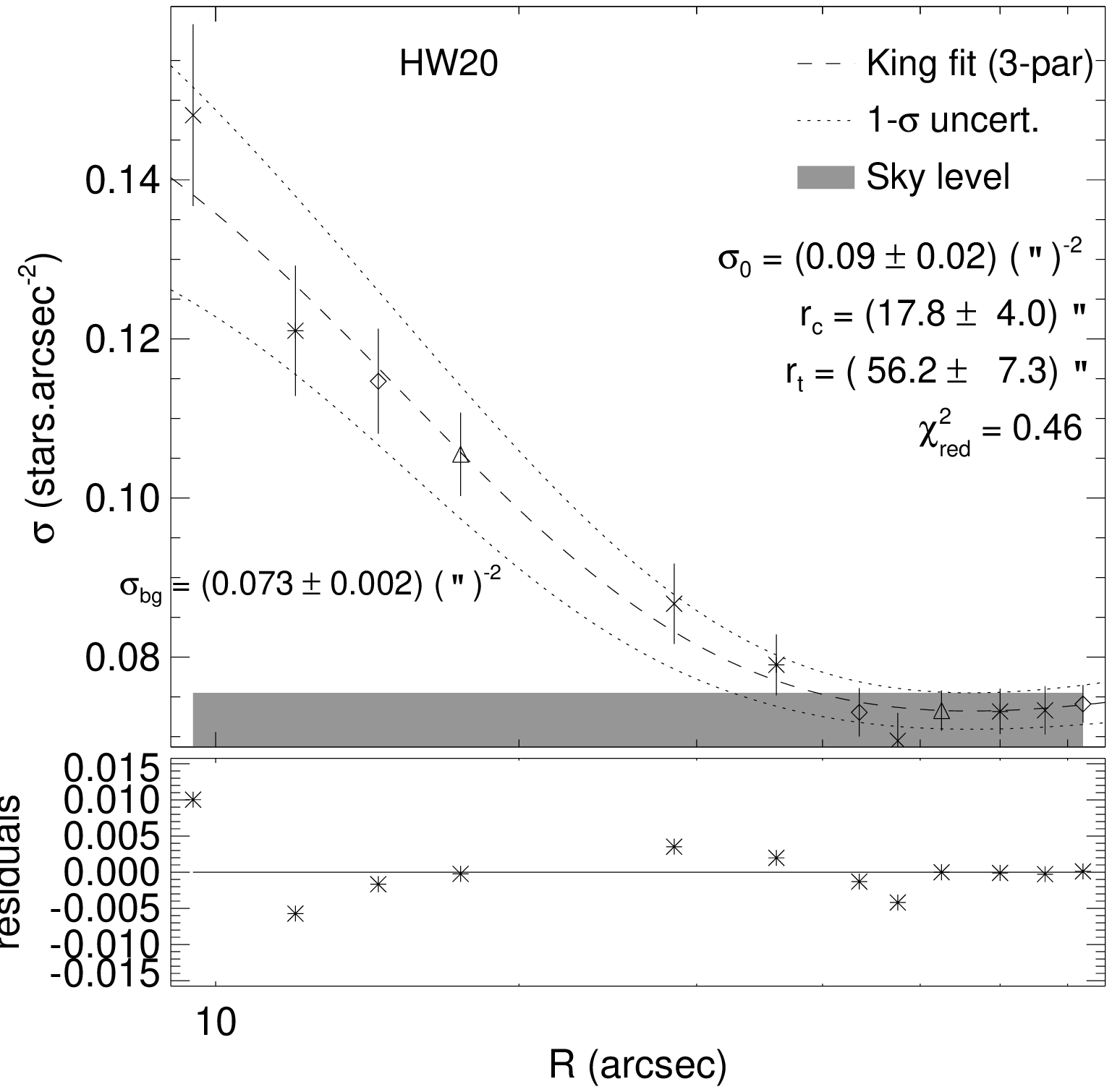}

\includegraphics[width=0.325\linewidth]{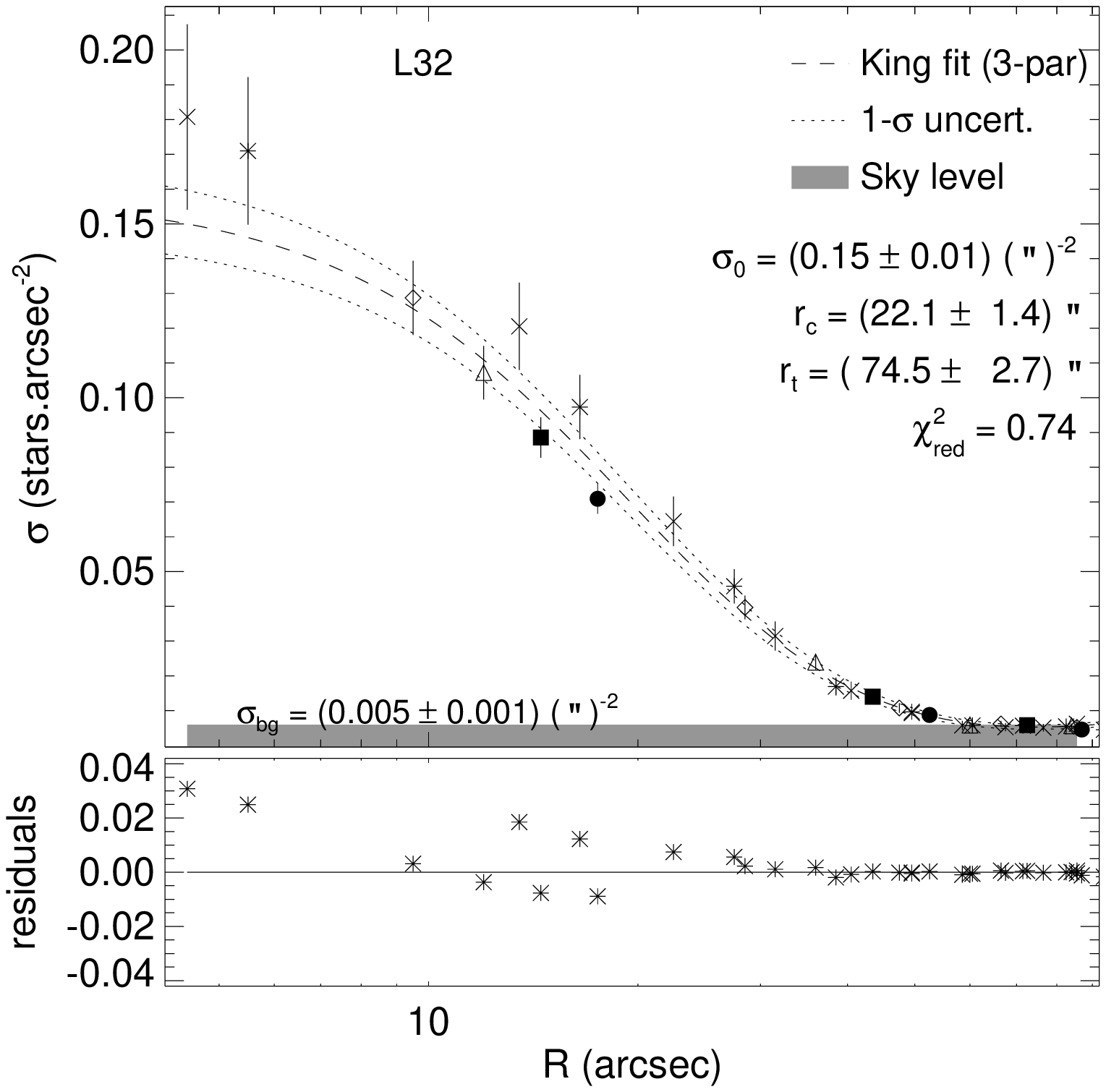}\includegraphics[width=0.325\linewidth]{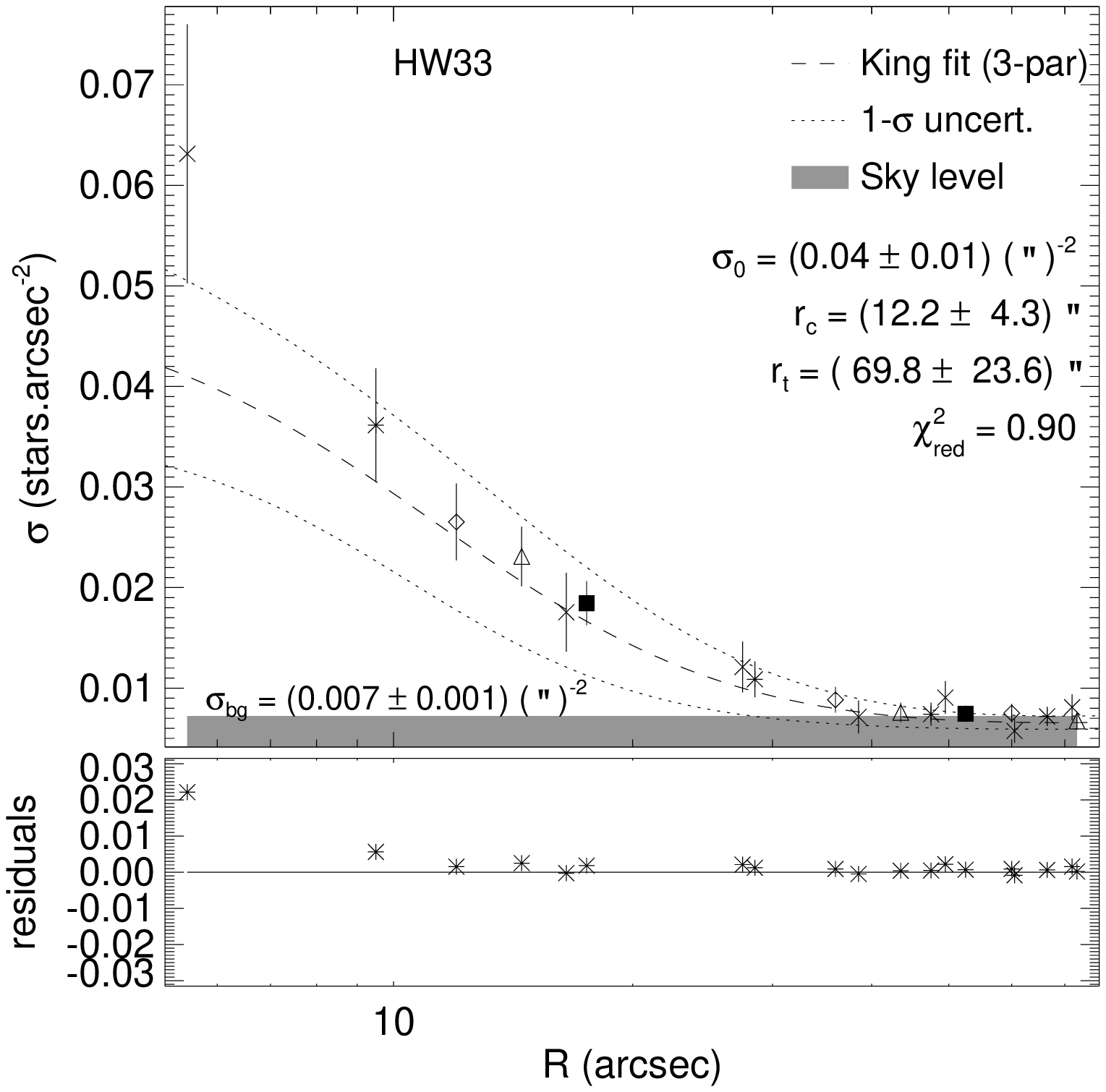}\includegraphics[width=0.325\linewidth]{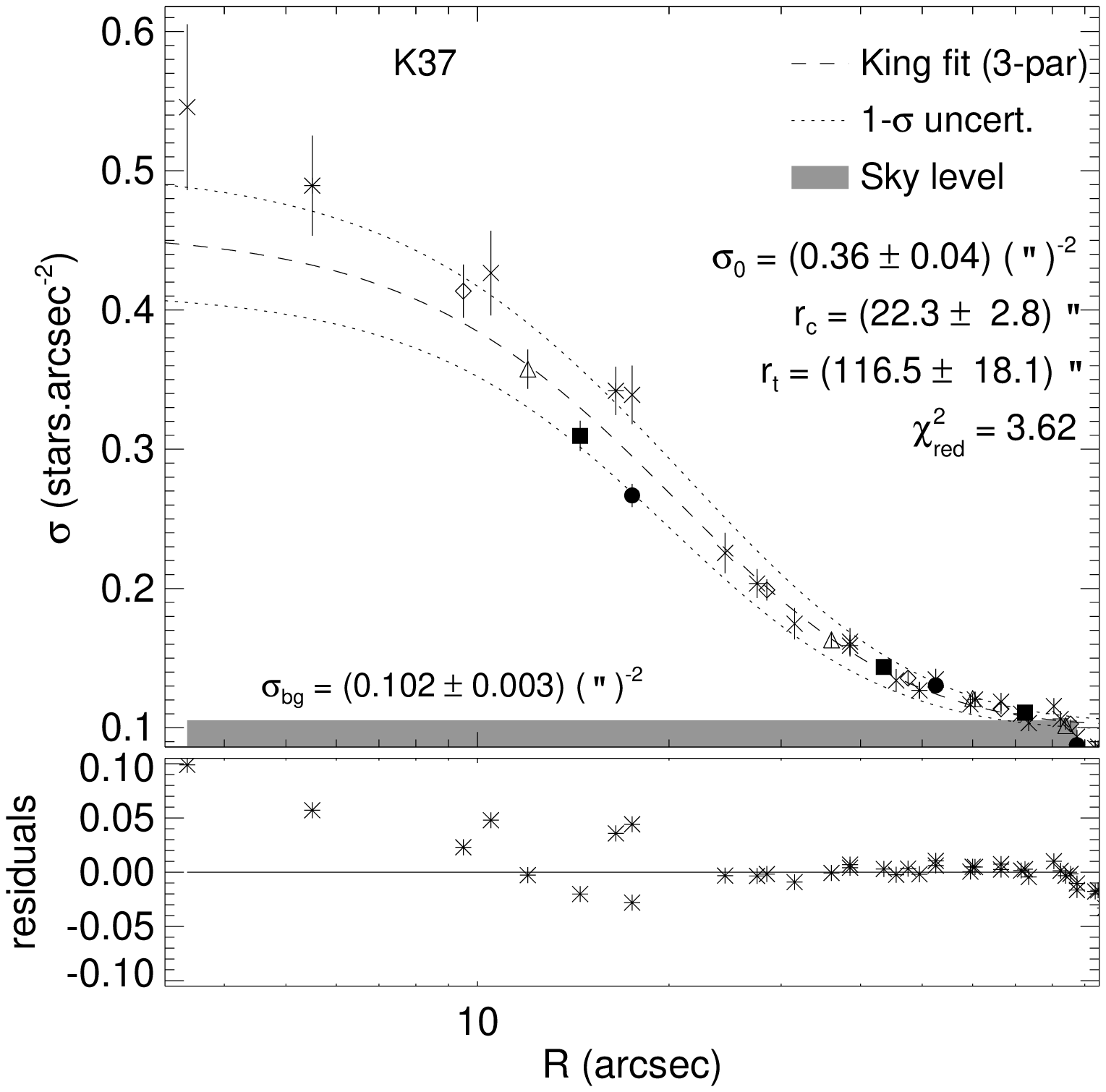}

\includegraphics[width=0.325\linewidth]{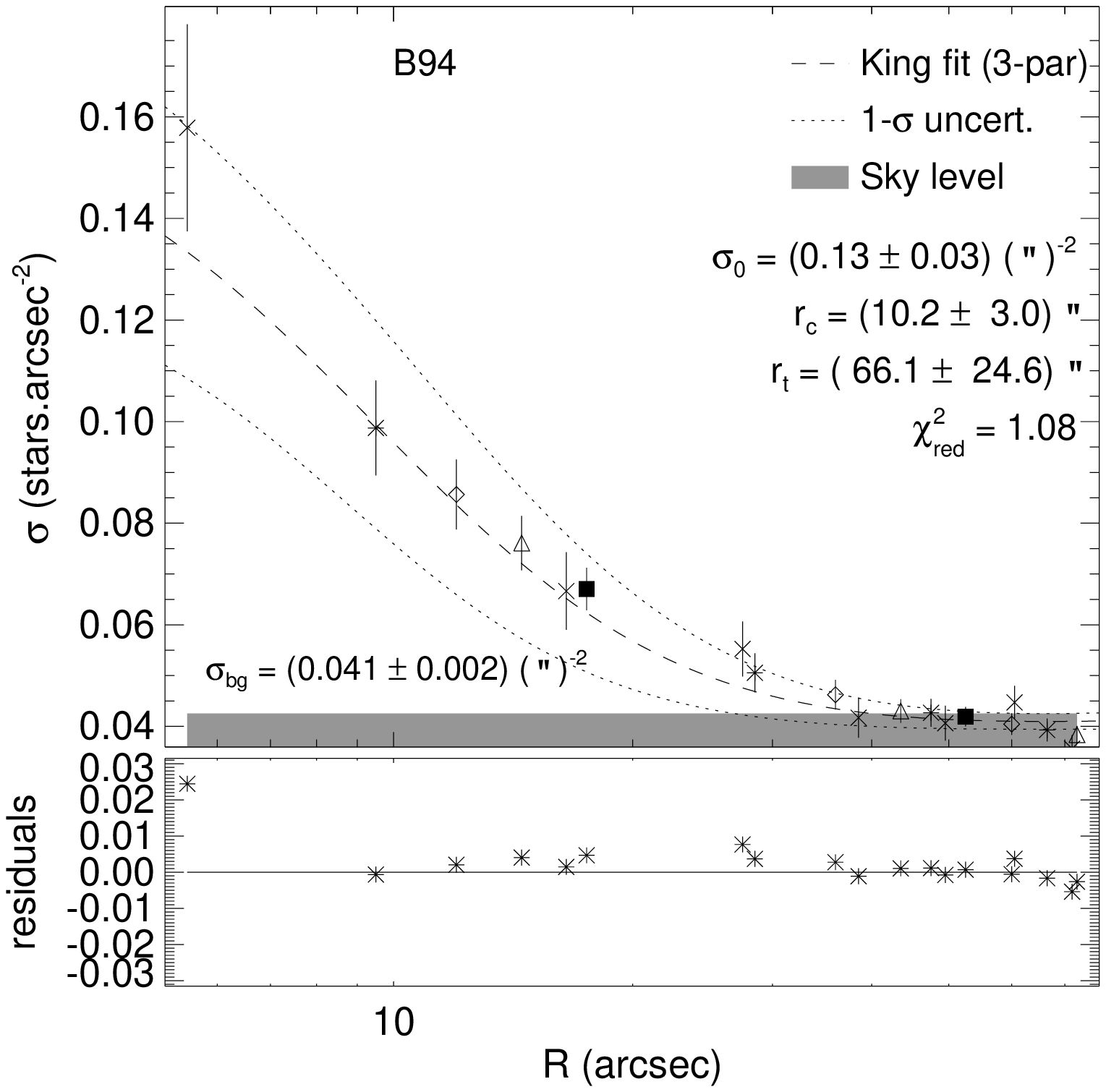}\includegraphics[width=0.325\linewidth]{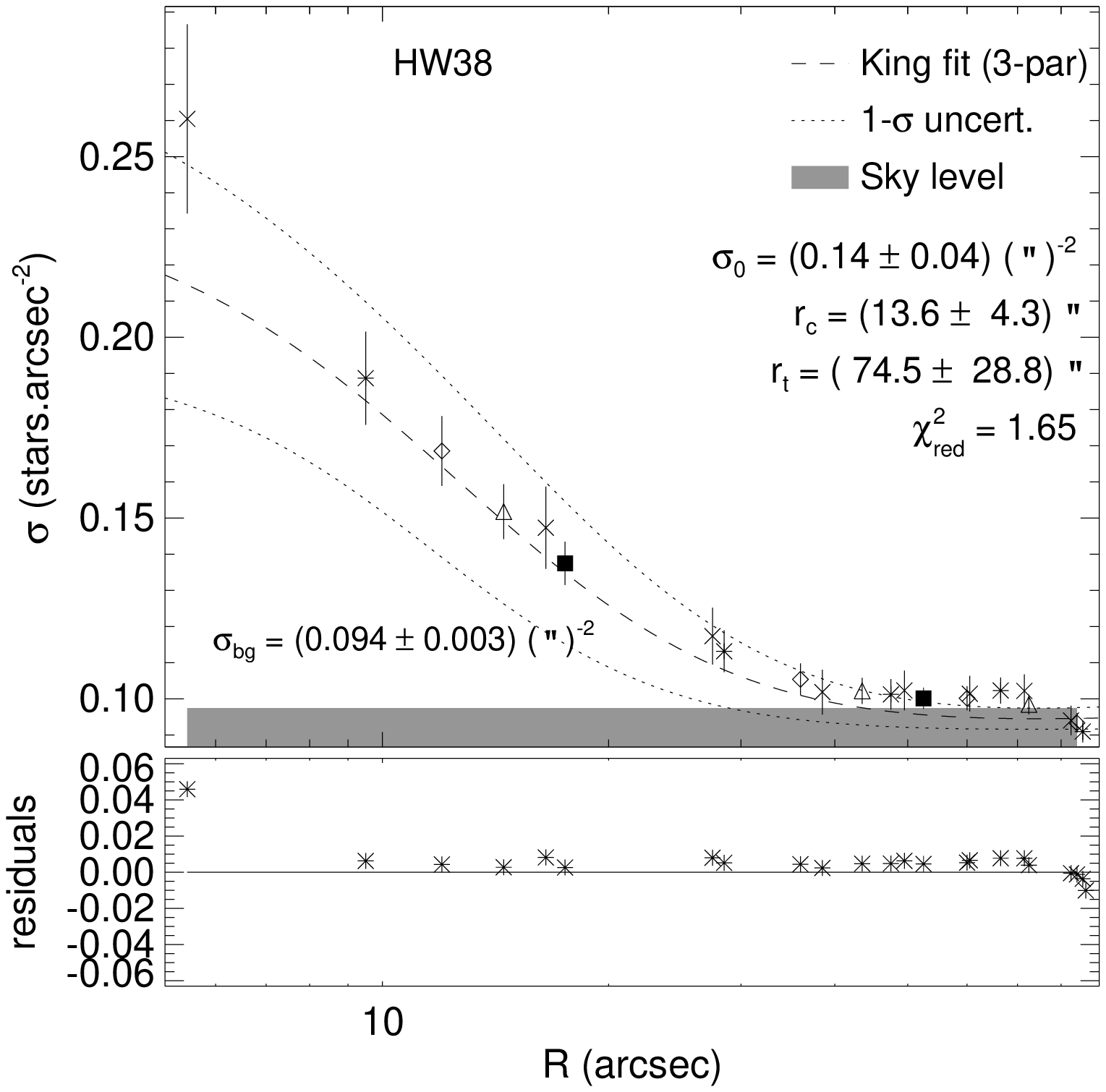}\includegraphics[width=0.325\linewidth]{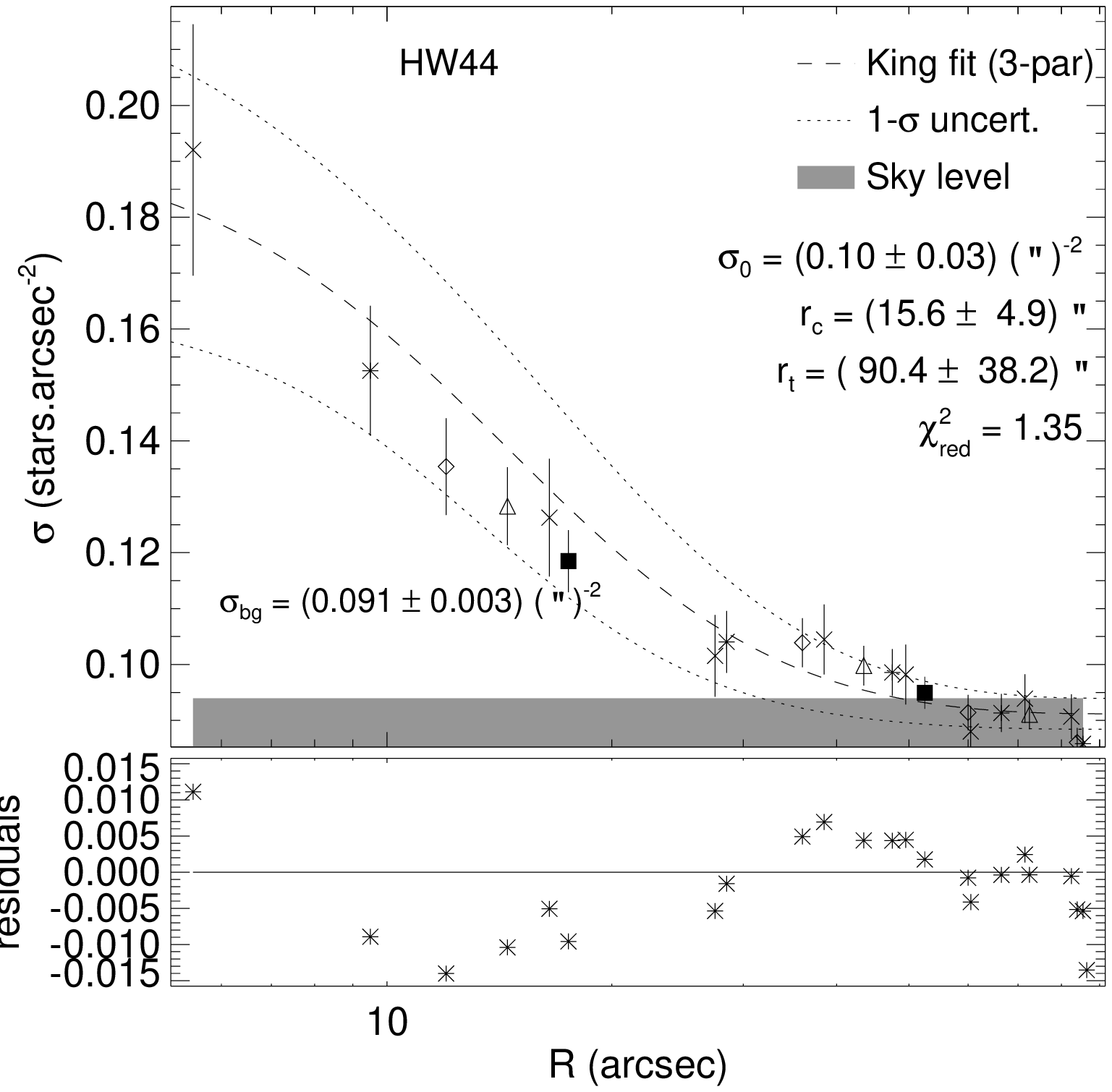}

\includegraphics[width=0.325\linewidth]{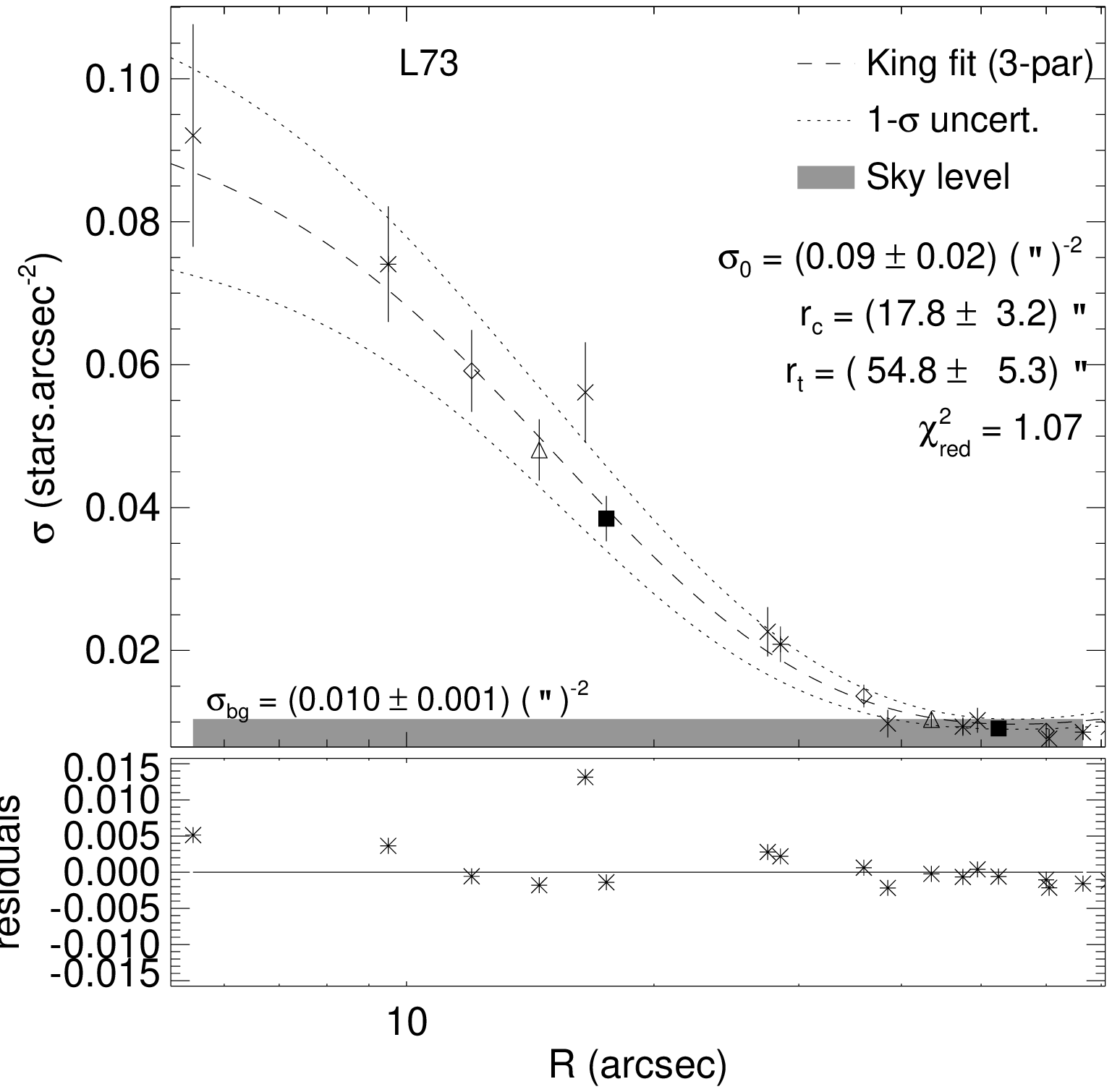}\includegraphics[width=0.325\linewidth]{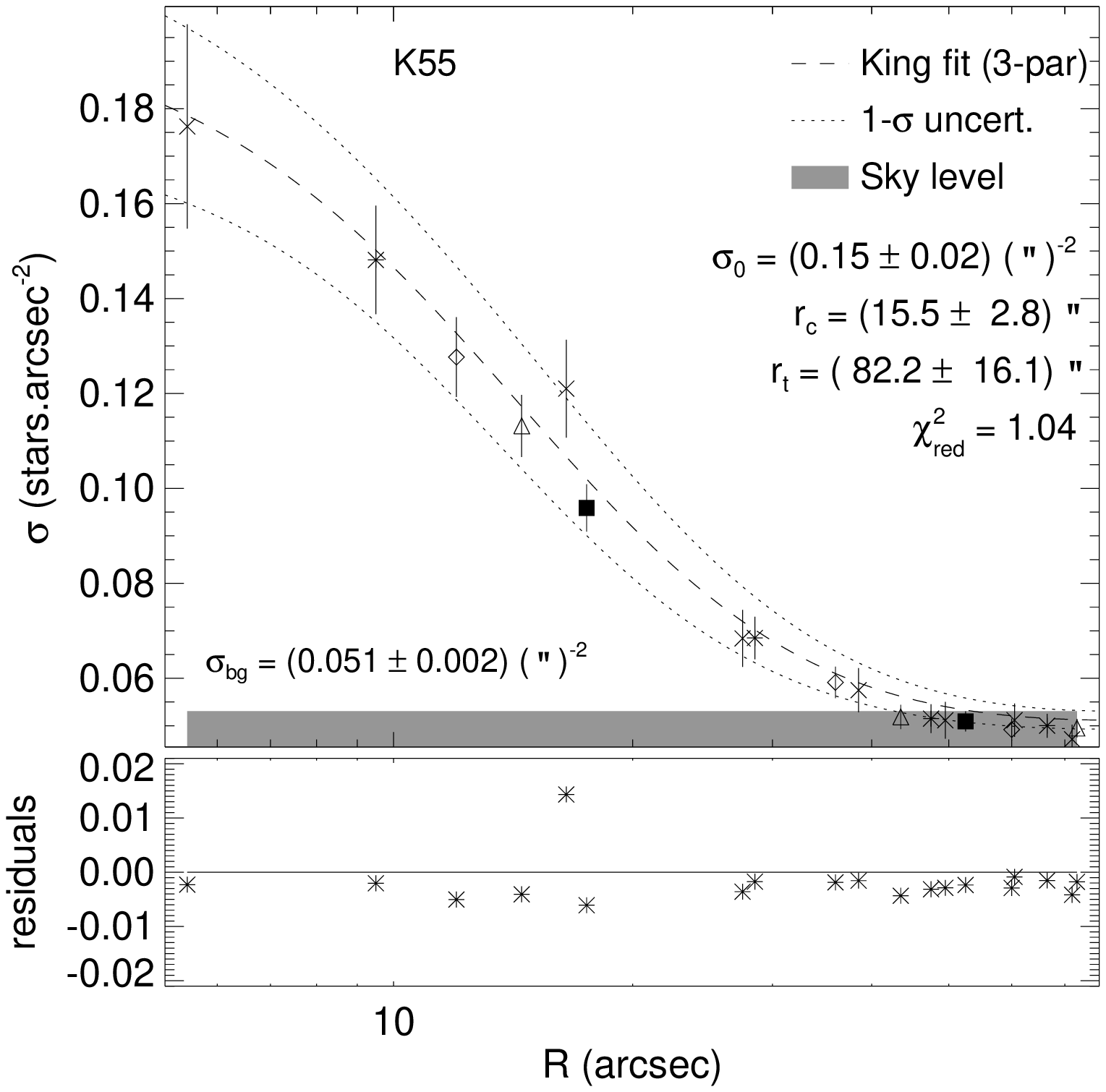}\includegraphics[width=0.325\linewidth]{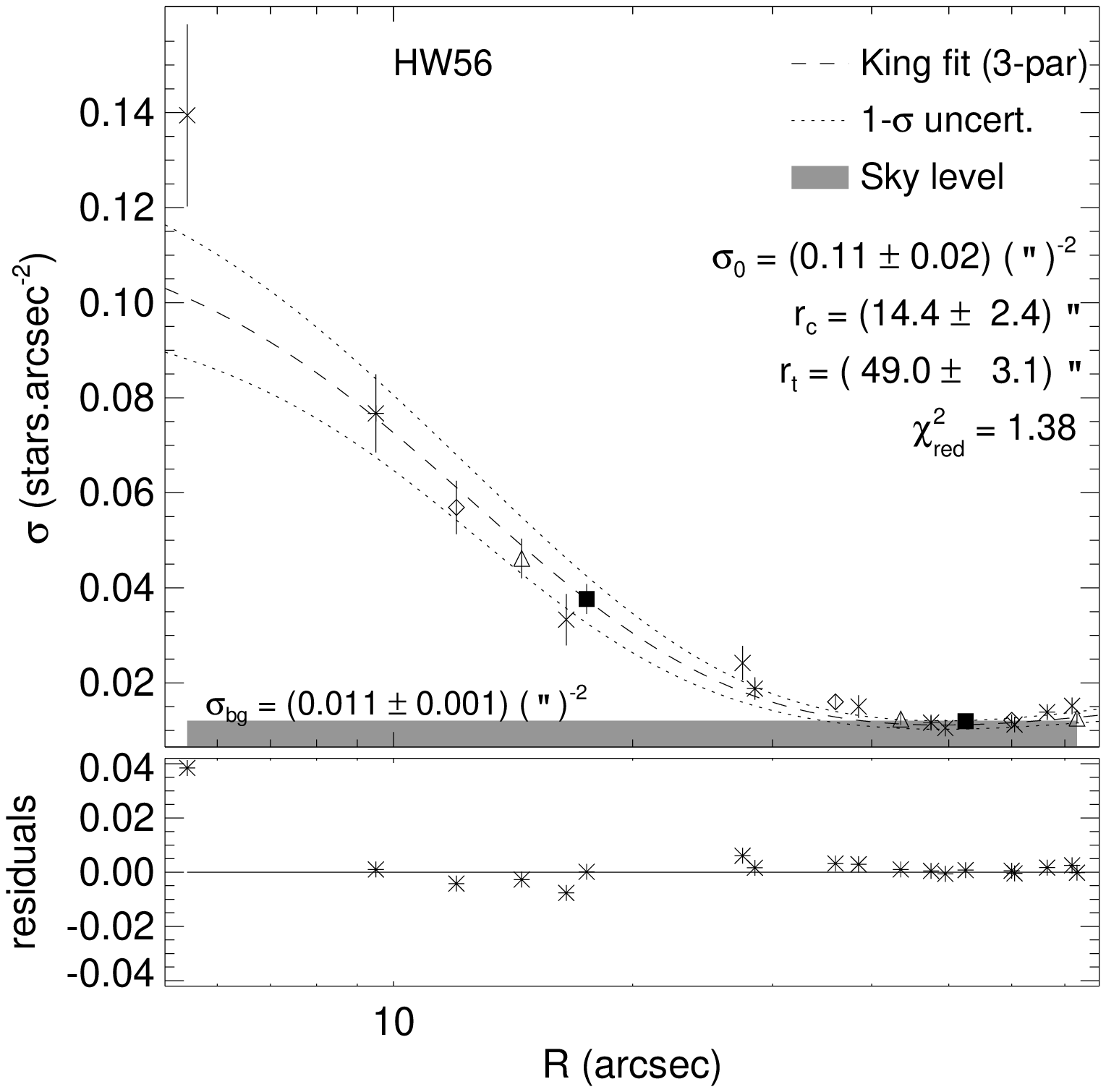}

\caption{Radial density profiles for additional SMC clusters complementing the sample presented in Fig.~\ref{fig:rdp_sbp}. The King model fits (dashed line) with envelopes of 1\,$\sigma$ uncertainty (dotted lines) are shown. Different symbols correspond to the various widths of the annular bins employed. The fitting residuals are also presented in the lower panel.}
\end{figure*}

\setcounter{figure}{0}

\begin{figure*}

\includegraphics[width=0.325\linewidth]{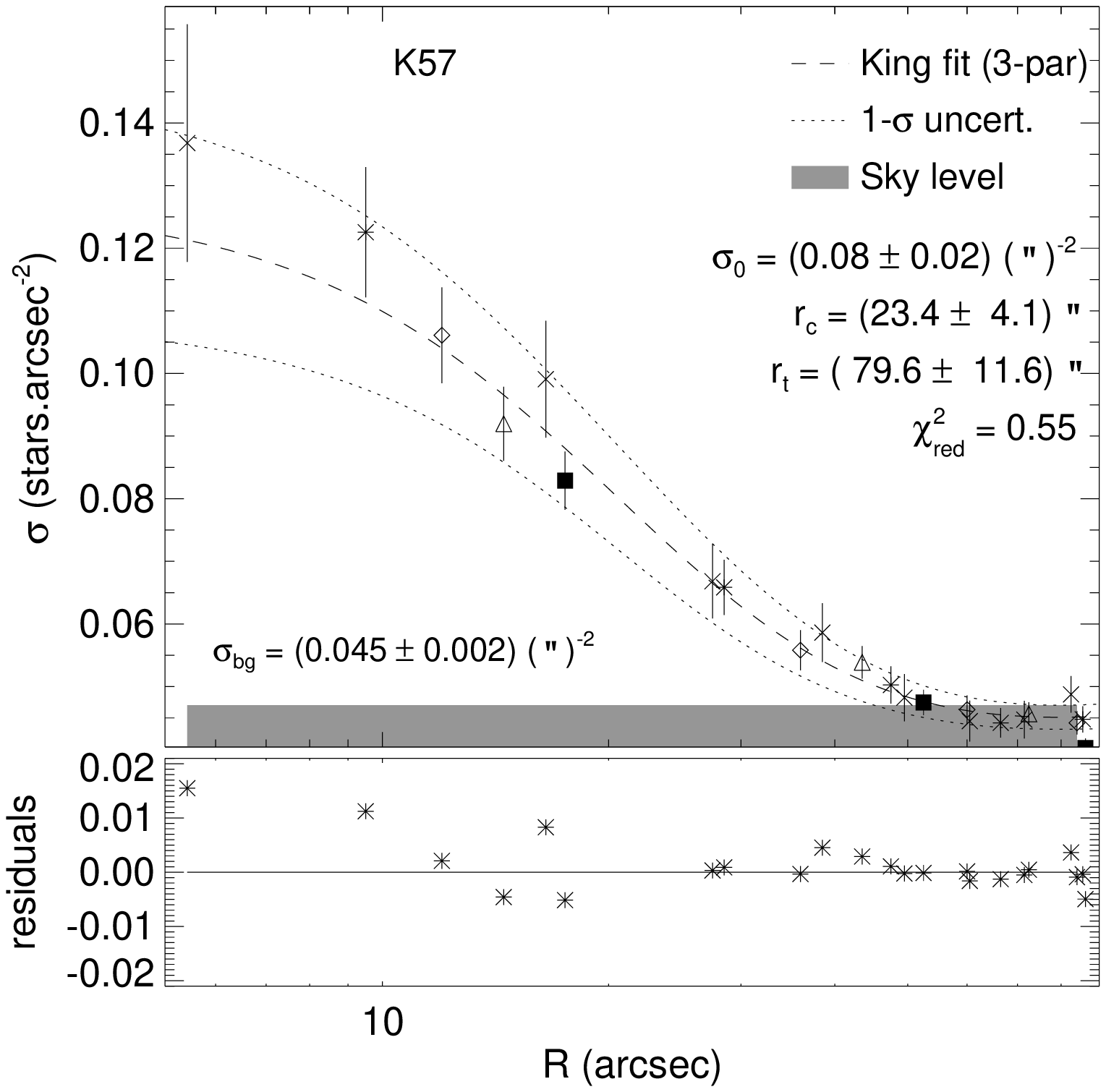}\includegraphics[width=0.325\linewidth]{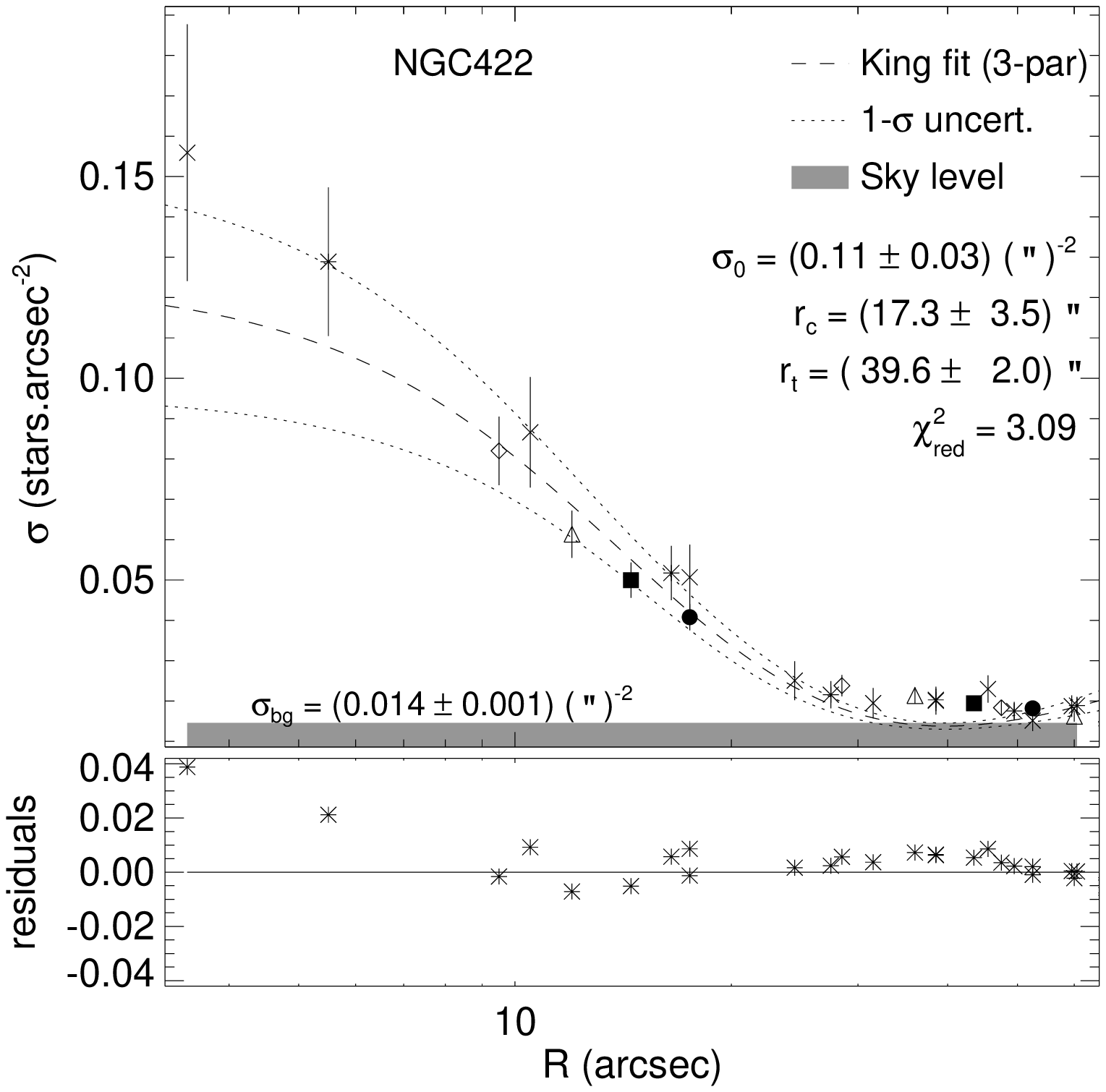}\includegraphics[width=0.325\linewidth]{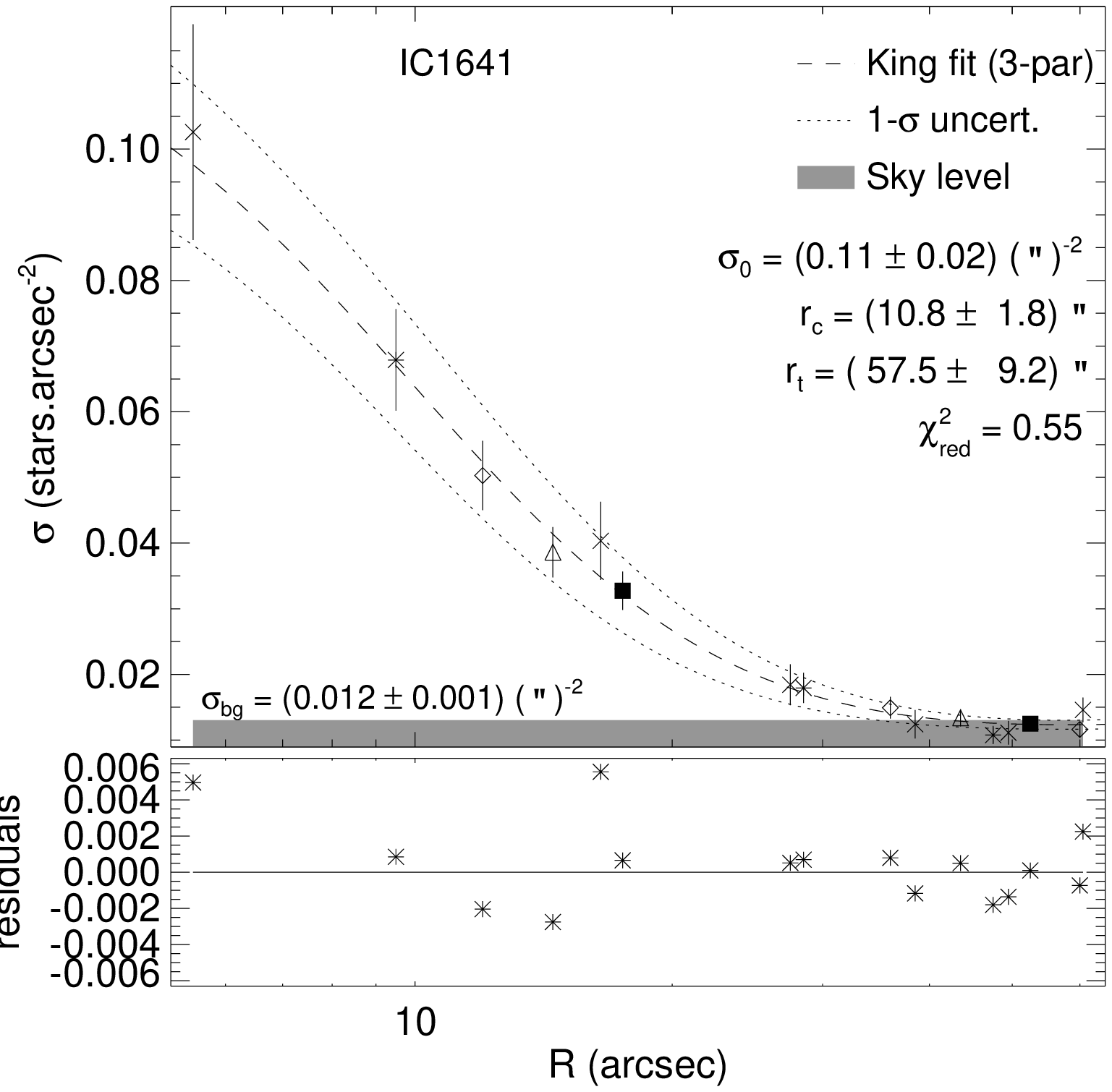}

\includegraphics[width=0.325\linewidth]{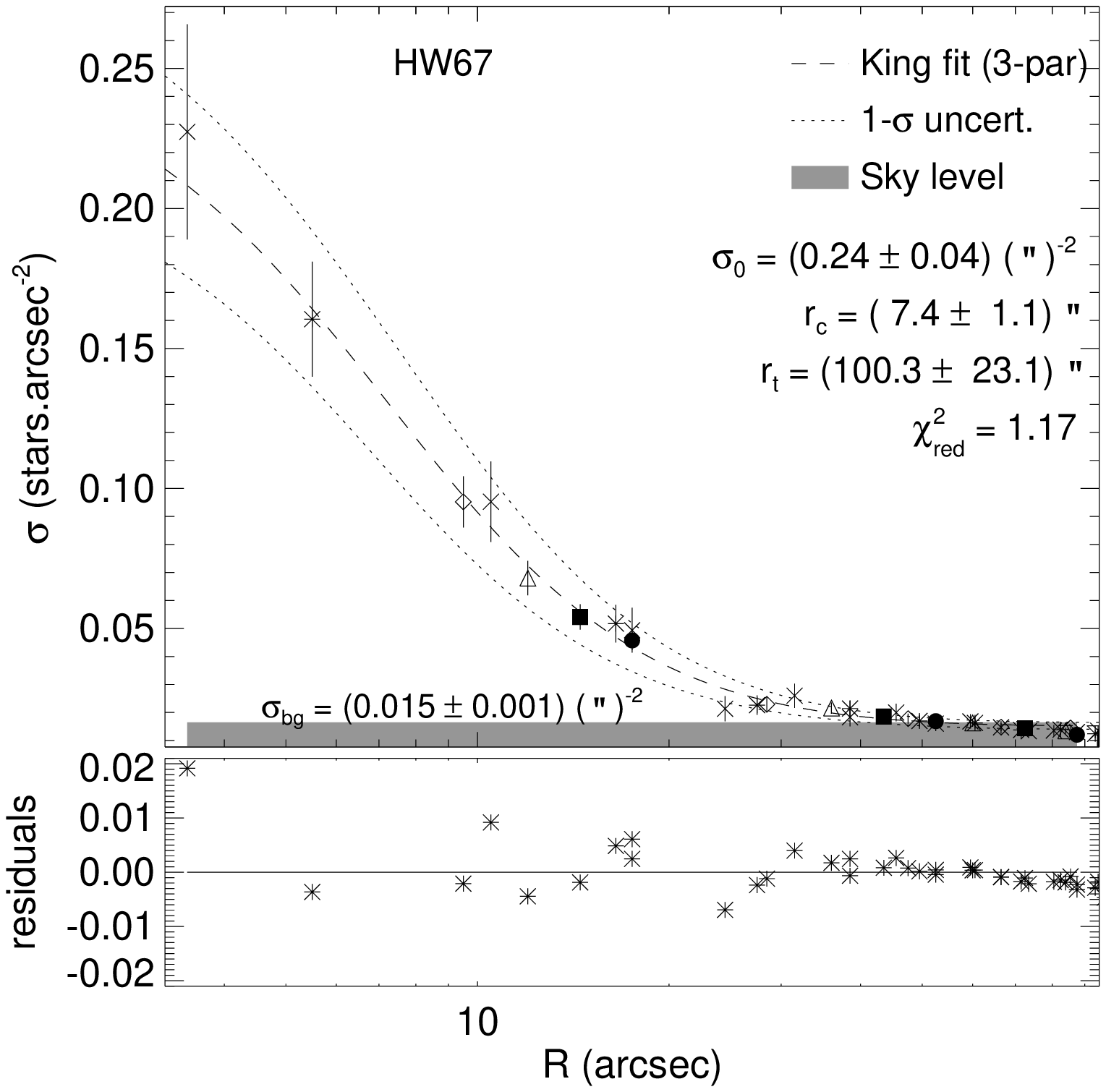}\includegraphics[width=0.325\linewidth]{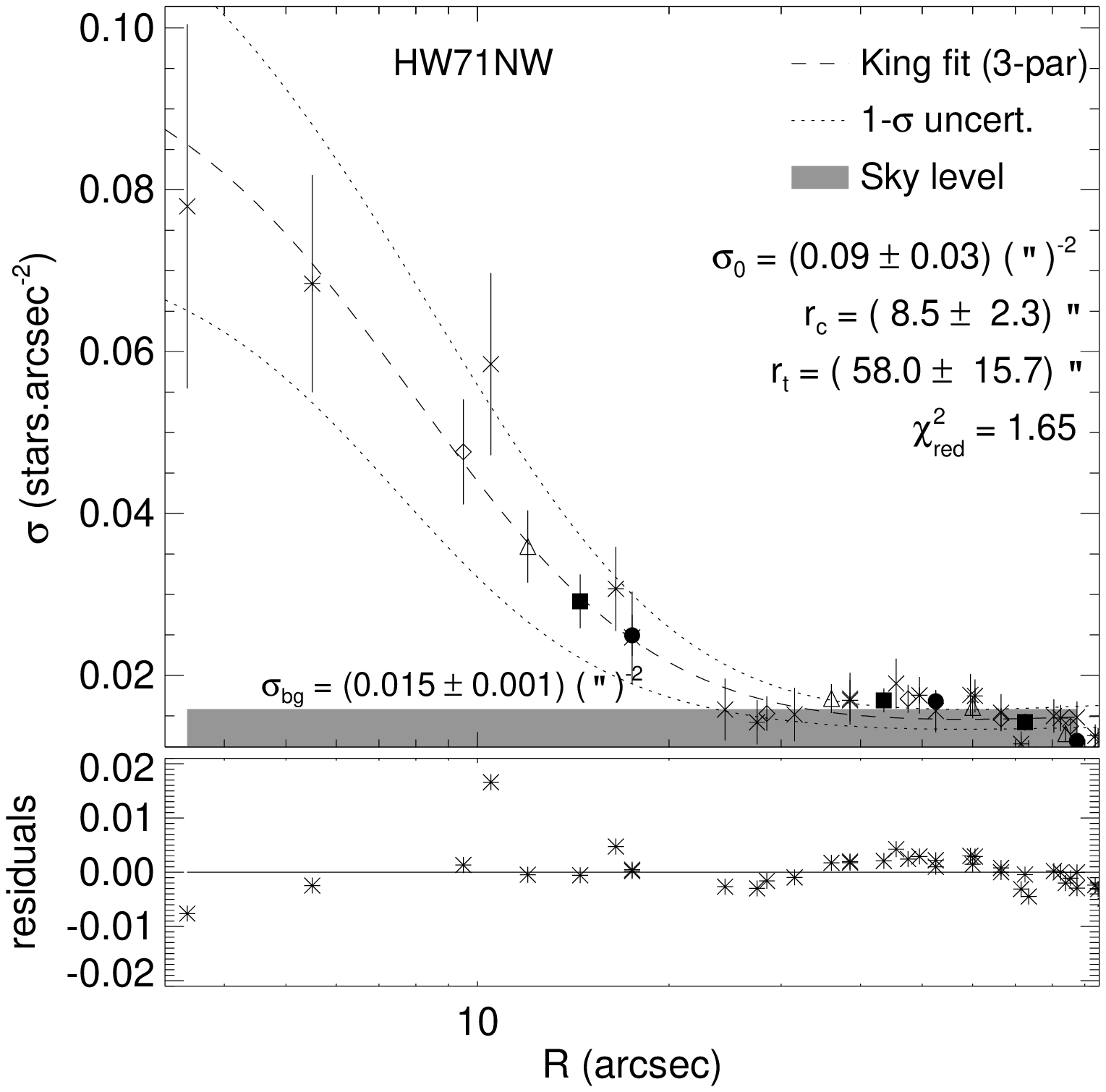}\includegraphics[width=0.325\linewidth]{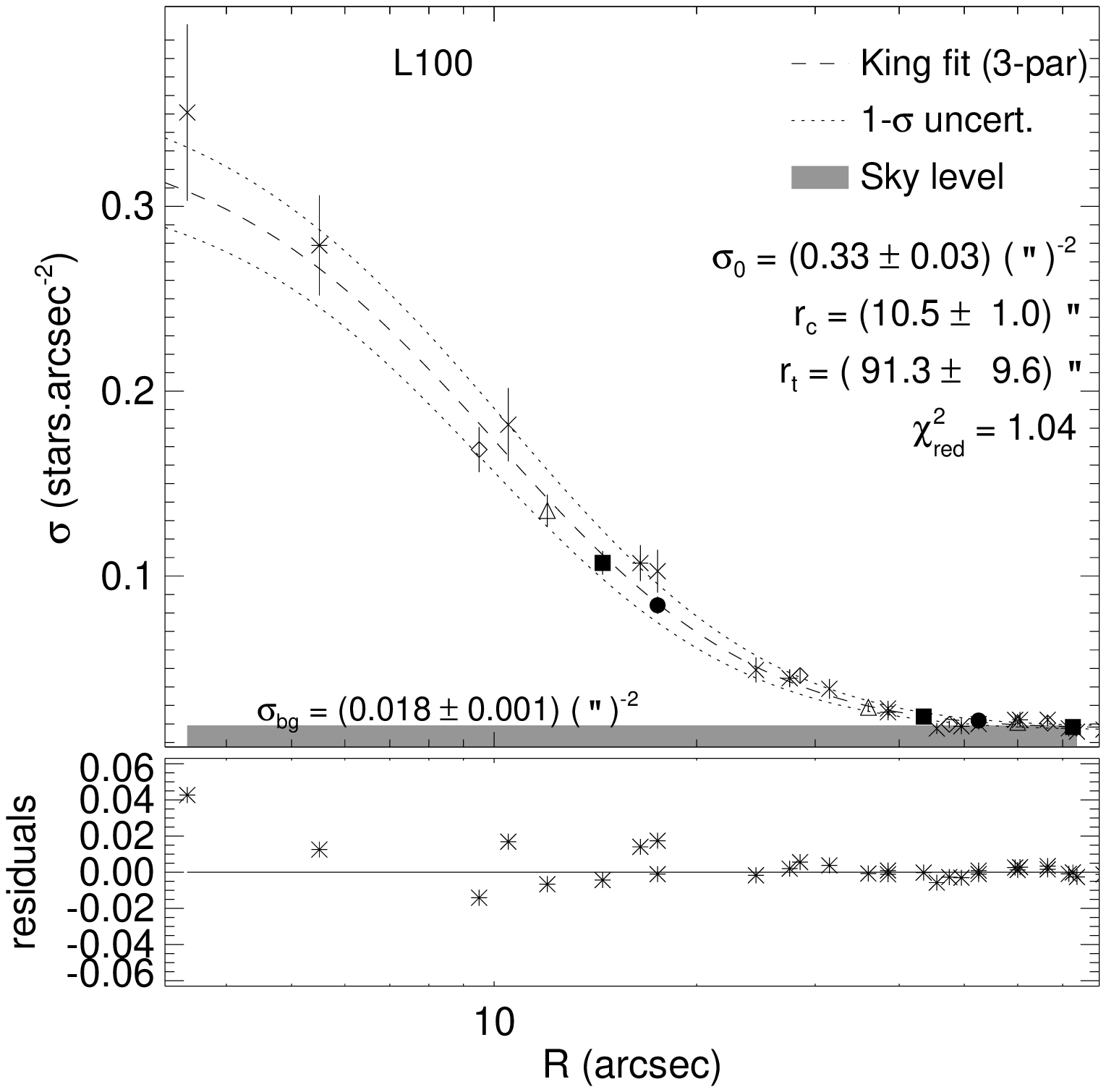}

\includegraphics[width=0.325\linewidth]{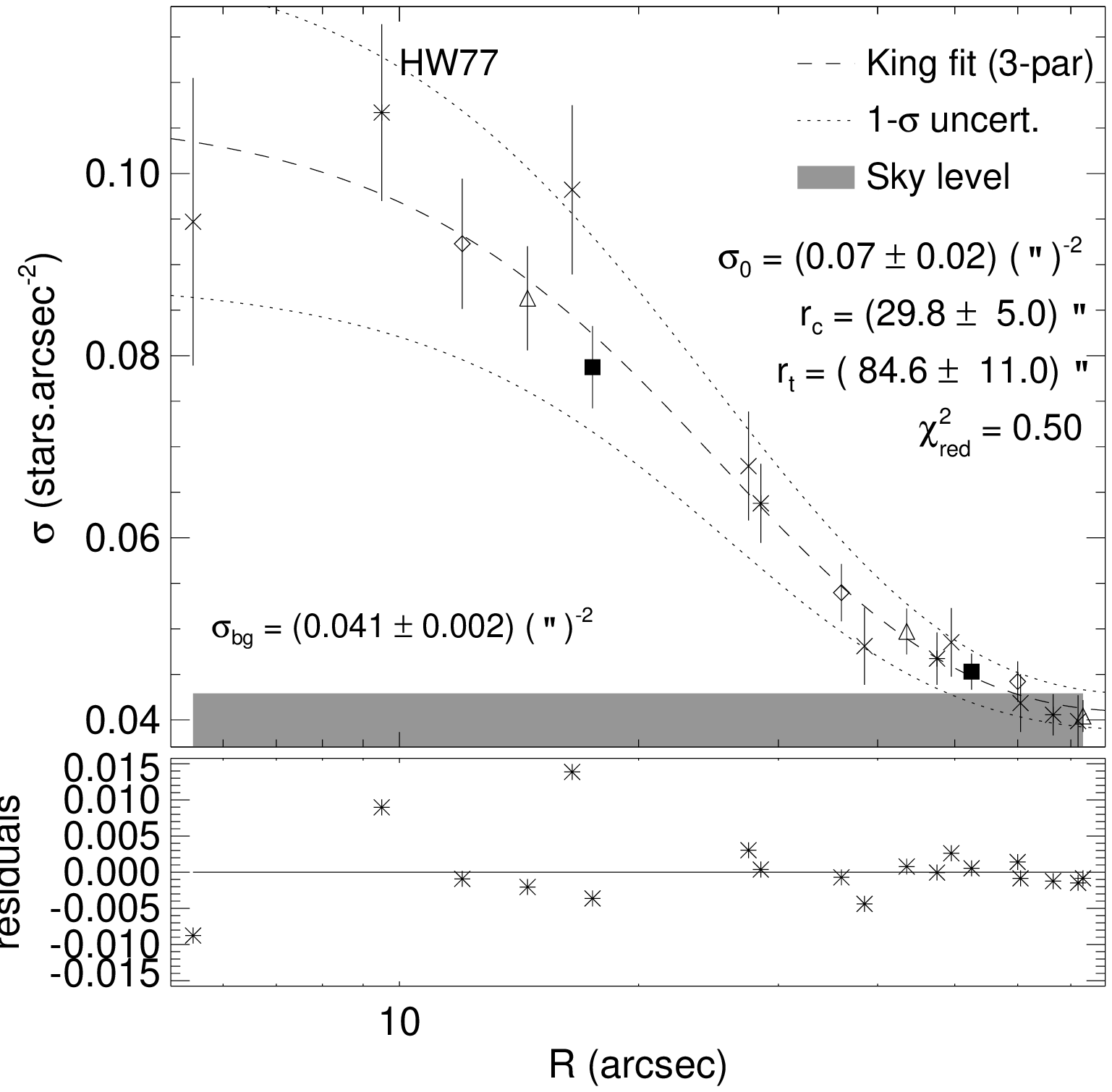}\includegraphics[width=0.325\linewidth]{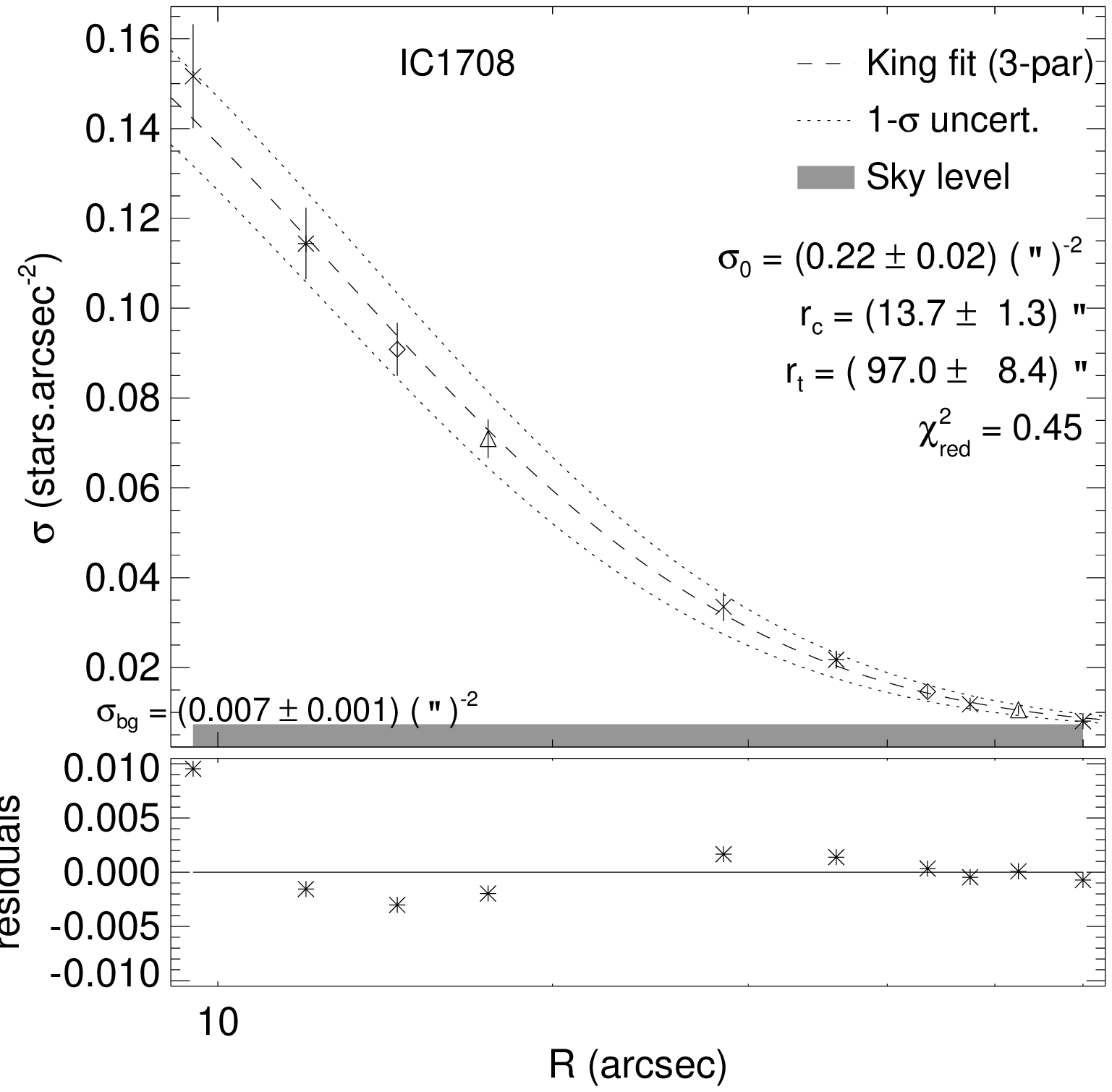}\includegraphics[width=0.325\linewidth]{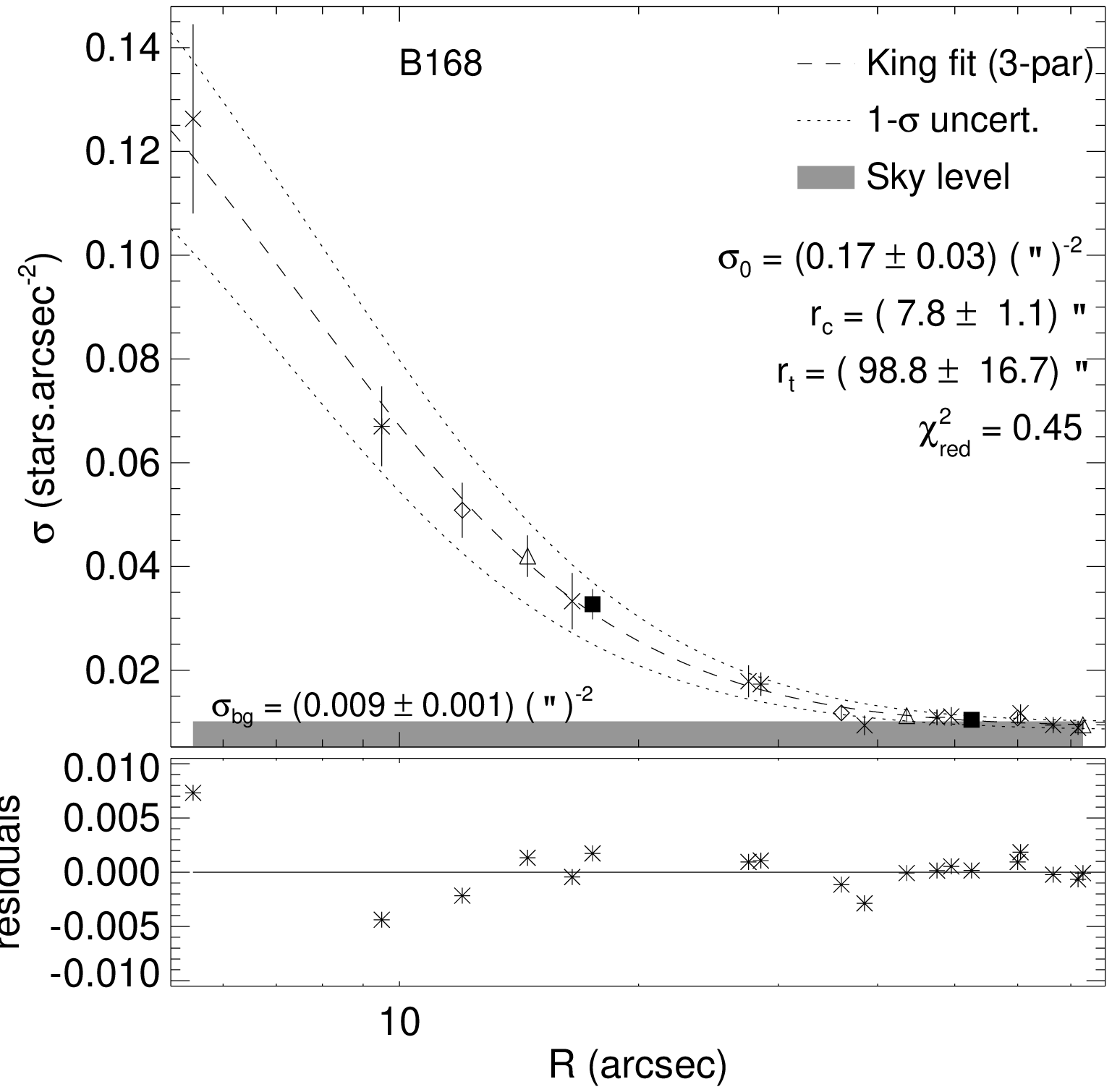}

\includegraphics[width=0.325\linewidth]{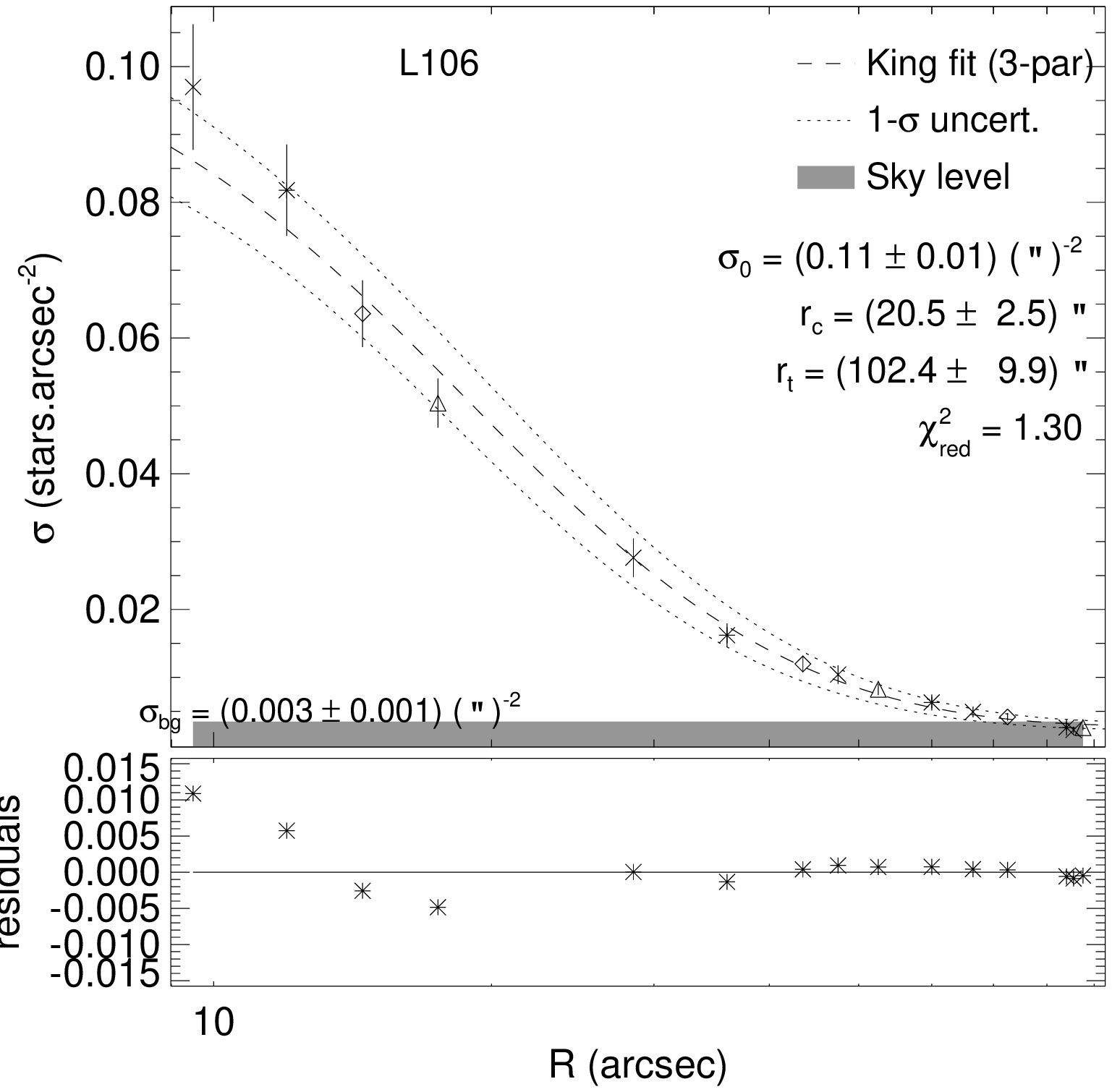}\includegraphics[width=0.325\linewidth]{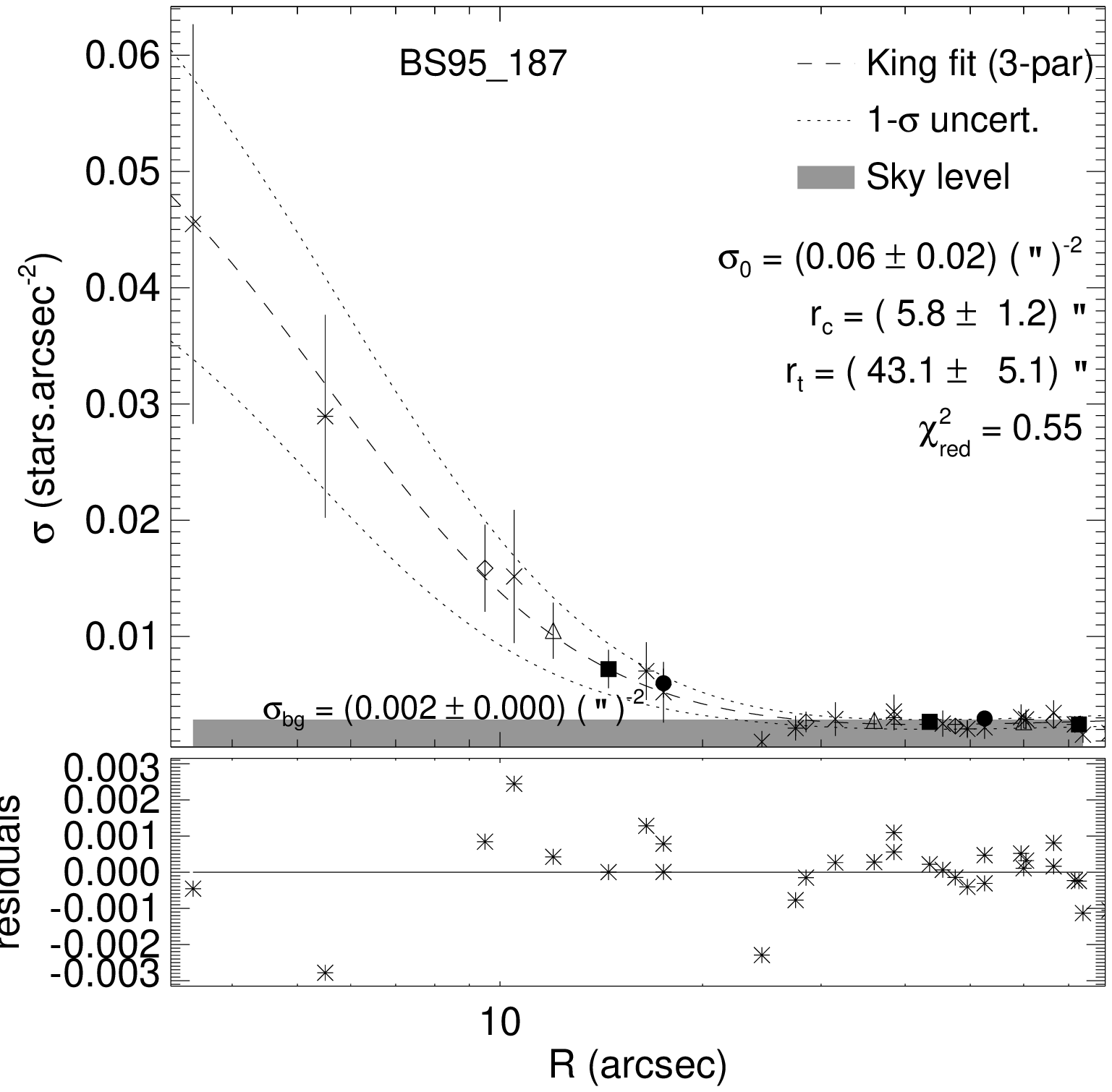}\includegraphics[width=0.325\linewidth]{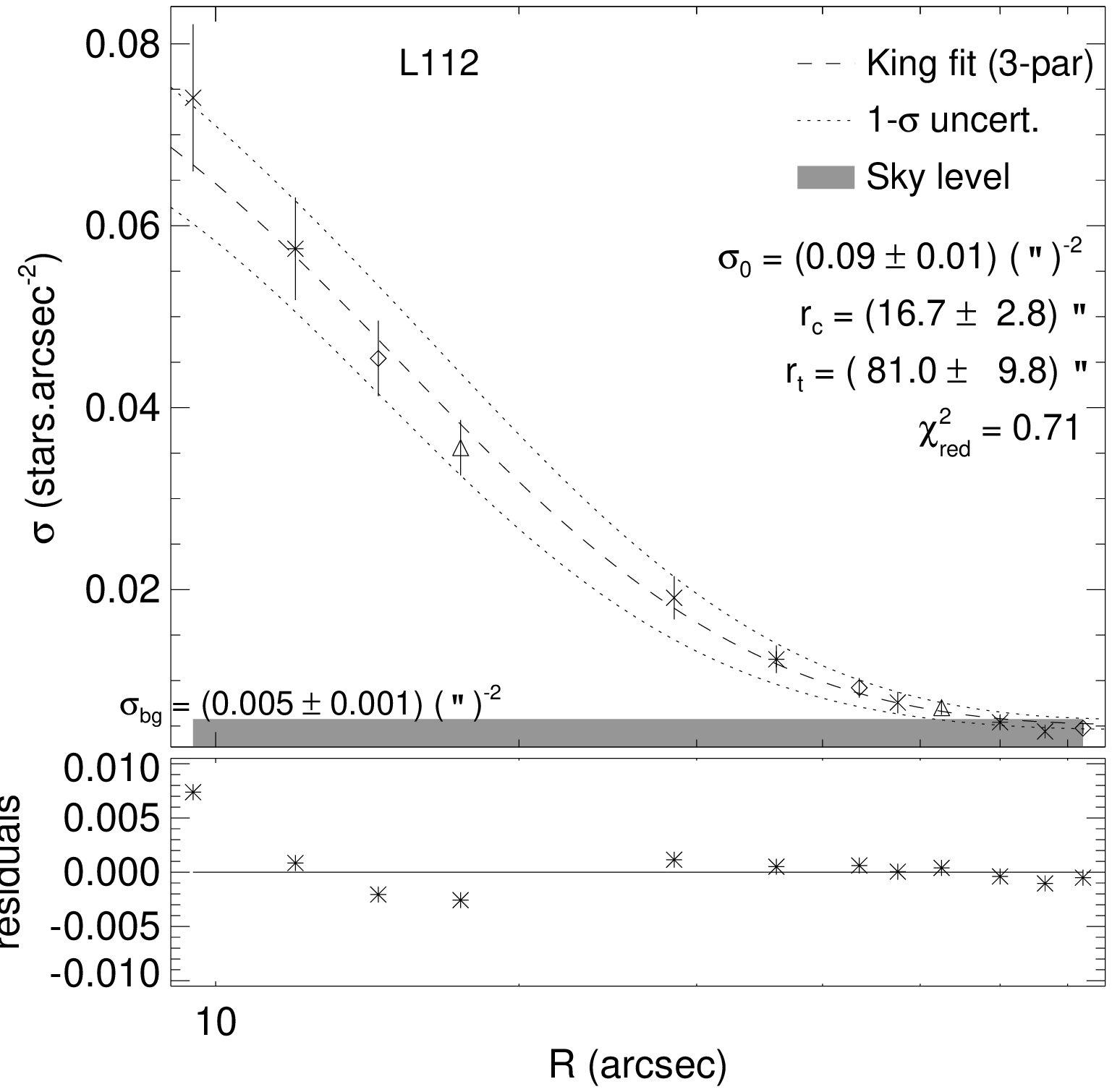}

\caption{cont.}

\end{figure*}

\setcounter{figure}{0}

\begin{figure*}
\includegraphics[width=0.325\linewidth]{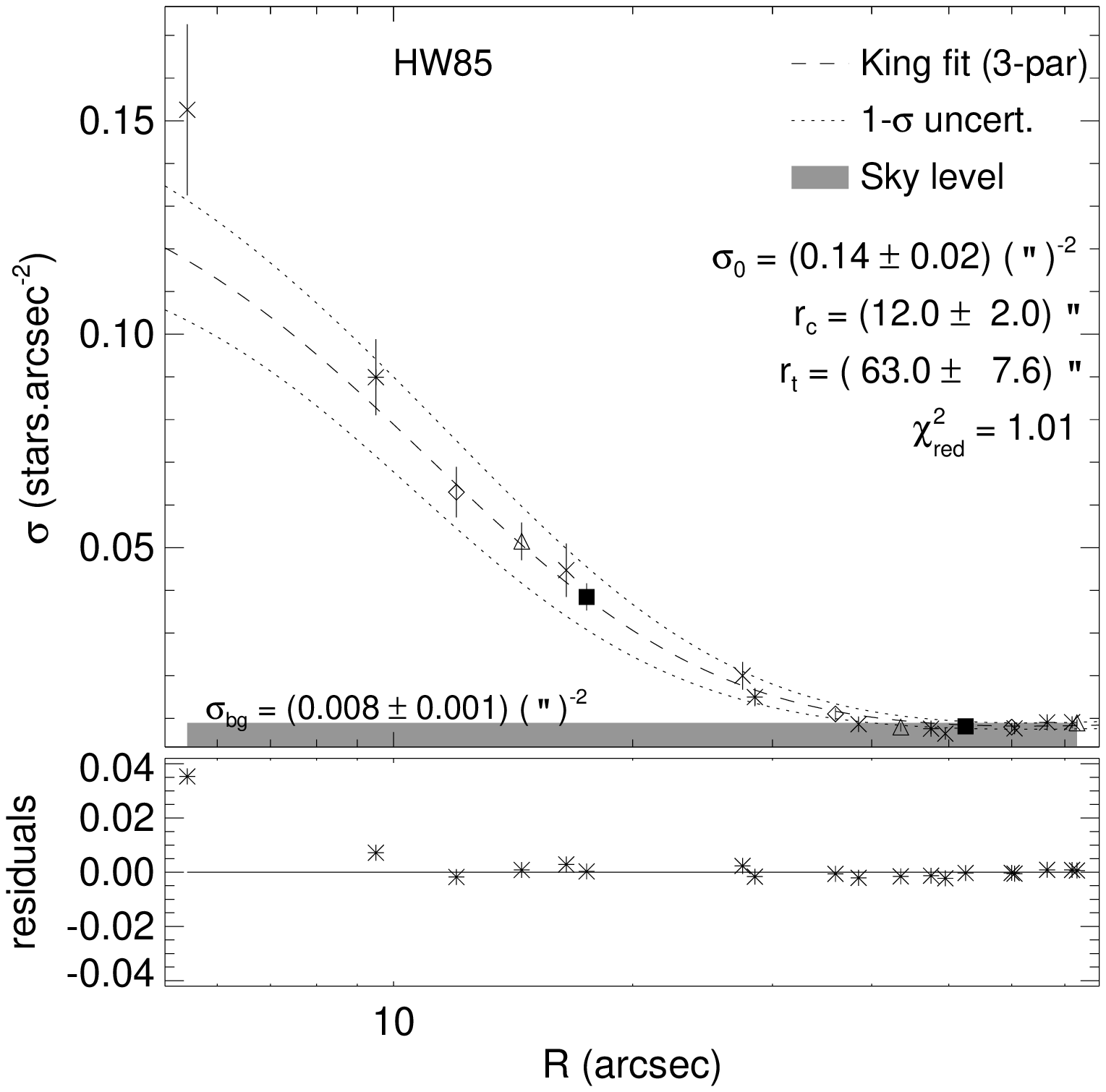}\includegraphics[width=0.325\linewidth]{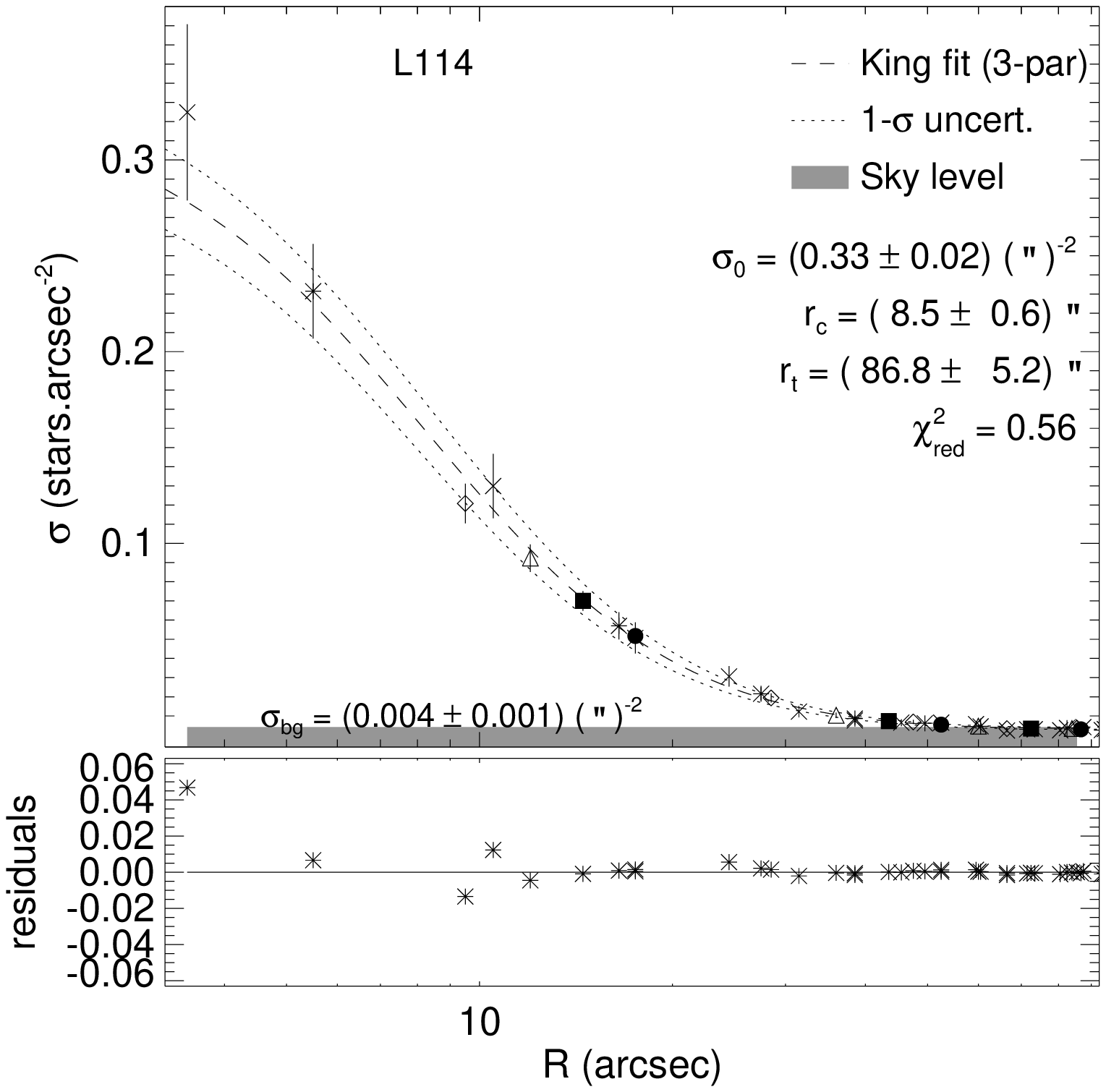}\includegraphics[width=0.325\linewidth]{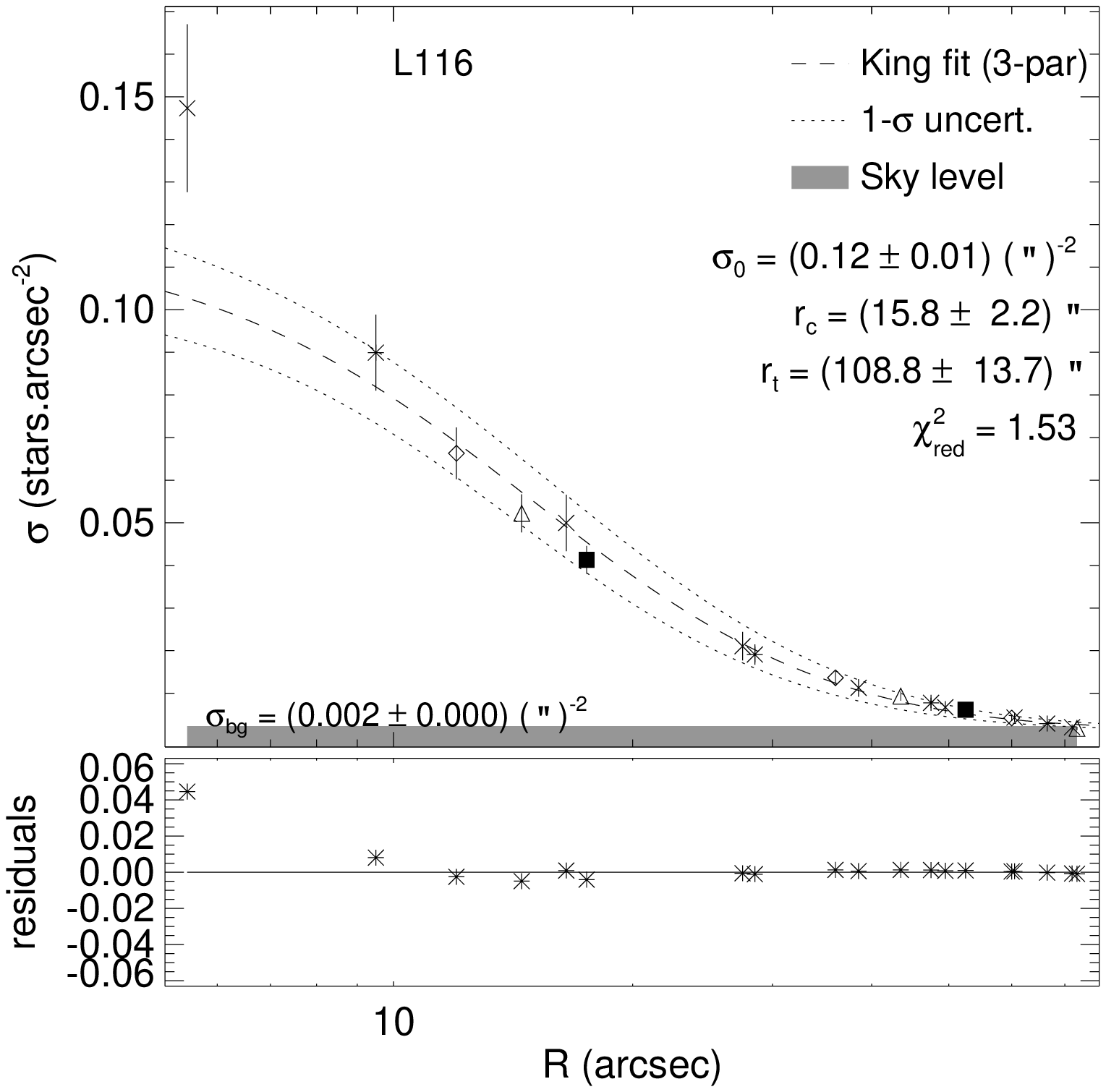}

\includegraphics[width=0.325\linewidth]{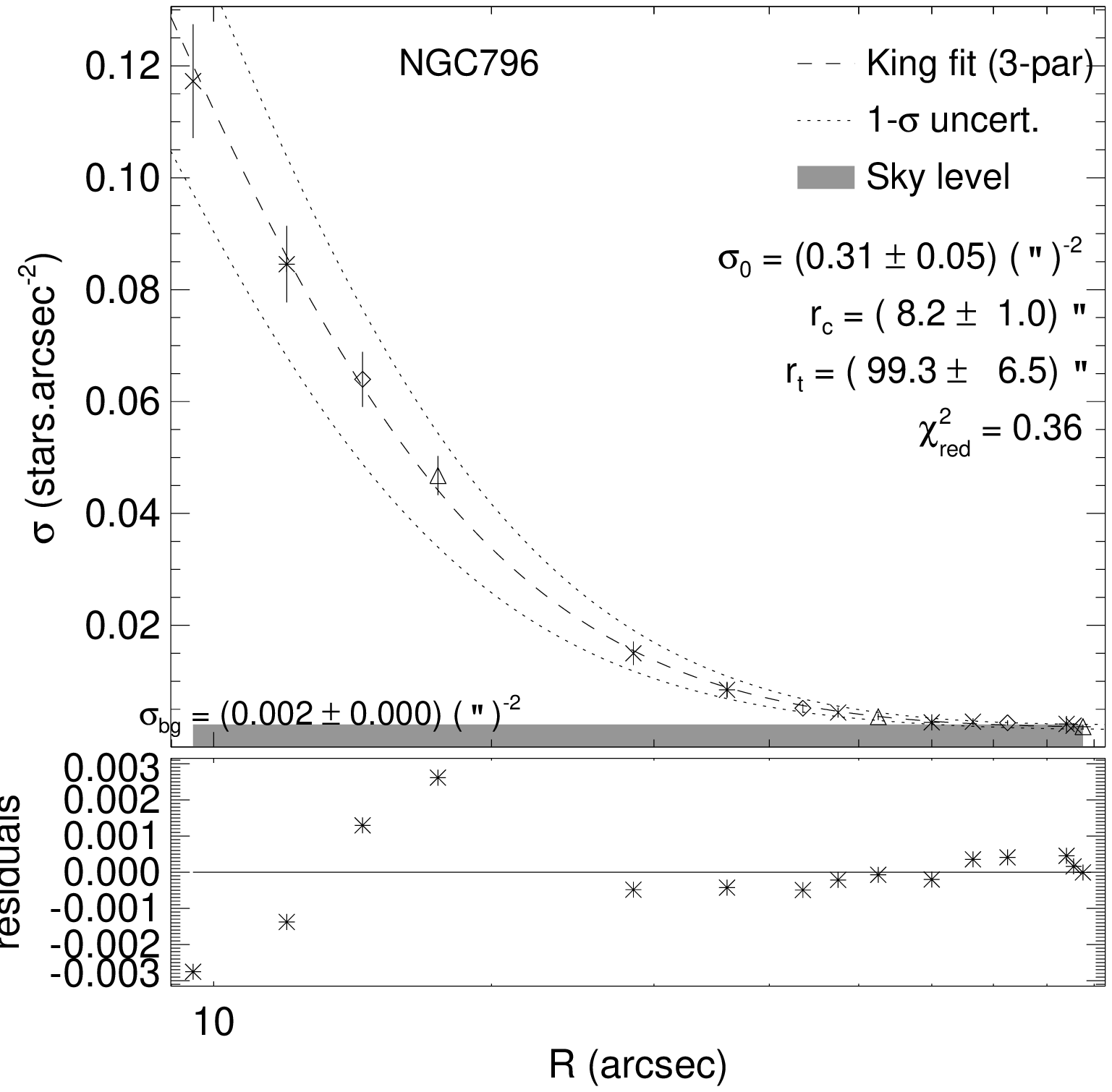}\includegraphics[width=0.325\linewidth]{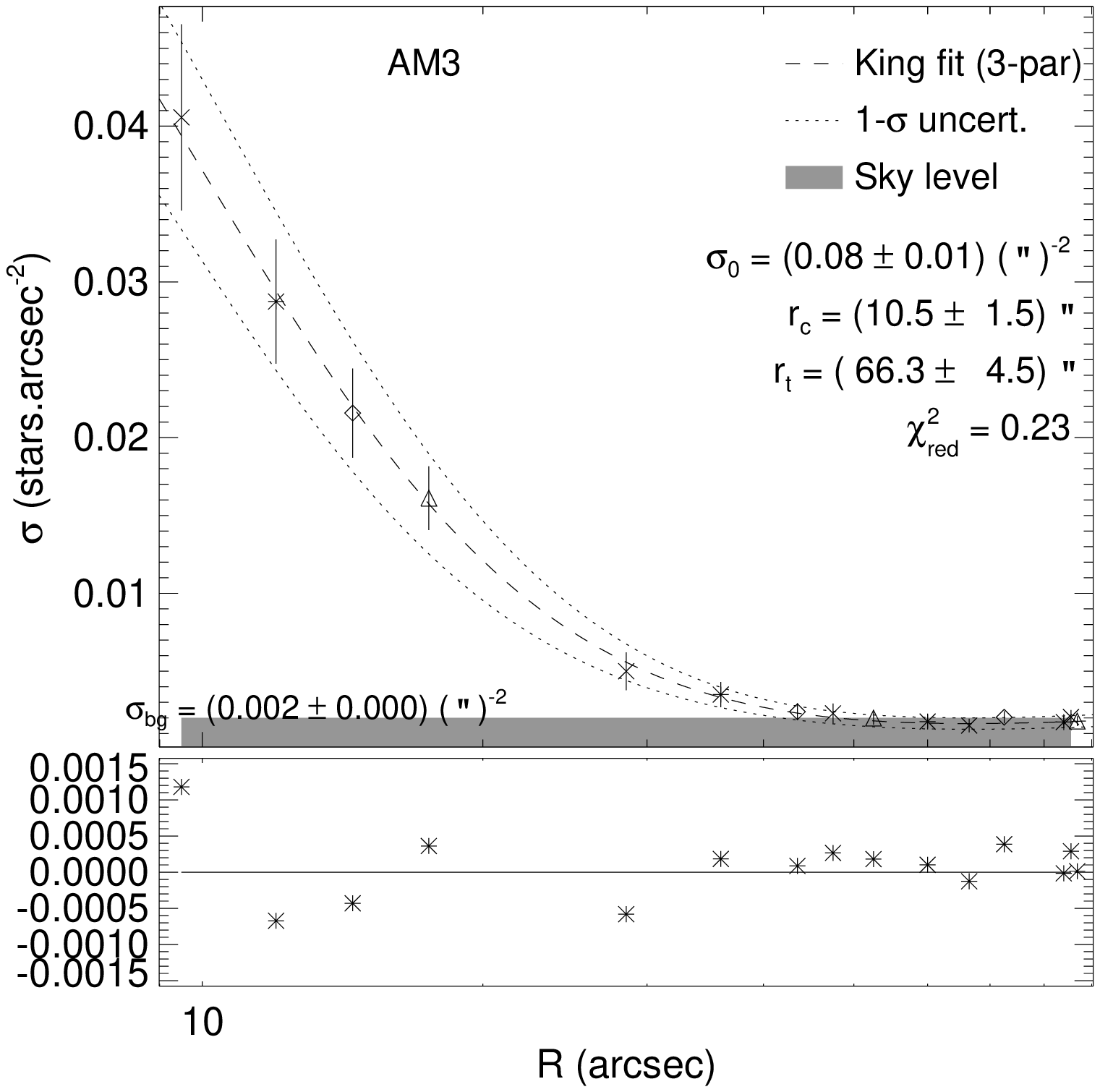}
\caption{cont.}

\end{figure*}

\begin{figure*}

\includegraphics[width=0.325\linewidth]{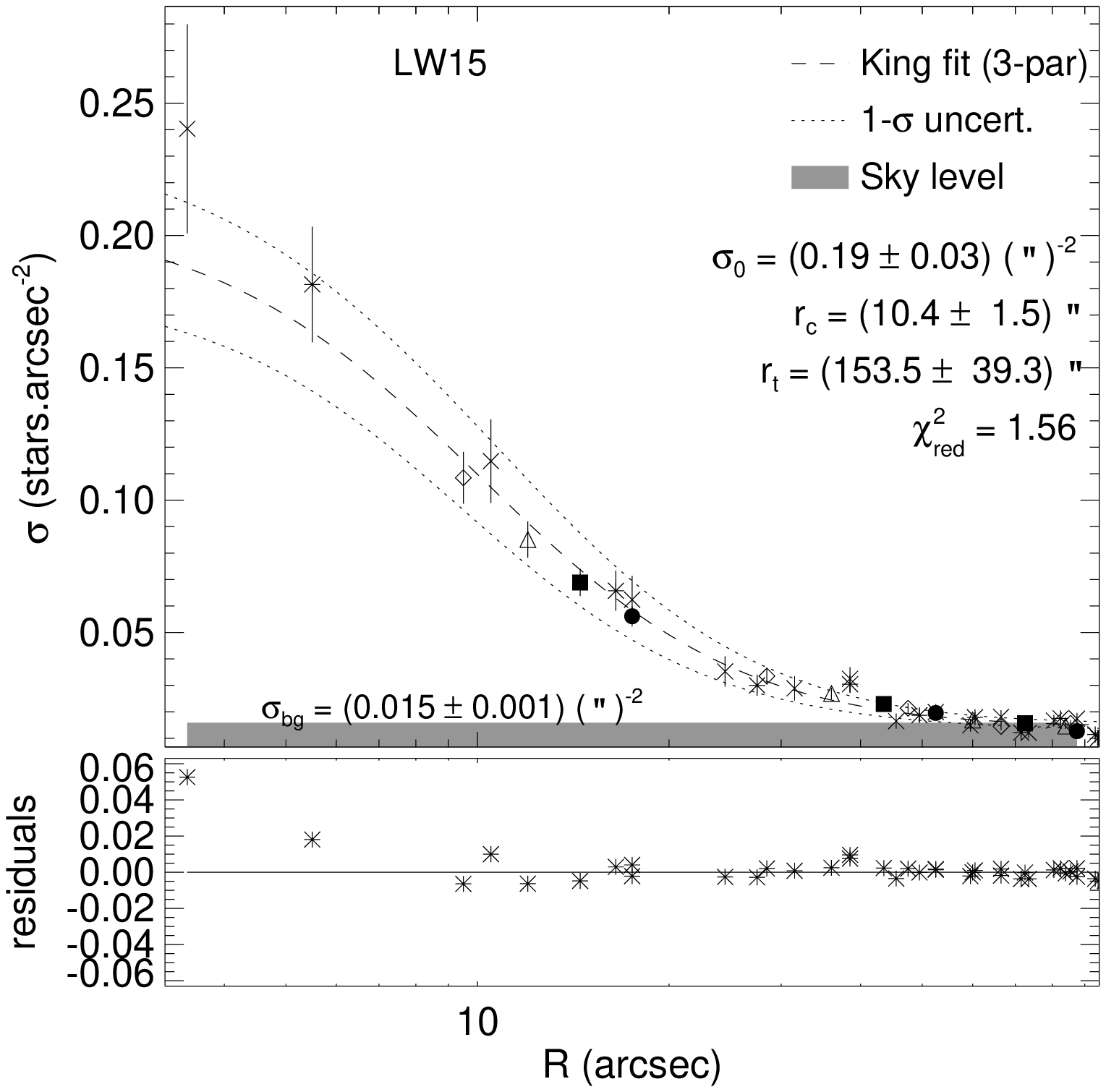}\includegraphics[width=0.325\linewidth]{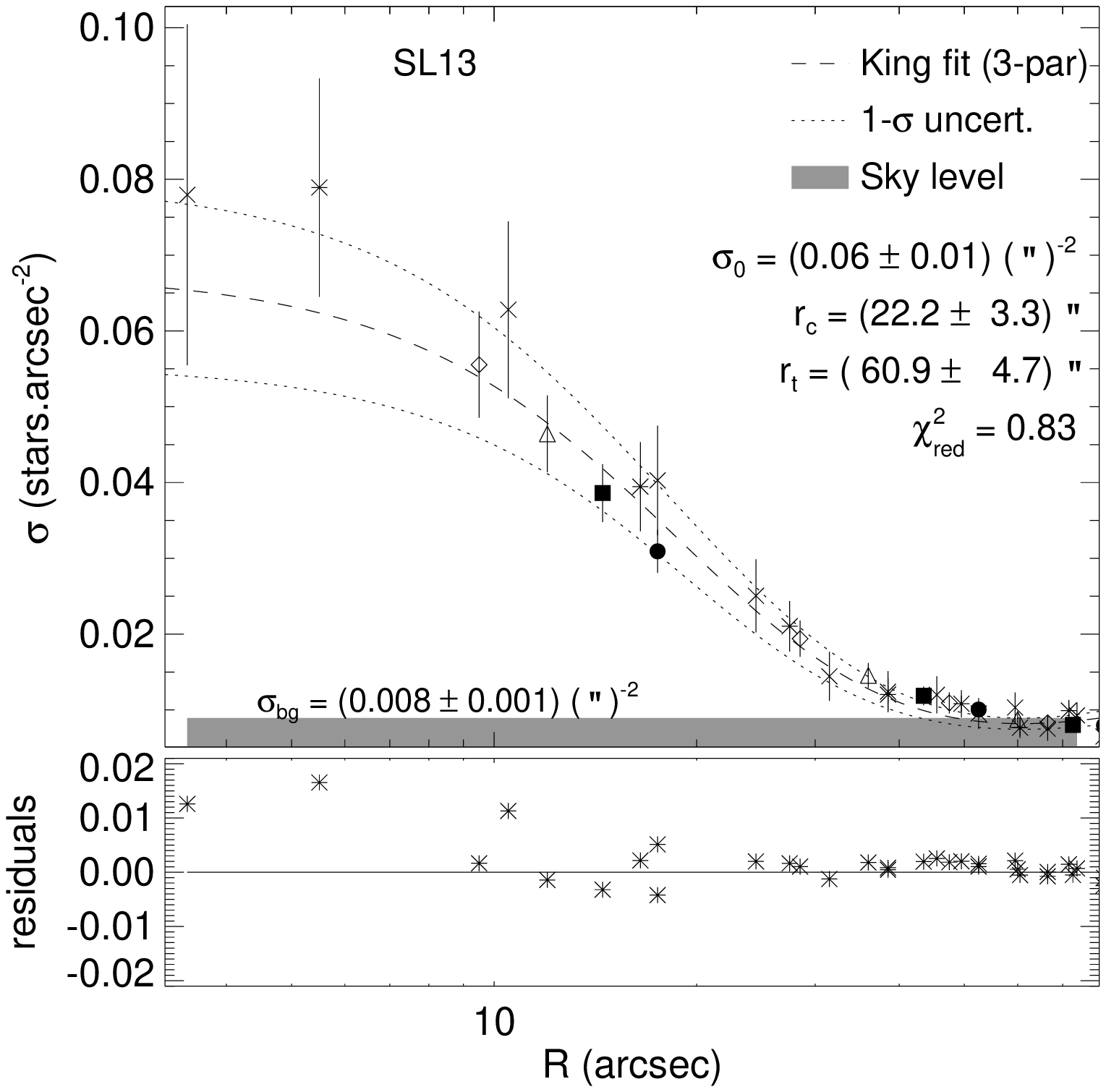}\includegraphics[width=0.325\linewidth]{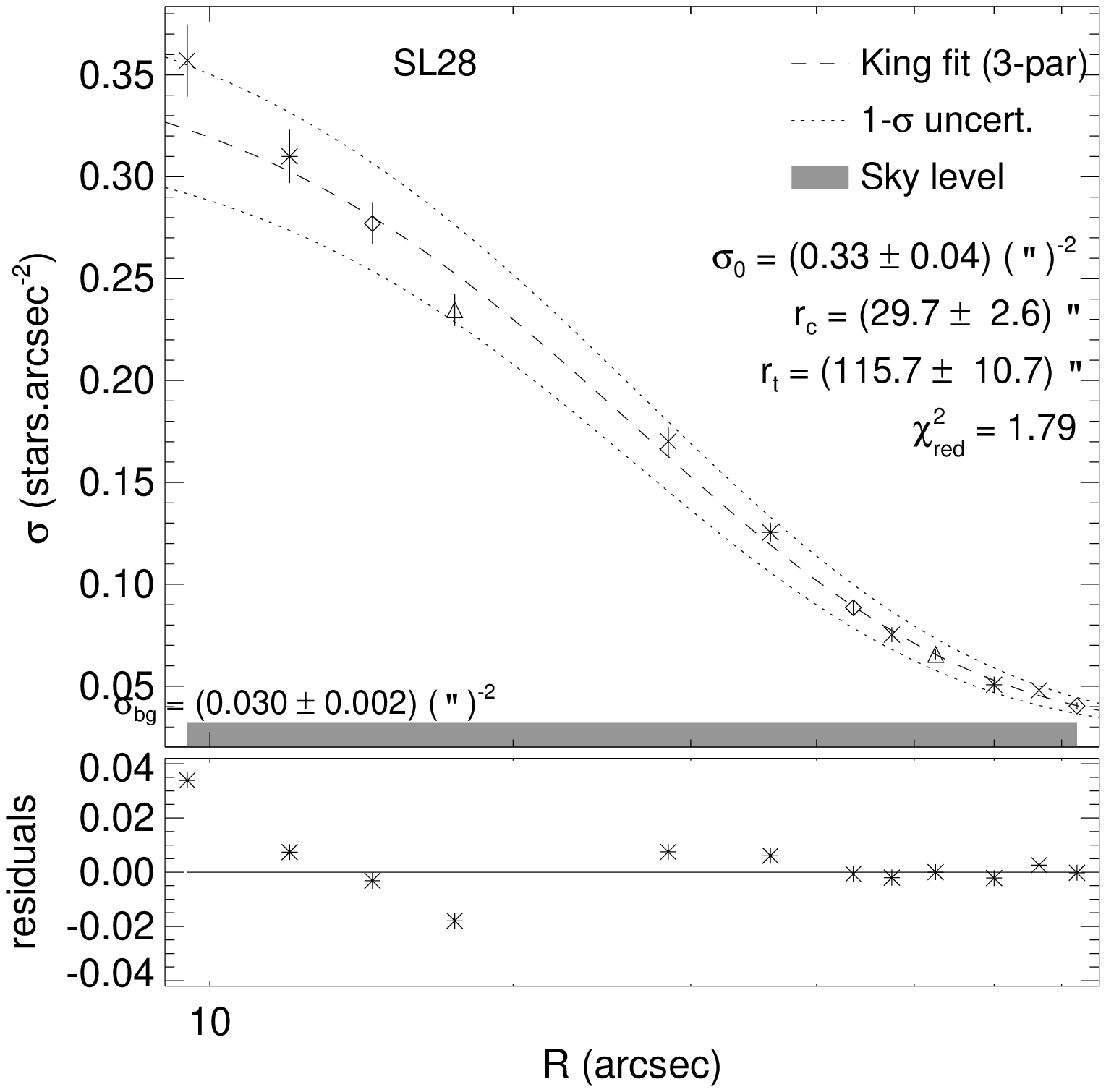}

\includegraphics[width=0.325\linewidth]{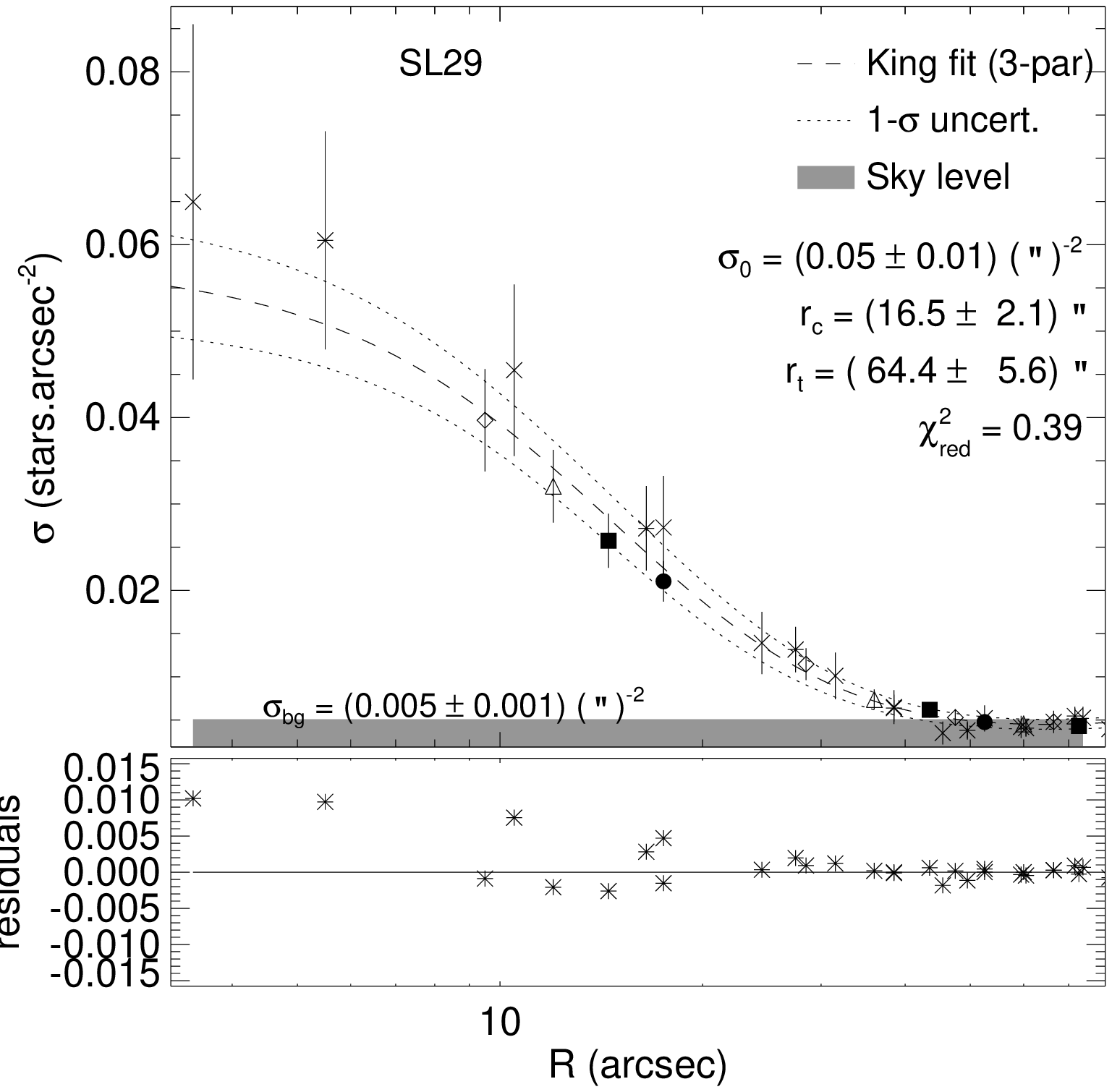}\includegraphics[width=0.325\linewidth]{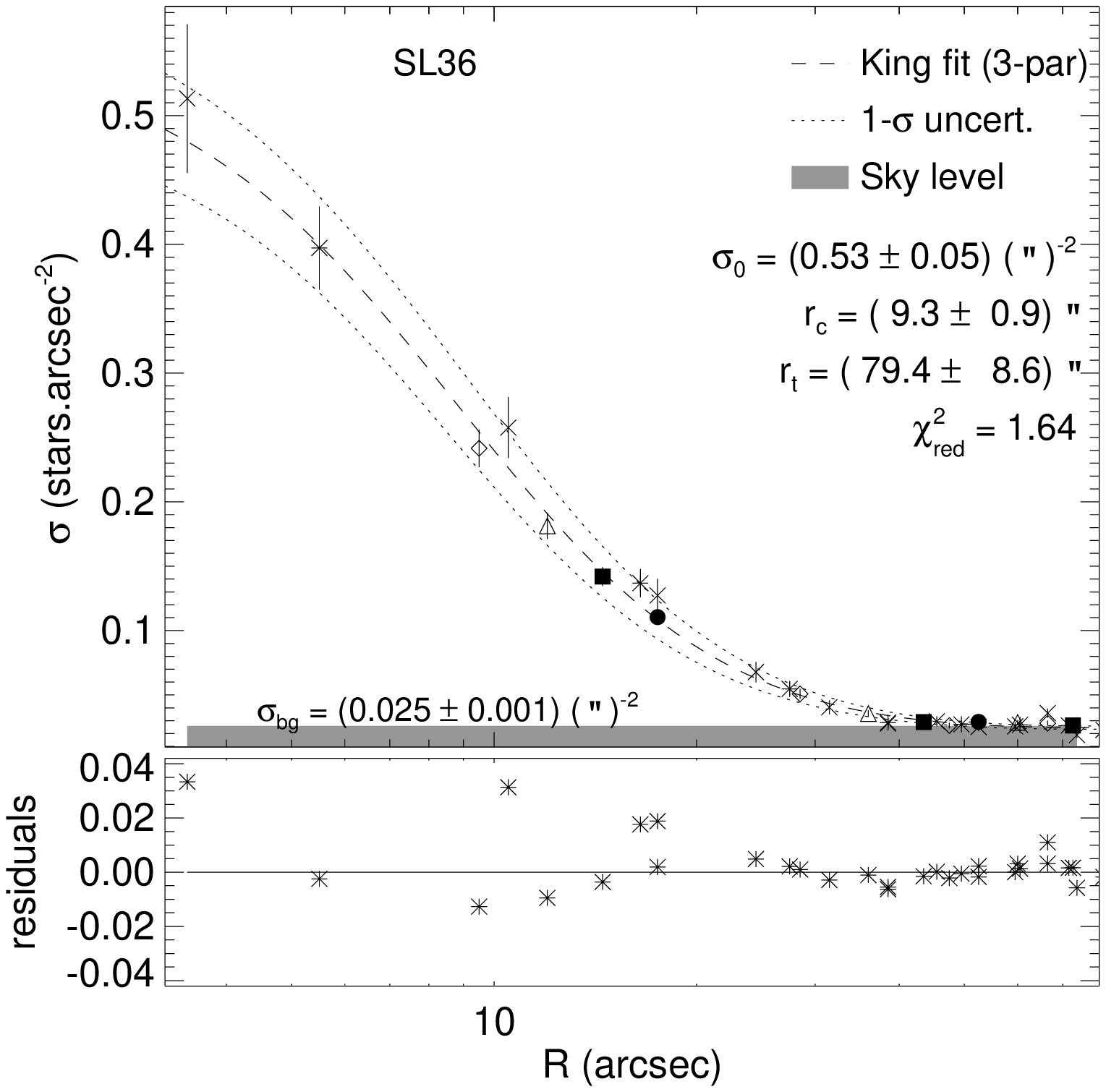}\includegraphics[width=0.325\linewidth]{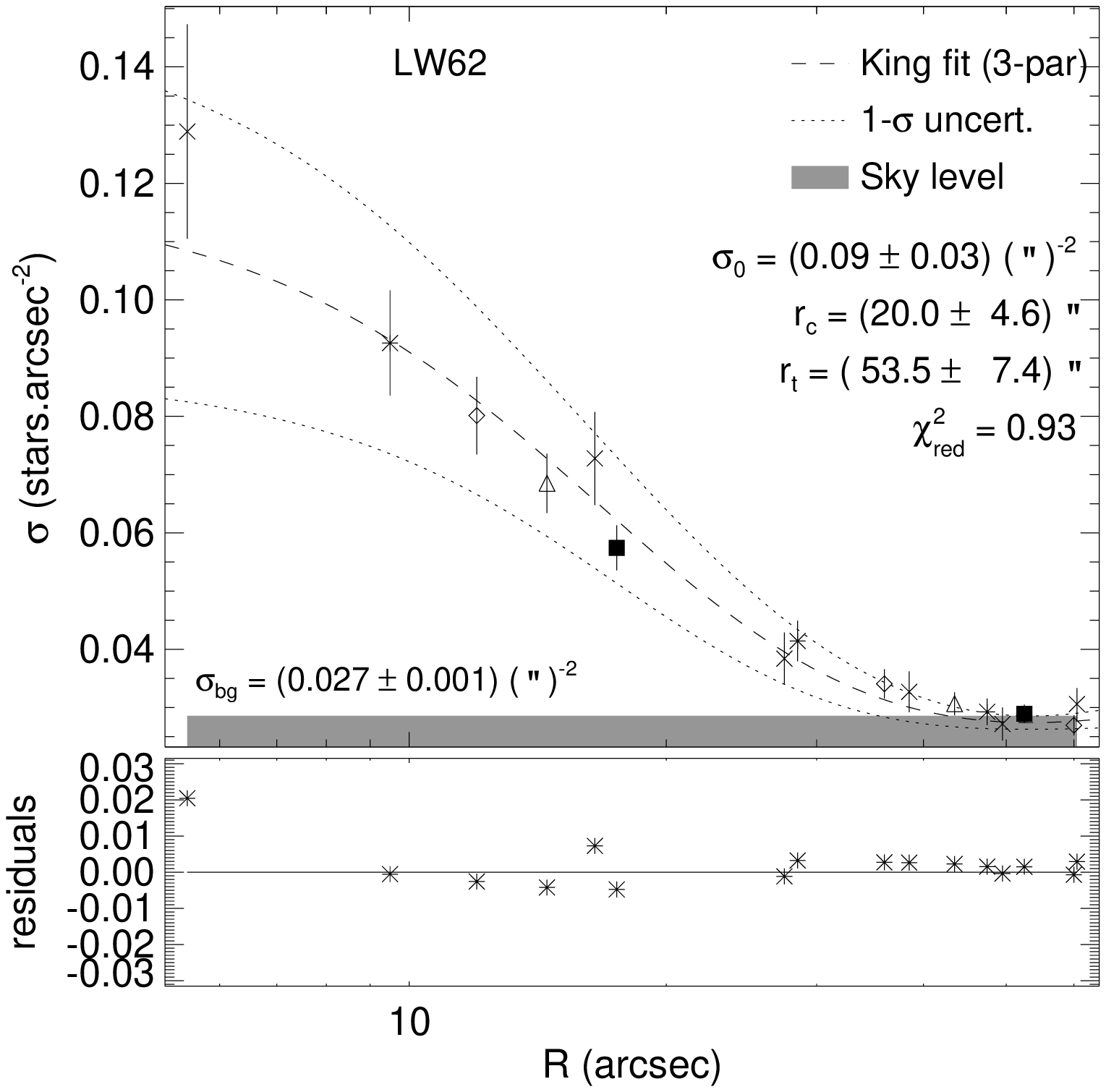}

\includegraphics[width=0.325\linewidth]{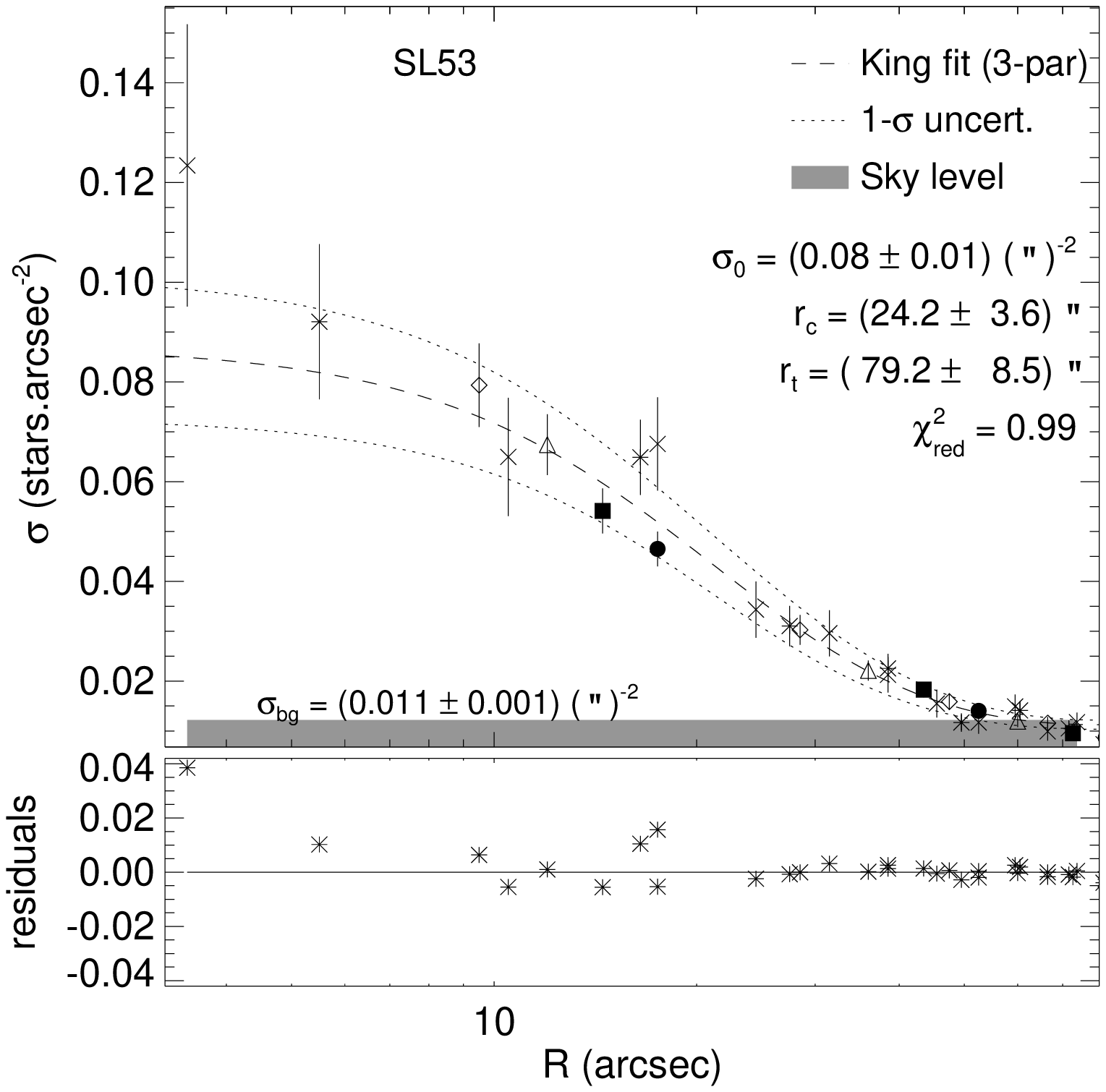}\includegraphics[width=0.325\linewidth]{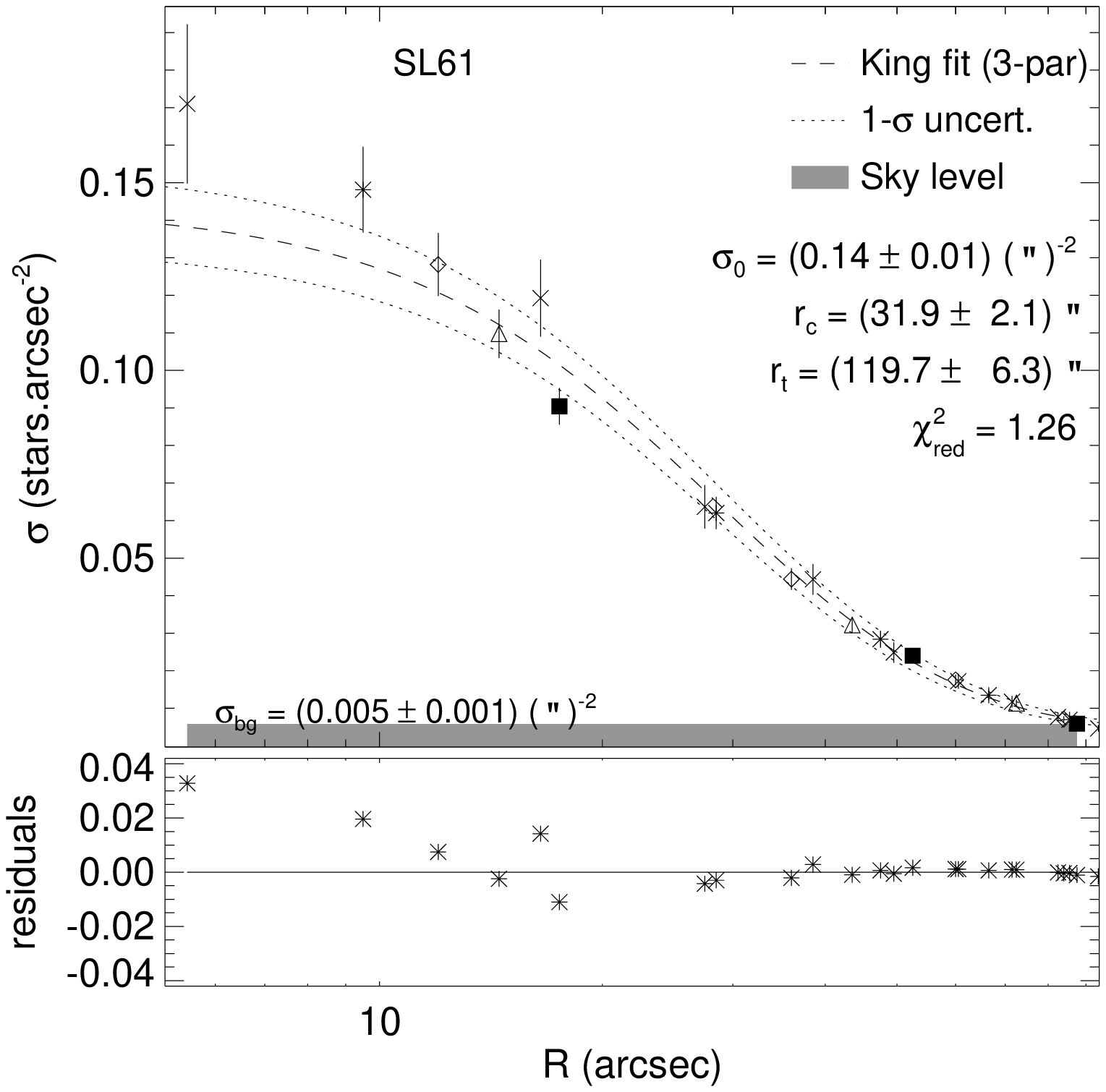}\includegraphics[width=0.325\linewidth]{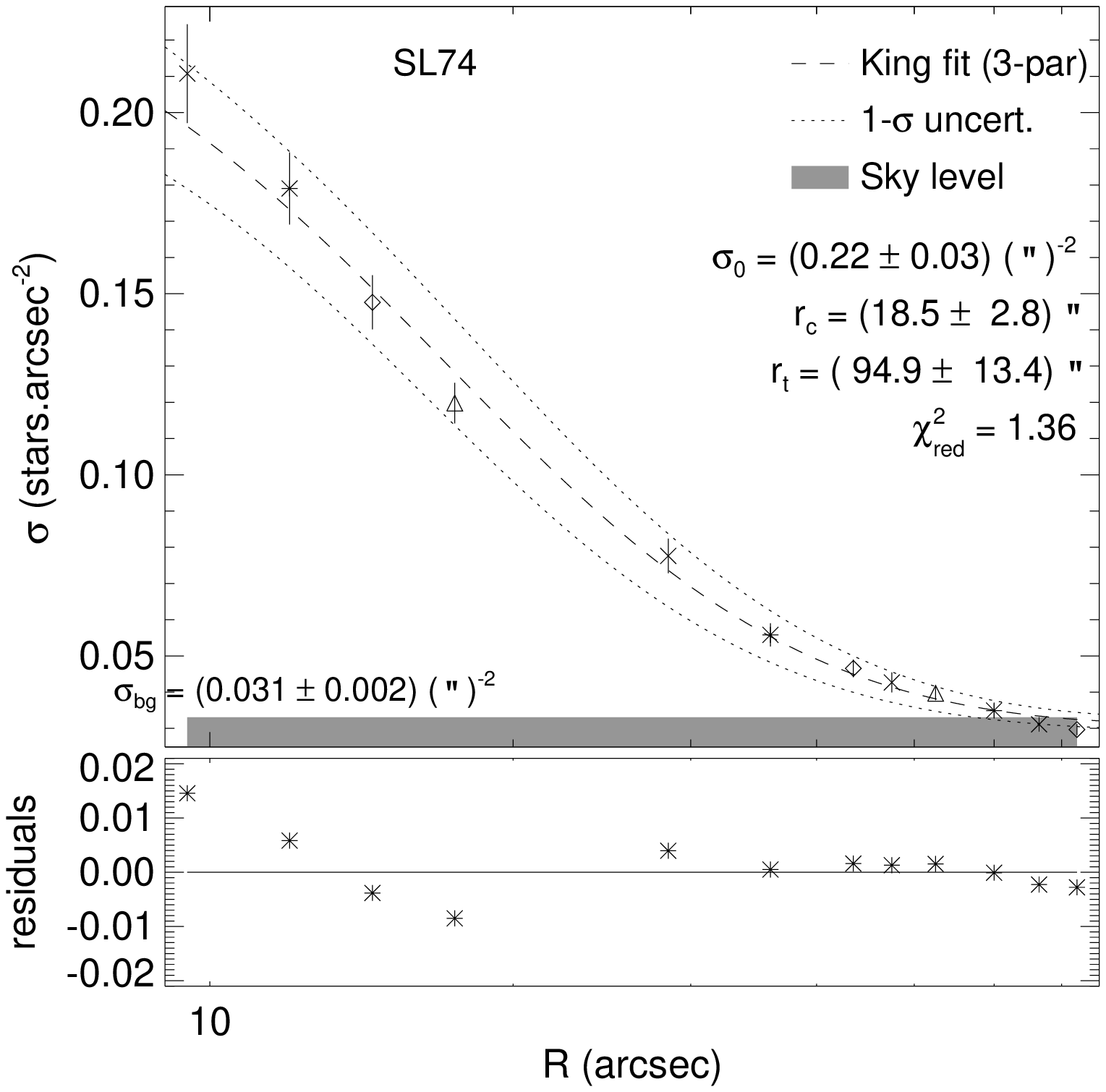}

\includegraphics[width=0.325\linewidth]{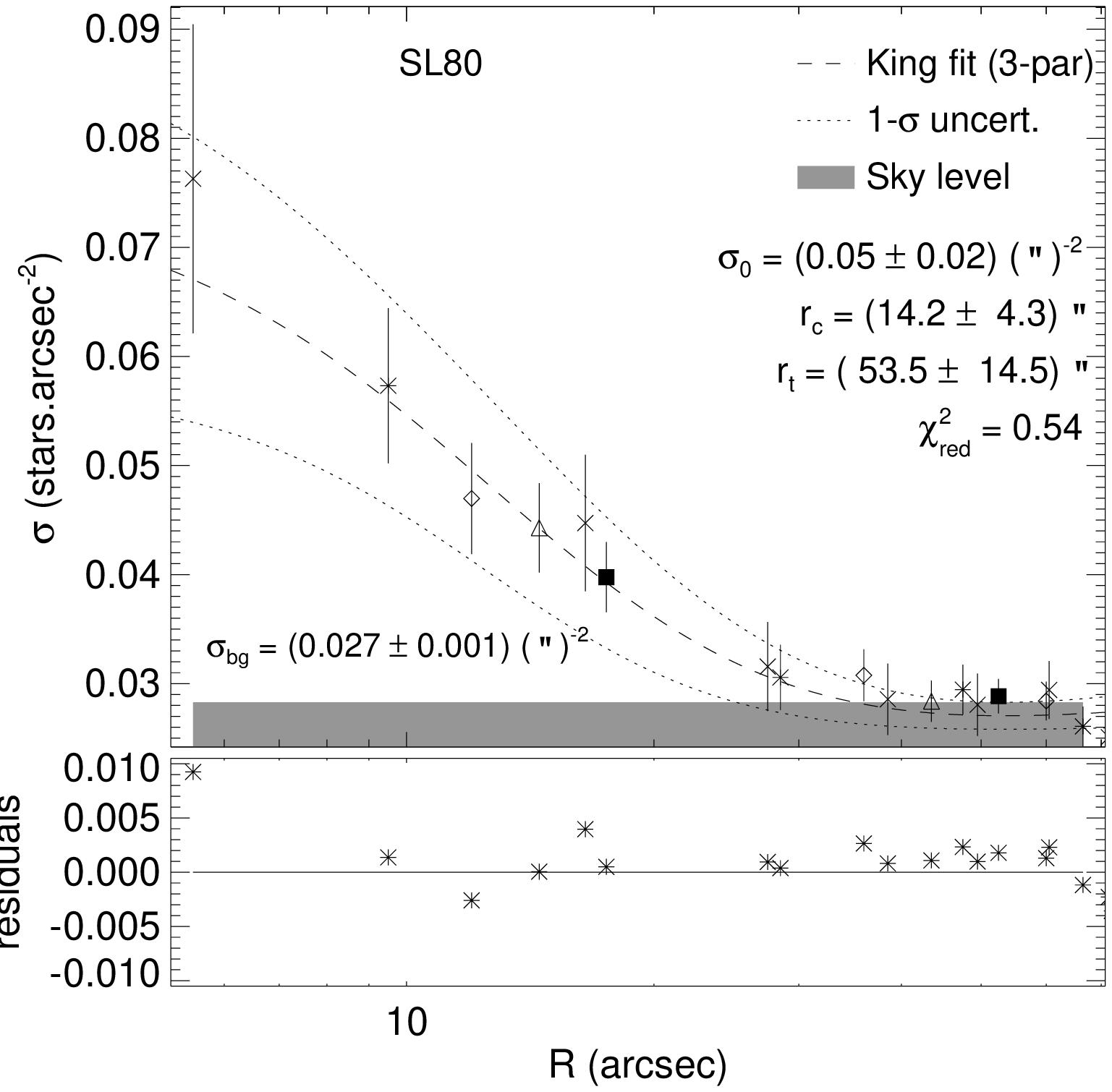}\includegraphics[width=0.325\linewidth]{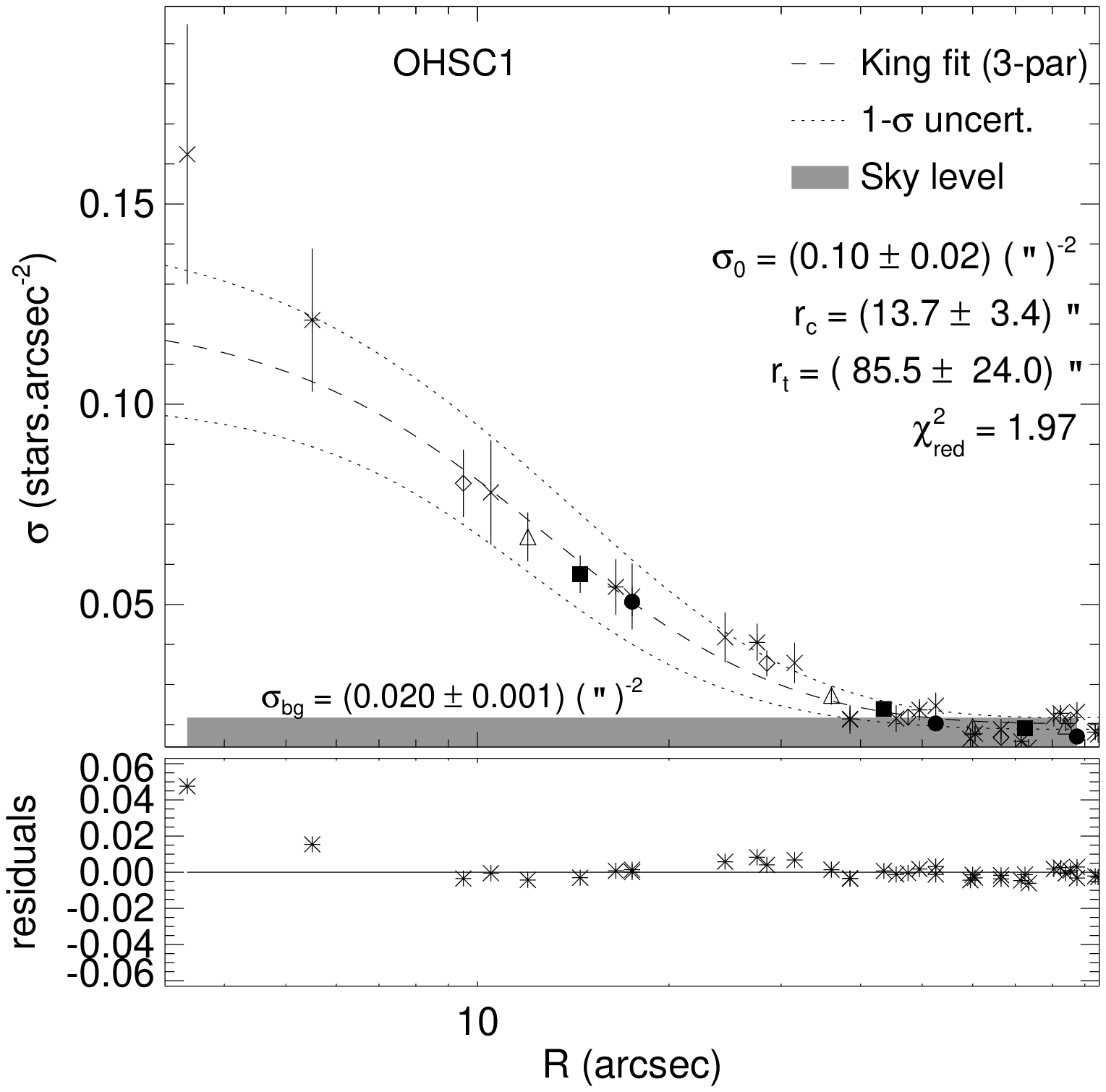}\includegraphics[width=0.325\linewidth]{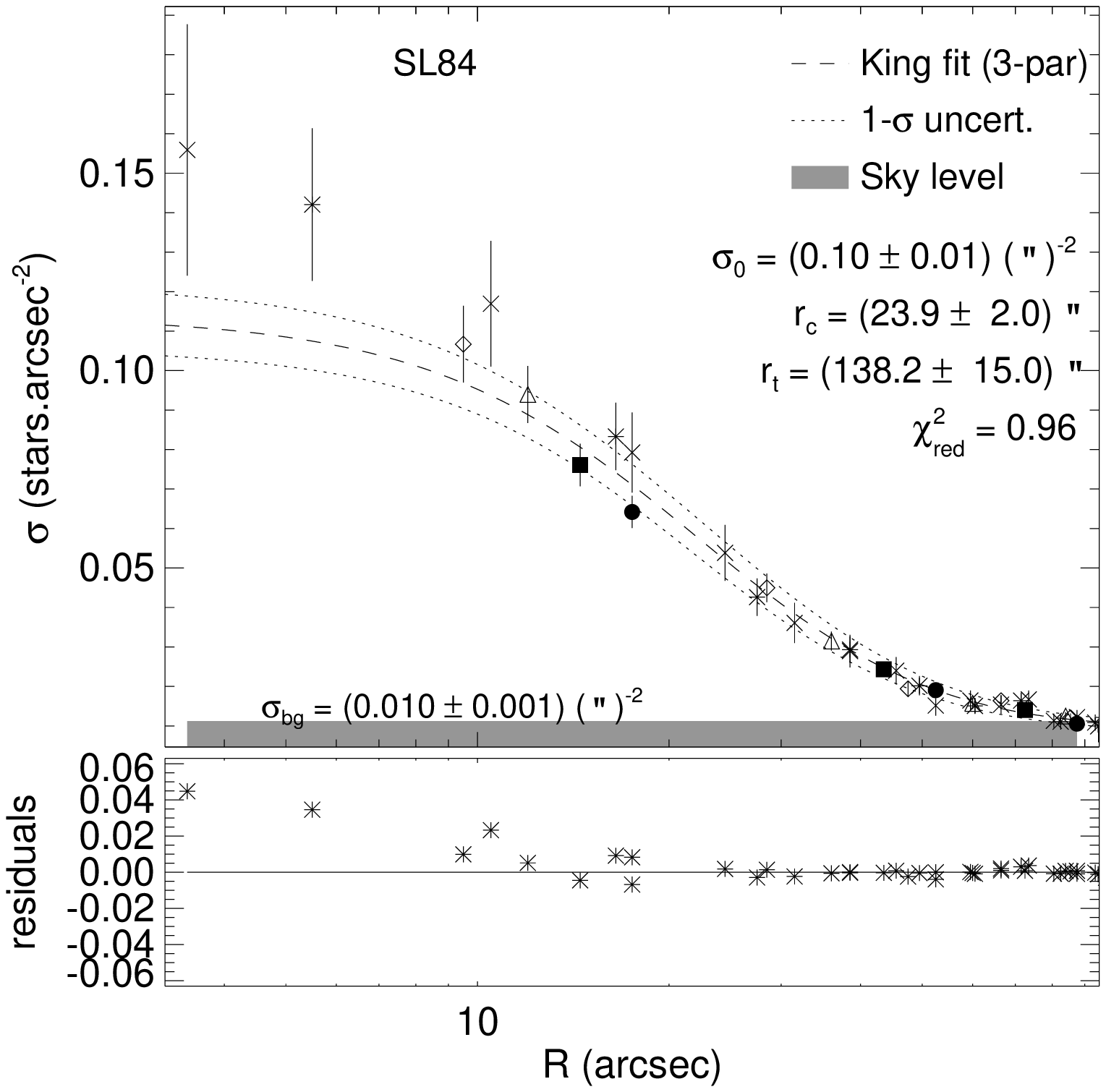}

\caption{Radial density profiles for additional LMC clusters complementing the sample presented in Fig.~\ref{fig:rdp_sbp}. The King model fits (dashed line) with envelopes of 1\,$\sigma$ uncertainty (dotted lines) are shown. Different symbols correspond to the various widths of the annular bins employed. The fitting residuals are also presented in the lower panel.}

\end{figure*}

\setcounter{figure}{1}
\begin{figure*}

\includegraphics[width=0.325\linewidth]{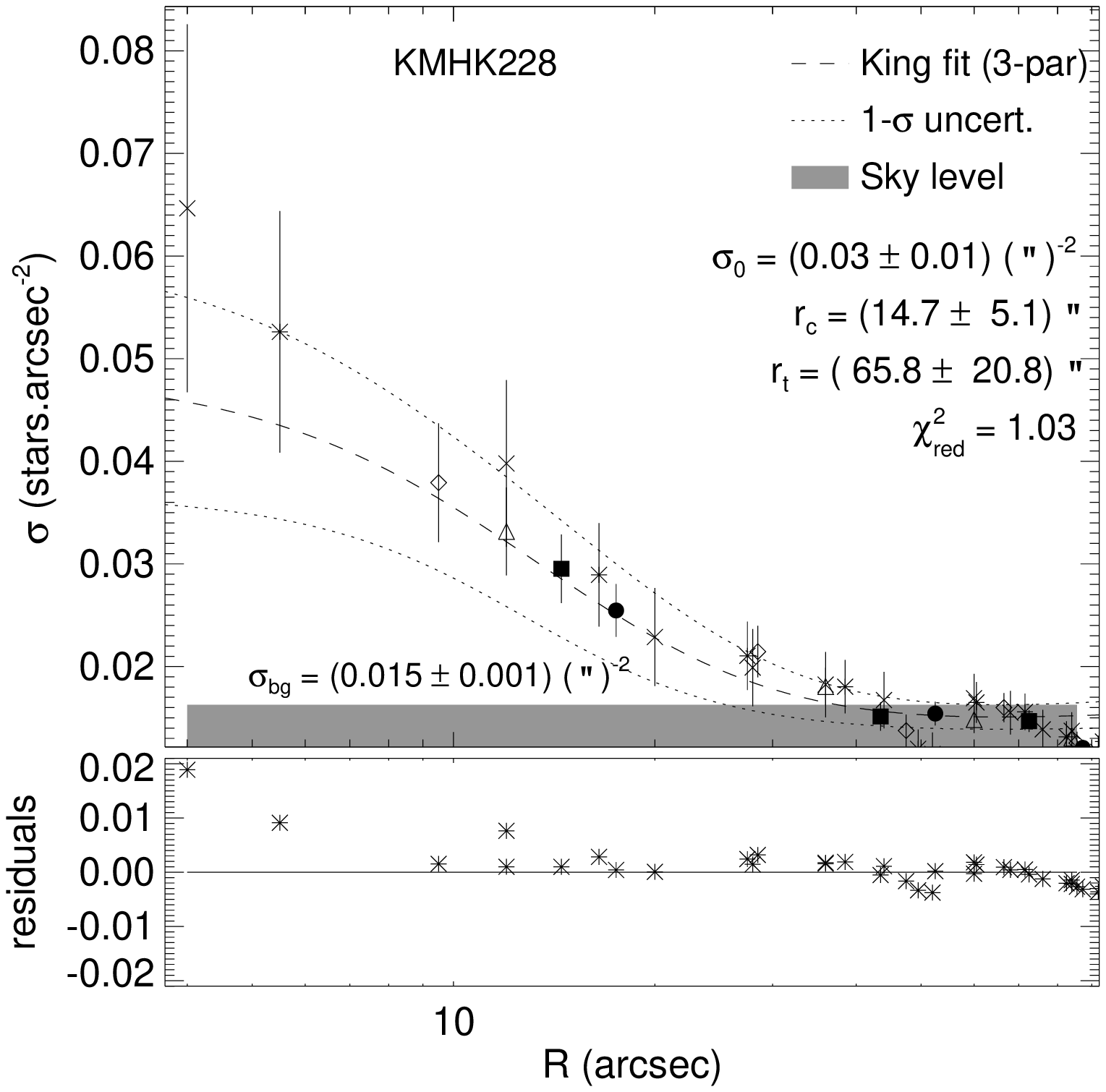}\includegraphics[width=0.325\linewidth]{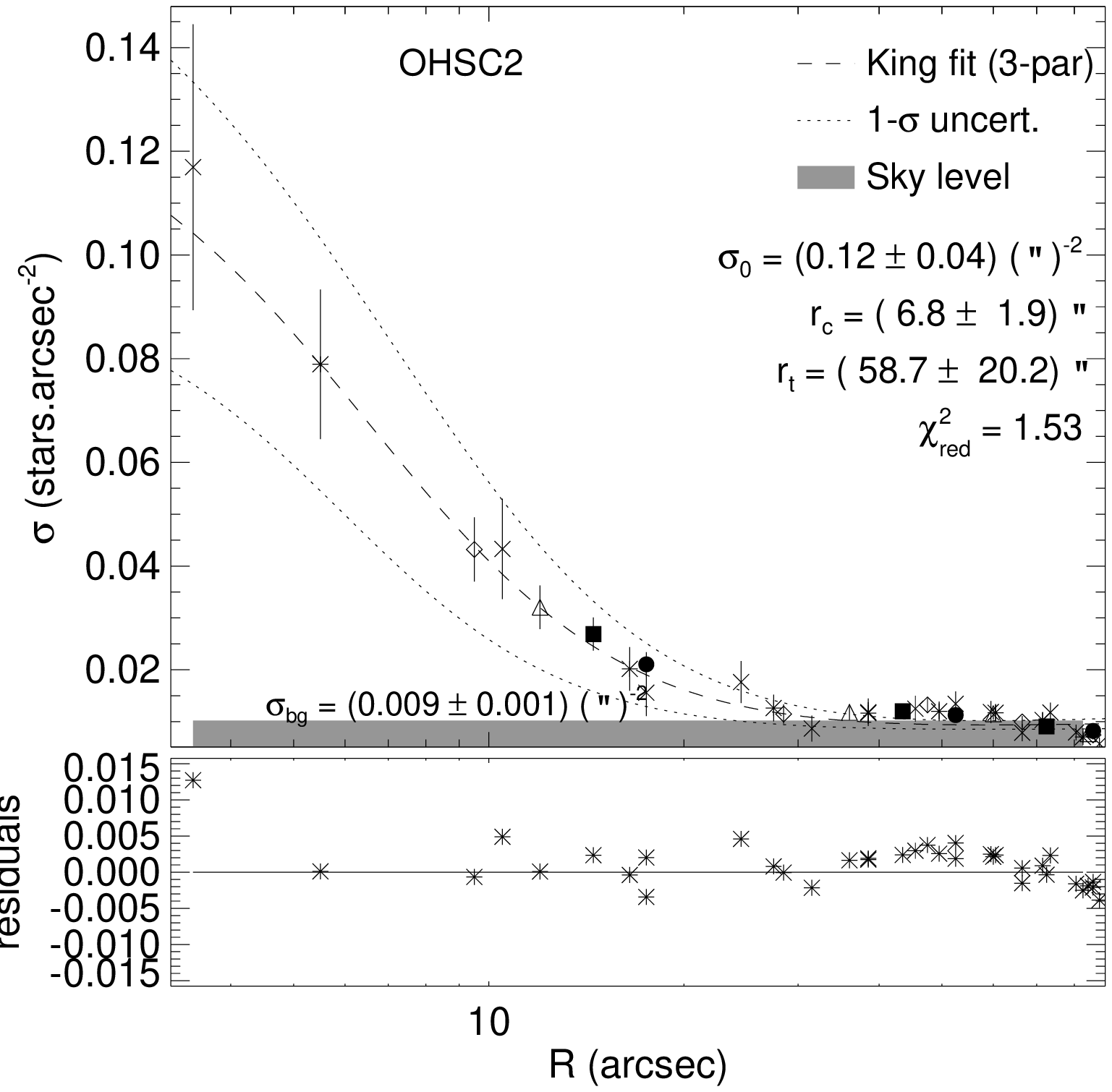}\includegraphics[width=0.325\linewidth]{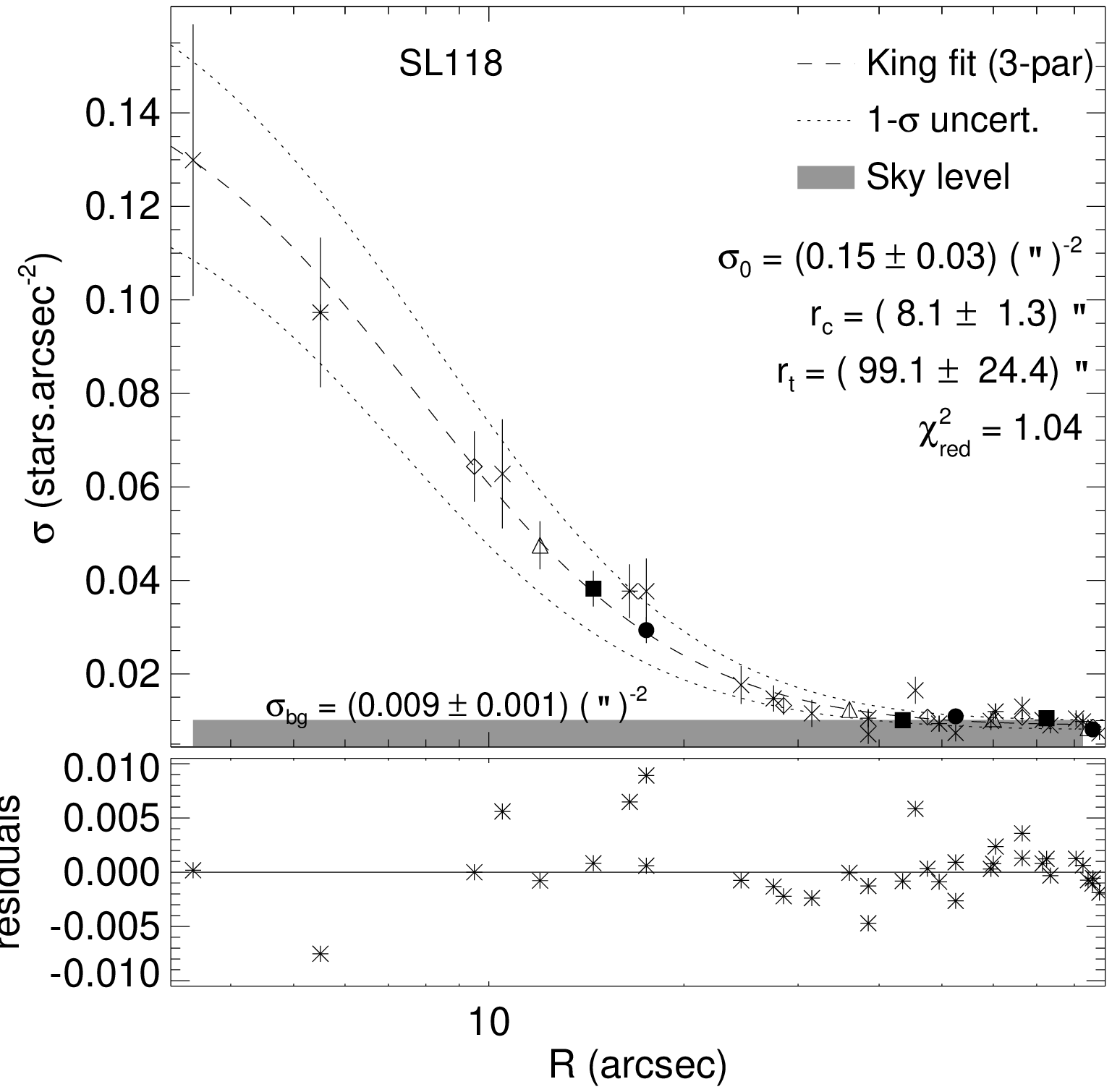}

\includegraphics[width=0.325\linewidth]{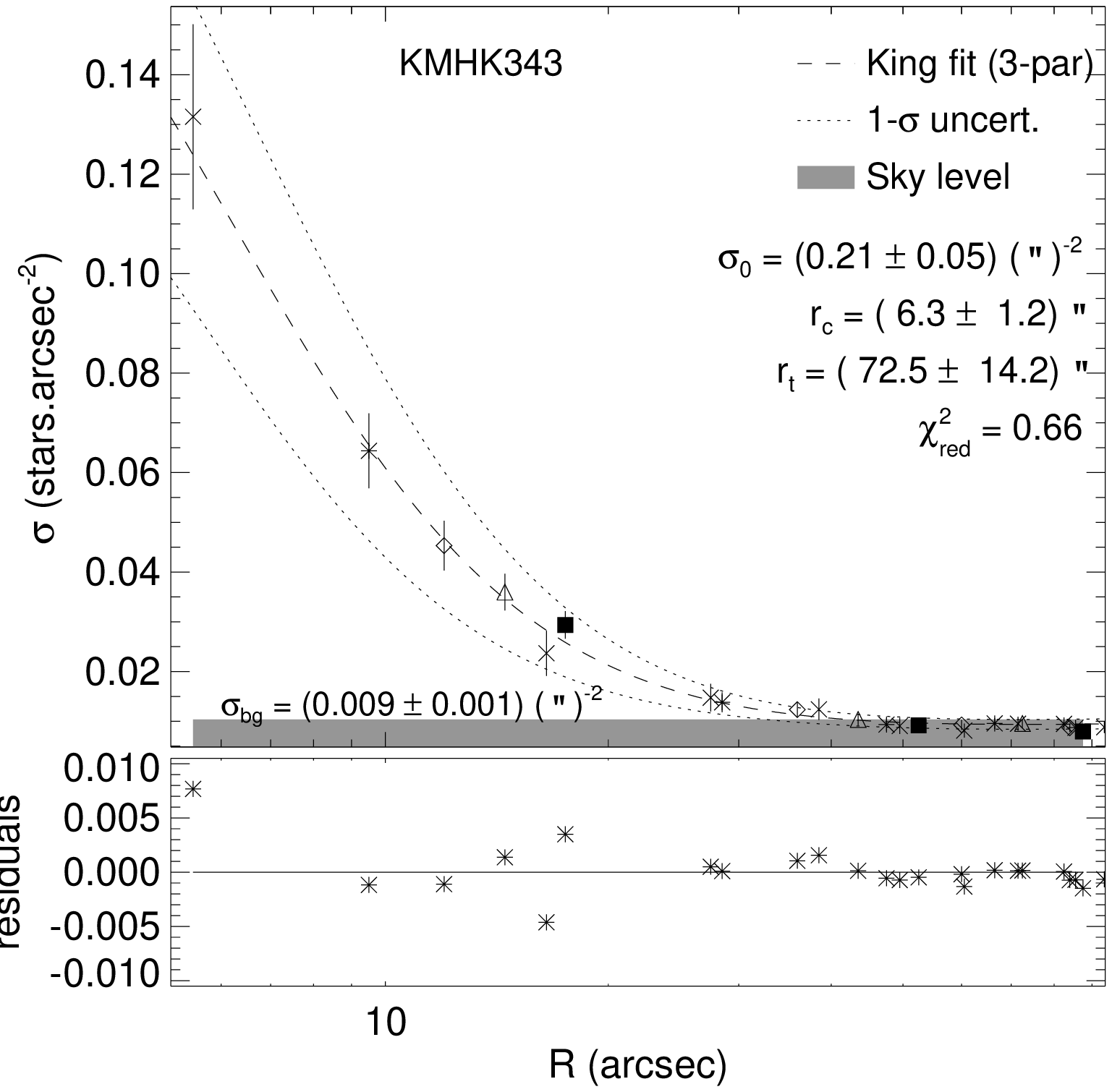}\includegraphics[width=0.325\linewidth]{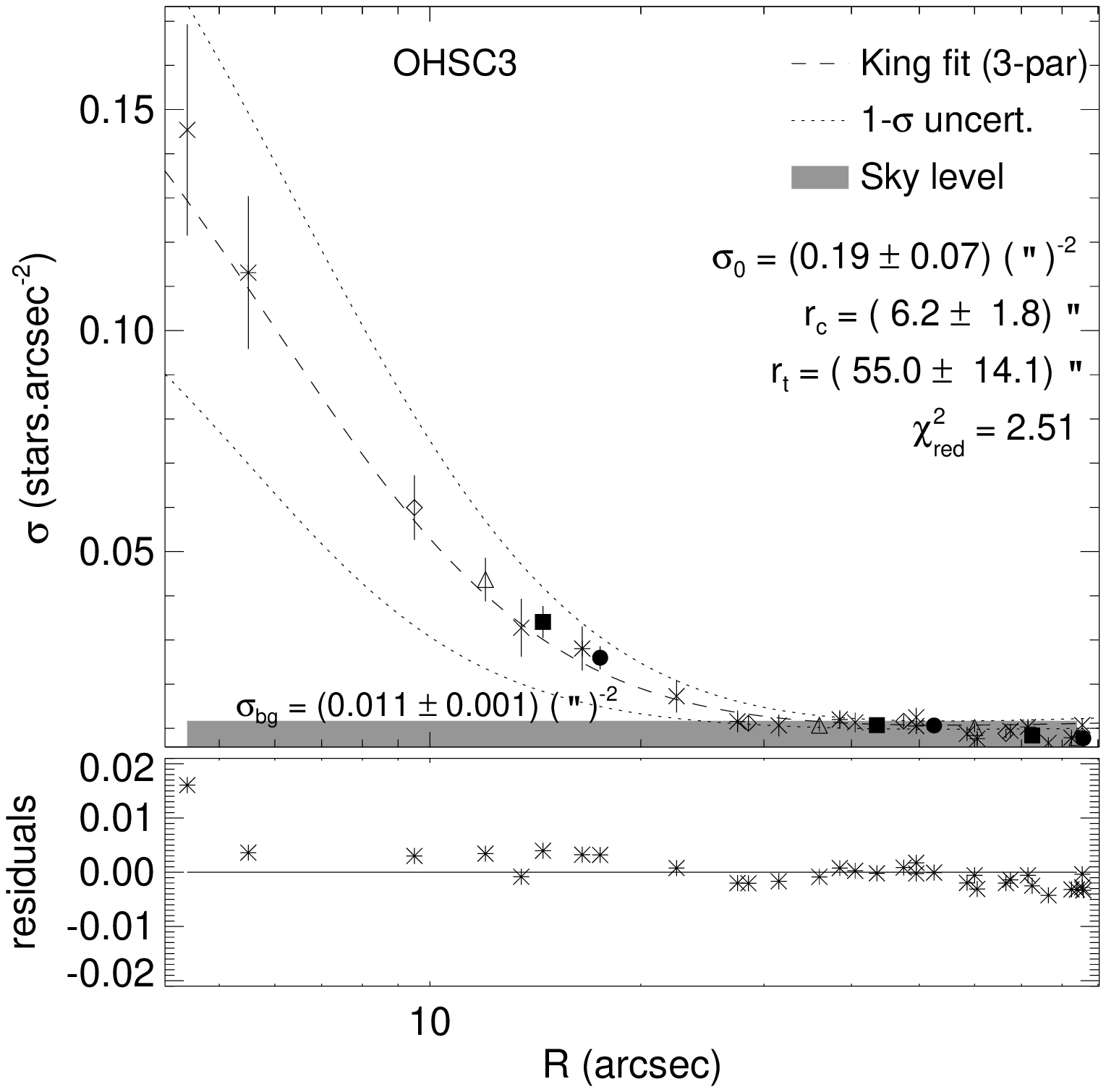}\includegraphics[width=0.325\linewidth]{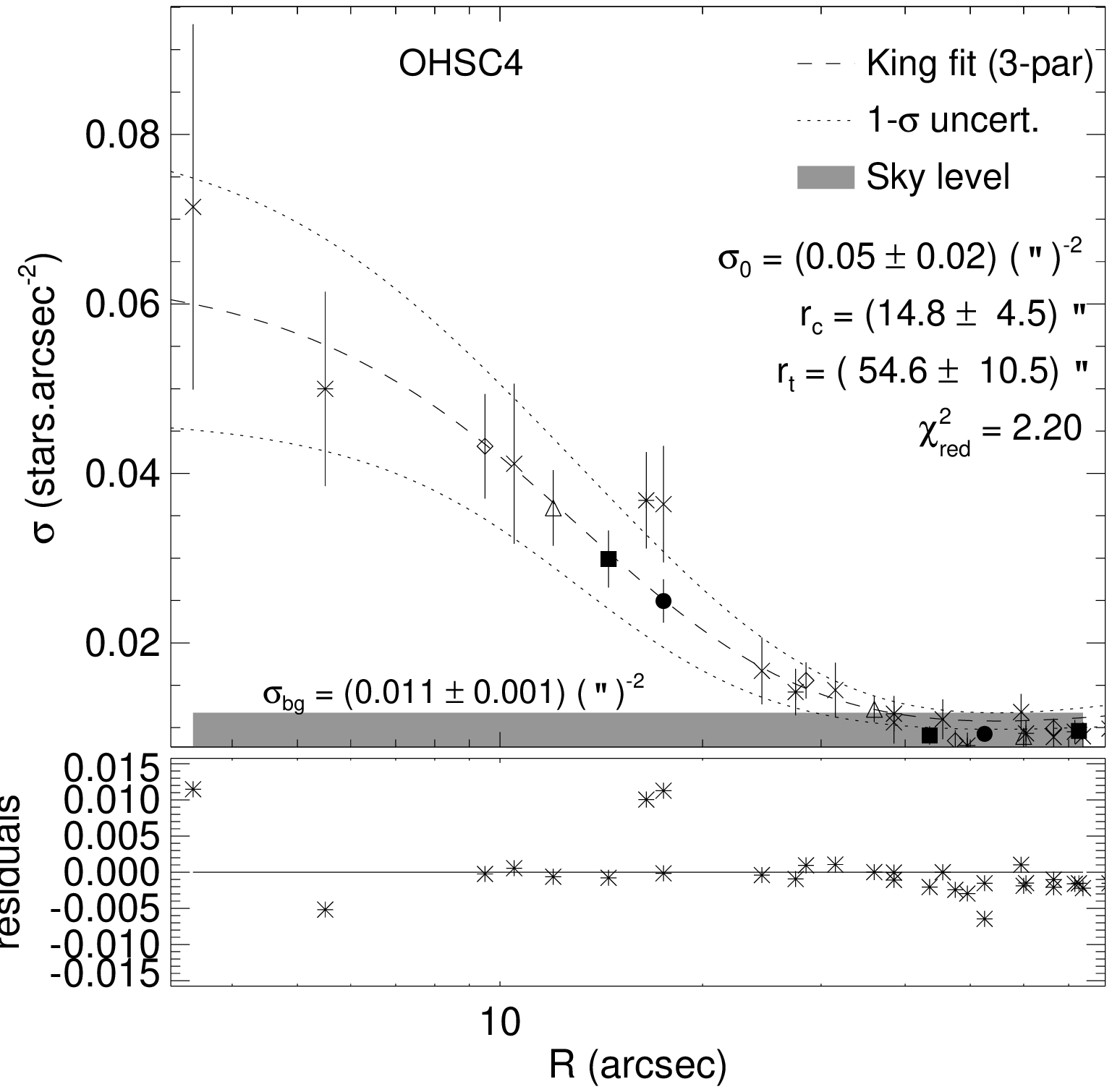}

\includegraphics[width=0.325\linewidth]{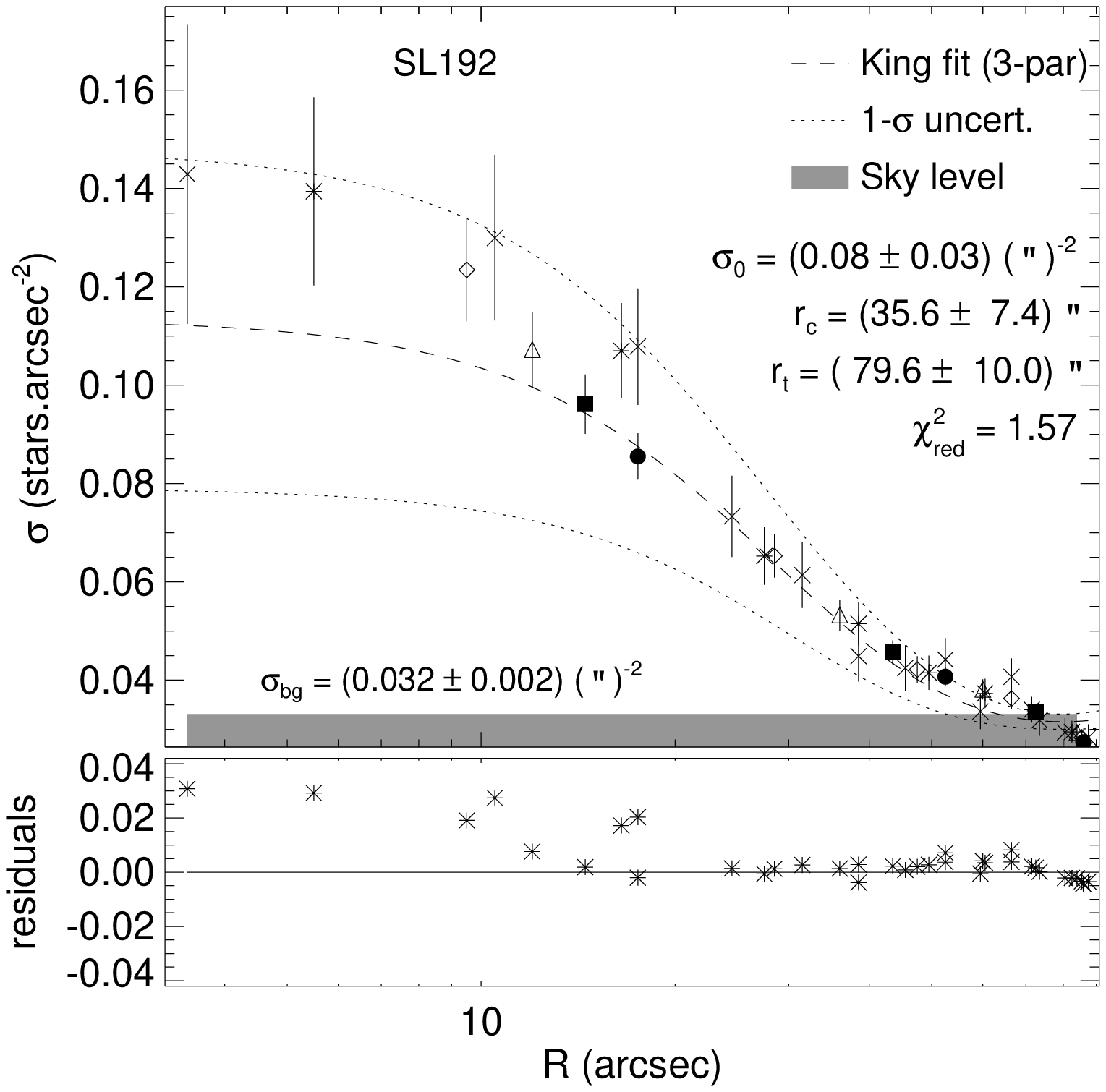}\includegraphics[width=0.325\linewidth]{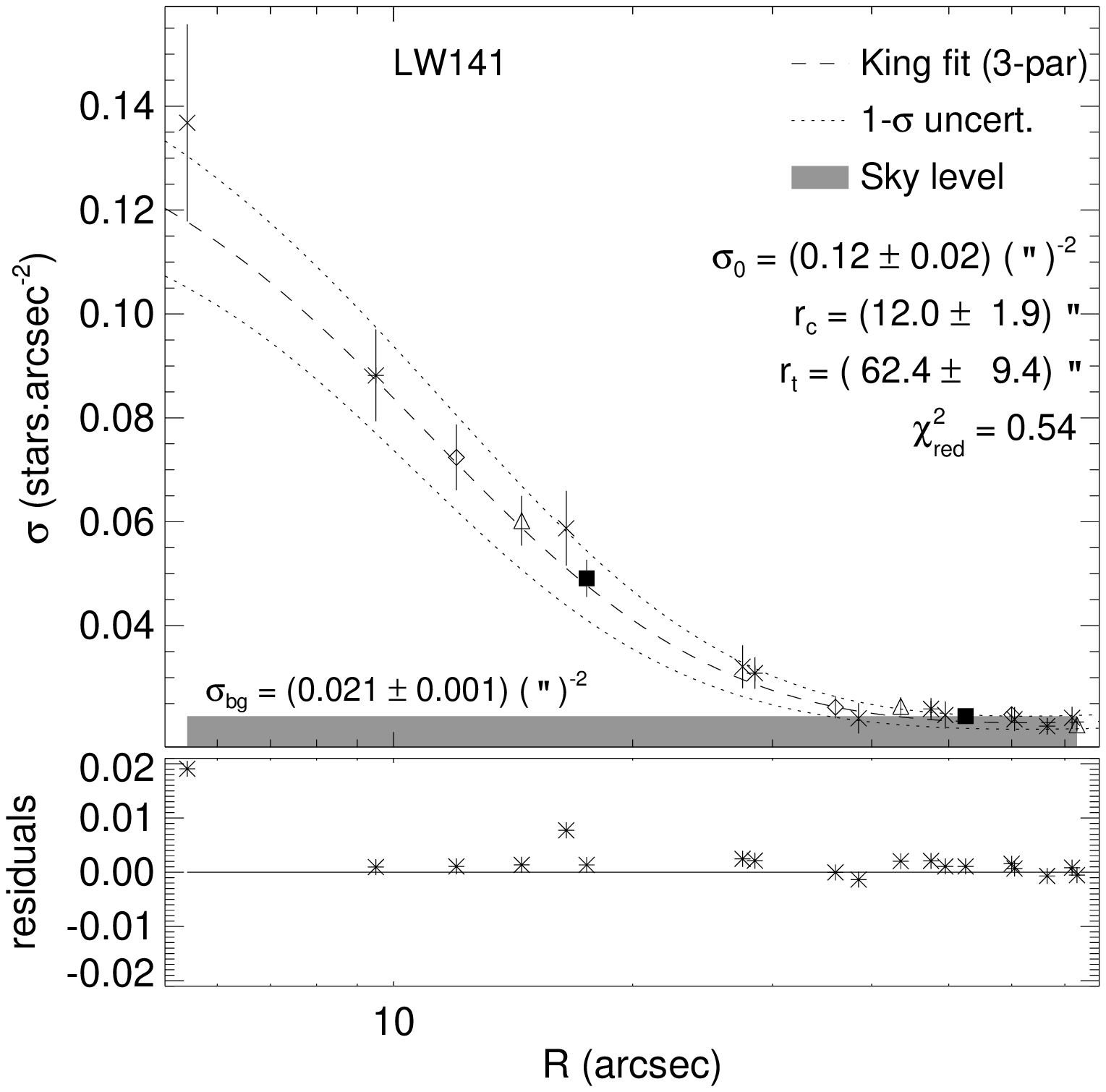}\includegraphics[width=0.325\linewidth]{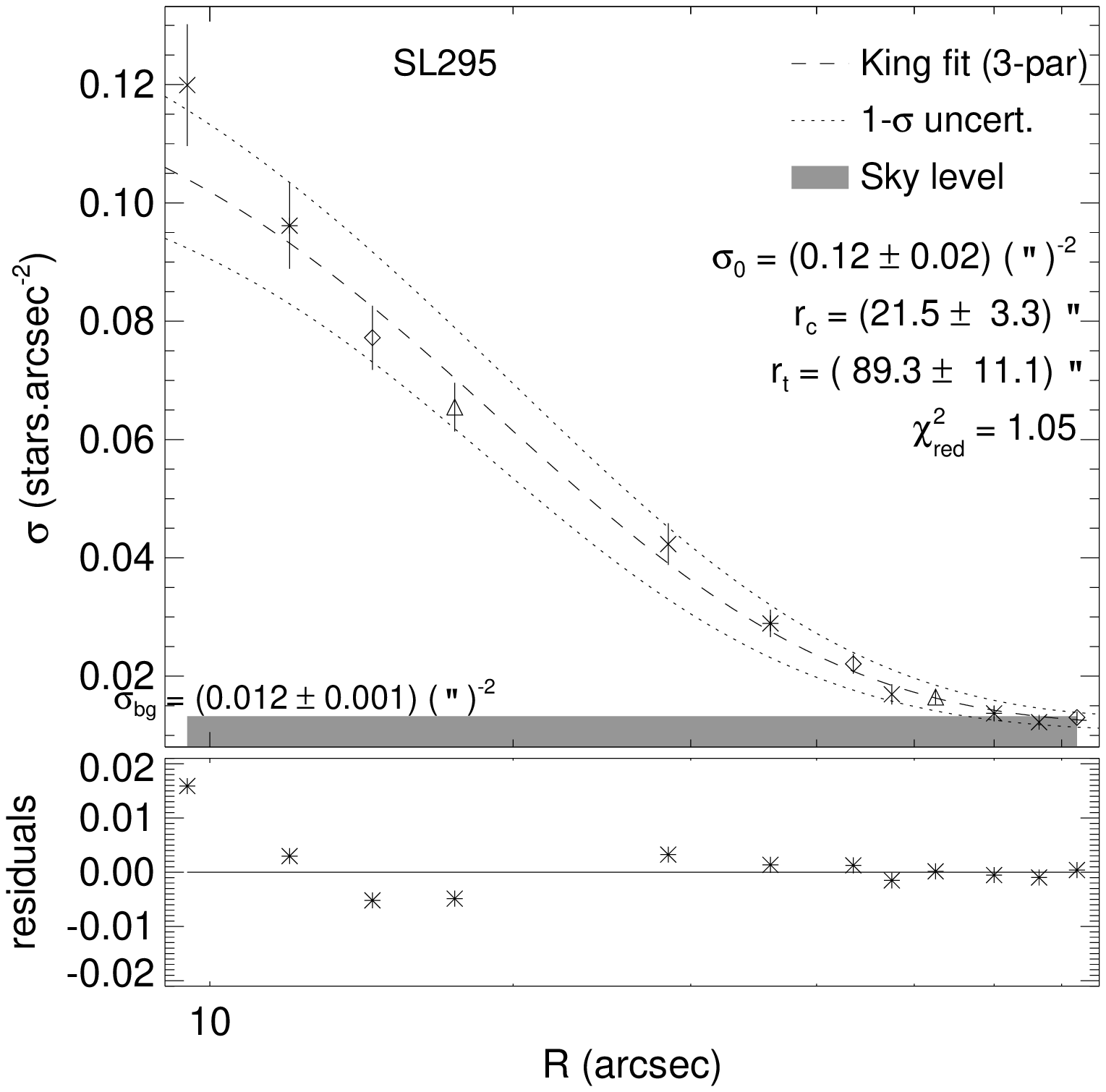}

\includegraphics[width=0.325\linewidth]{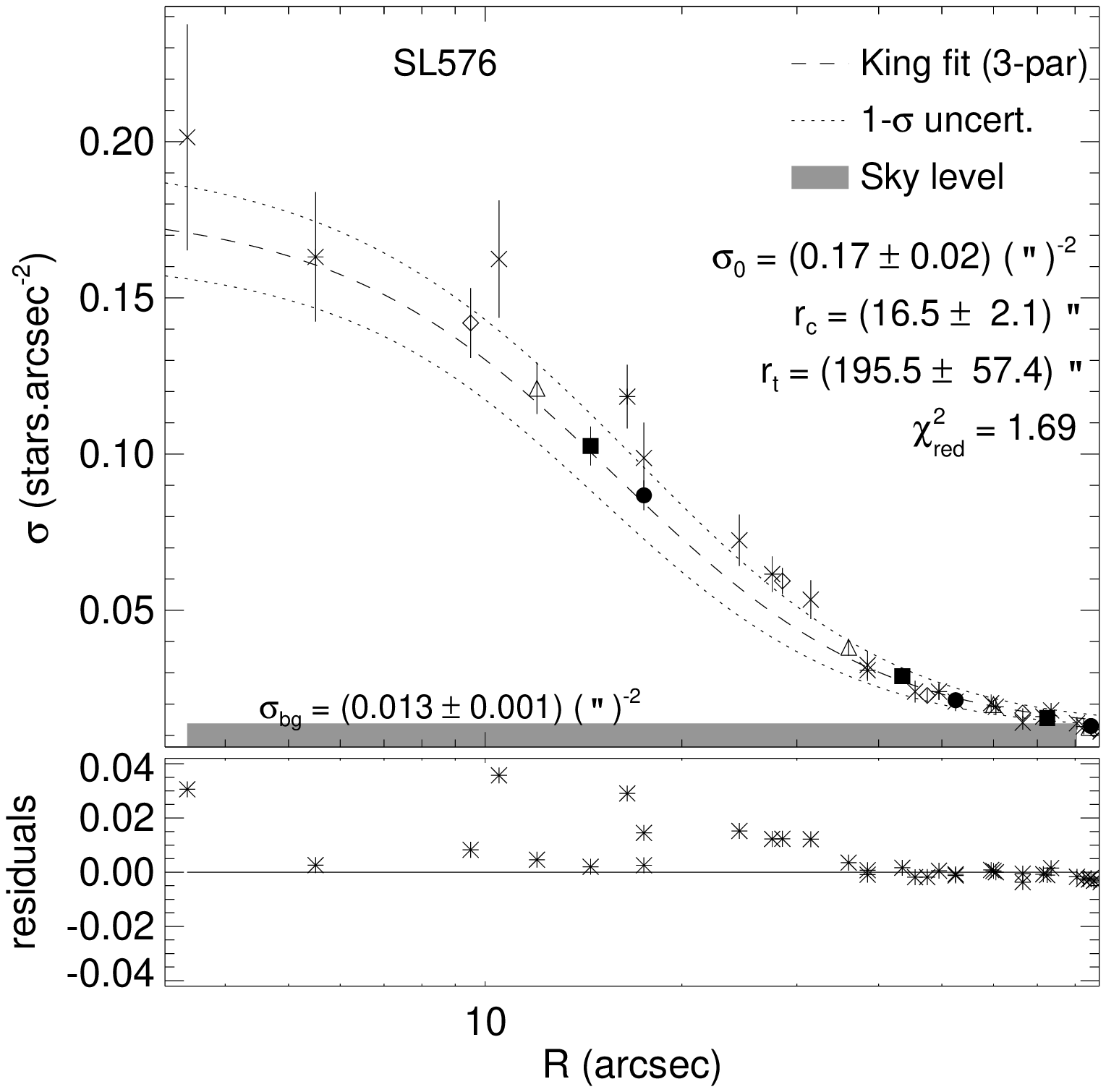}\includegraphics[width=0.325\linewidth]{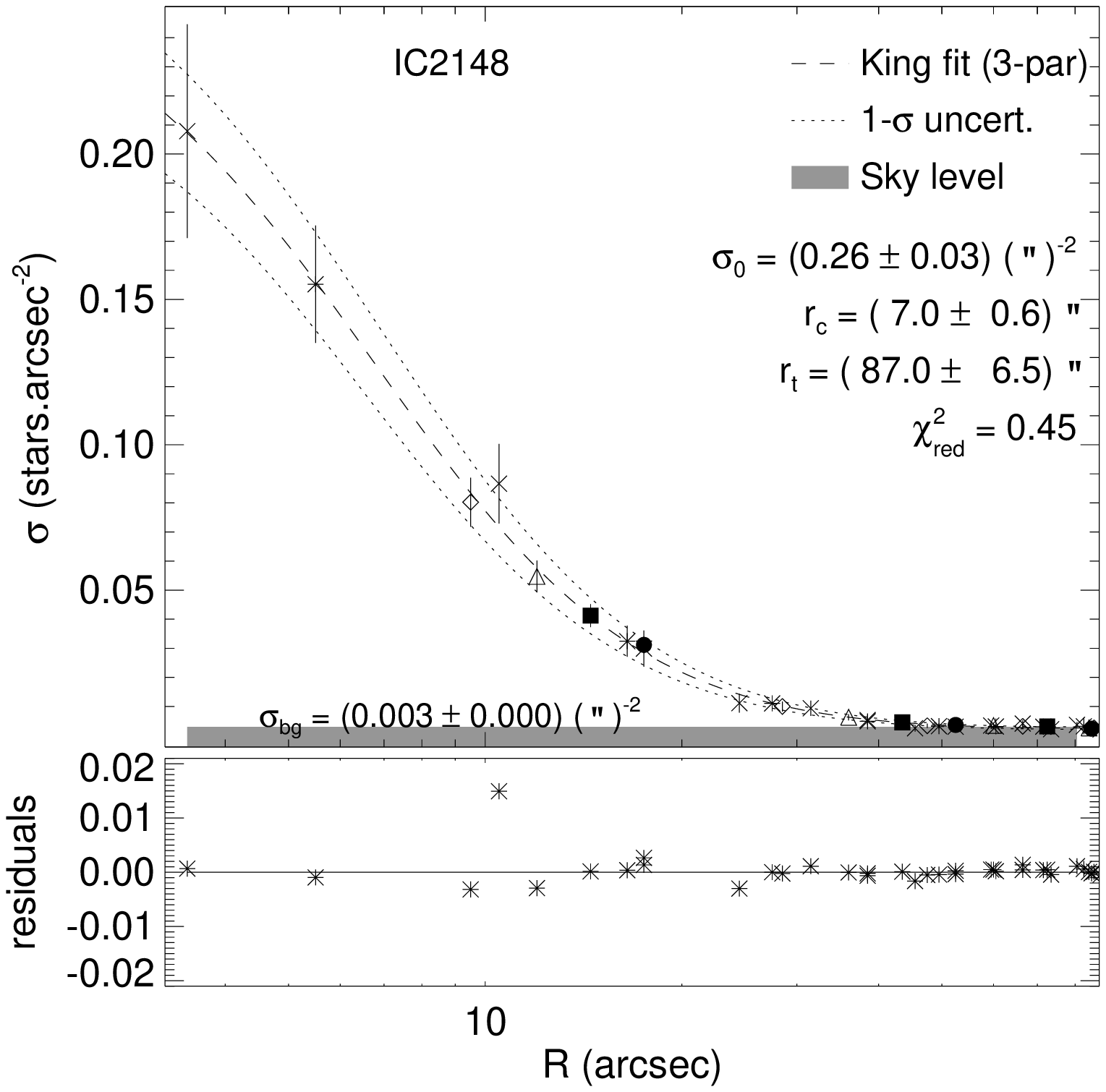}\includegraphics[width=0.325\linewidth]{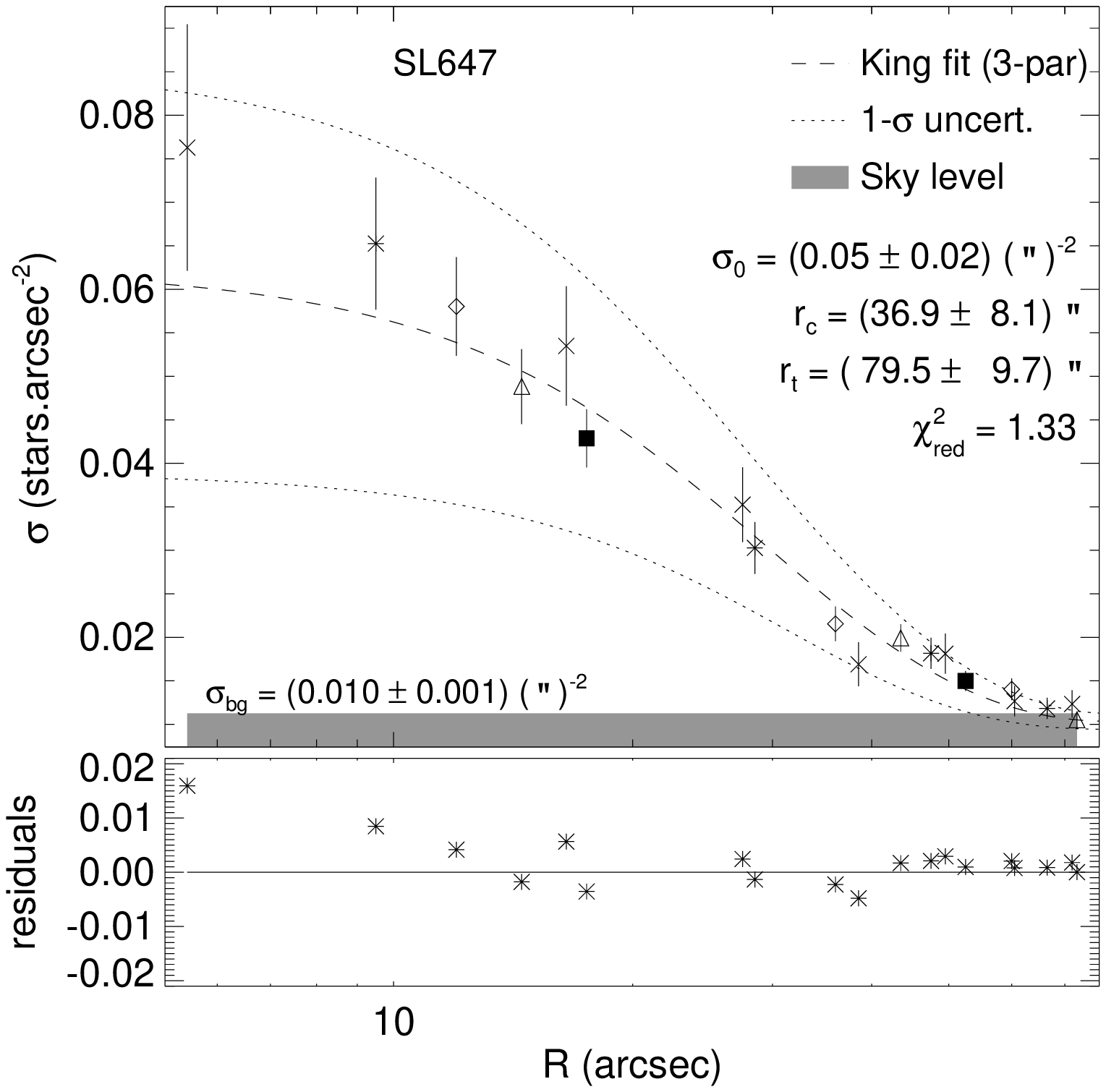}

\caption{cont.}

\end{figure*}

\setcounter{figure}{1}
\begin{figure*}

\includegraphics[width=0.325\linewidth]{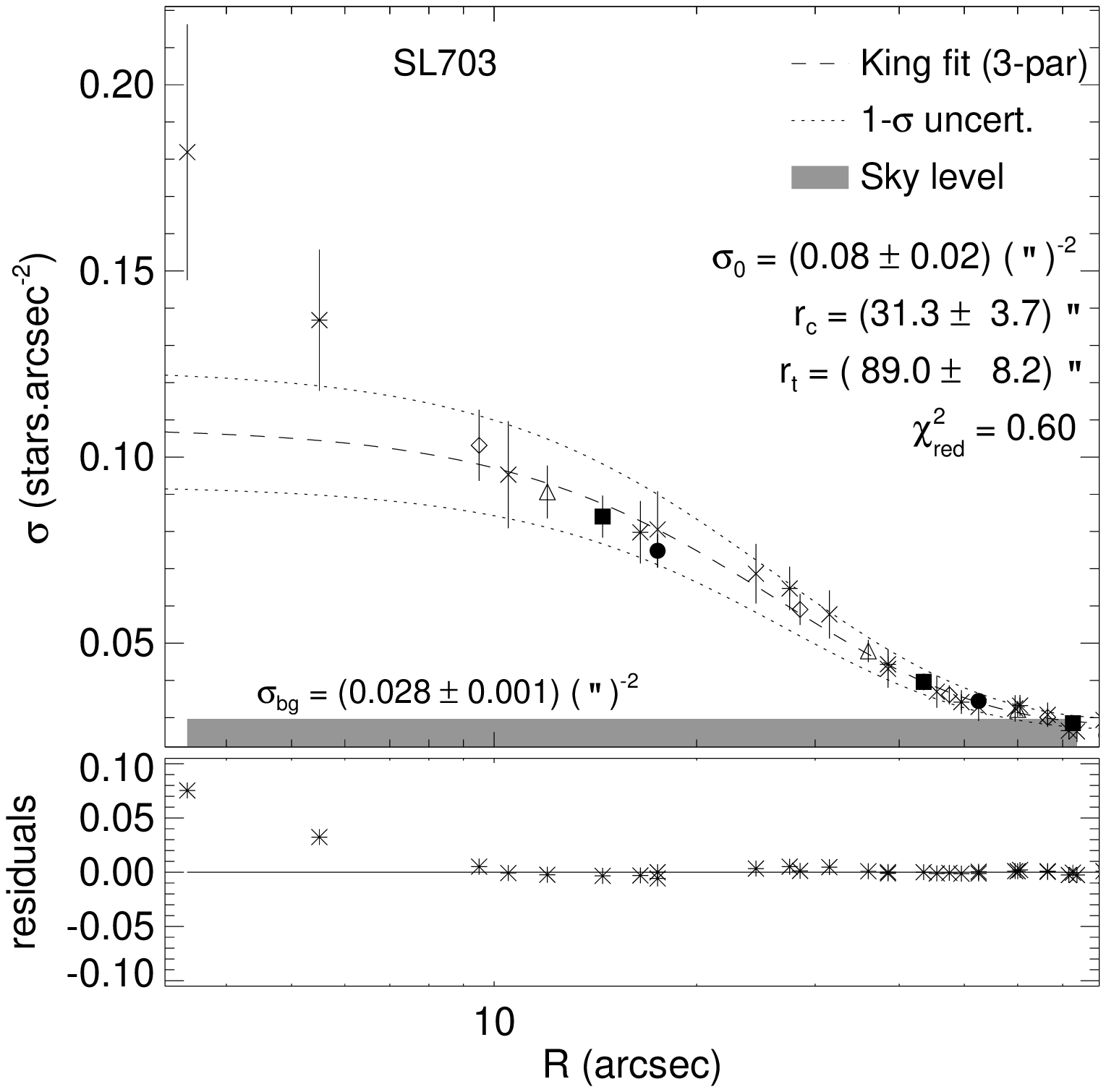}\includegraphics[width=0.325\linewidth]{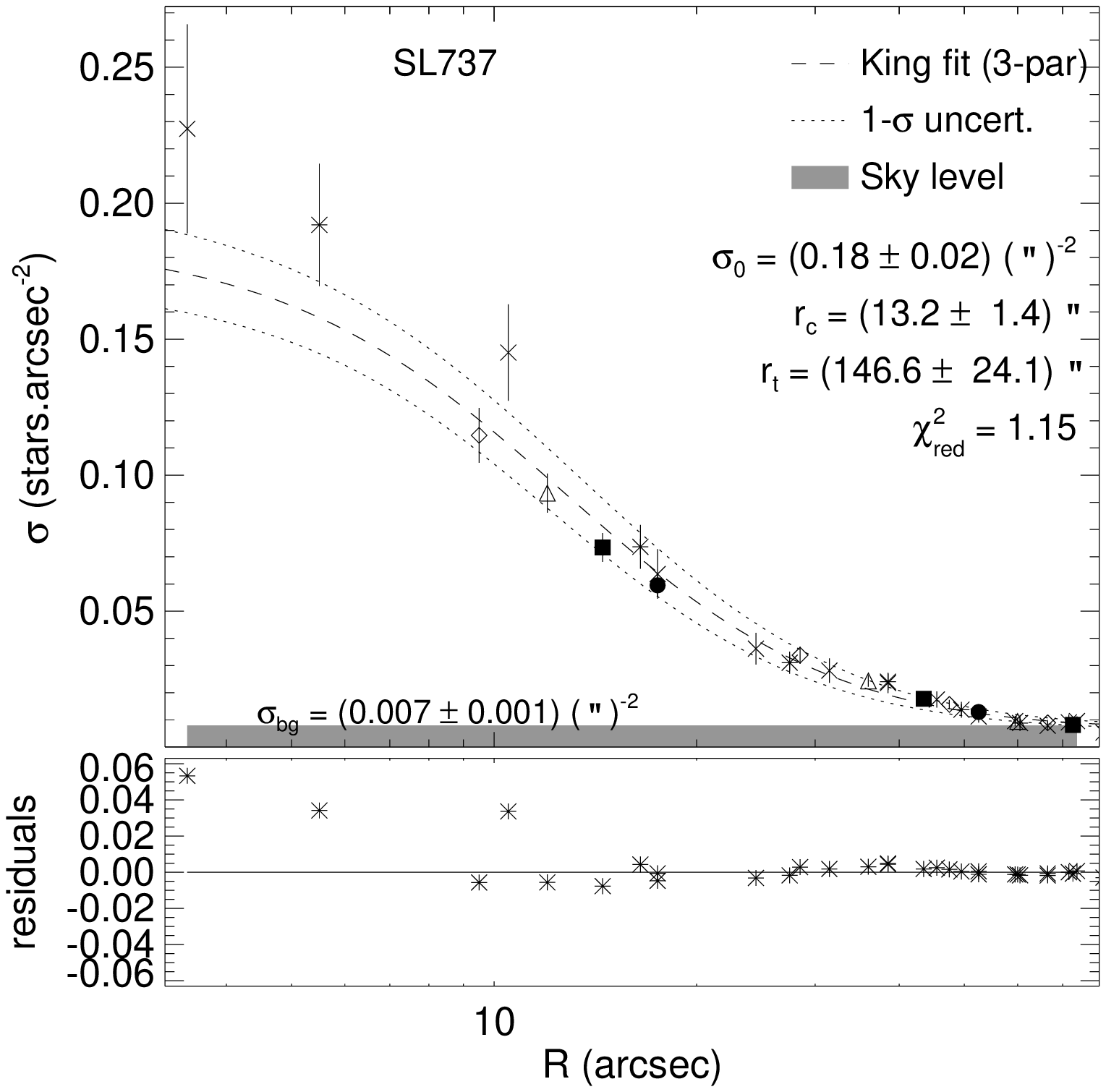}\includegraphics[width=0.325\linewidth]{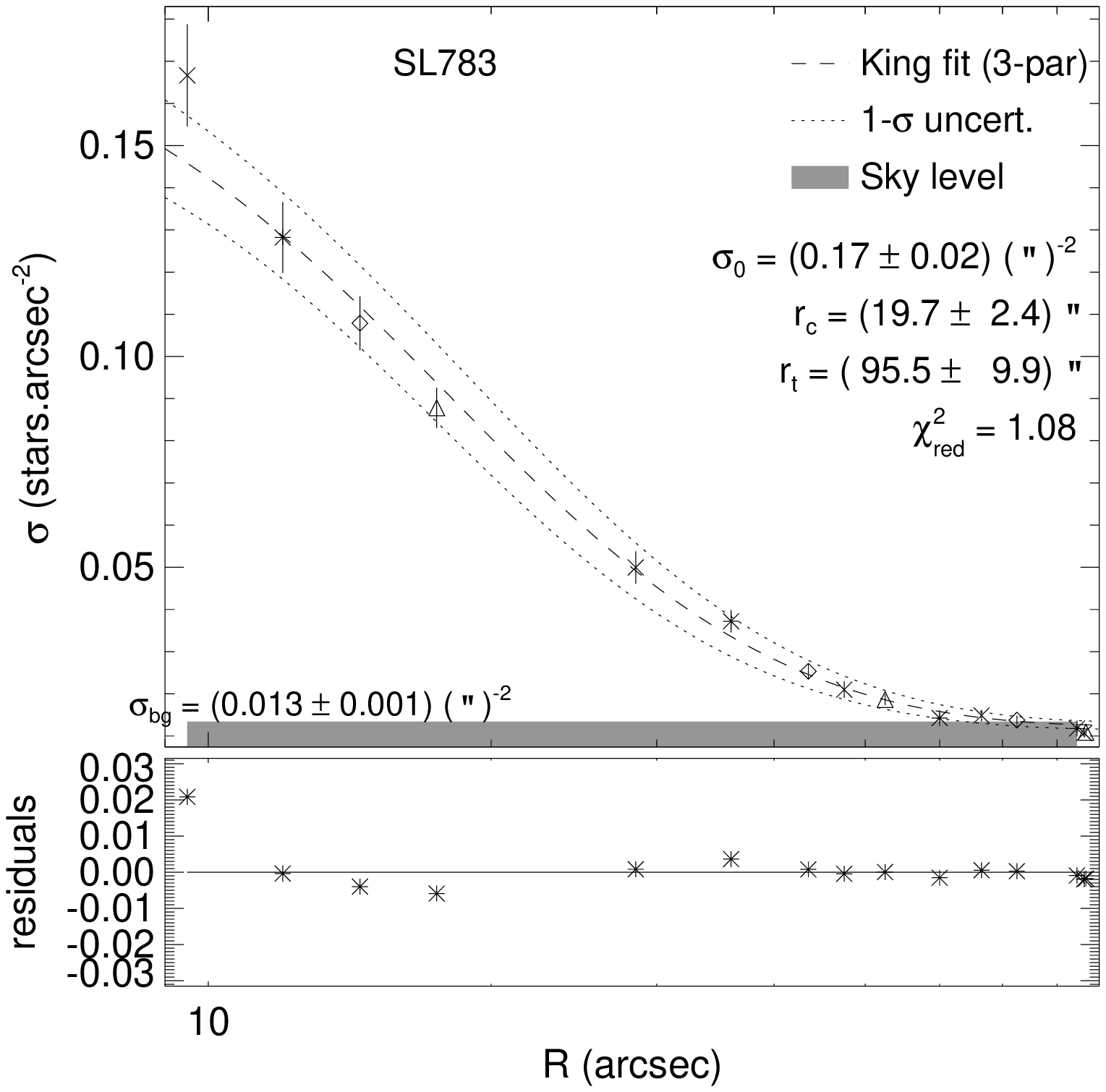}

\includegraphics[width=0.325\linewidth]{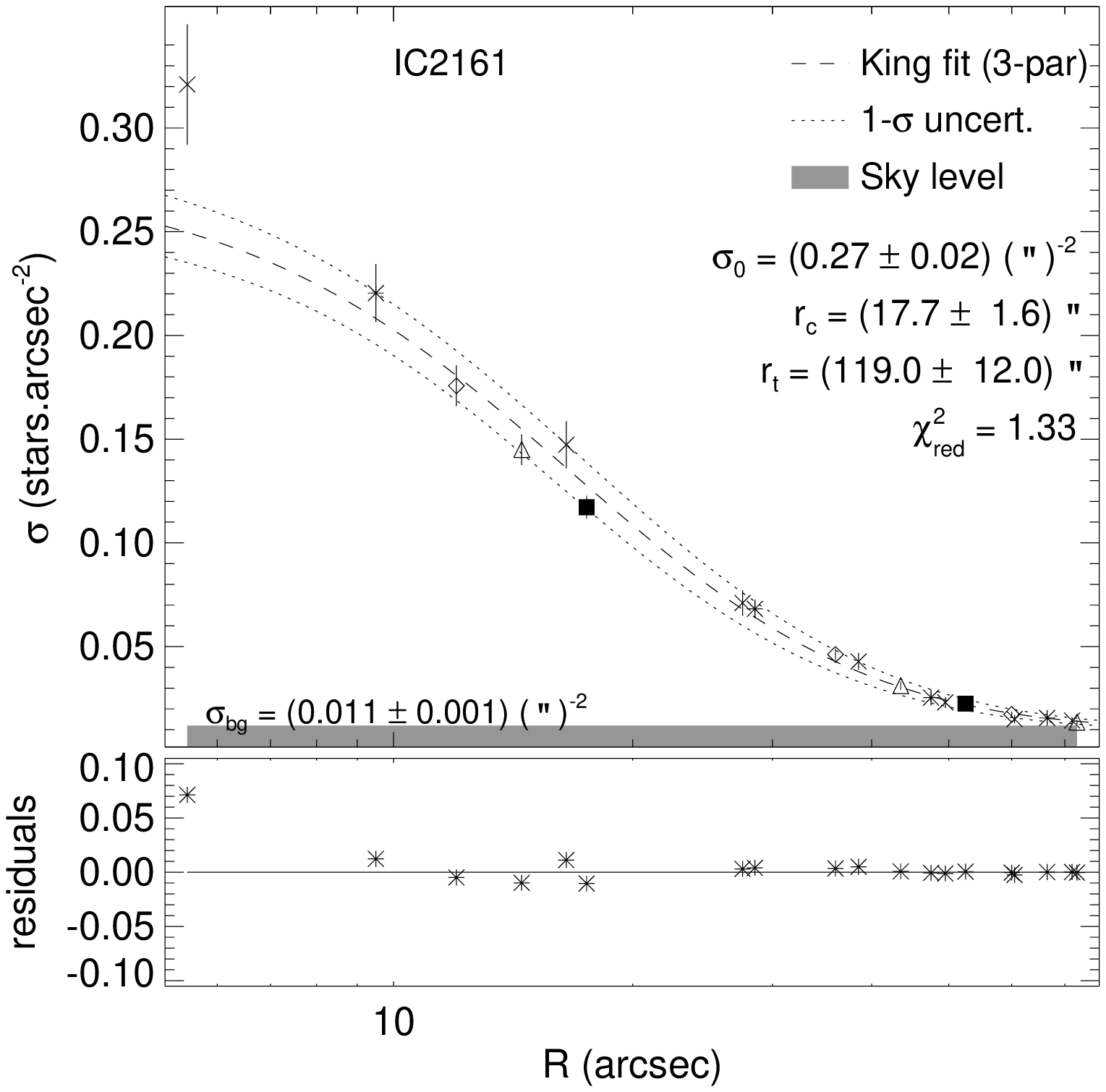}\includegraphics[width=0.325\linewidth]{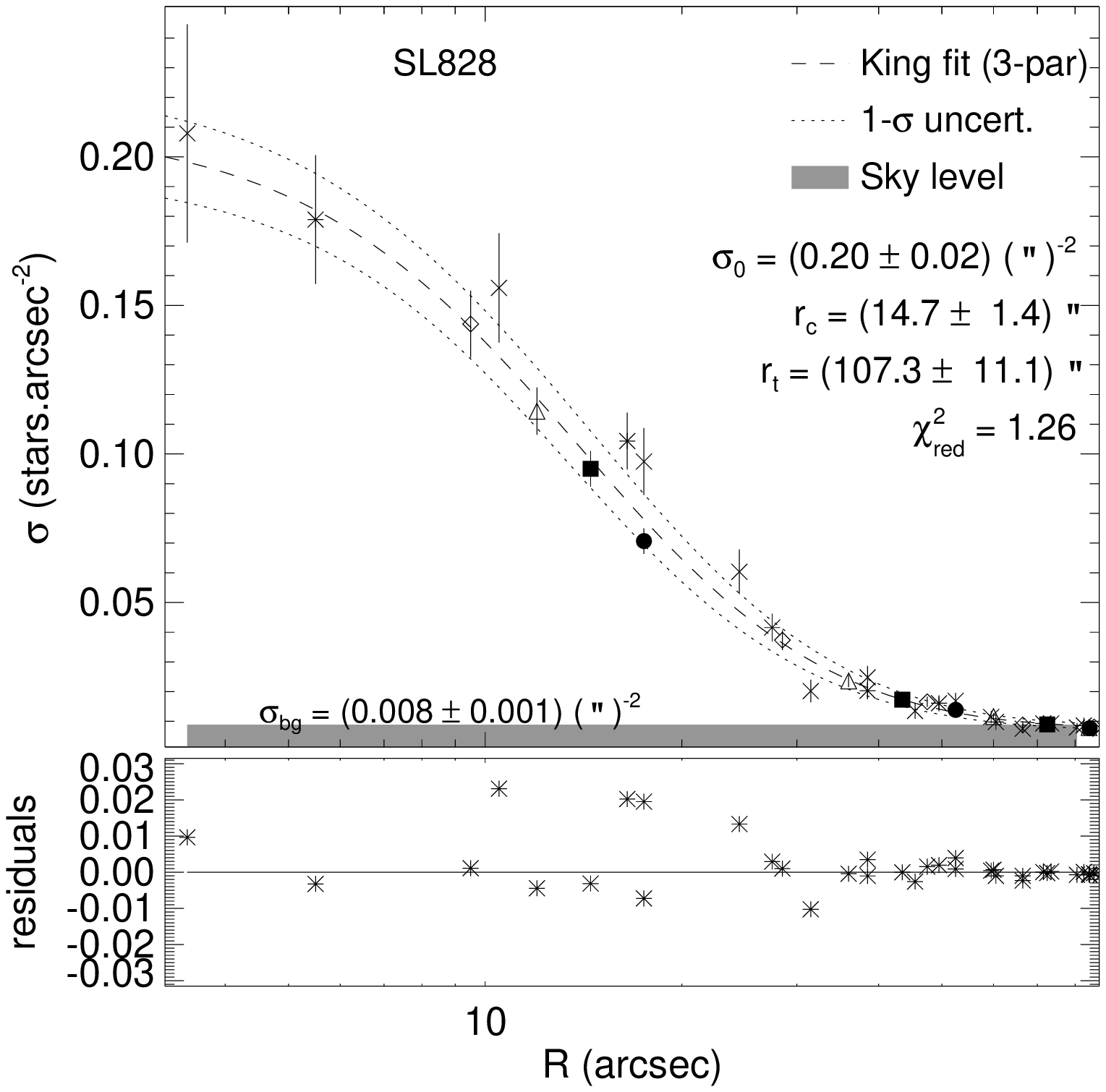}\includegraphics[width=0.325\linewidth]{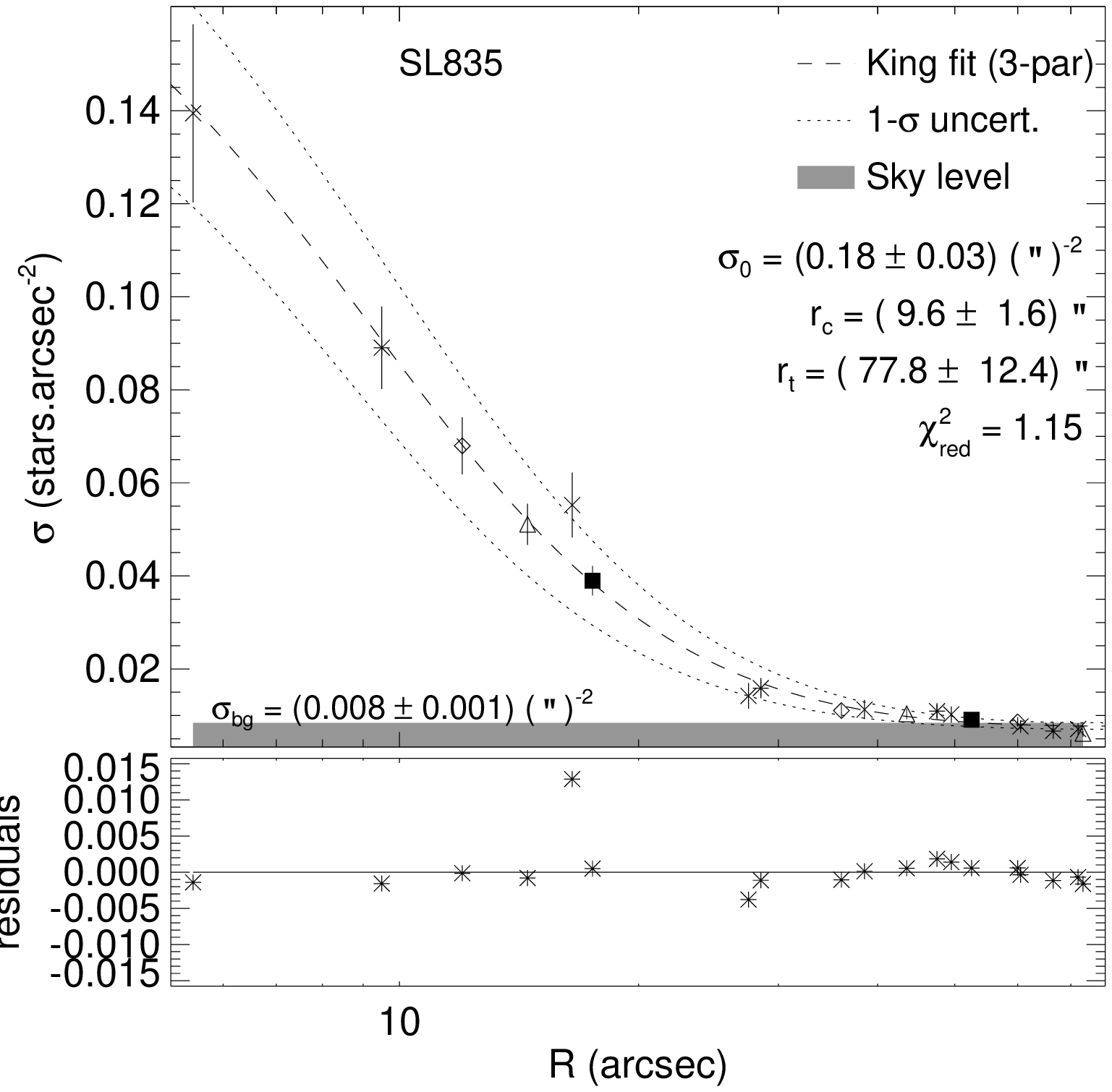}

\includegraphics[width=0.325\linewidth]{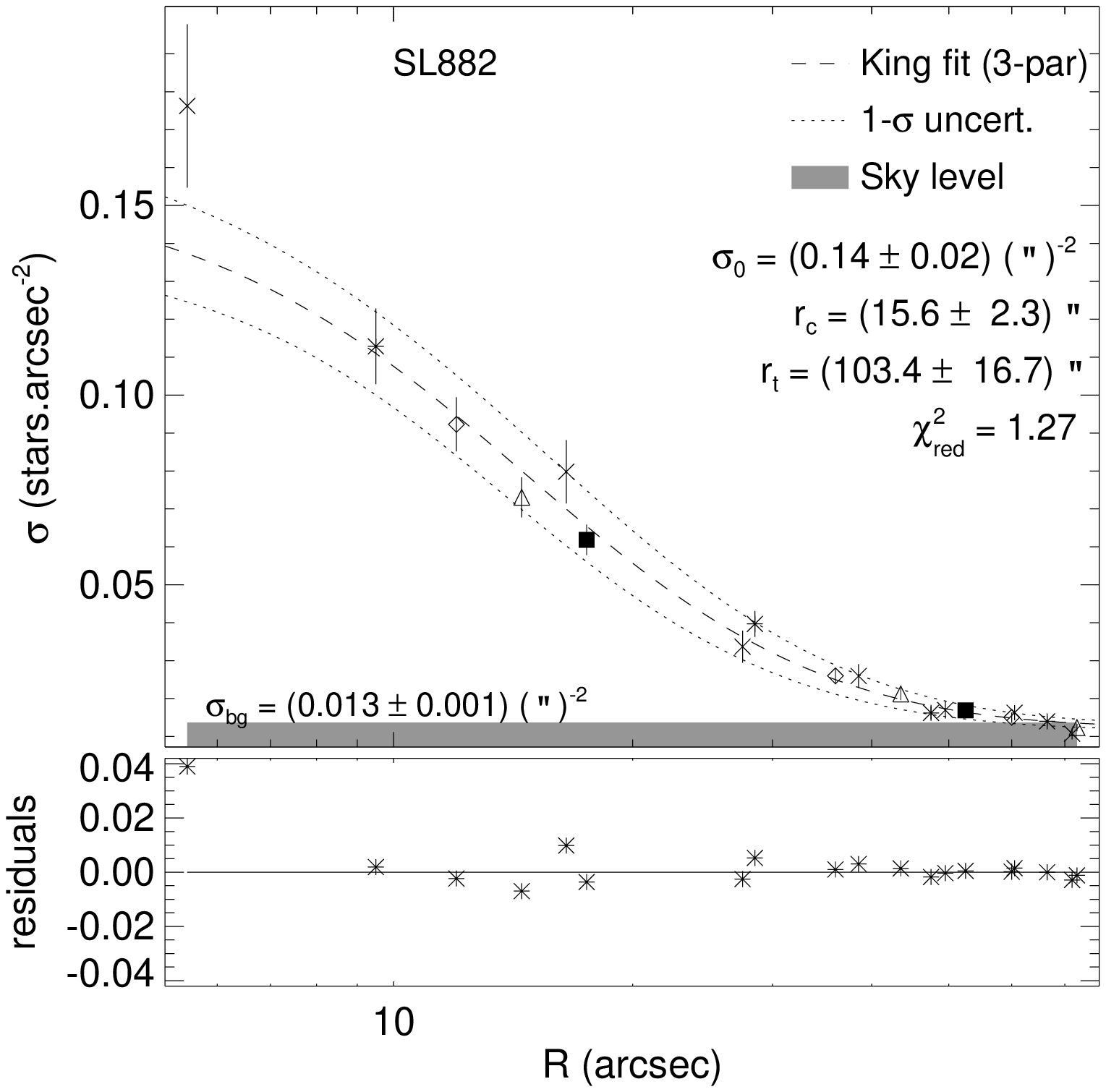}\includegraphics[width=0.325\linewidth]{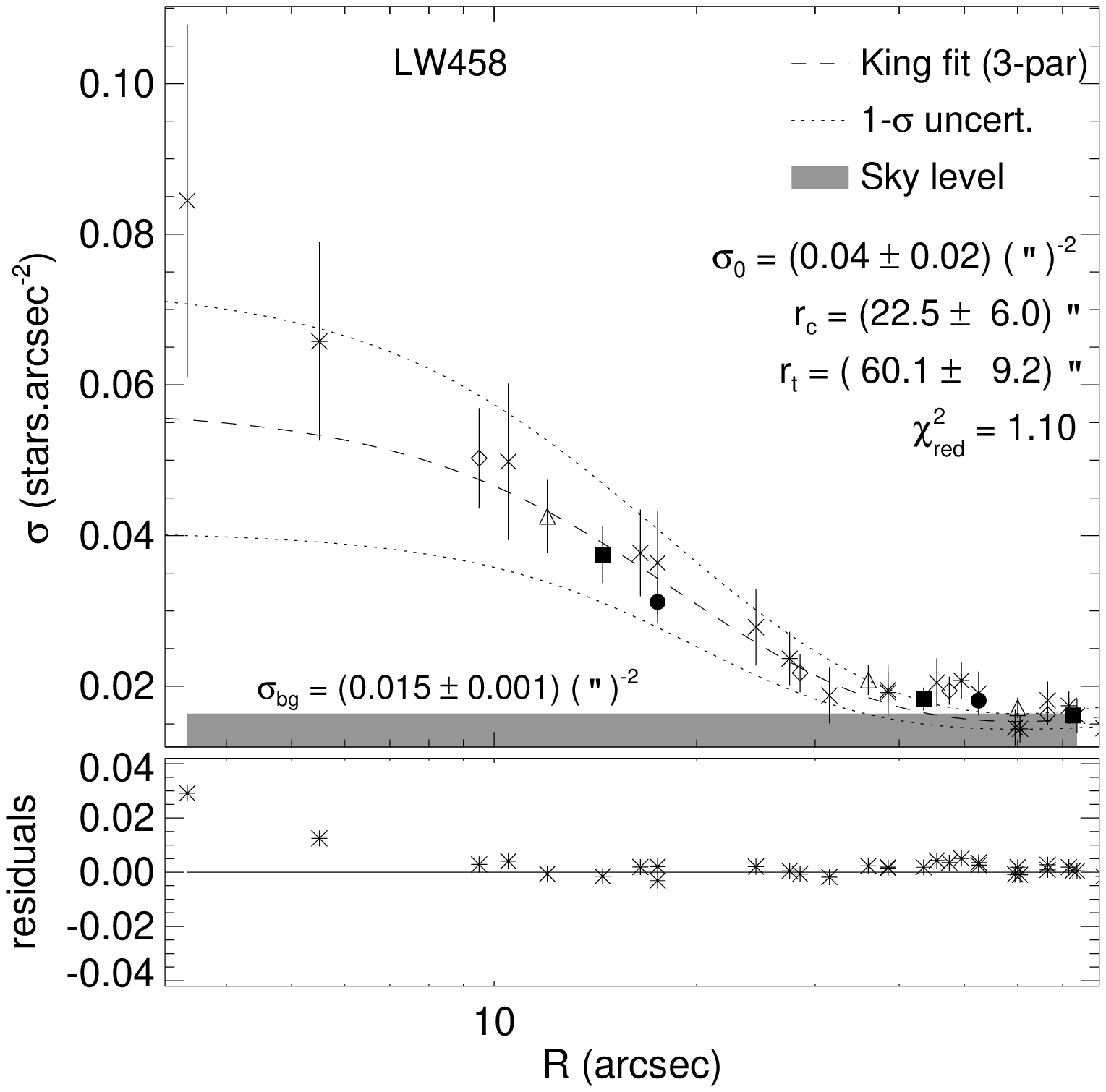}\includegraphics[width=0.325\linewidth]{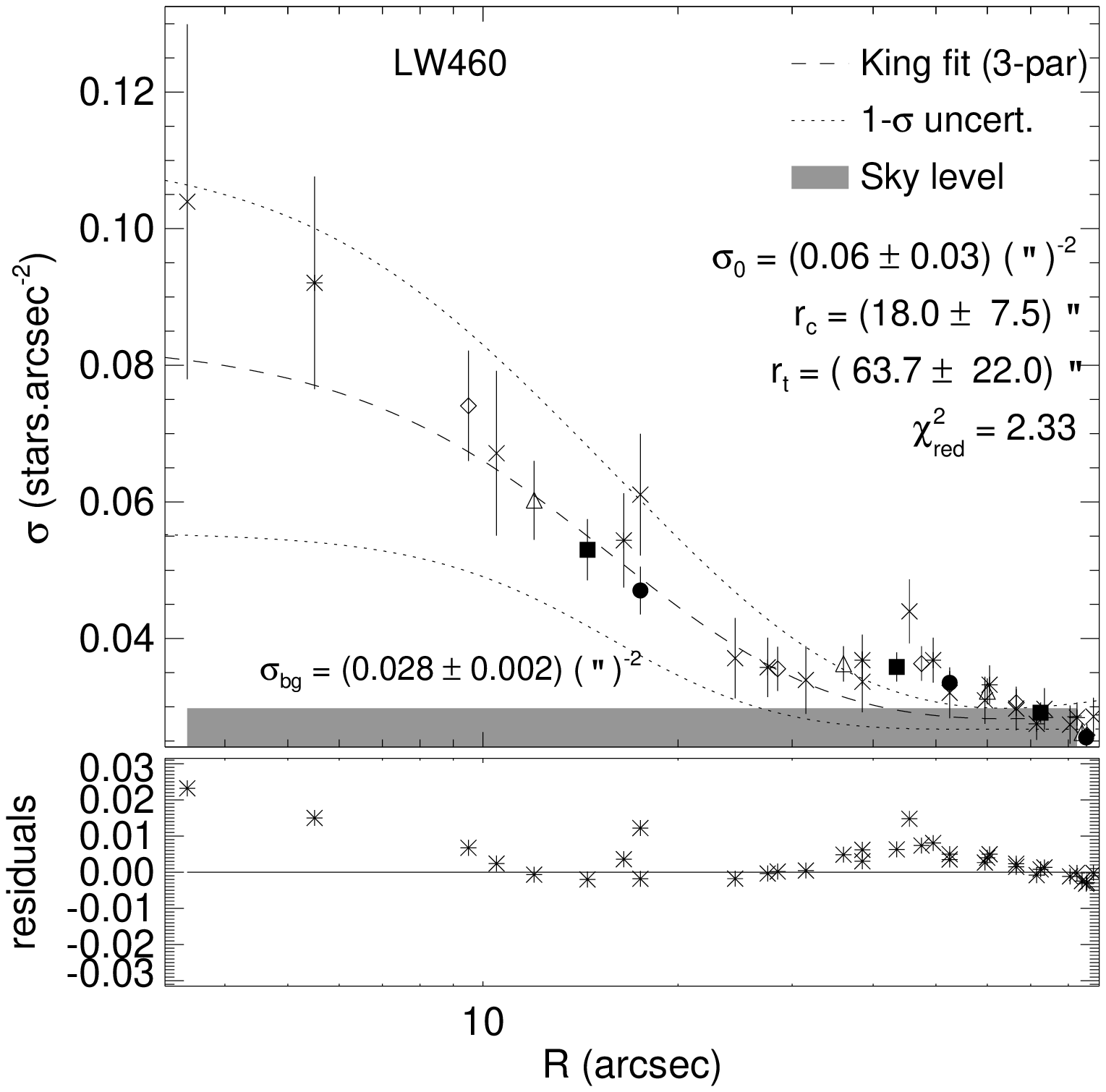}

\includegraphics[width=0.325\linewidth]{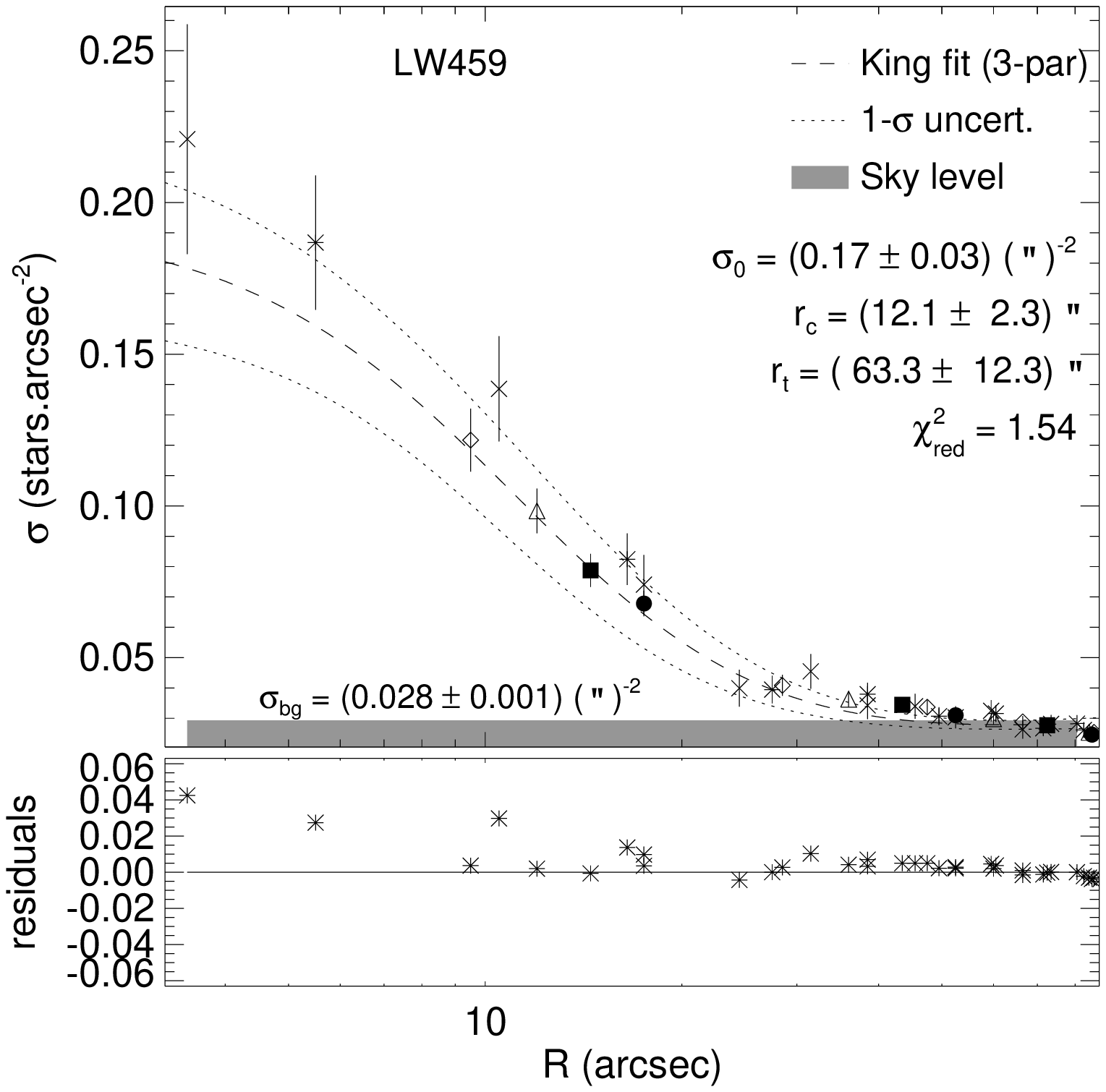}\includegraphics[width=0.325\linewidth]{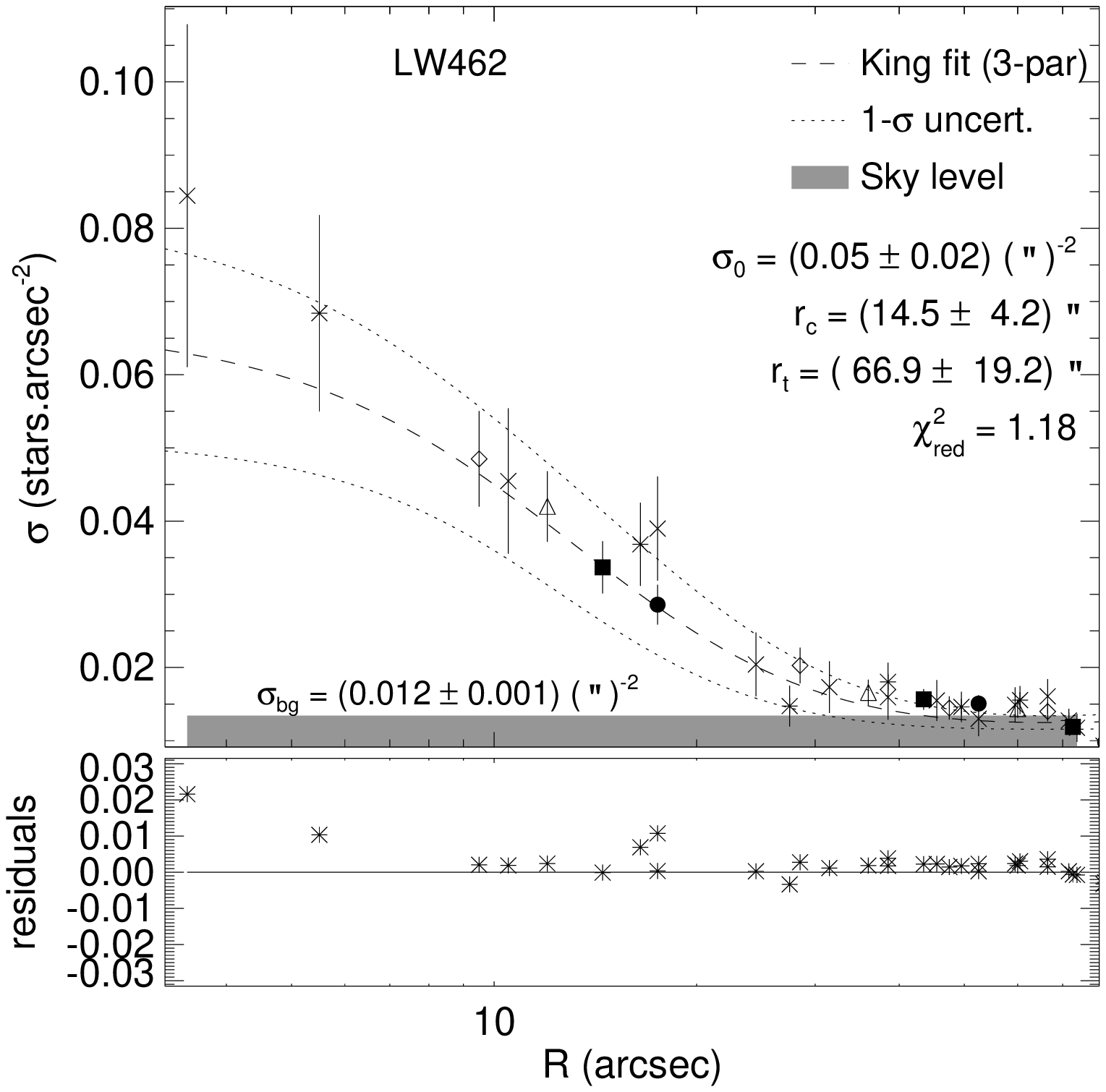}\includegraphics[width=0.325\linewidth]{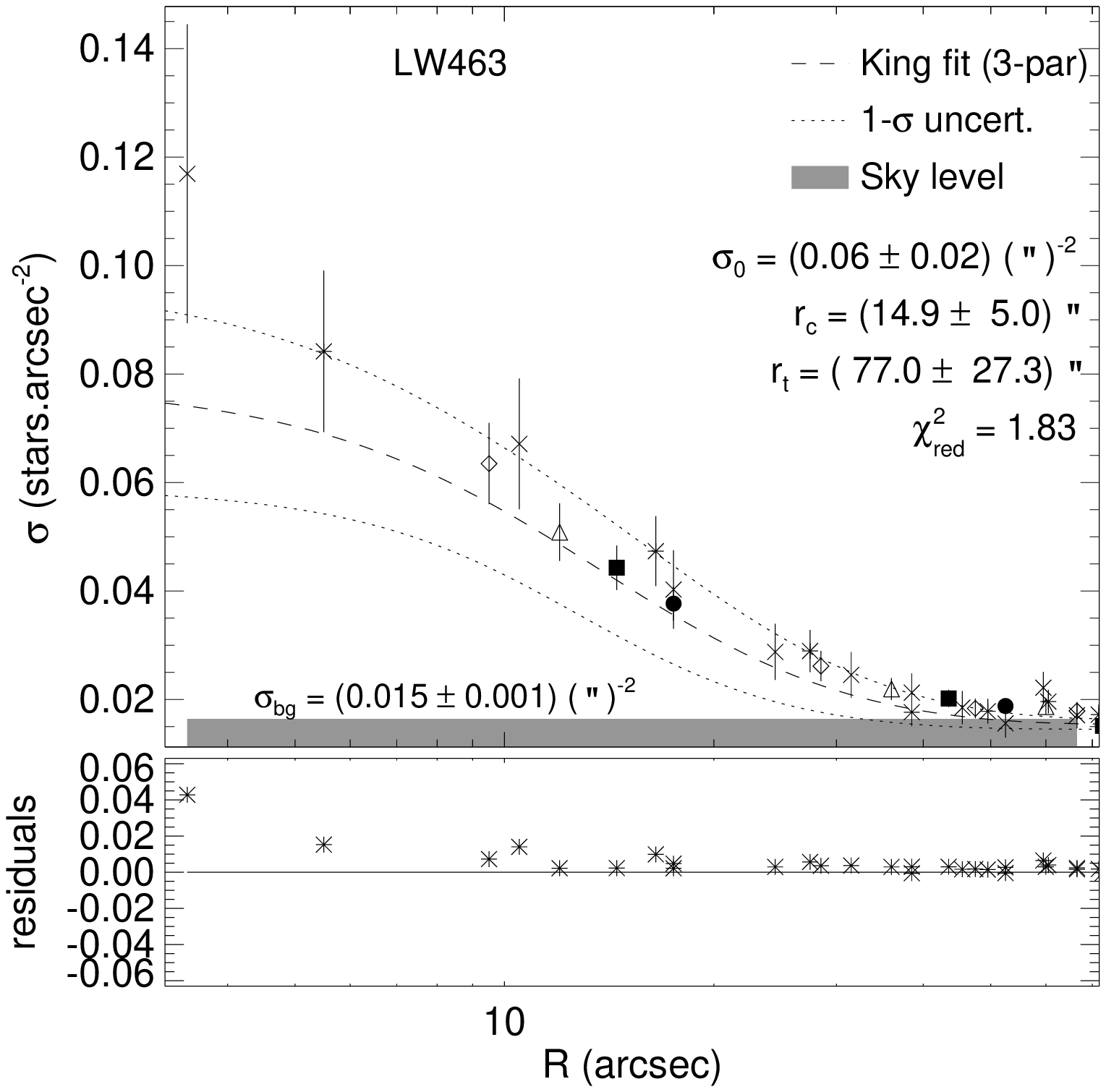}

\caption{cont.}

\end{figure*}

\setcounter{figure}{1}
\begin{figure*}

\includegraphics[width=0.325\linewidth]{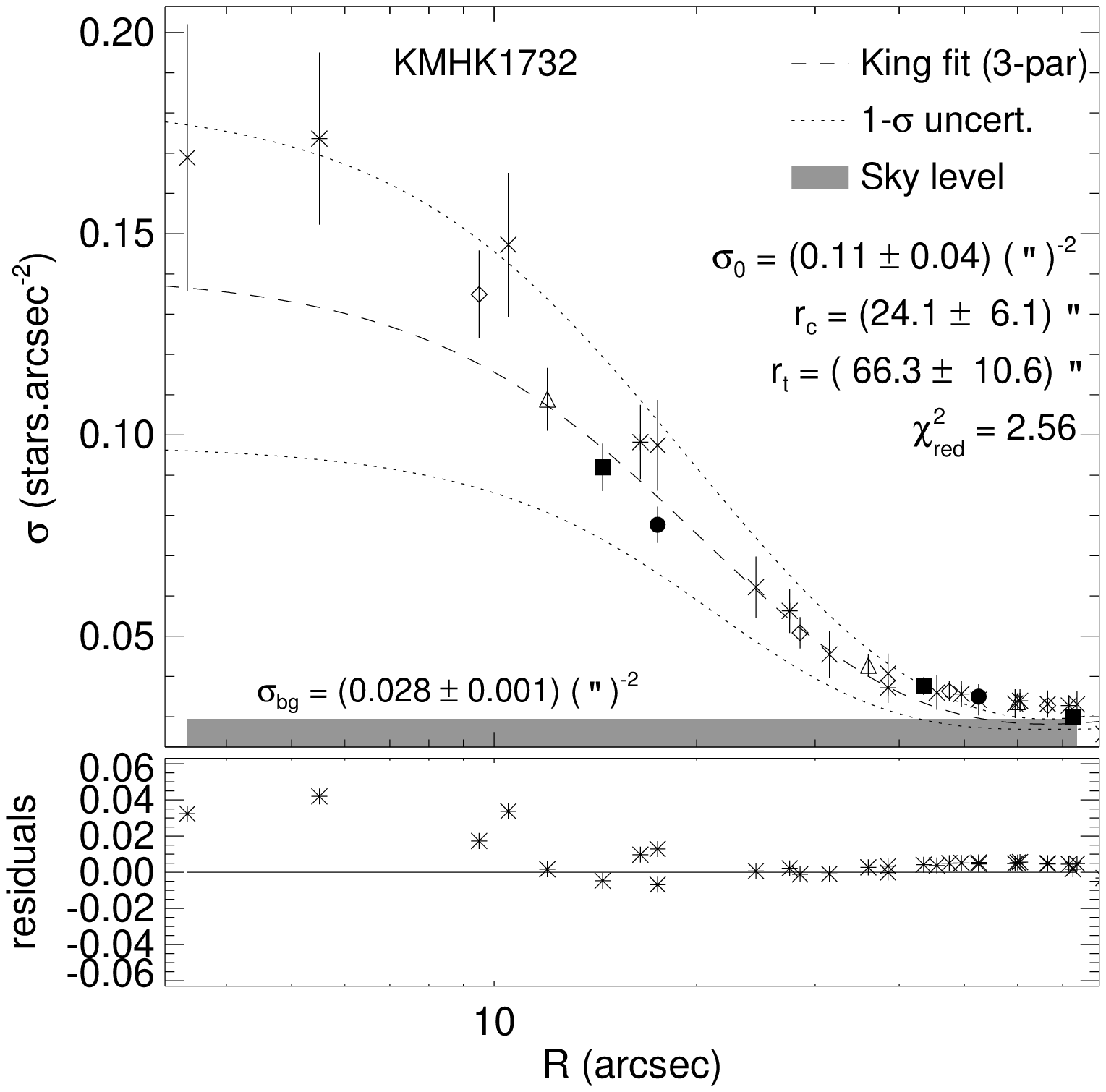}\includegraphics[width=0.325\linewidth]{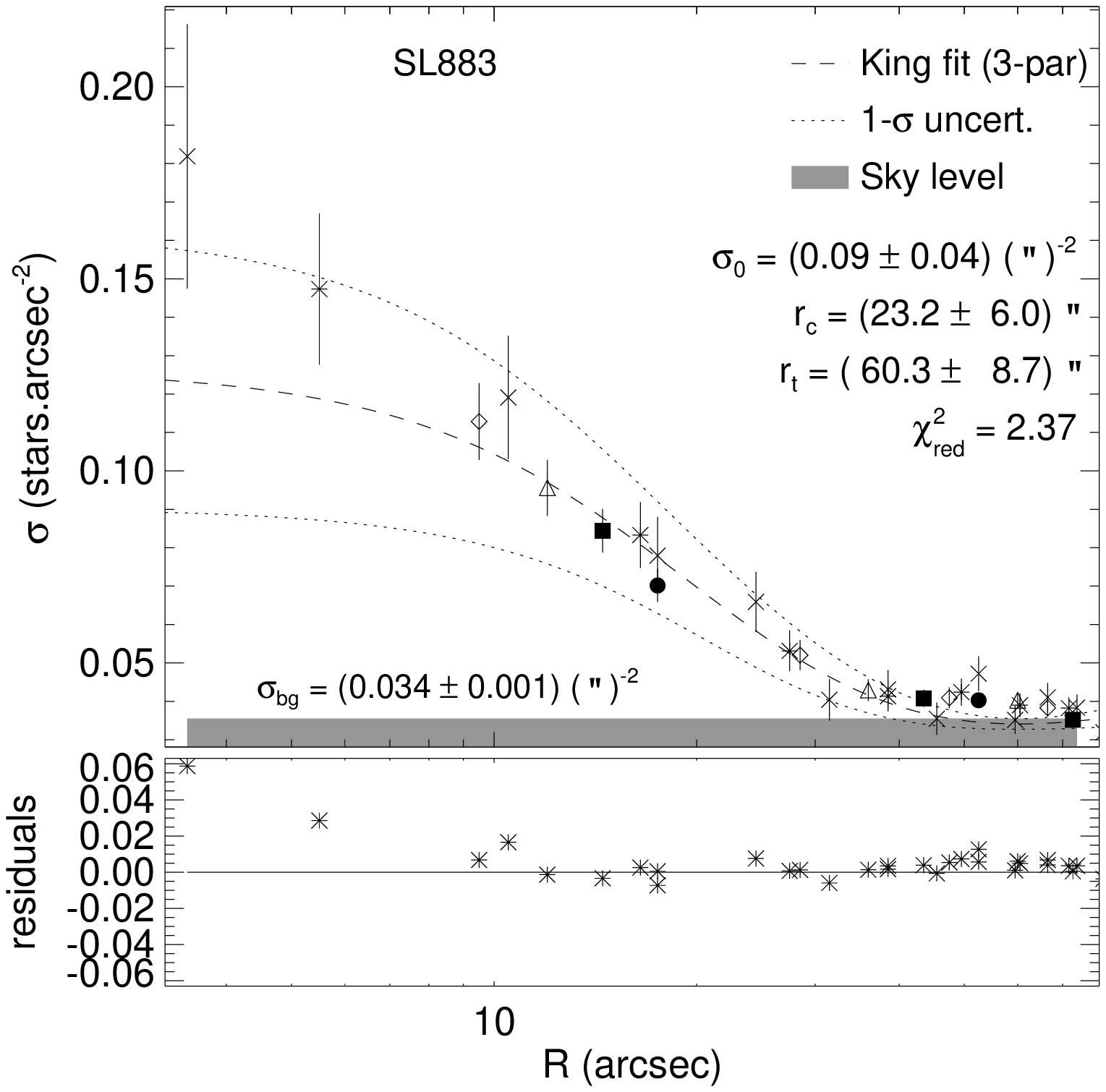}\includegraphics[width=0.325\linewidth]{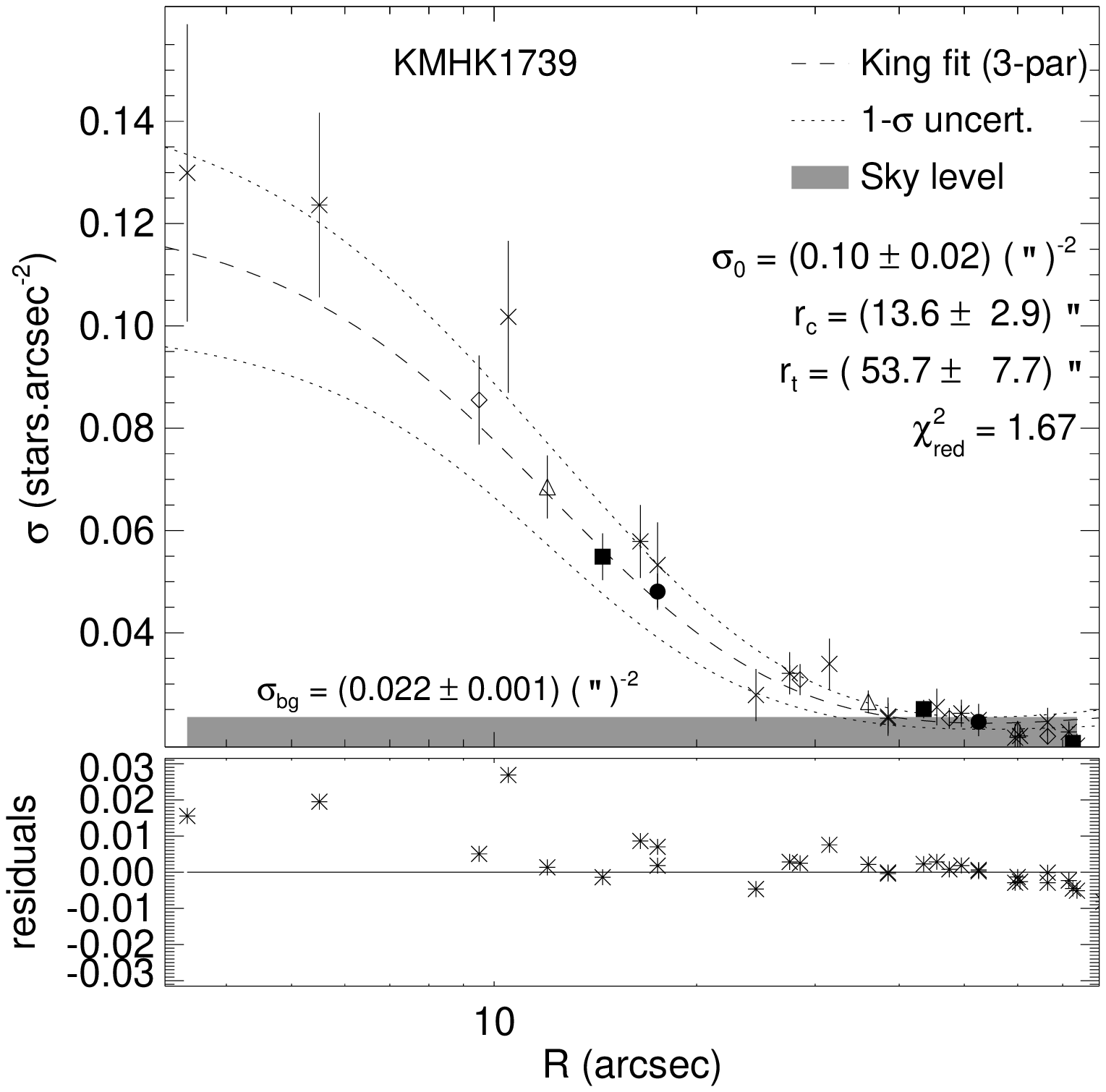}

\includegraphics[width=0.325\linewidth]{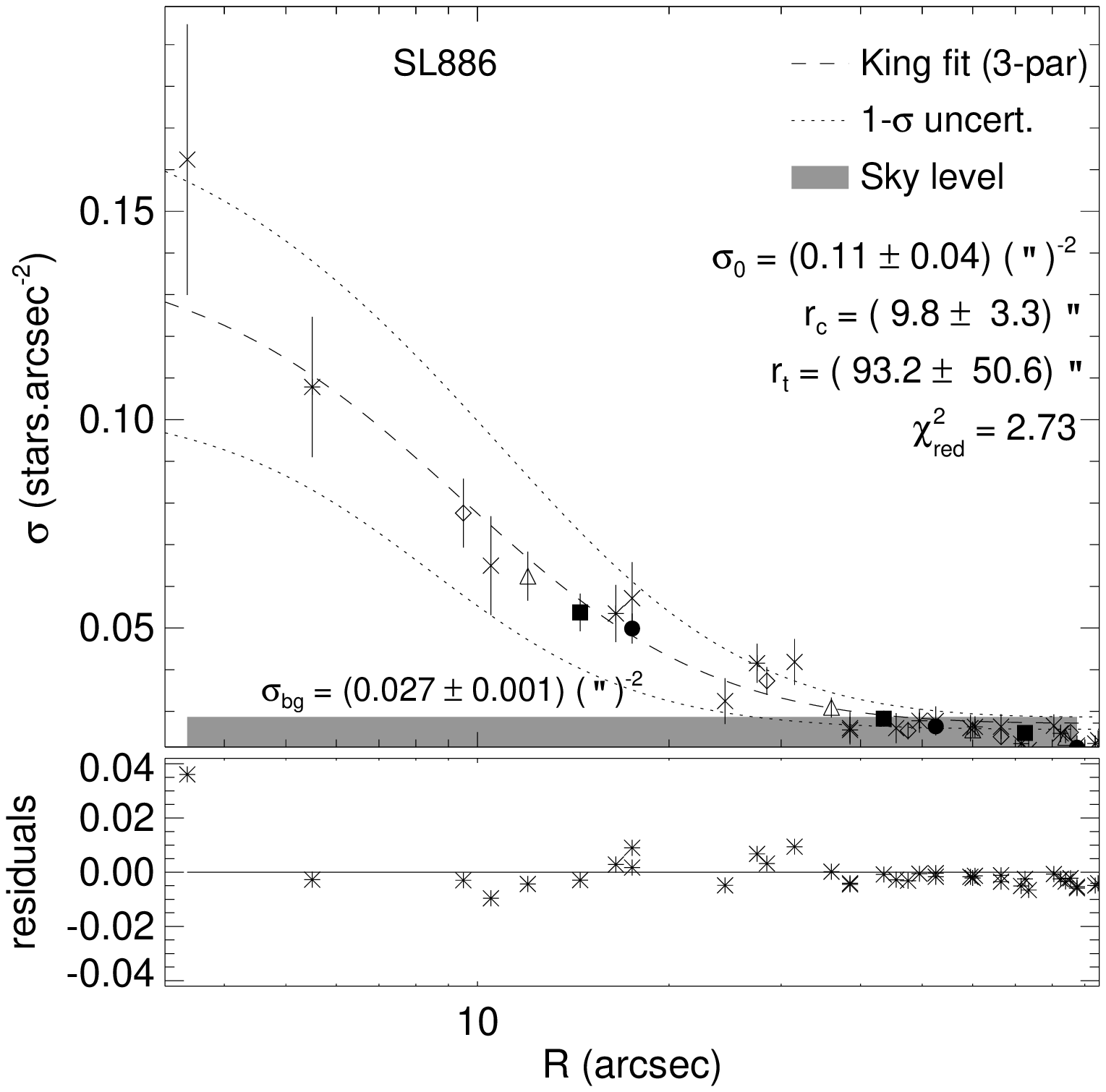}\includegraphics[width=0.325\linewidth]{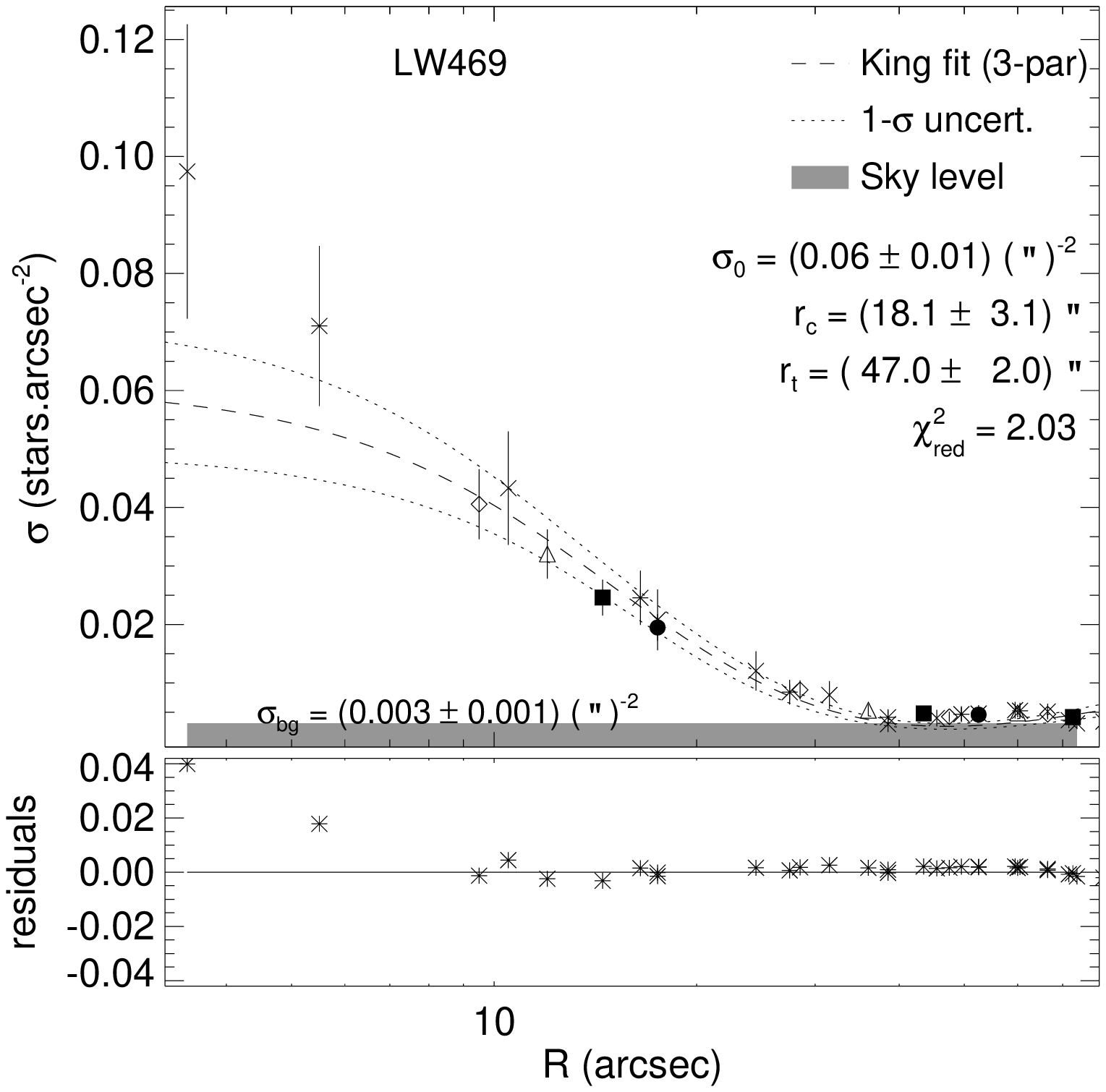}\includegraphics[width=0.325\linewidth]{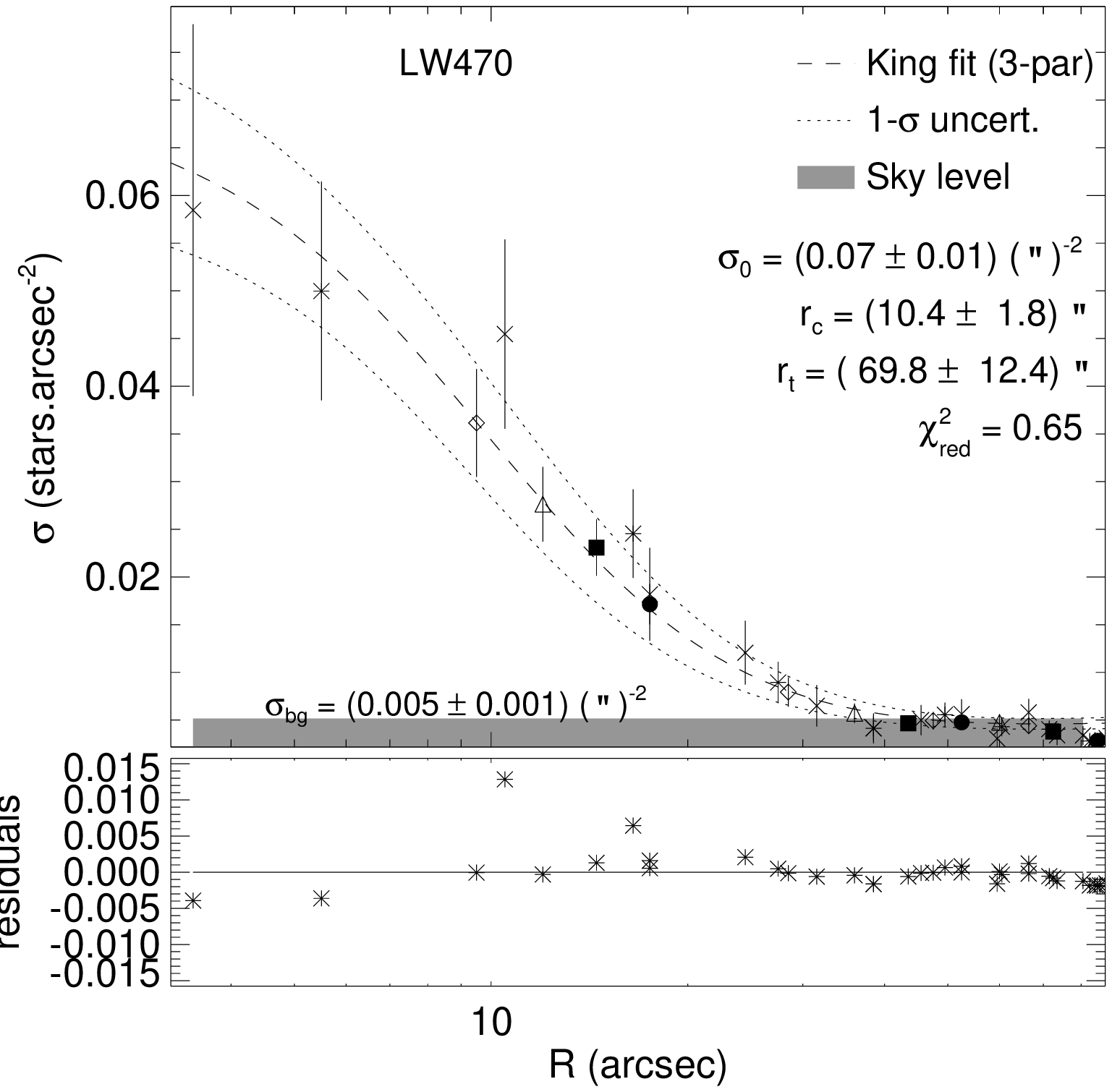}

\includegraphics[width=0.325\linewidth]{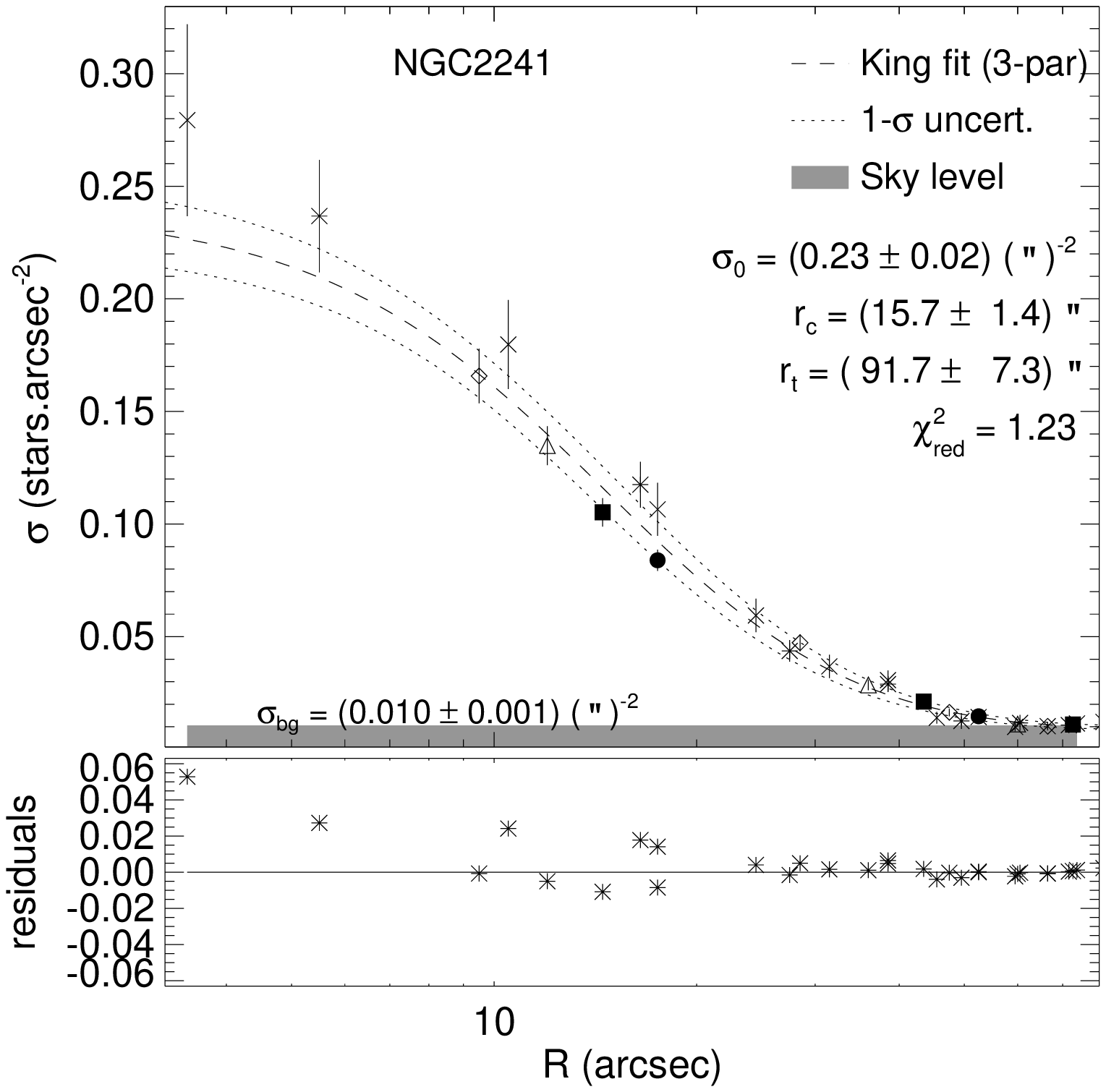}\includegraphics[width=0.325\linewidth]{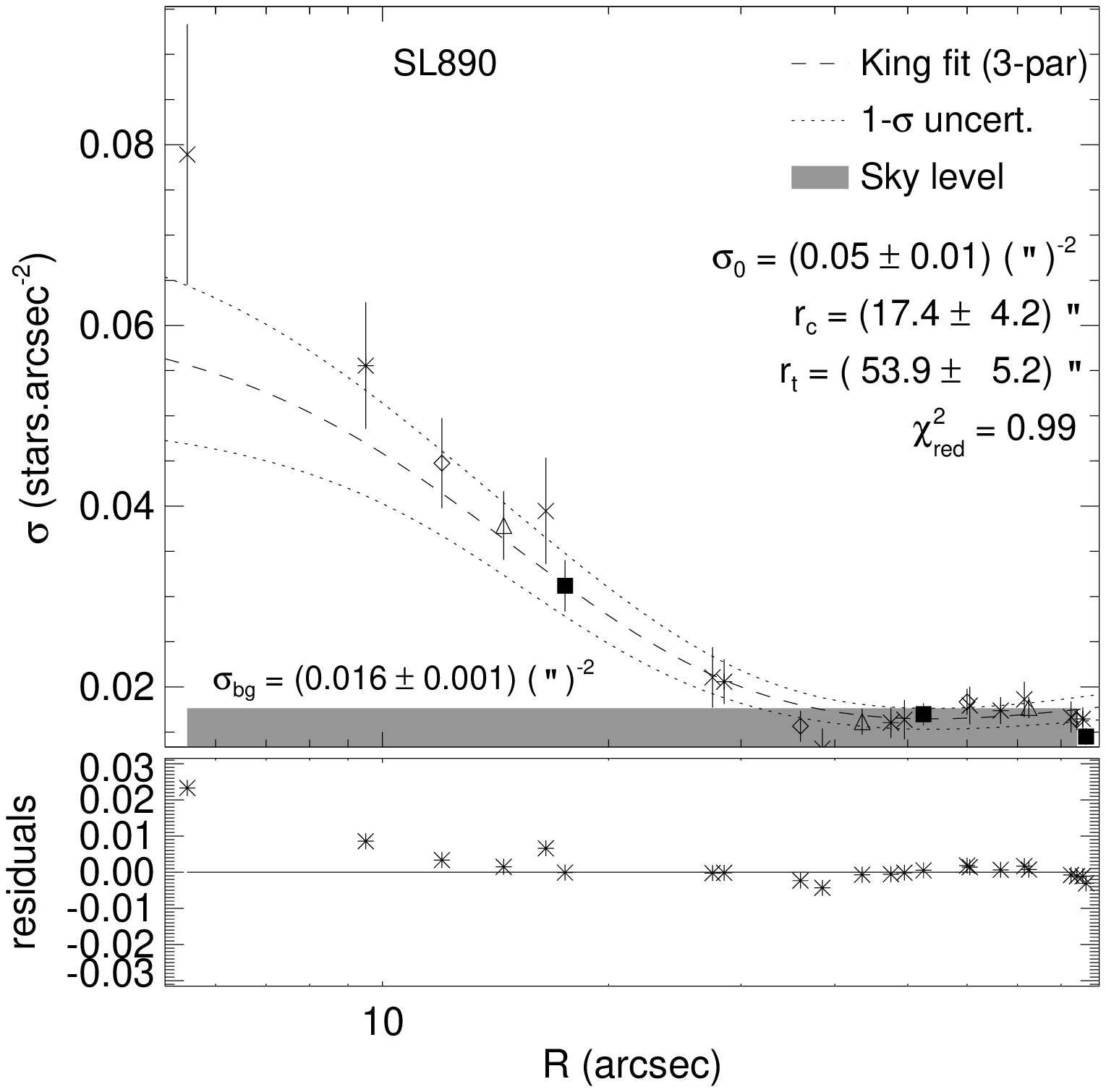}\includegraphics[width=0.325\linewidth]{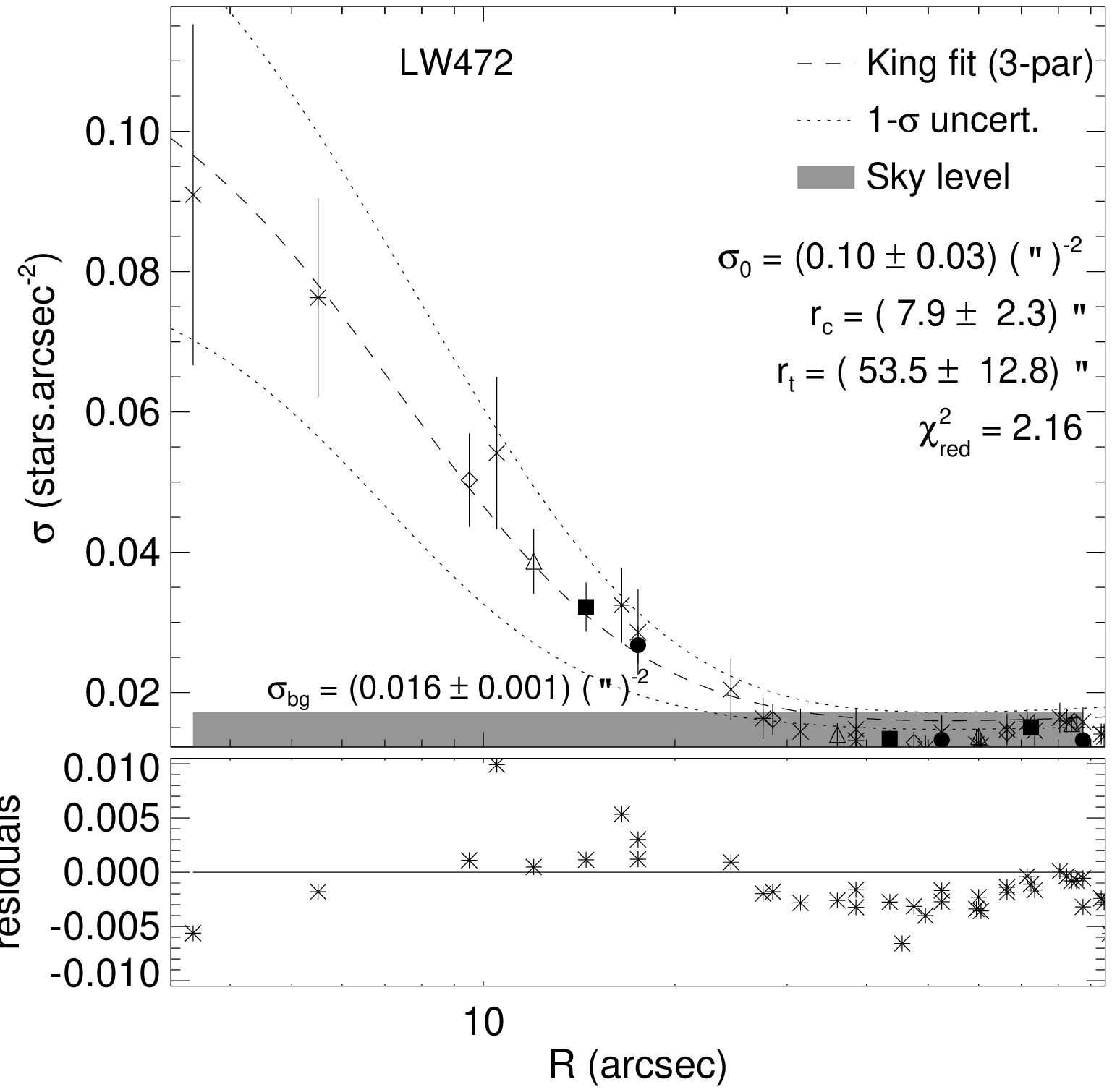}

\includegraphics[width=0.325\linewidth]{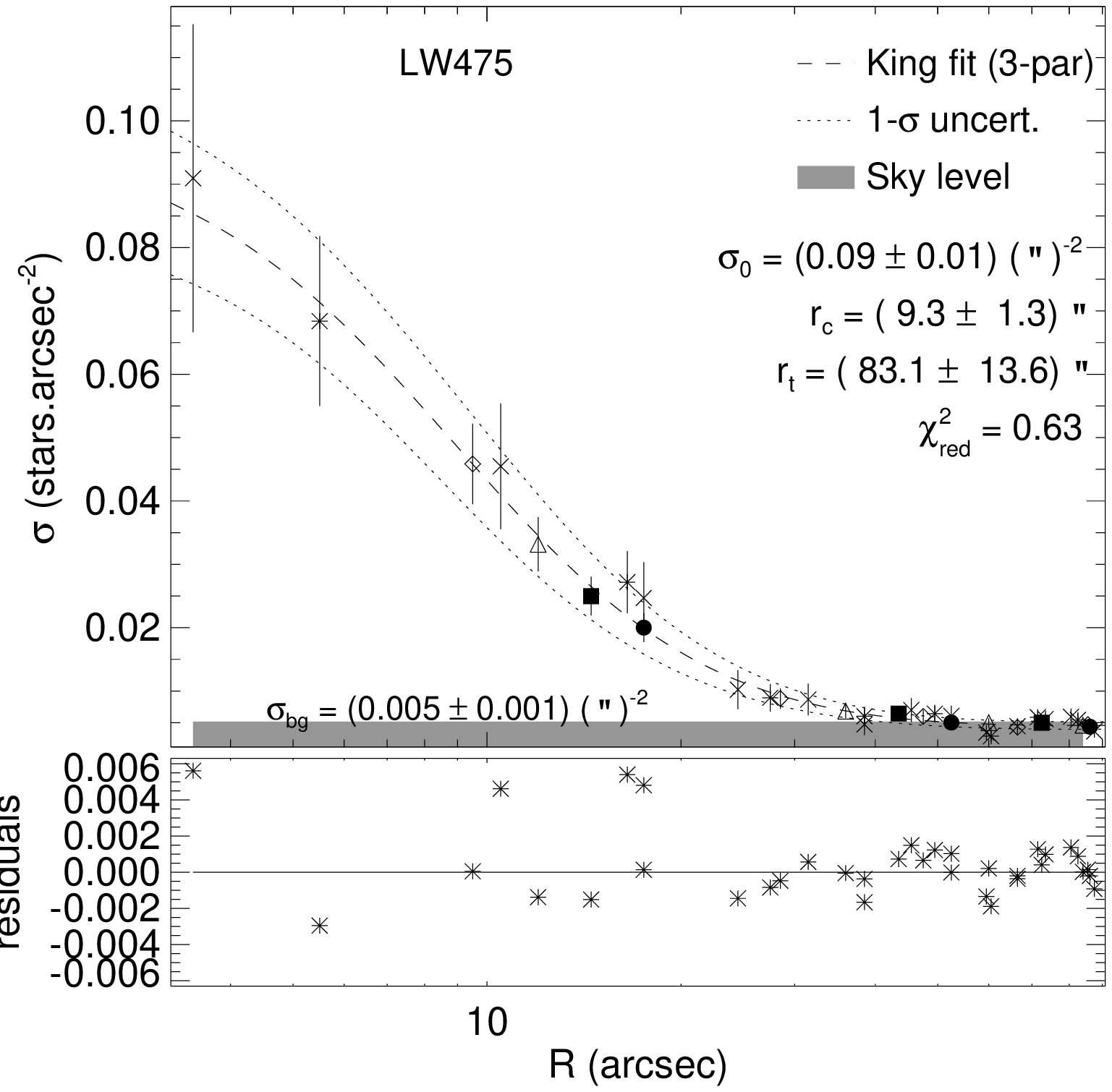}\includegraphics[width=0.325\linewidth]{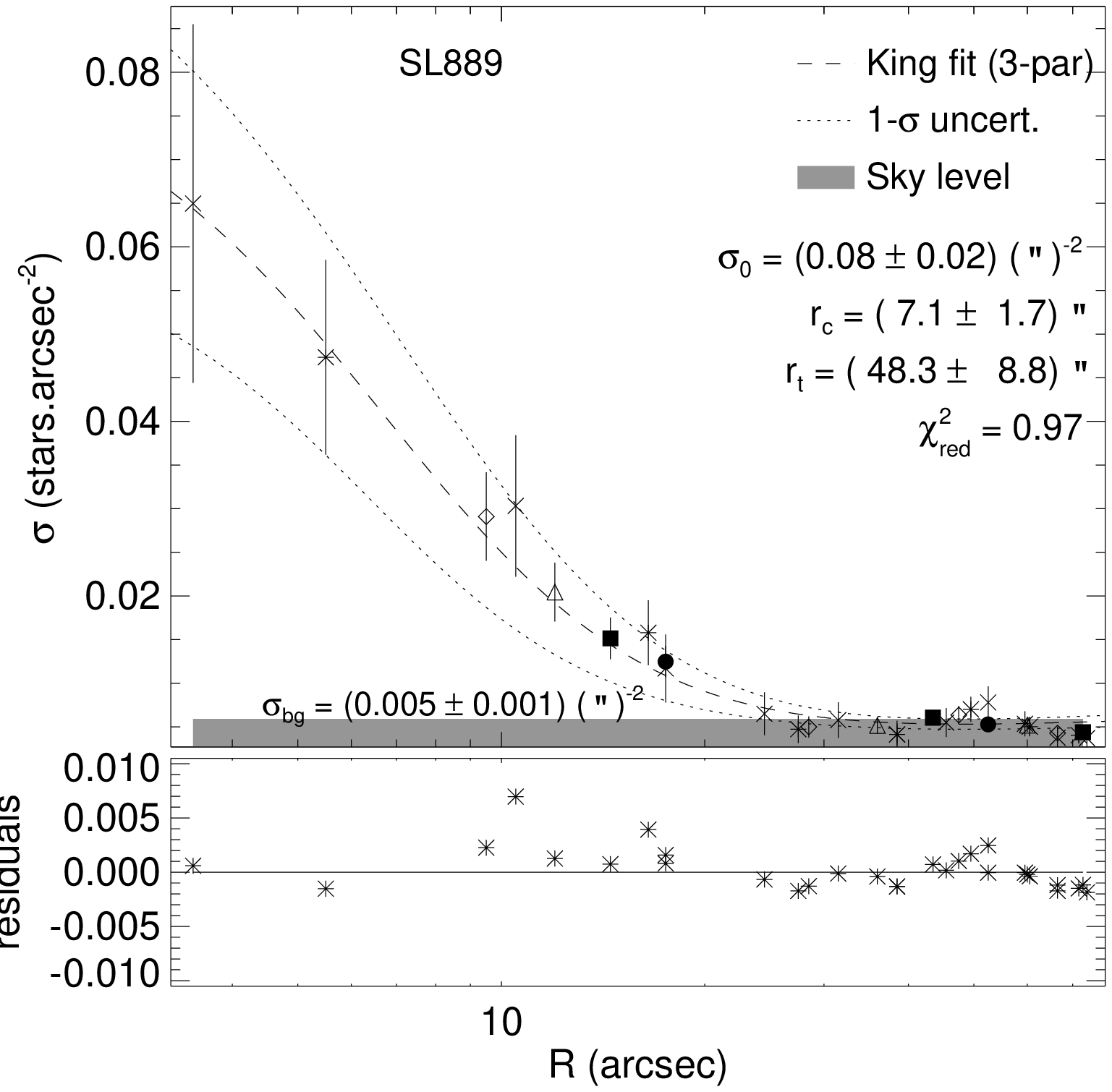}\includegraphics[width=0.325\linewidth]{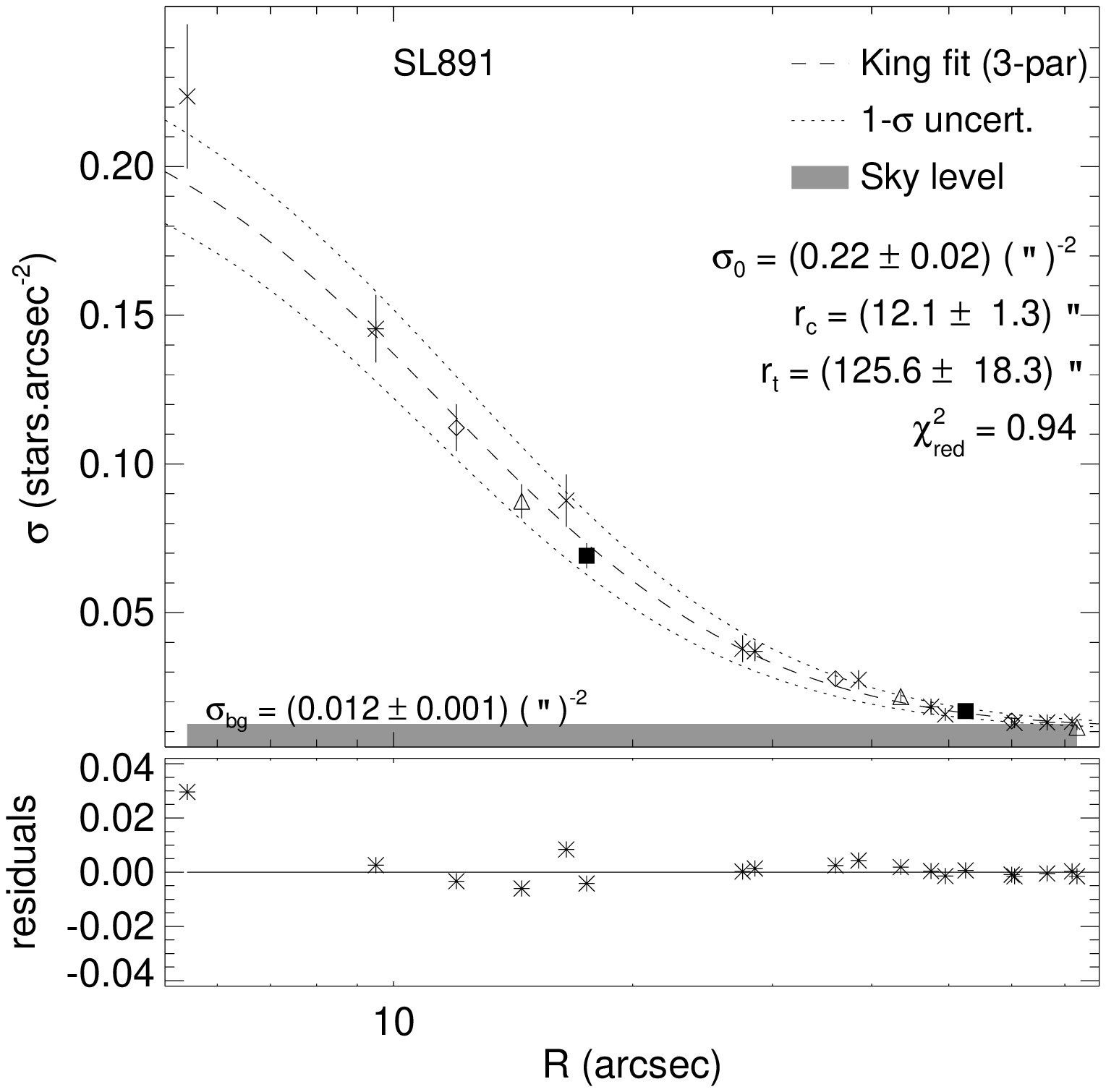}

\caption{cont.}

\end{figure*}

\setcounter{figure}{1}
\begin{figure*}

\includegraphics[width=0.325\linewidth]{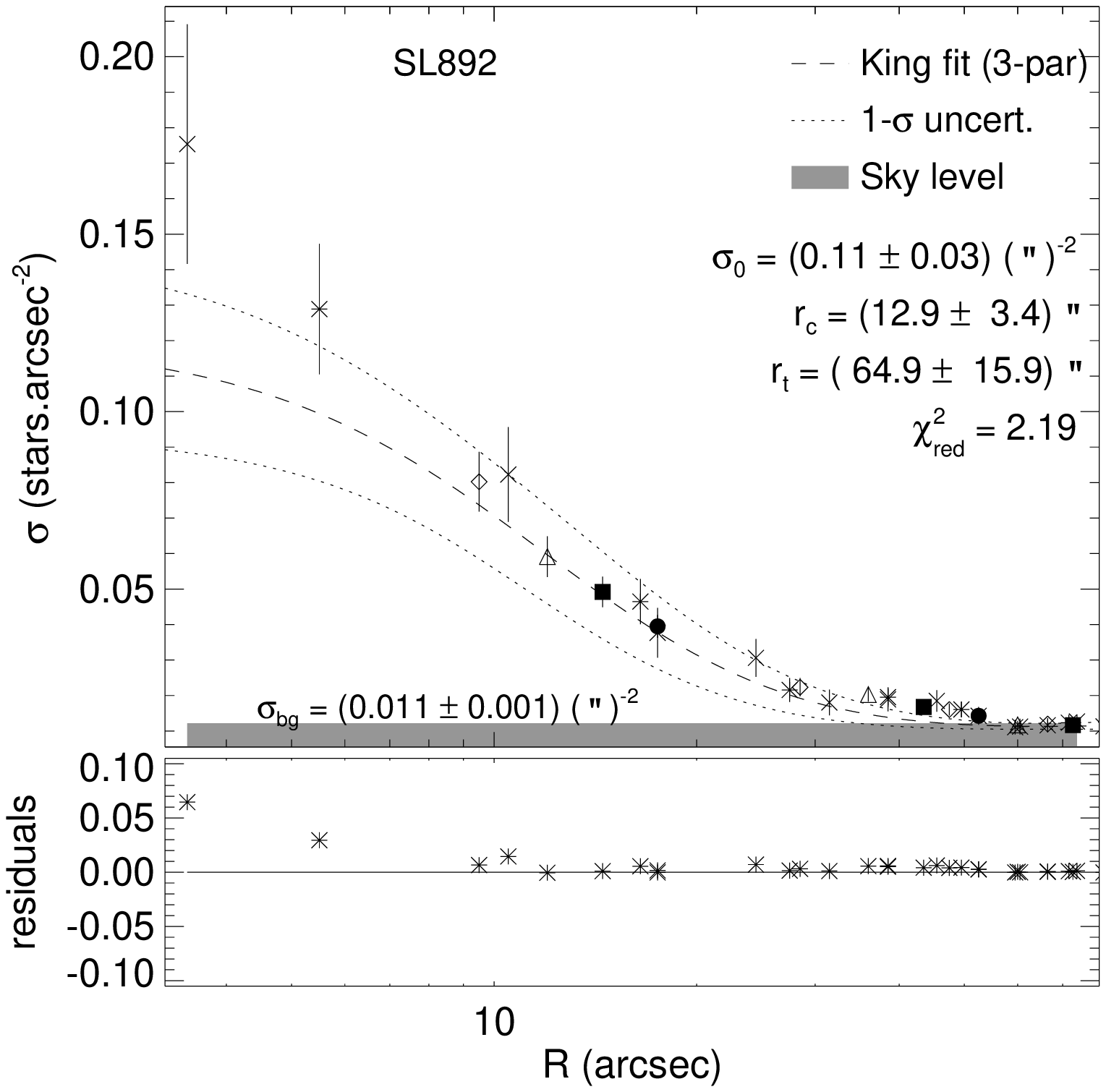}\includegraphics[width=0.325\linewidth]{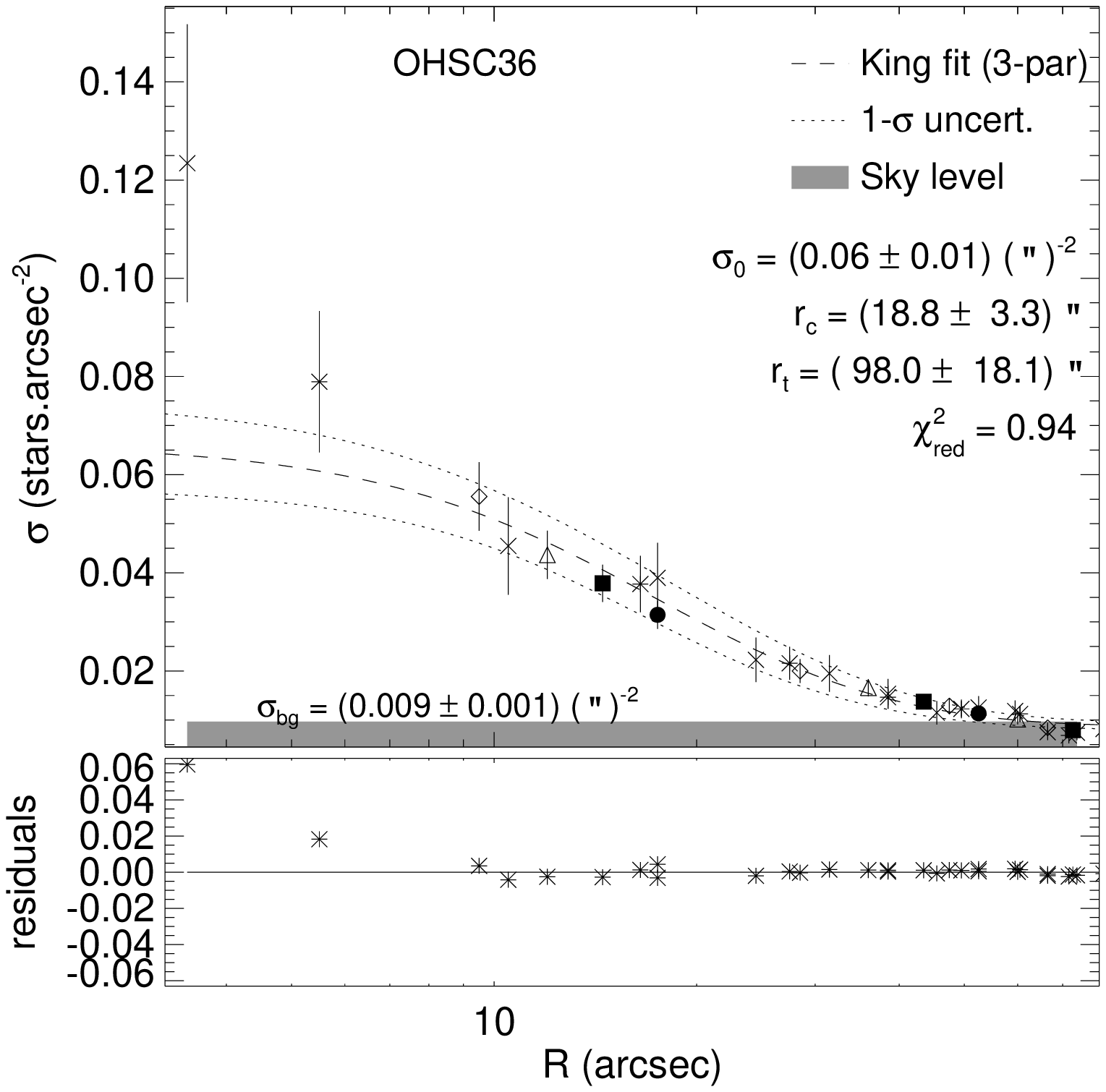}\includegraphics[width=0.325\linewidth]{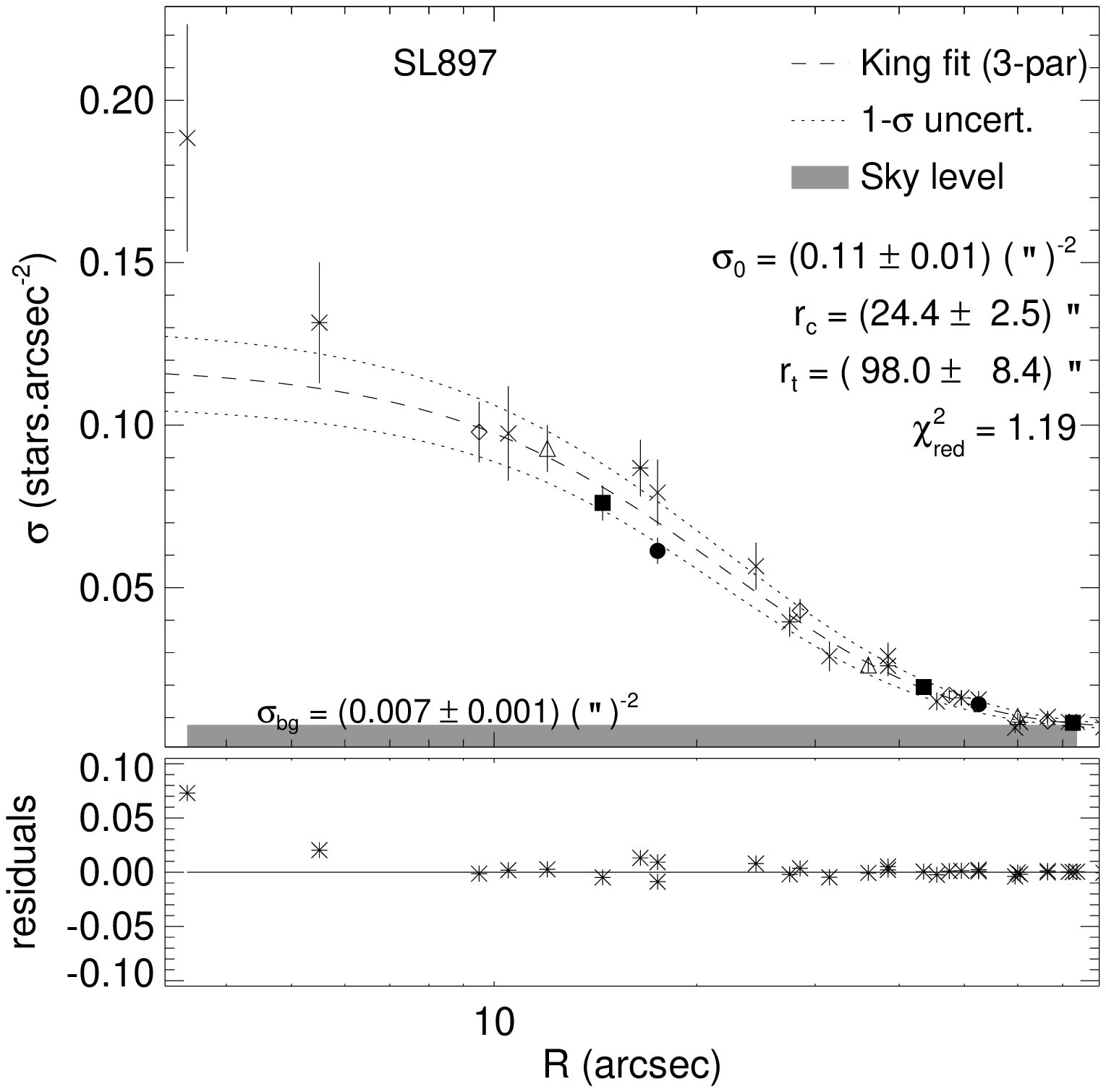}
\caption{cont.}

\end{figure*}

\begin{figure*}
\includegraphics[width=0.325\linewidth]{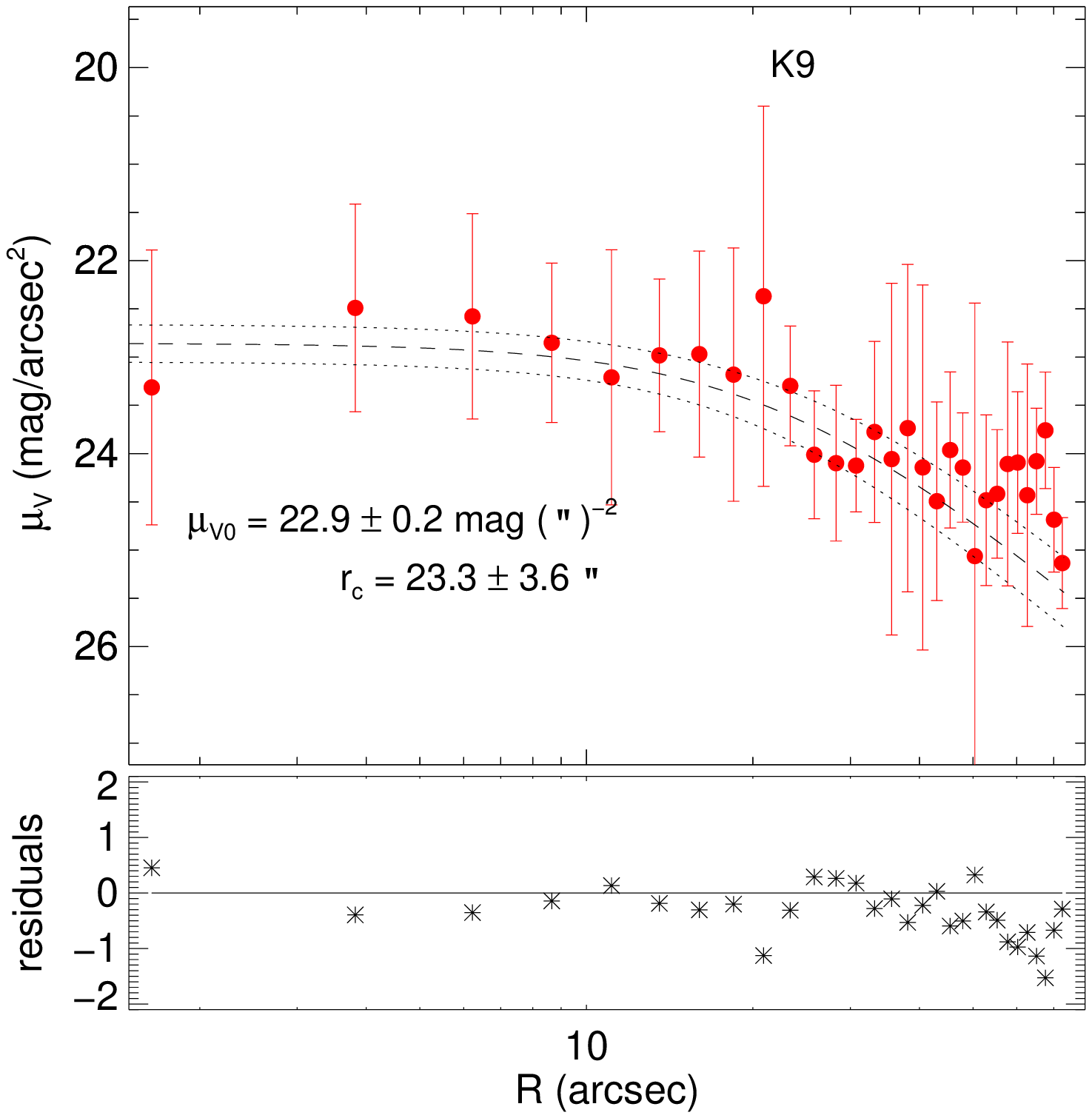}\includegraphics[width=0.325\linewidth]{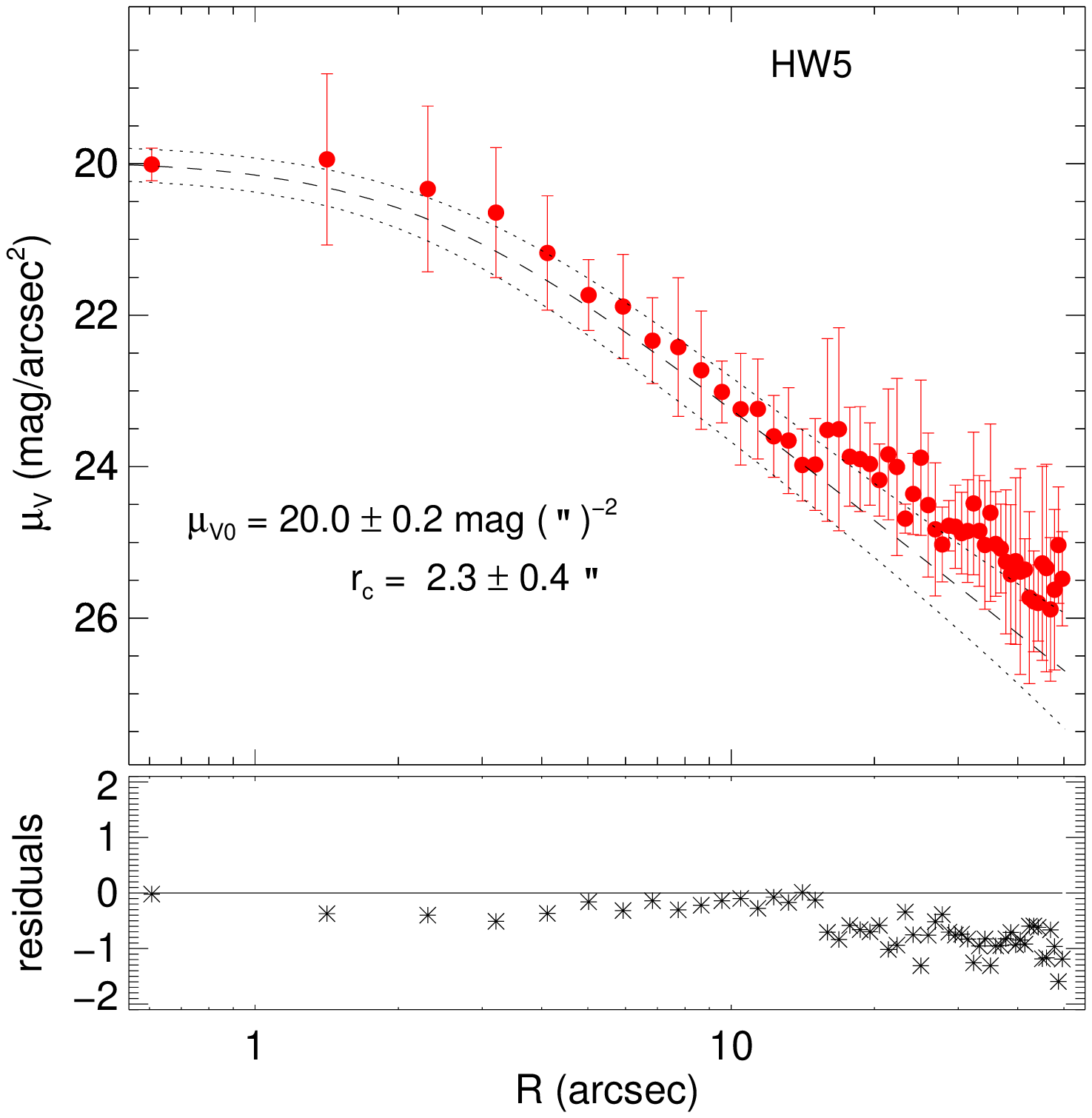}\includegraphics[width=0.325\linewidth]{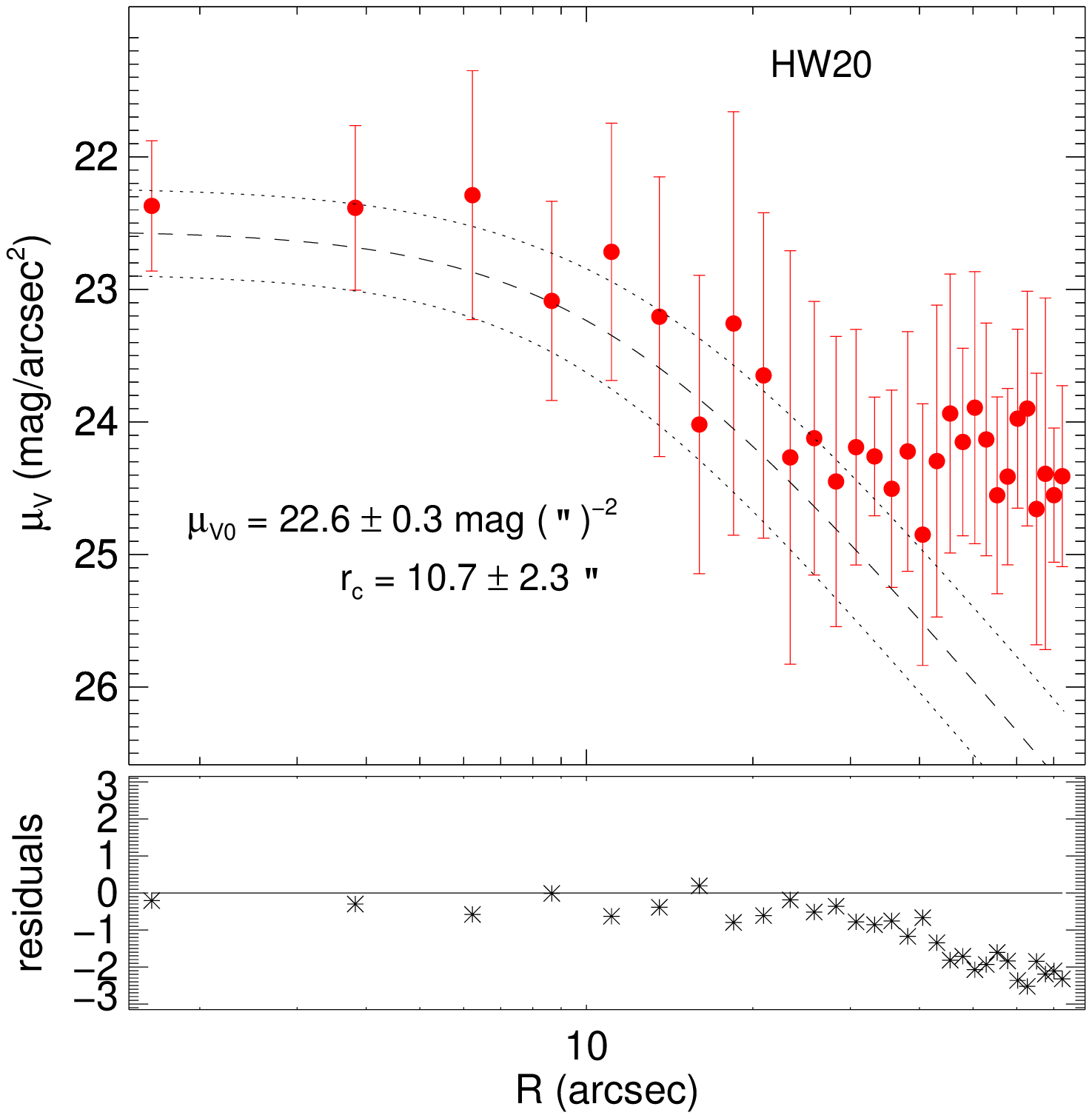}

\includegraphics[width=0.325\linewidth]{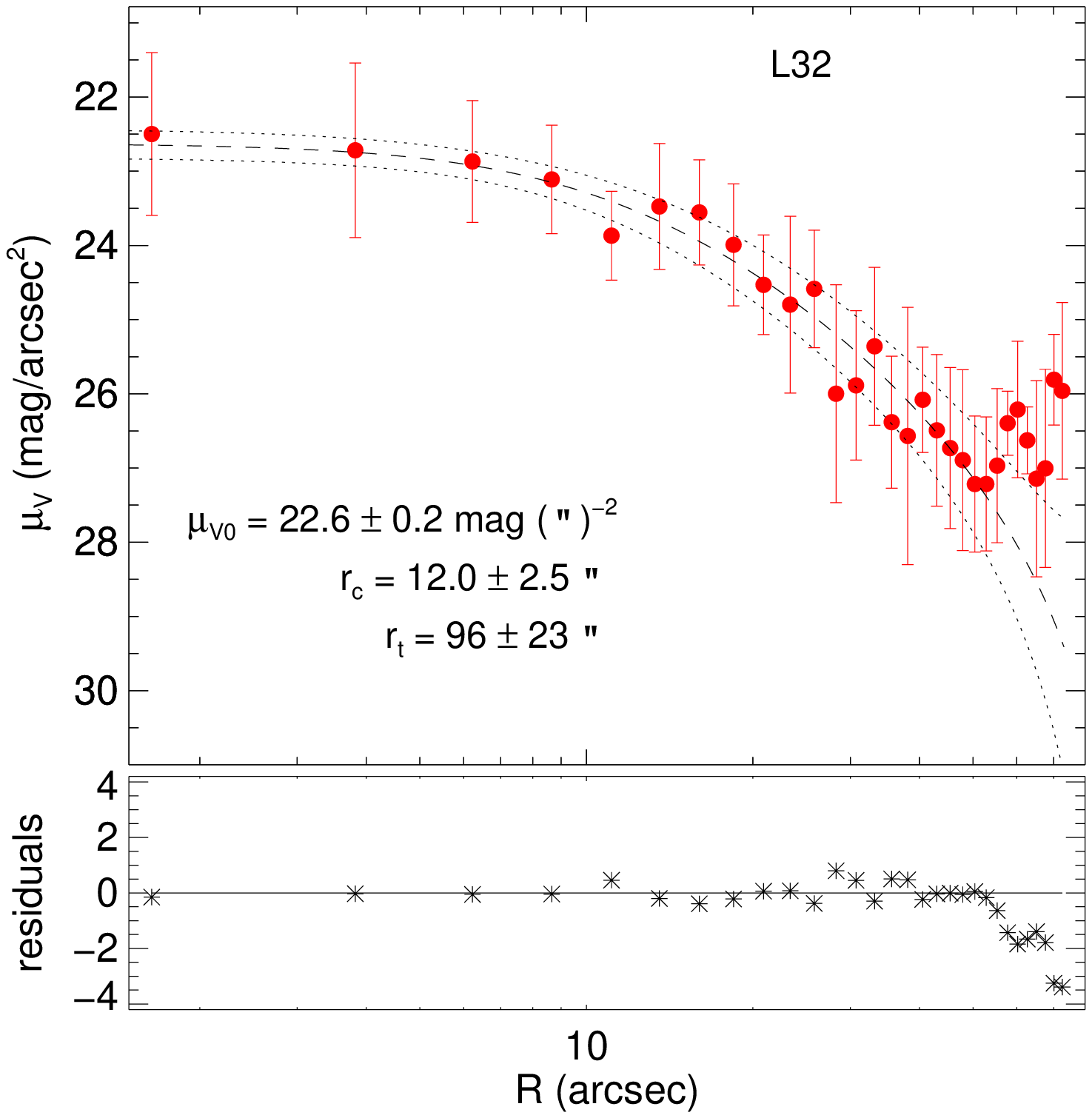}\includegraphics[width=0.325\linewidth]{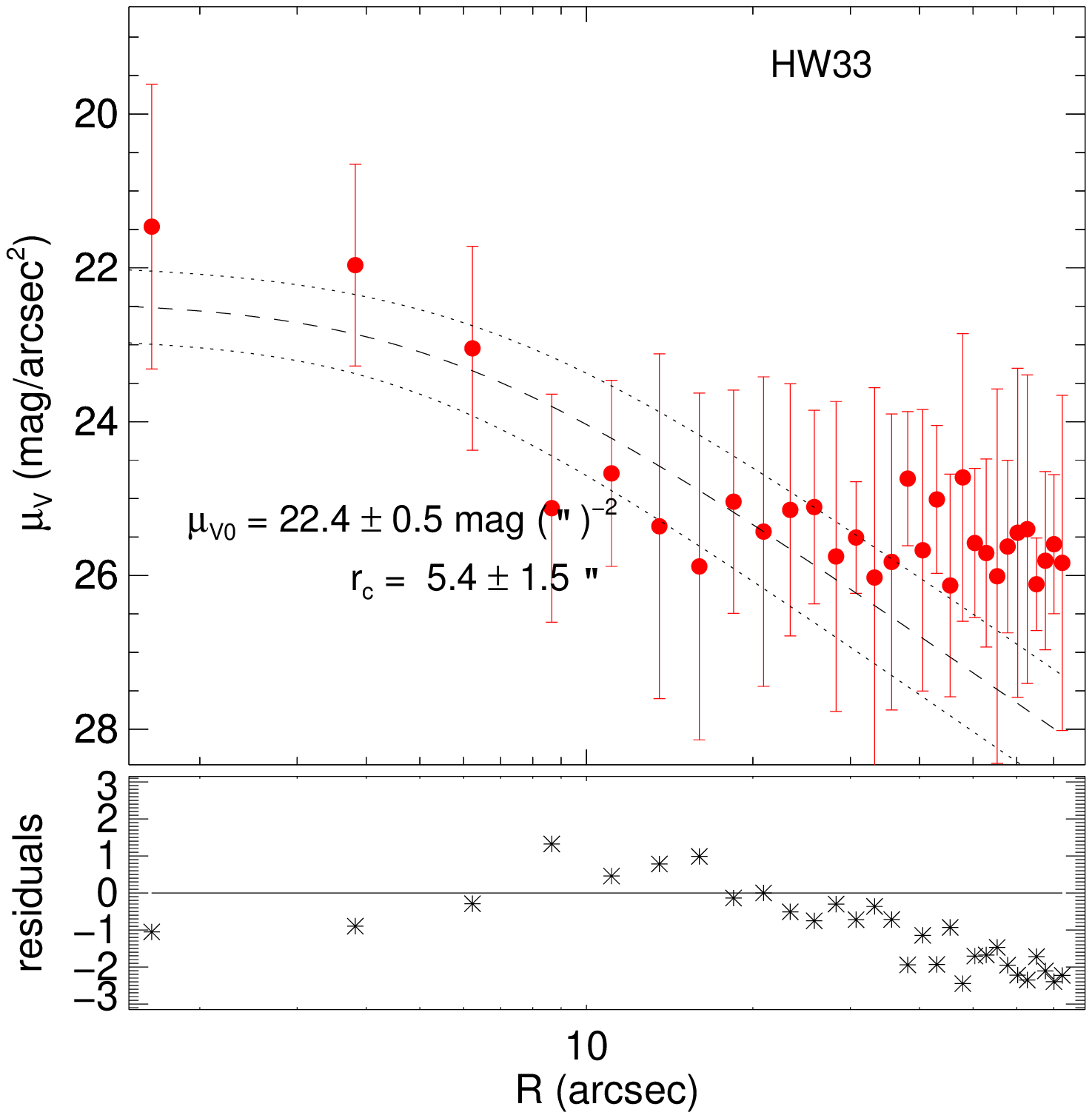}\includegraphics[width=0.325\linewidth]{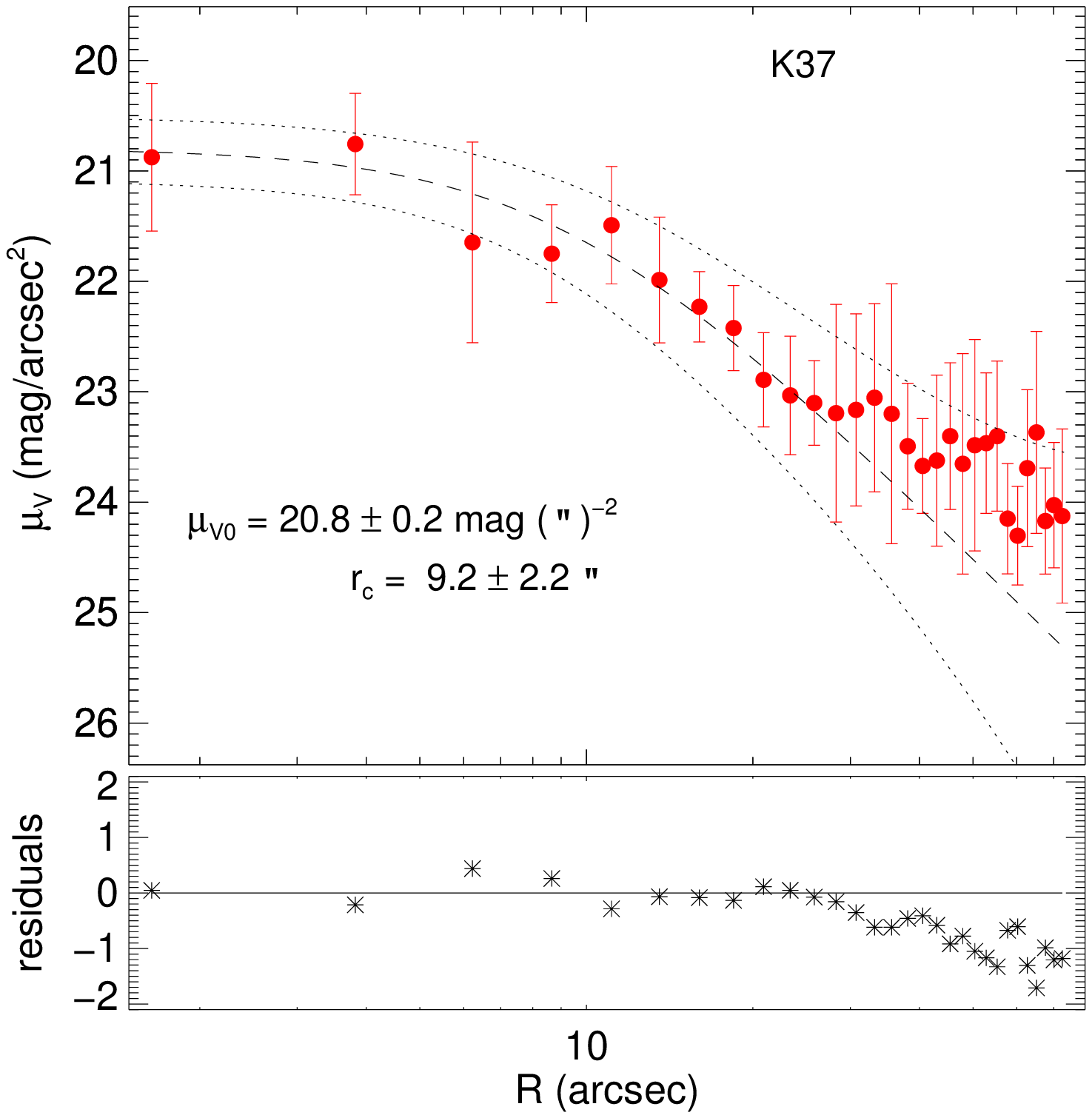}

\includegraphics[width=0.325\linewidth]{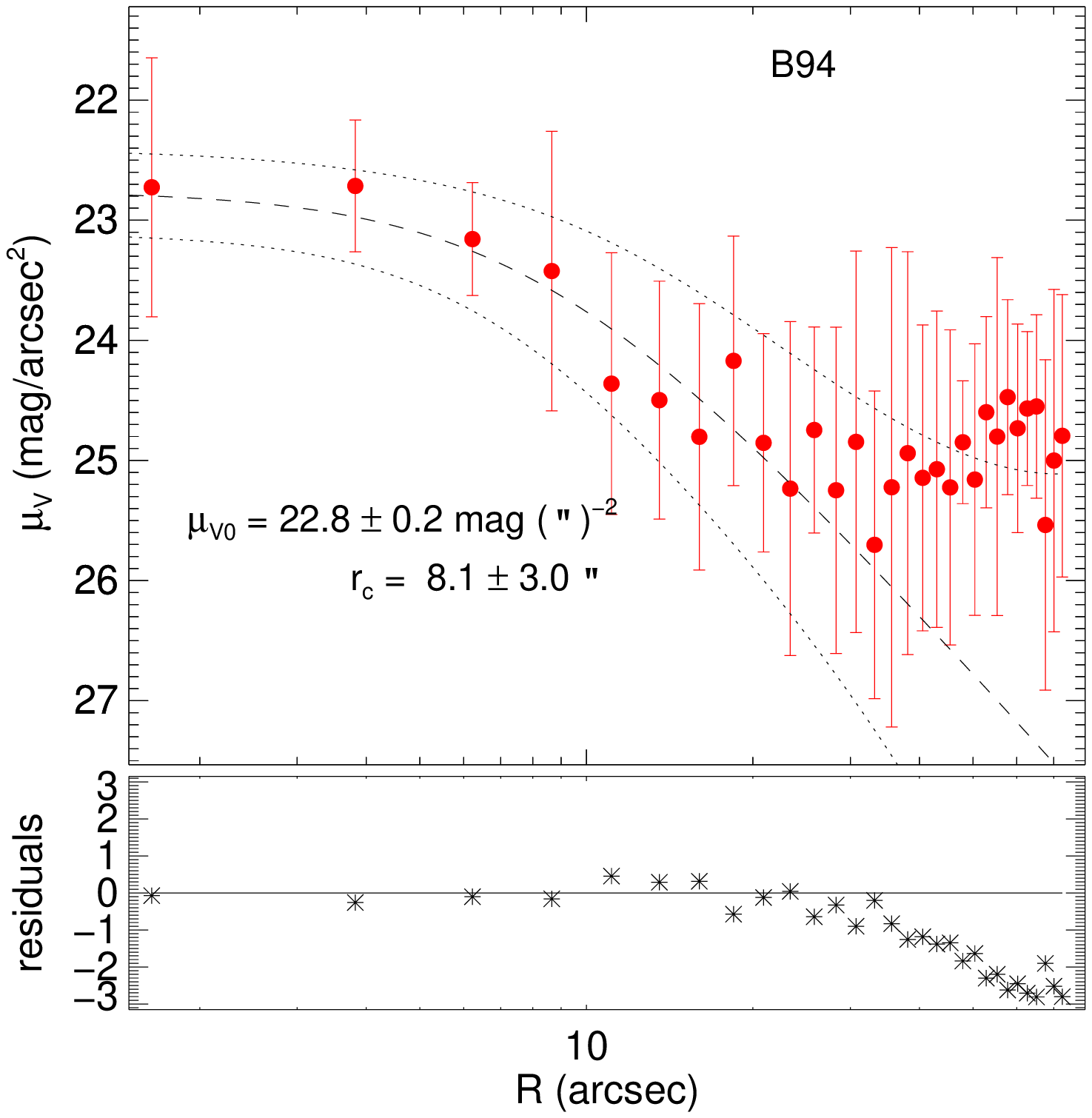}\includegraphics[width=0.325\linewidth]{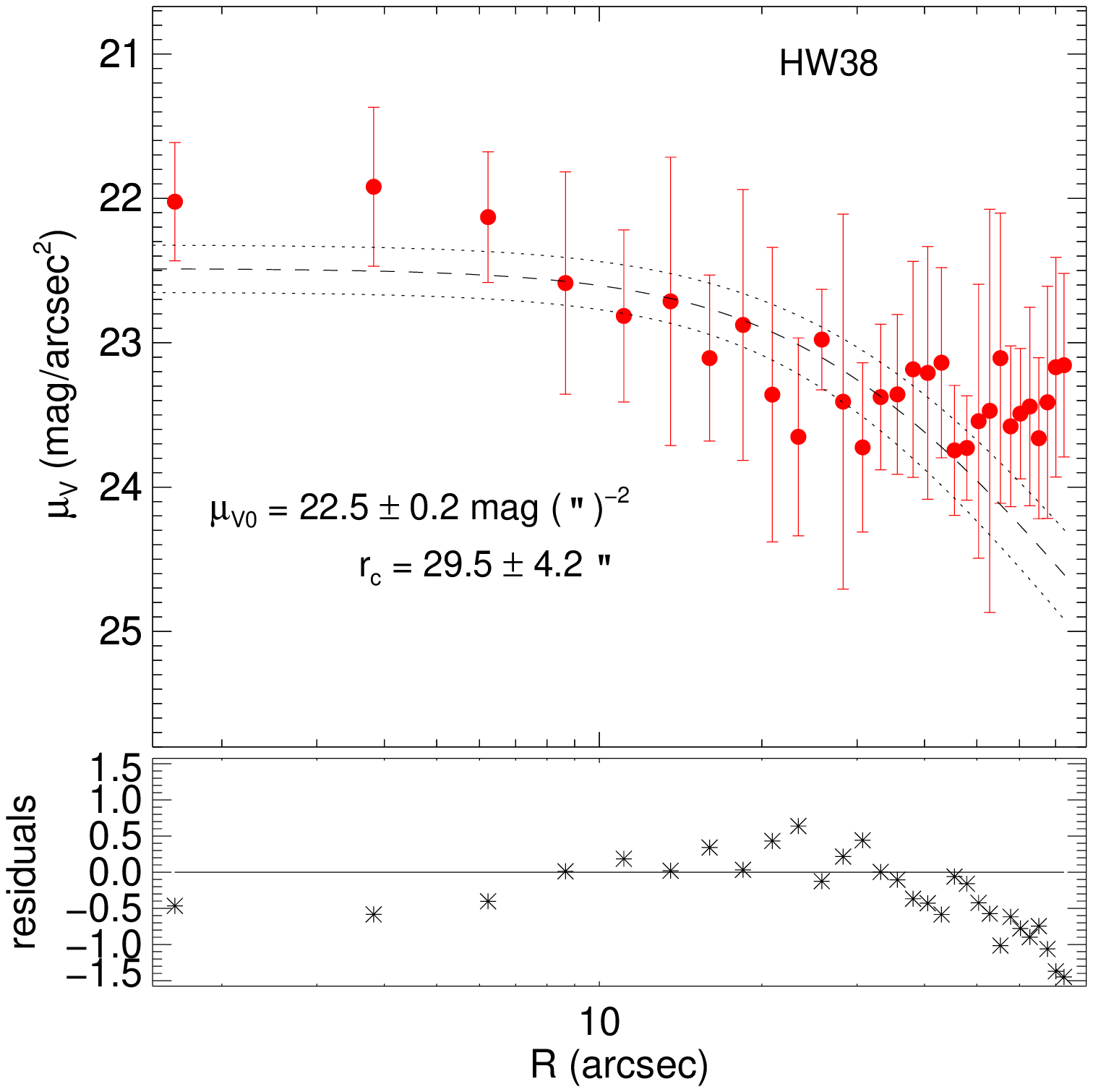}\includegraphics[width=0.325\linewidth]{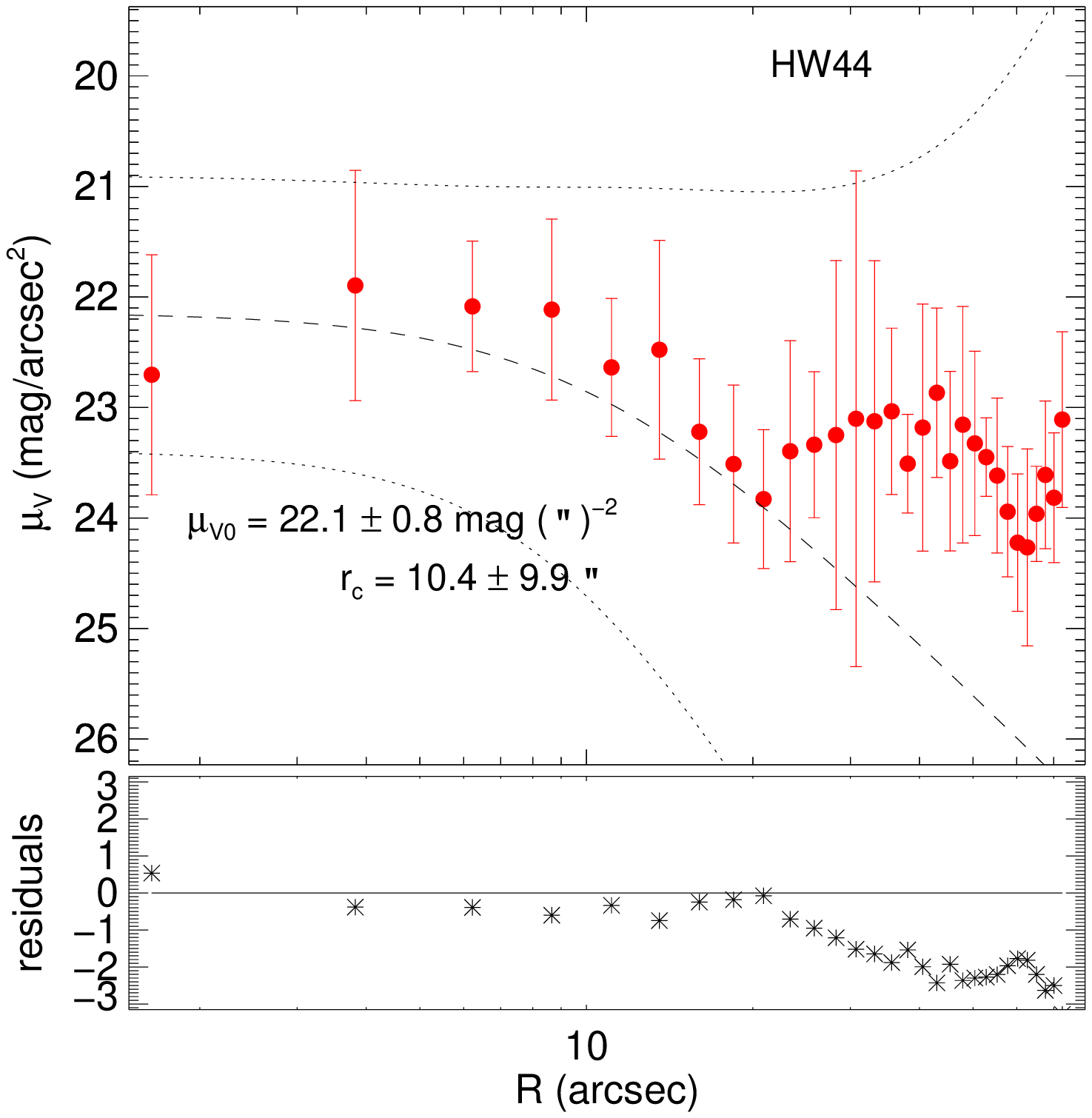}

\includegraphics[width=0.325\linewidth]{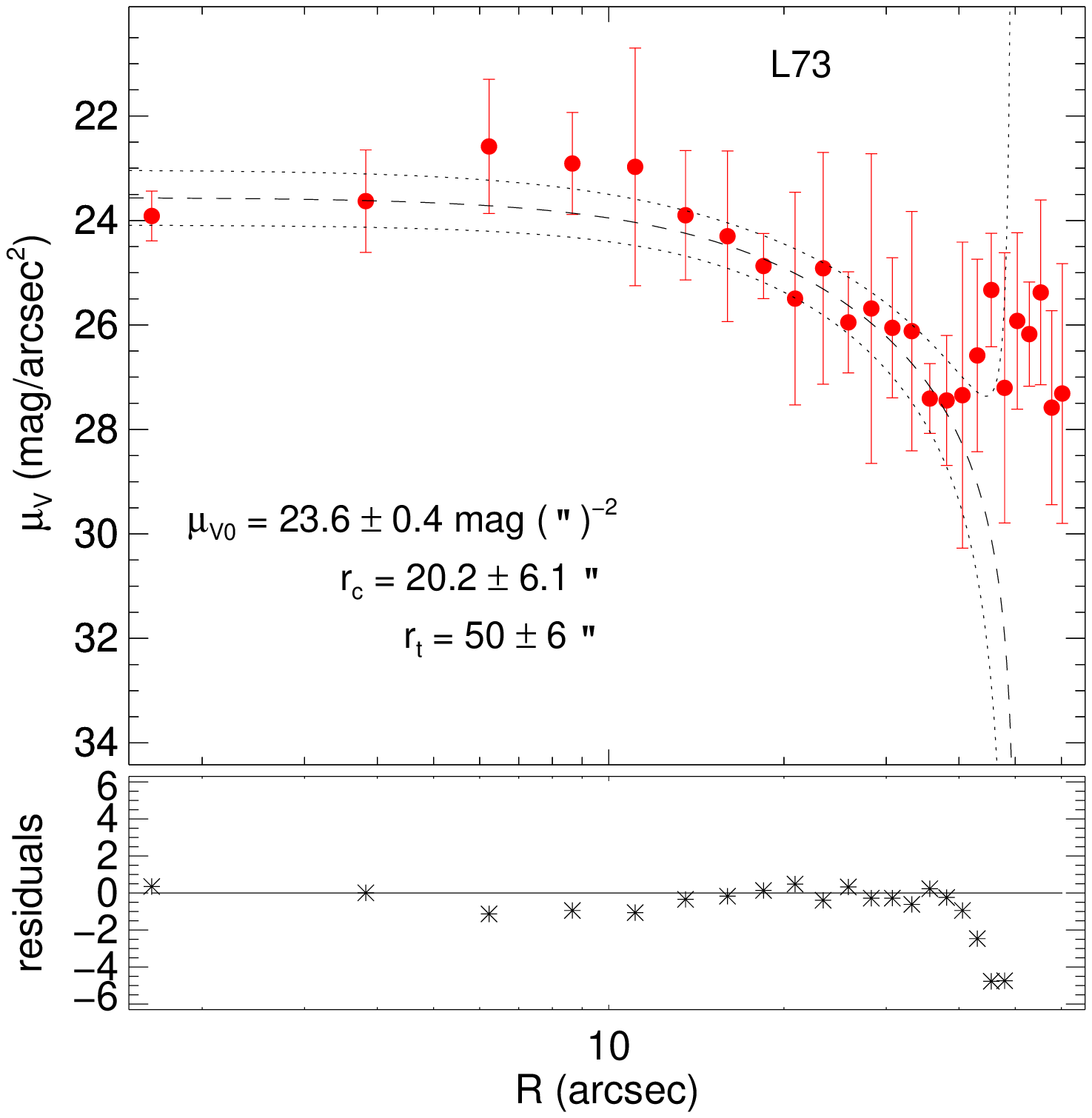}\includegraphics[width=0.325\linewidth]{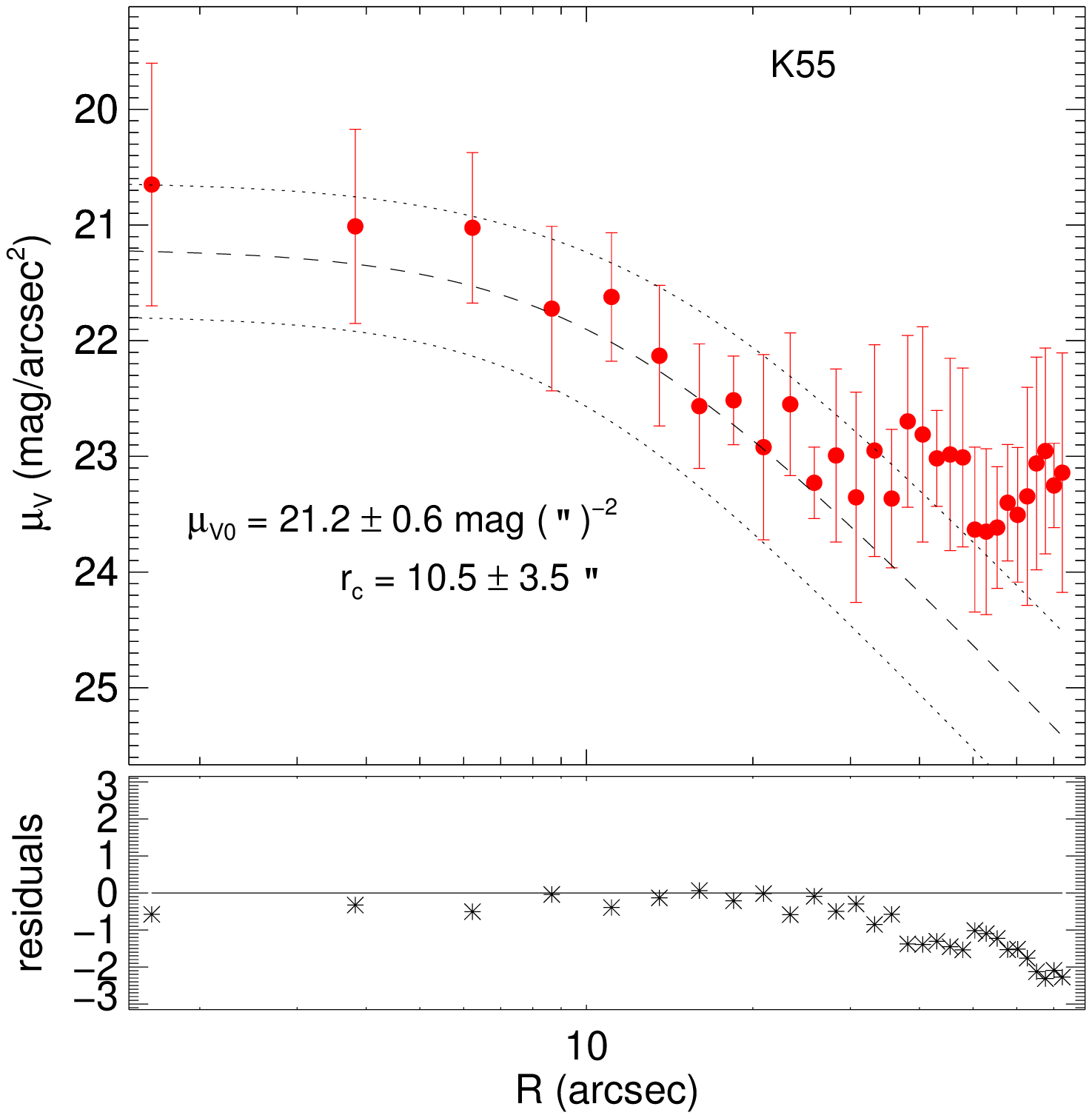}\includegraphics[width=0.325\linewidth]{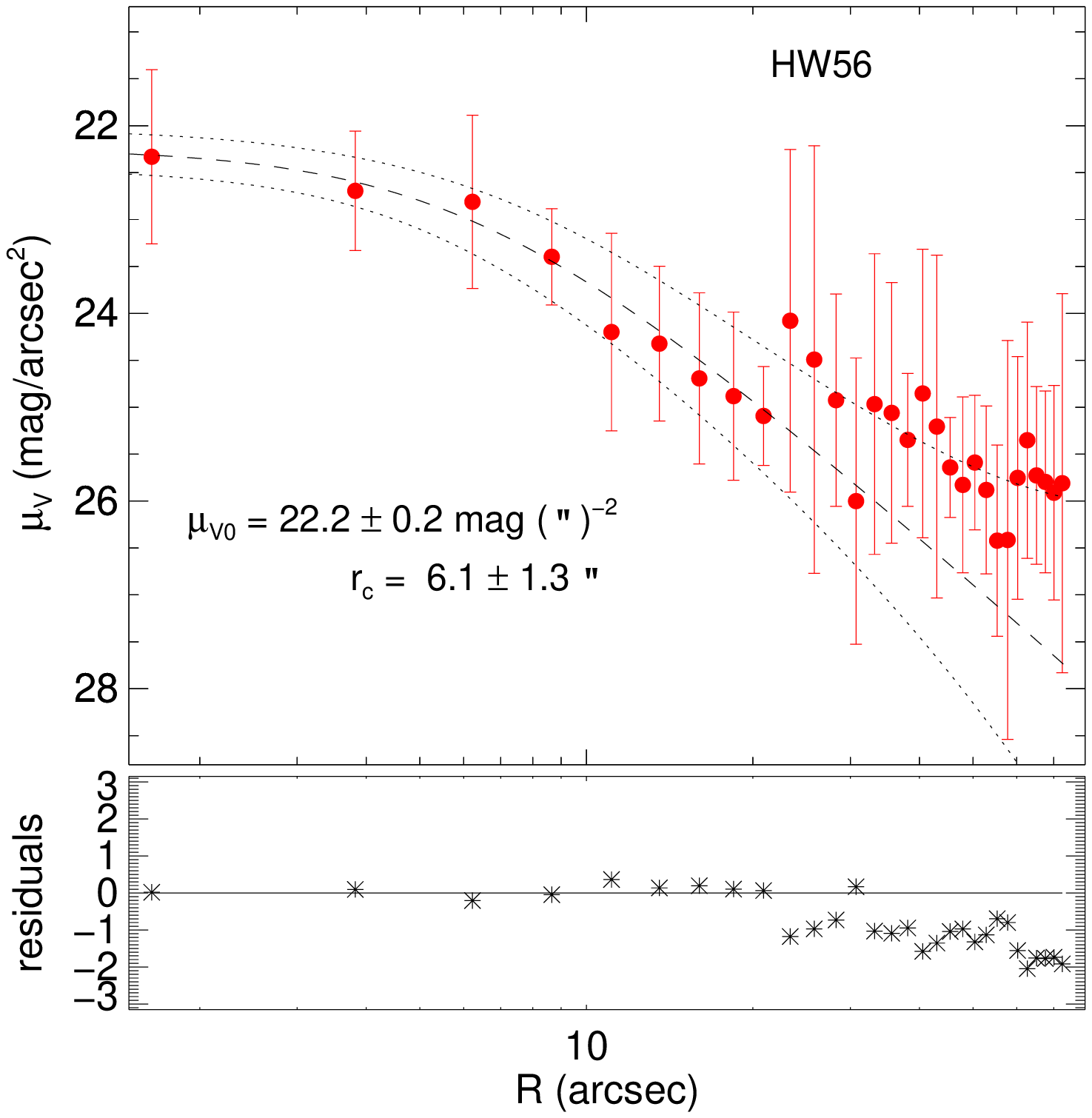}

\caption{Surface brightness profiles of additional SMC clusters complementing the sample presented in Fig.~\ref{fig:rdp_sbp}. The King model fits (dashed lines)  and 1\,$\sigma$ uncertainties (dotted lines) are shown. The best-fitting  parameters are indicated and the fit residuals are plotted in the lower panel.}

\end{figure*}

\setcounter{figure}{2}

\begin{figure*}
\includegraphics[width=0.325\linewidth]{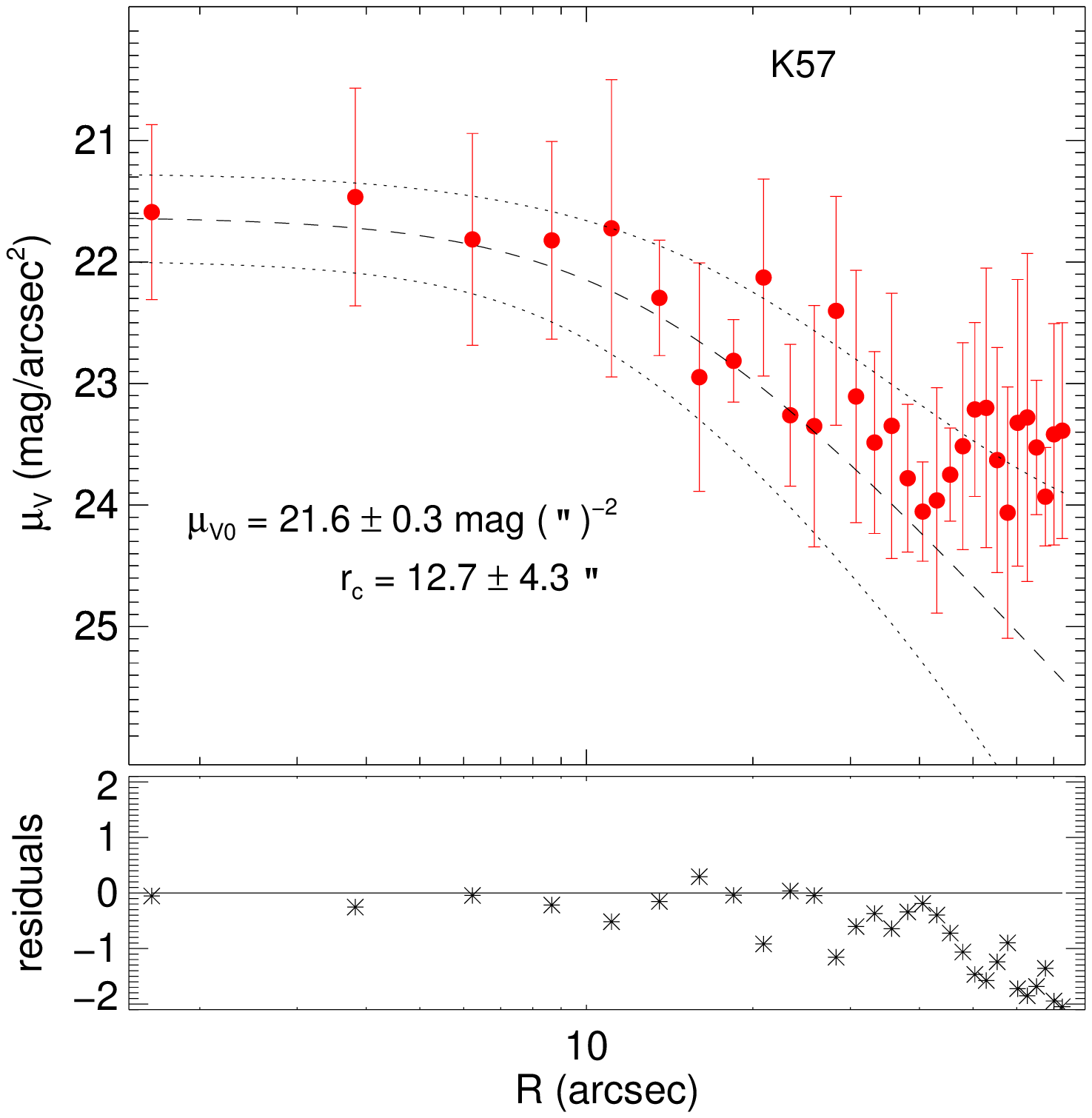}\includegraphics[width=0.325\linewidth]{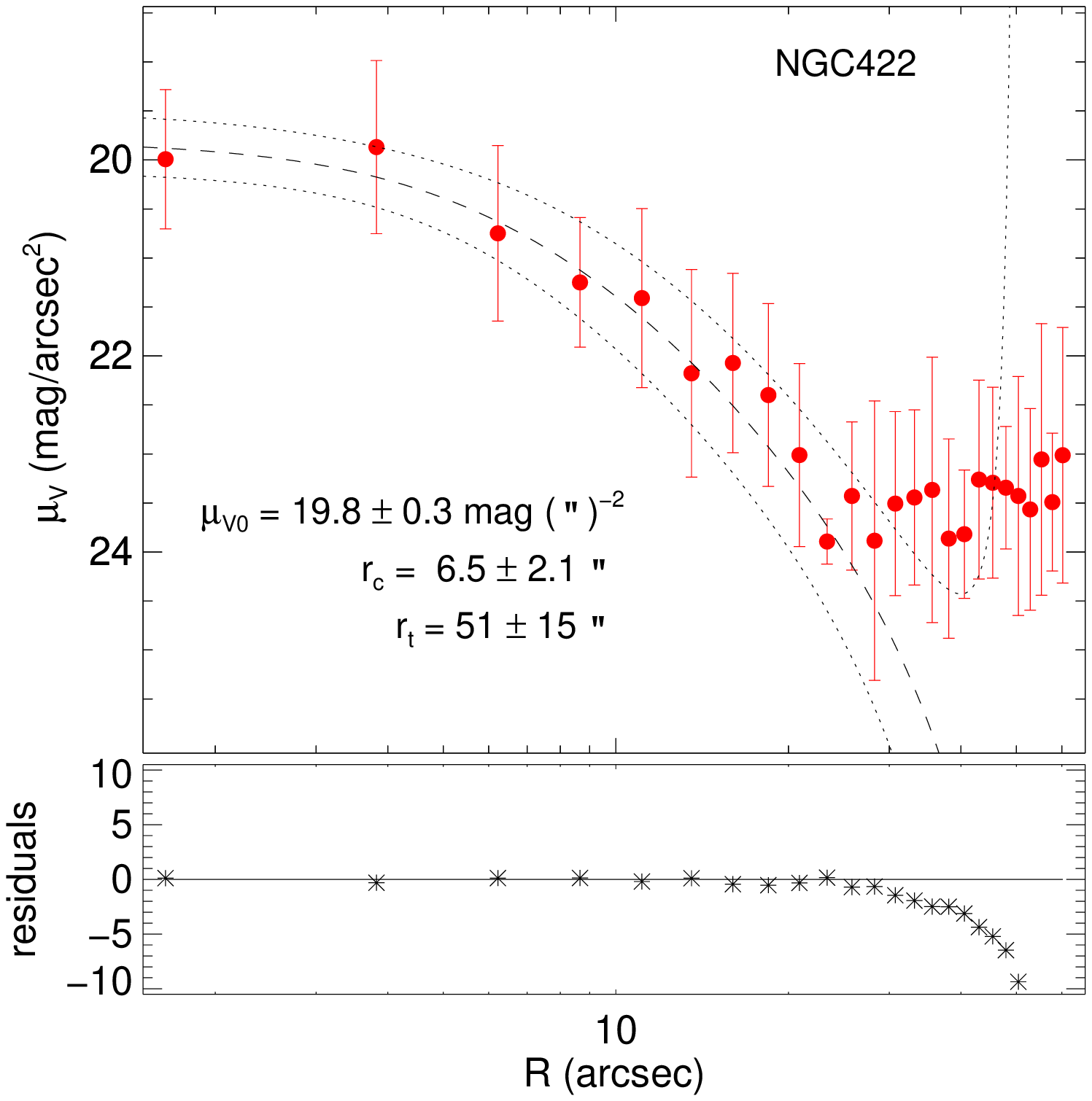}\includegraphics[width=0.325\linewidth]{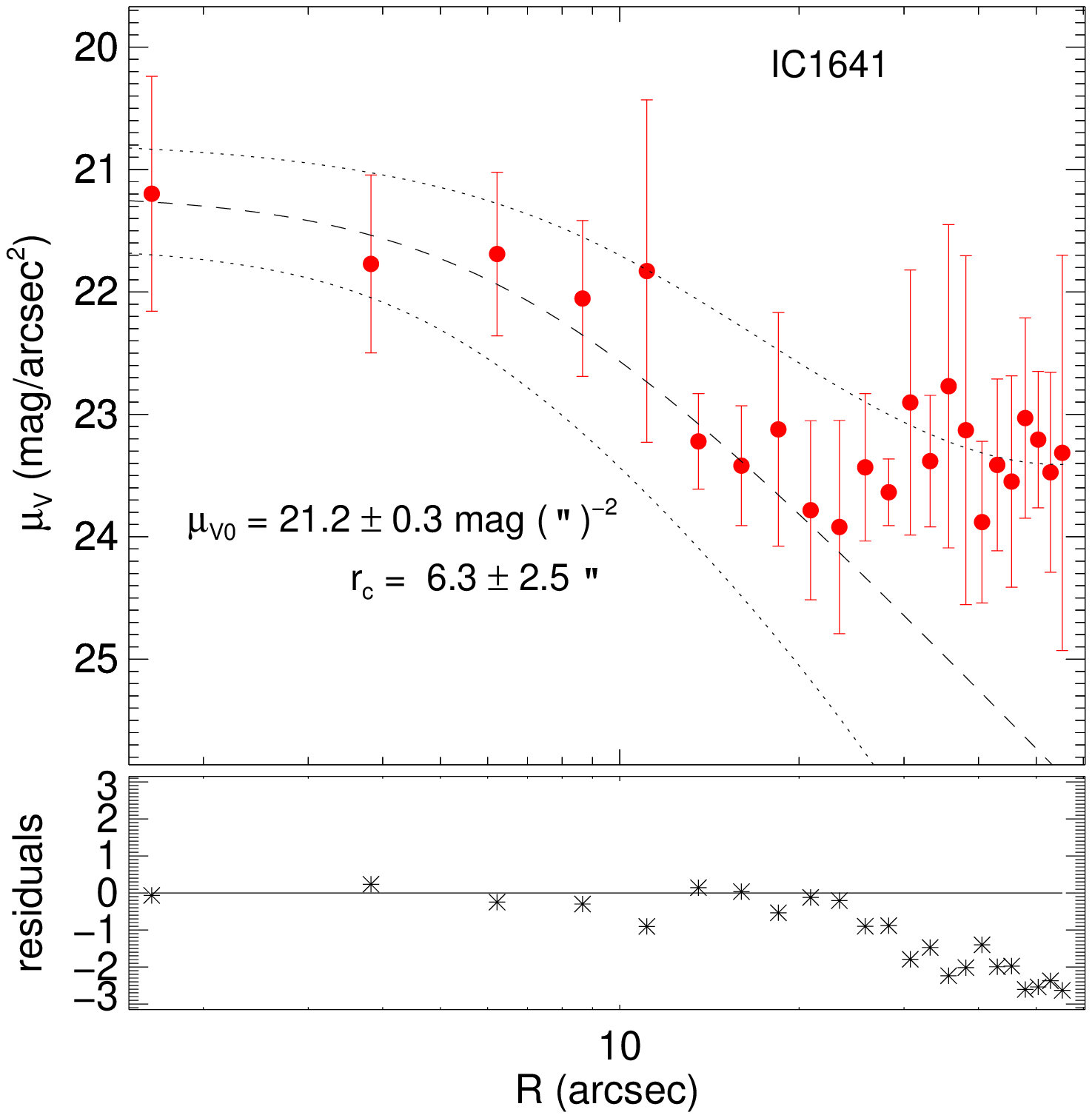}

\includegraphics[width=0.325\linewidth]{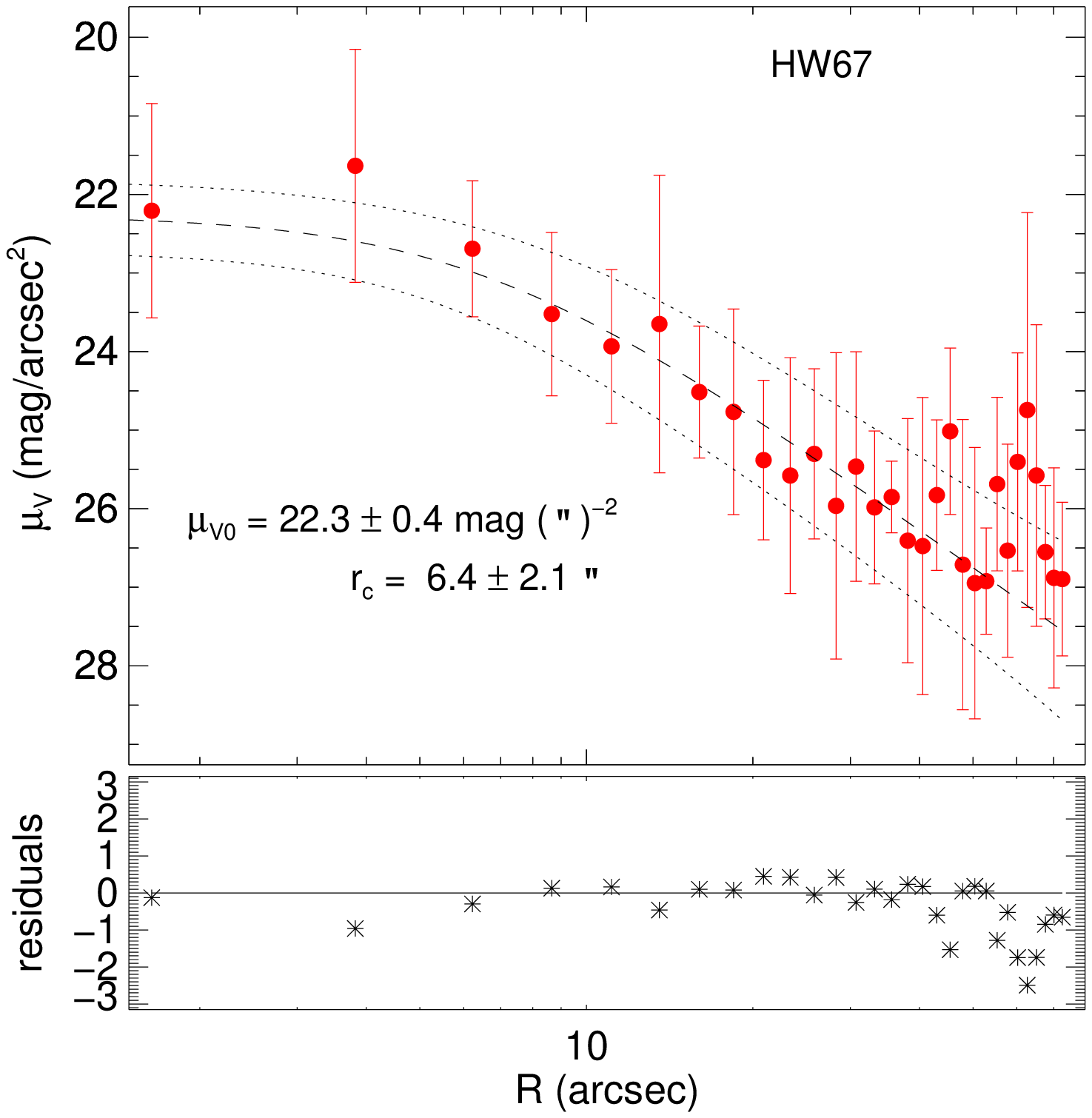}\includegraphics[width=0.325\linewidth]{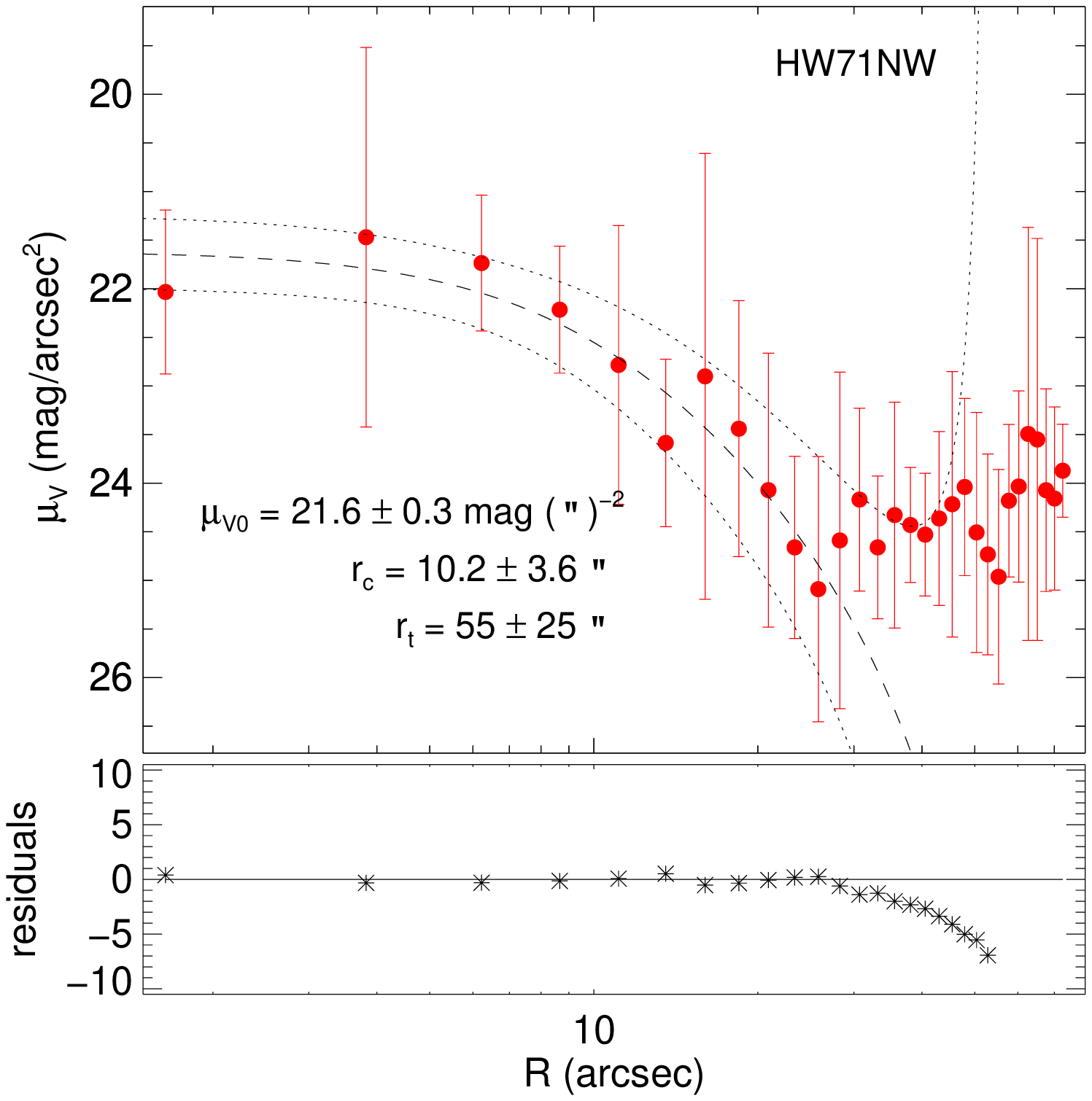}\includegraphics[width=0.325\linewidth]{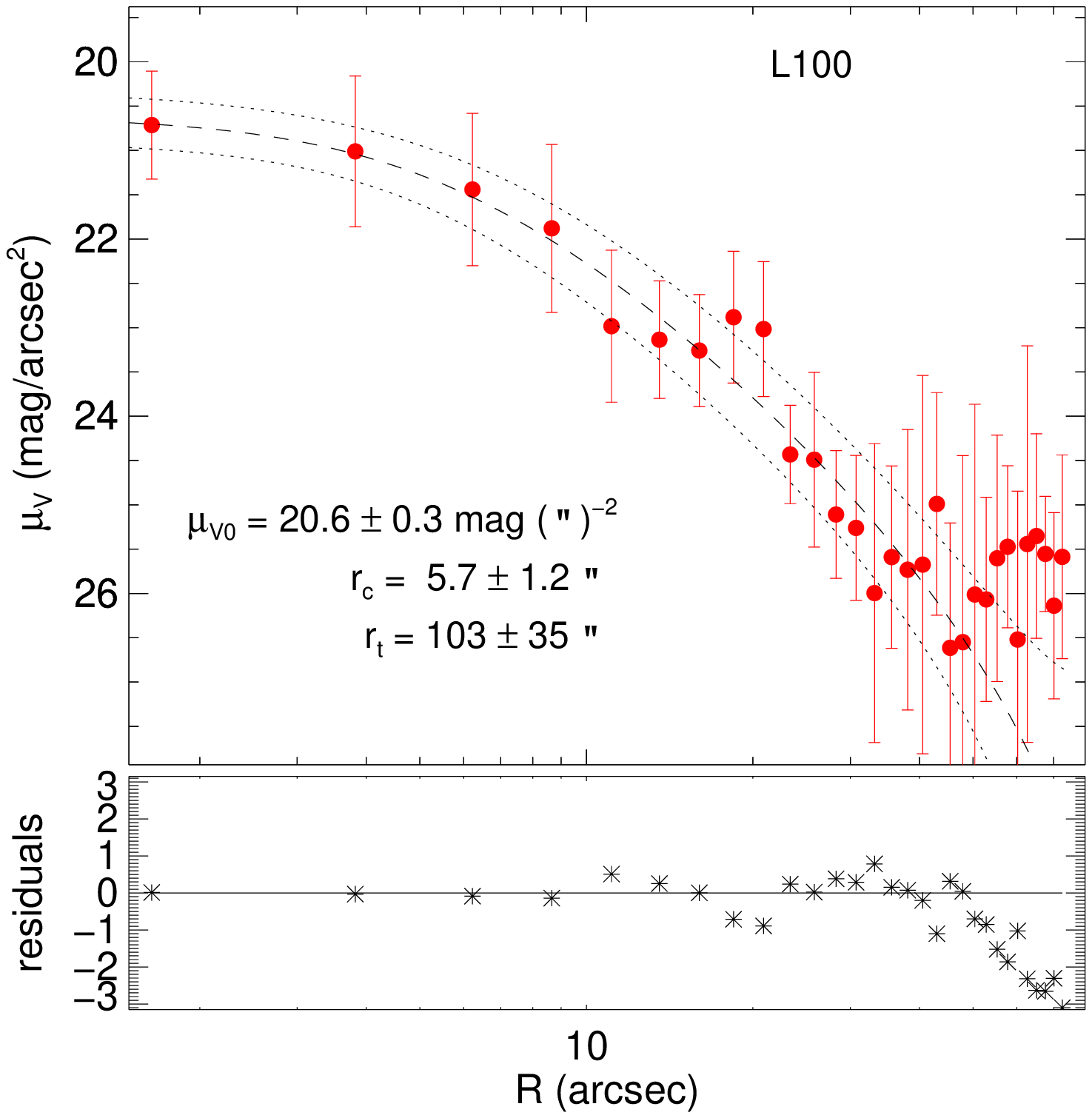}

\includegraphics[width=0.325\linewidth]{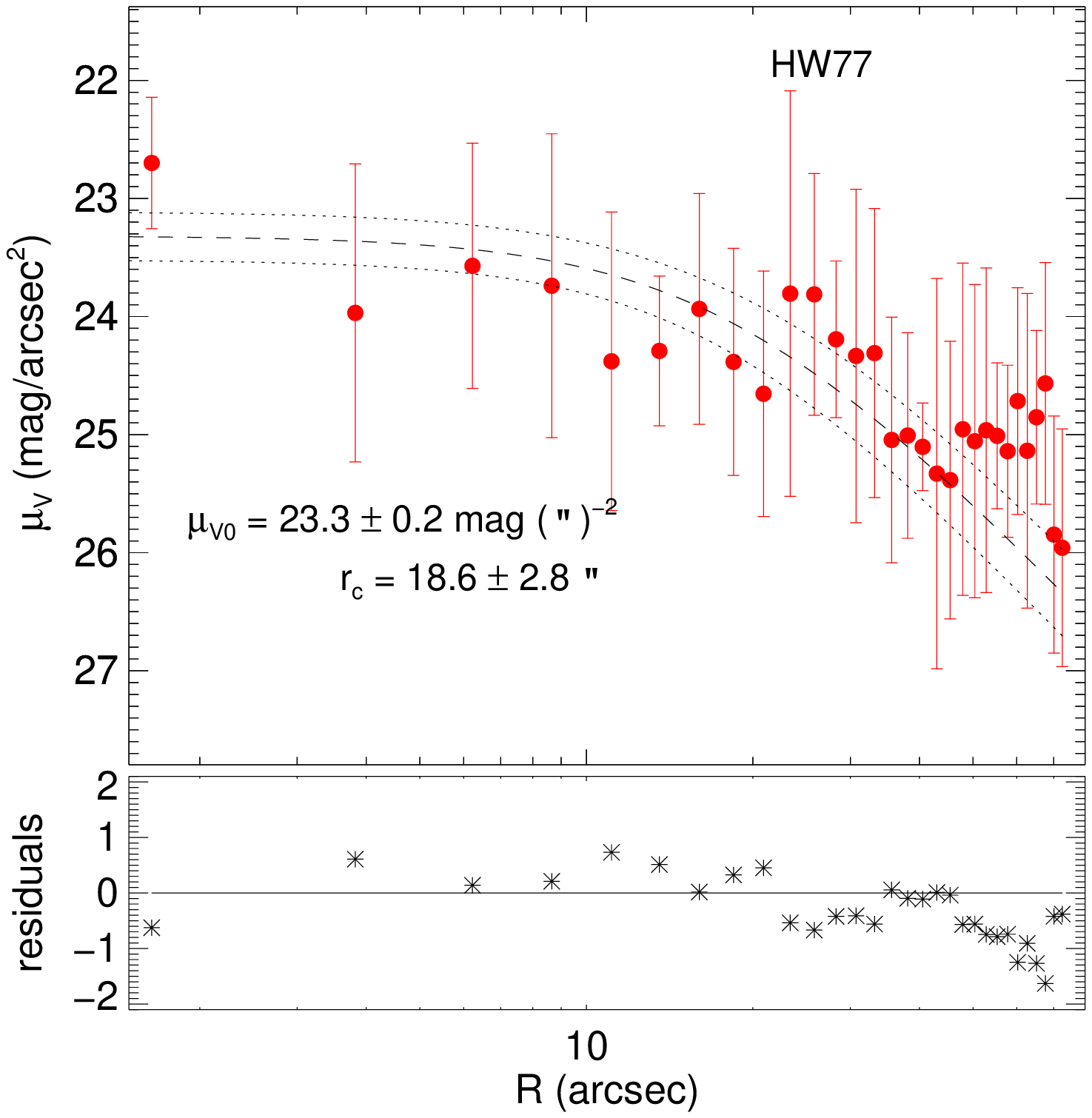}\includegraphics[width=0.325\linewidth]{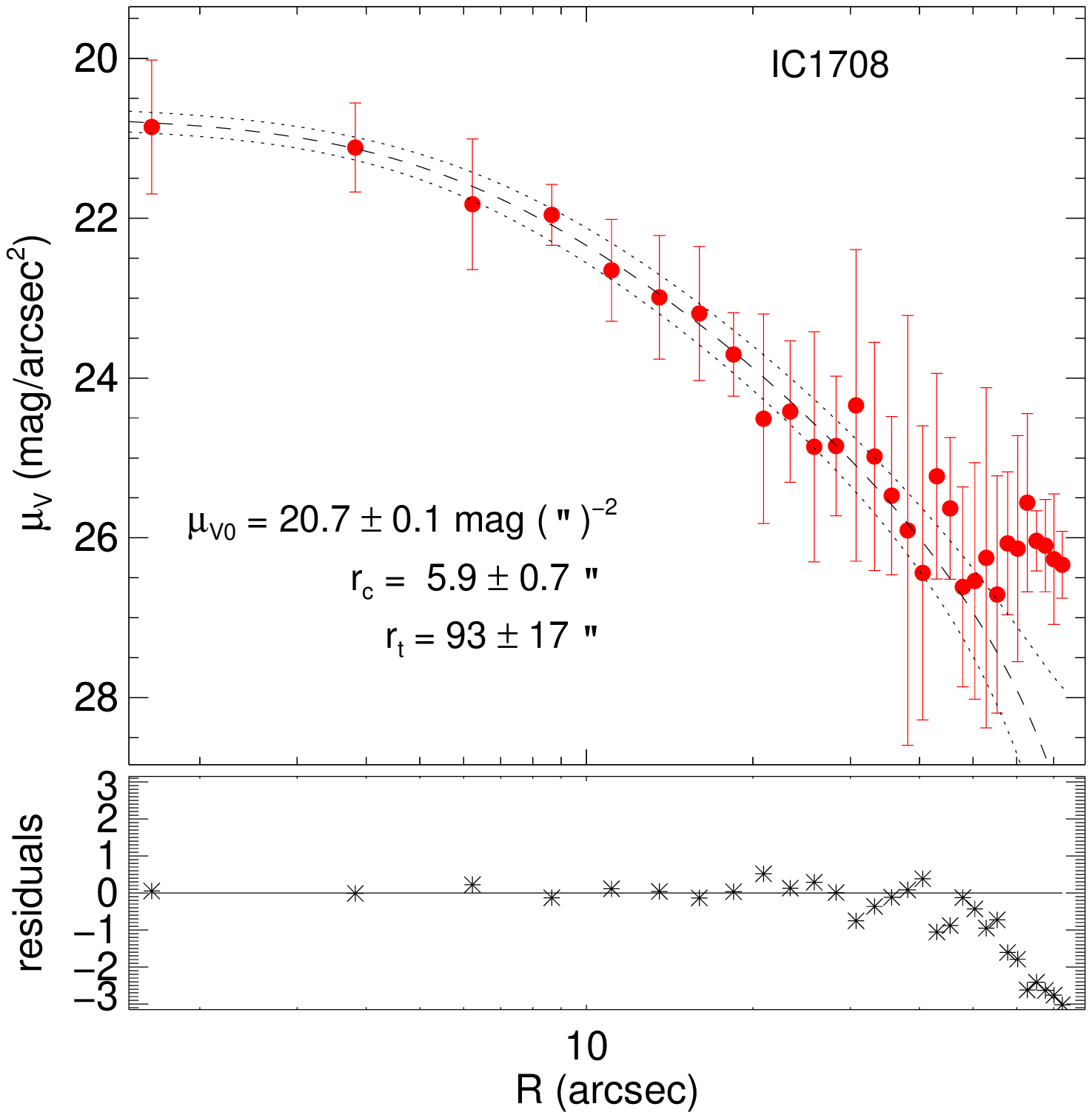}\includegraphics[width=0.325\linewidth]{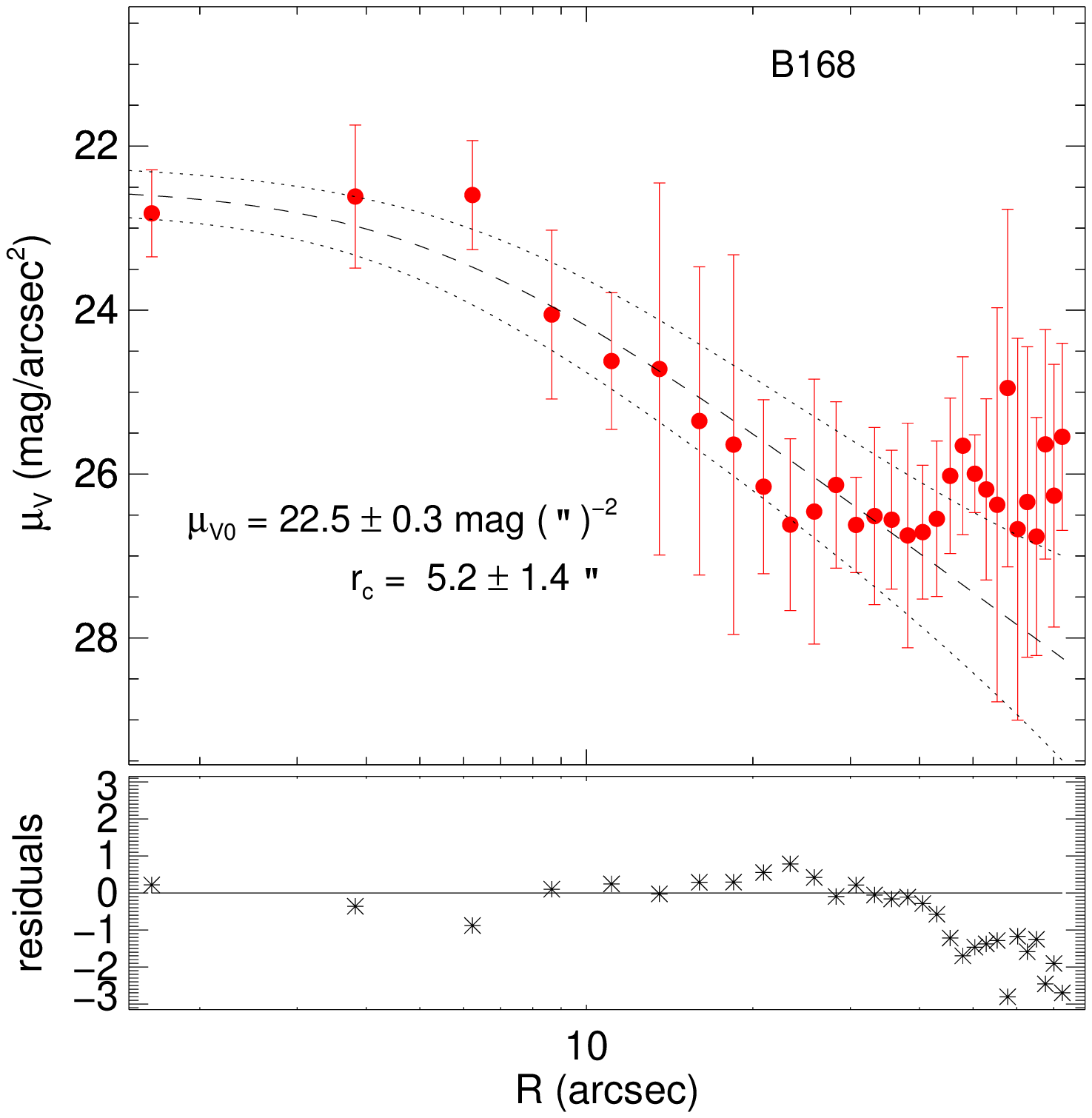}

\includegraphics[width=0.325\linewidth]{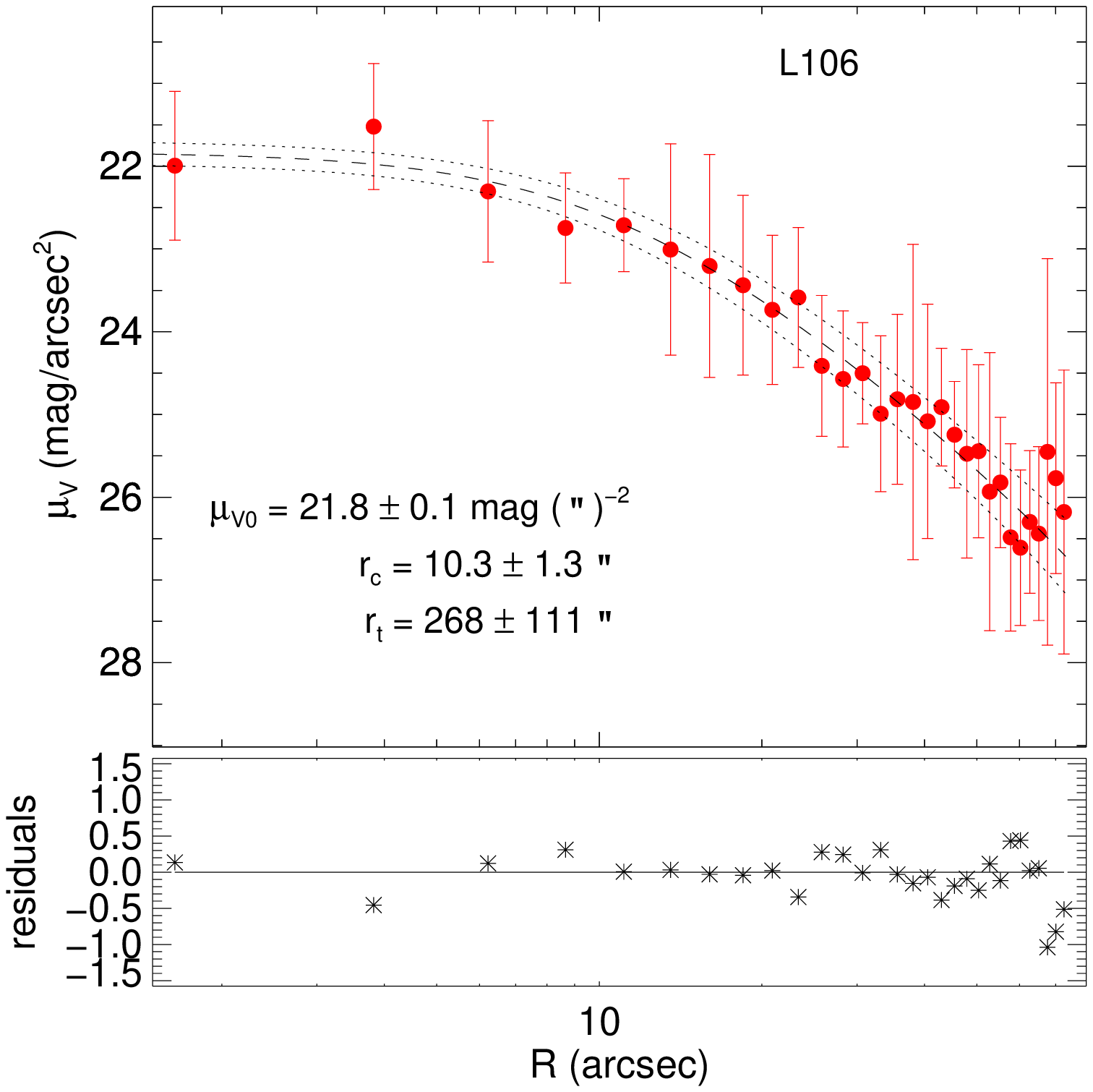}\includegraphics[width=0.325\linewidth]{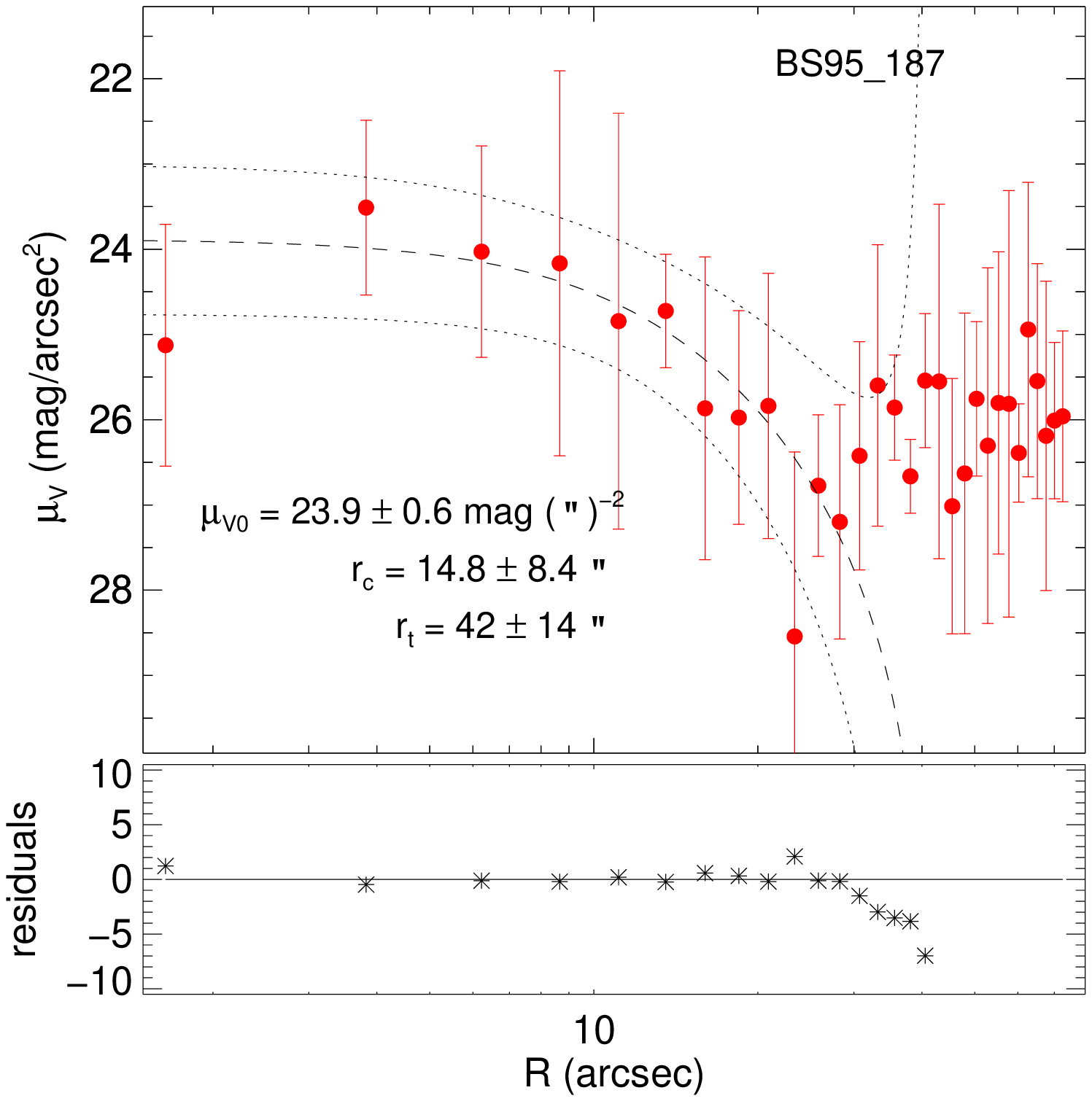}\includegraphics[width=0.325\linewidth]{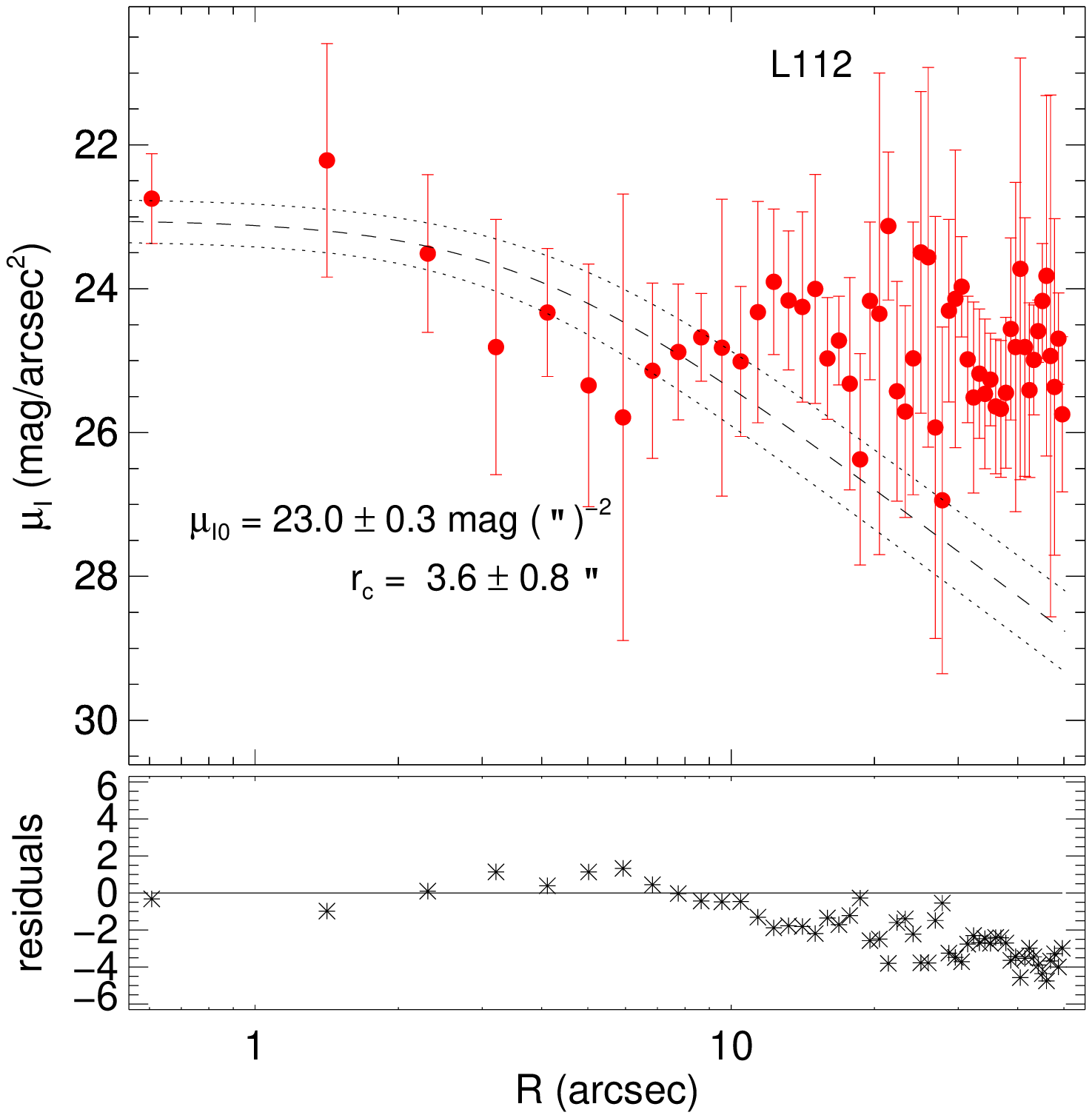}

\caption{cont.}

\end{figure*}

\setcounter{figure}{2}

\begin{figure*}

\includegraphics[width=0.325\linewidth]{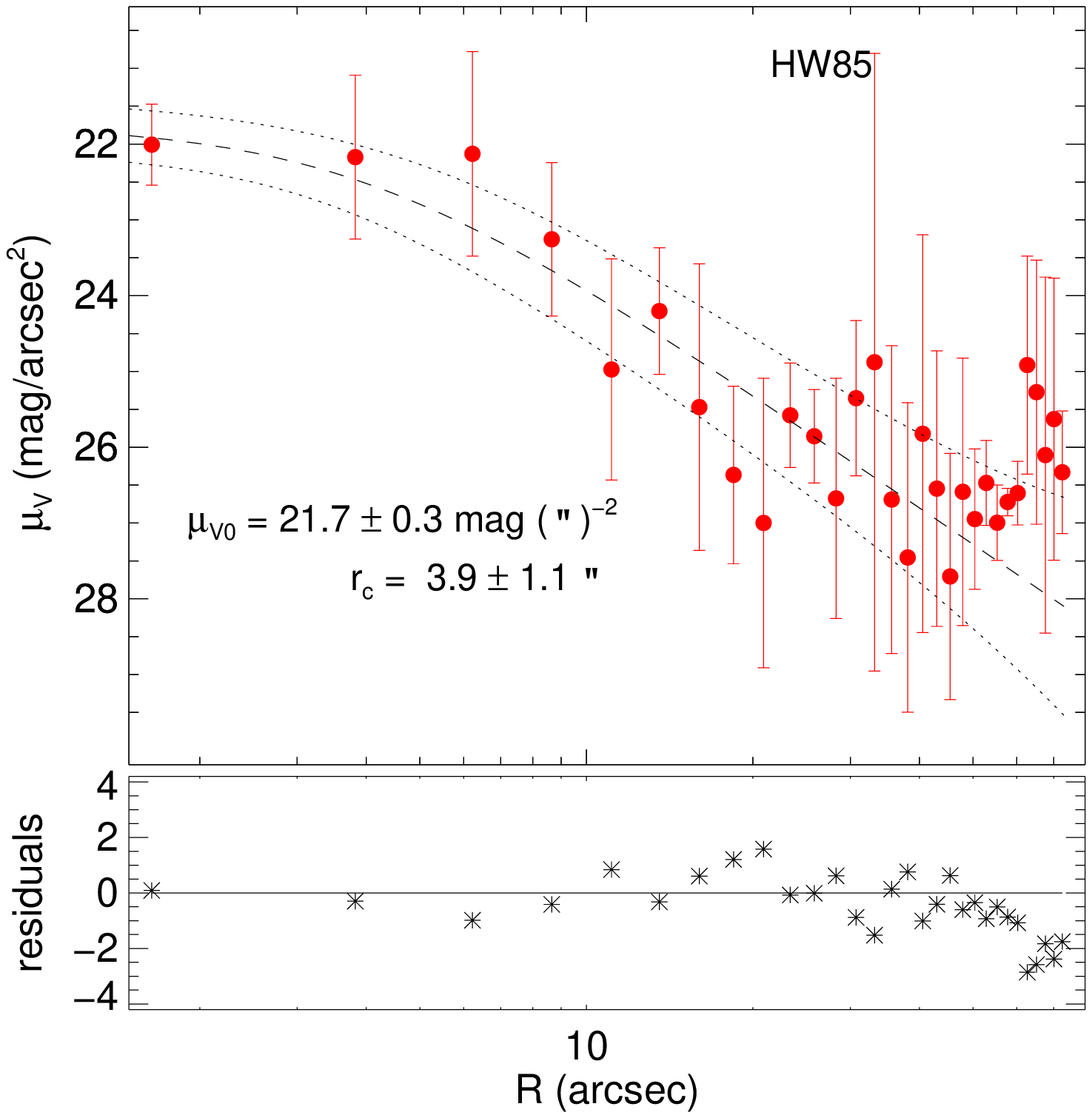}\includegraphics[width=0.325\linewidth]{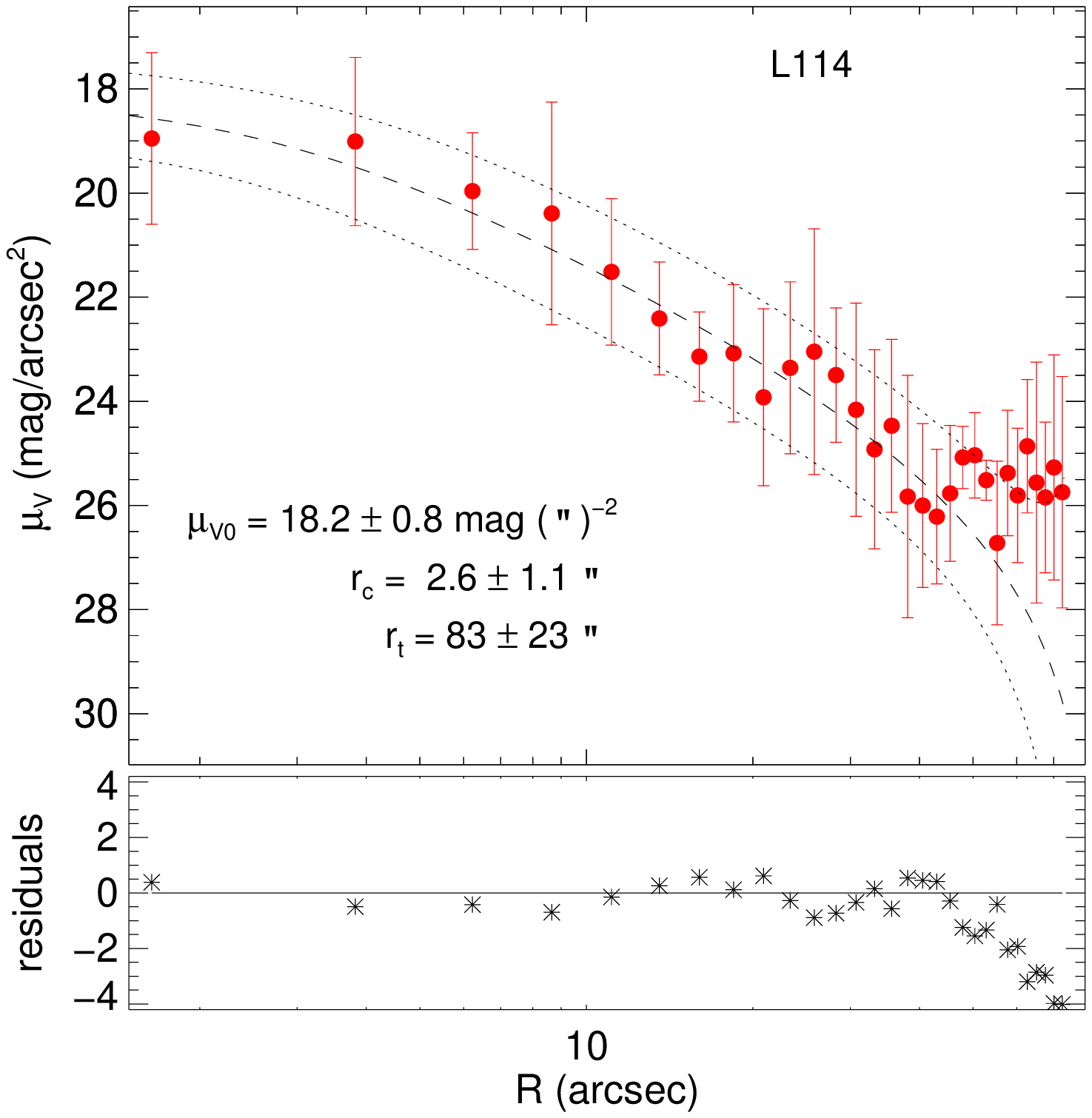}\includegraphics[width=0.325\linewidth]{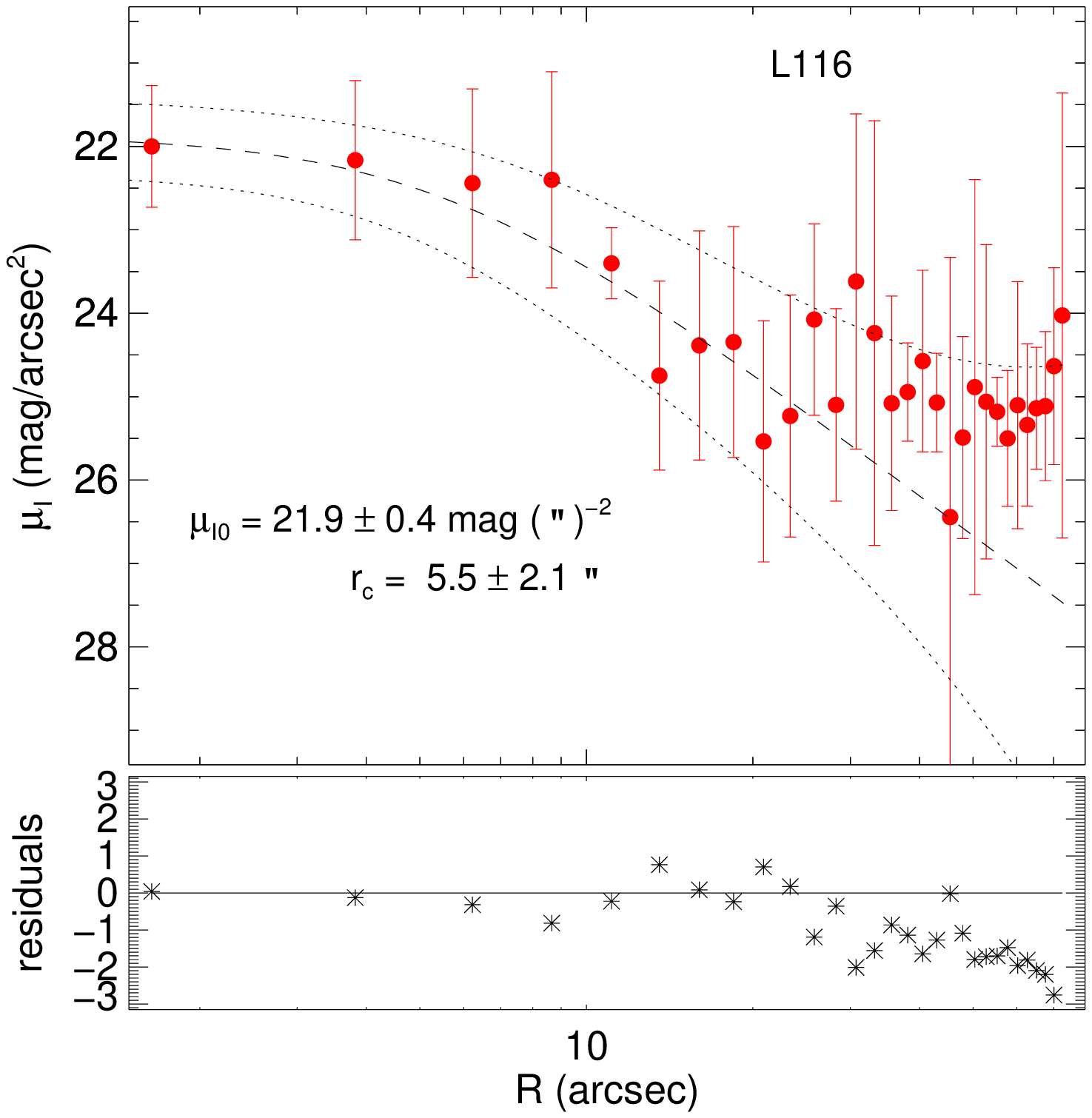}

\includegraphics[width=0.325\linewidth]{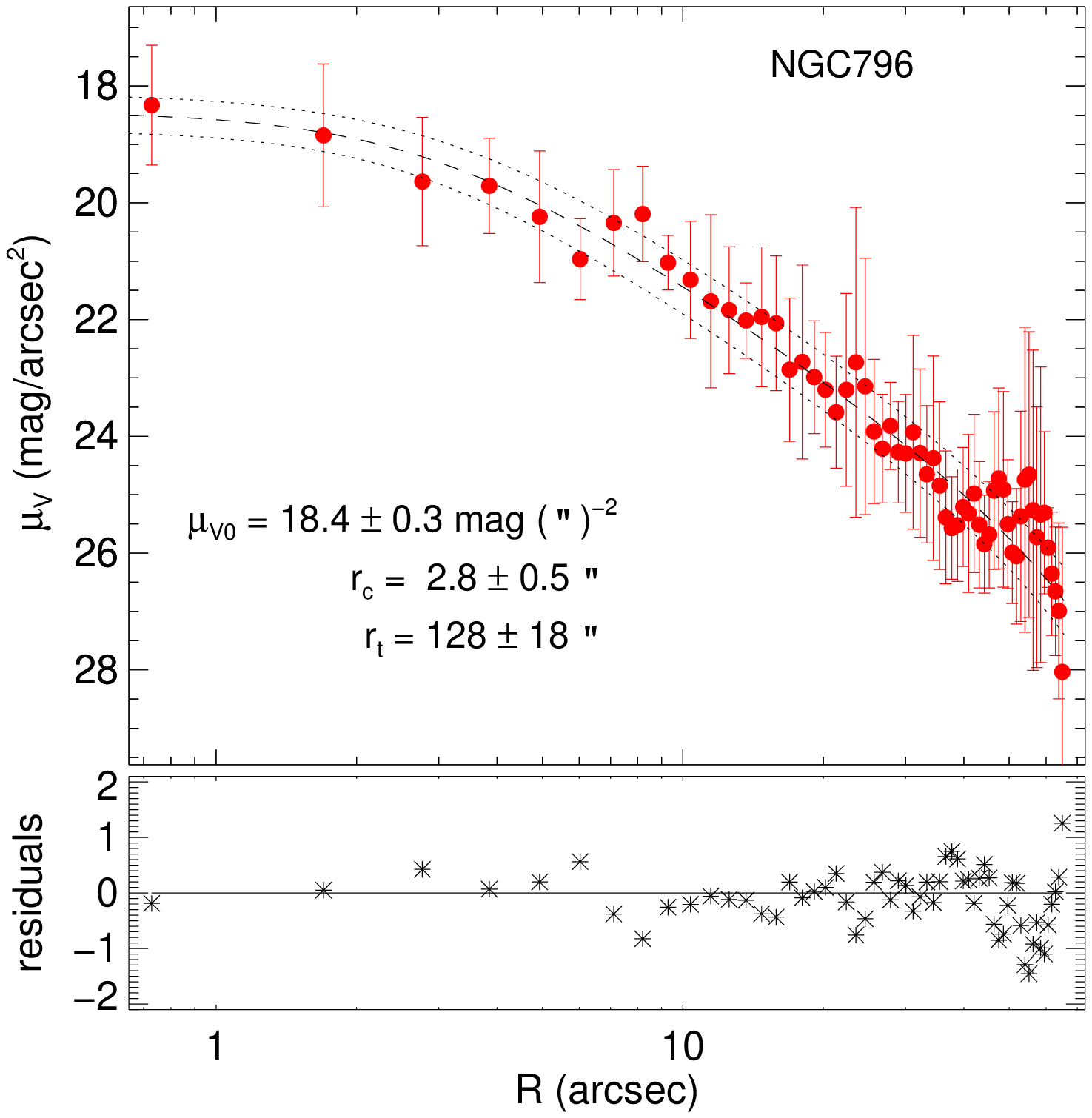}\includegraphics[width=0.325\linewidth]{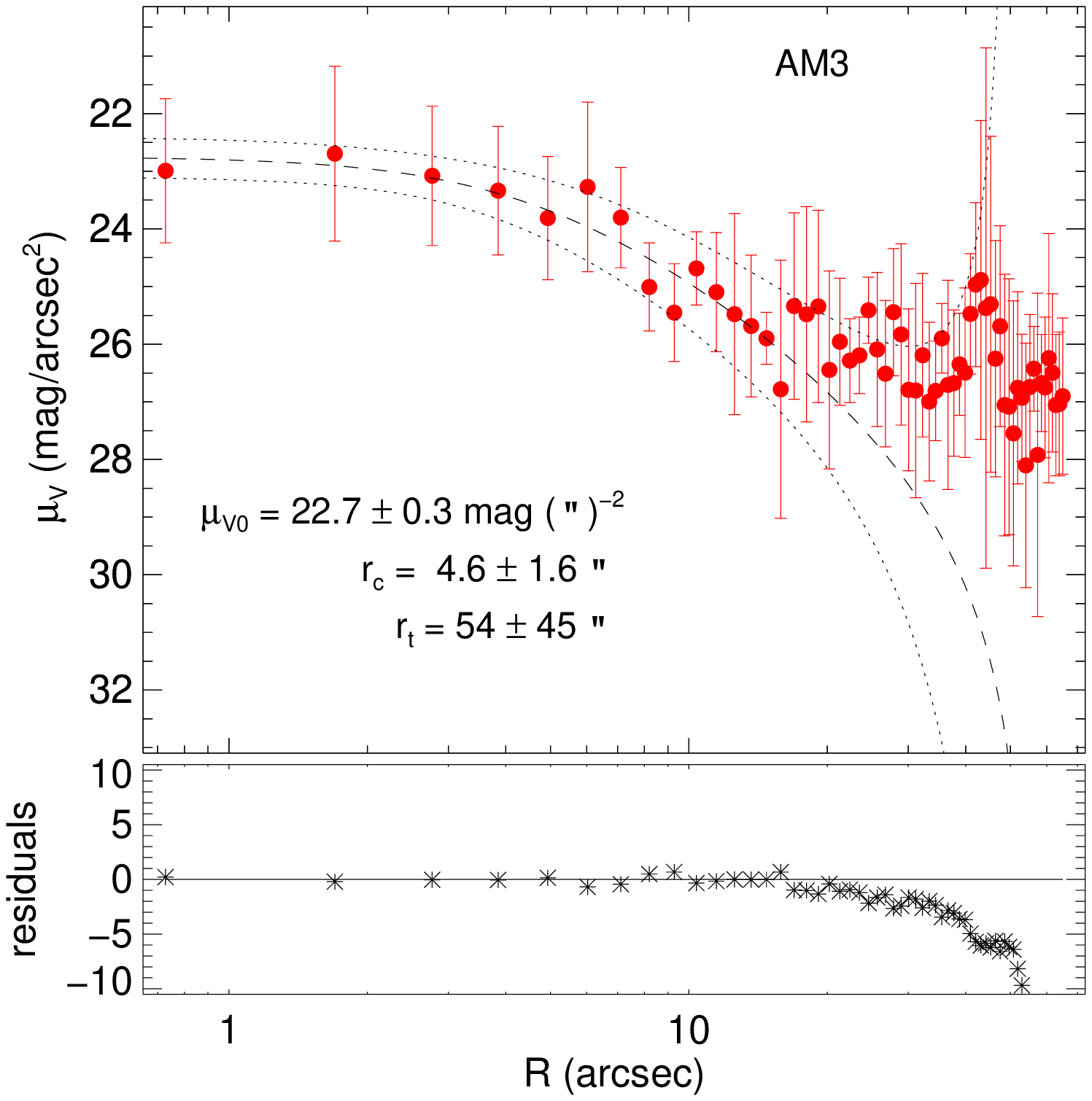}
\caption{cont.}

\end{figure*}

\begin{figure*}
\includegraphics[width=0.325\linewidth]{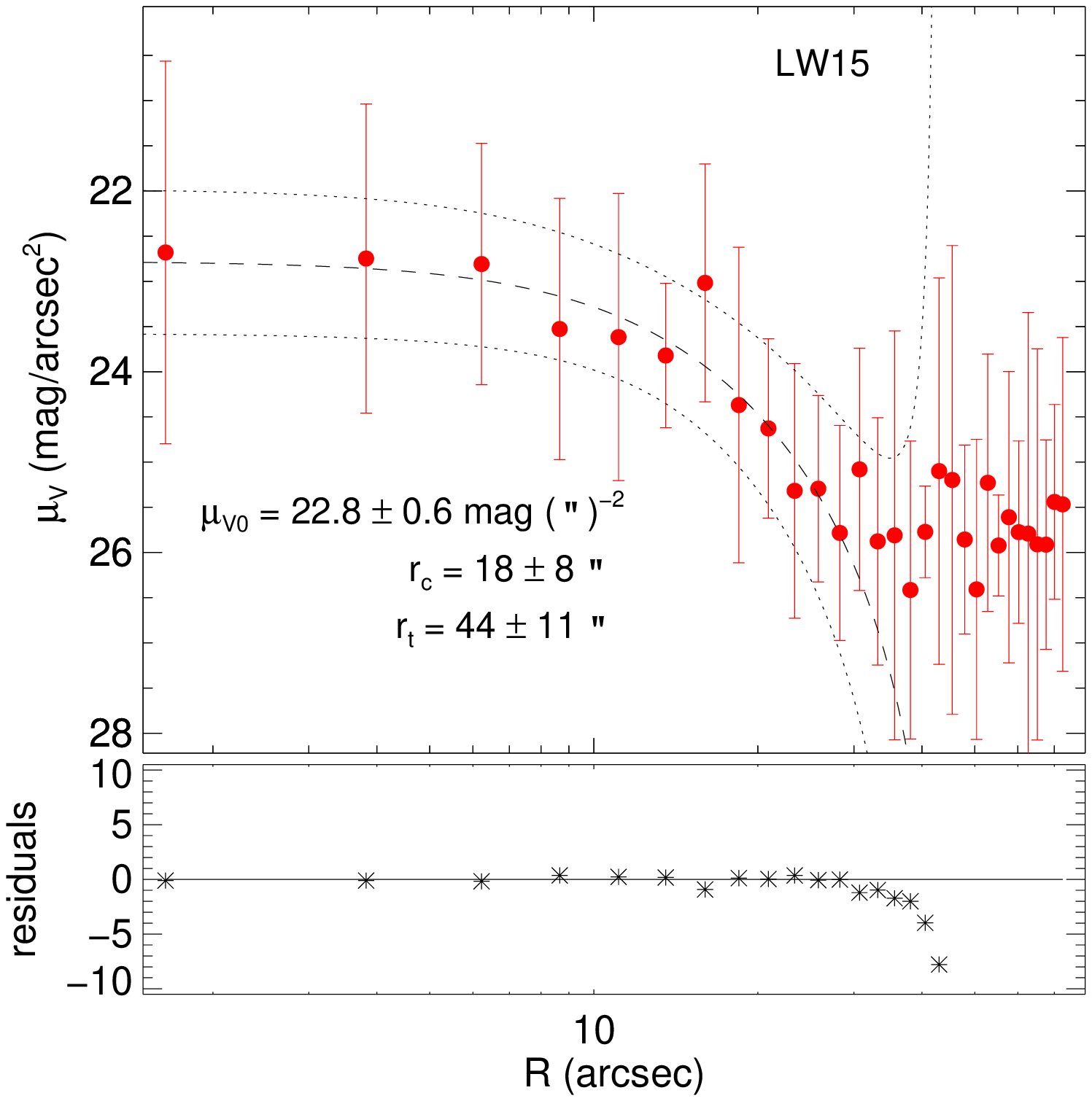}\includegraphics[width=0.325\linewidth]{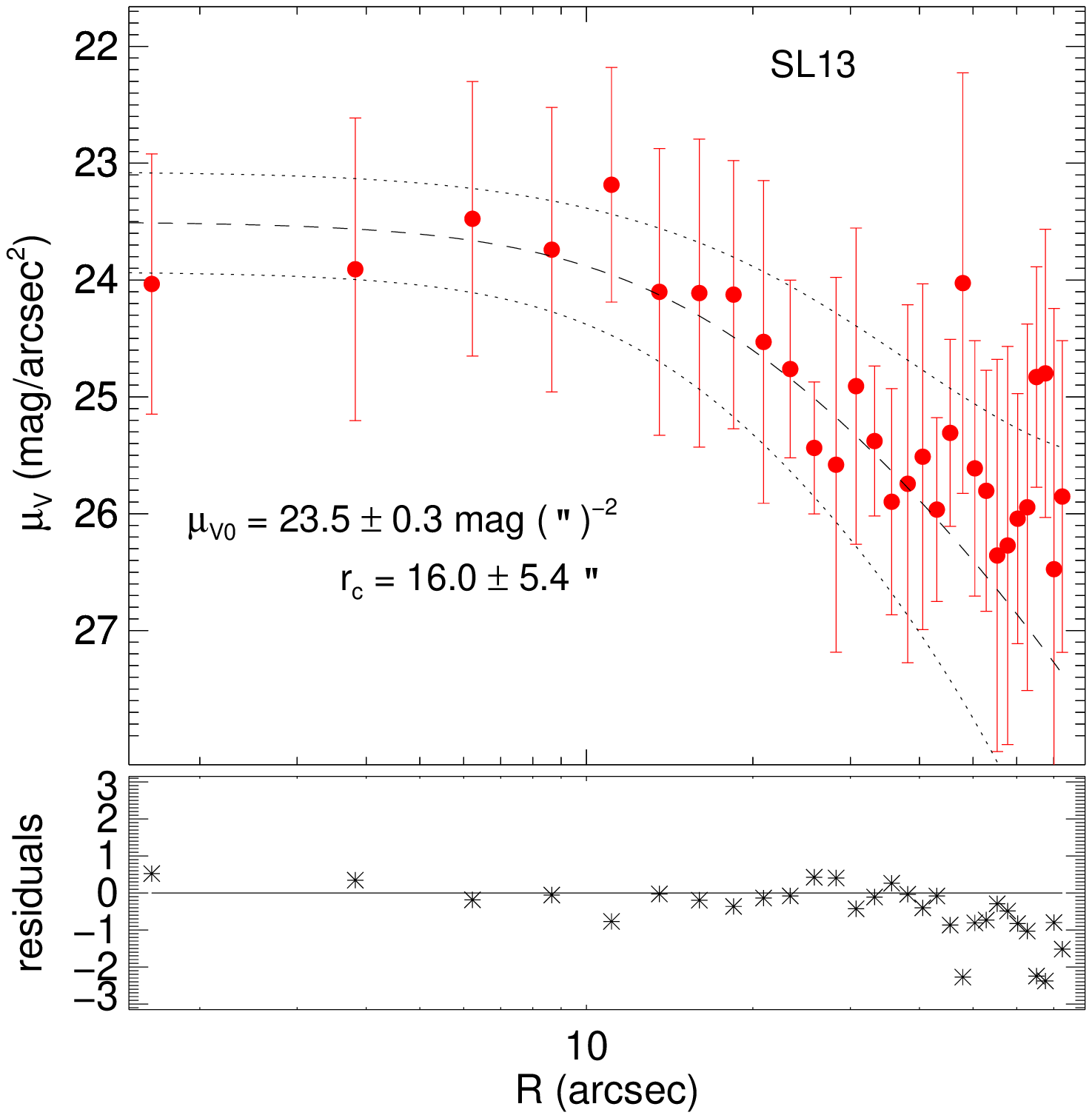}\includegraphics[width=0.325\linewidth]{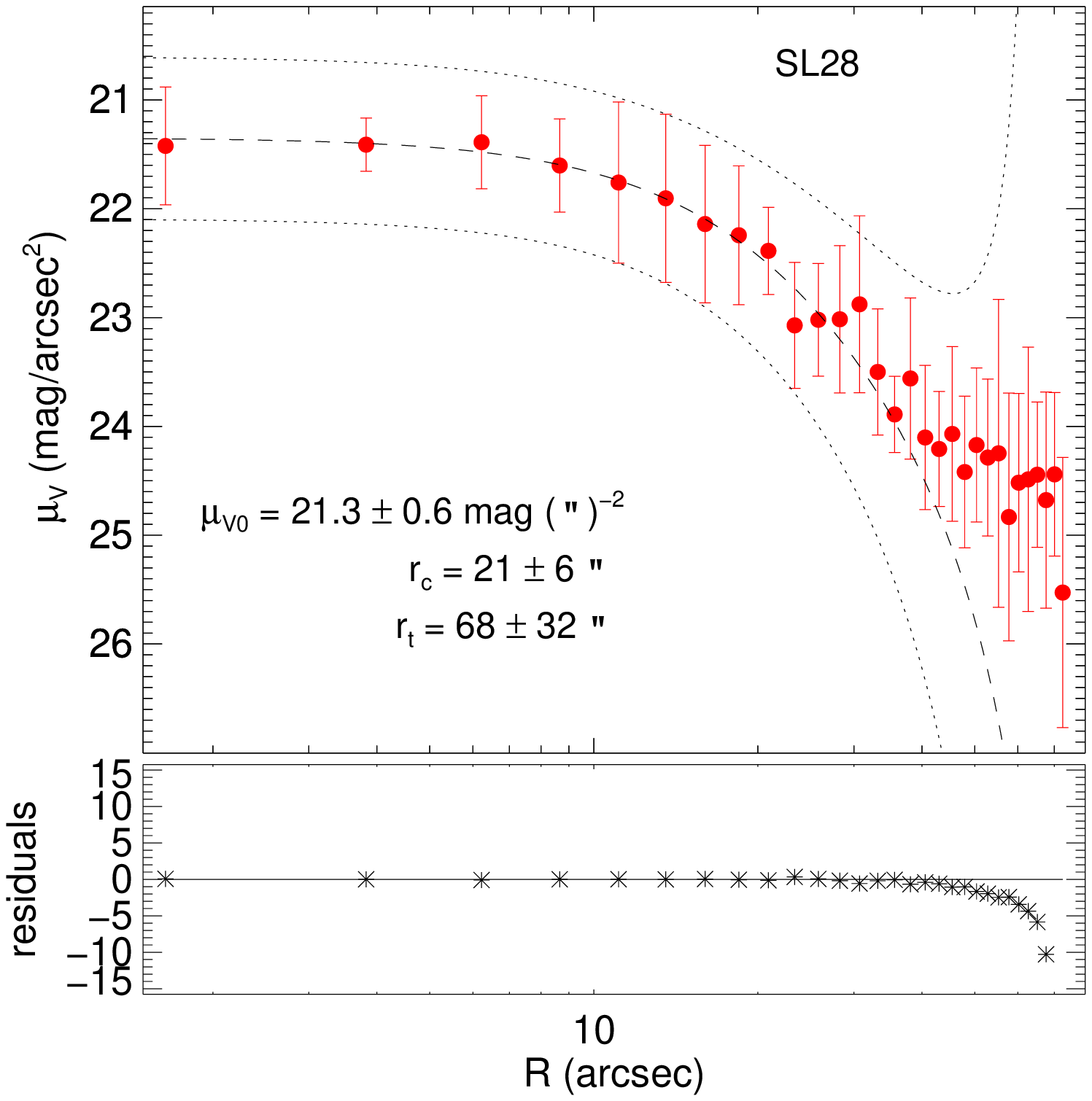}

\includegraphics[width=0.325\linewidth]{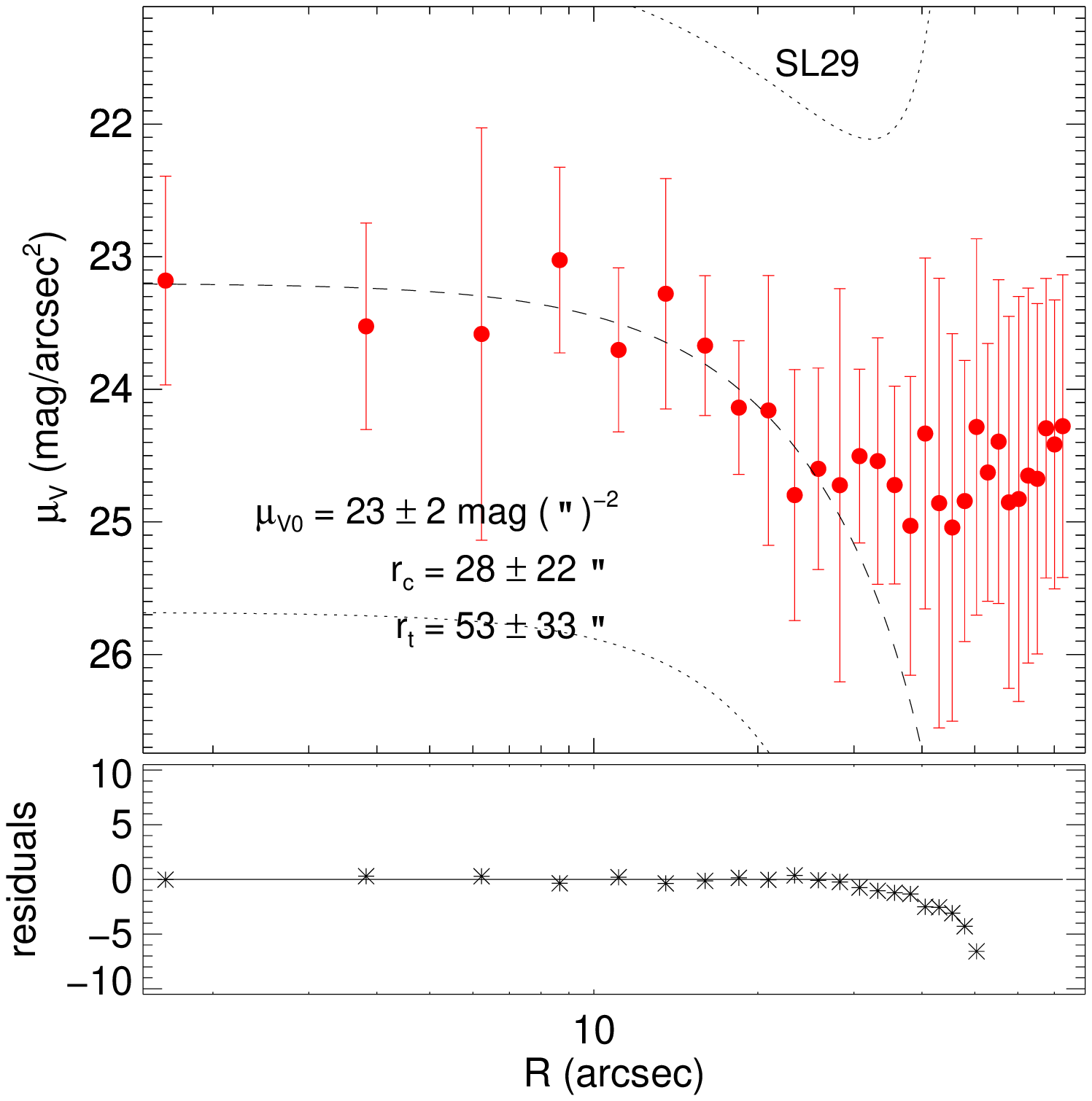}\includegraphics[width=0.325\linewidth]{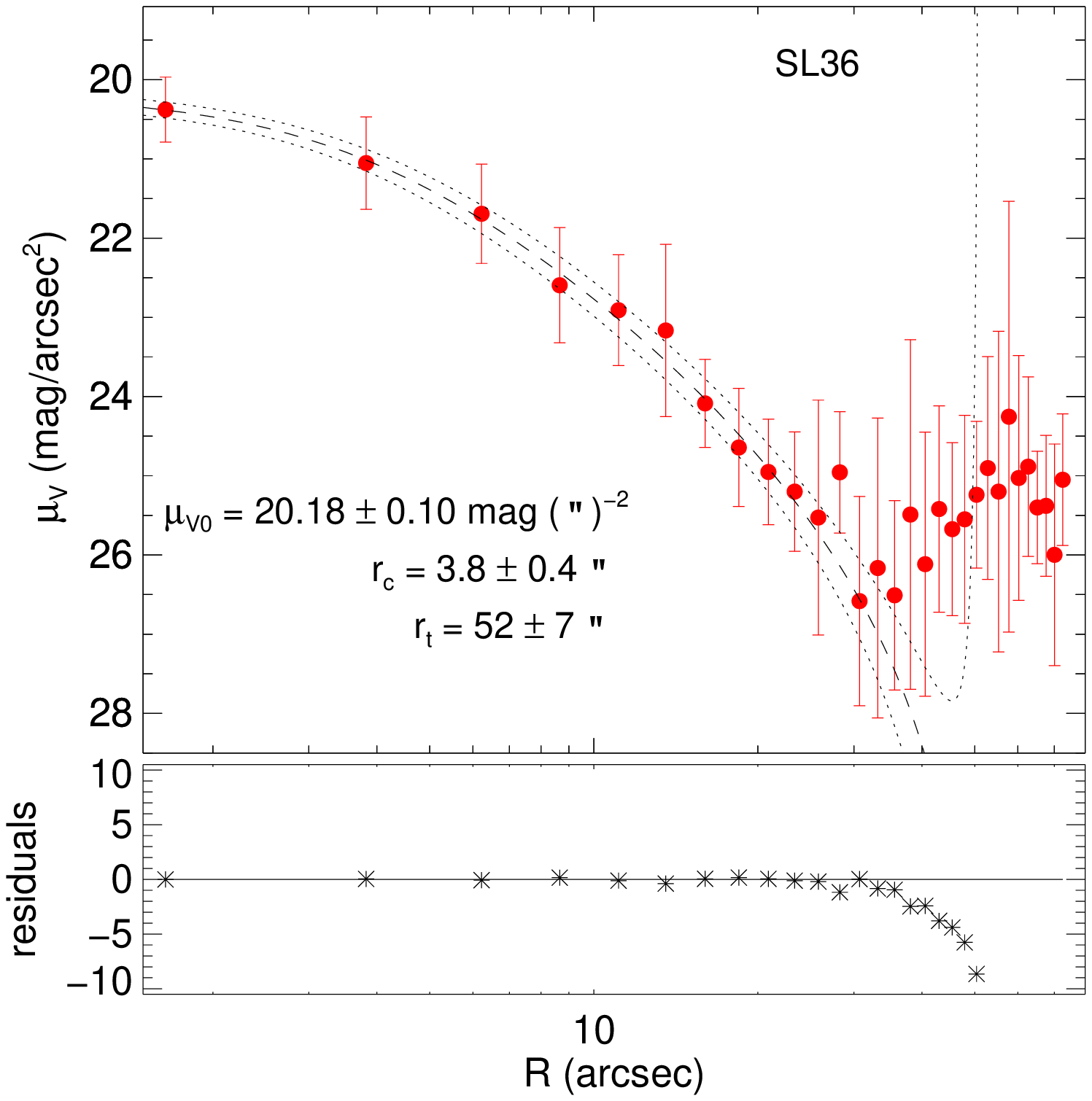}\includegraphics[width=0.325\linewidth]{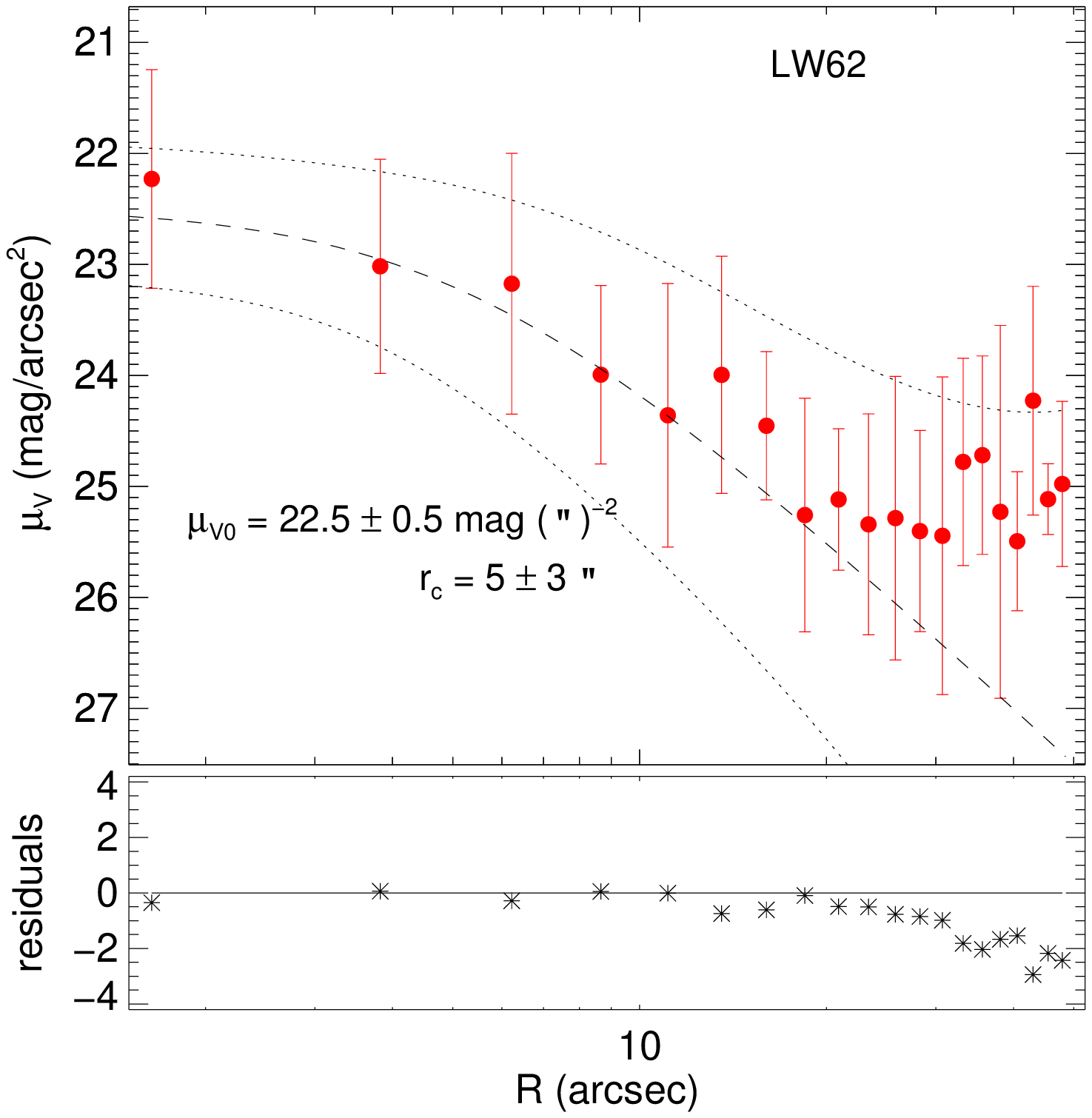}

\includegraphics[width=0.325\linewidth]{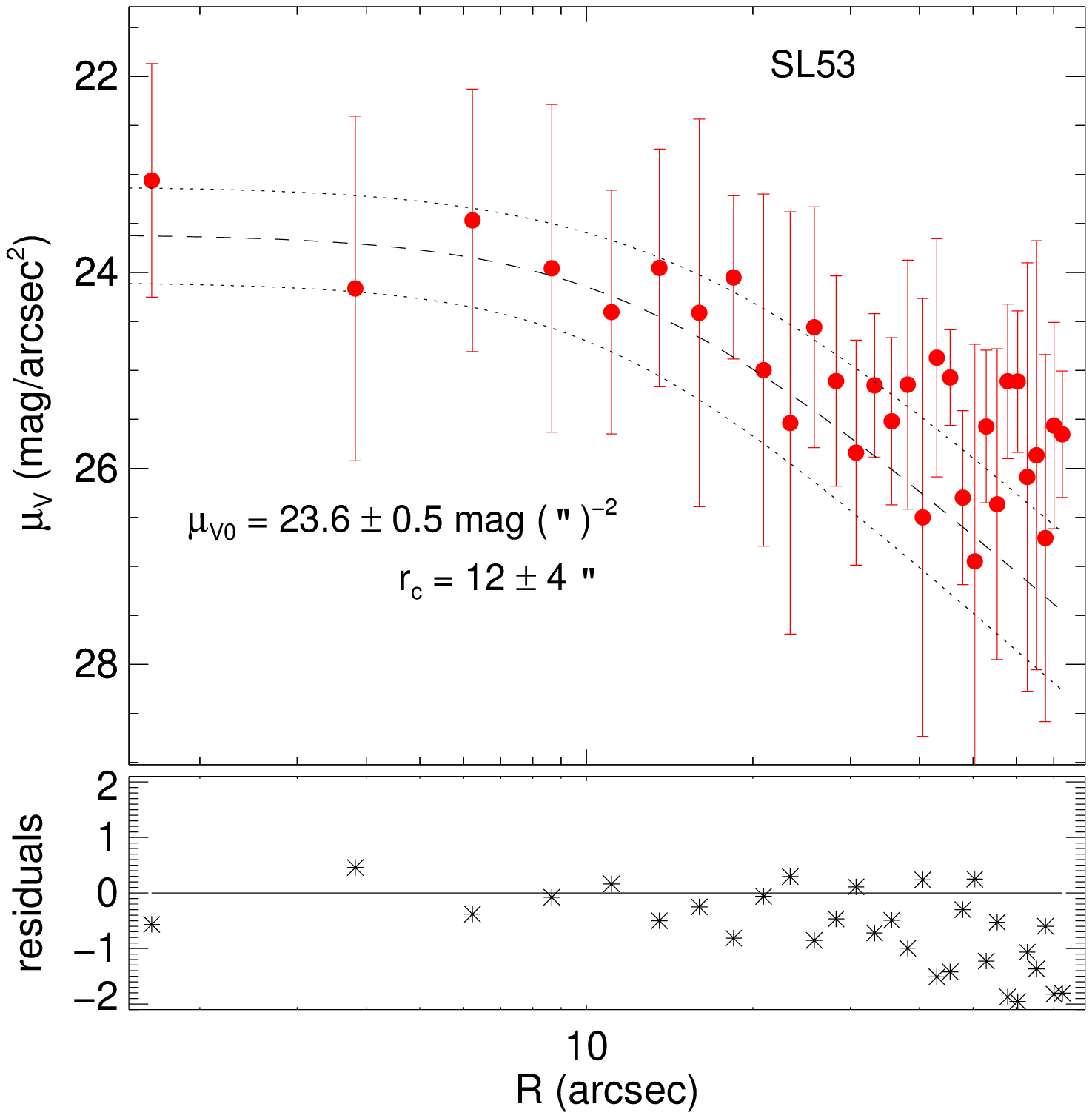}\includegraphics[width=0.325\linewidth]{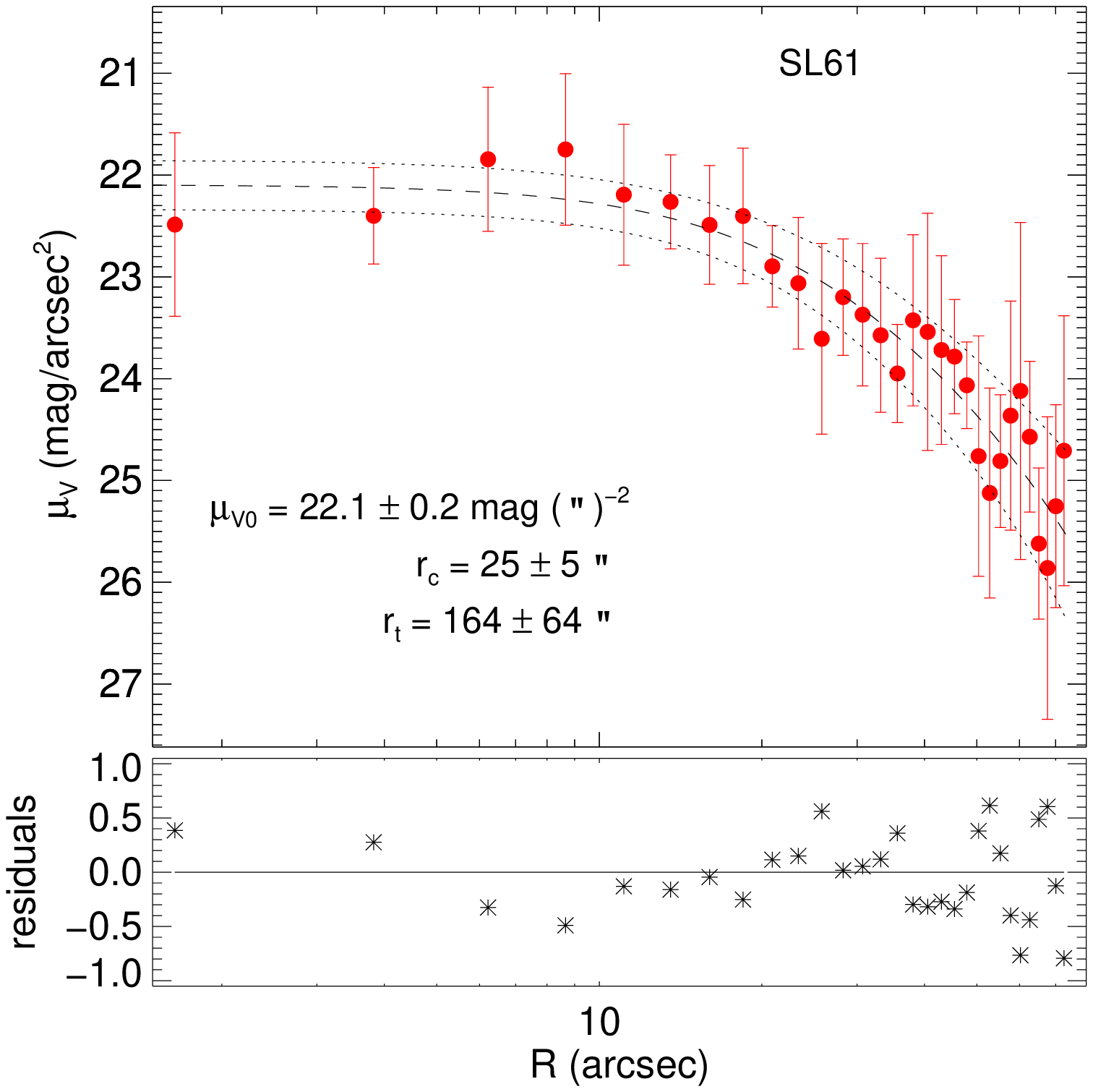}\includegraphics[width=0.325\linewidth]{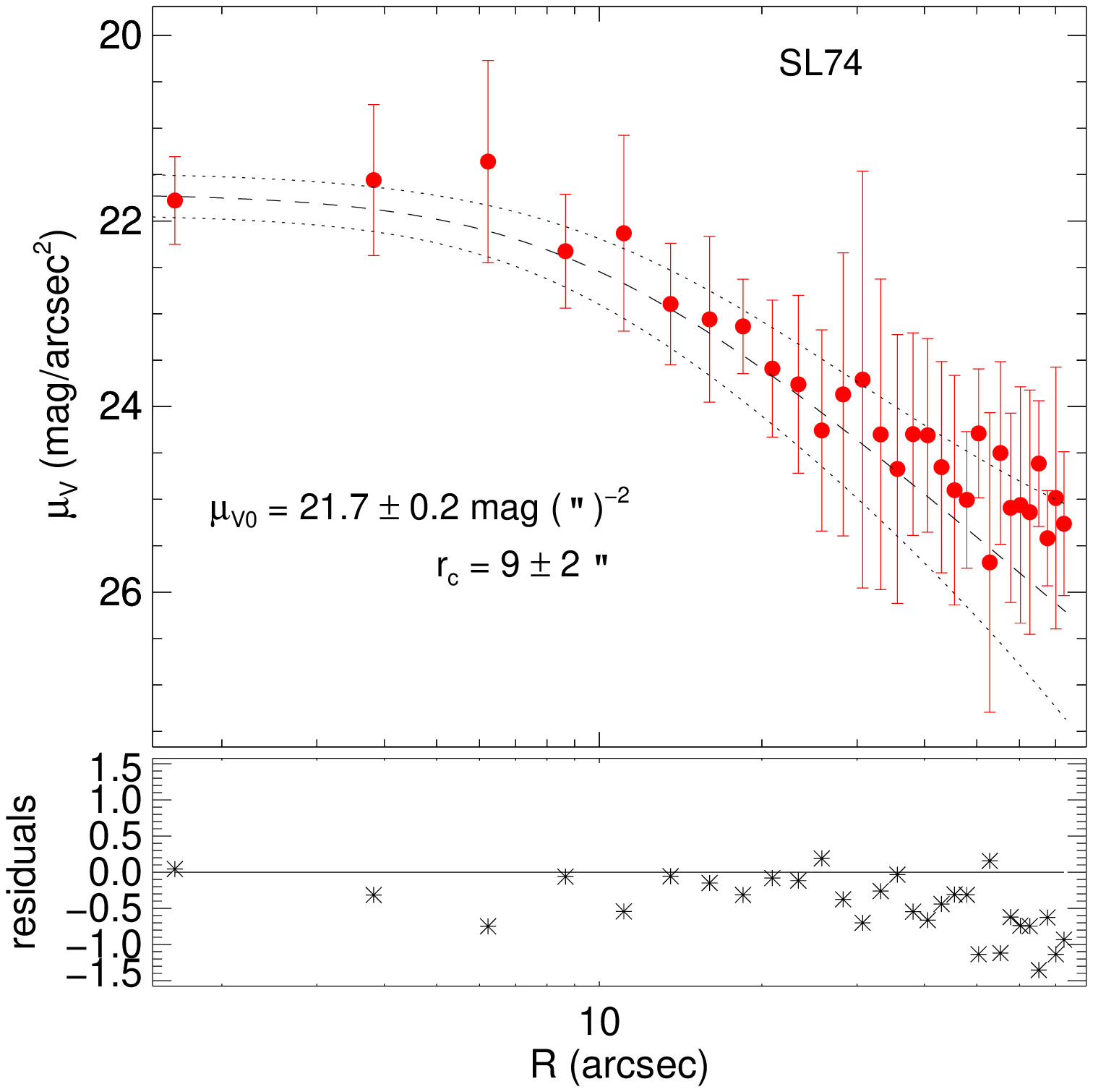}

\includegraphics[width=0.325\linewidth]{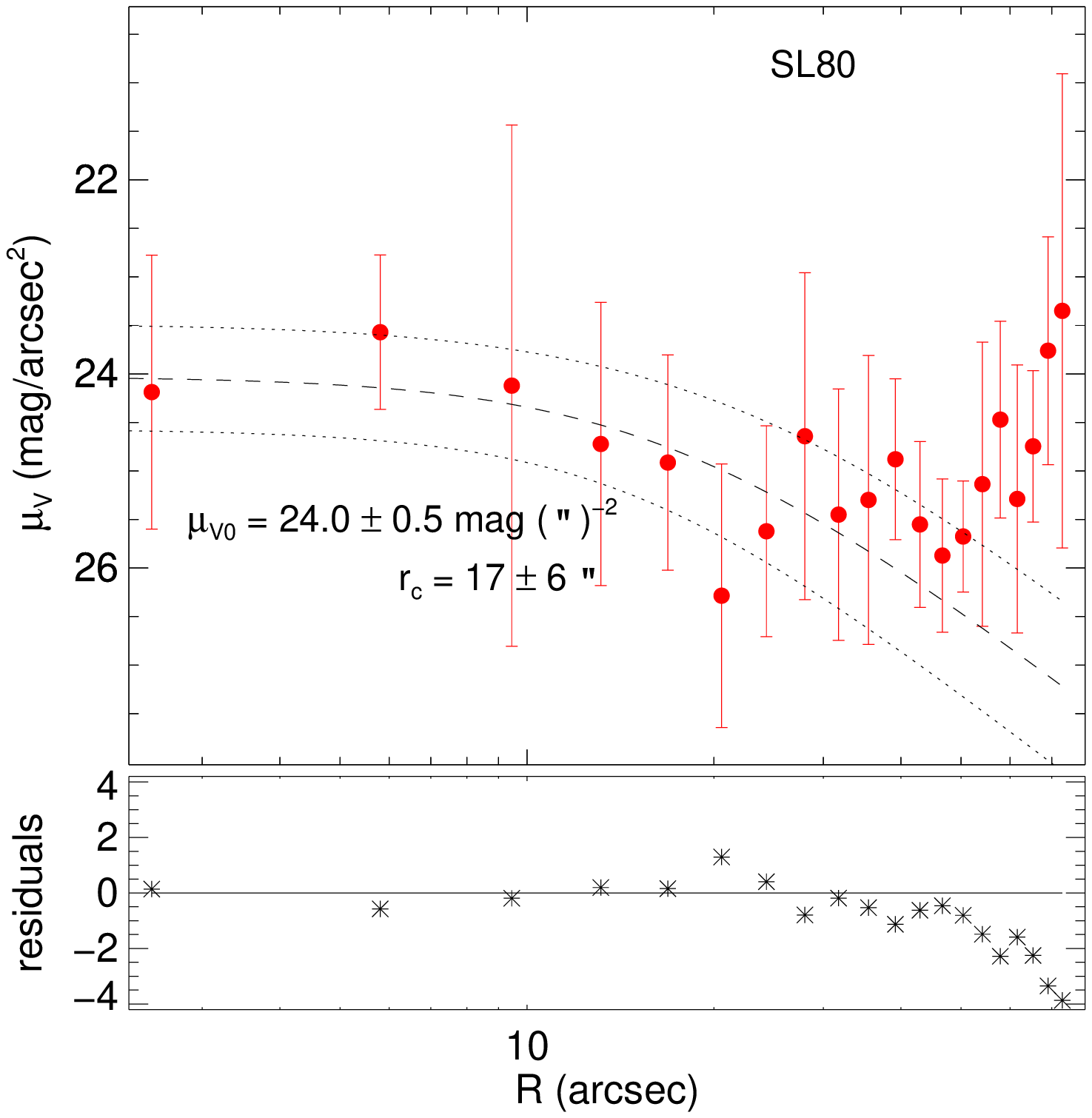}\includegraphics[width=0.325\linewidth]{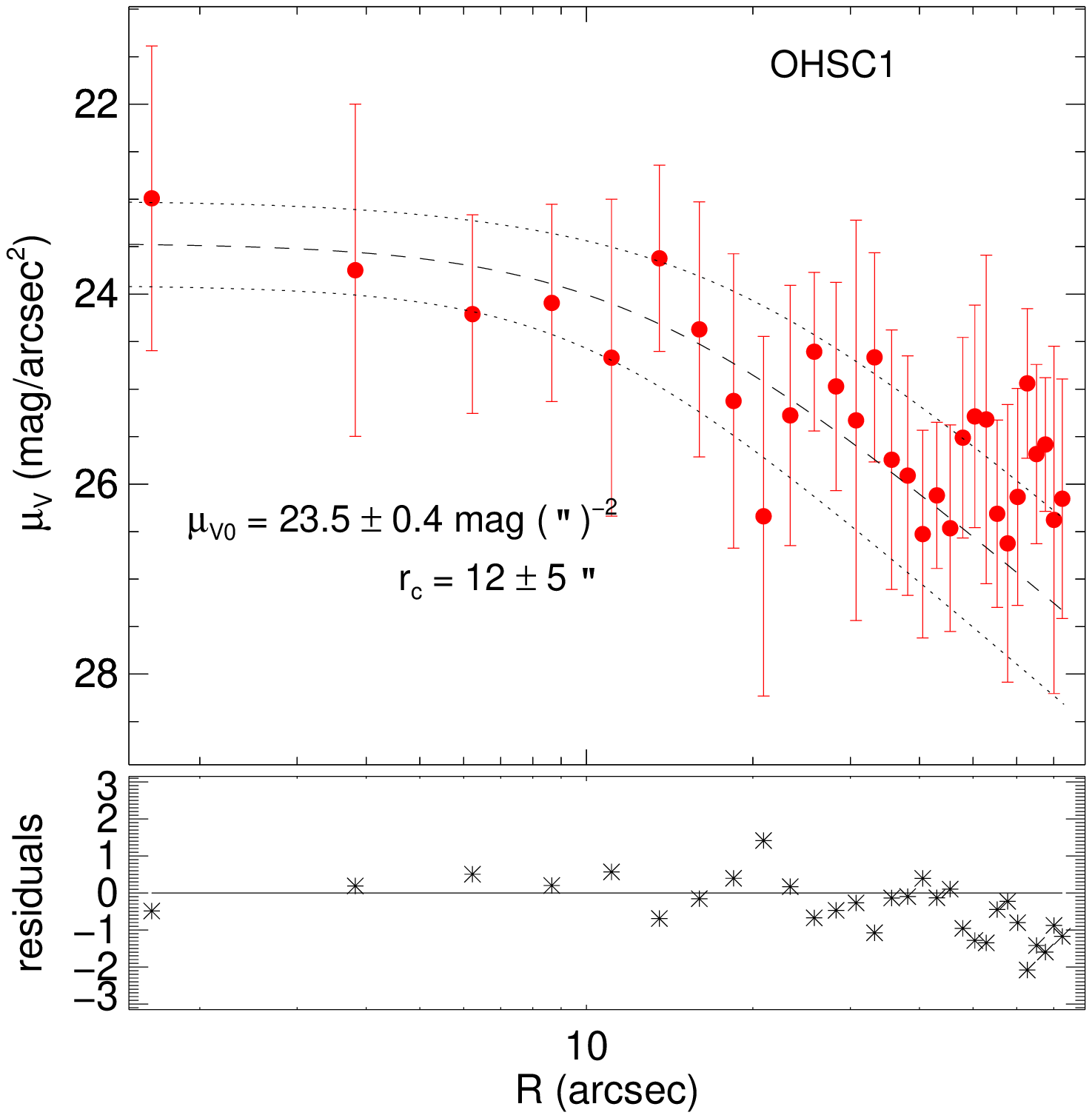}\includegraphics[width=0.325\linewidth]{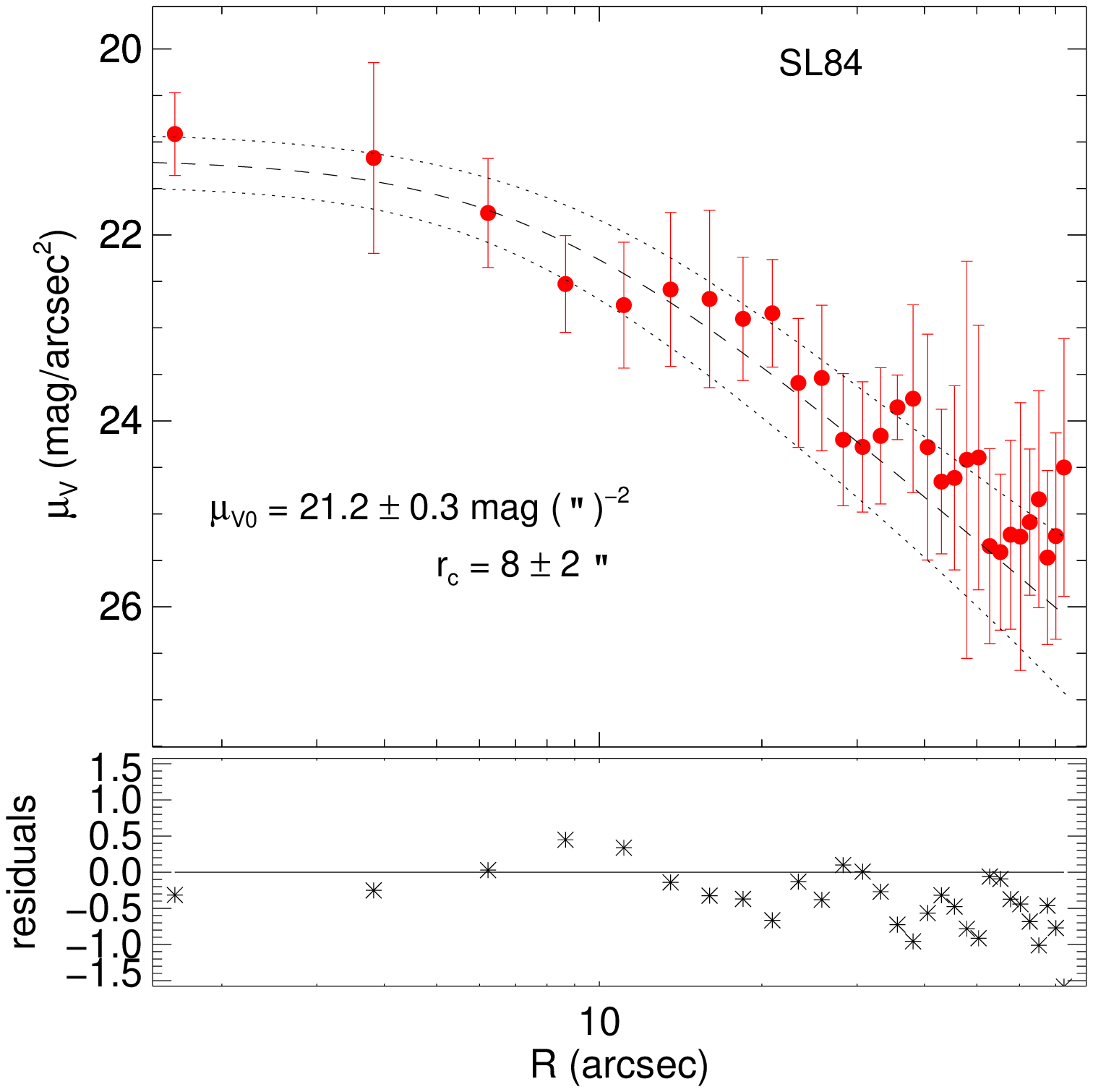}

\caption{Surface brightness profiles of additional LMC clusters complementing the sample presented in Fig.~\ref{fig:rdp_sbp}. The King model fits (dashed lines)  and 1\,$\sigma$ uncertainties (dotted lines) are shown. The best-fitting  parameters are indicated and the fit residuals are plotted in the lower panel.}

\end{figure*}

\setcounter{figure}{3}

\begin{figure*}
\includegraphics[width=0.325\linewidth]{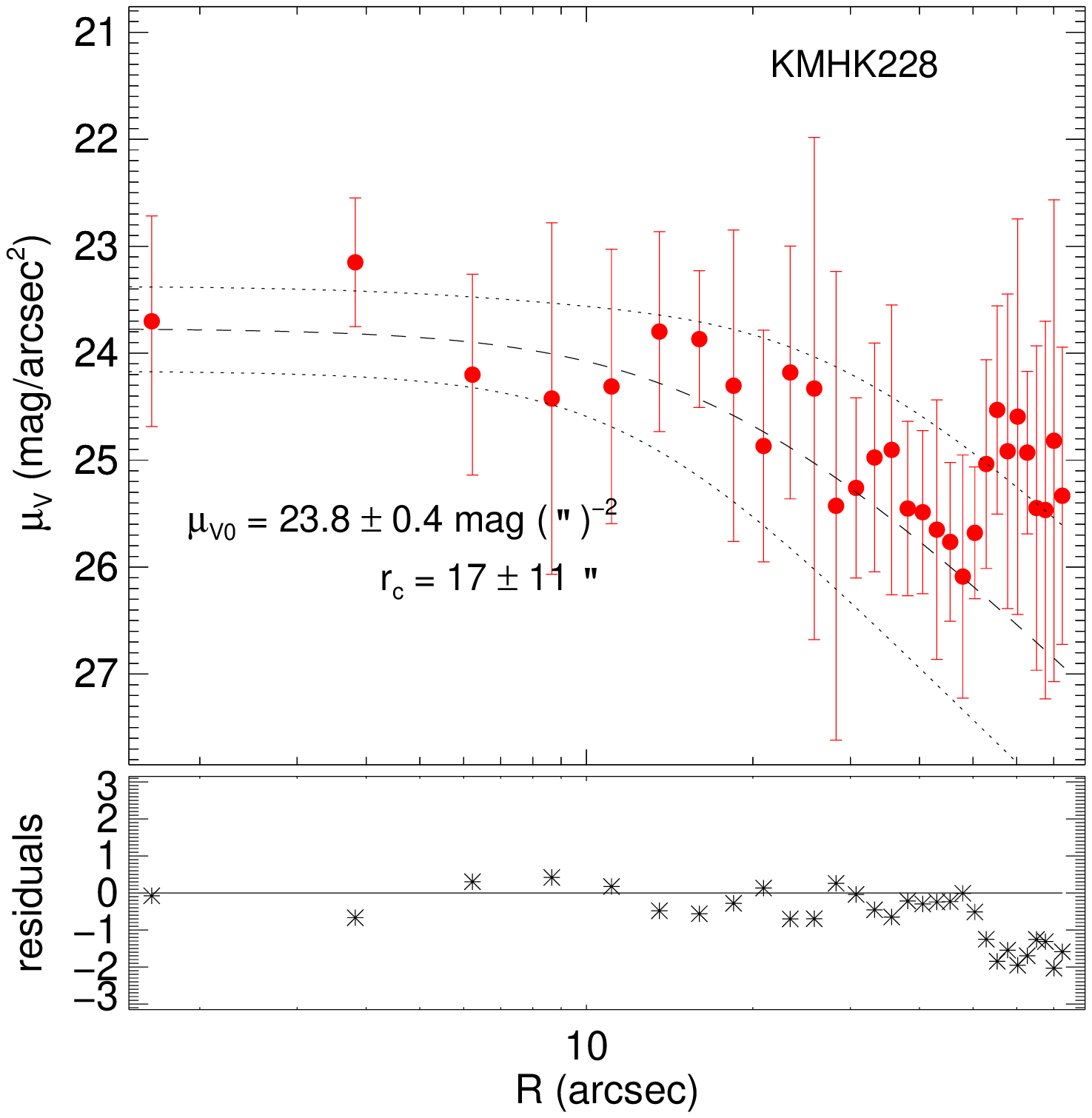}\includegraphics[width=0.325\linewidth]{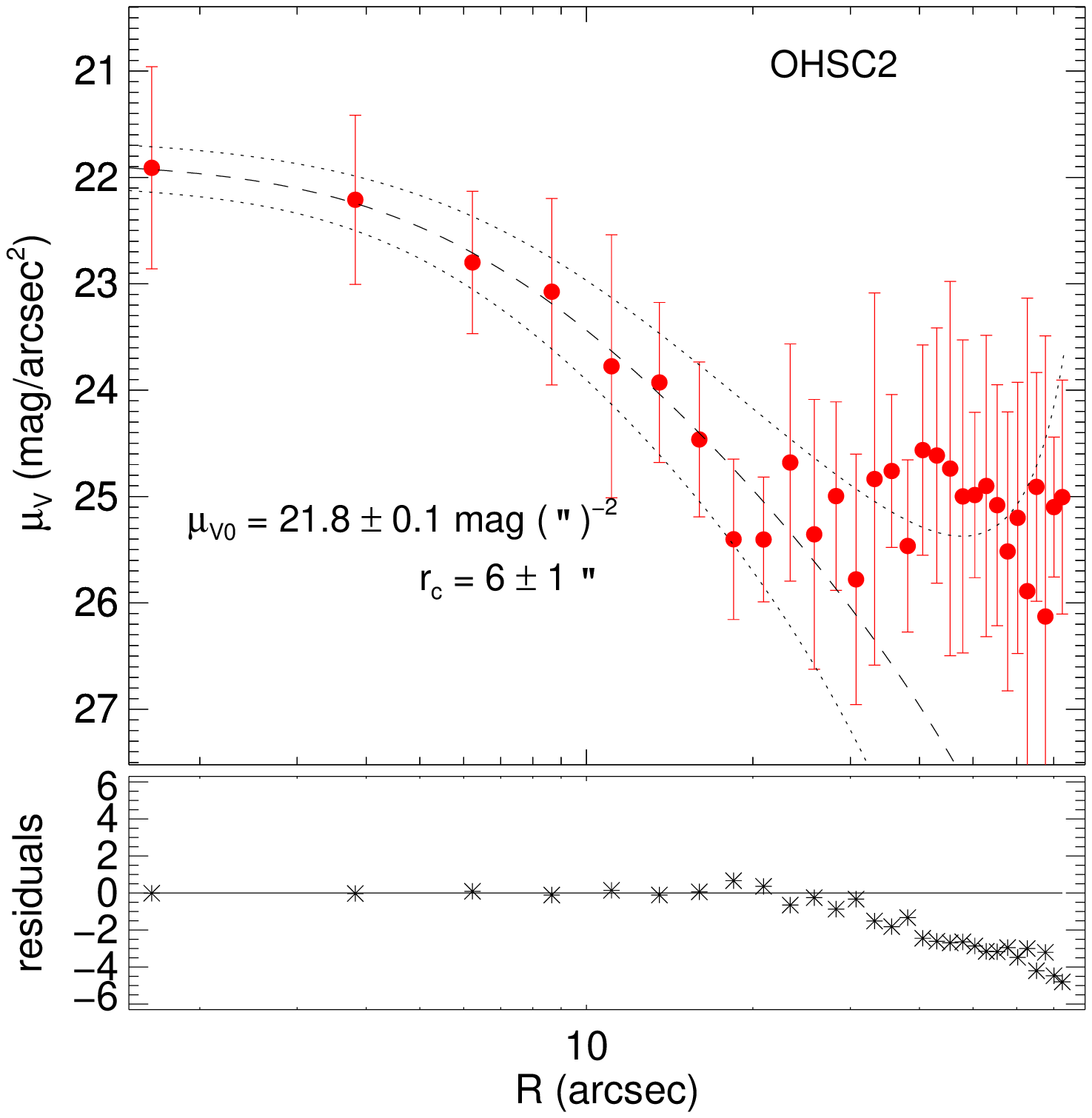}\includegraphics[width=0.325\linewidth]{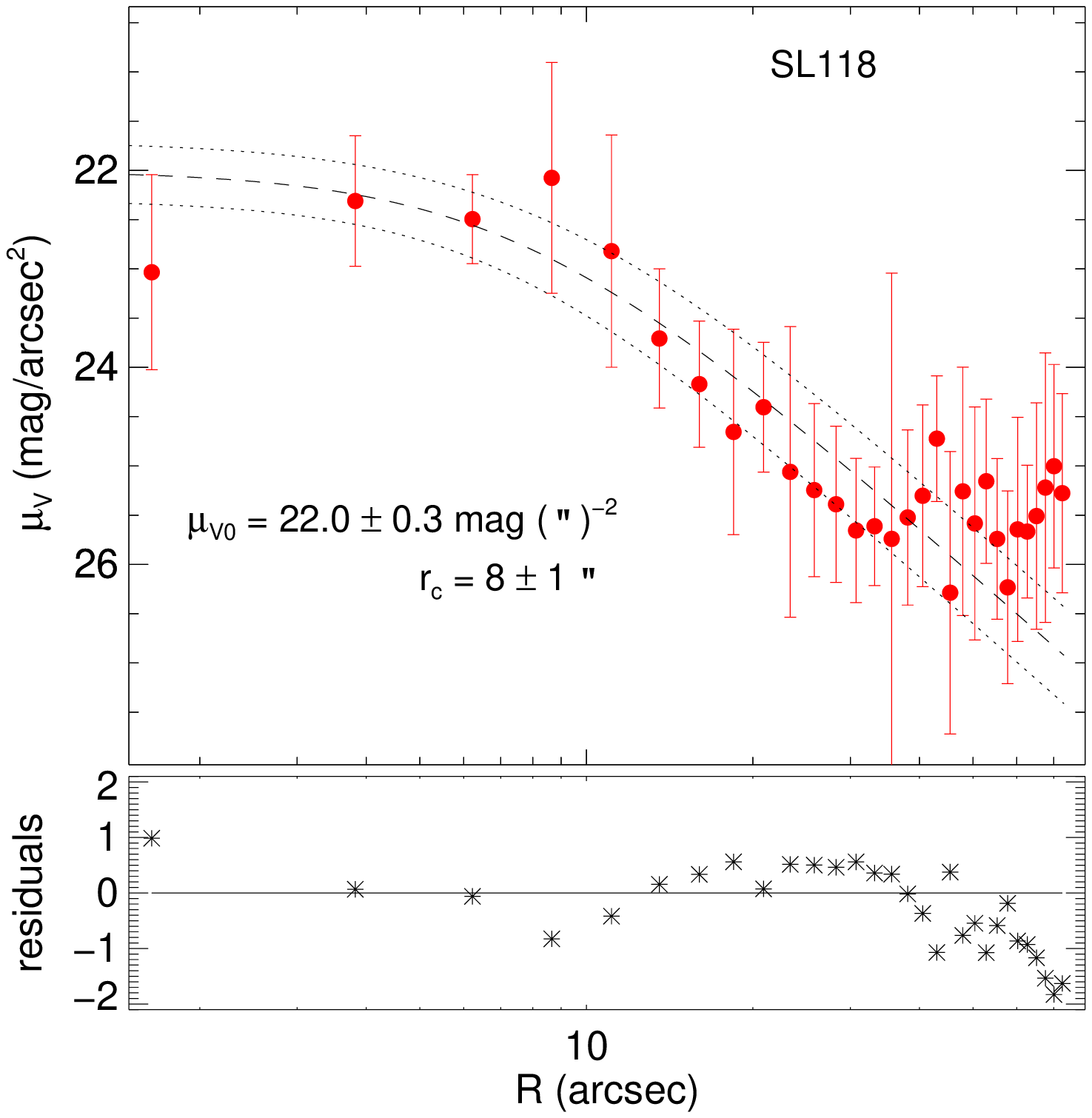}

\includegraphics[width=0.325\linewidth]{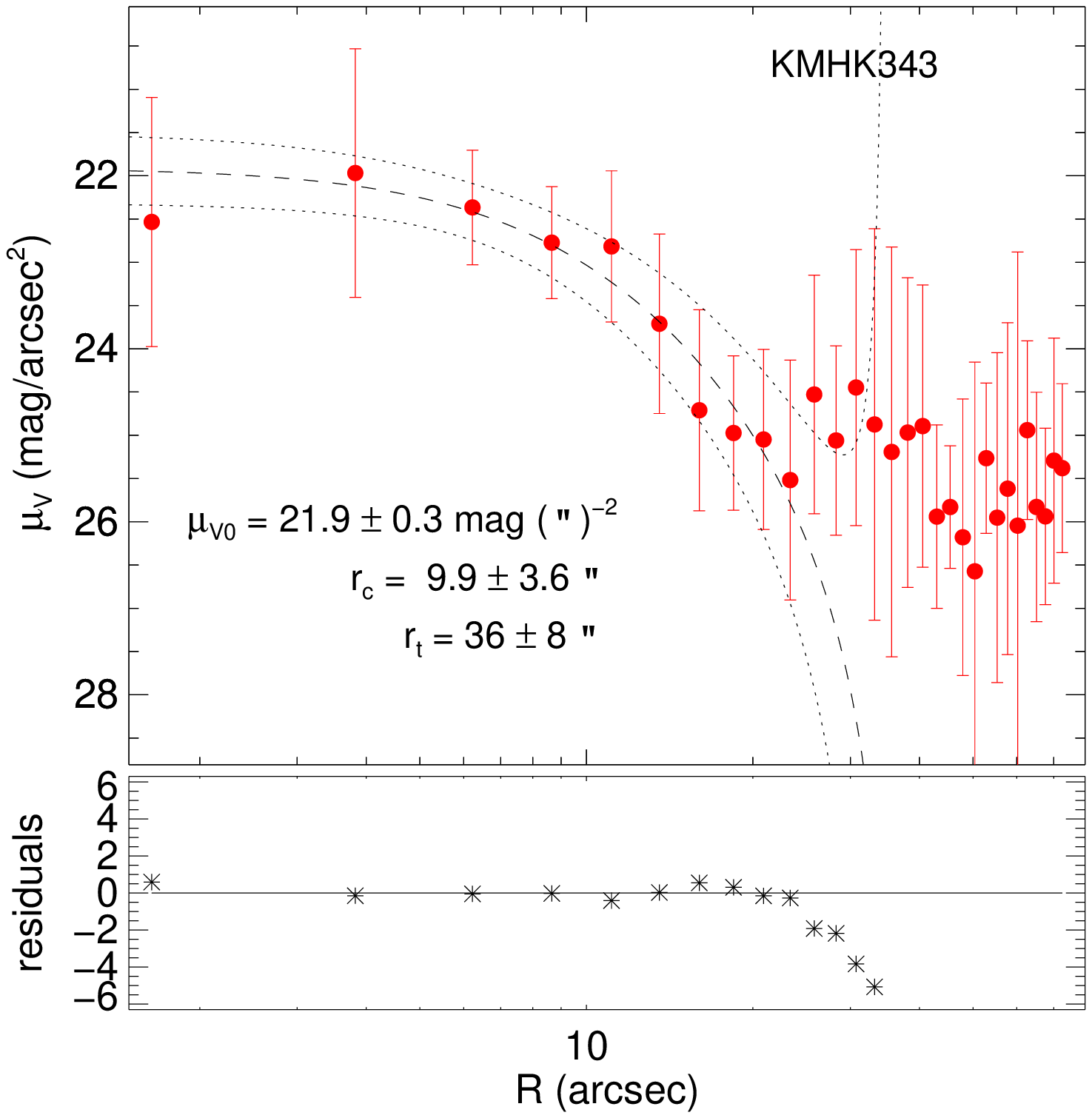}\includegraphics[width=0.325\linewidth]{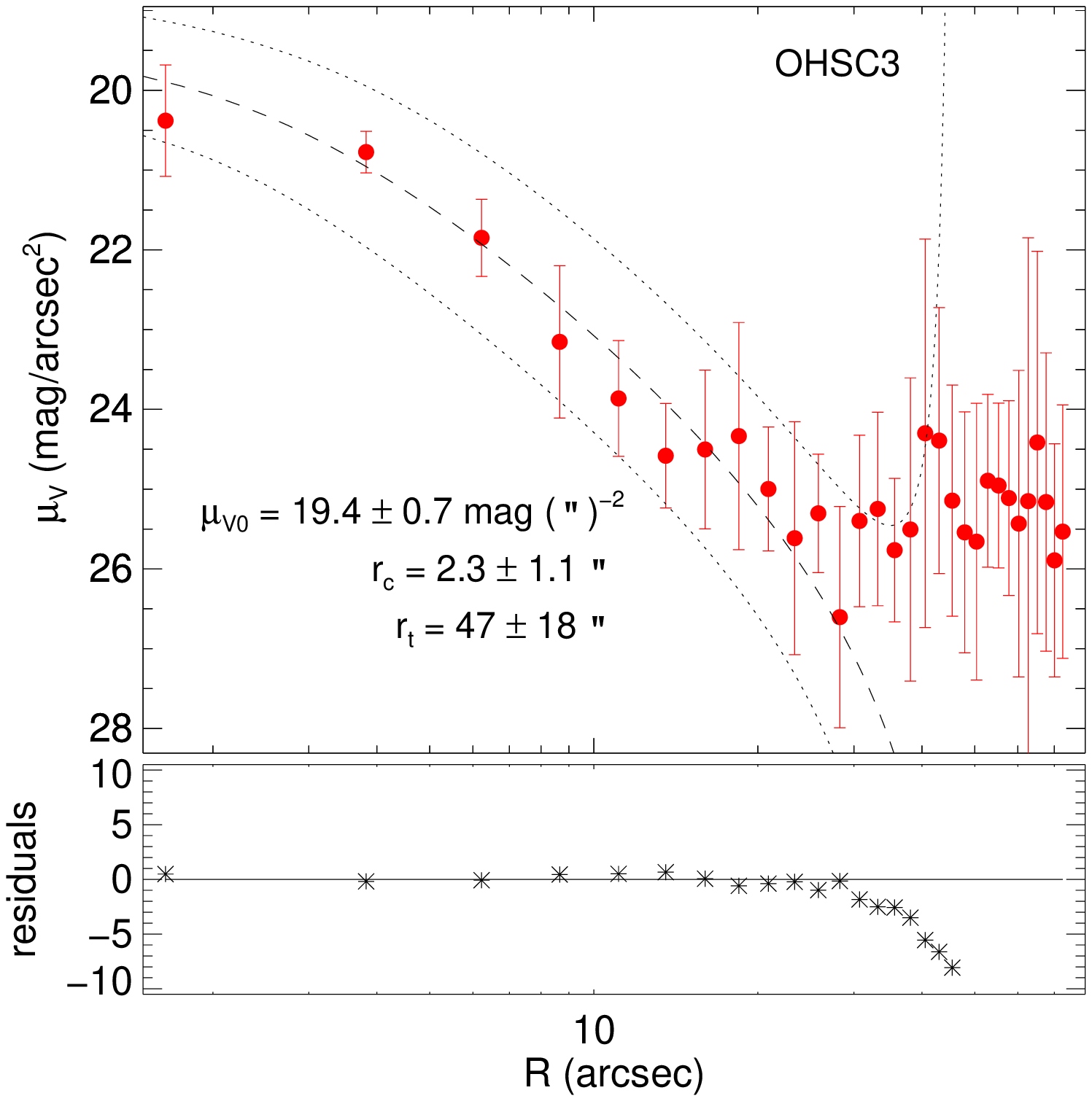}\includegraphics[width=0.325\linewidth]{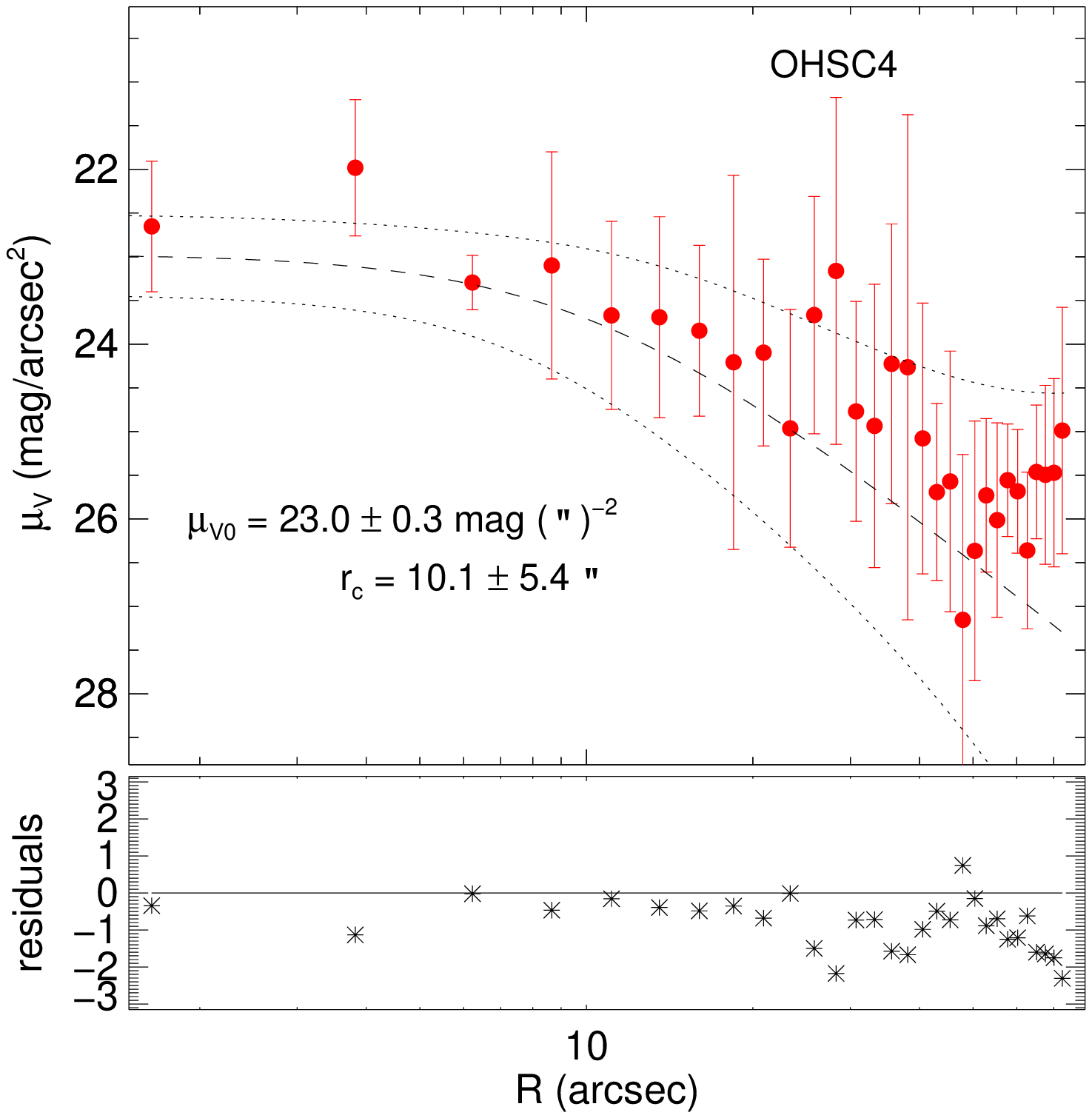}

\includegraphics[width=0.325\linewidth]{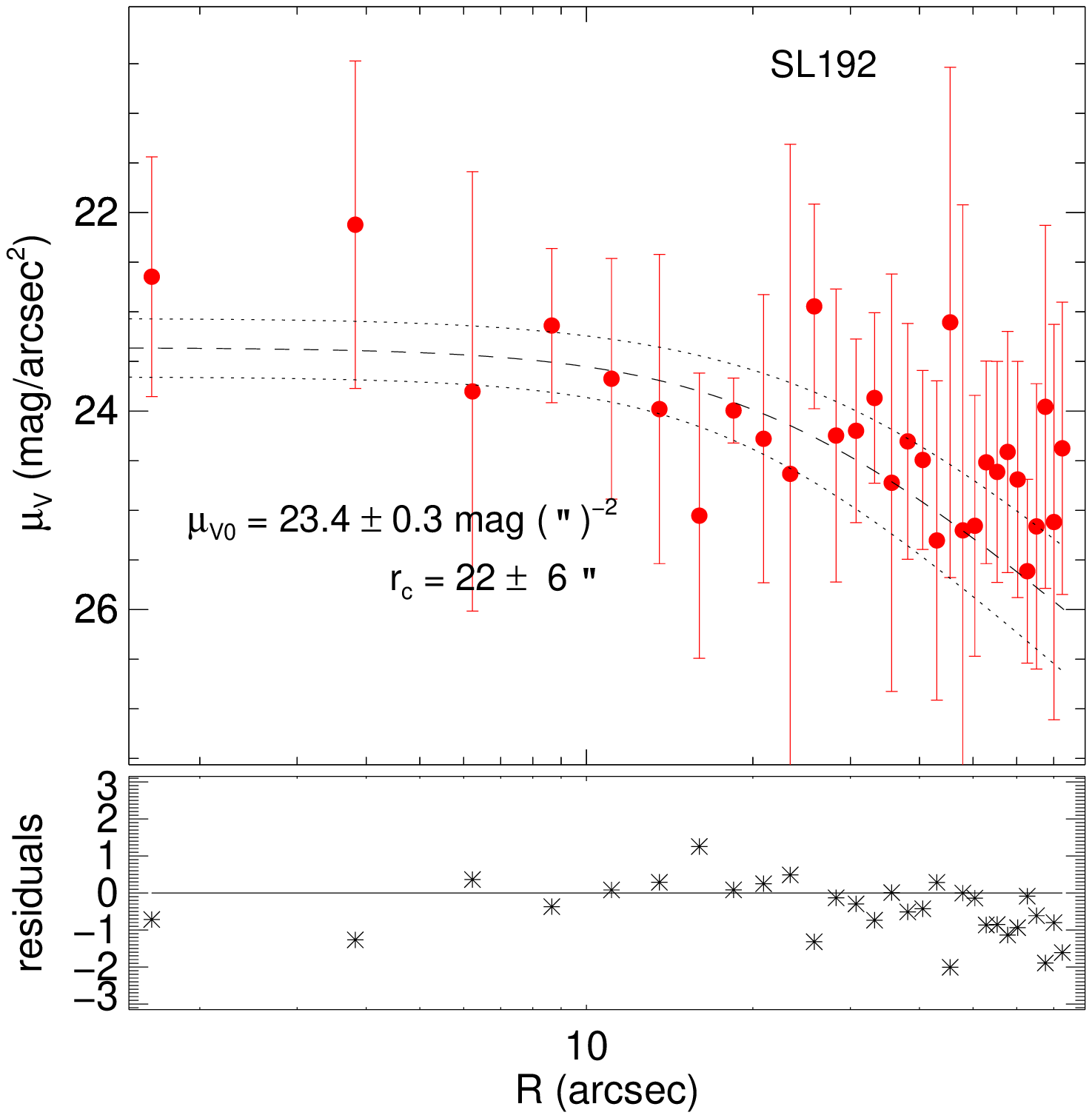}\includegraphics[width=0.325\linewidth]{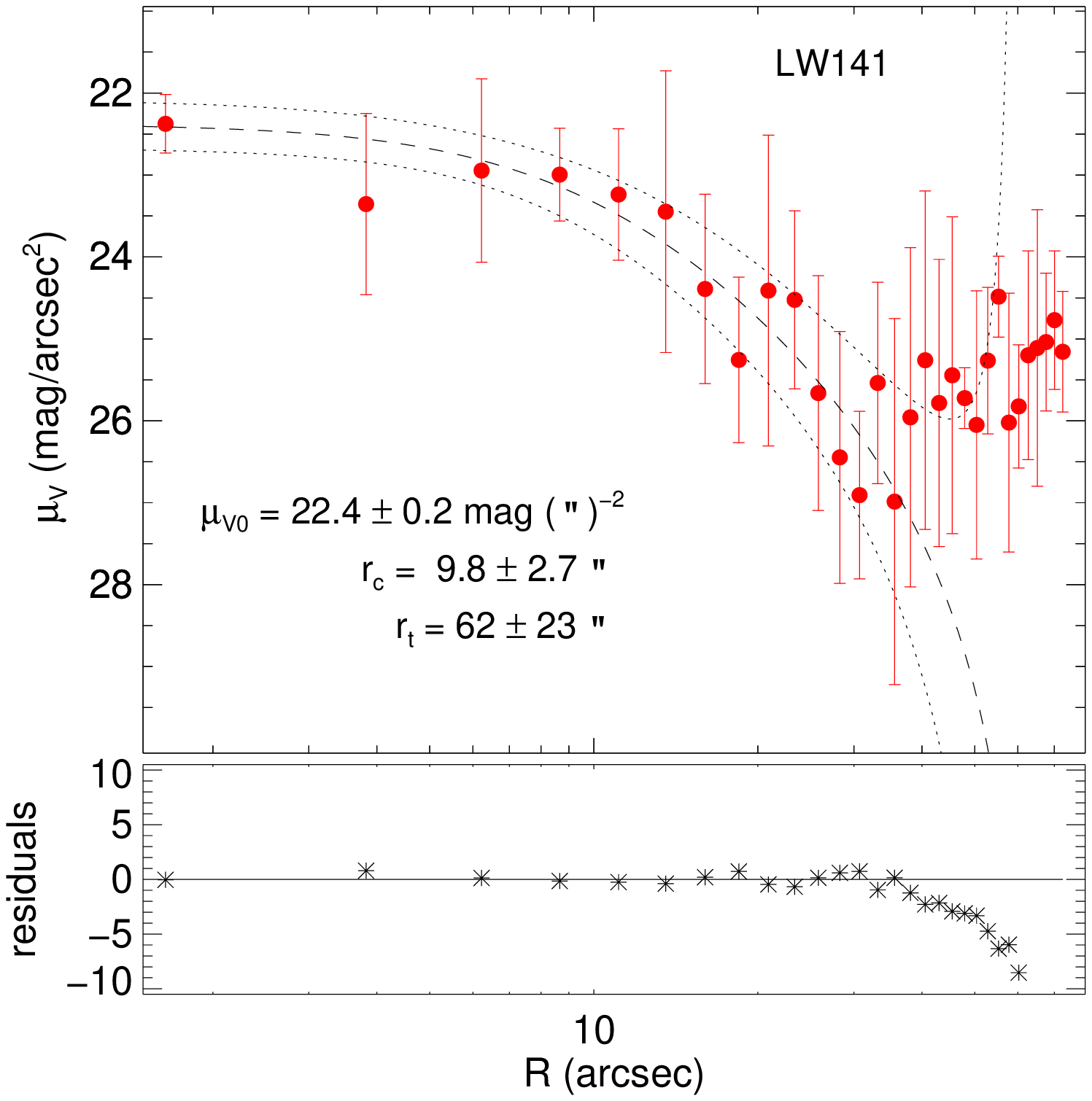}\includegraphics[width=0.325\linewidth]{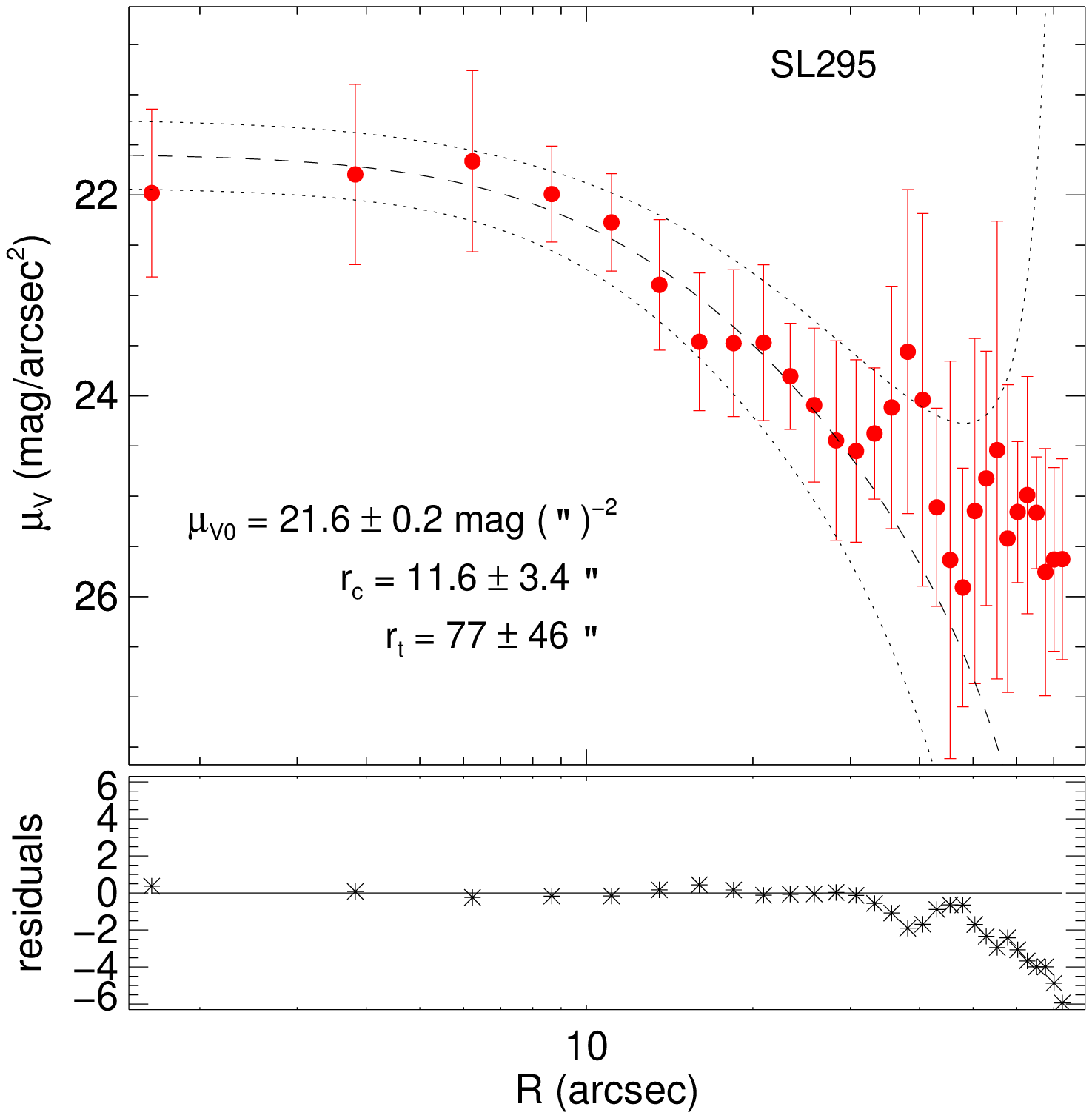}

\includegraphics[width=0.325\linewidth]{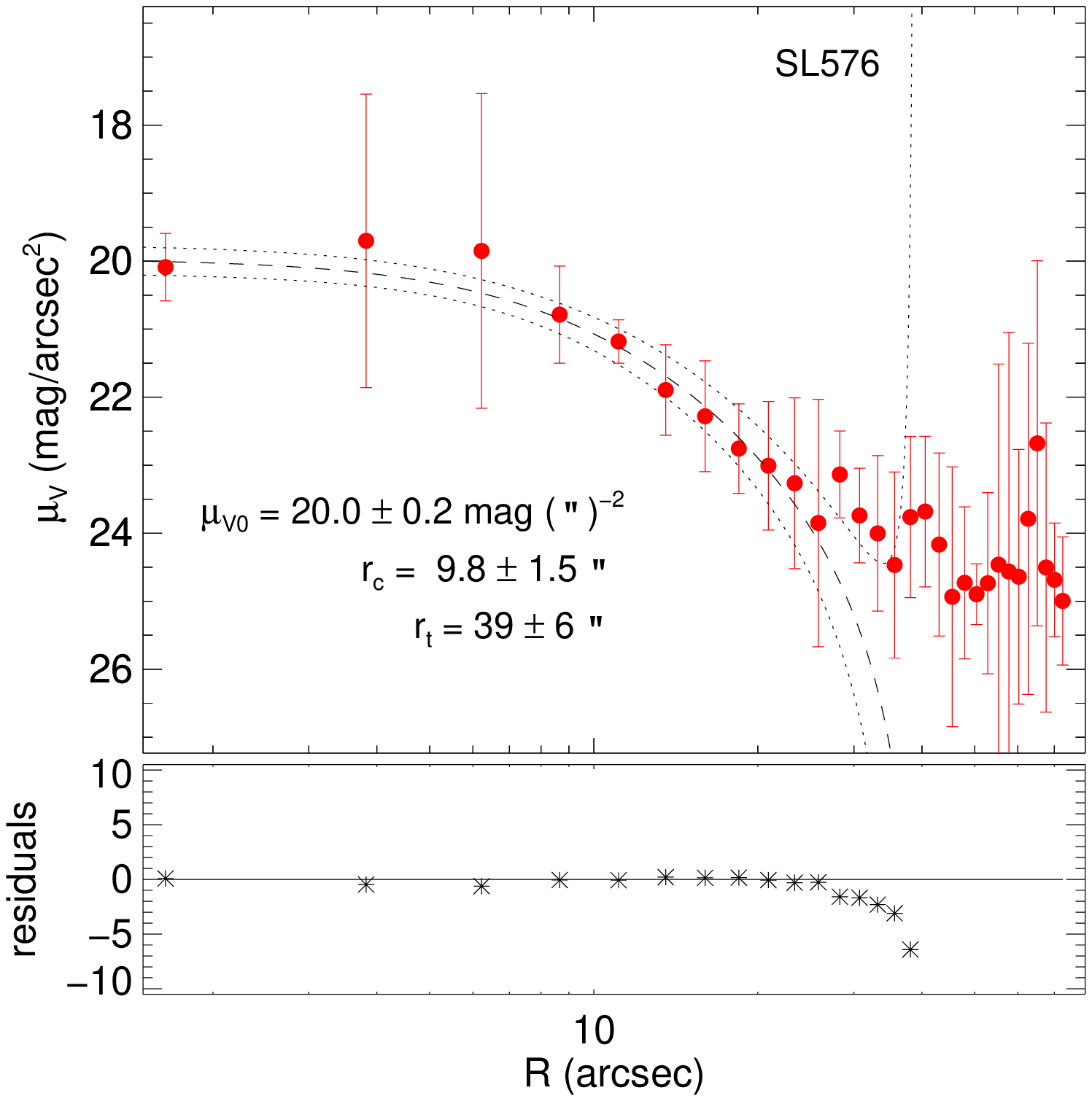}\includegraphics[width=0.325\linewidth]{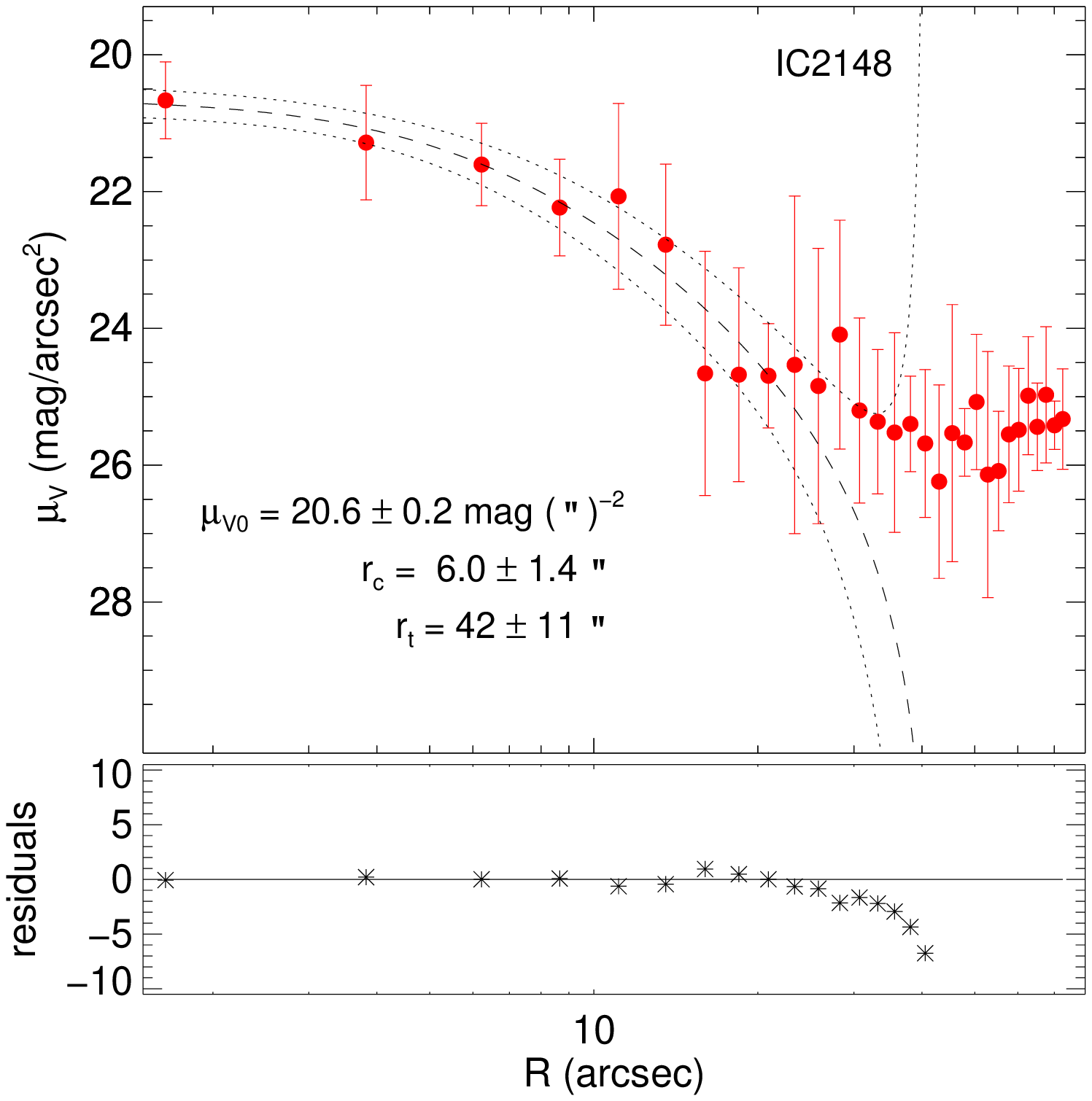}\includegraphics[width=0.325\linewidth]{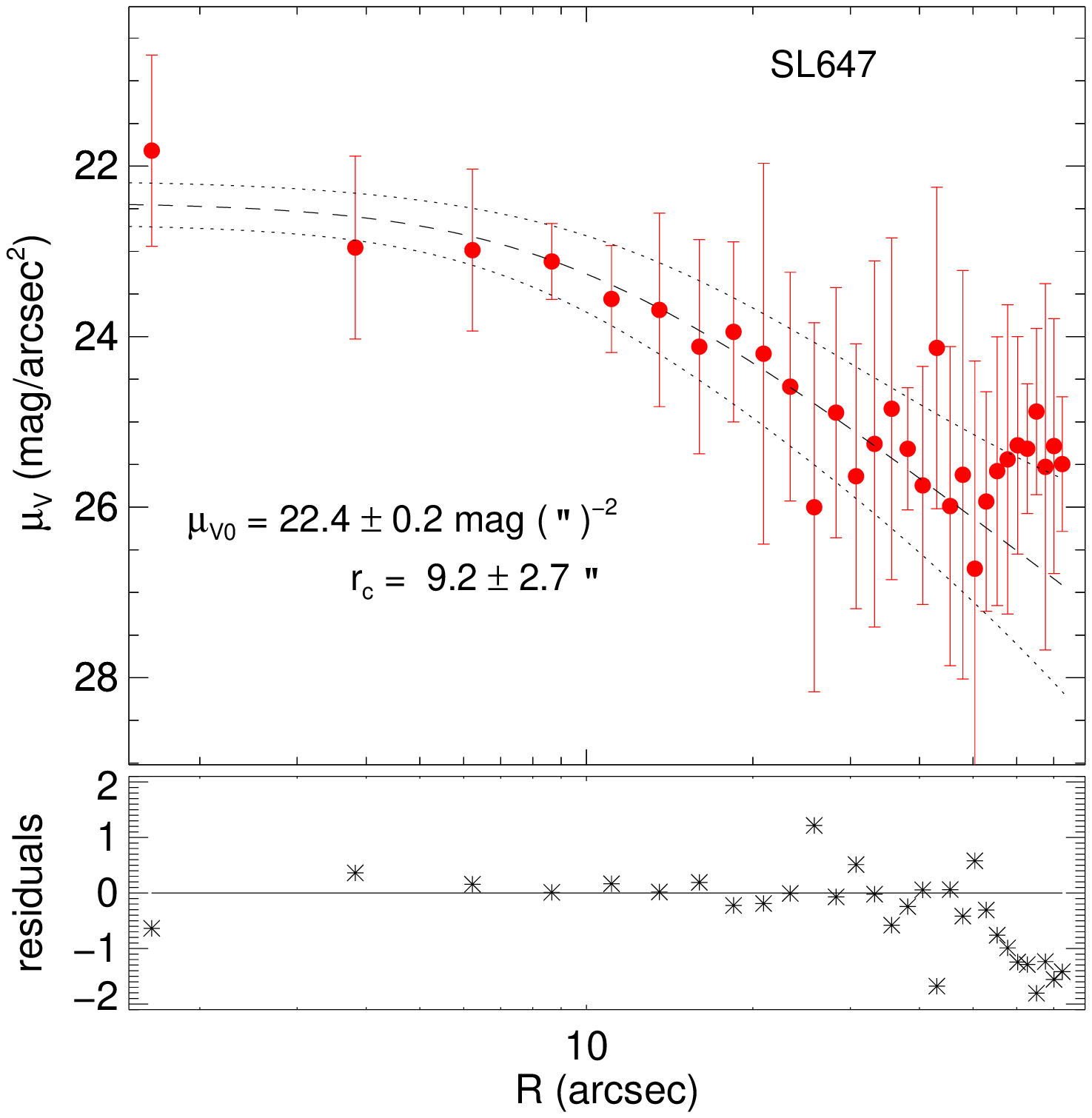}

\caption{cont.}

\end{figure*}

\setcounter{figure}{3}

\begin{figure*}
\includegraphics[width=0.325\linewidth]{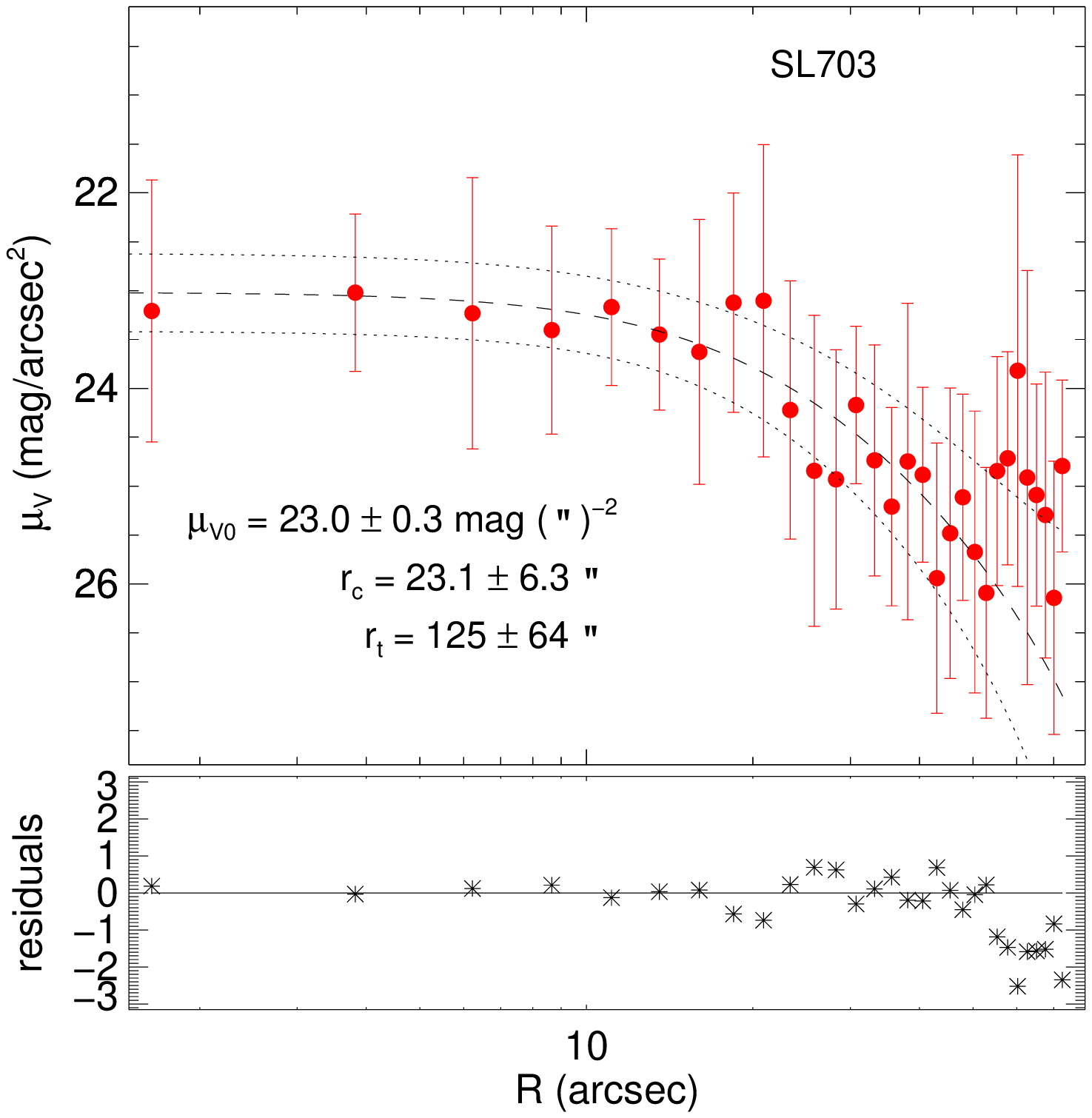}\includegraphics[width=0.325\linewidth]{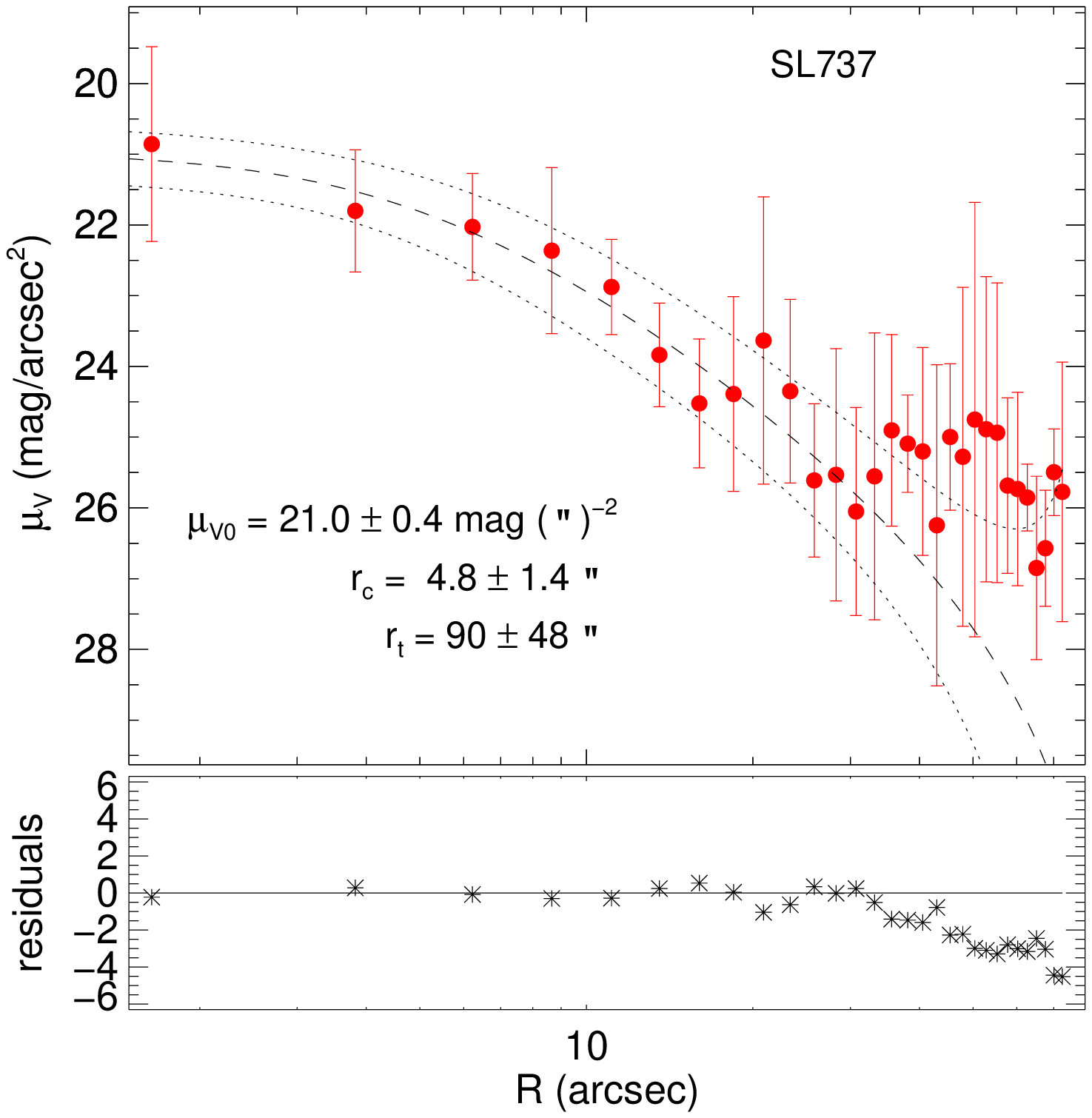}\includegraphics[width=0.325\linewidth]{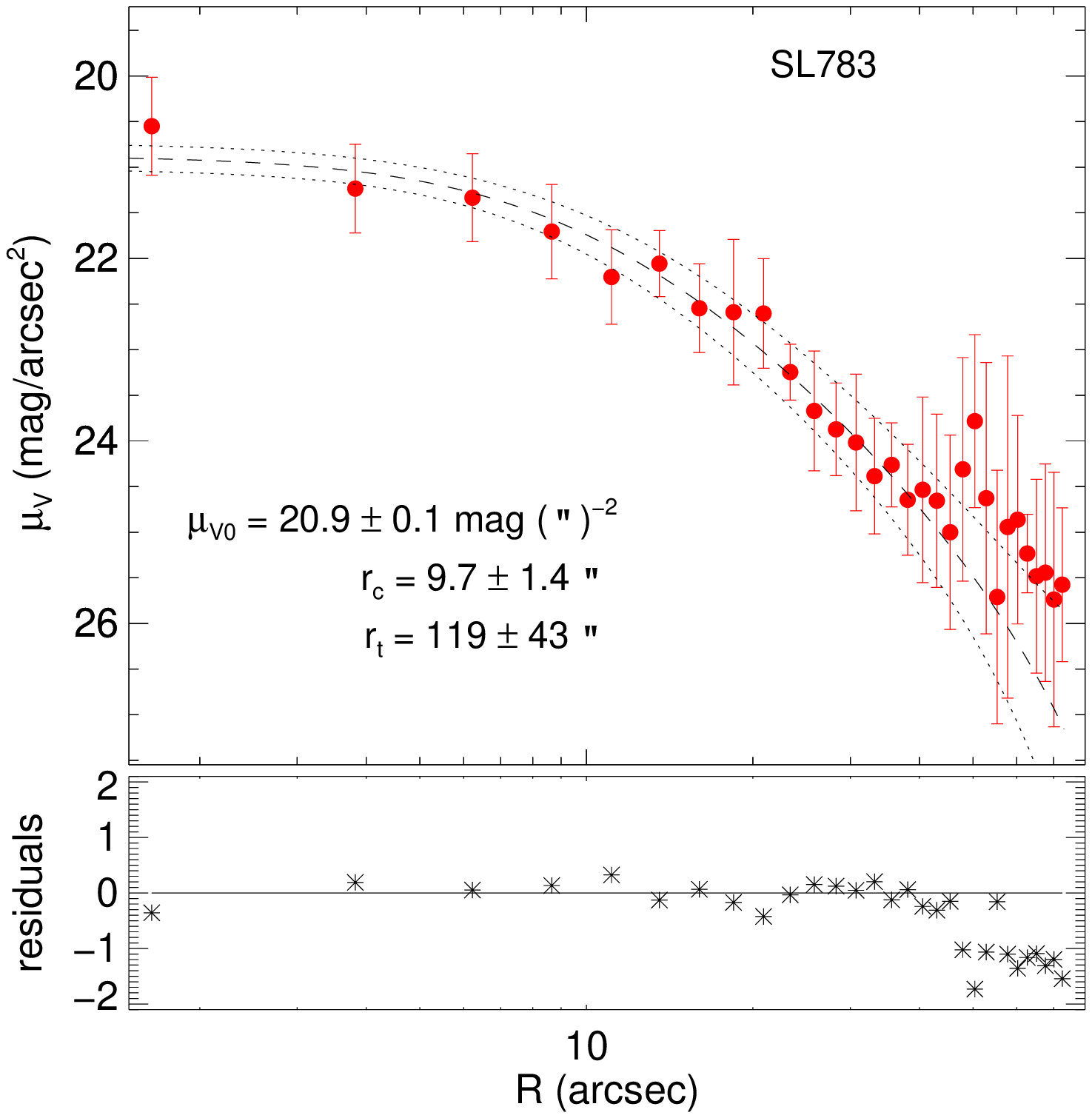}

\includegraphics[width=0.325\linewidth]{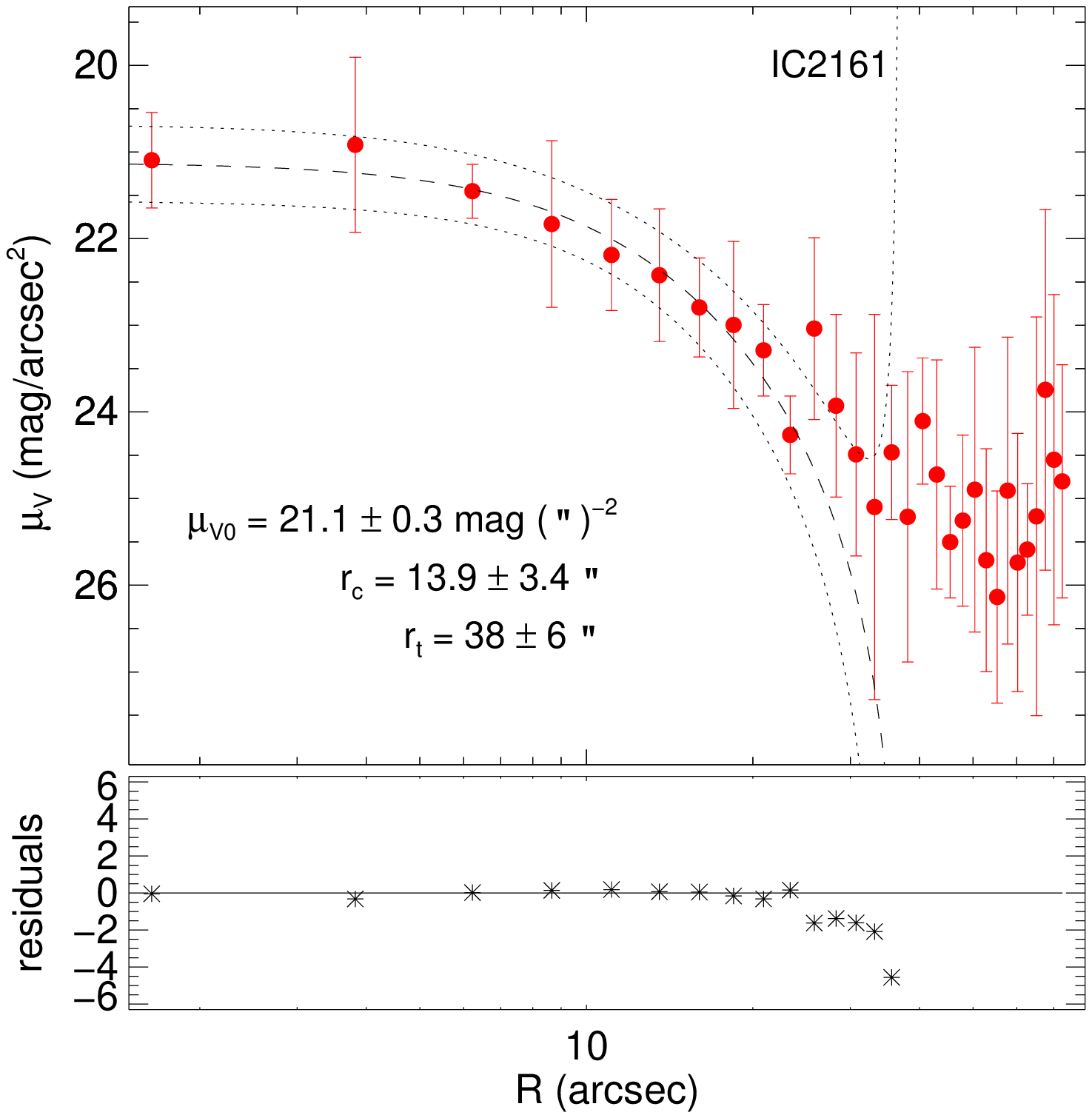}\includegraphics[width=0.325\linewidth]{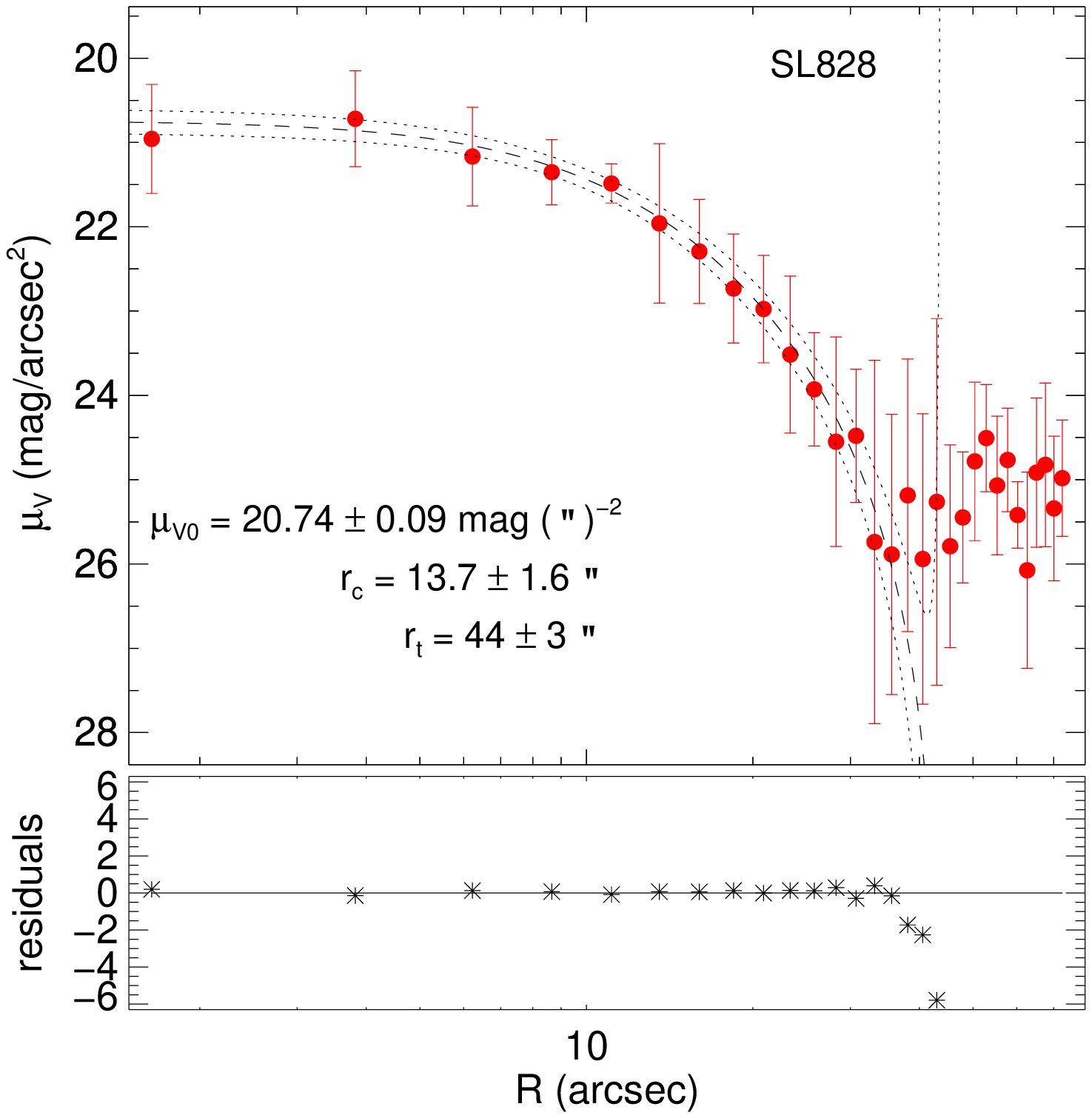}\includegraphics[width=0.325\linewidth]{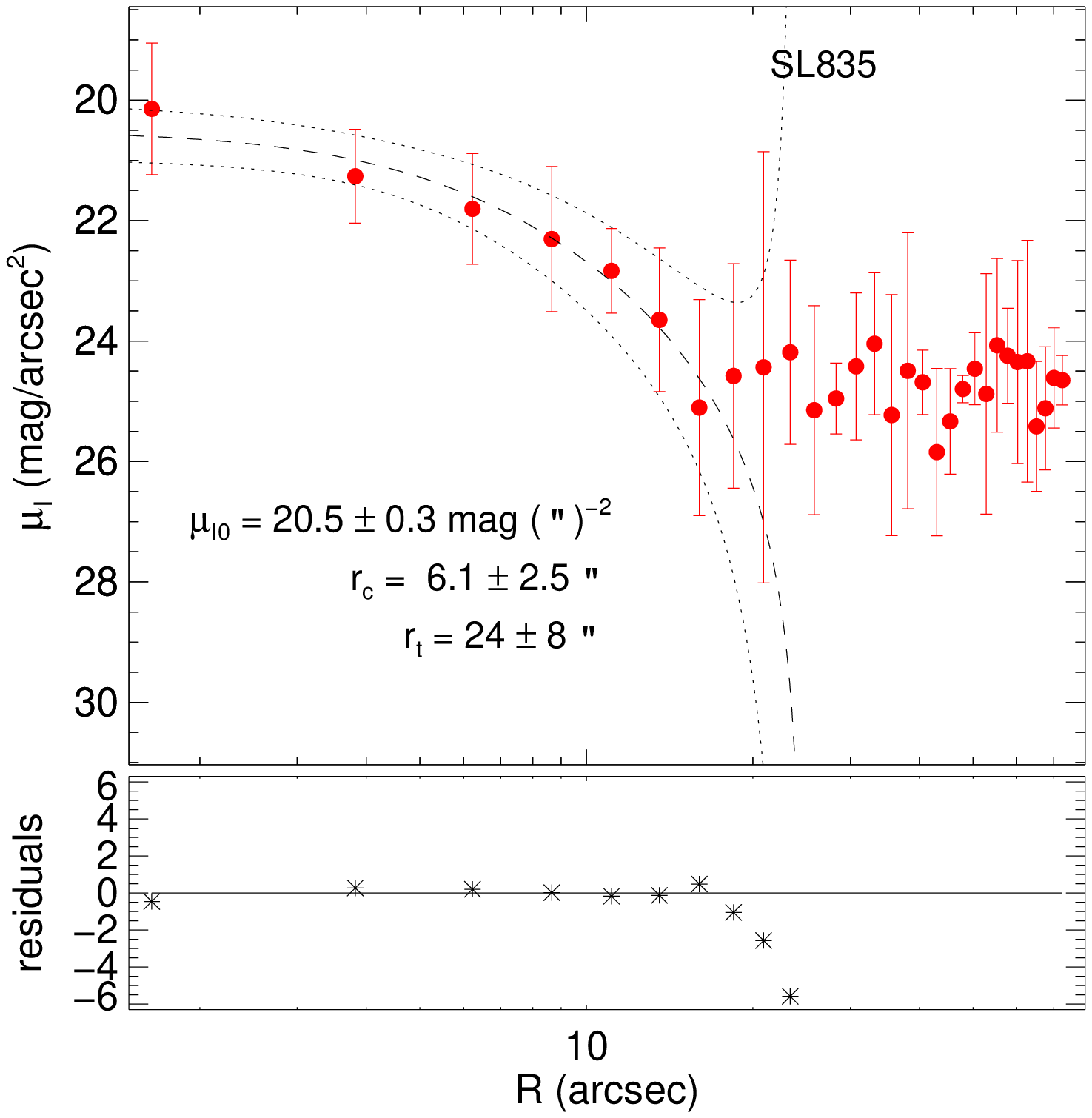}

\includegraphics[width=0.325\linewidth]{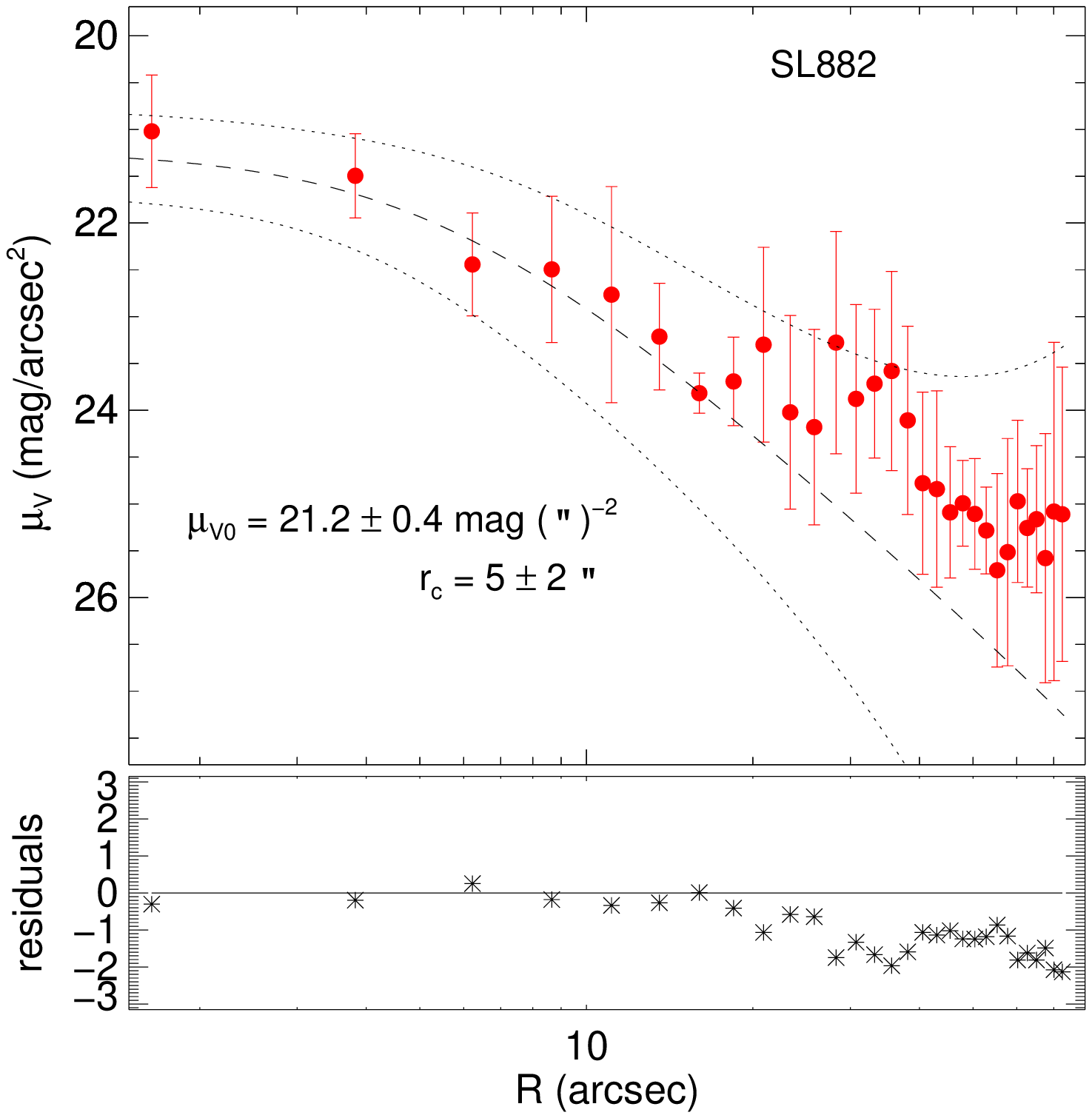}\includegraphics[width=0.325\linewidth]{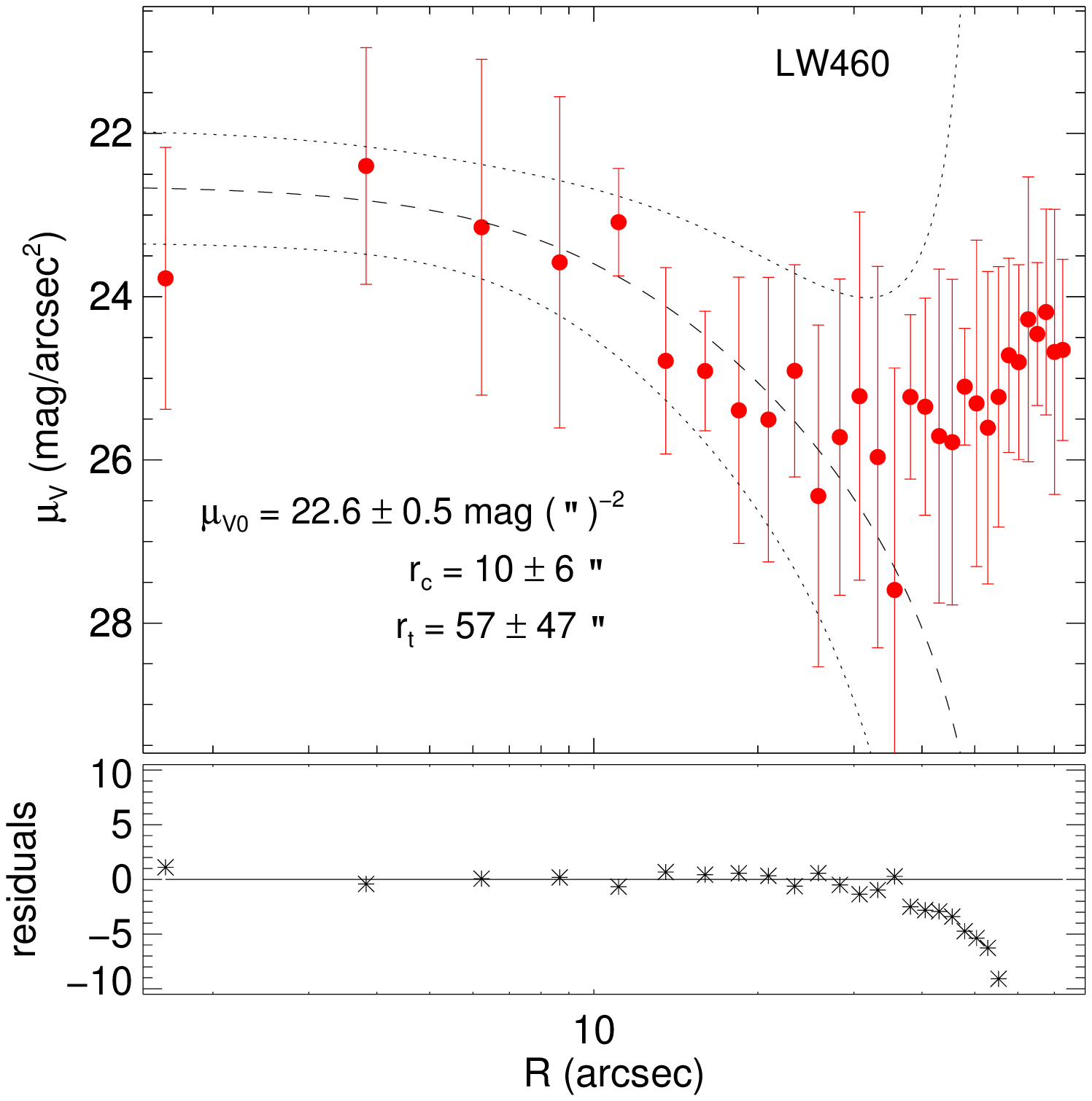}\includegraphics[width=0.325\linewidth]{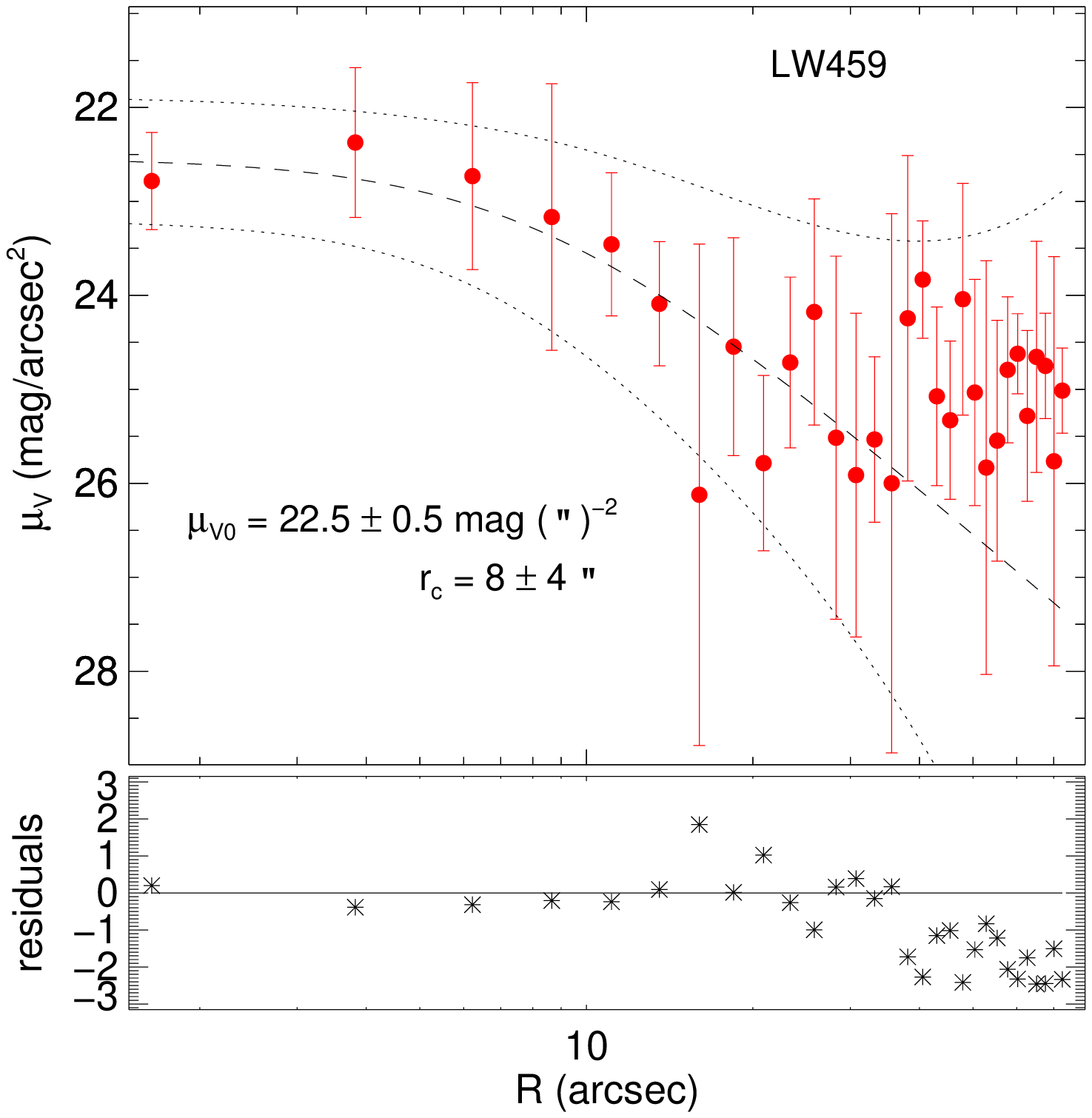}

\includegraphics[width=0.325\linewidth]{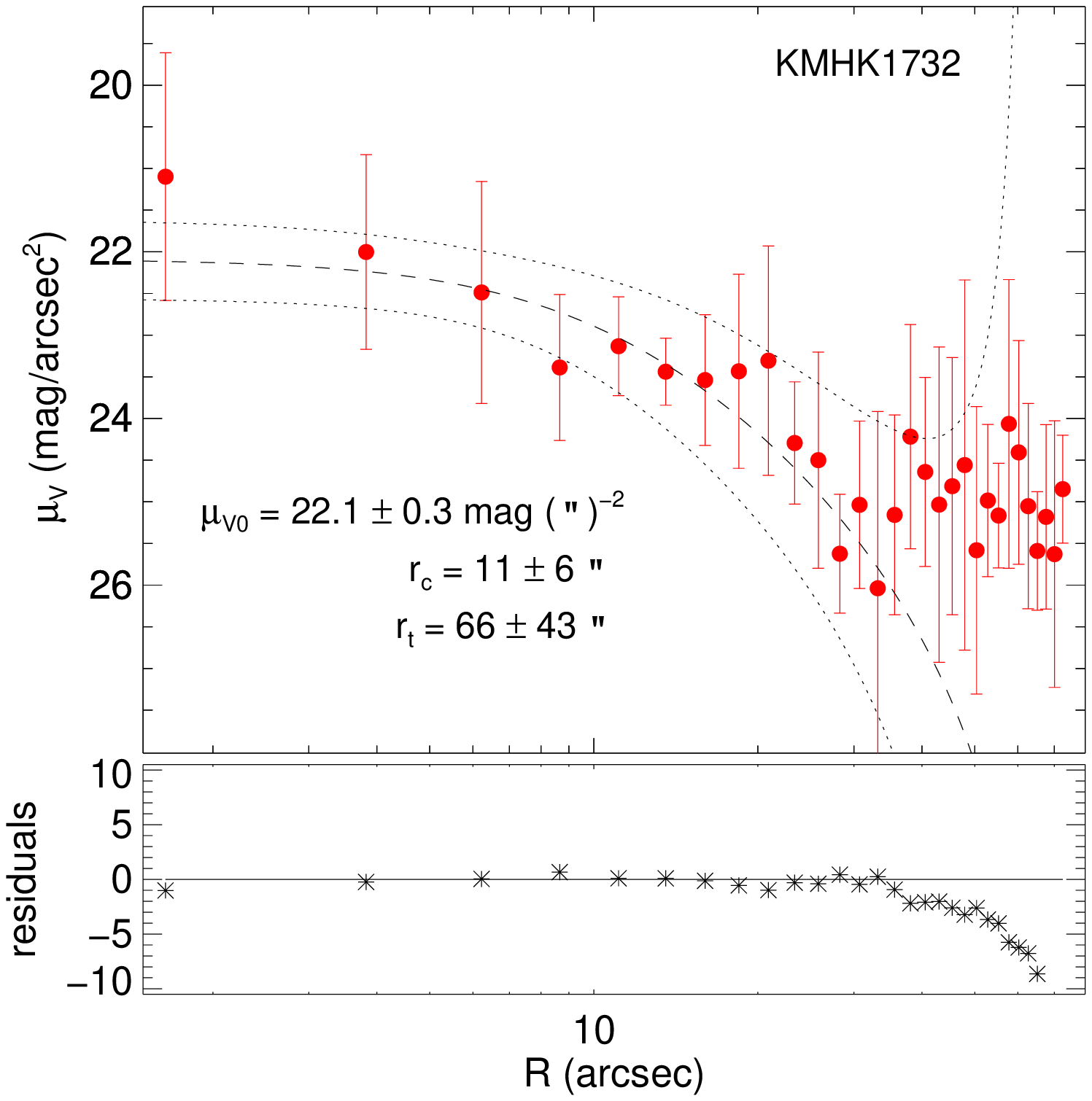}\includegraphics[width=0.325\linewidth]{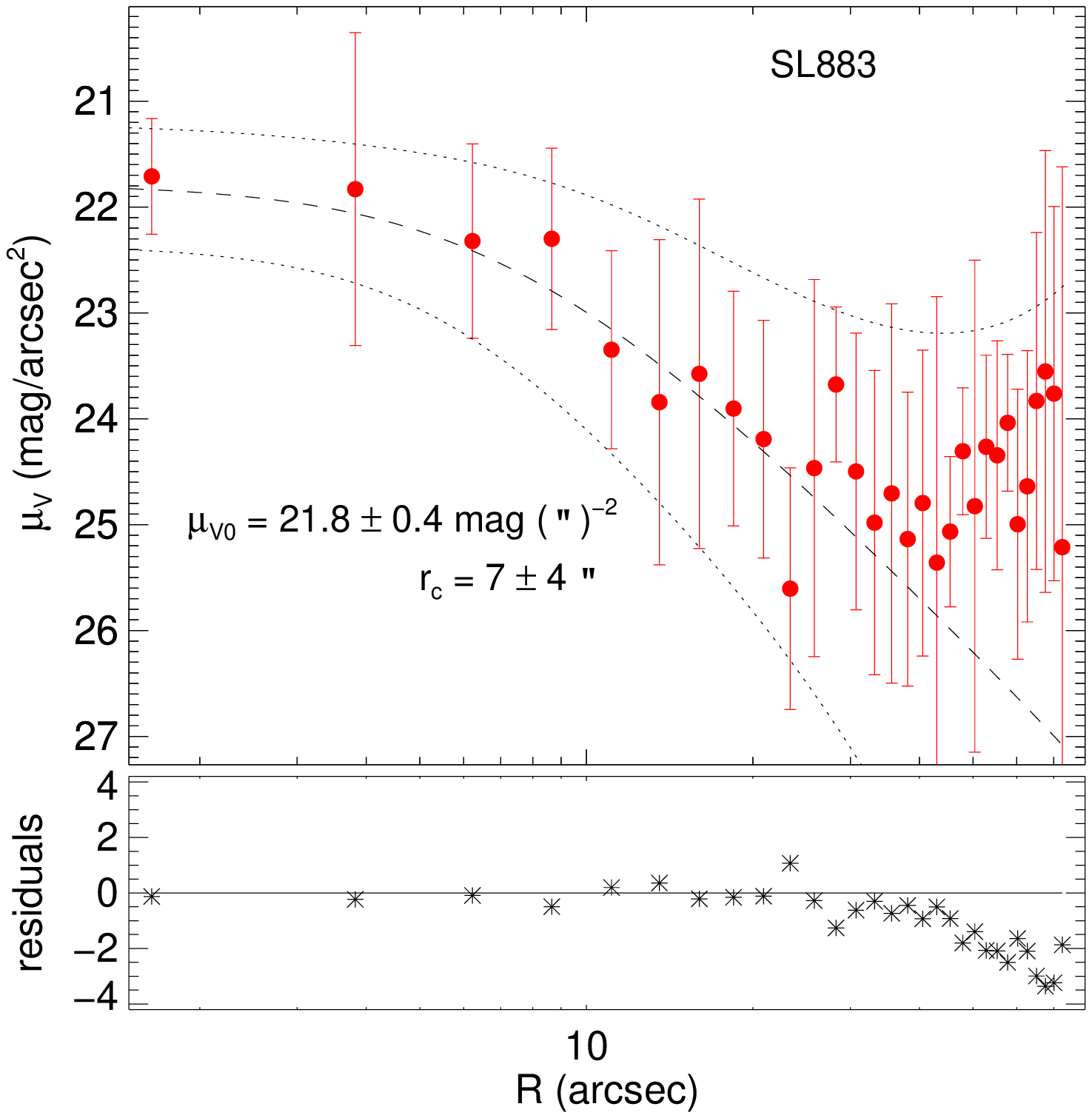}\includegraphics[width=0.325\linewidth]{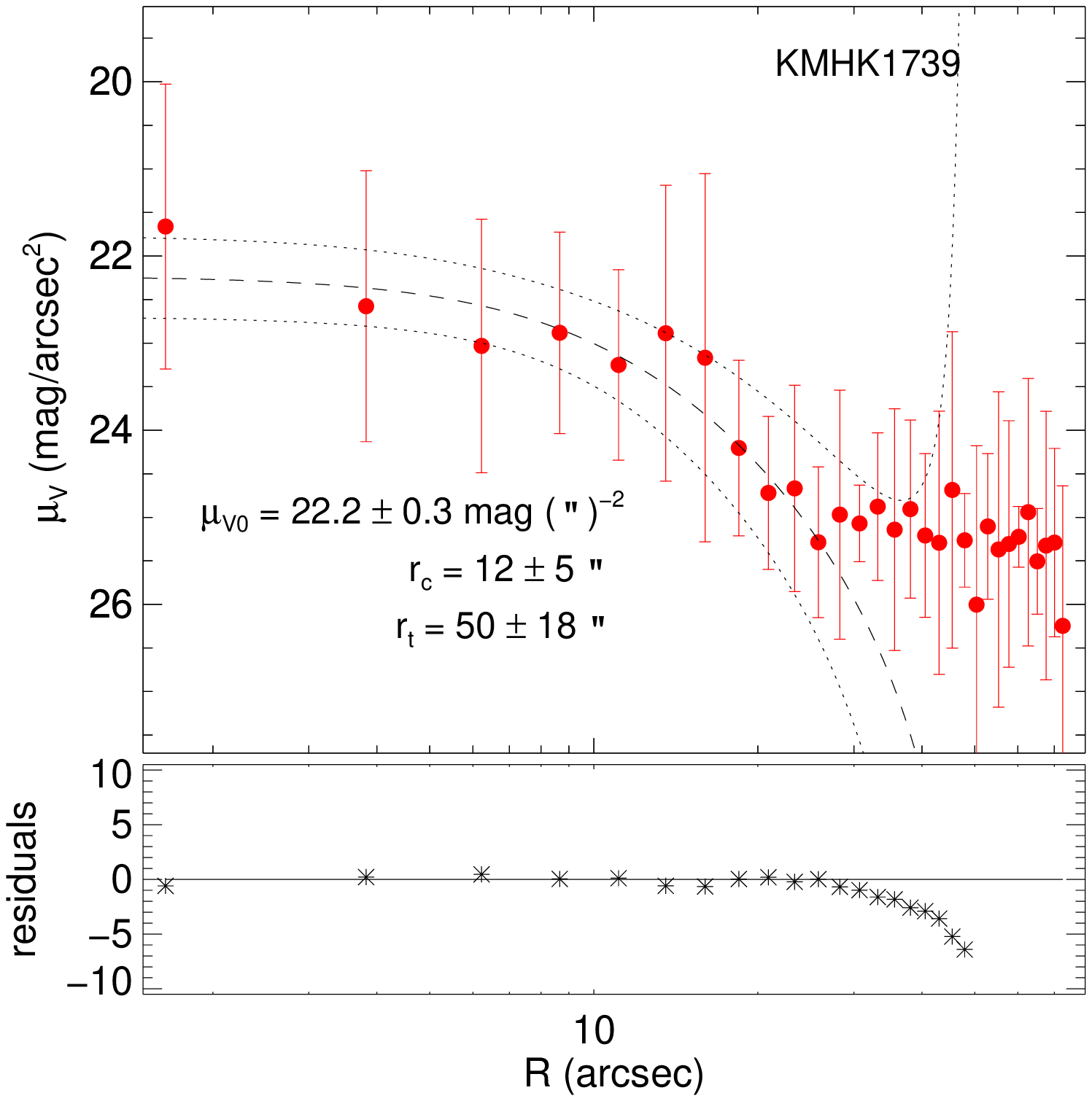}

\caption{cont.}

\end{figure*}

\setcounter{figure}{3}

\begin{figure*}
\includegraphics[width=0.325\linewidth]{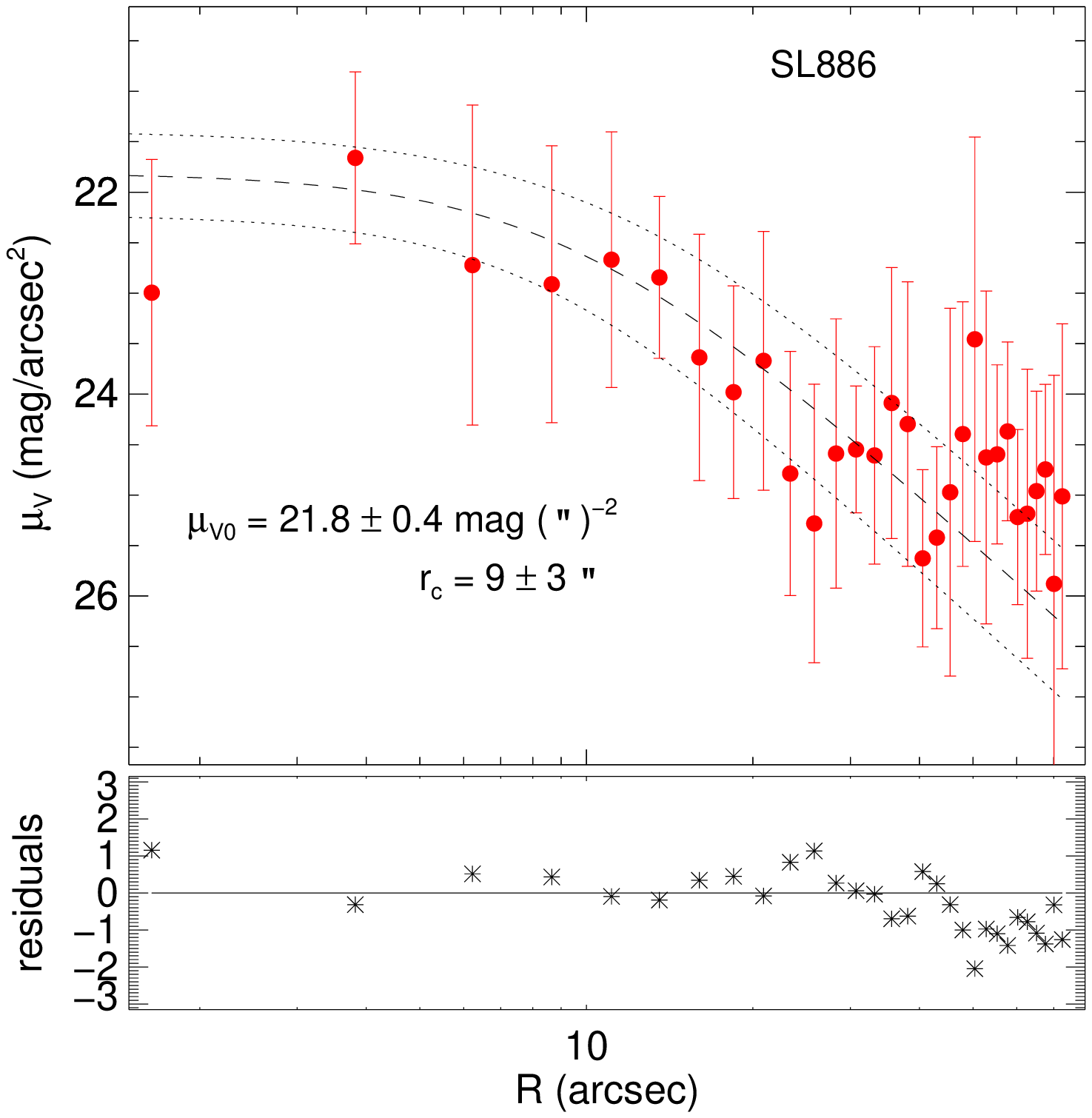}\includegraphics[width=0.325\linewidth]{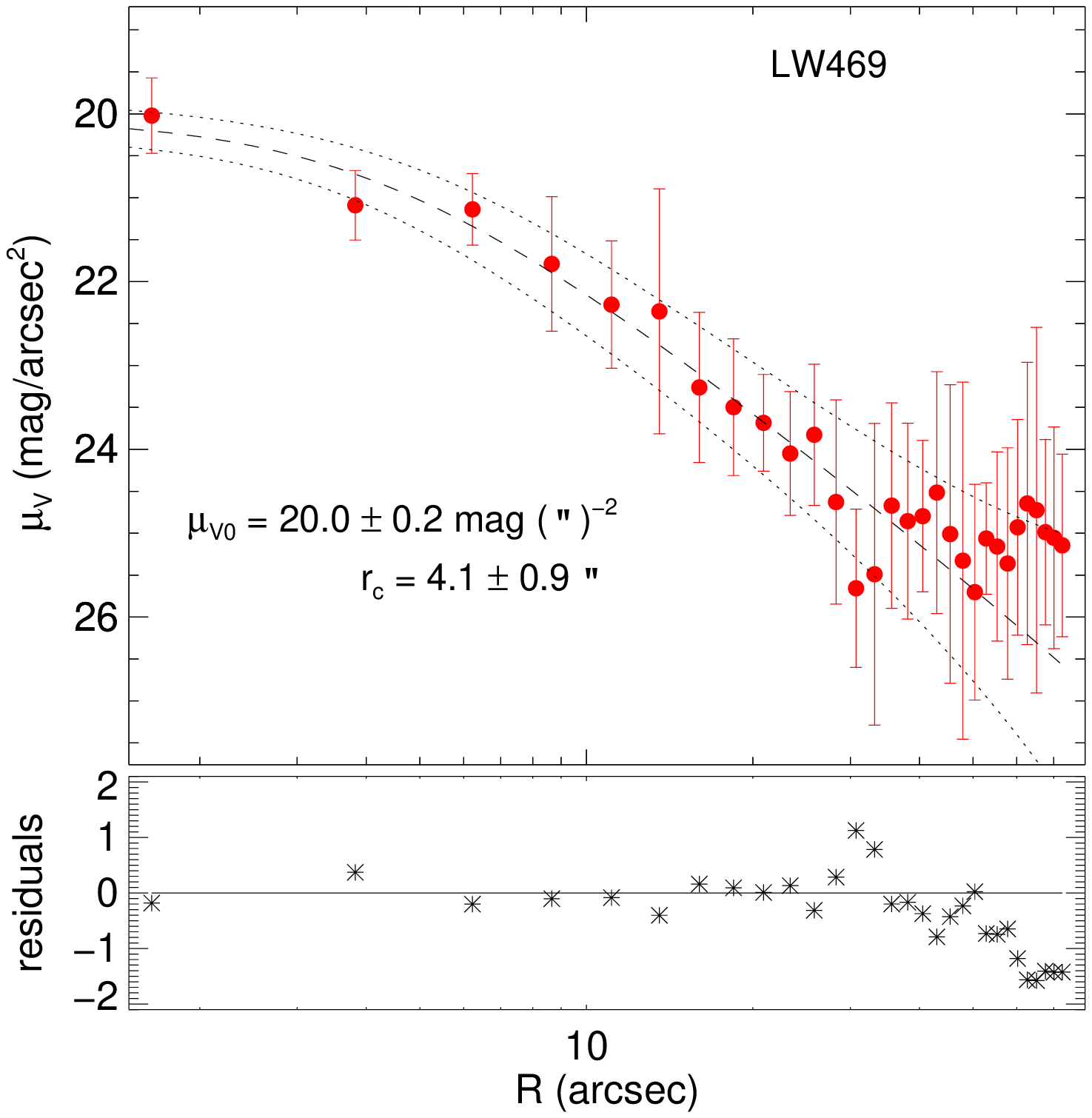}\includegraphics[width=0.325\linewidth]{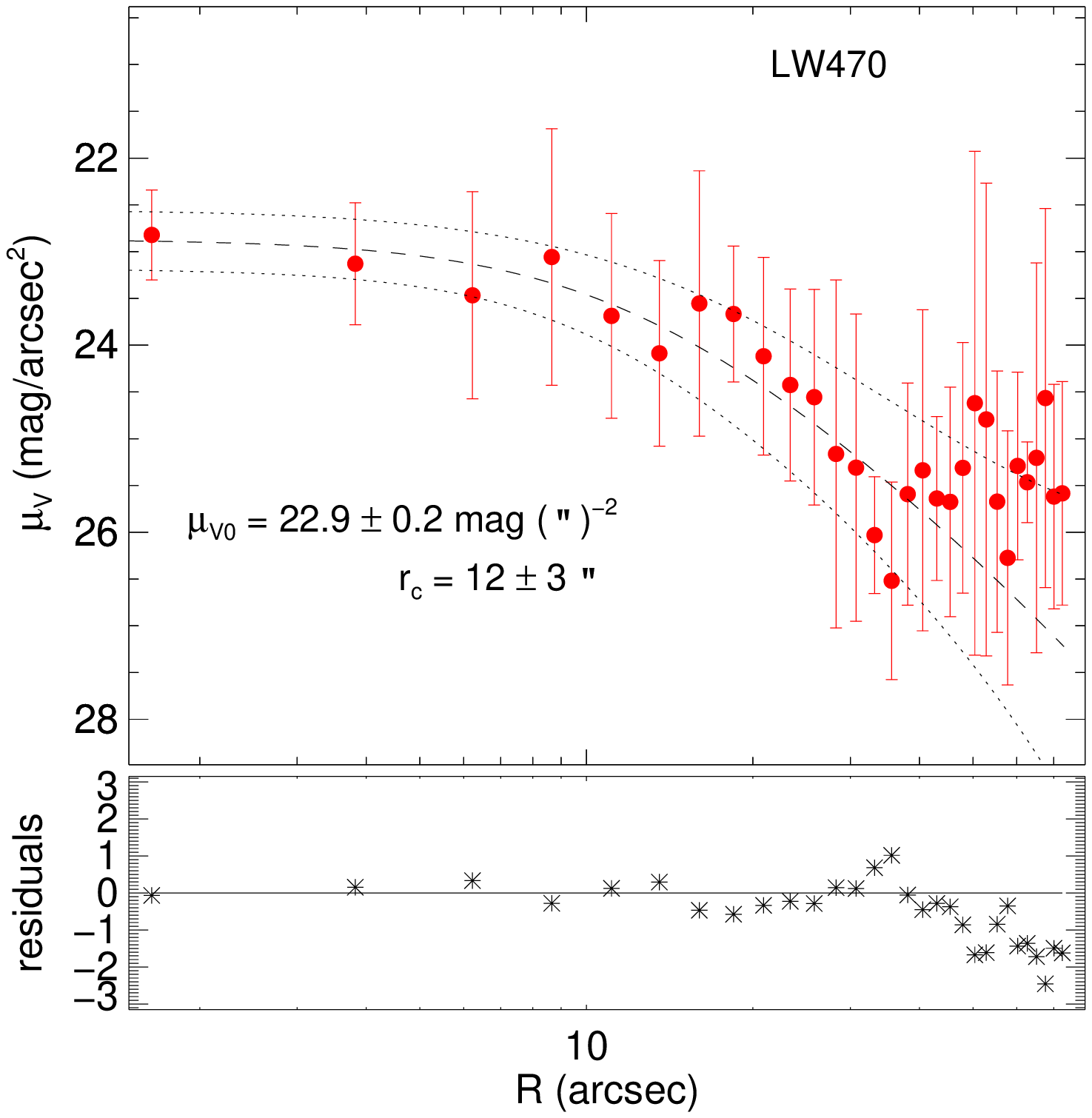}

\includegraphics[width=0.325\linewidth]{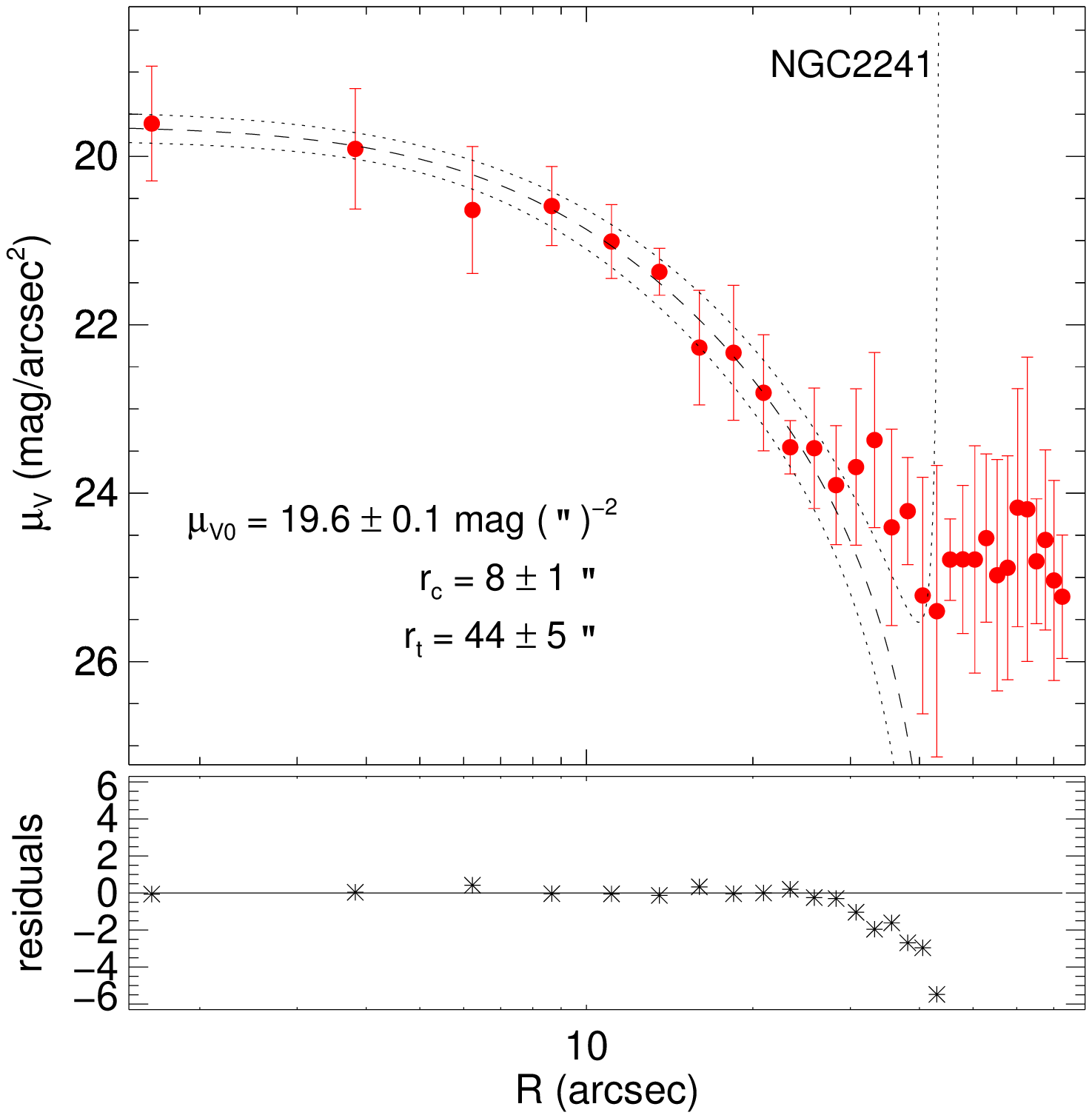}\includegraphics[width=0.325\linewidth]{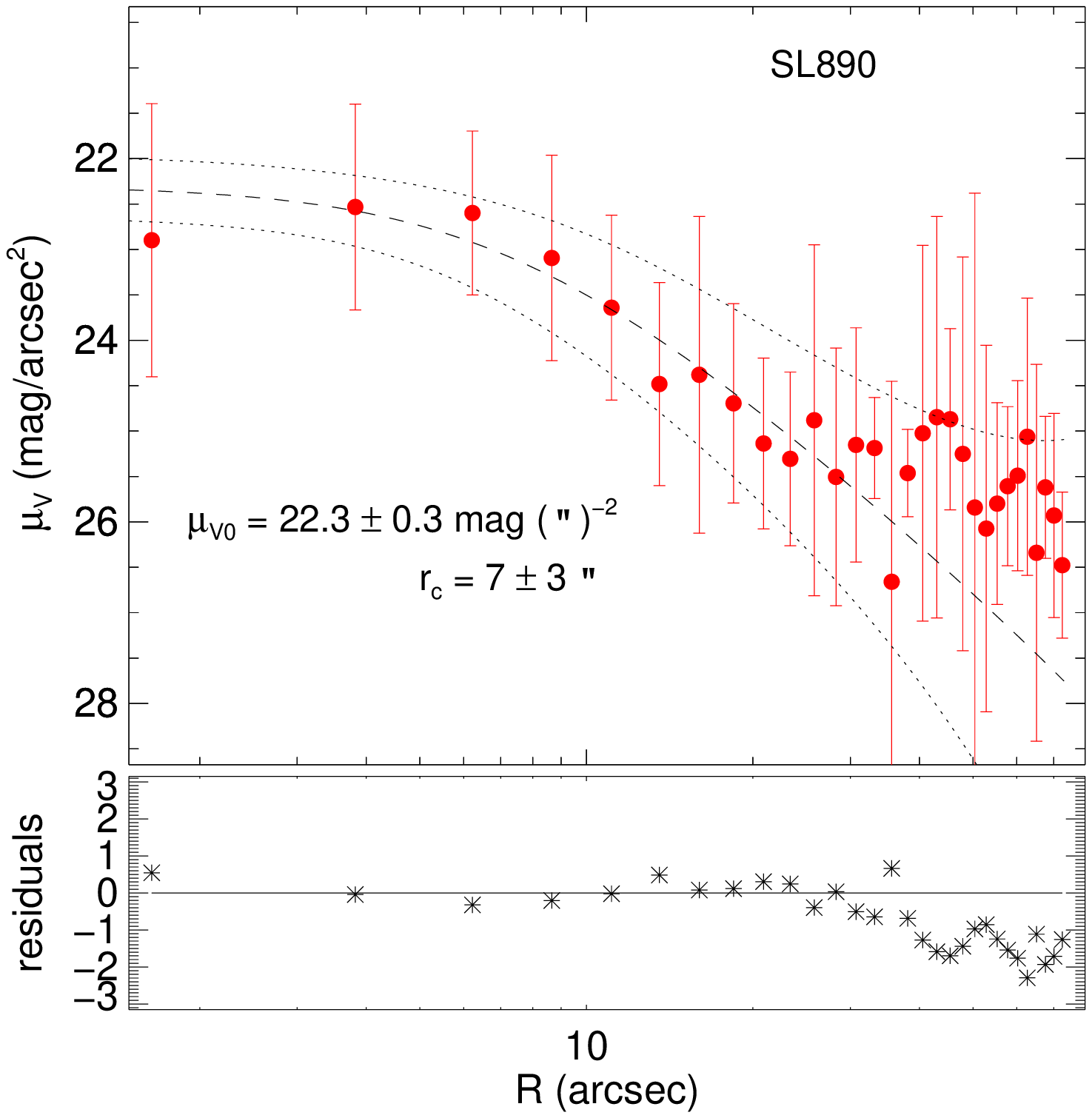}\includegraphics[width=0.325\linewidth]{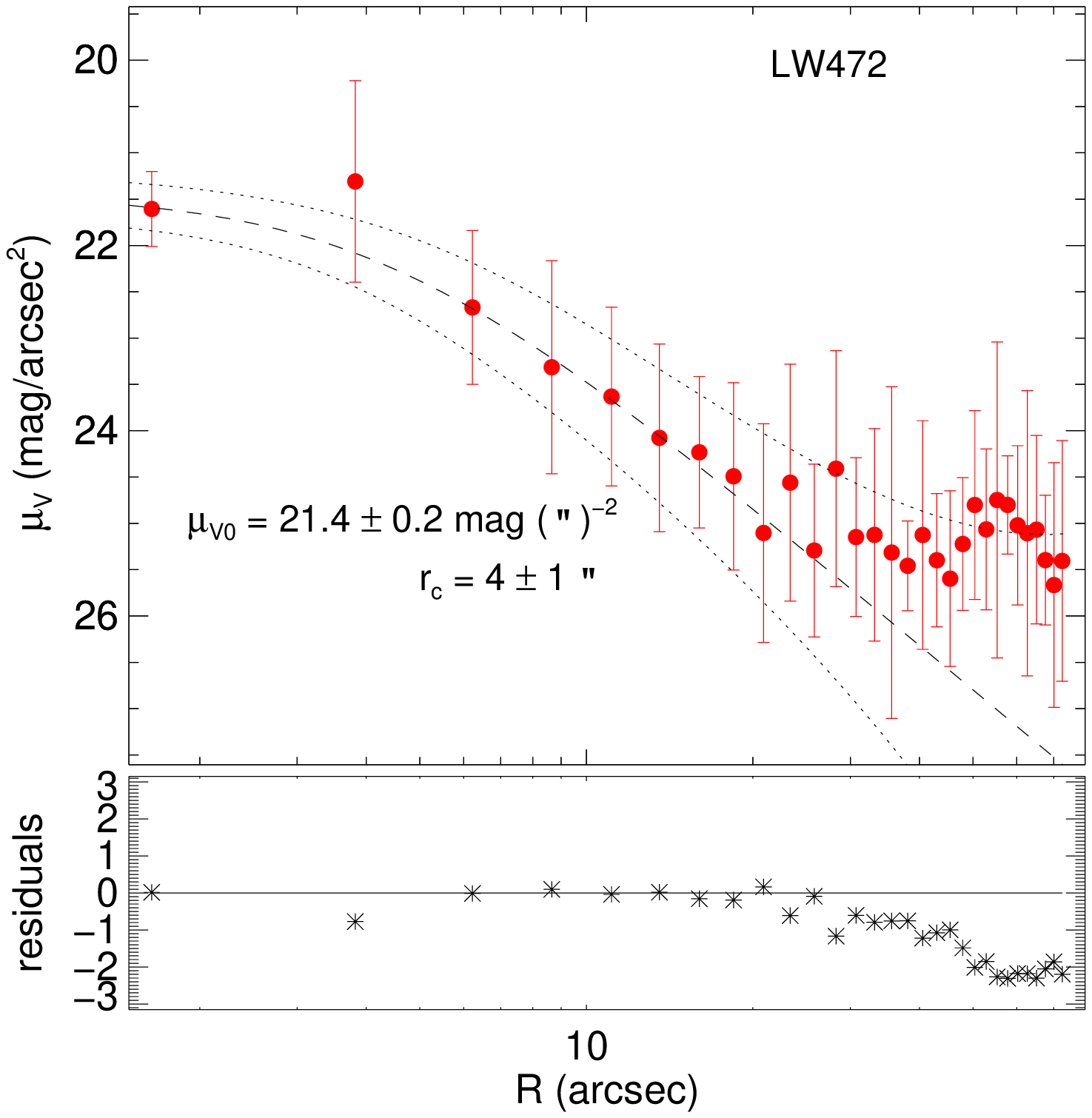}

\includegraphics[width=0.325\linewidth]{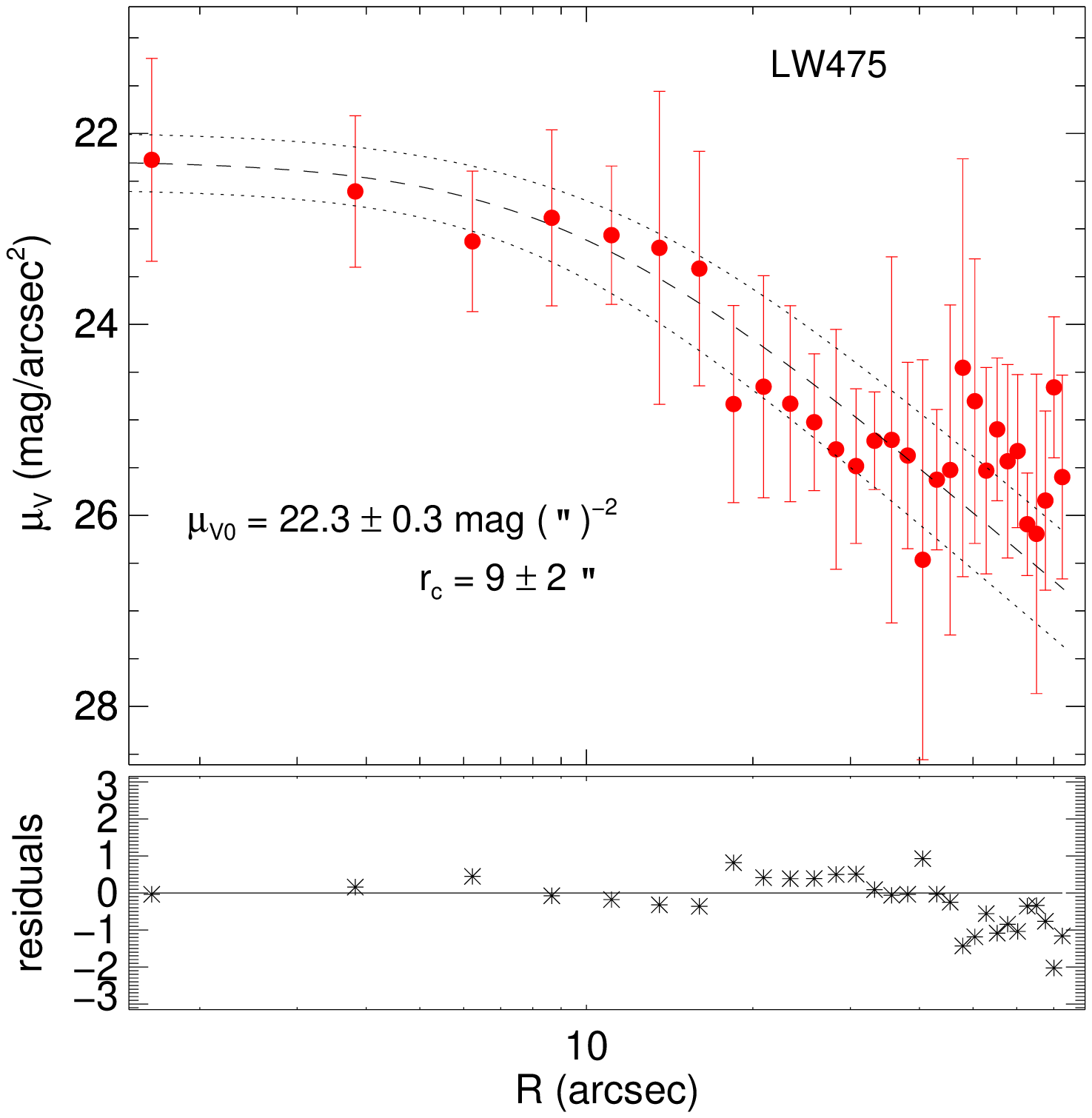}\includegraphics[width=0.325\linewidth]{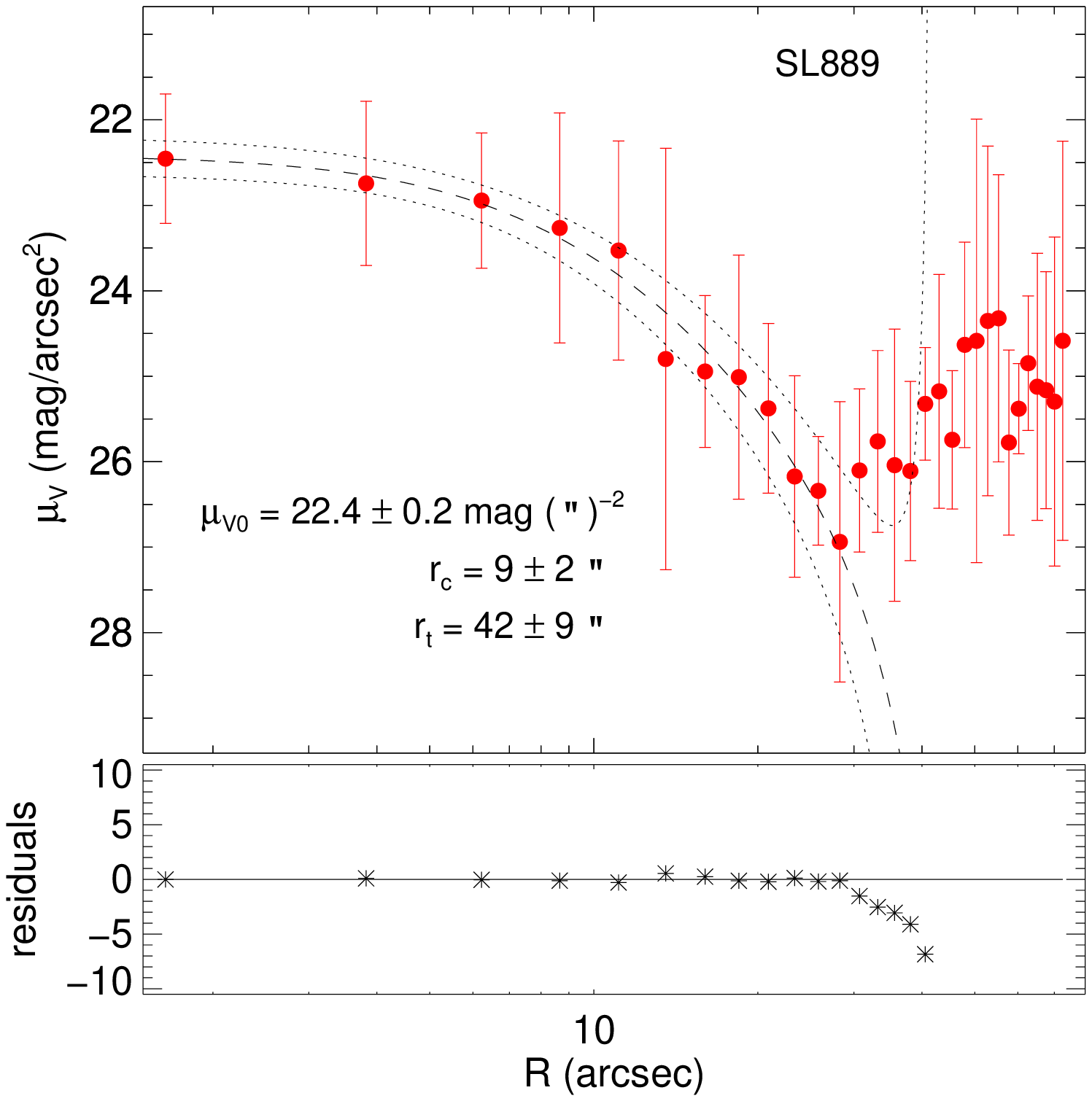}\includegraphics[width=0.325\linewidth]{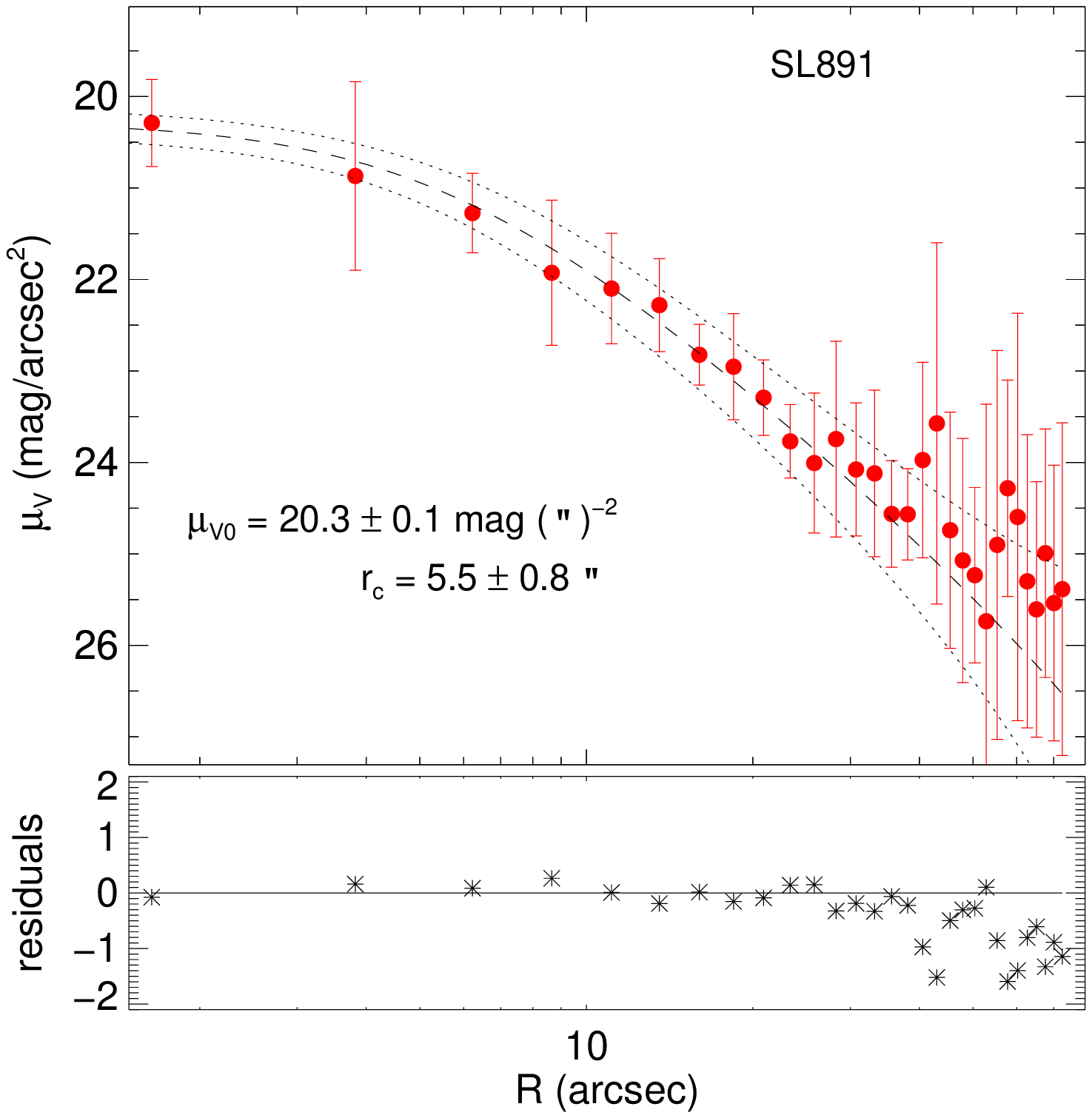}

\includegraphics[width=0.325\linewidth]{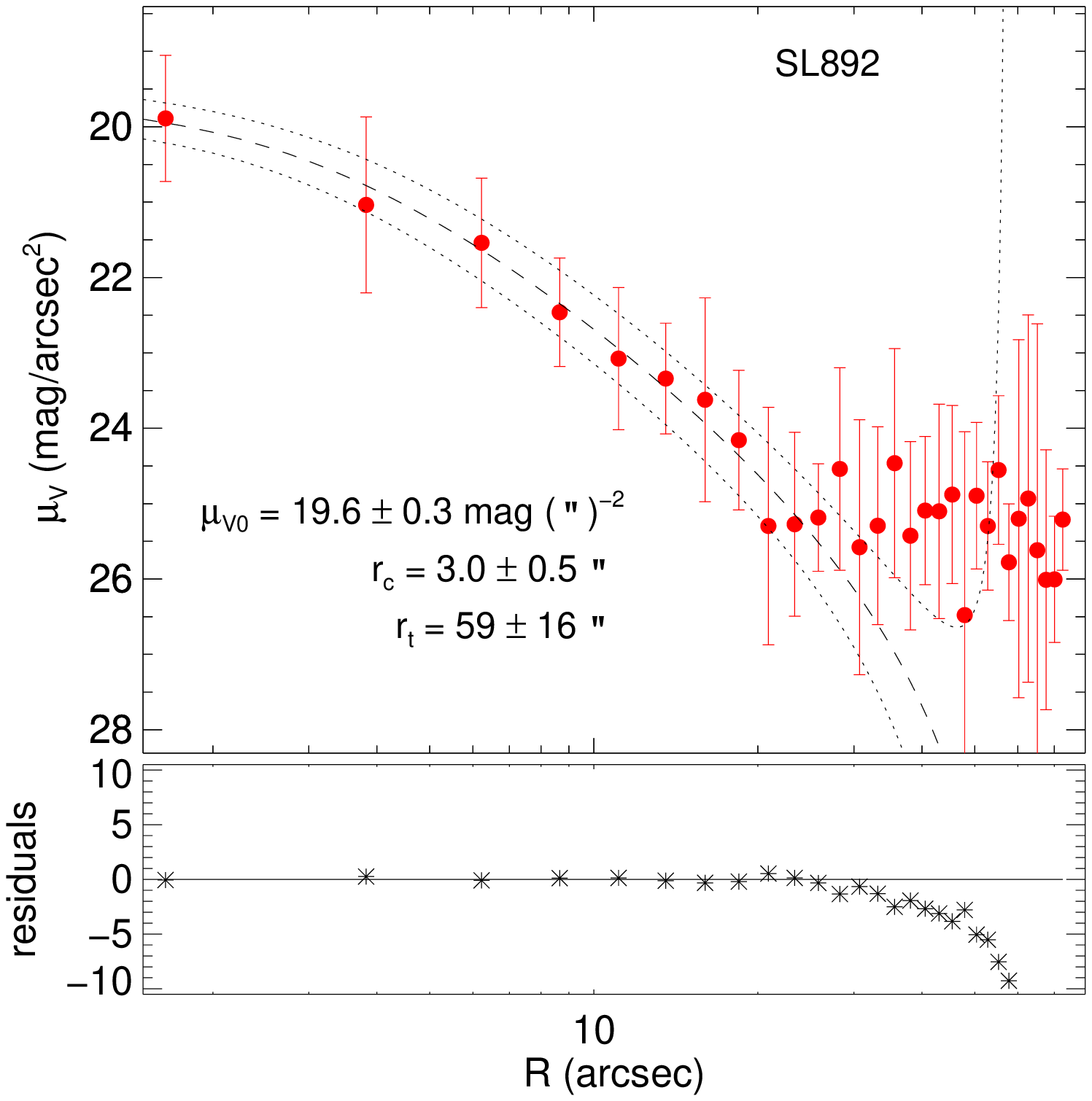}\includegraphics[width=0.325\linewidth]{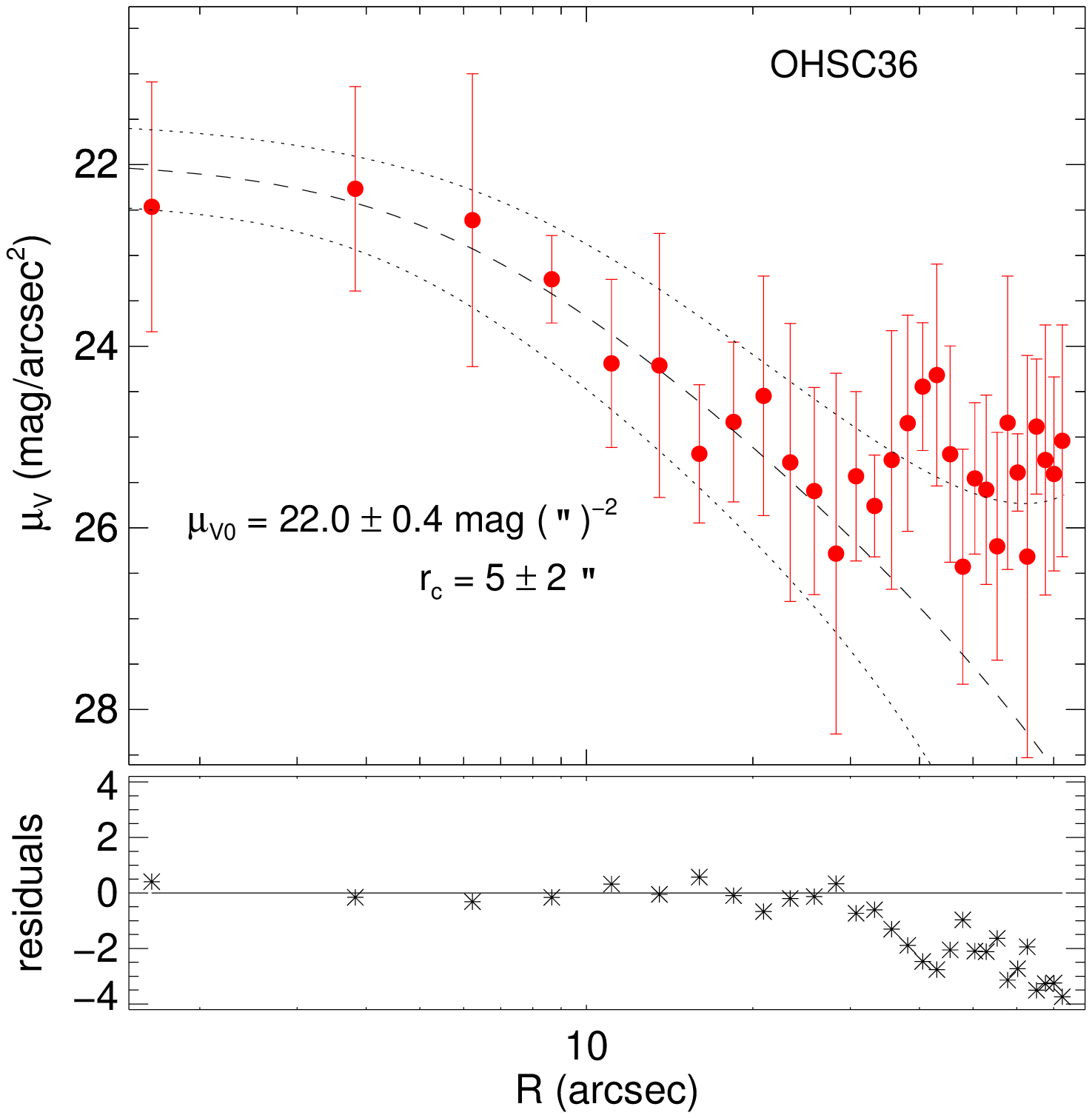}\includegraphics[width=0.325\linewidth]{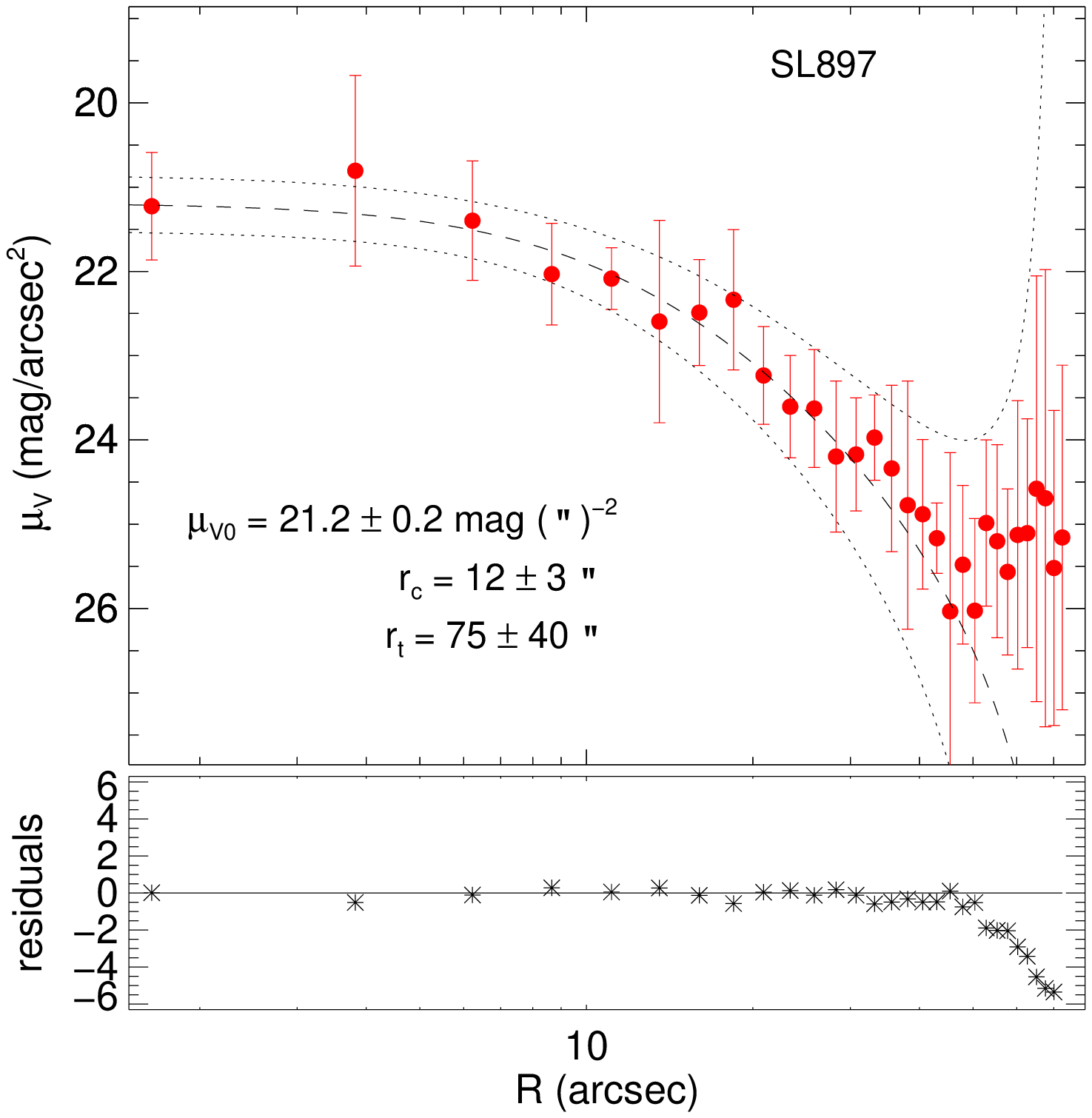}

\caption{cont.}

\end{figure*}

\clearpage


\begin{figure*}

\includegraphics[width=0.325\linewidth]{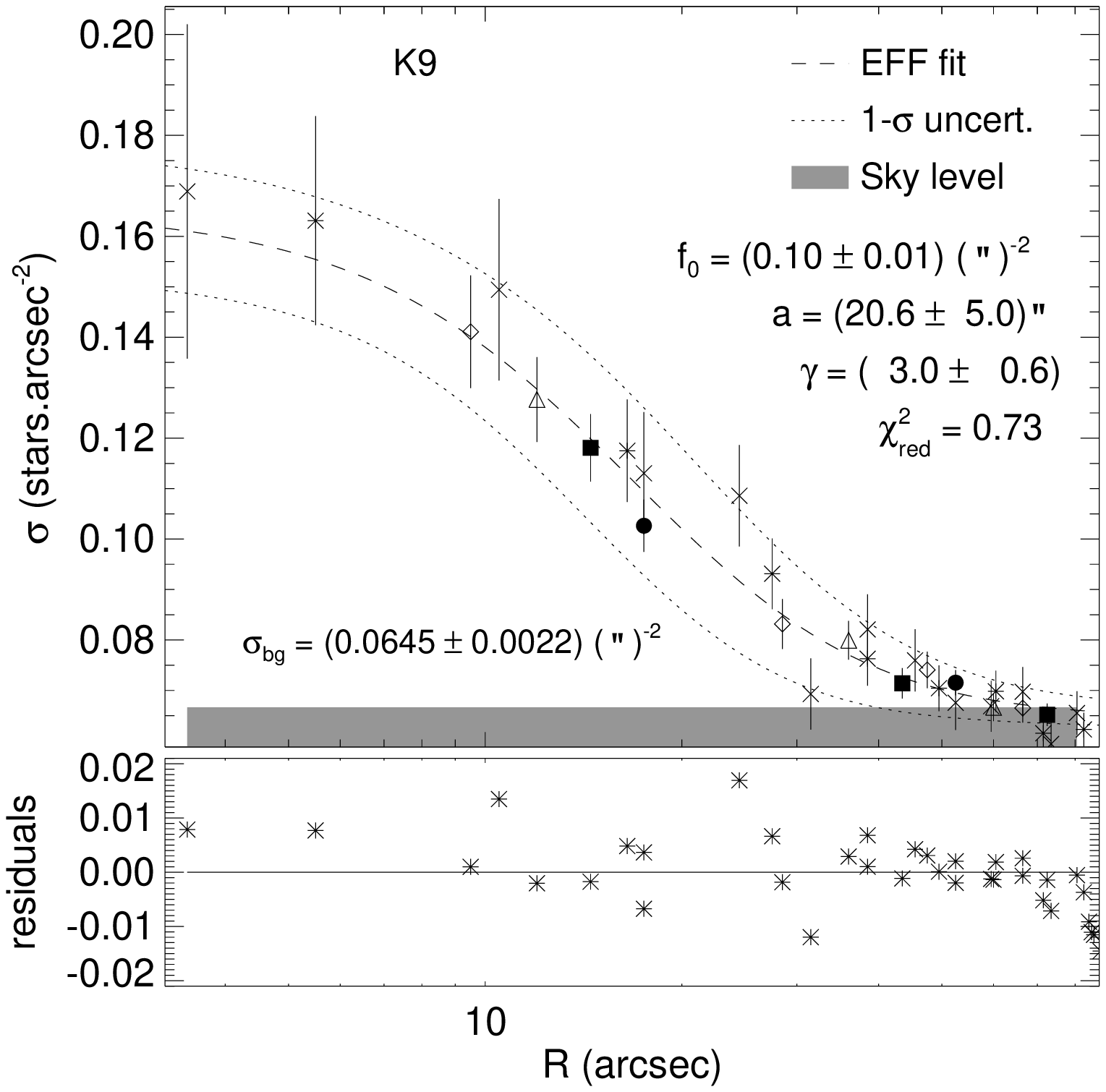}\includegraphics[width=0.325\linewidth]{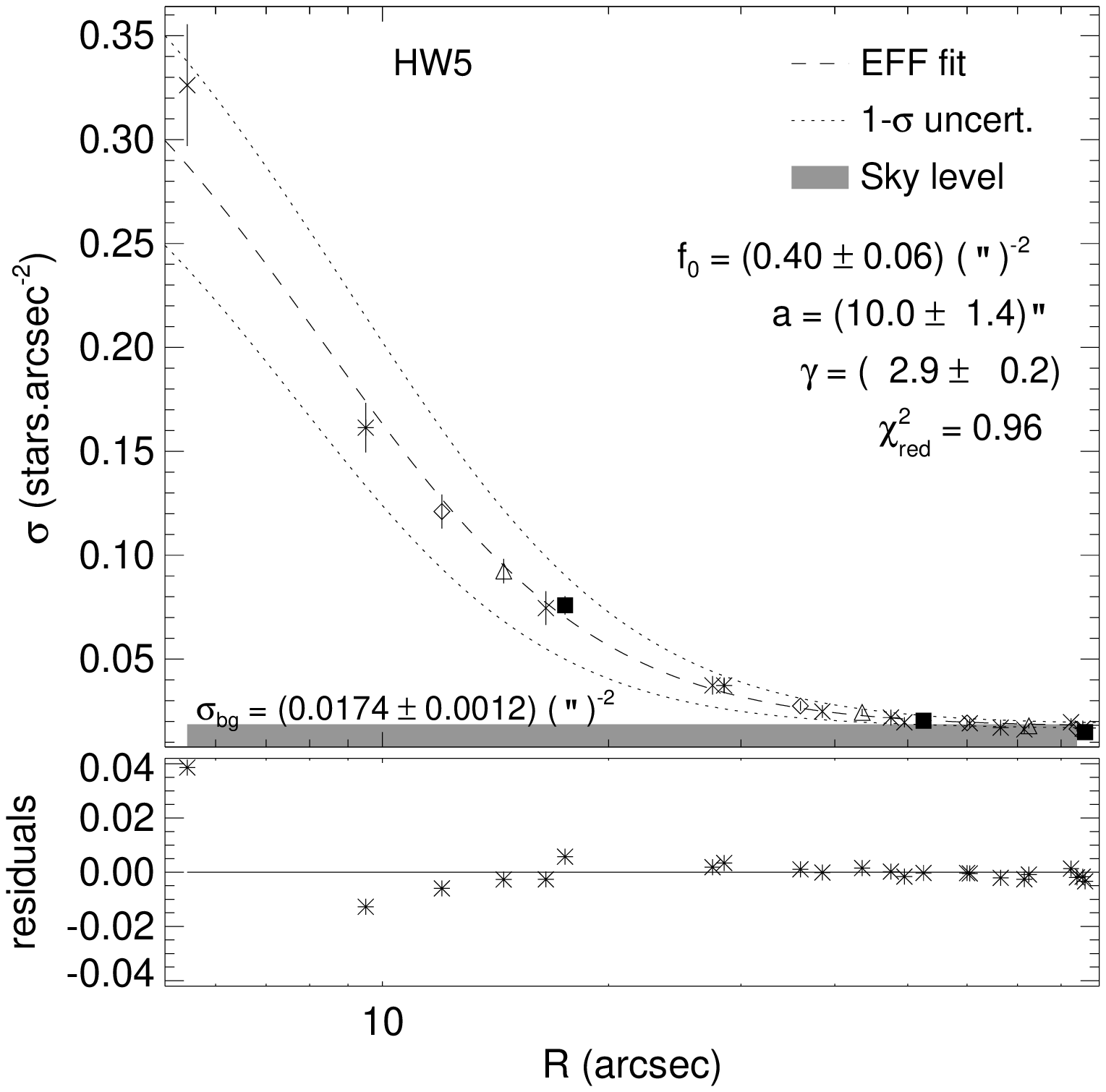}\includegraphics[width=0.325\linewidth]{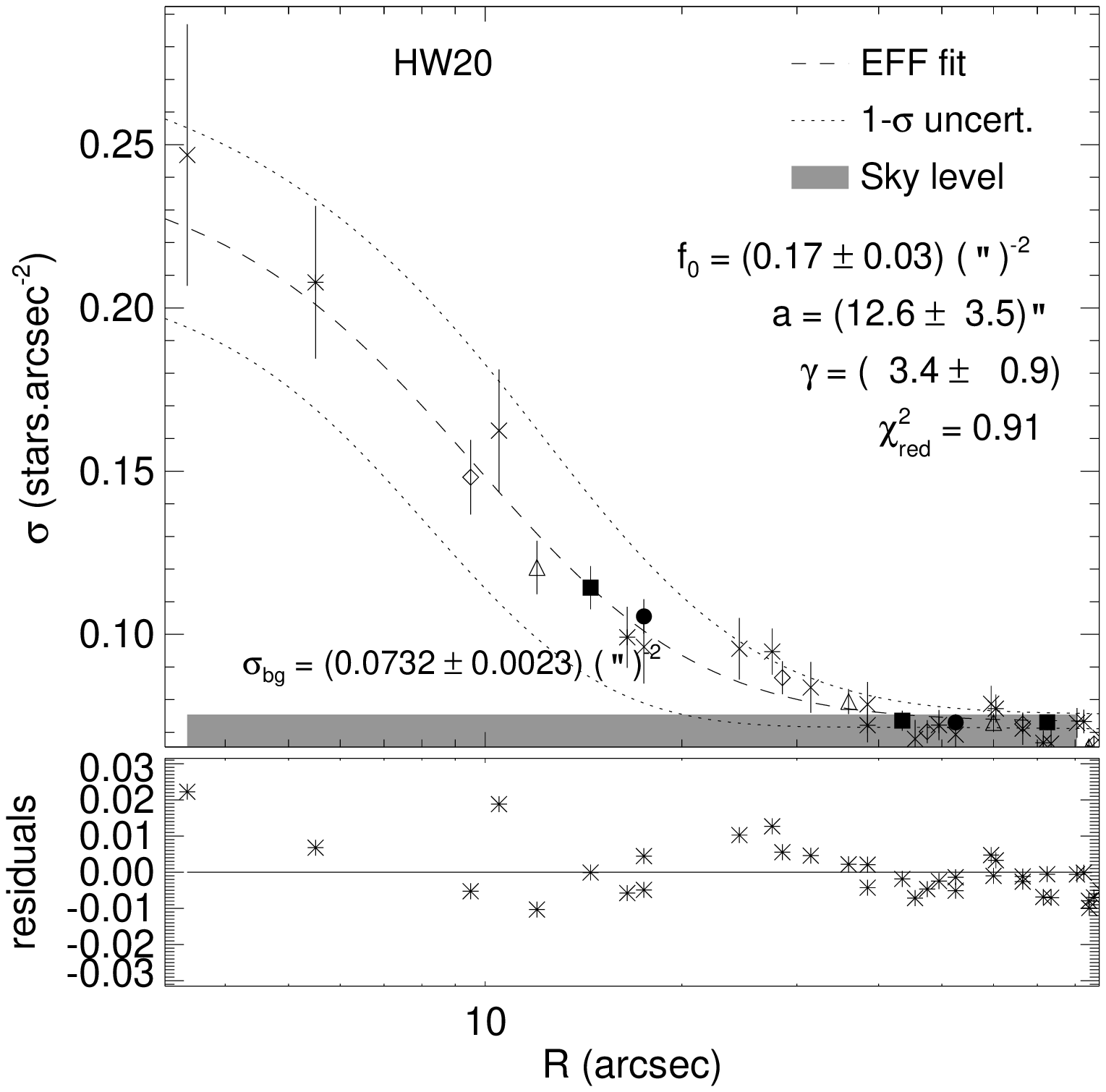}

\includegraphics[width=0.325\linewidth]{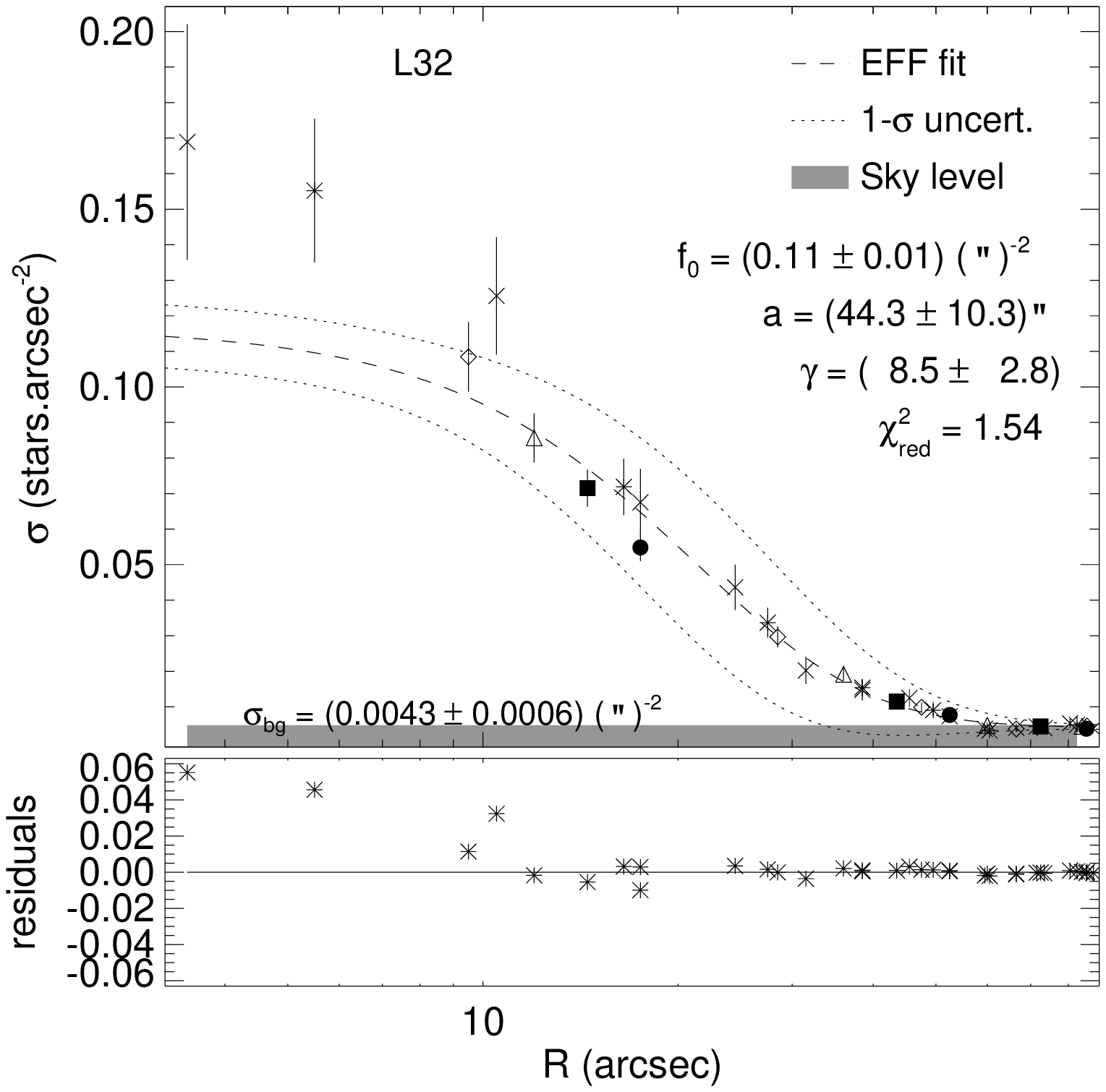}\includegraphics[width=0.325\linewidth]{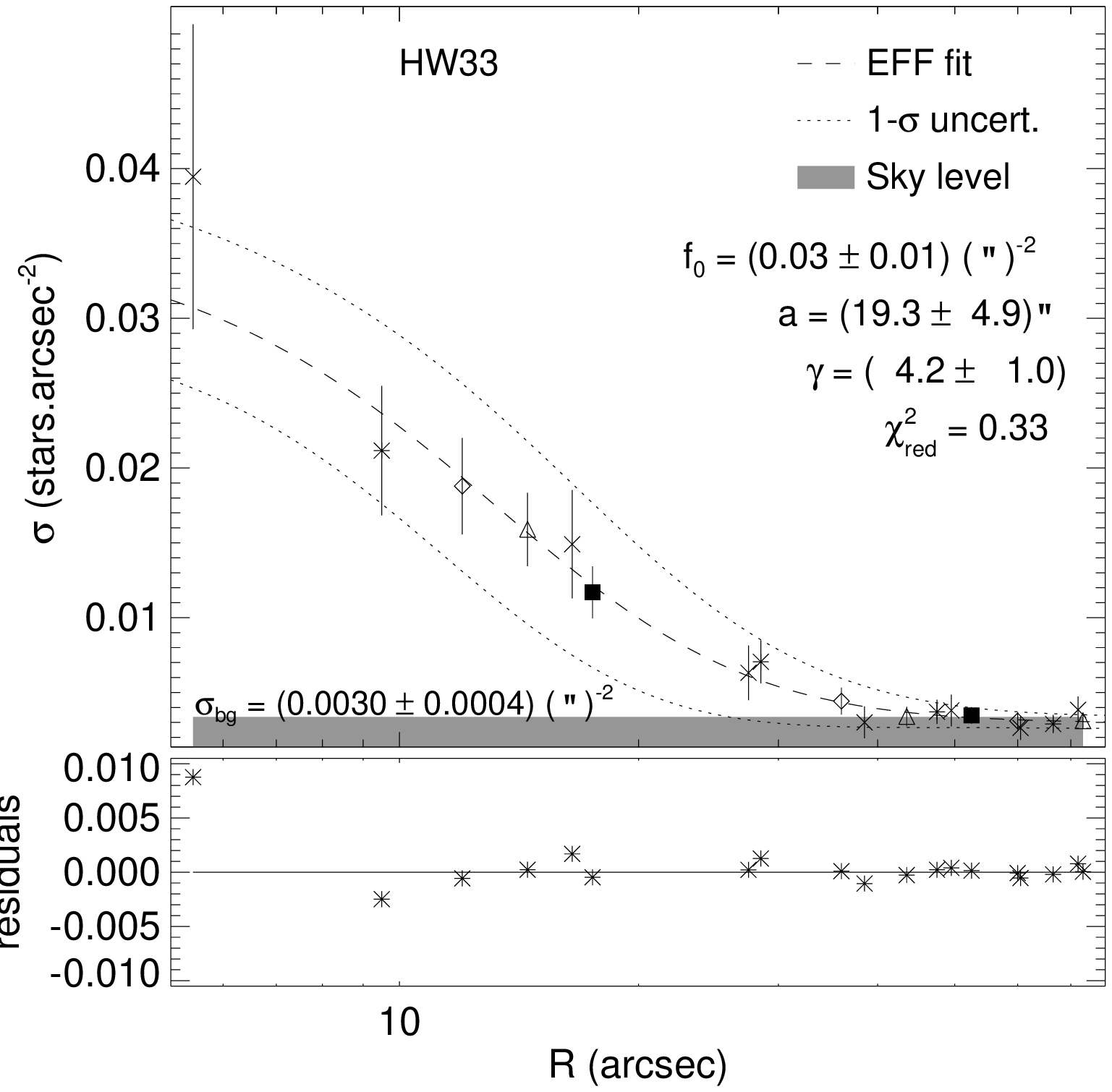}\includegraphics[width=0.325\linewidth]{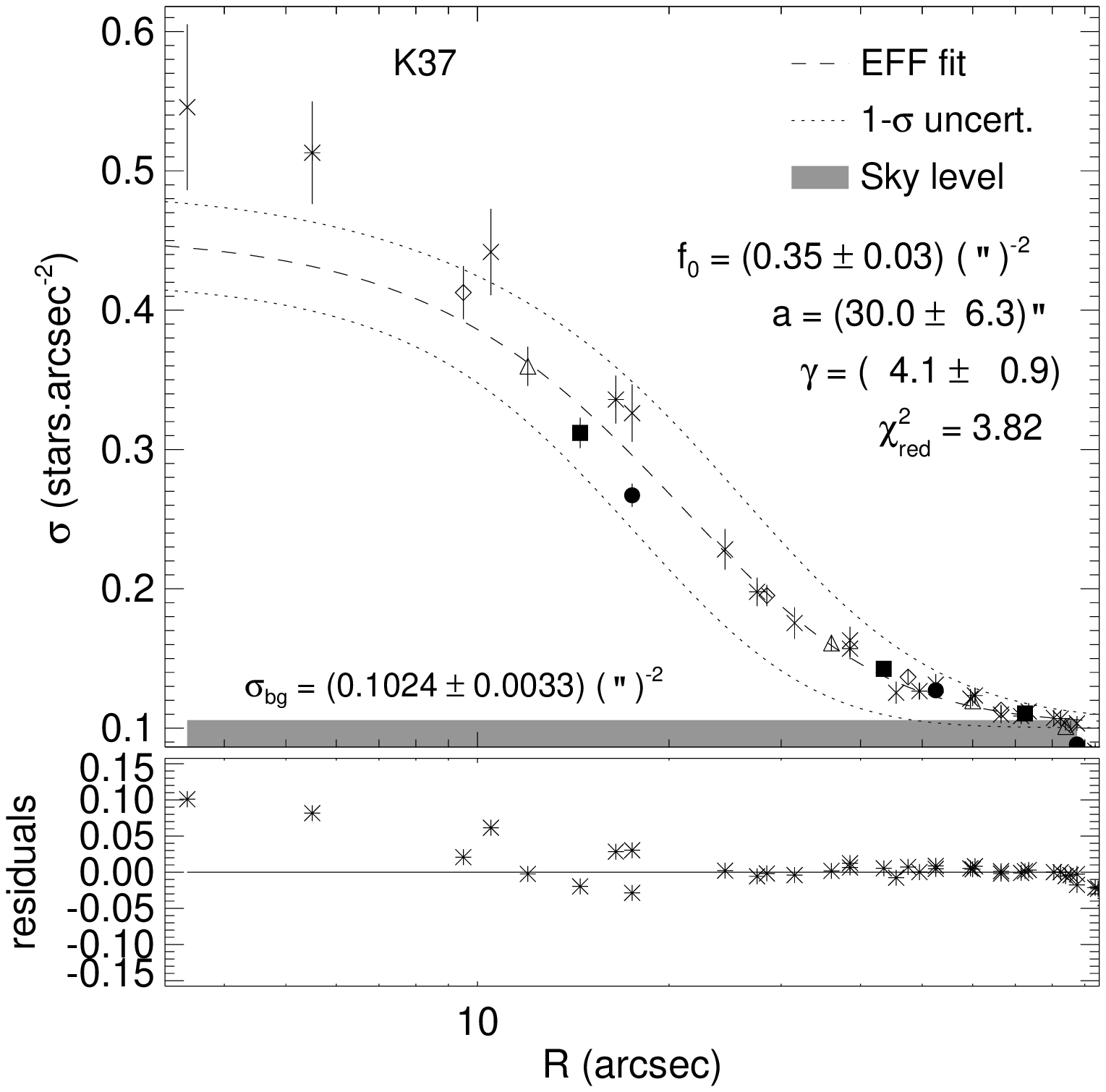}

\includegraphics[width=0.325\linewidth]{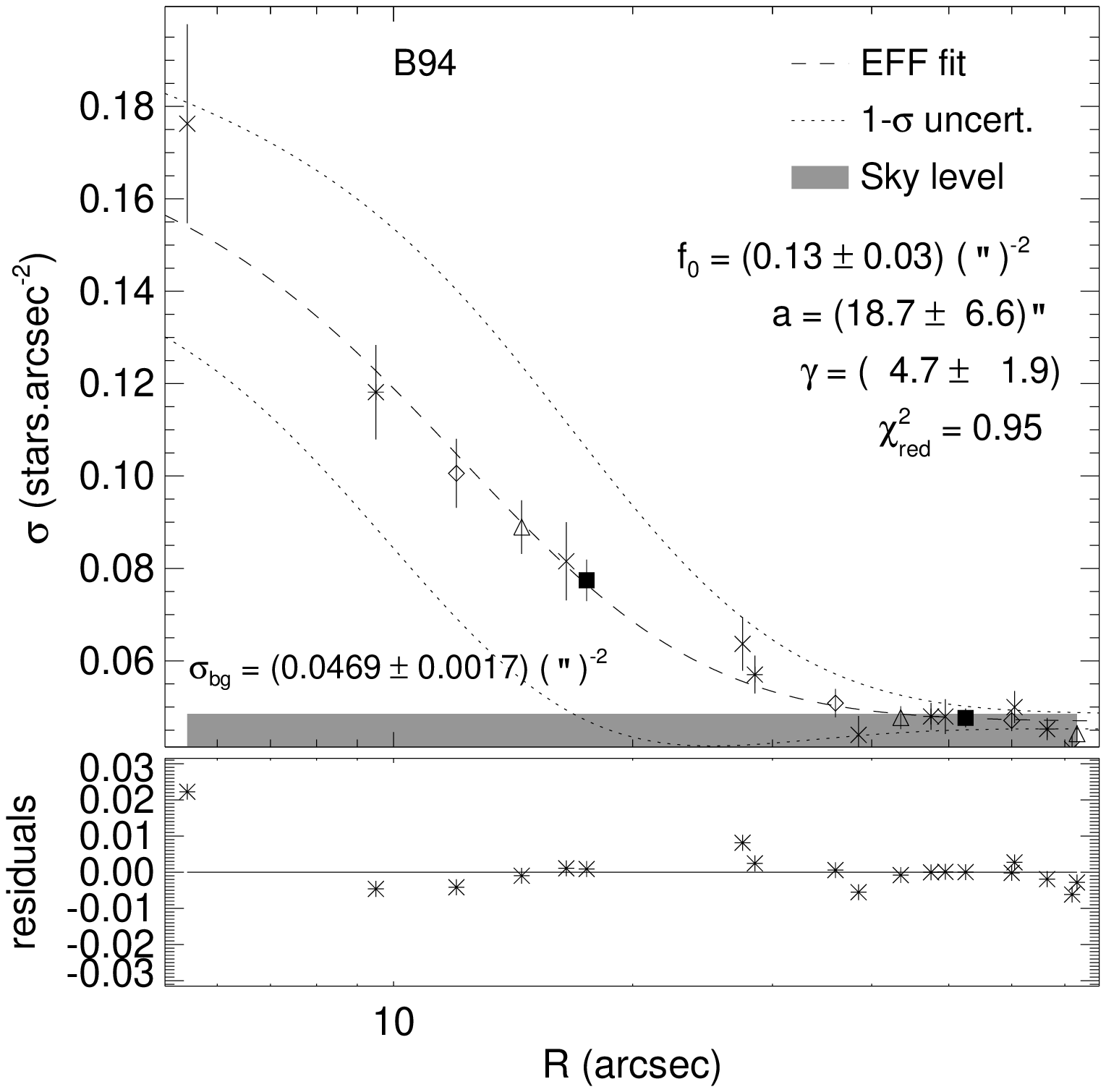}\includegraphics[width=0.325\linewidth]{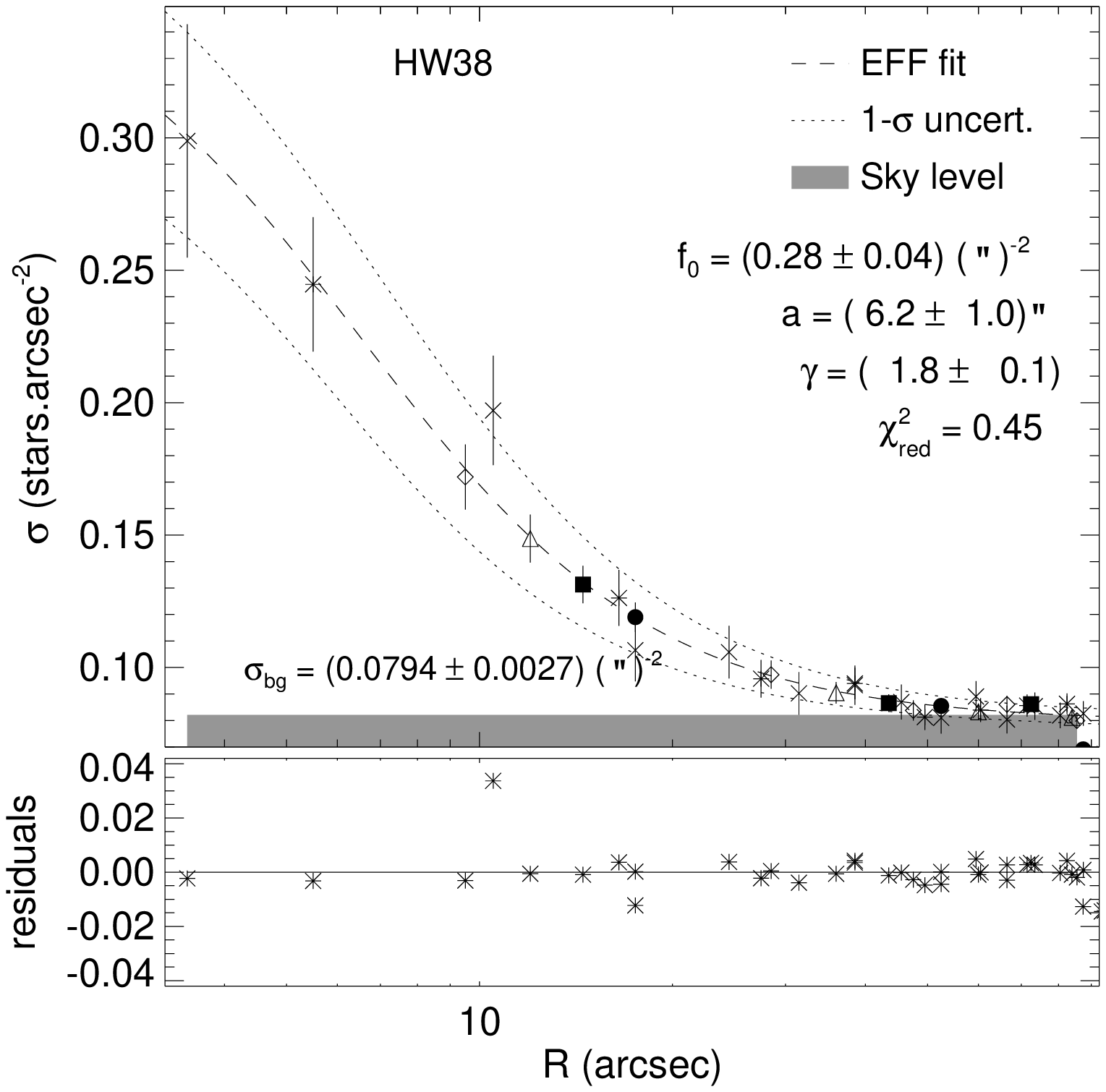}\includegraphics[width=0.325\linewidth]{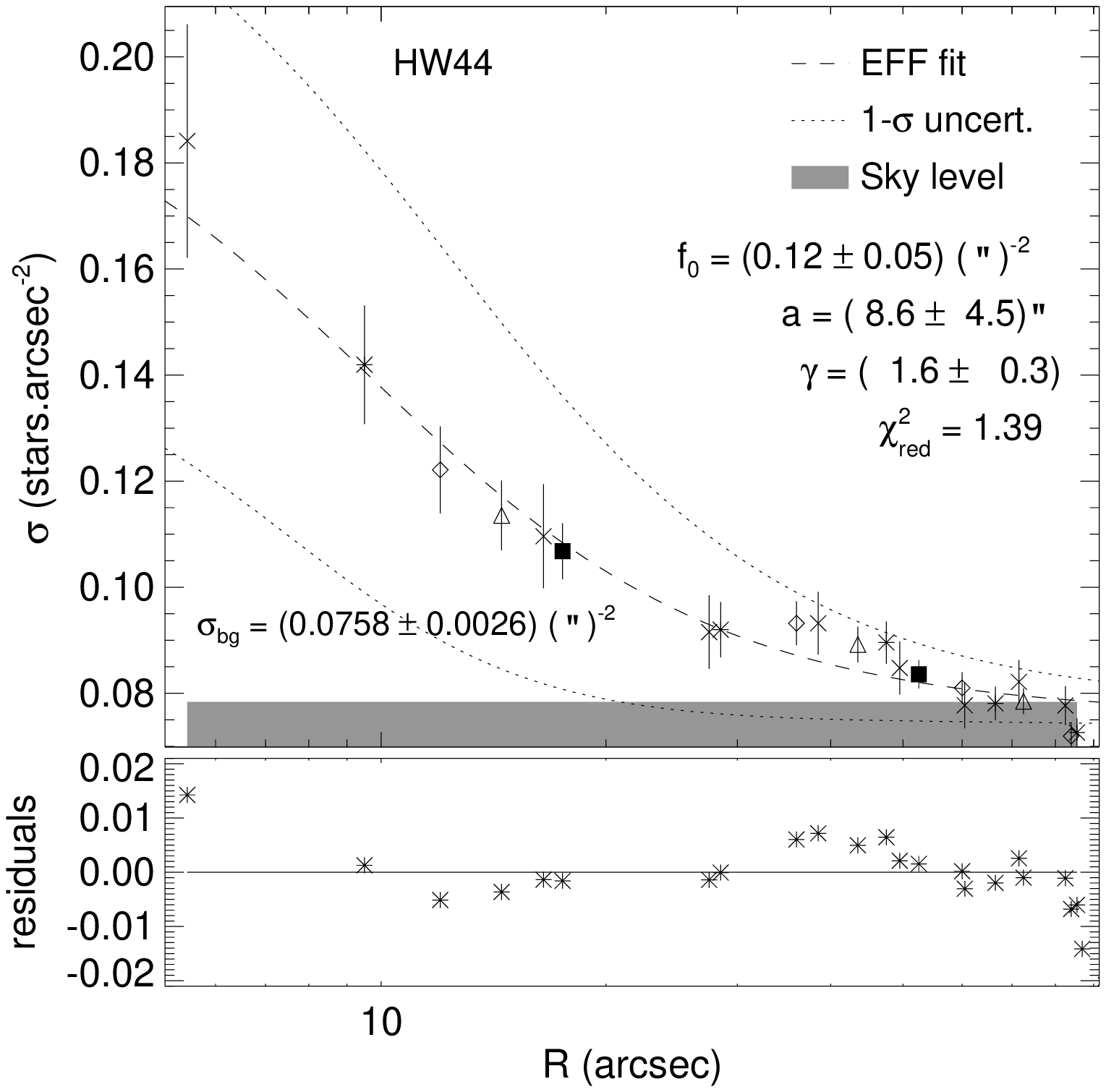}

\includegraphics[width=0.325\linewidth]{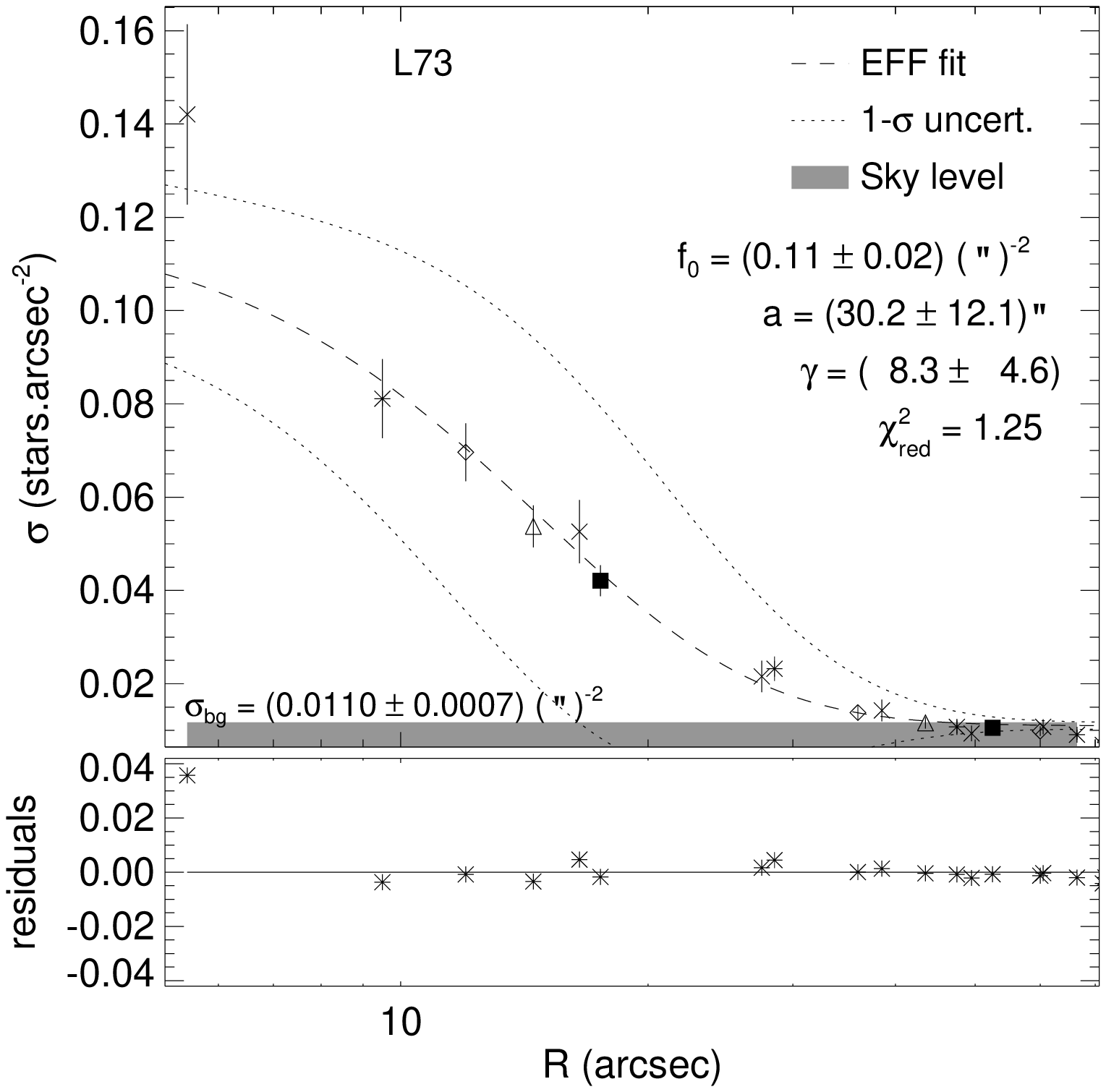}\includegraphics[width=0.325\linewidth]{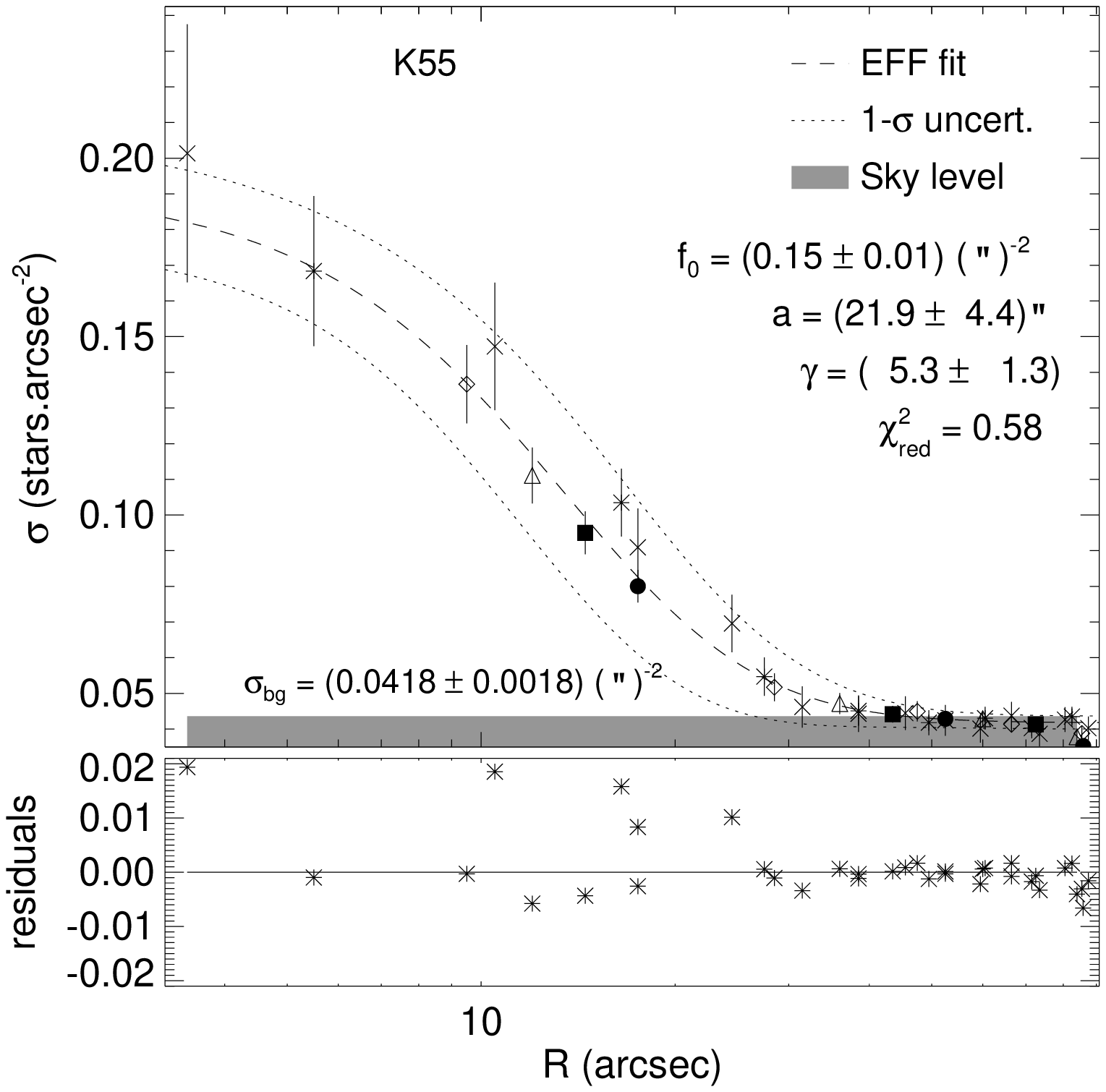}\includegraphics[width=0.325\linewidth]{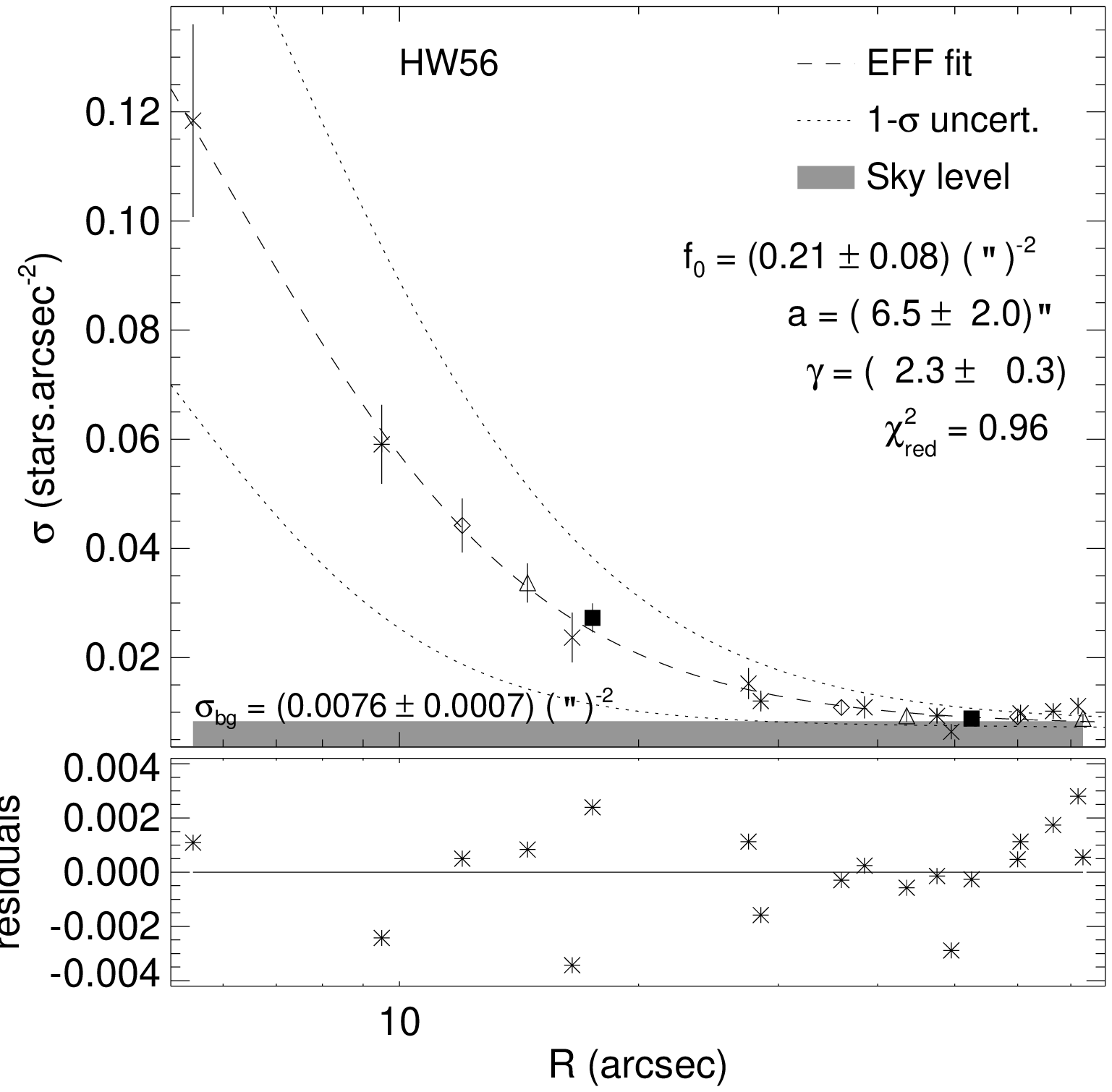}

\caption{Radial density profiles for additional SMC clusters complementing the sample presented in Fig.~\ref{fig:rdp_sbp}. The EFF model fits (dashed line) with envelopes of 1\,$\sigma$ uncertainty (dotted lines) are shown. Different symbols correspond to the various widths of the annular bins employed. The fitting residuals are also presented in the lower panel.}
\end{figure*}

\setcounter{figure}{4}

\begin{figure*}

\includegraphics[width=0.325\linewidth]{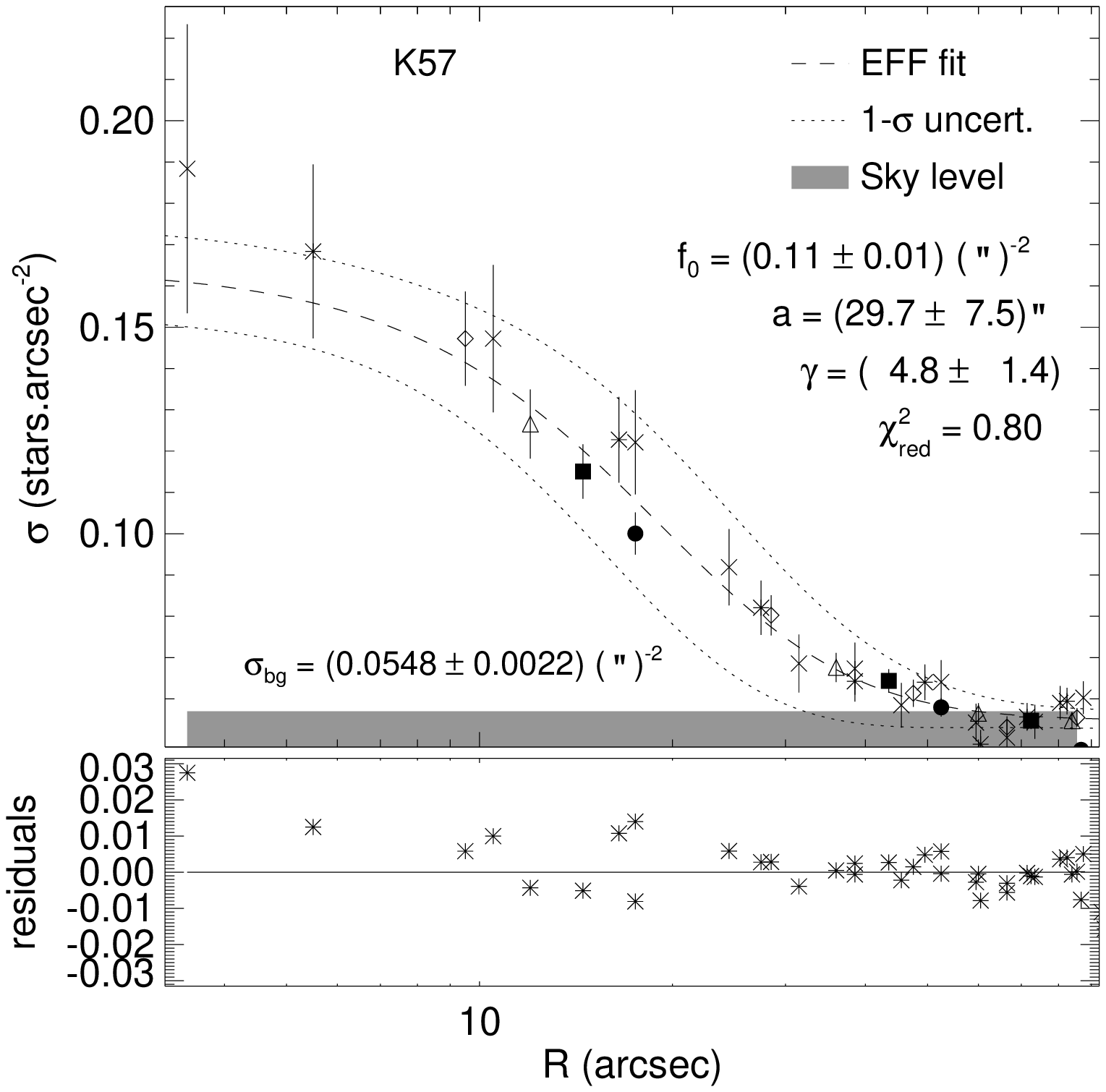}\includegraphics[width=0.325\linewidth]{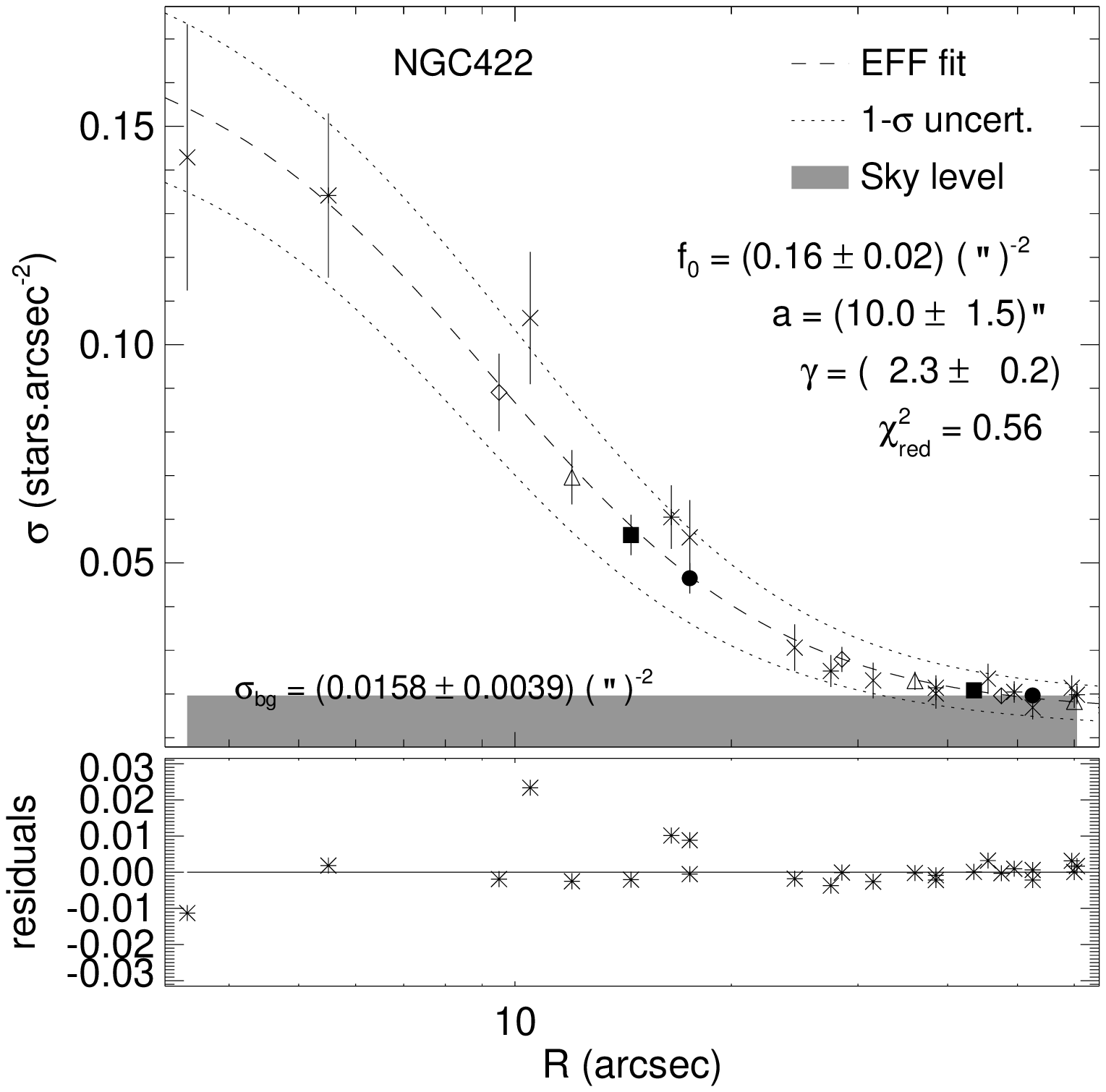}\includegraphics[width=0.325\linewidth]{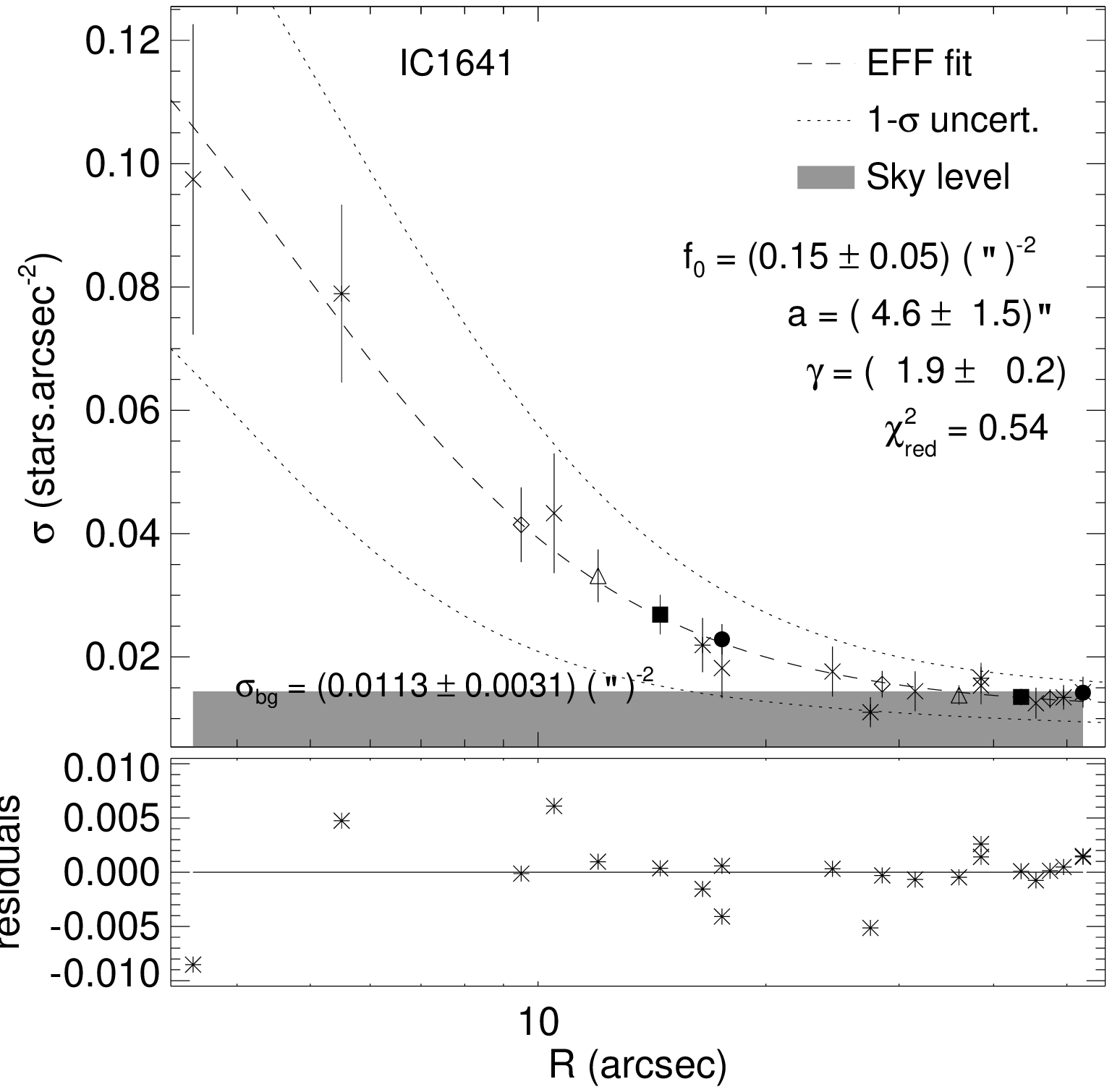}

\includegraphics[width=0.325\linewidth]{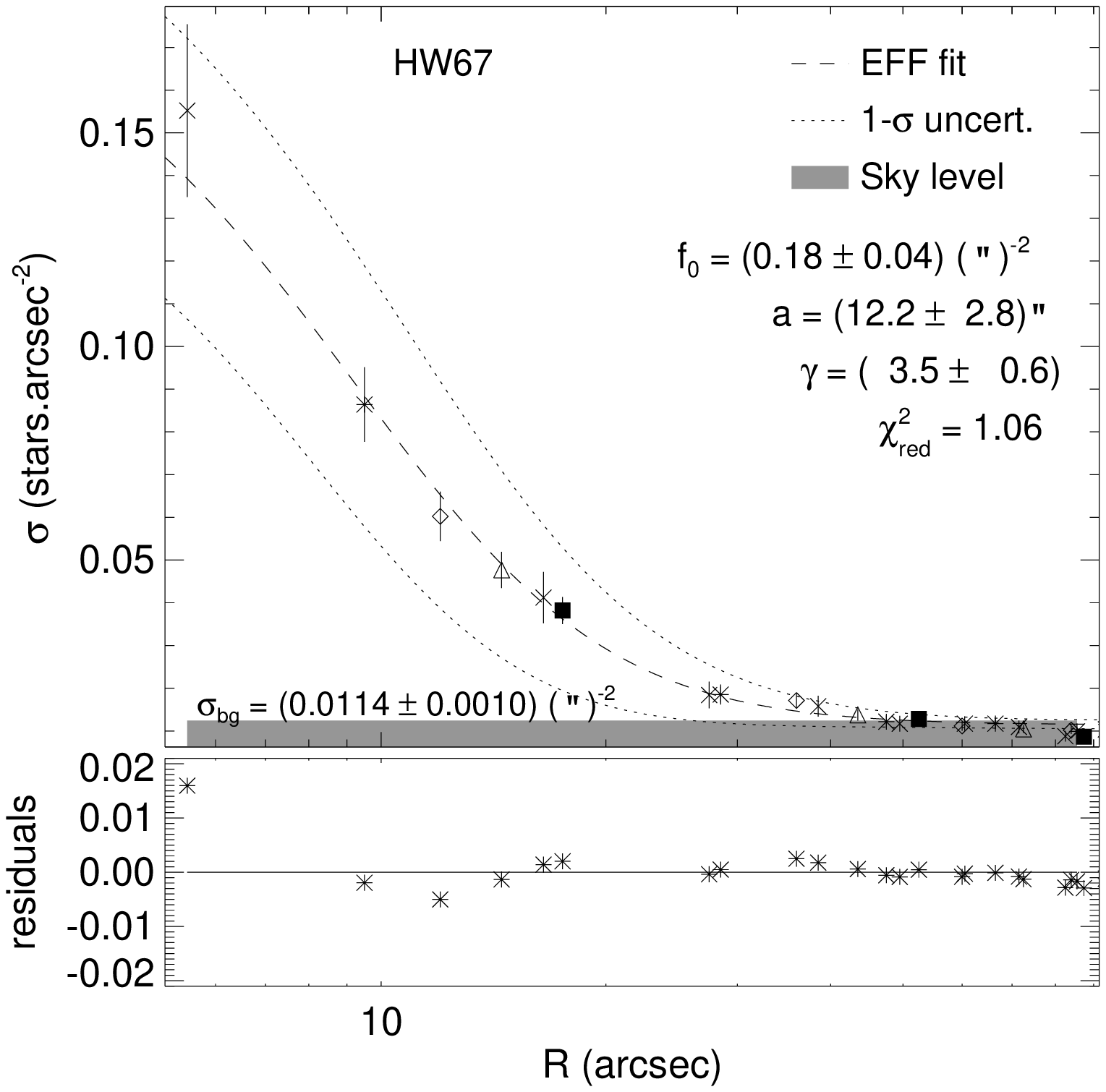}\includegraphics[width=0.325\linewidth]{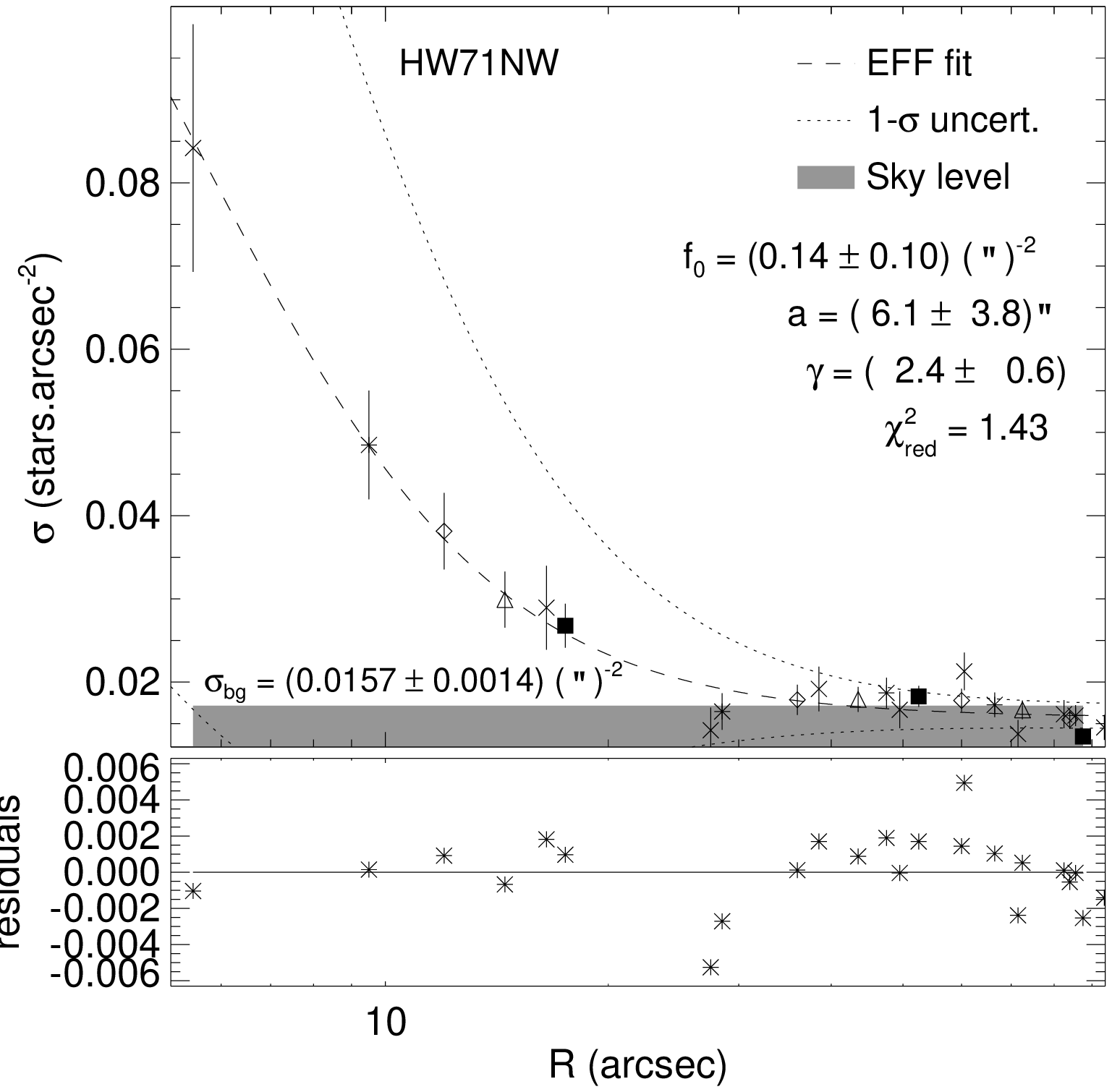}\includegraphics[width=0.325\linewidth]{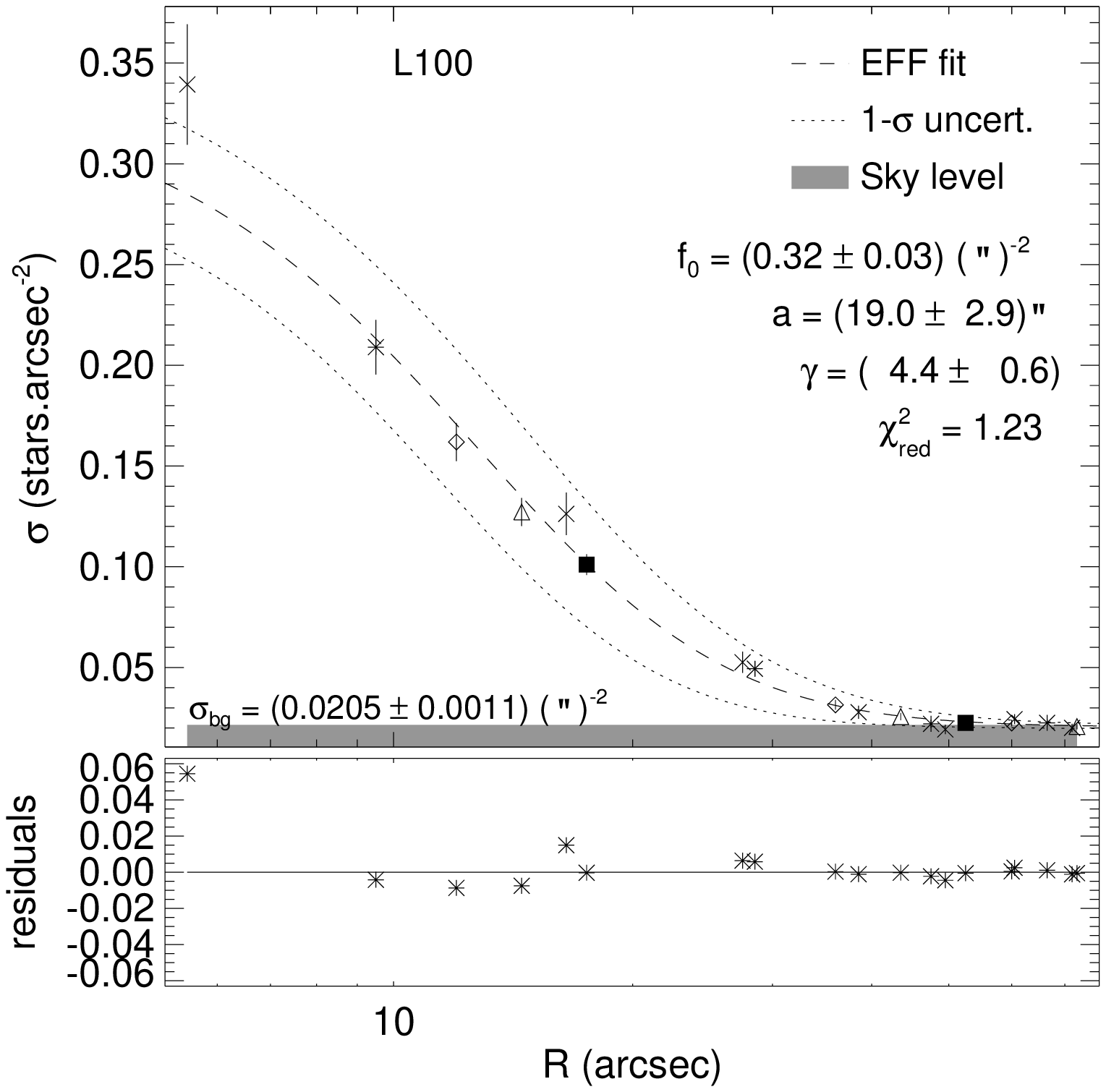}

\includegraphics[width=0.325\linewidth]{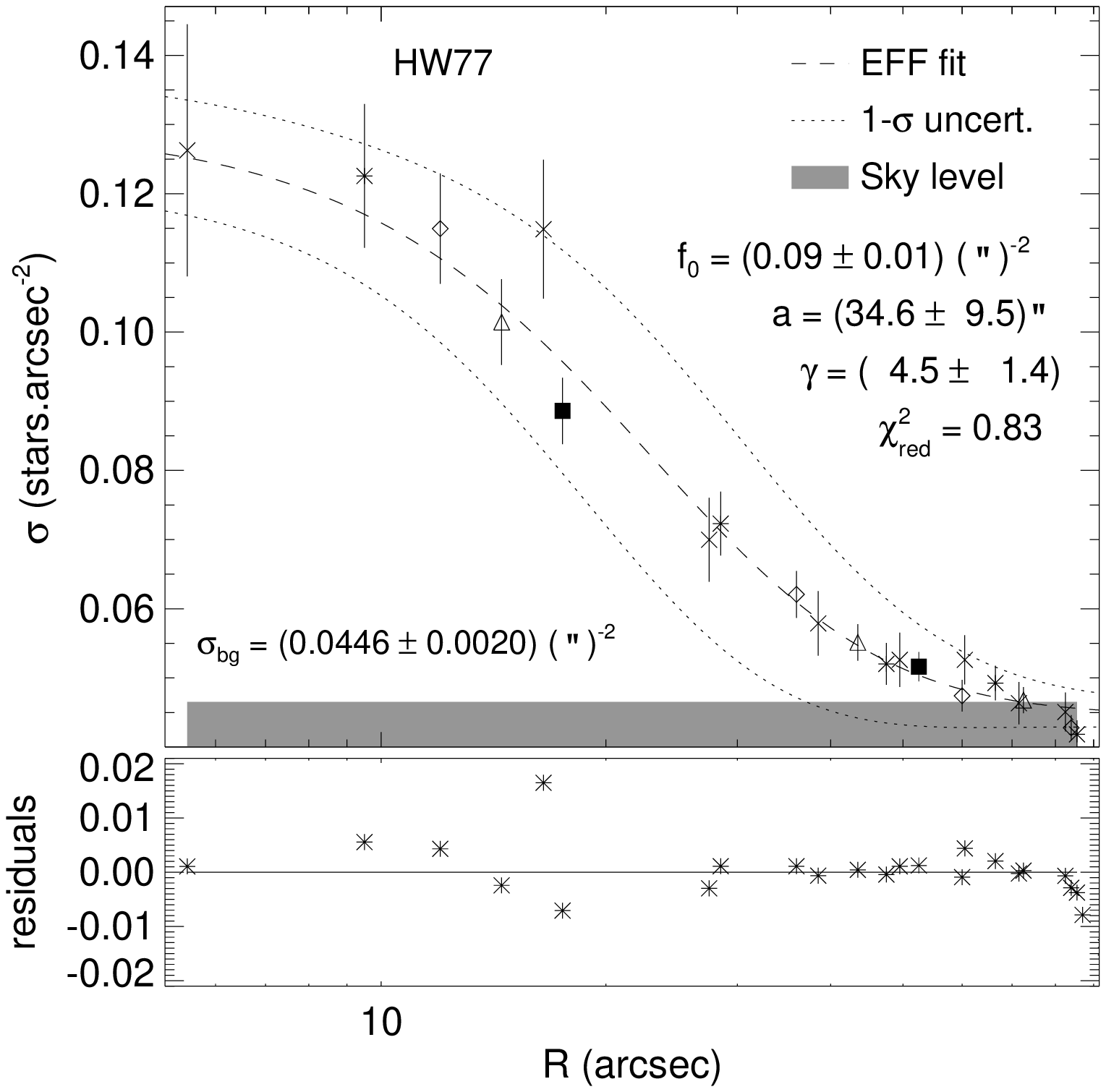}\includegraphics[width=0.325\linewidth]{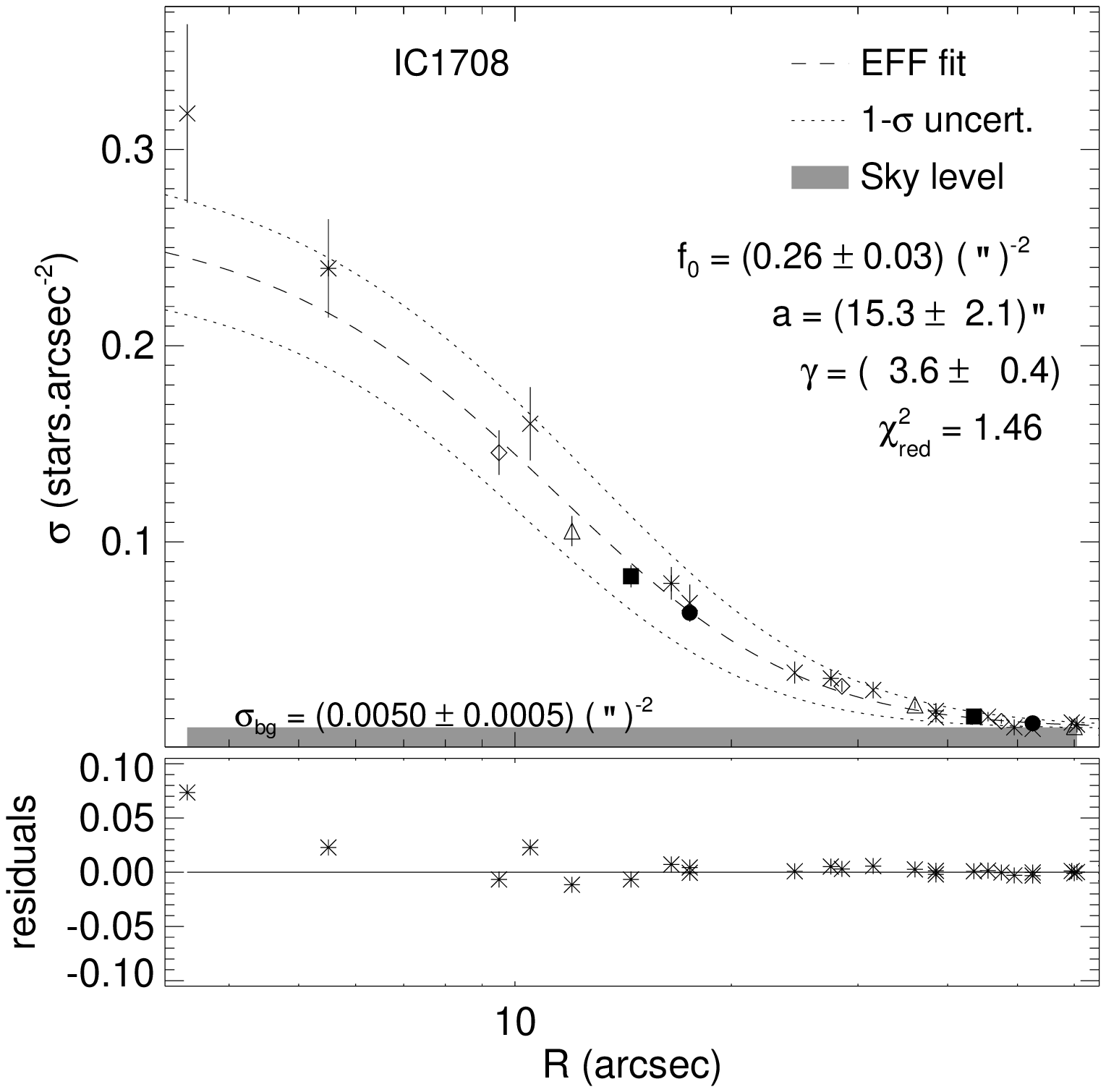}\includegraphics[width=0.325\linewidth]{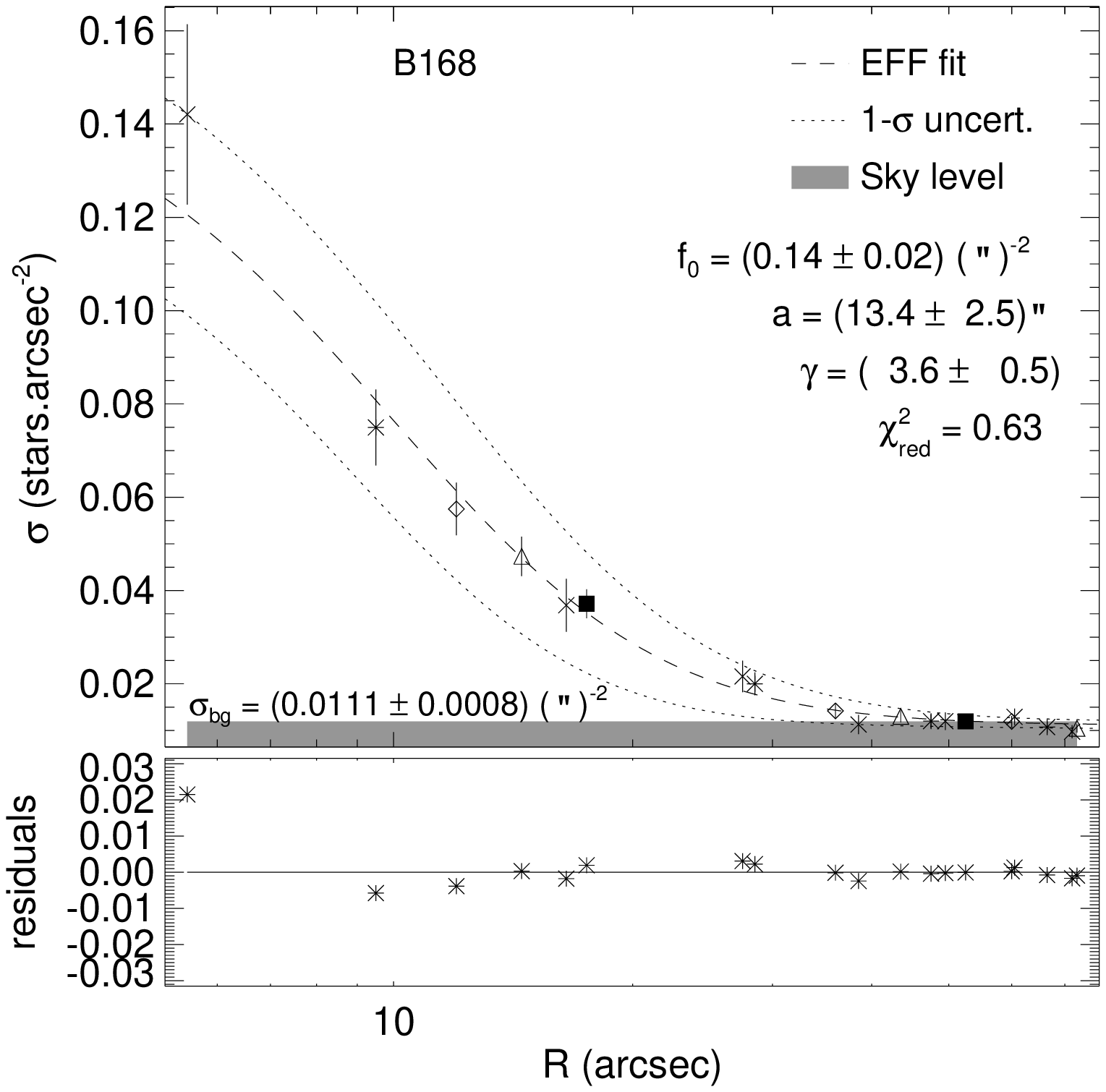}

\includegraphics[width=0.325\linewidth]{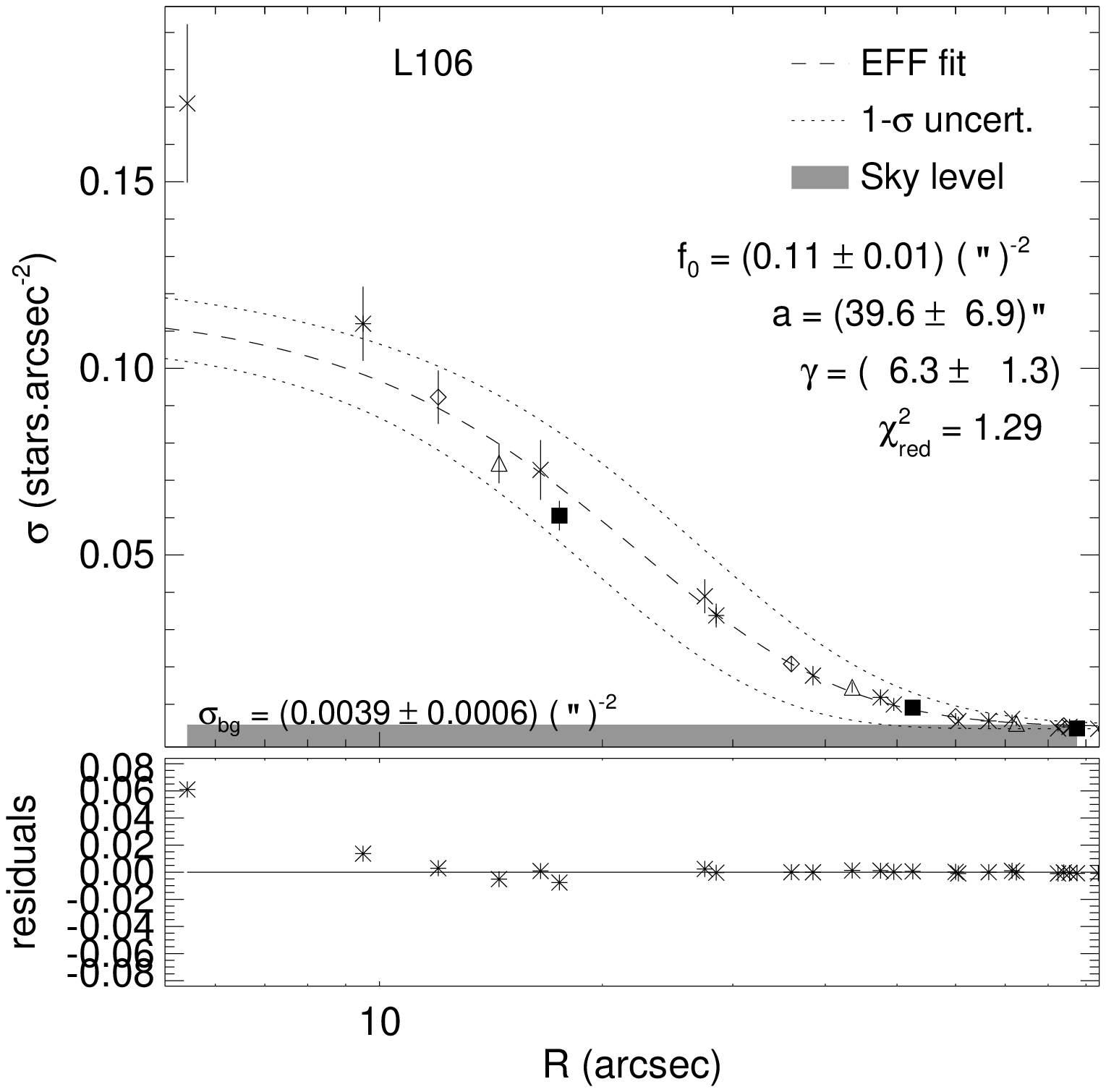}\includegraphics[width=0.325\linewidth]{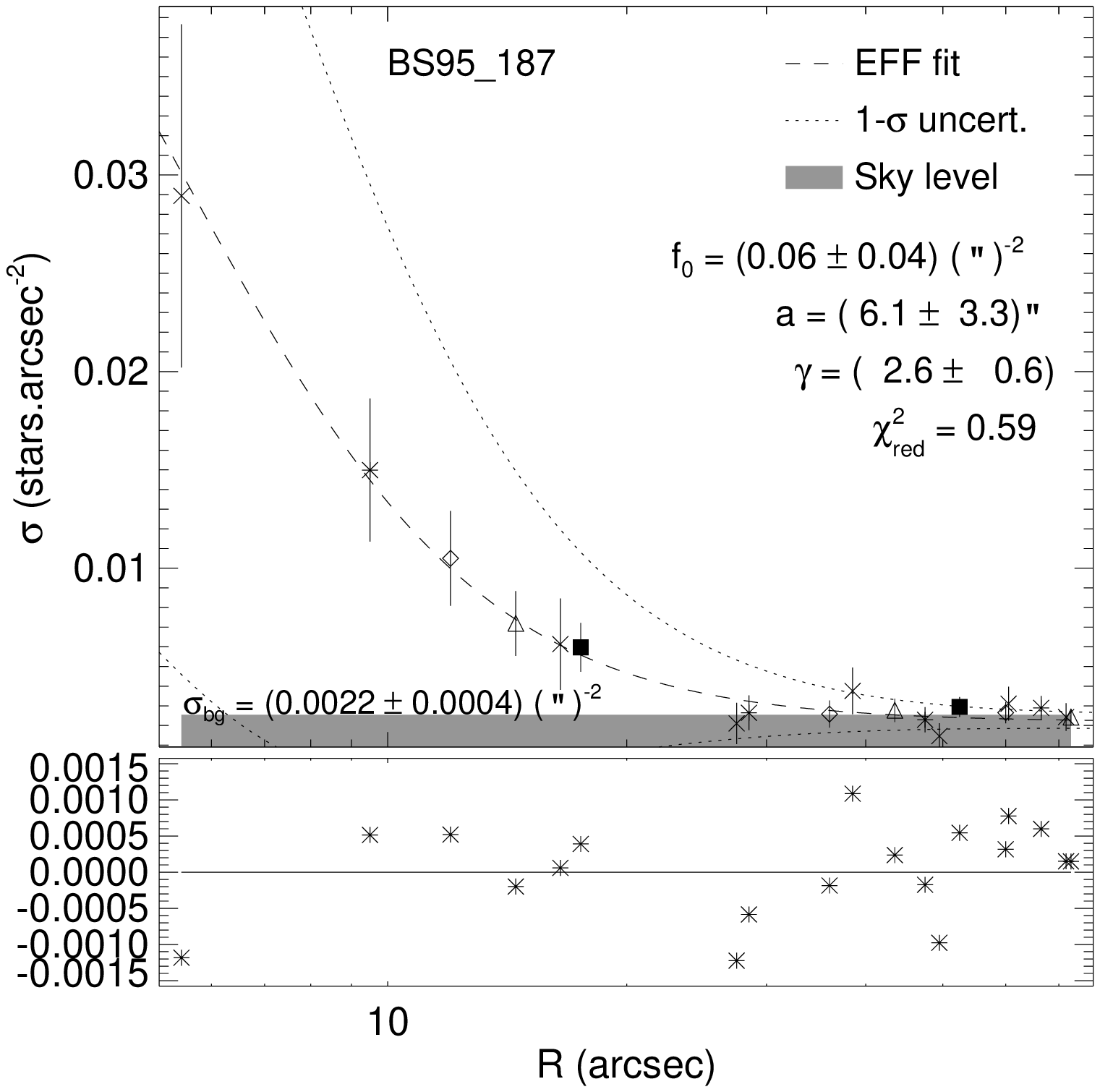}\includegraphics[width=0.325\linewidth]{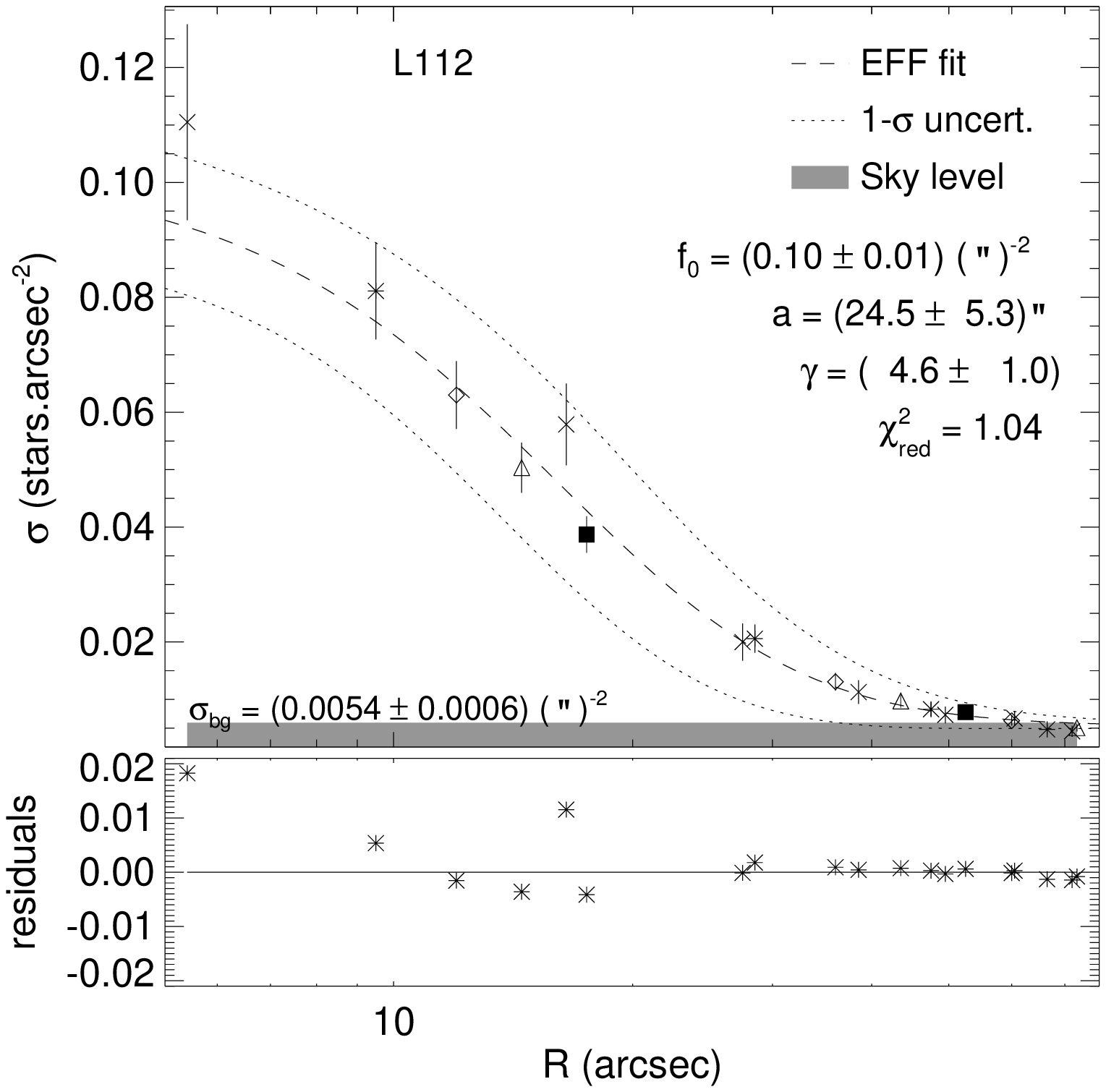}

\caption{cont.}

\end{figure*}

\setcounter{figure}{4}

\begin{figure*}
\includegraphics[width=0.325\linewidth]{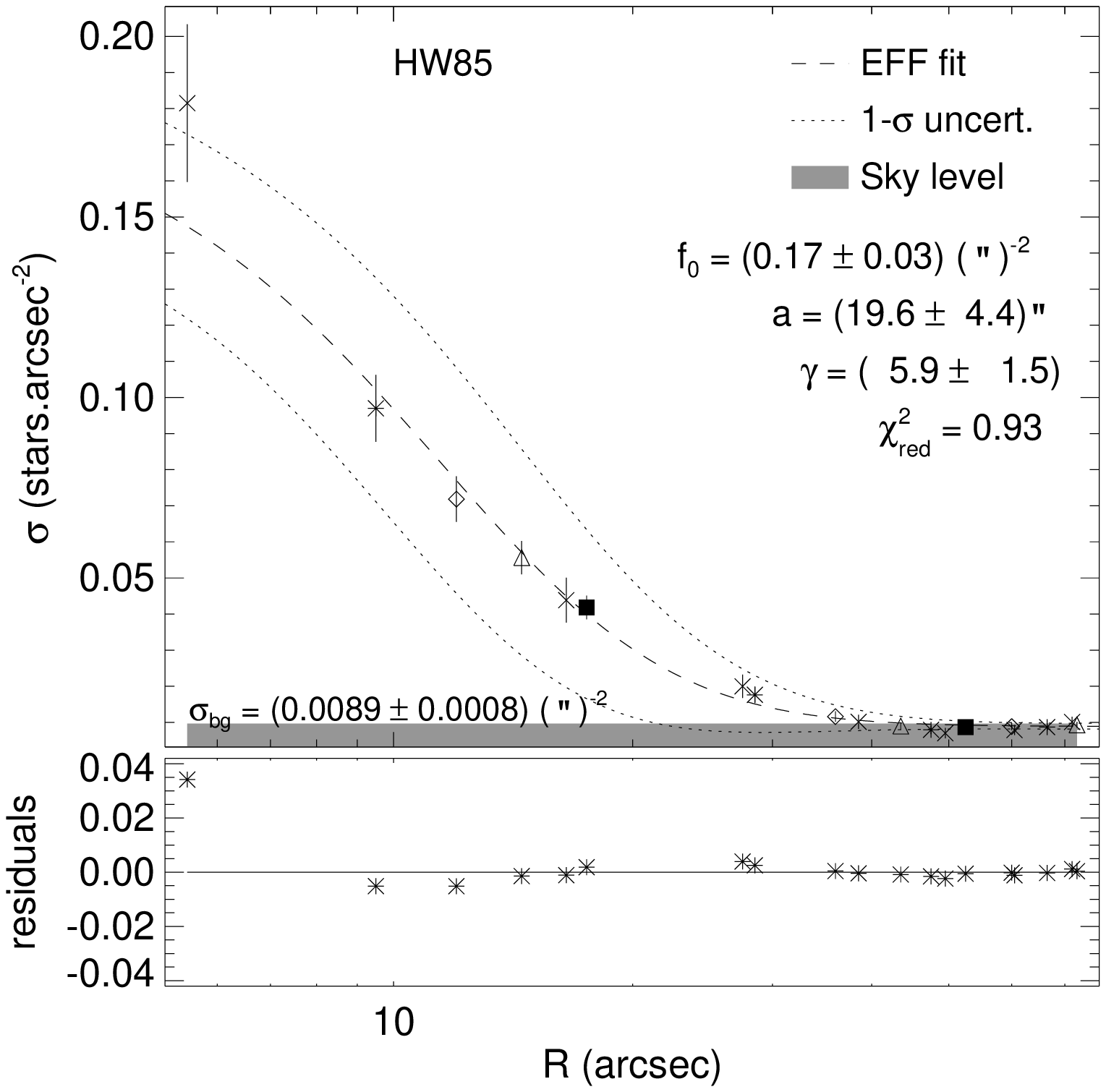}\includegraphics[width=0.325\linewidth]{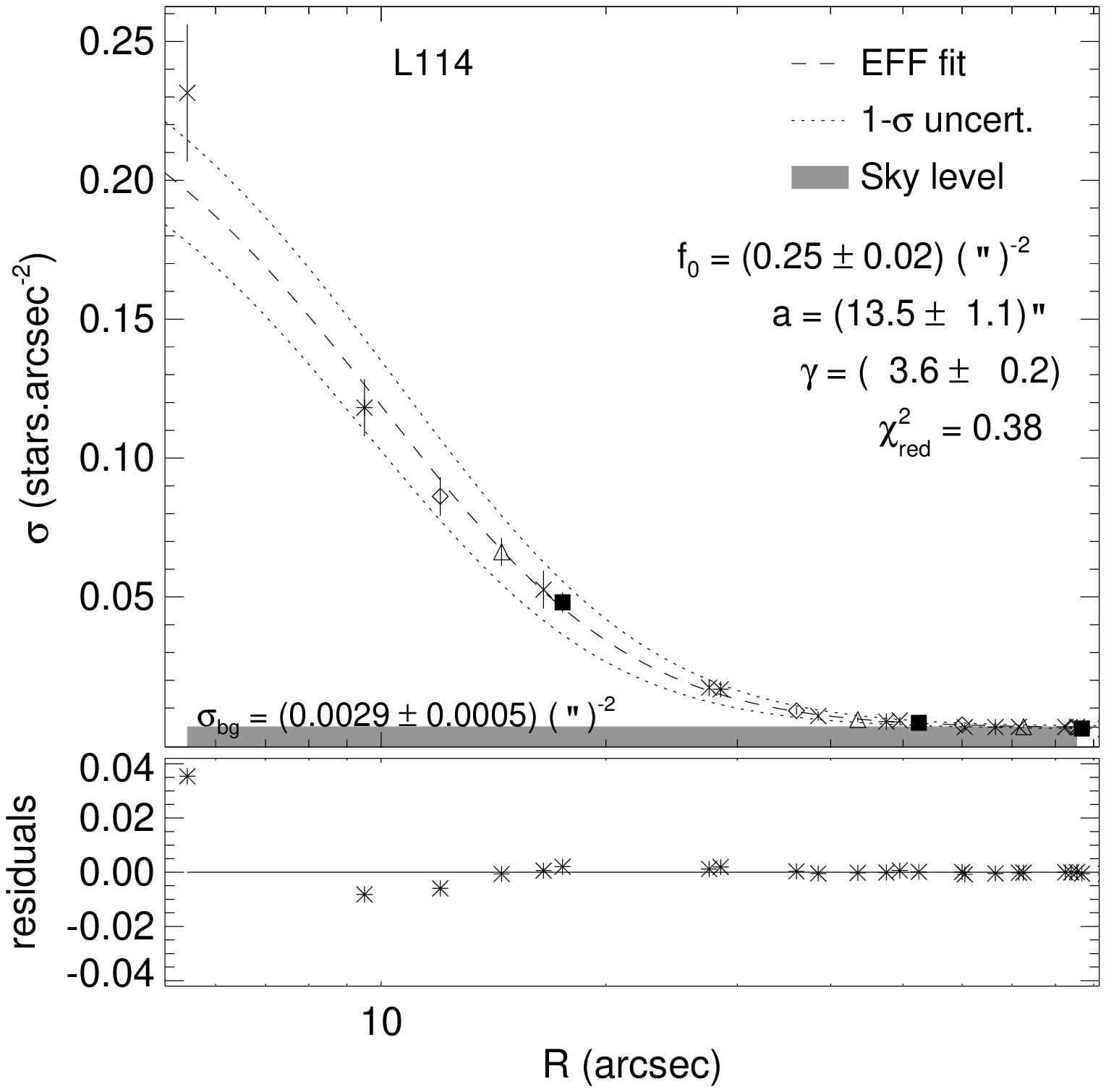}\includegraphics[width=0.325\linewidth]{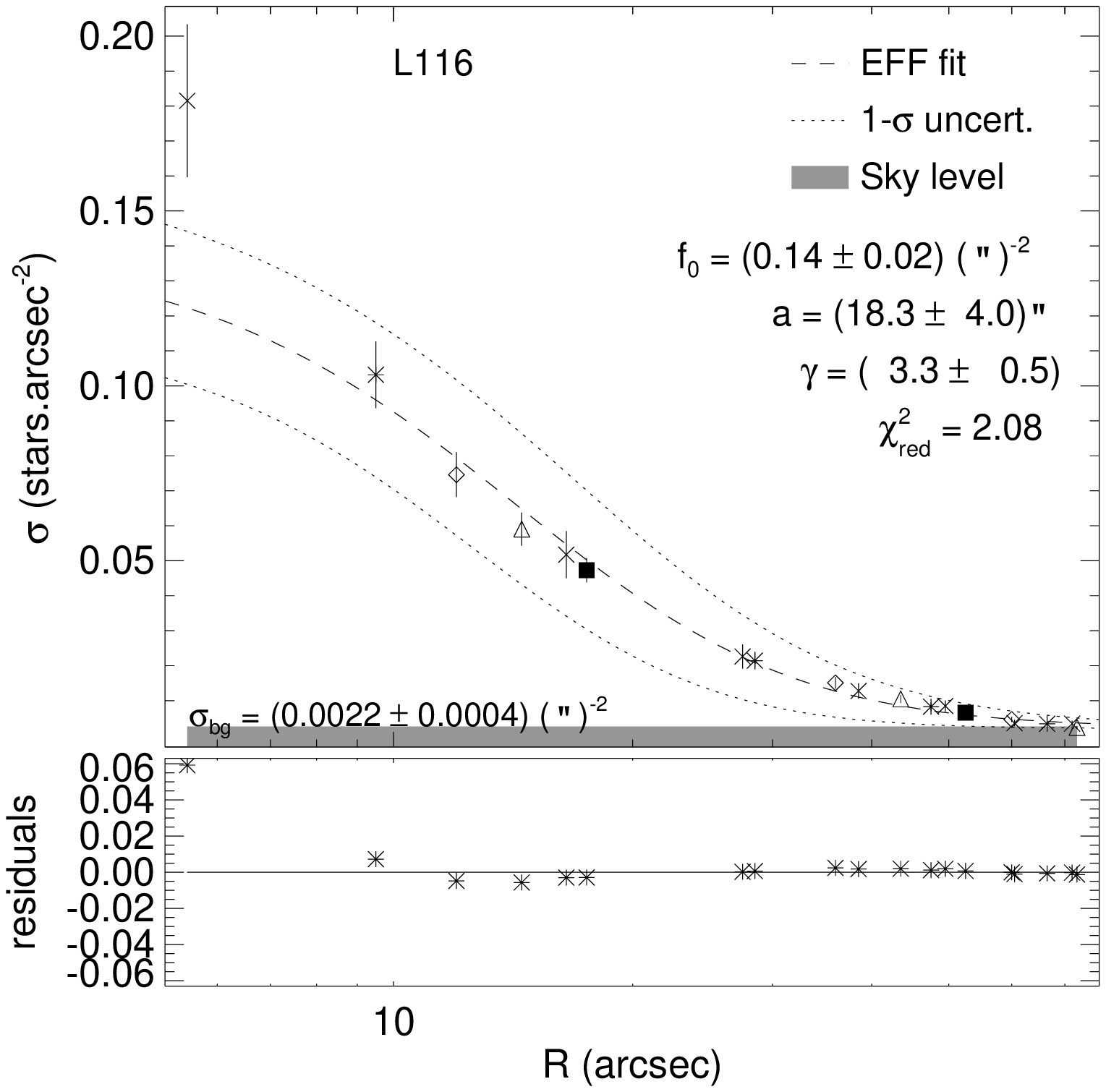}

\includegraphics[width=0.325\linewidth]{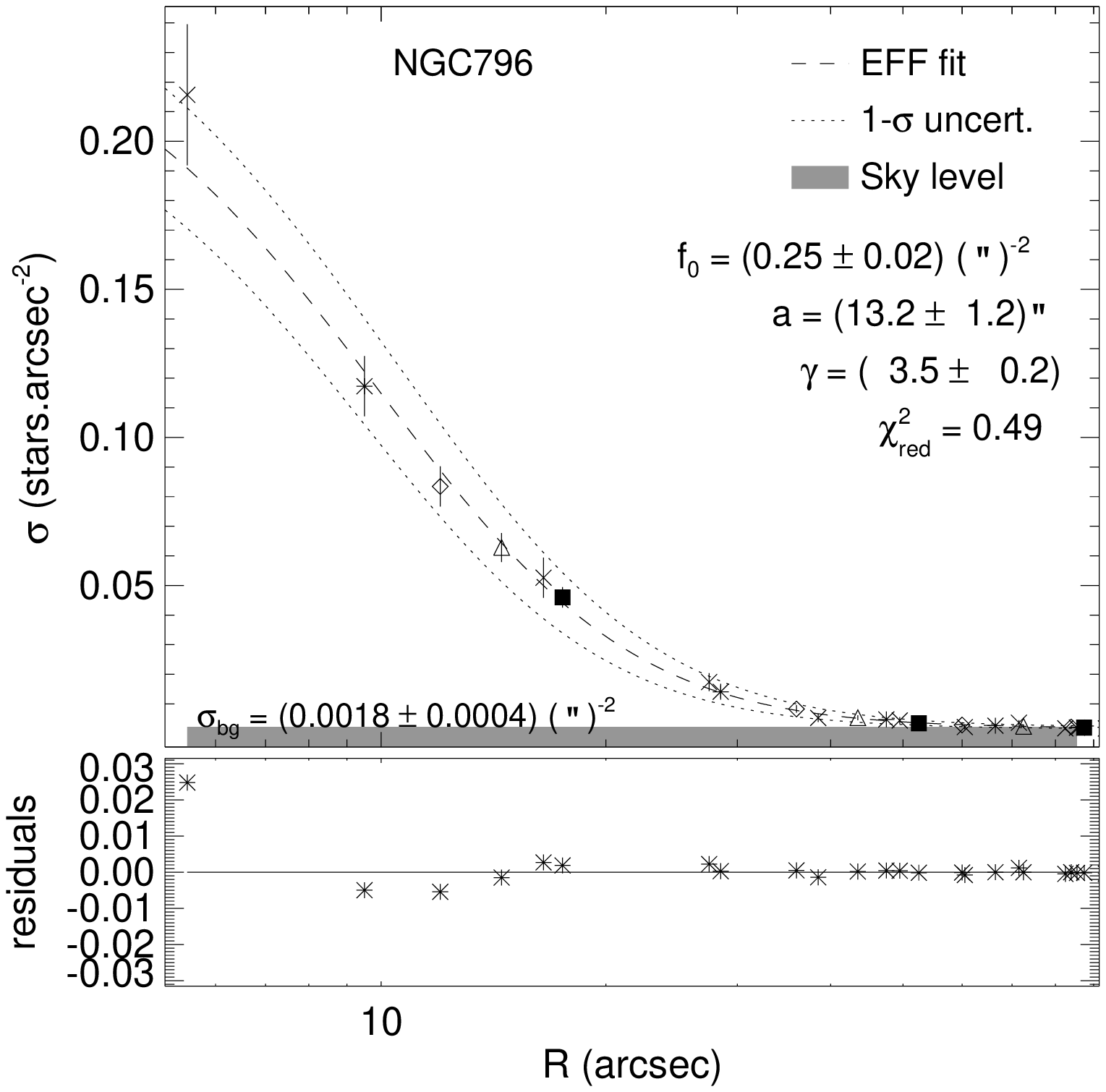}\includegraphics[width=0.325\linewidth]{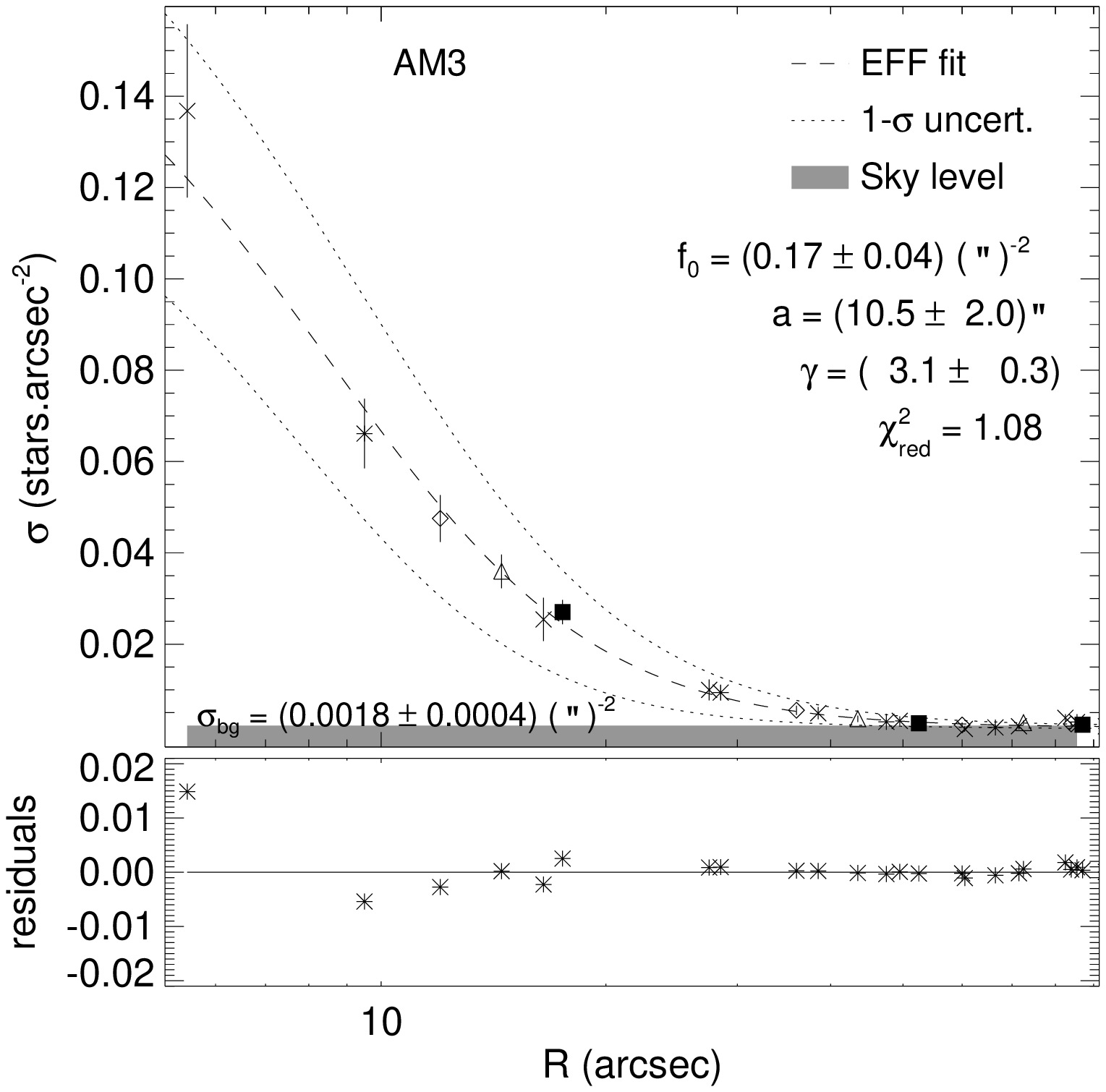}
\caption{cont.}

\end{figure*}

\begin{figure*}

\includegraphics[width=0.325\linewidth]{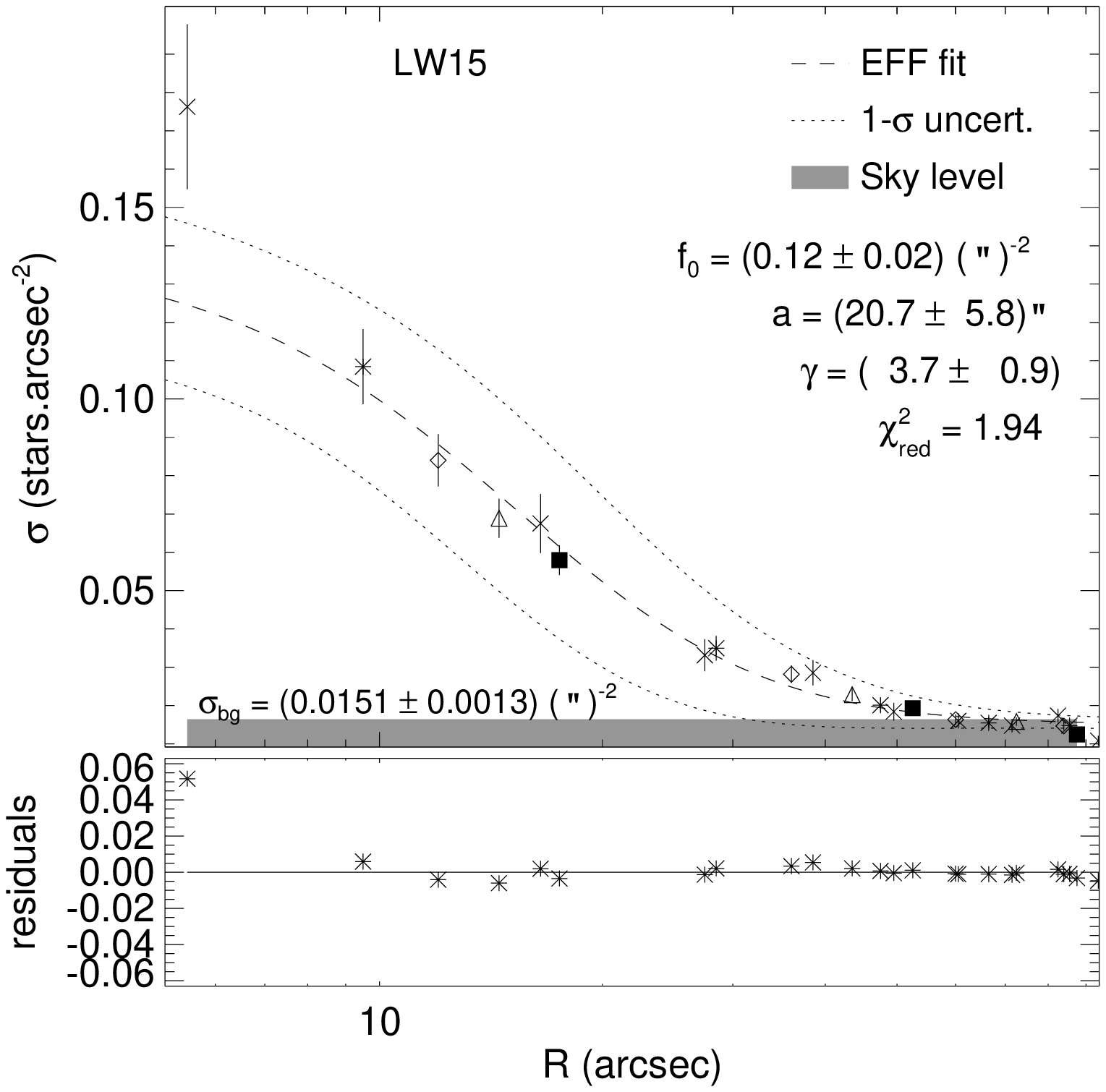}\includegraphics[width=0.325\linewidth]{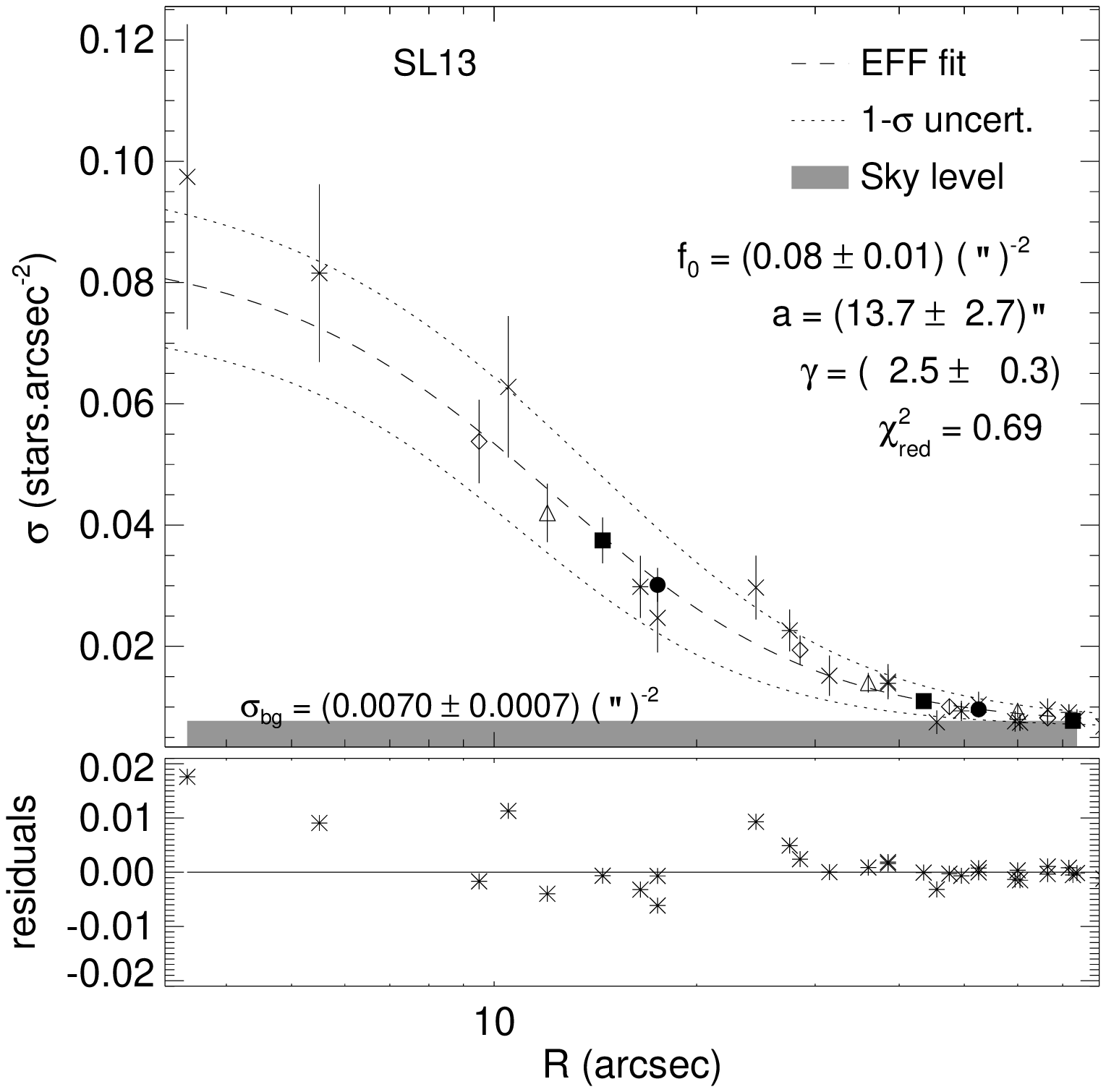}\includegraphics[width=0.325\linewidth]{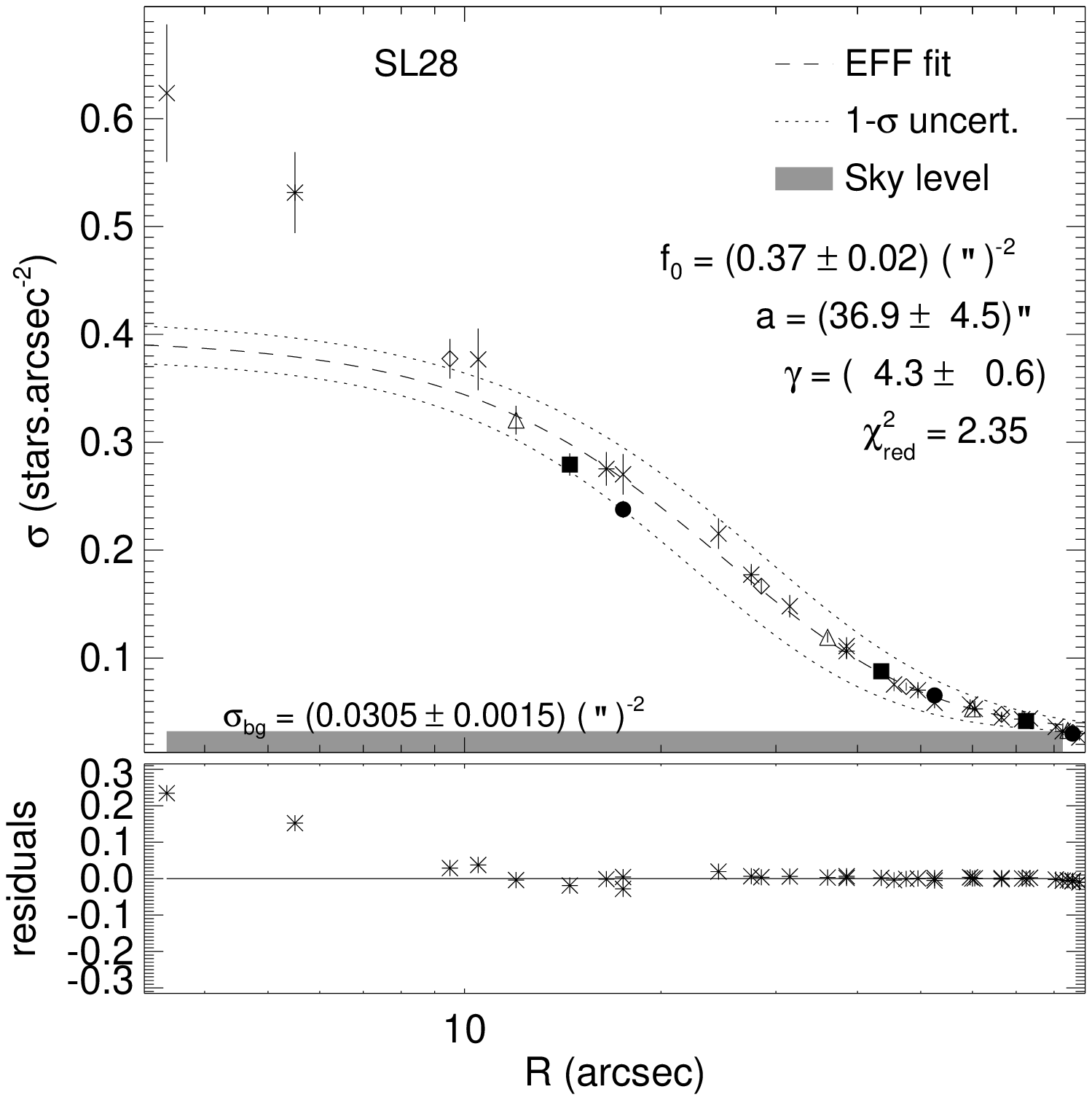}

\includegraphics[width=0.325\linewidth]{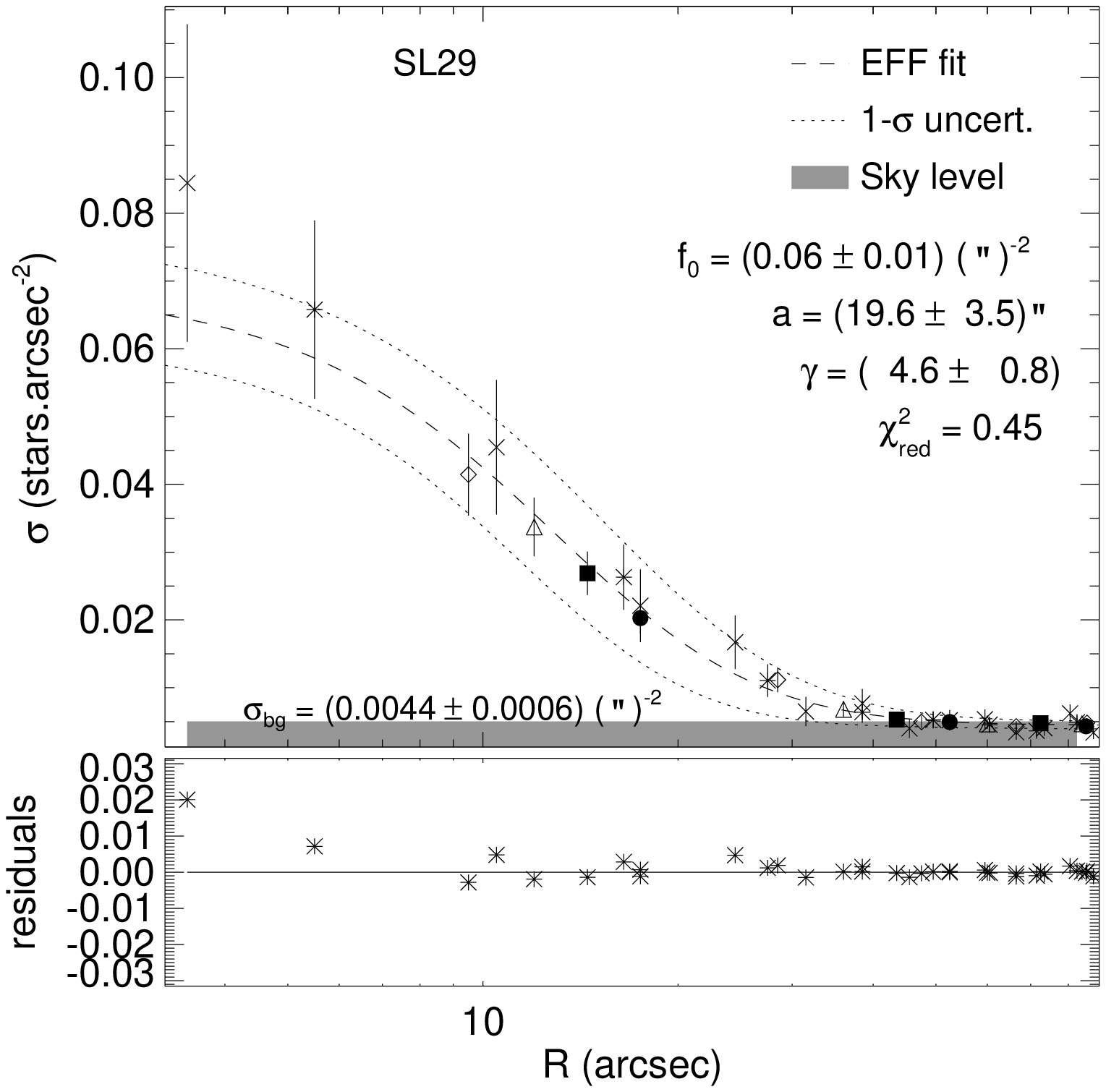}\includegraphics[width=0.325\linewidth]{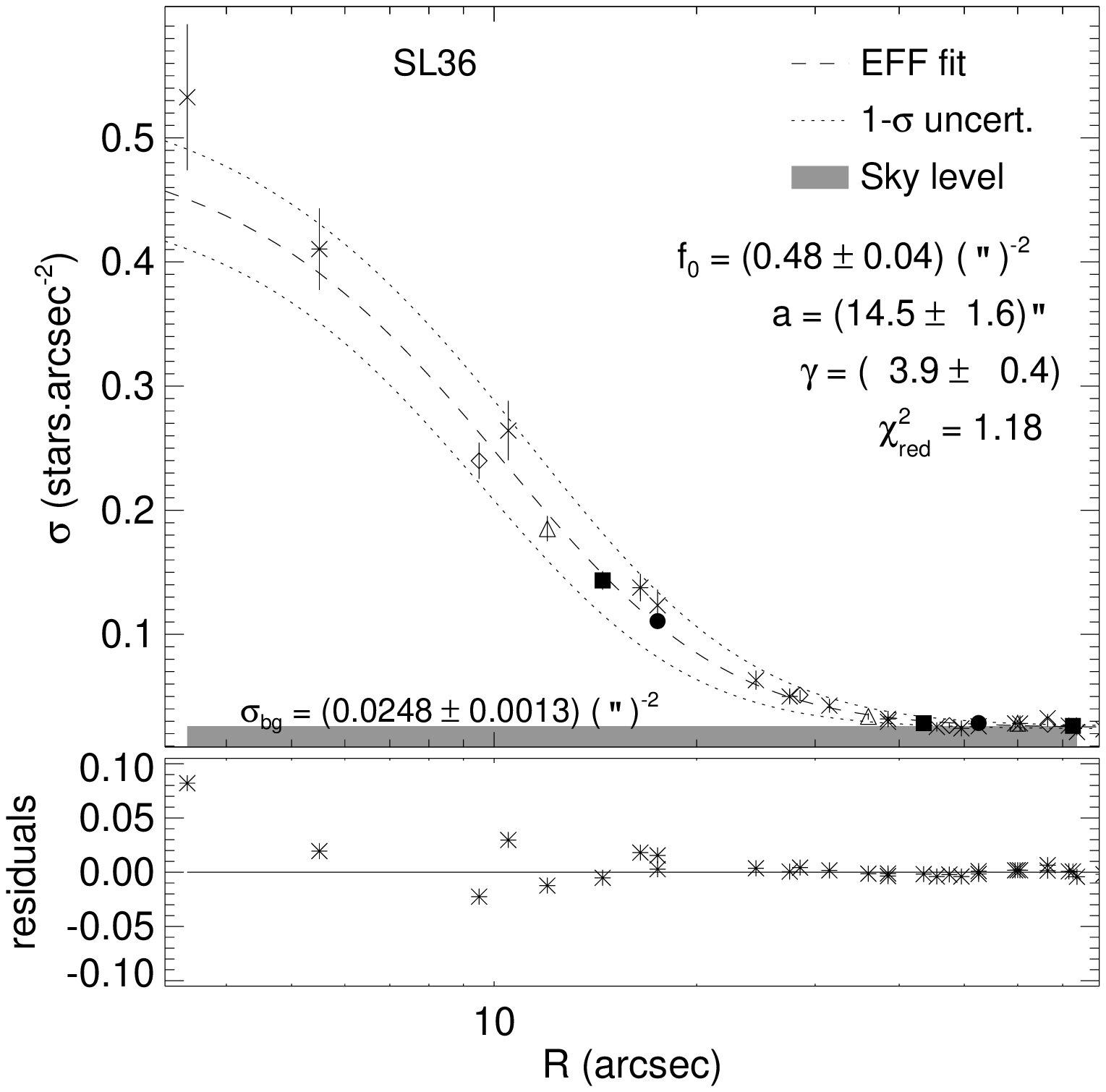}\includegraphics[width=0.325\linewidth]{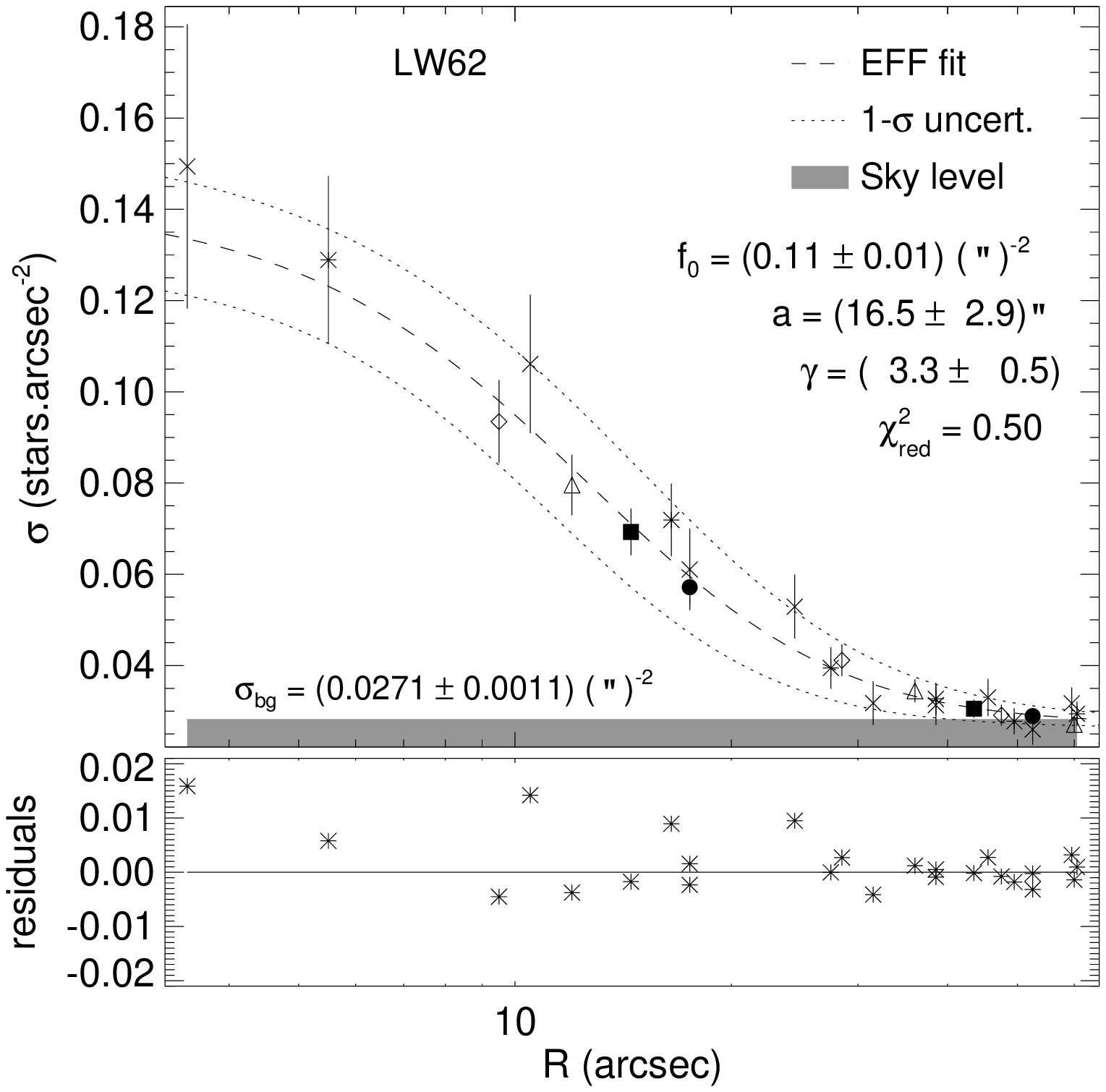}

\includegraphics[width=0.325\linewidth]{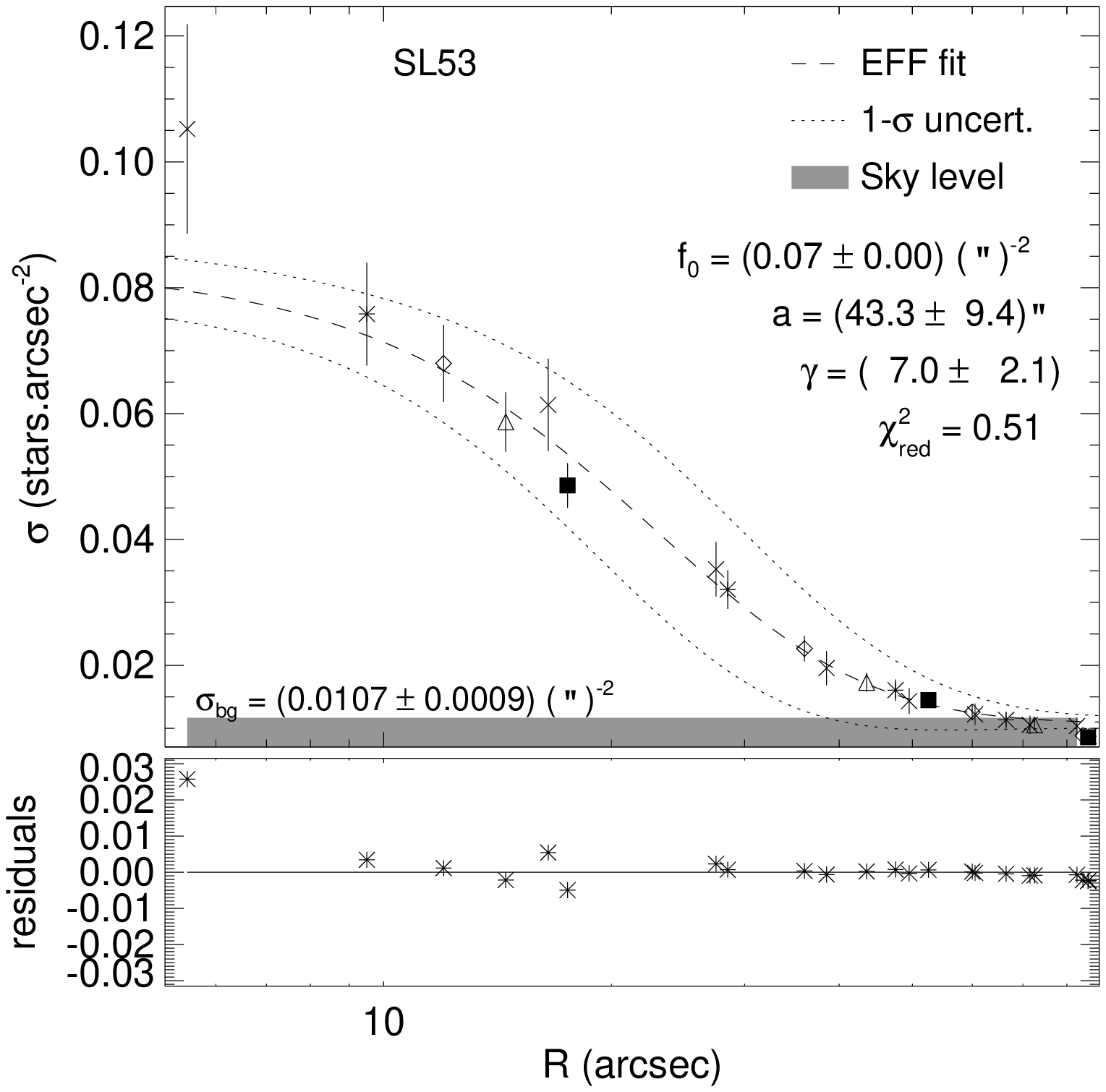}\includegraphics[width=0.325\linewidth]{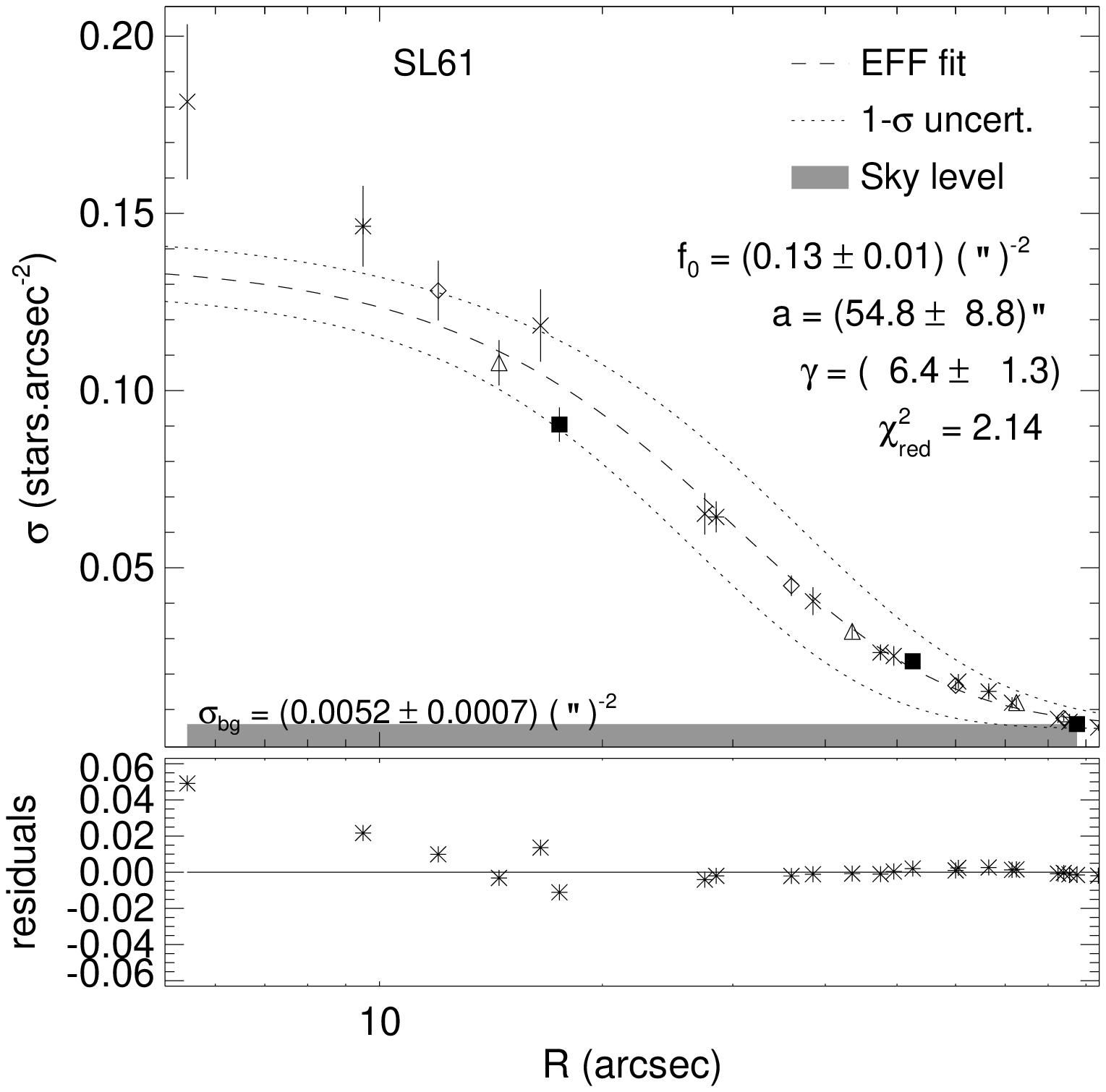}\includegraphics[width=0.325\linewidth]{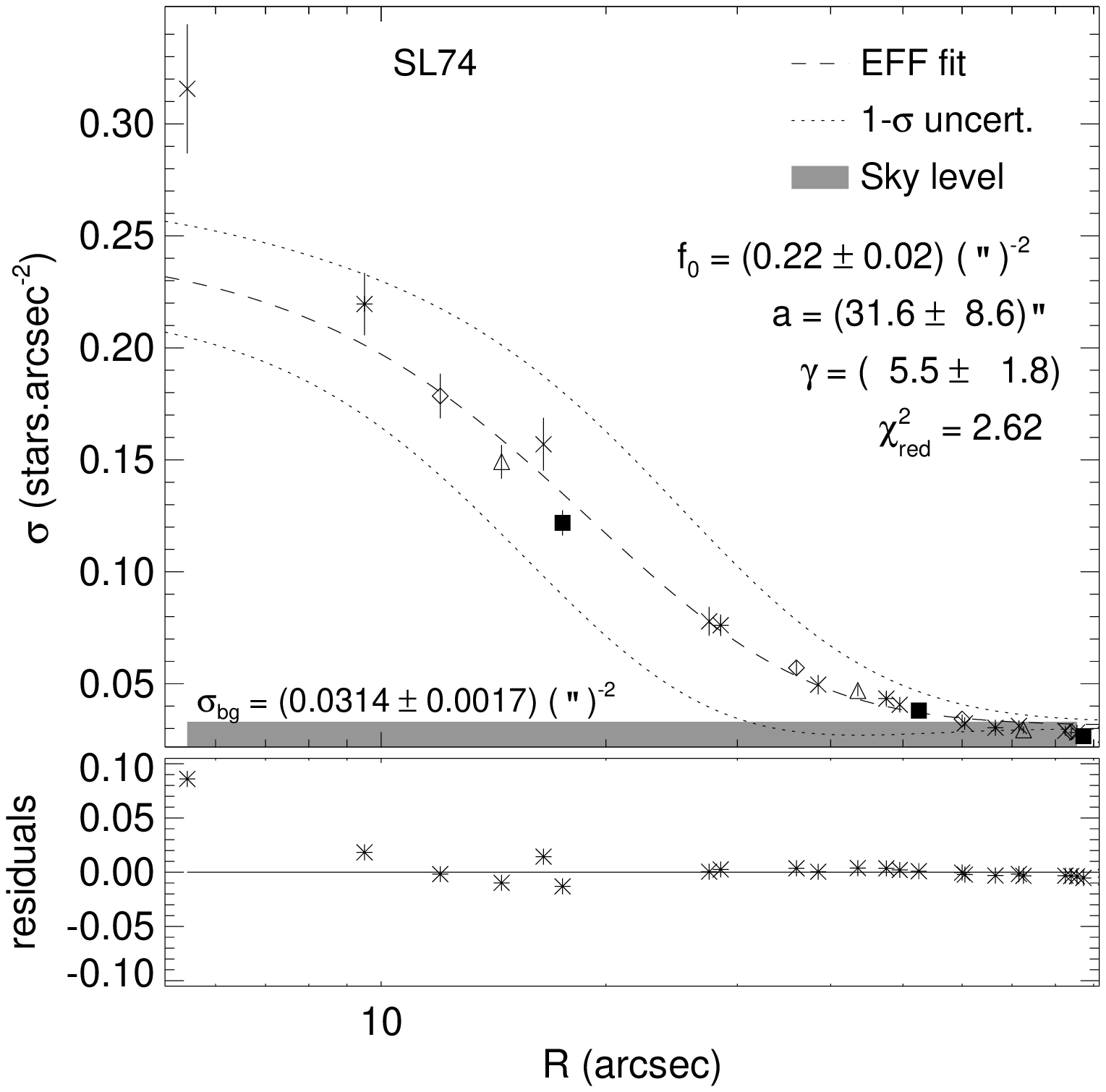}

\includegraphics[width=0.325\linewidth]{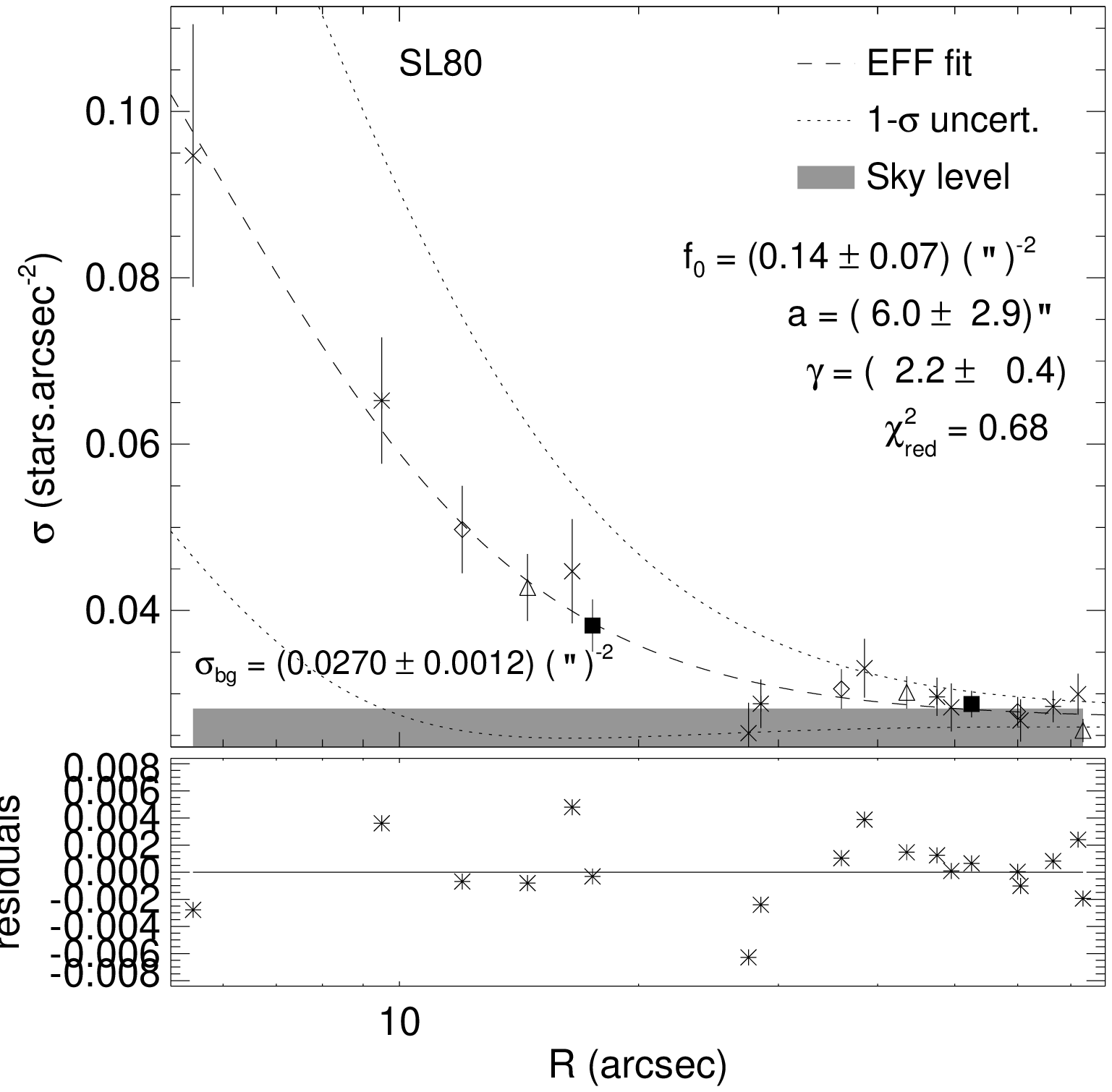}\includegraphics[width=0.325\linewidth]{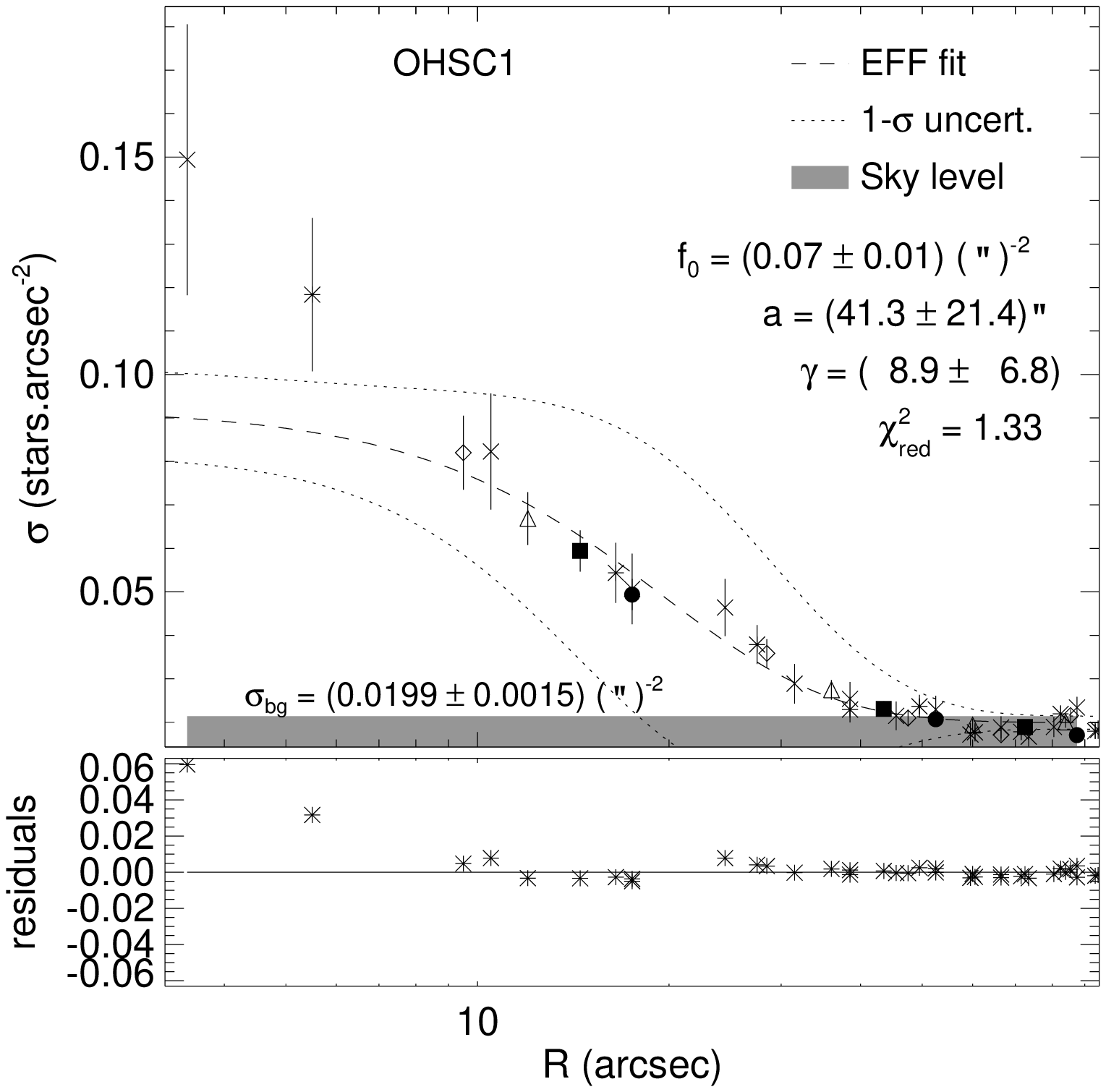}\includegraphics[width=0.325\linewidth]{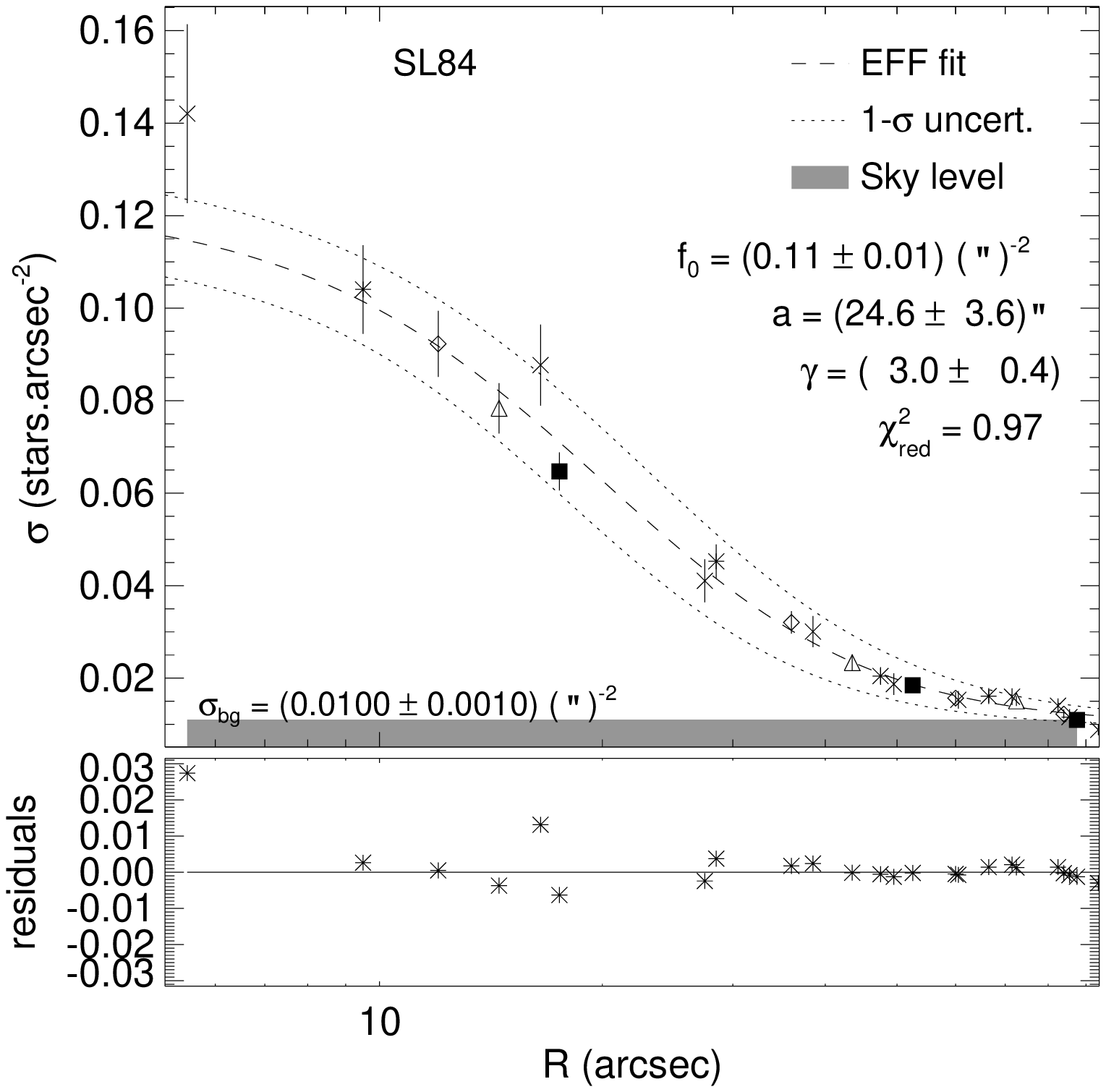}

\caption{Radial density profiles for additional LMC clusters complementing the sample presented in Fig.~\ref{fig:rdp_sbp}. The EFF model fits (dashed line) with envelopes of 1\,$\sigma$ uncertainty (dotted lines) are shown. Different symbols correspond to the various widths of the annular bins employed. The fitting residuals are also presented in the lower panel.}

\end{figure*}

\setcounter{figure}{5}
\begin{figure*}

\includegraphics[width=0.325\linewidth]{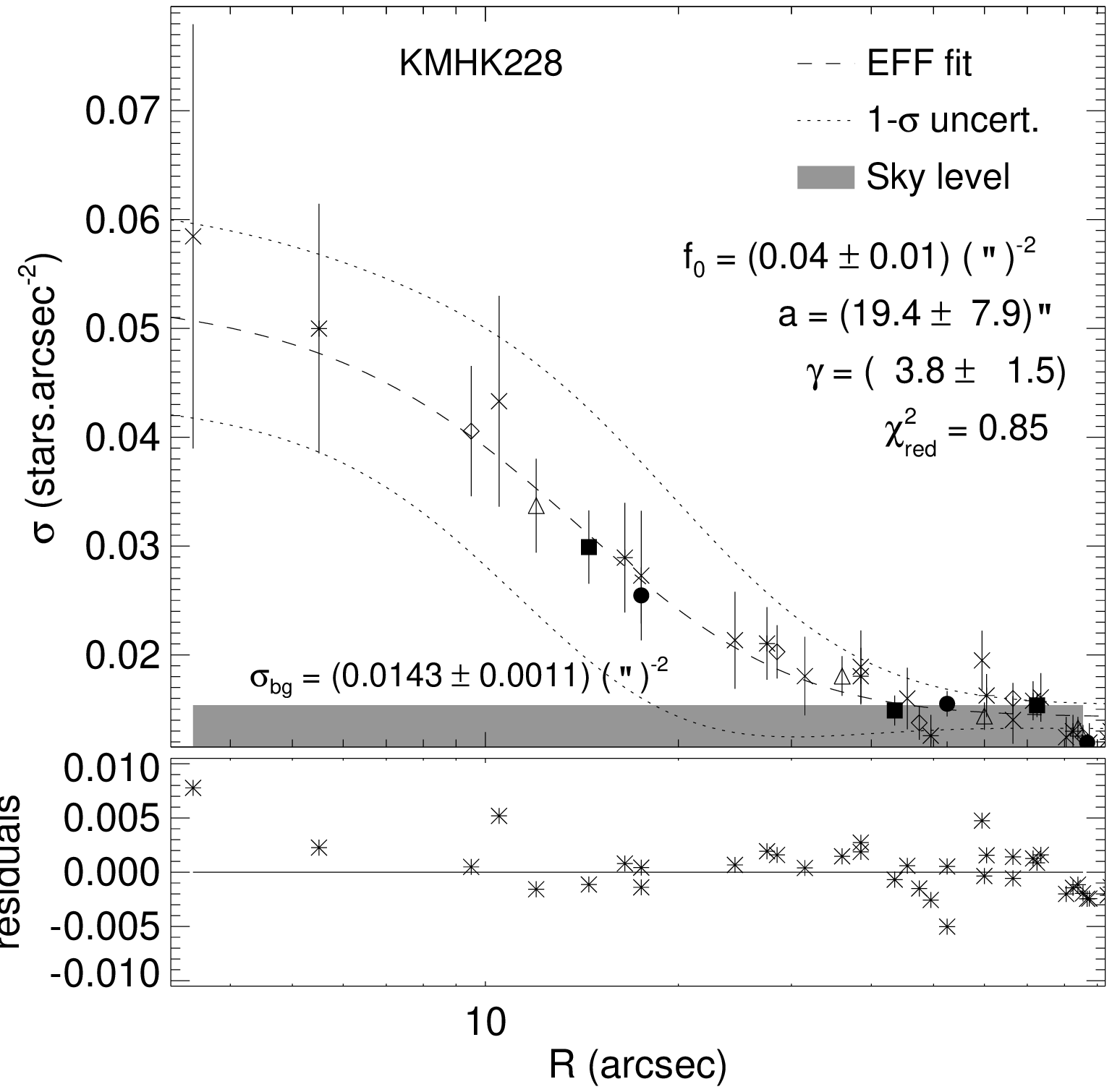}\includegraphics[width=0.325\linewidth]{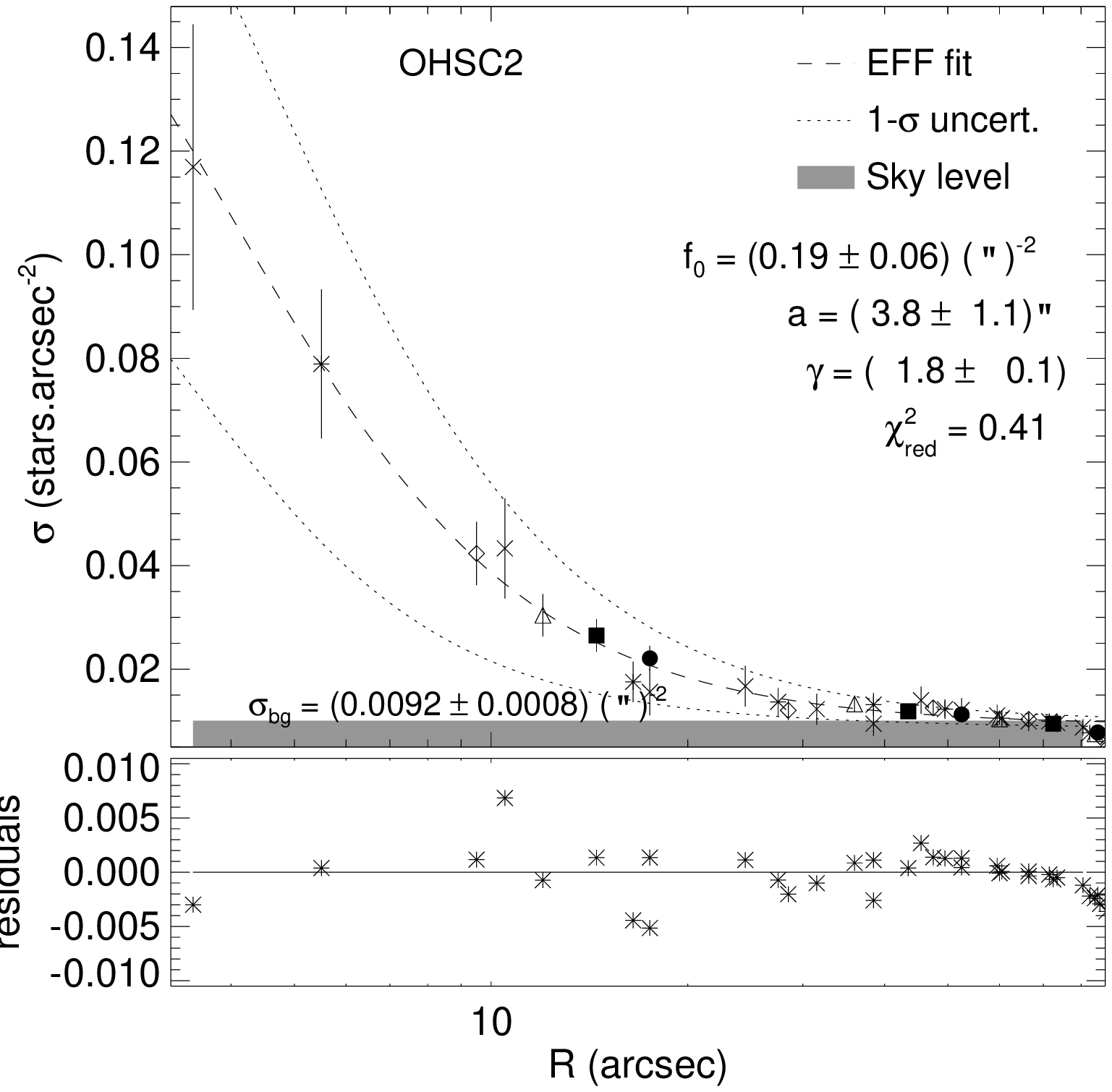}\includegraphics[width=0.325\linewidth]{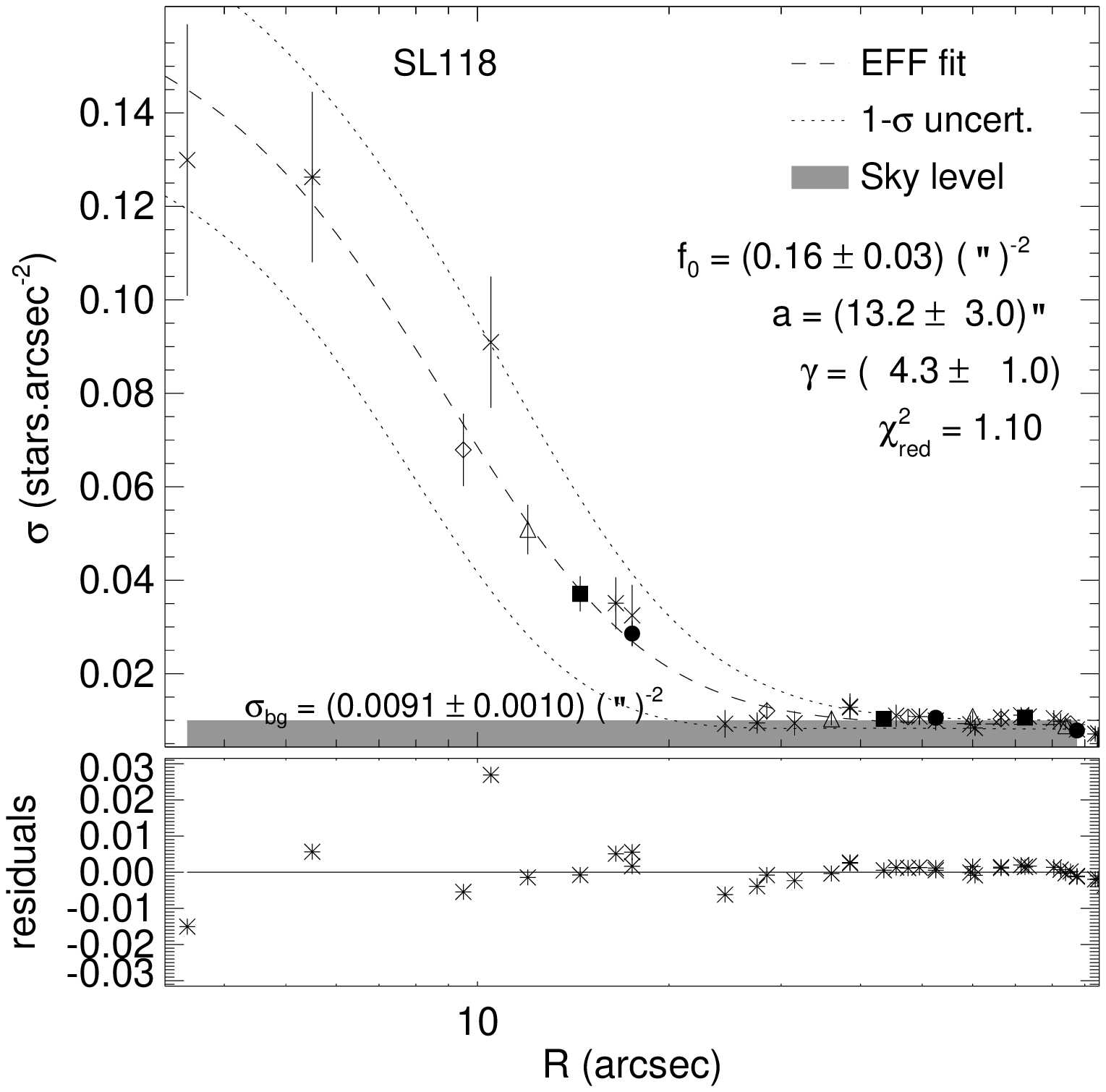}

\includegraphics[width=0.325\linewidth]{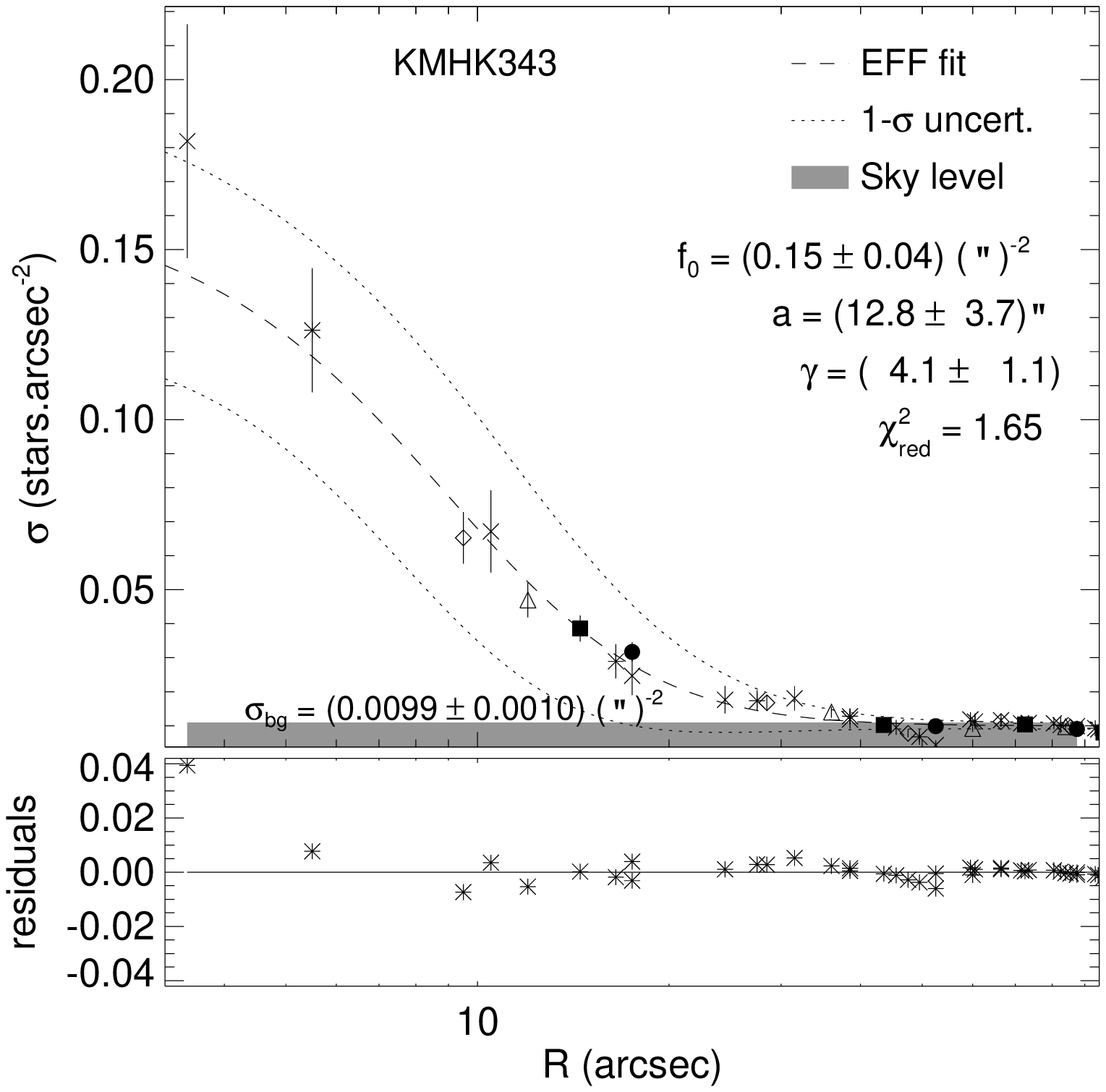}\includegraphics[width=0.325\linewidth]{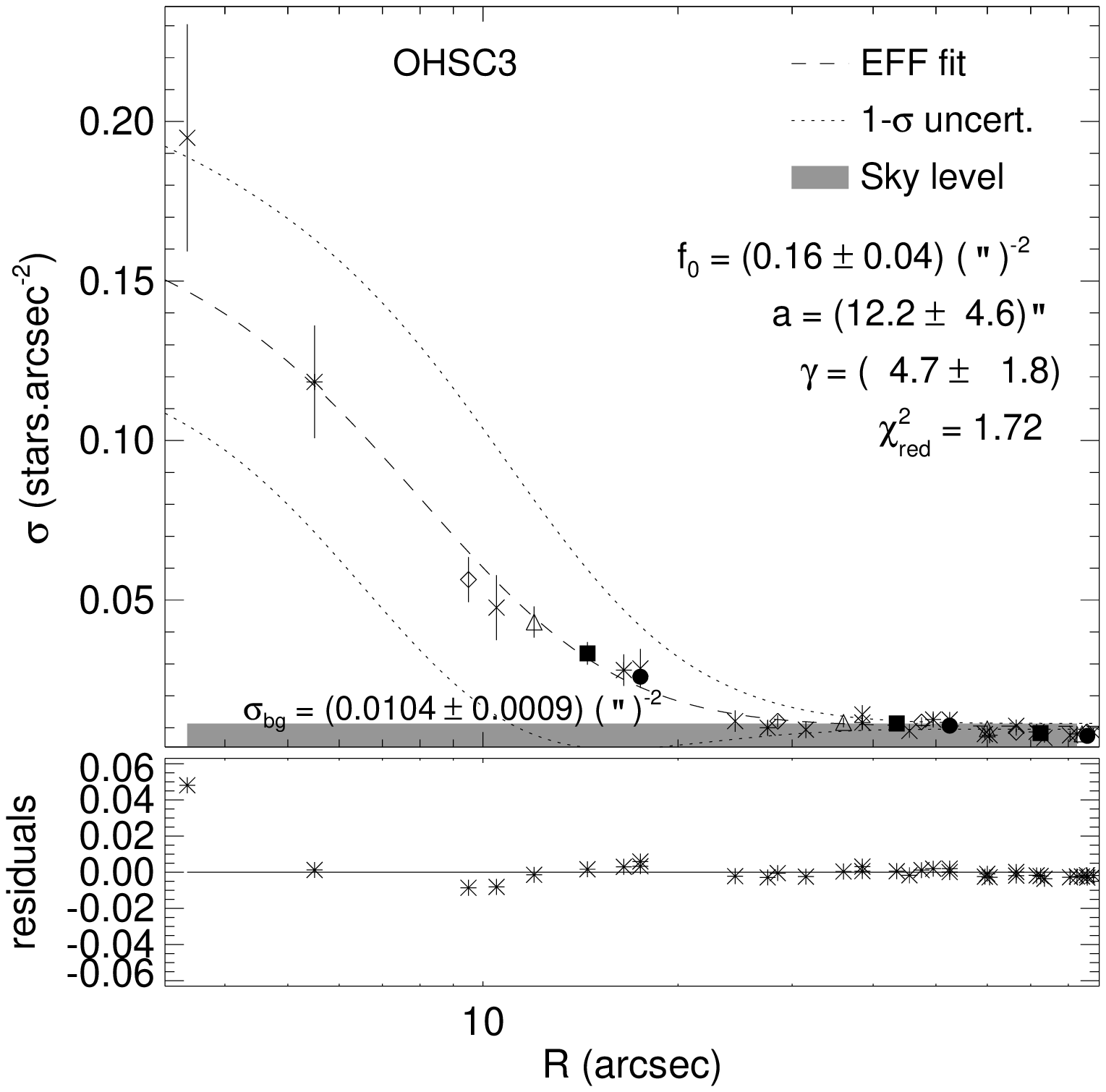}\includegraphics[width=0.325\linewidth]{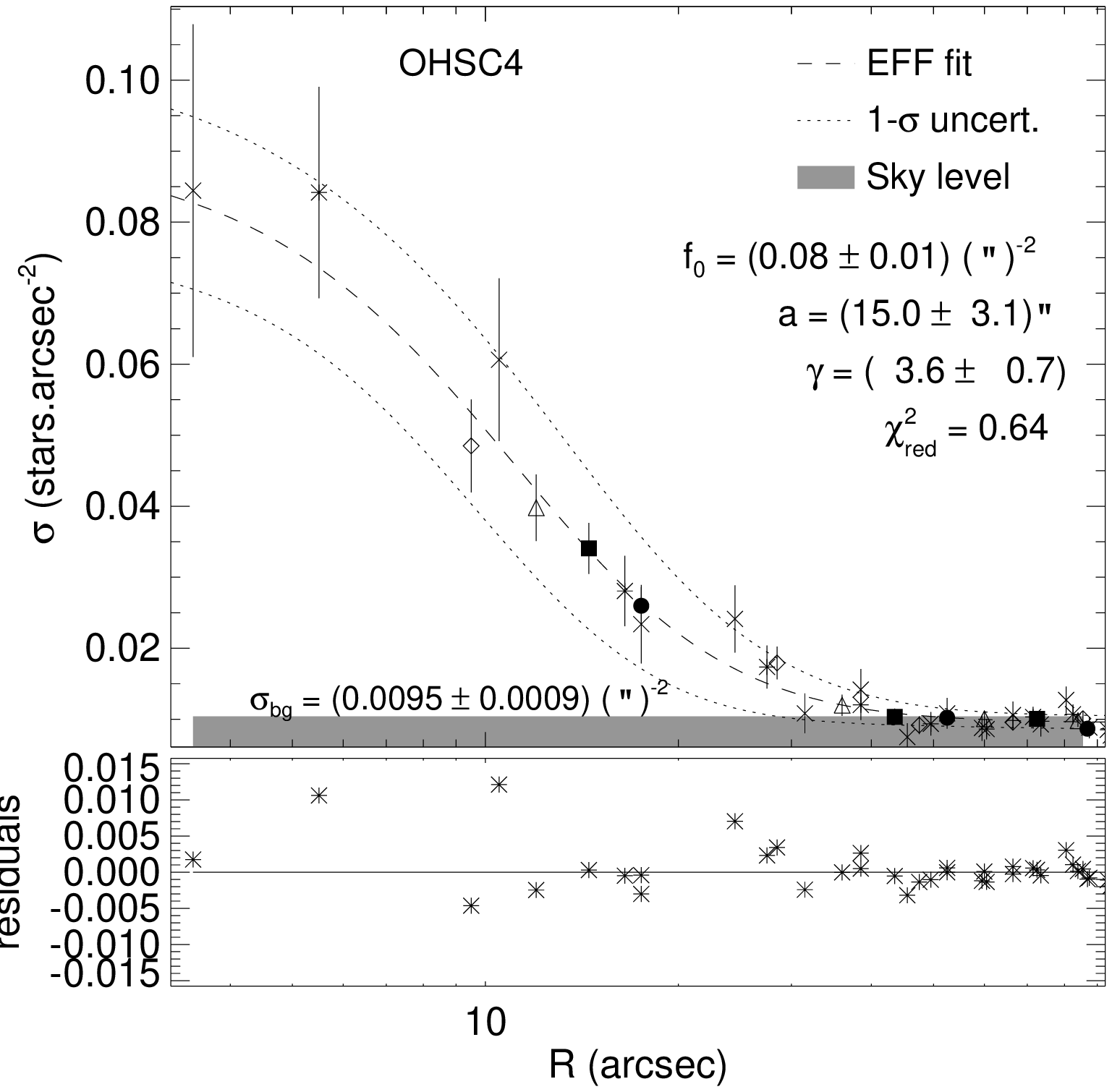}

\includegraphics[width=0.325\linewidth]{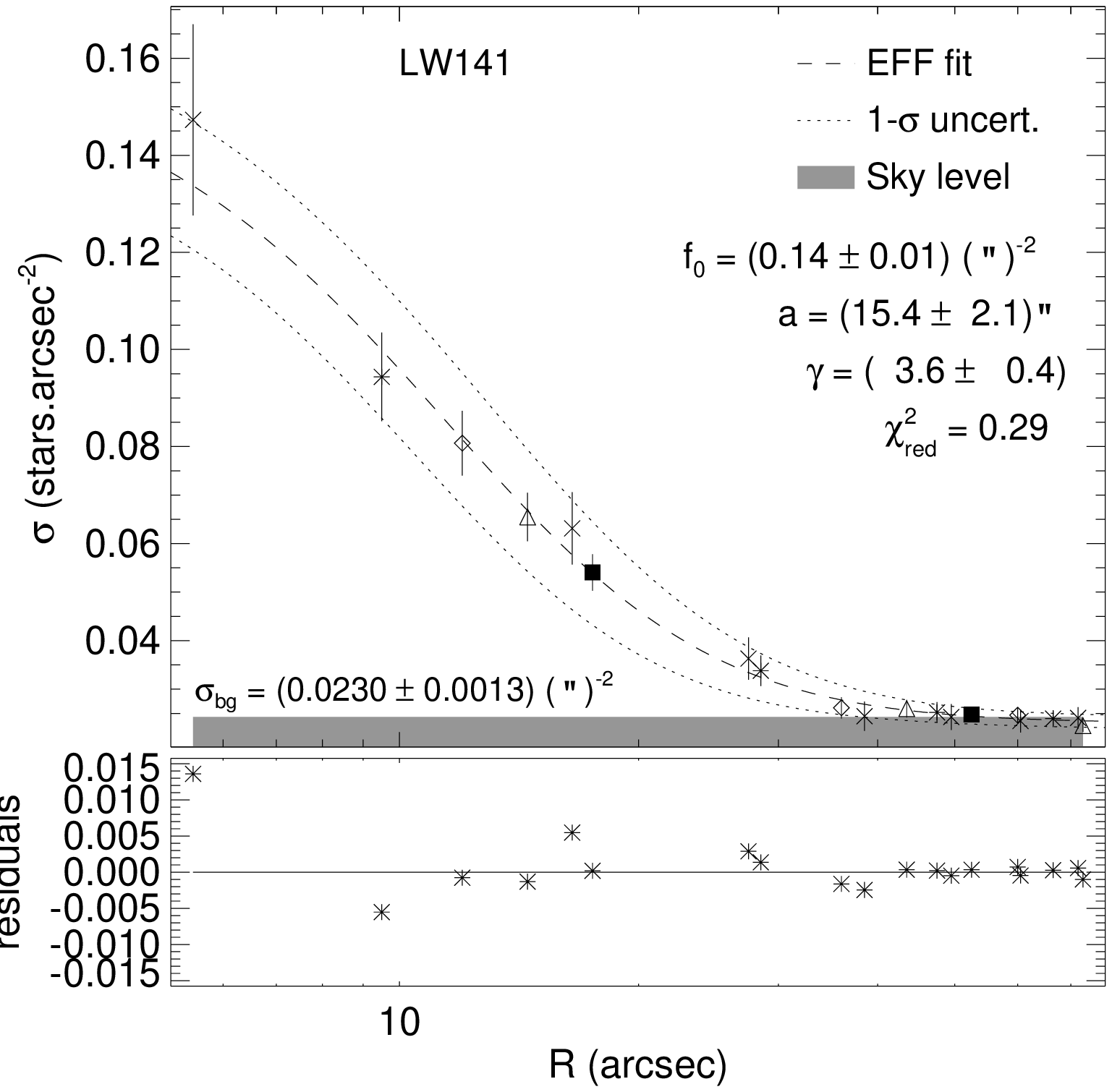}\includegraphics[width=0.325\linewidth]{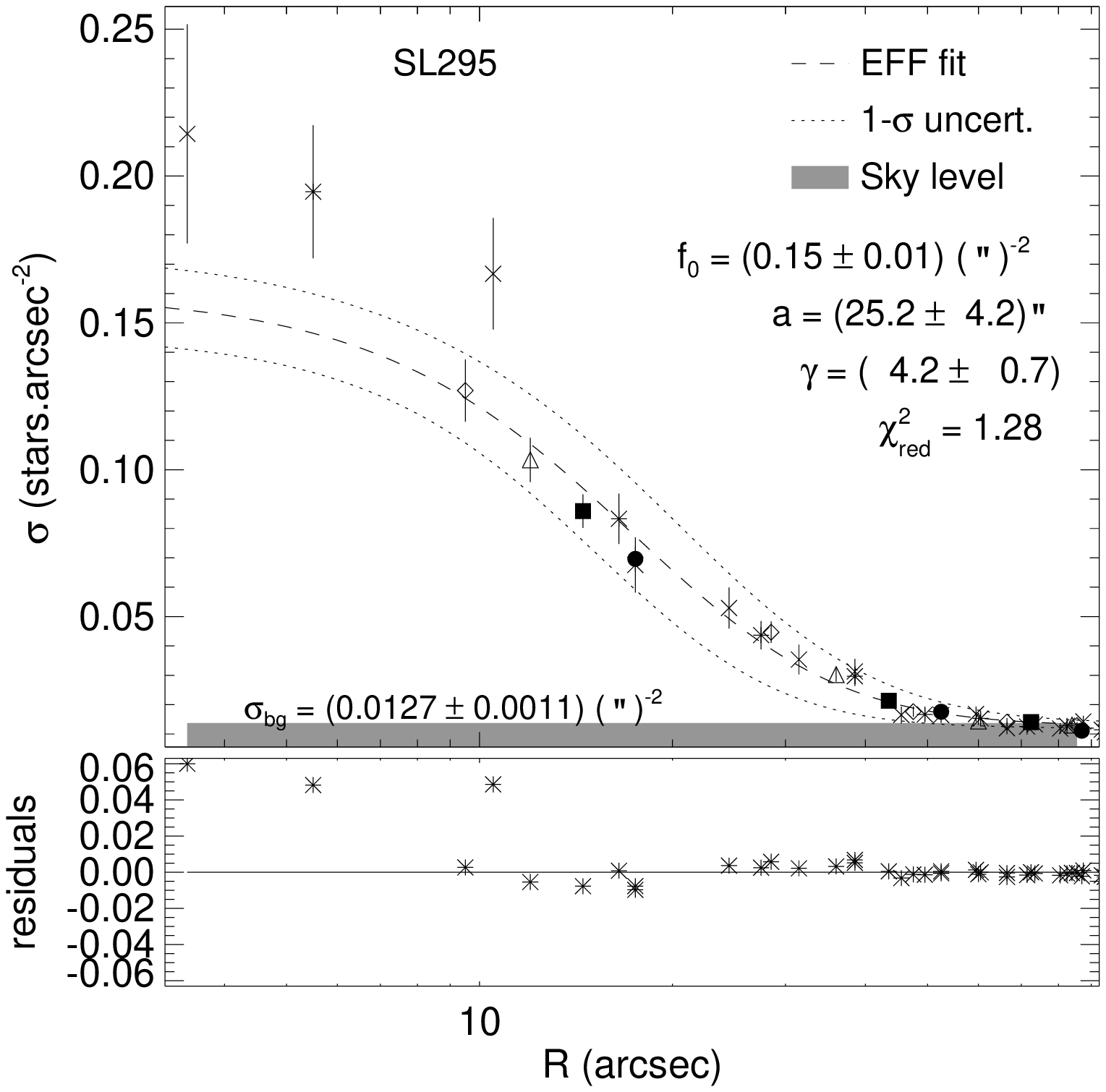}\includegraphics[width=0.325\linewidth]{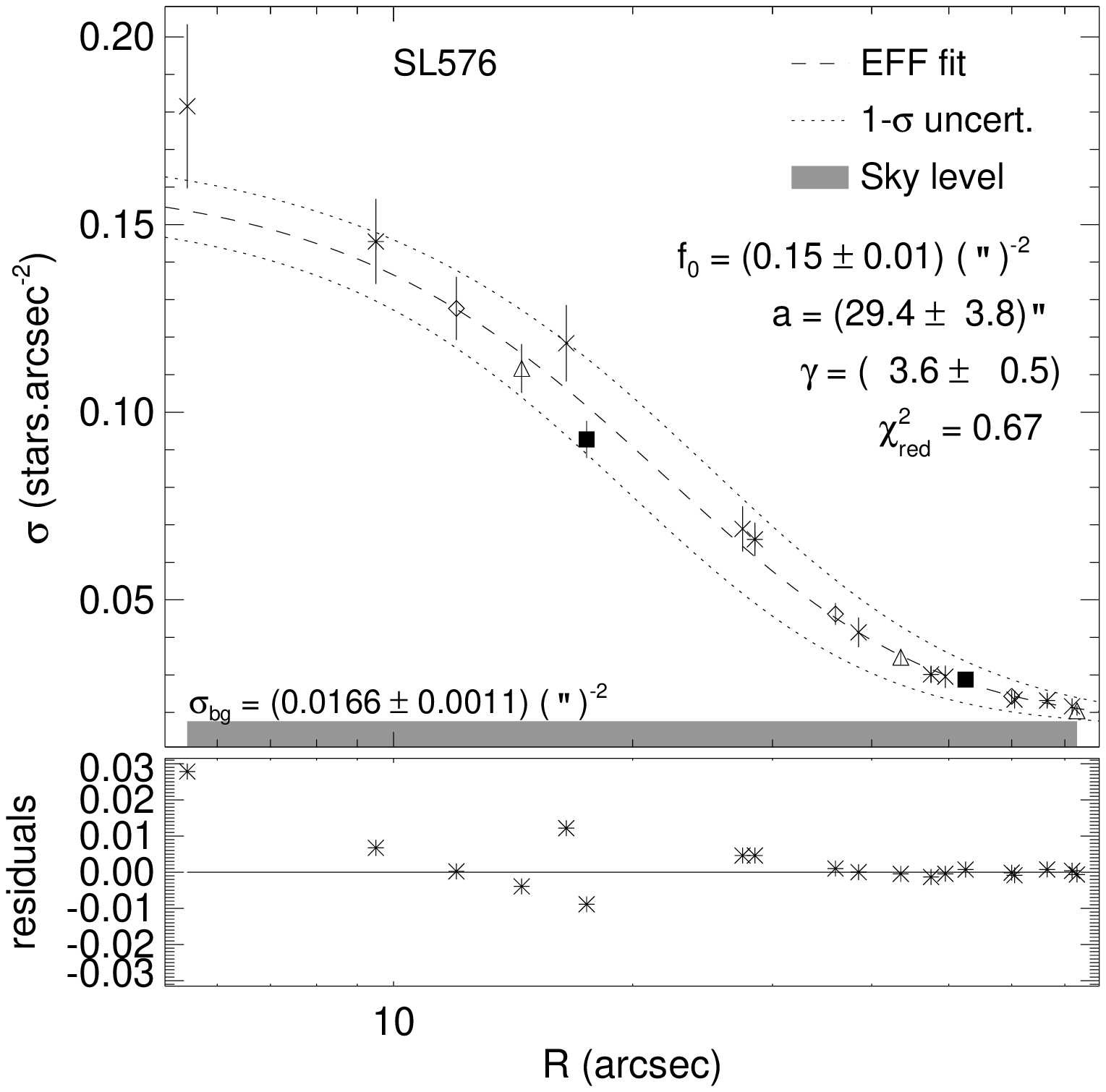}

\includegraphics[width=0.325\linewidth]{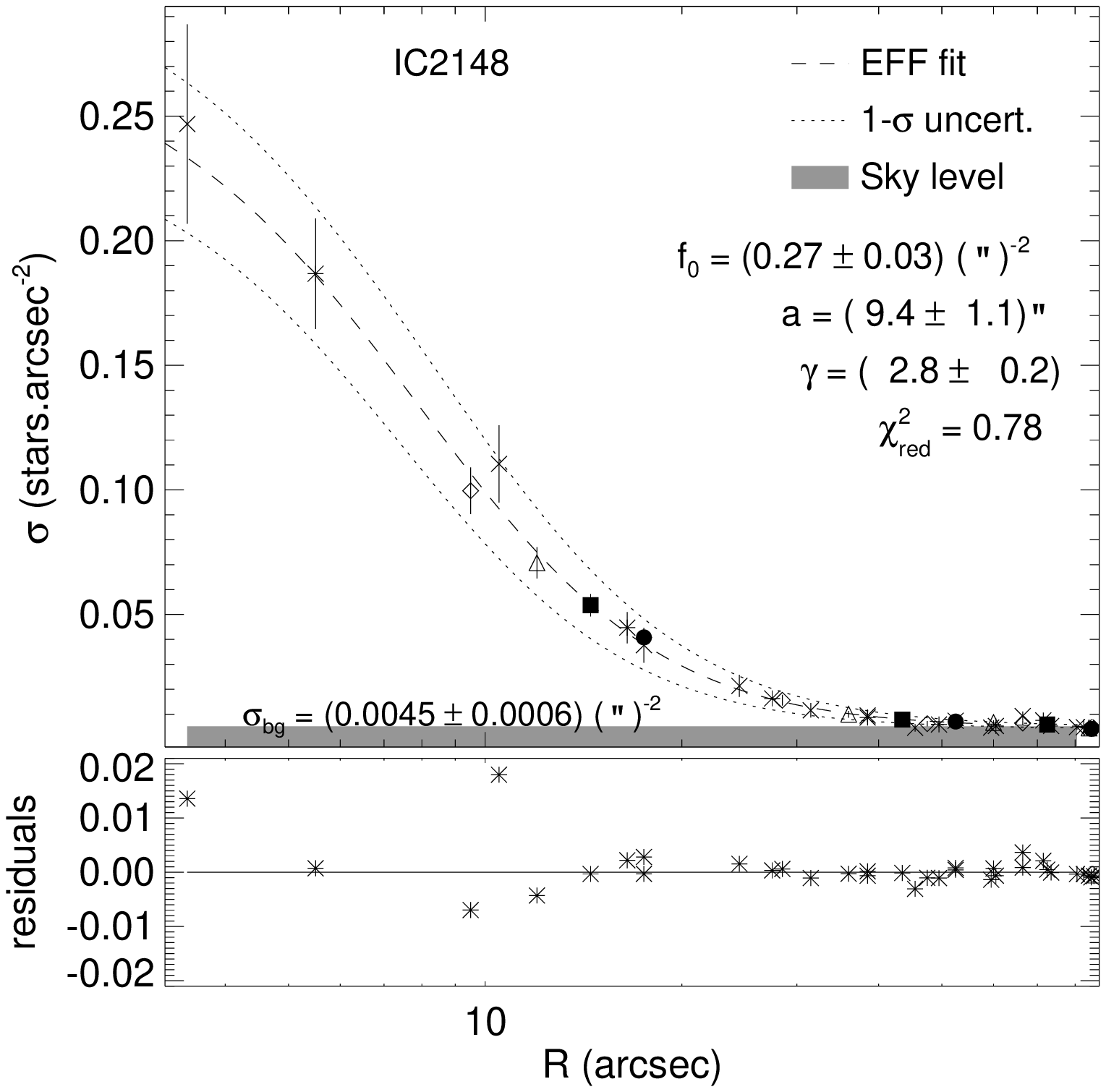}\includegraphics[width=0.325\linewidth]{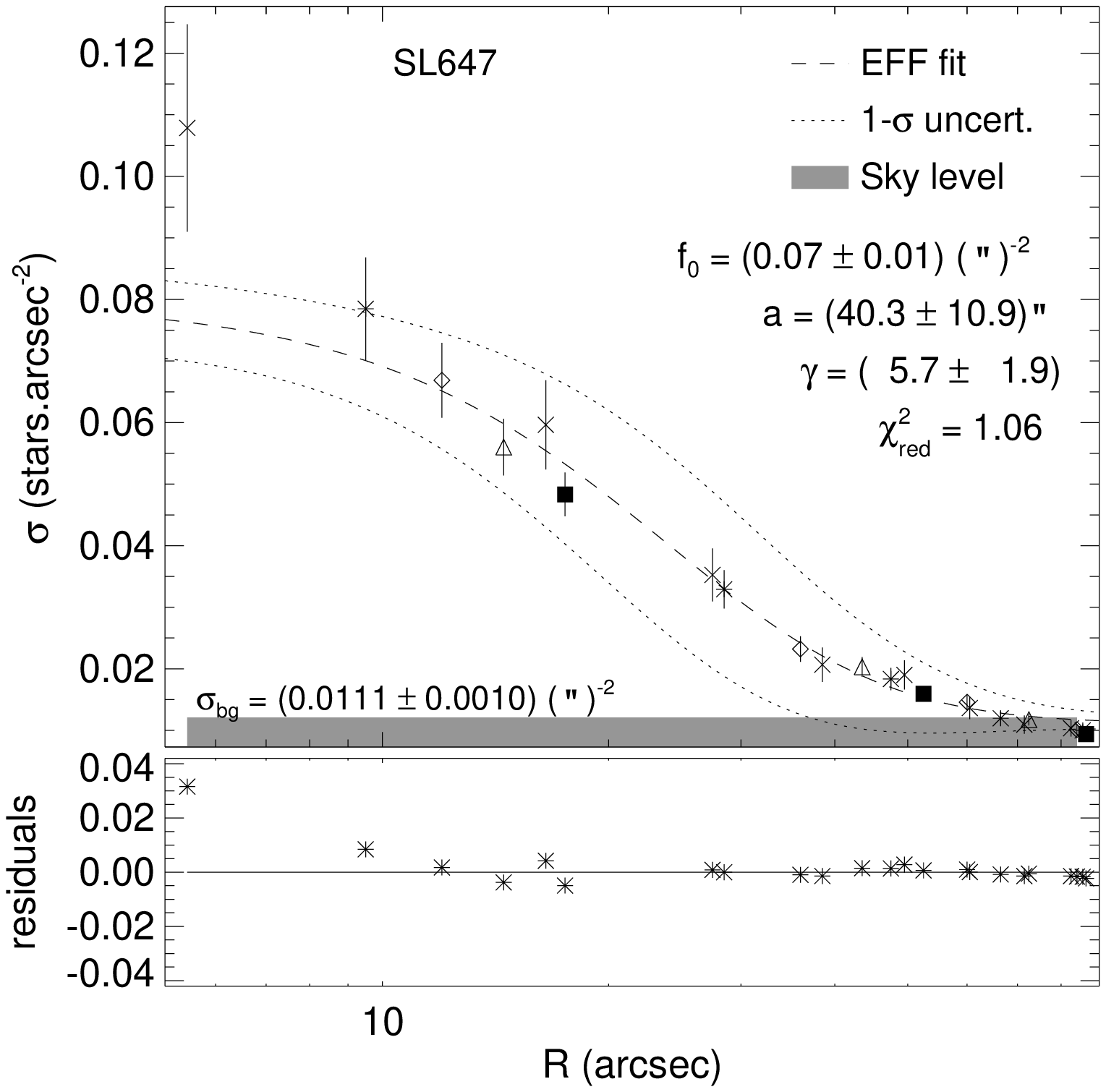}\includegraphics[width=0.325\linewidth]{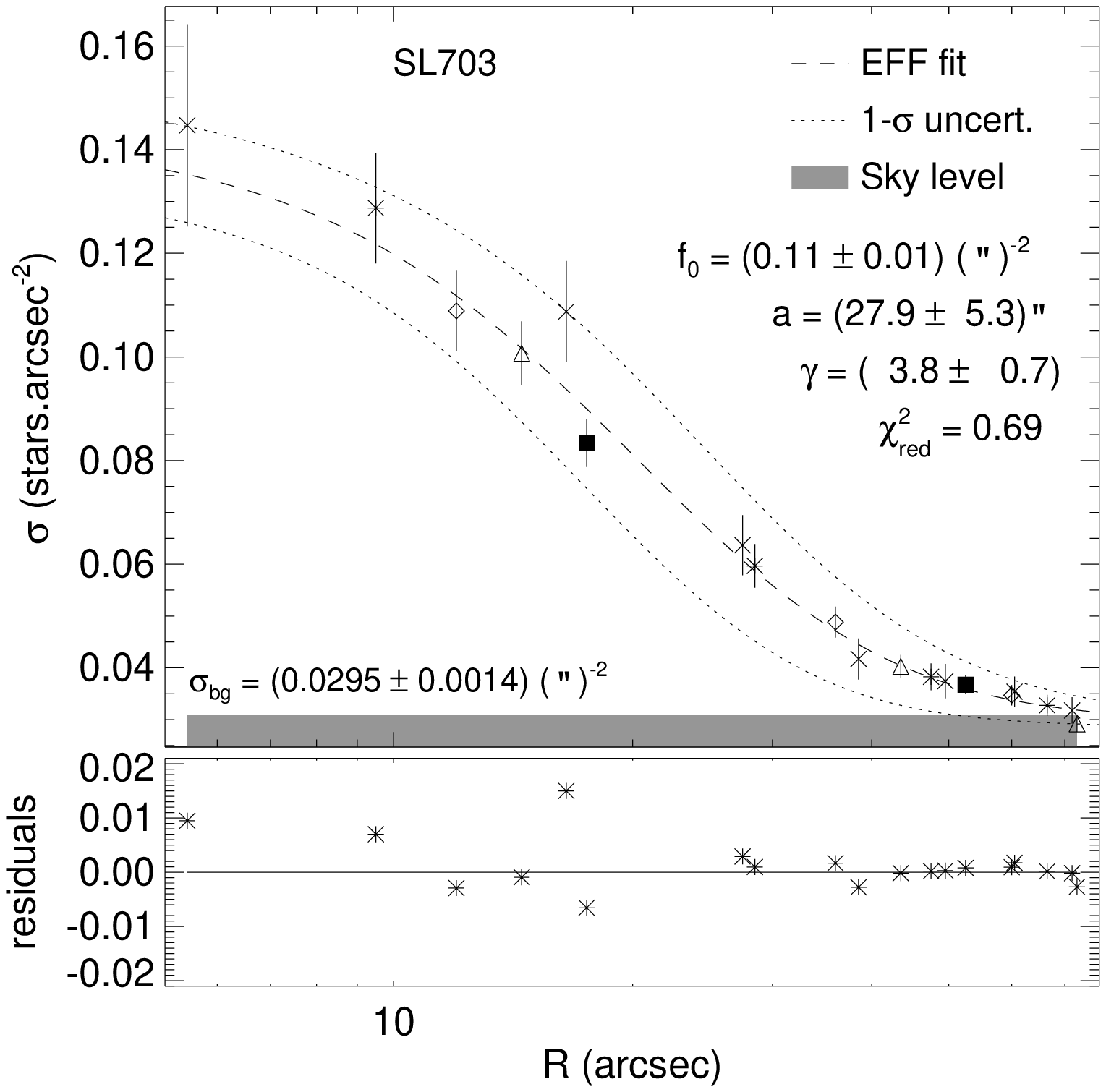}

\caption{cont.}

\end{figure*}

\setcounter{figure}{5}
\begin{figure*}

\includegraphics[width=0.325\linewidth]{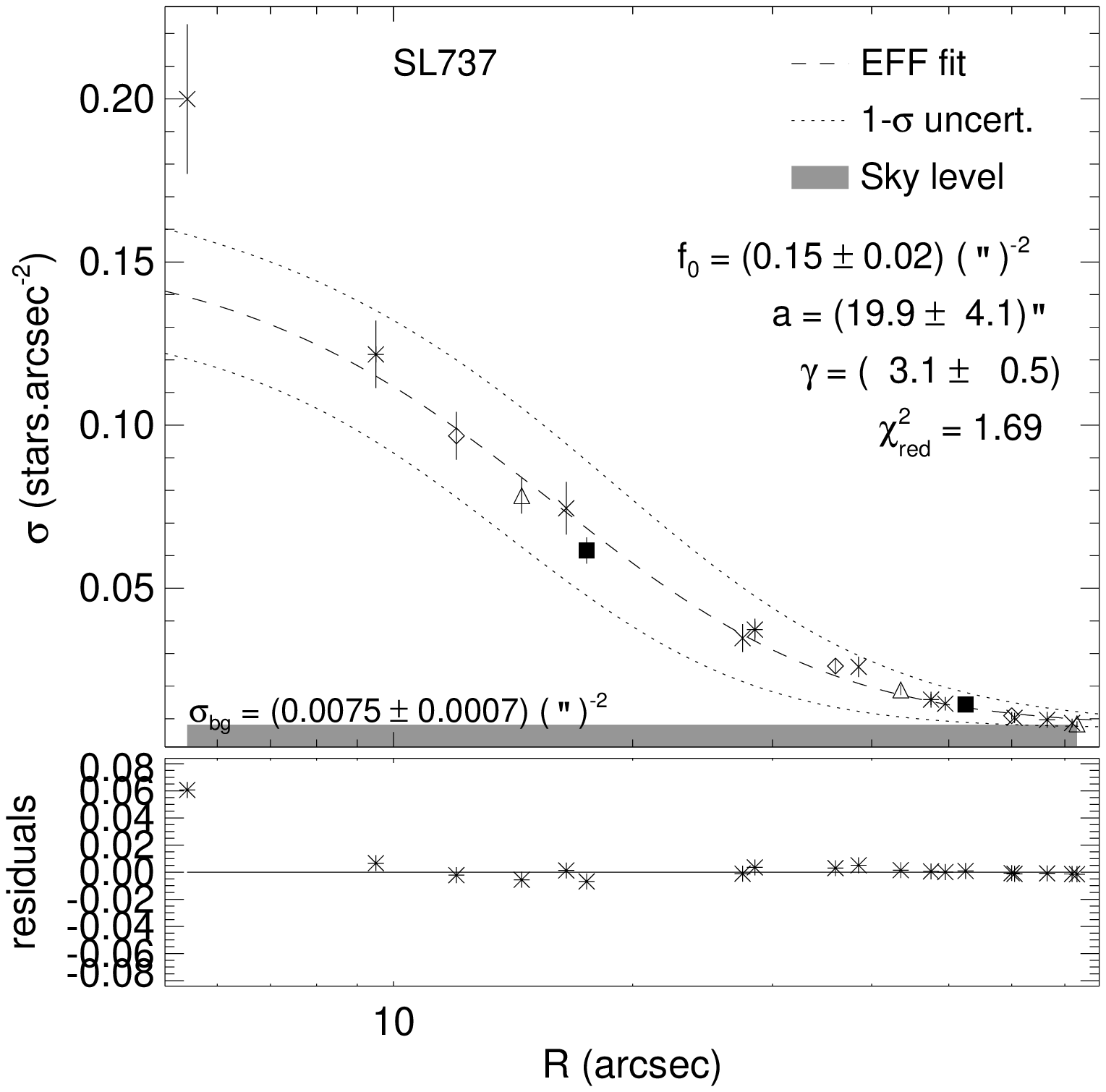}\includegraphics[width=0.325\linewidth]{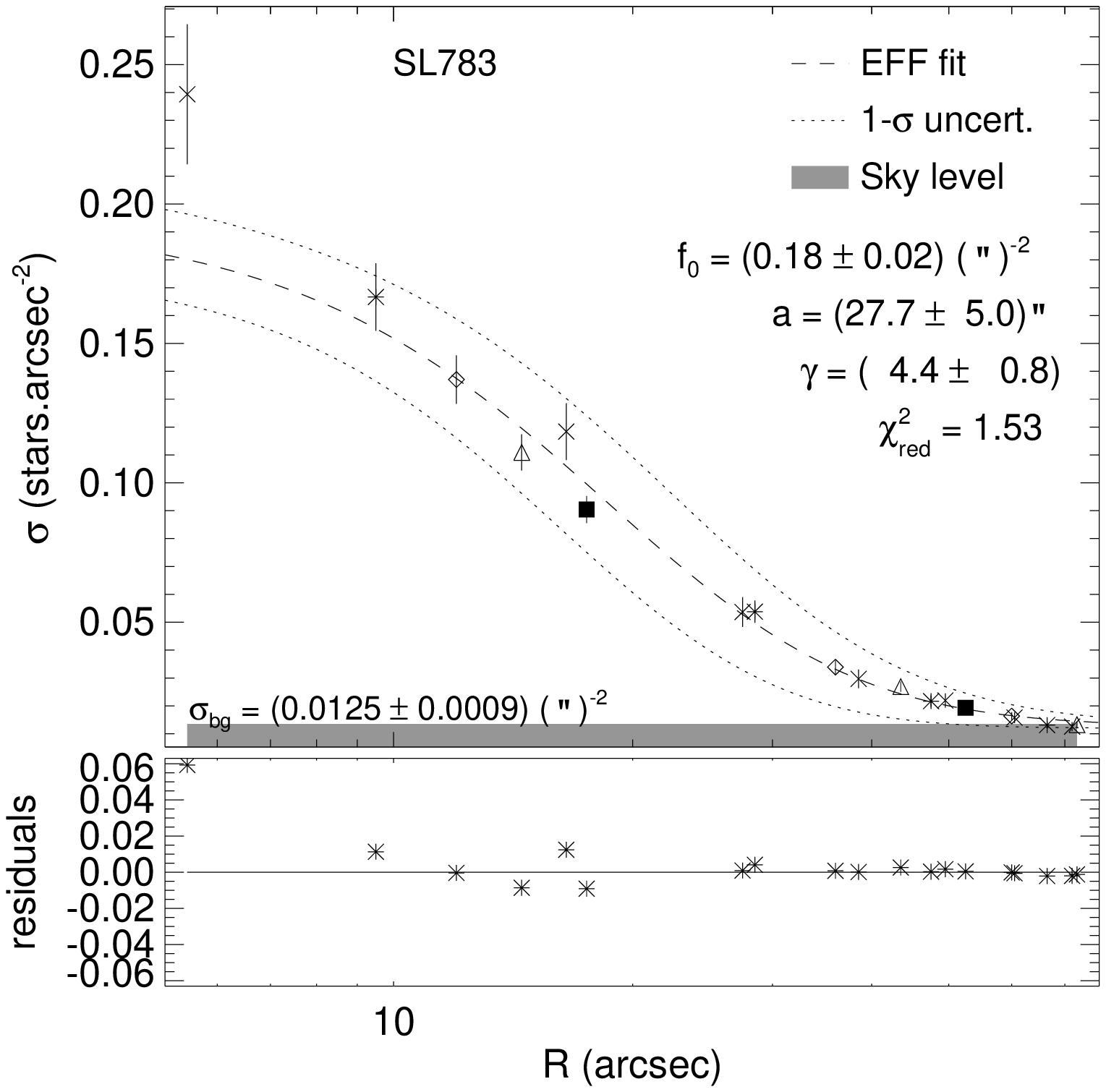}\includegraphics[width=0.325\linewidth]{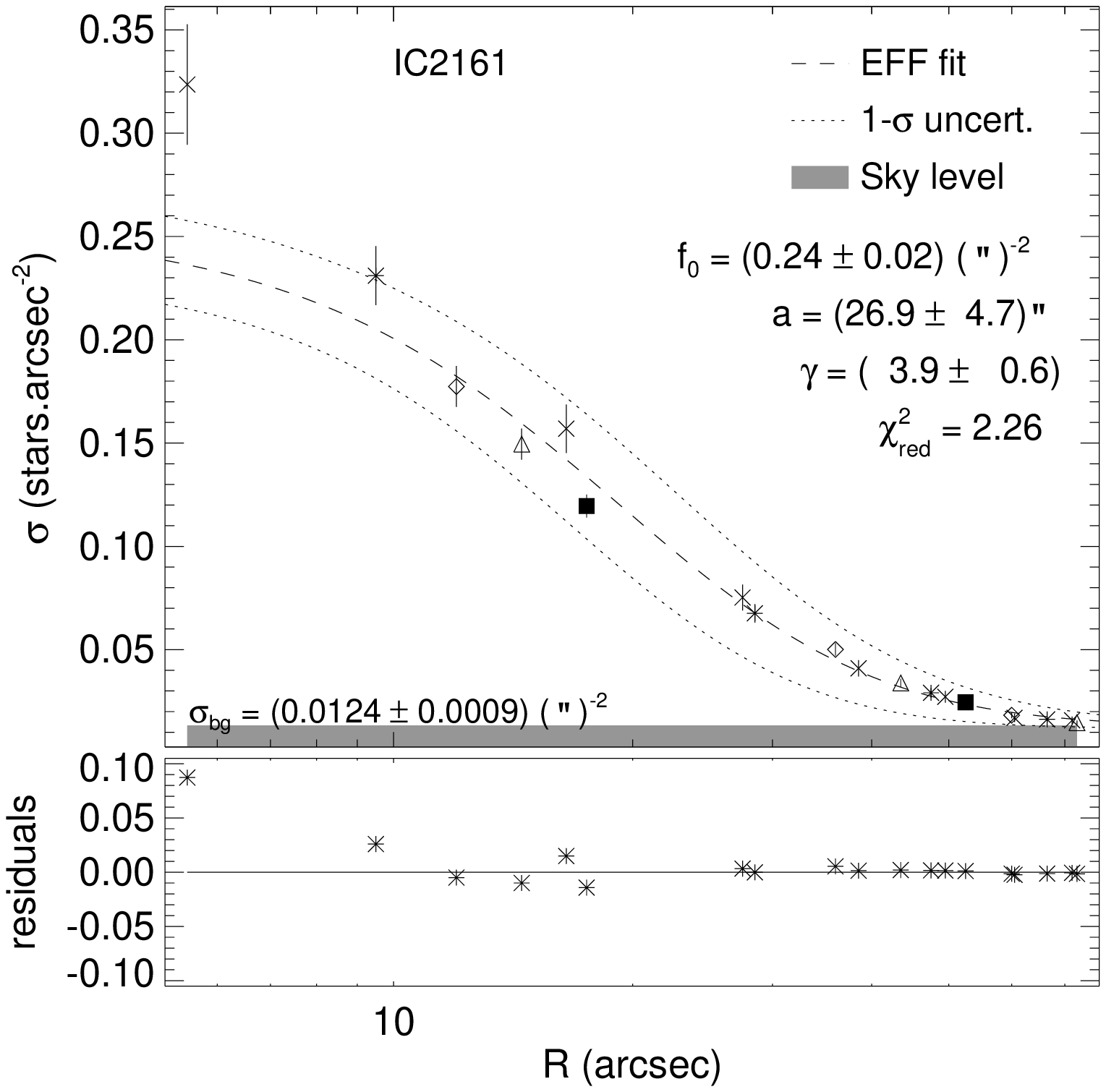}

\includegraphics[width=0.325\linewidth]{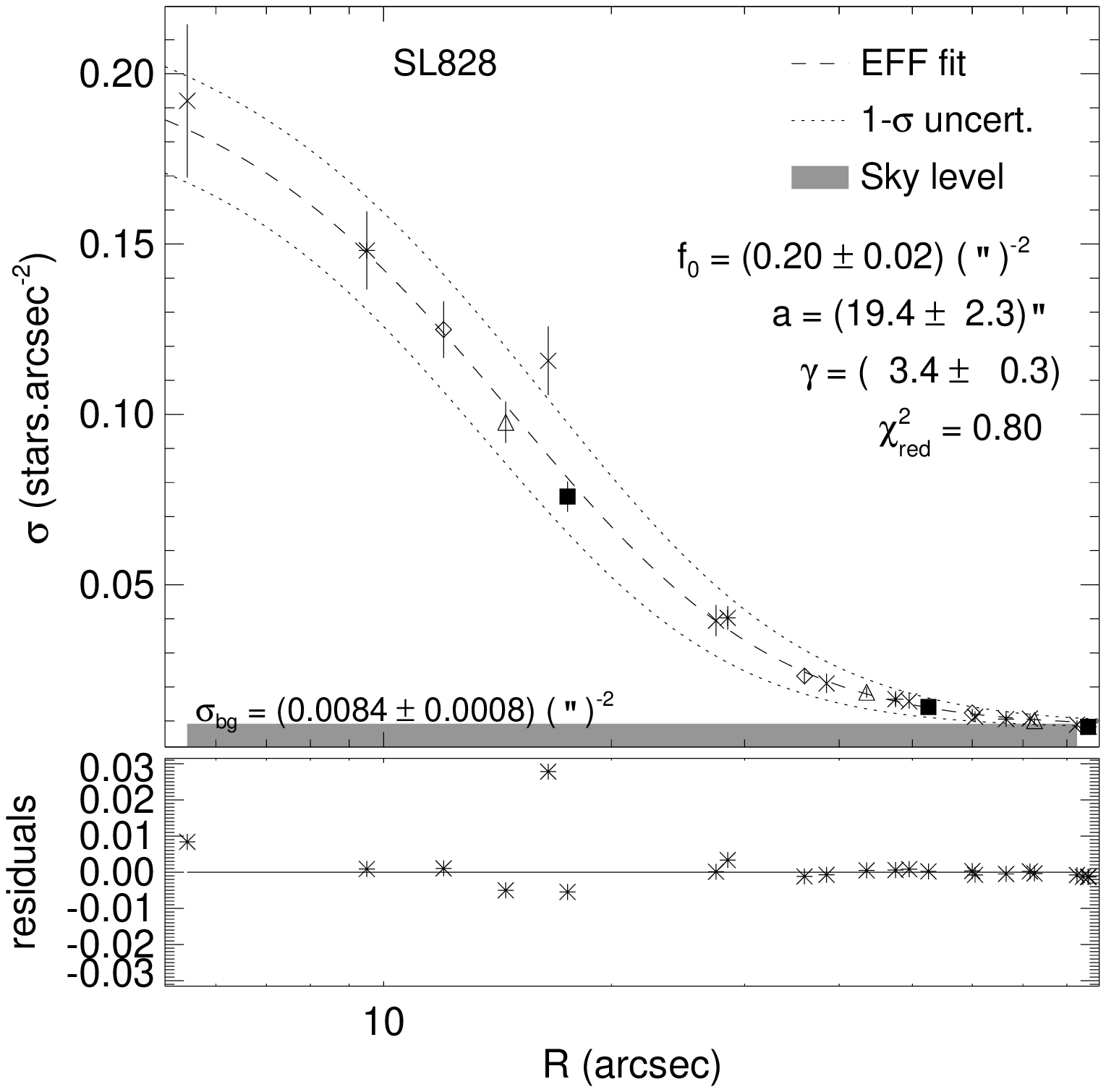}\includegraphics[width=0.325\linewidth]{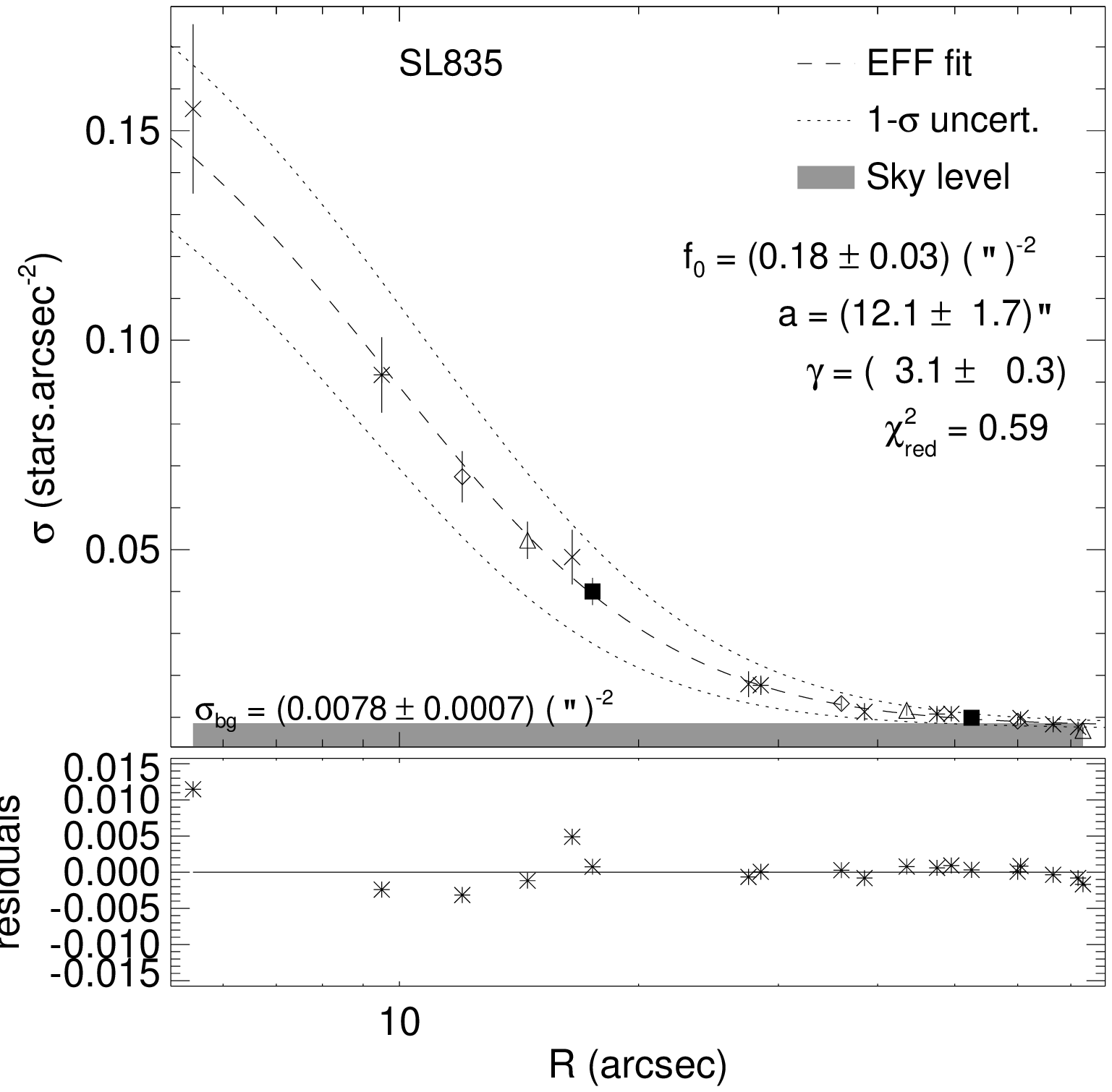}\includegraphics[width=0.325\linewidth]{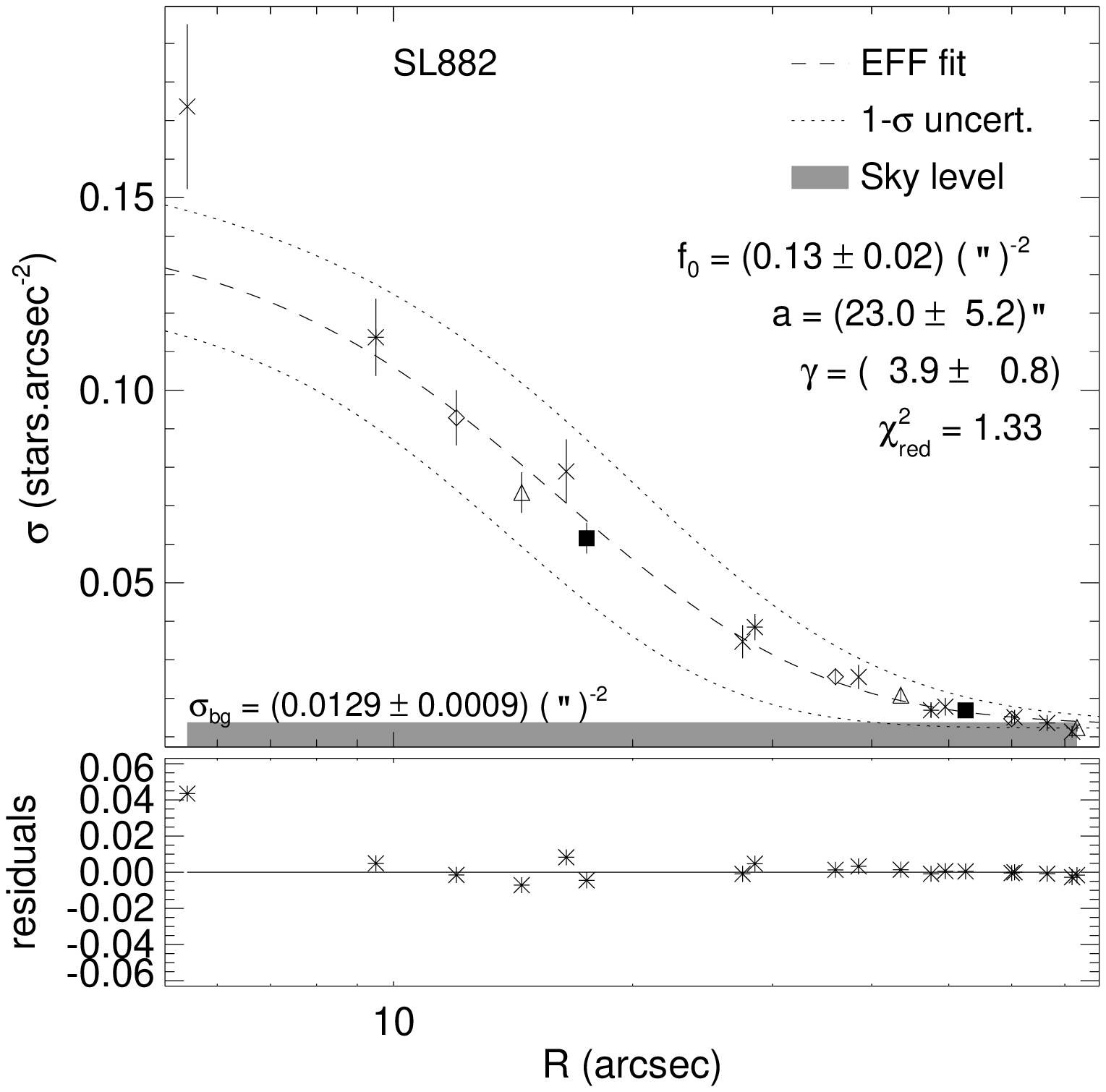}

\includegraphics[width=0.325\linewidth]{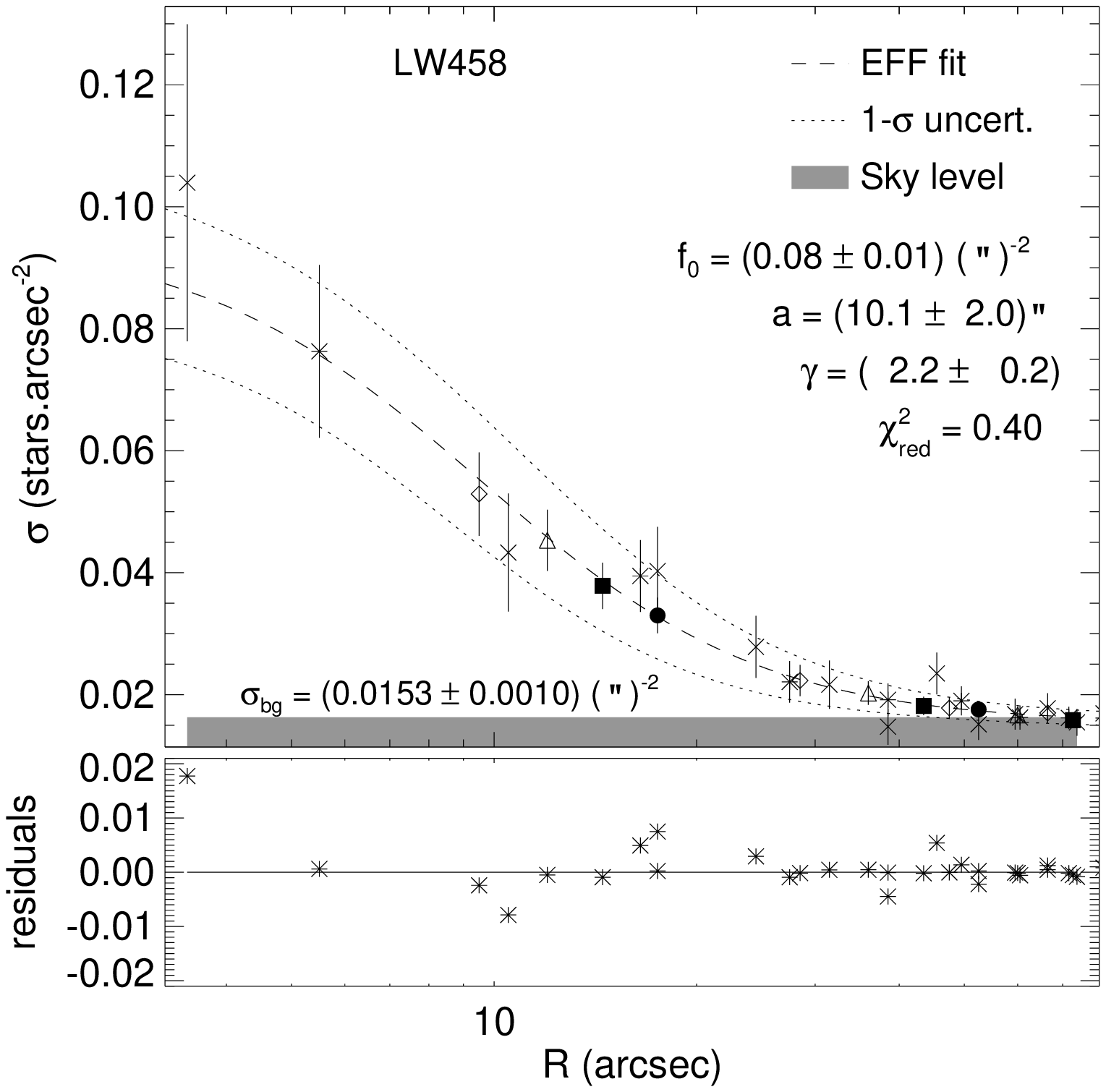}\includegraphics[width=0.325\linewidth]{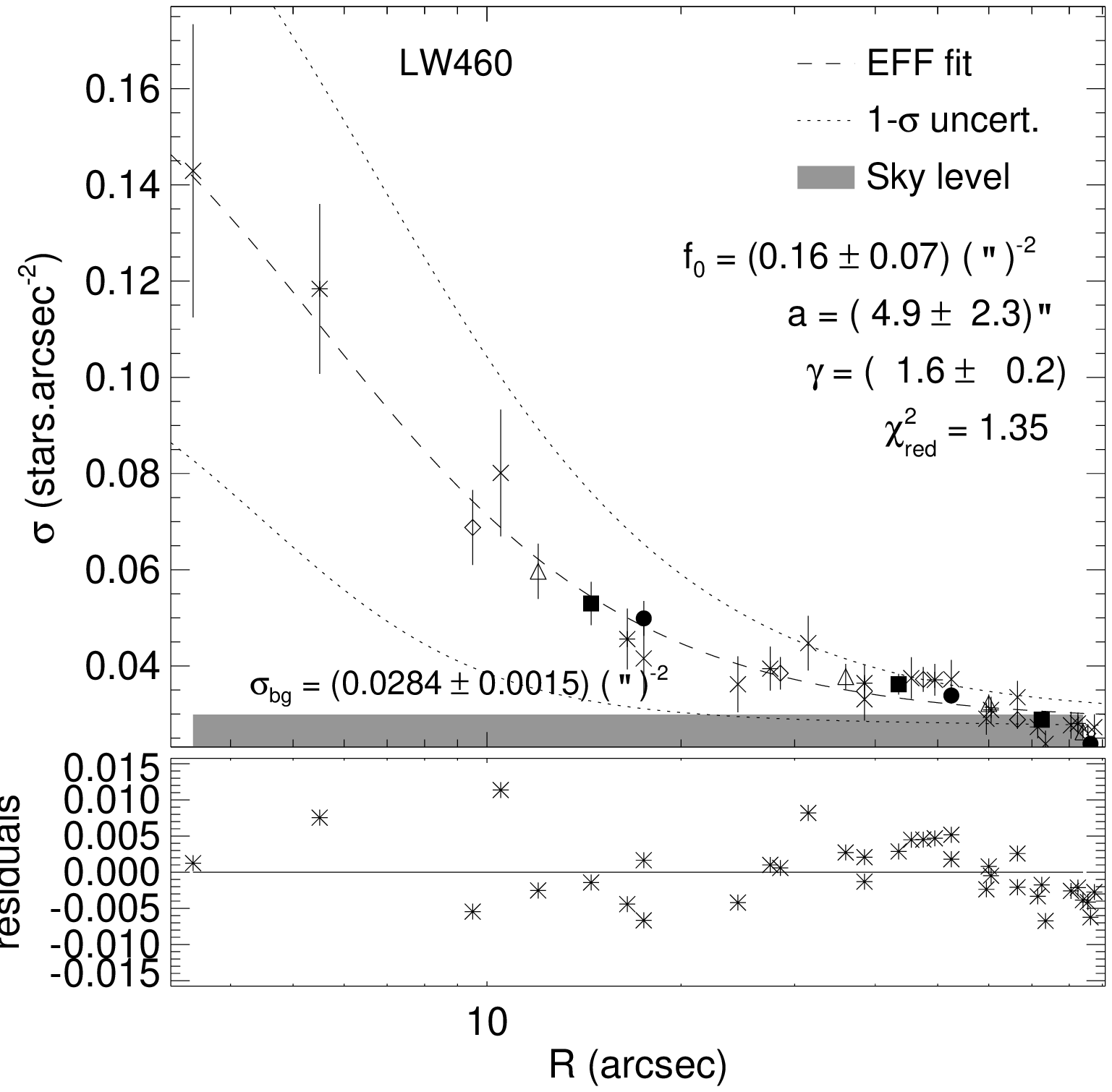}\includegraphics[width=0.325\linewidth]{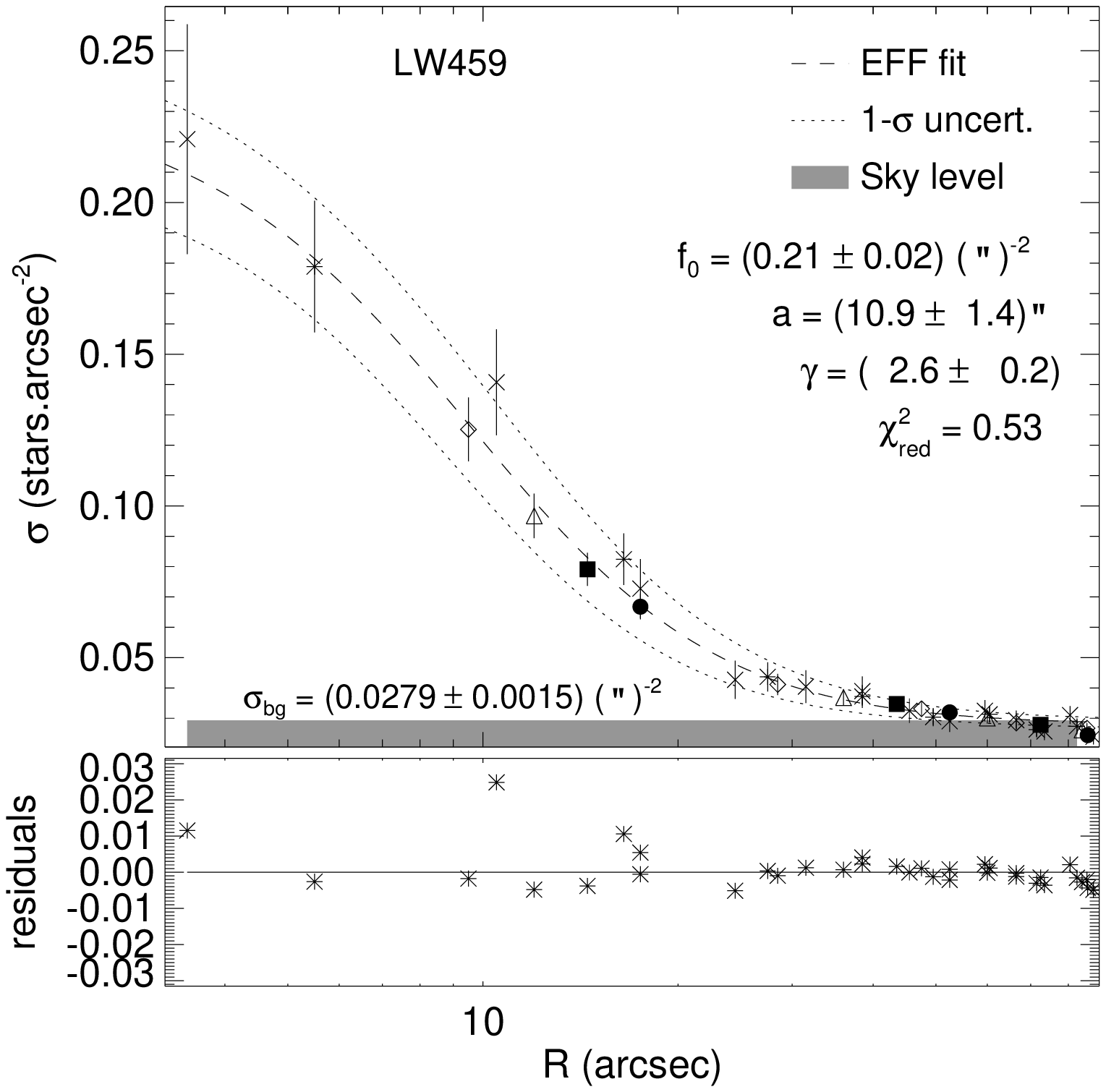}

\includegraphics[width=0.325\linewidth]{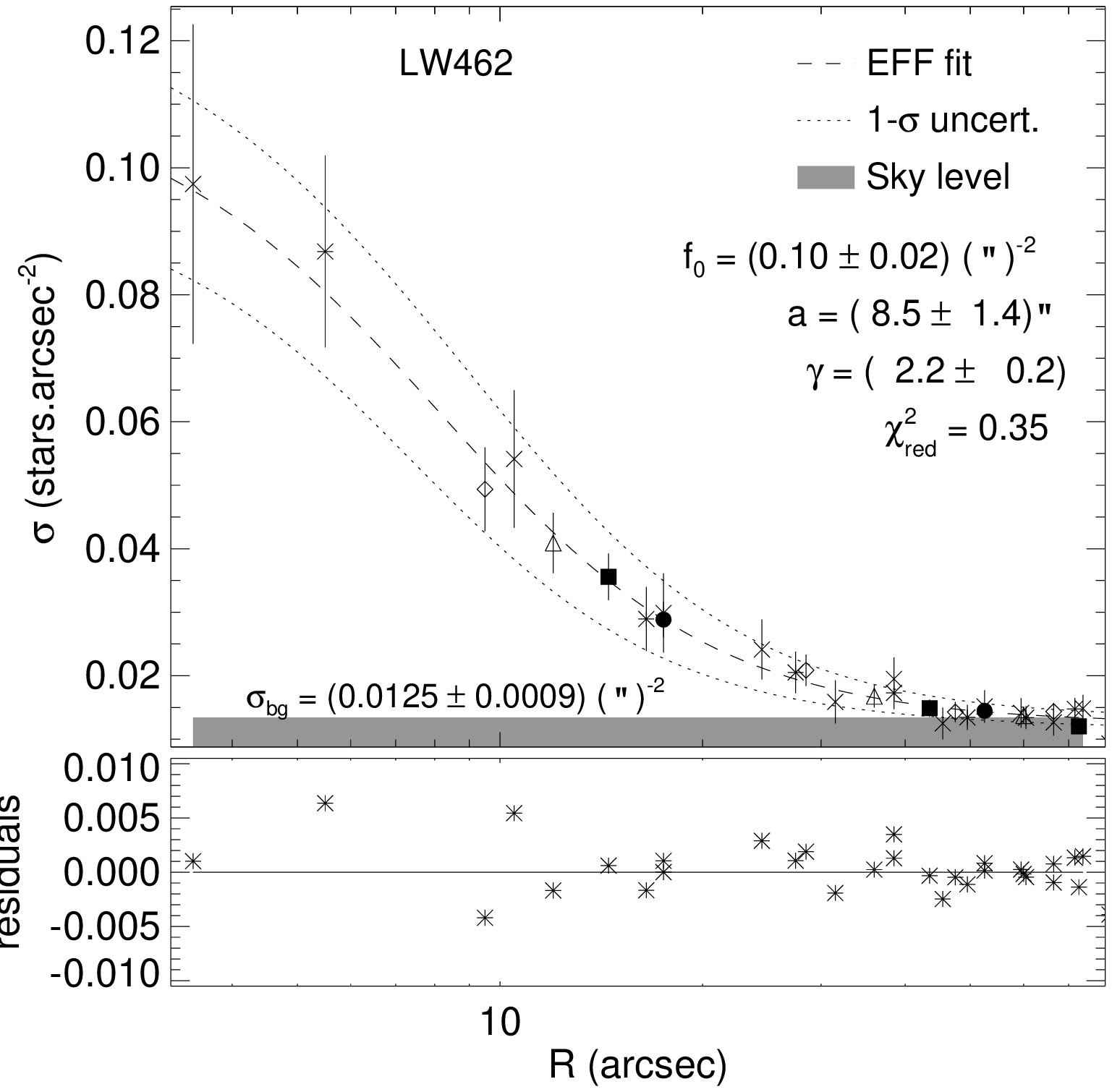}\includegraphics[width=0.325\linewidth]{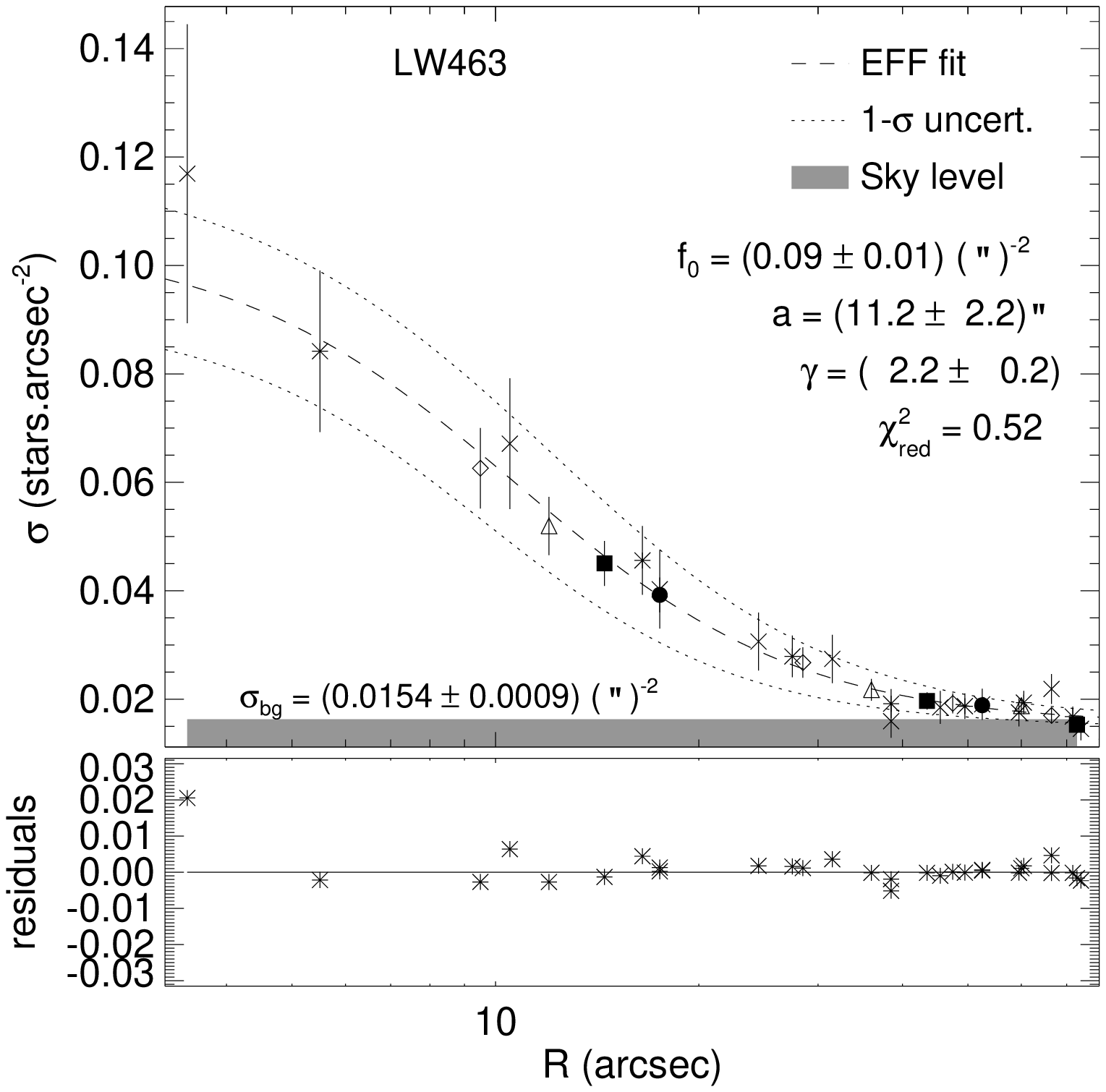}\includegraphics[width=0.325\linewidth]{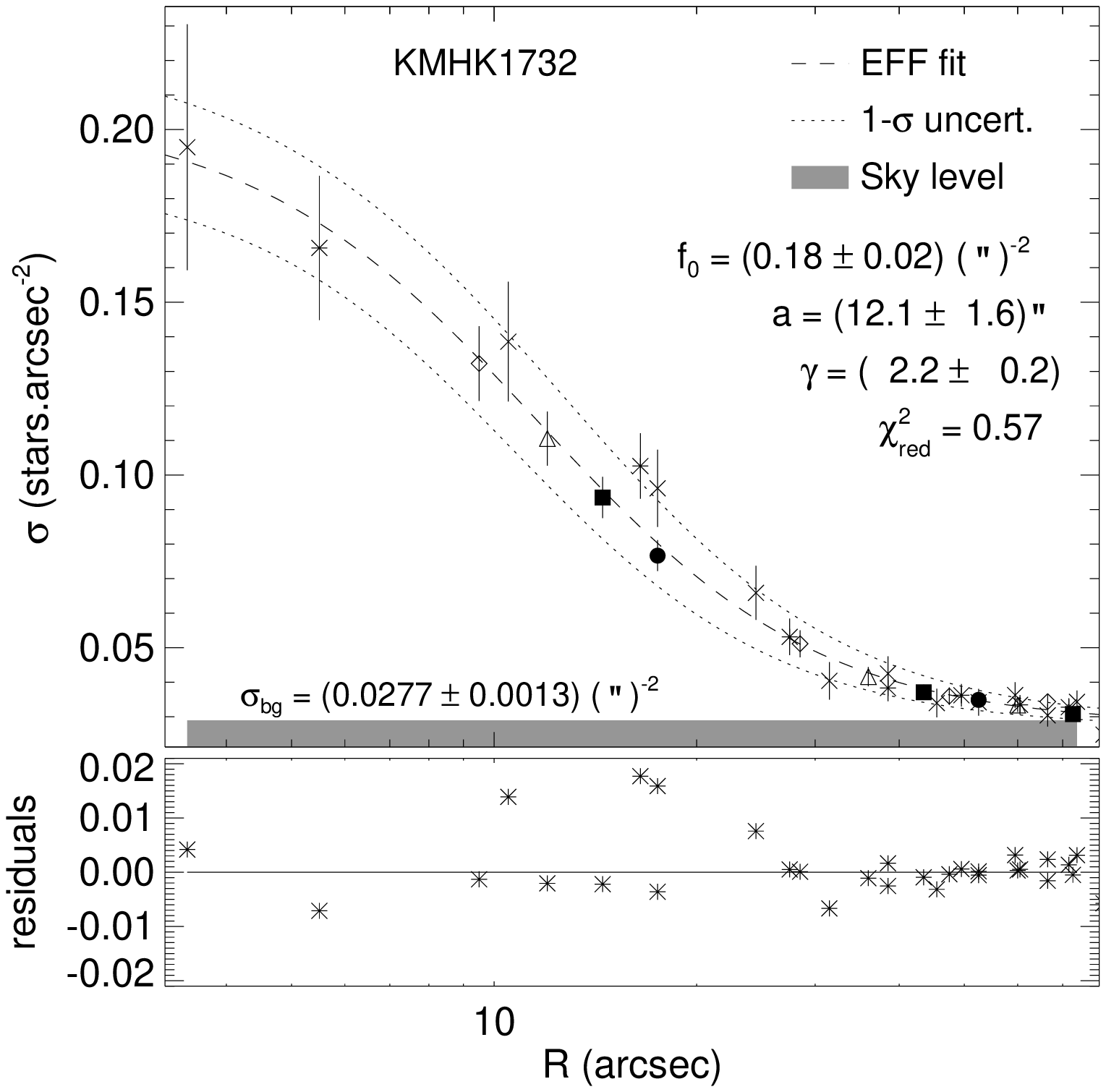}

\caption{cont.}

\end{figure*}

\setcounter{figure}{5}
\begin{figure*}

\includegraphics[width=0.325\linewidth]{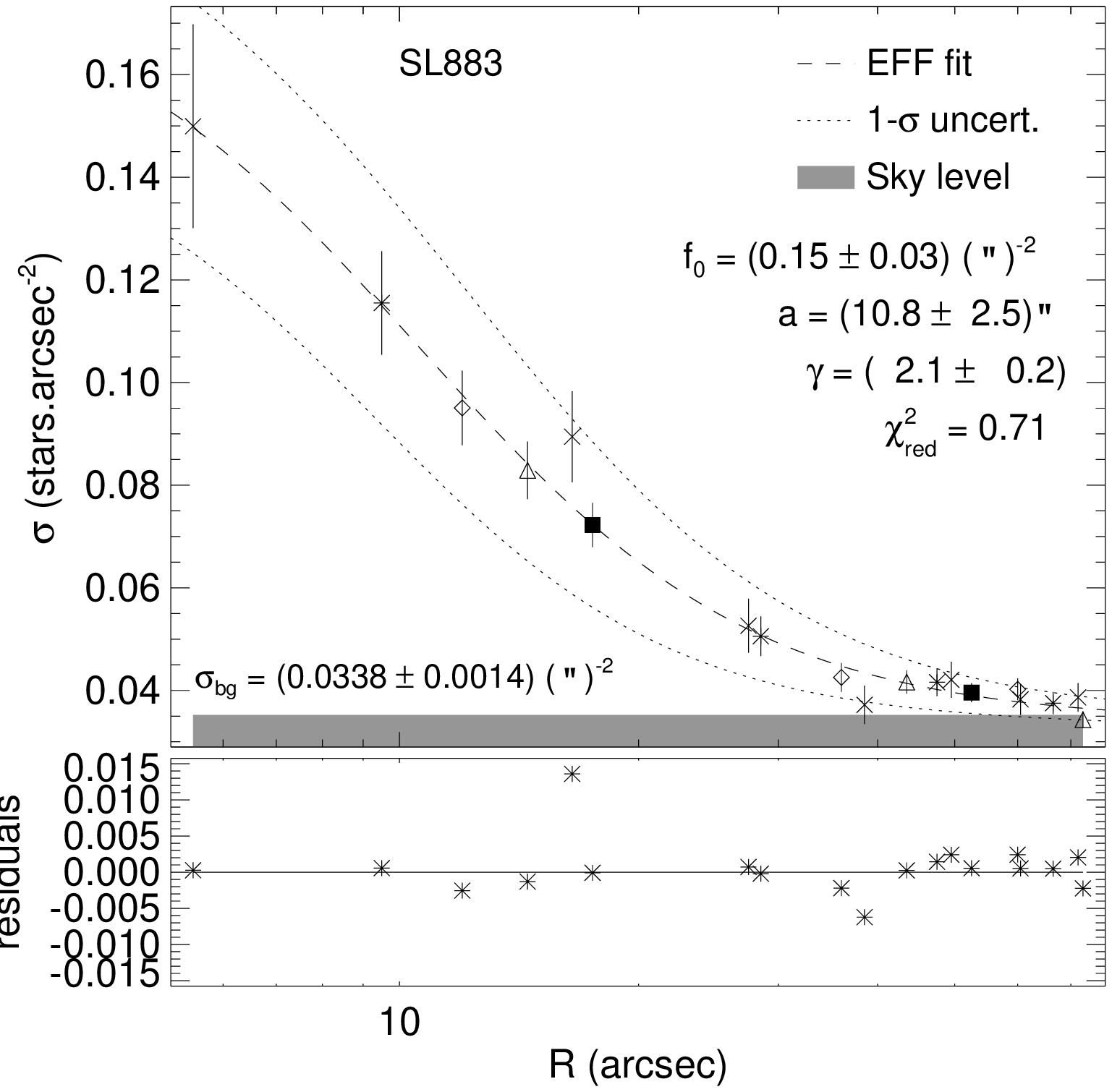}\includegraphics[width=0.325\linewidth]{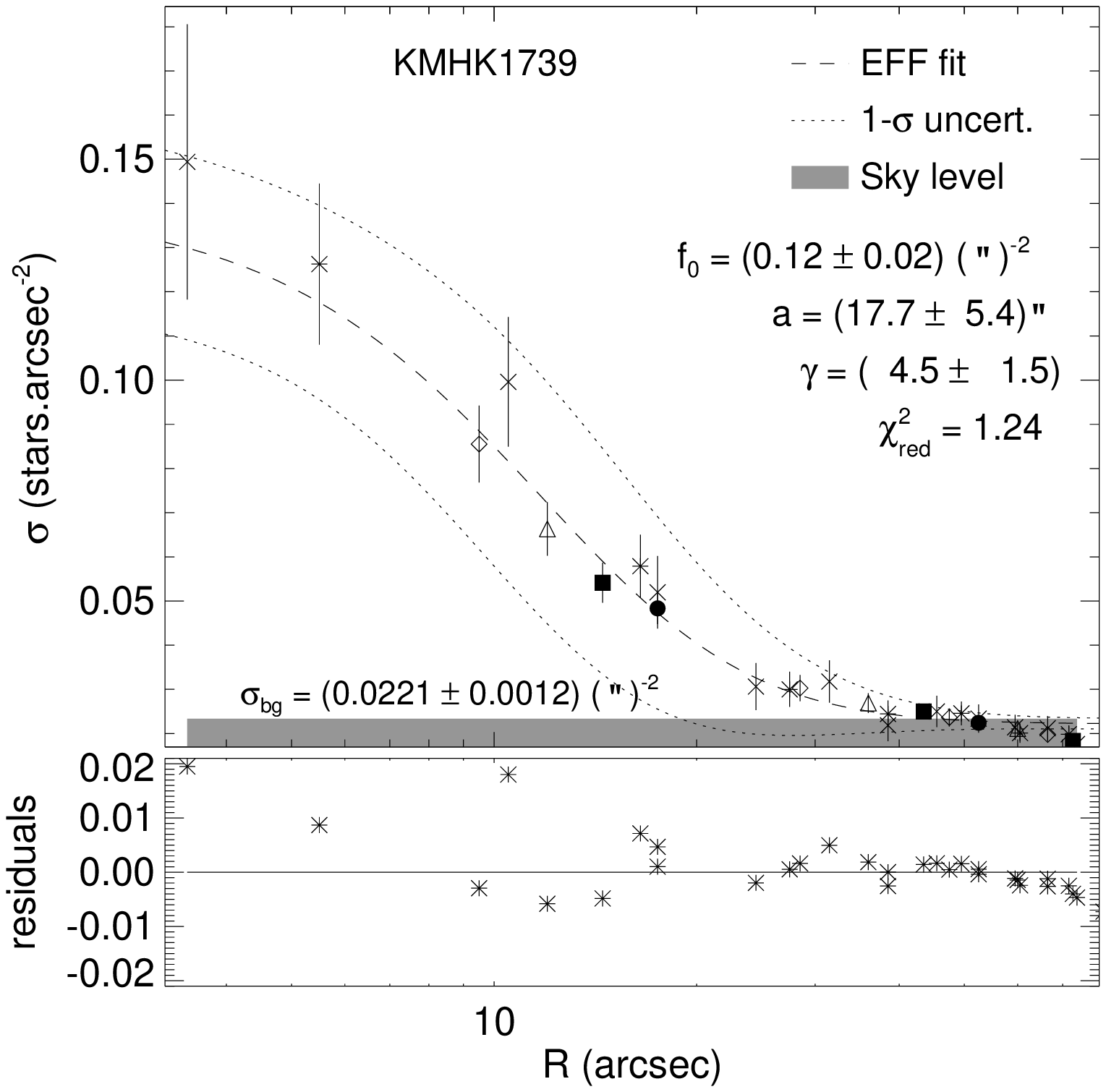}\includegraphics[width=0.325\linewidth]{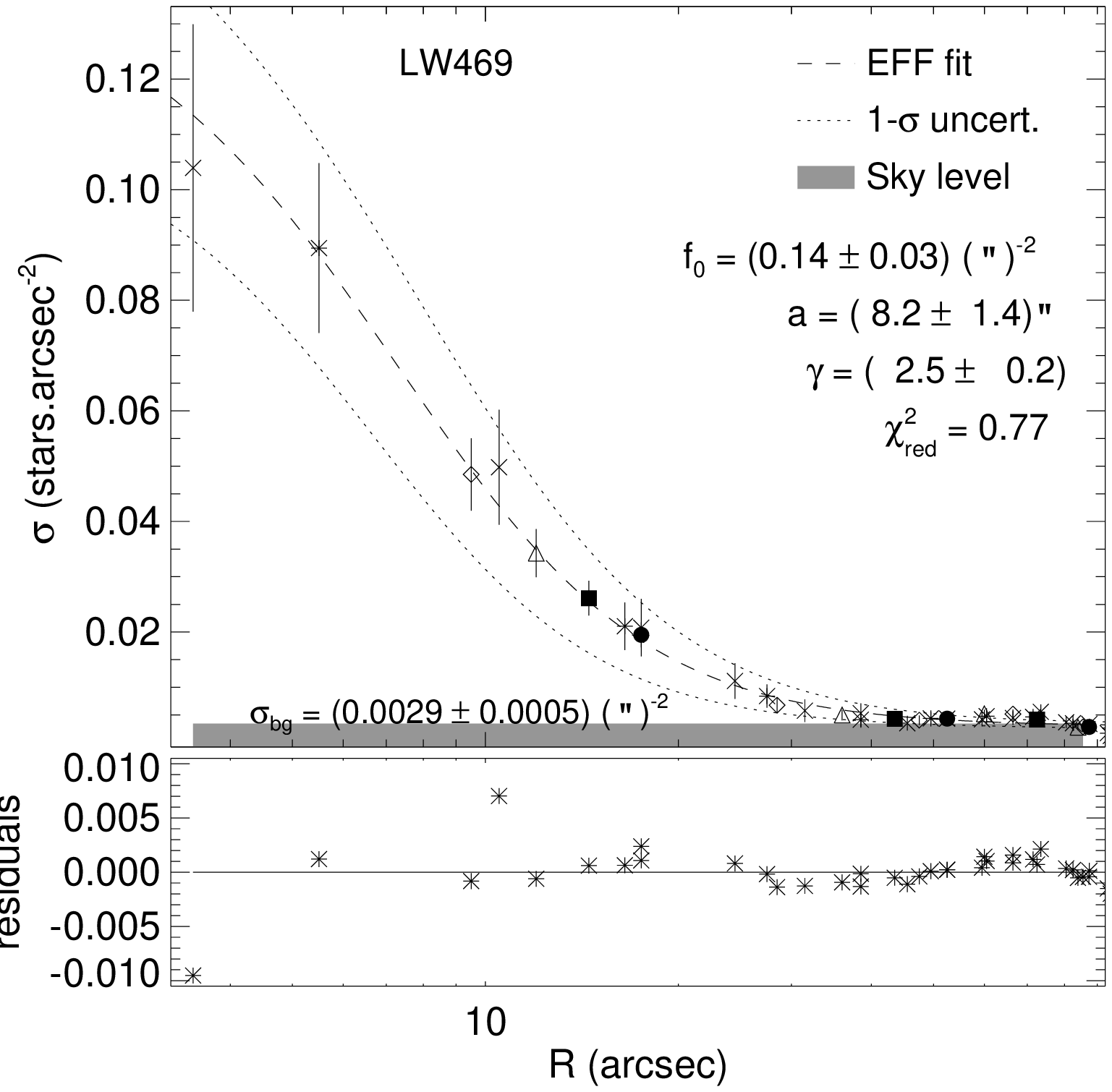}

\includegraphics[width=0.325\linewidth]{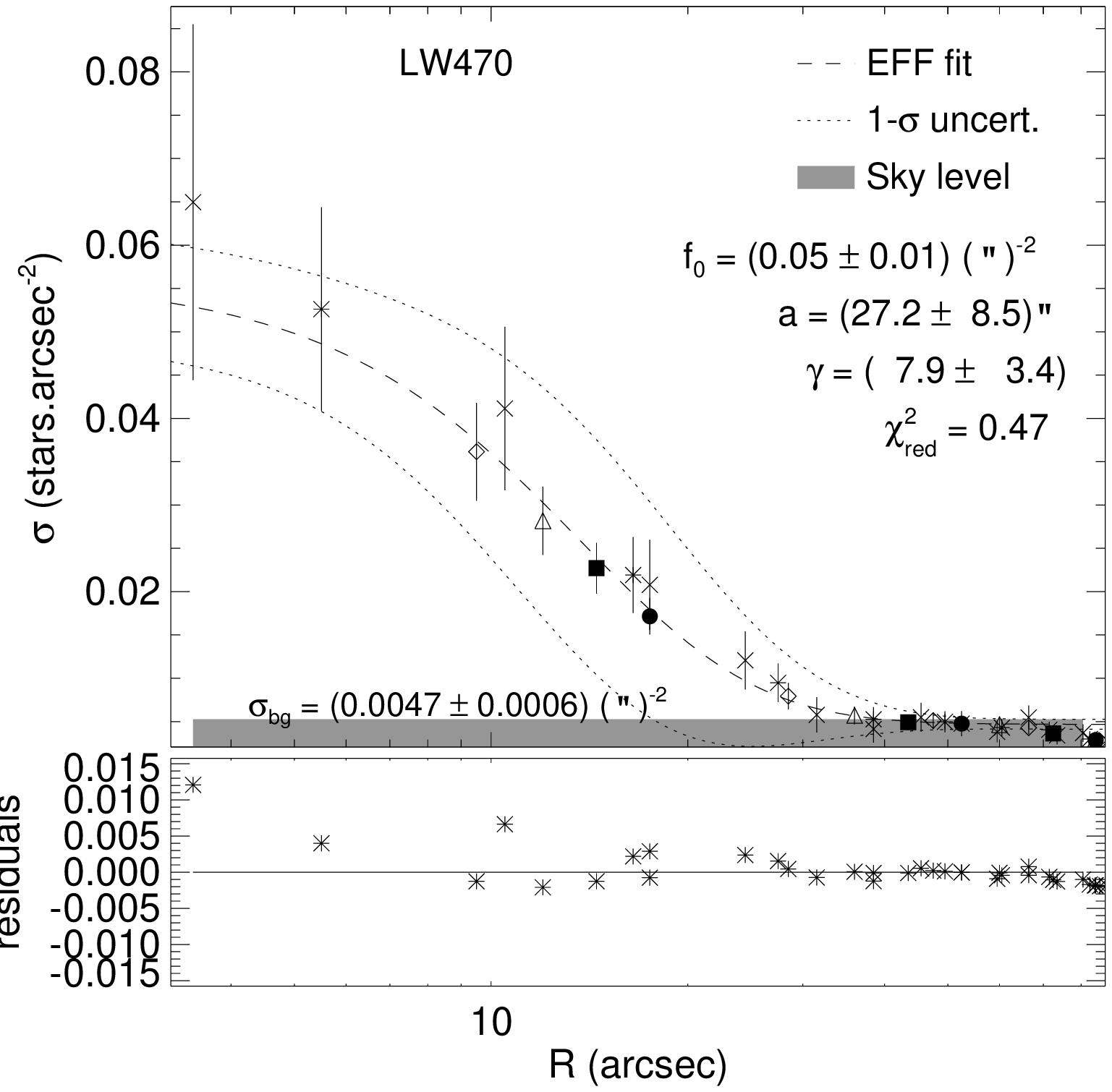}\includegraphics[width=0.325\linewidth]{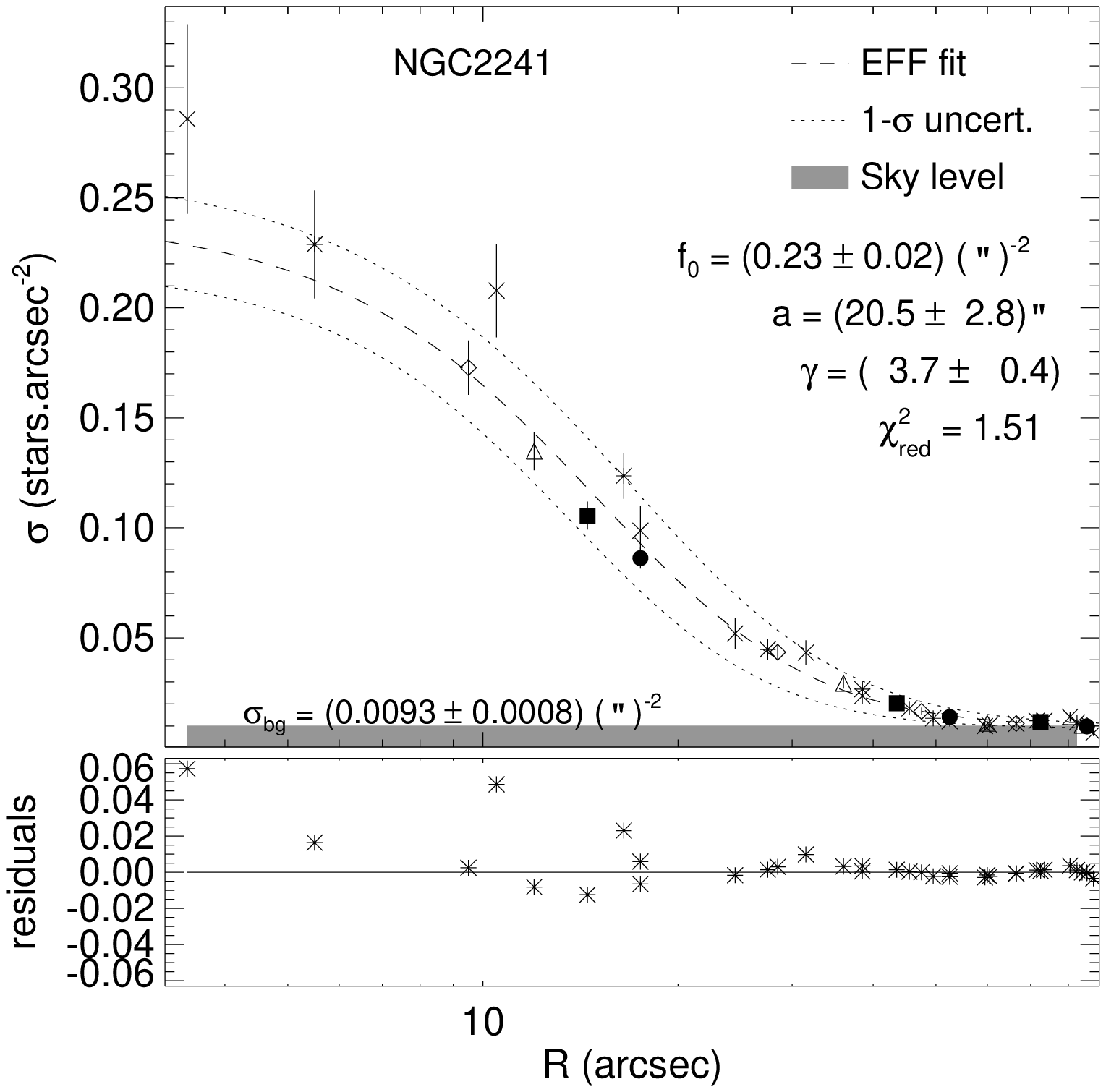}\includegraphics[width=0.325\linewidth]{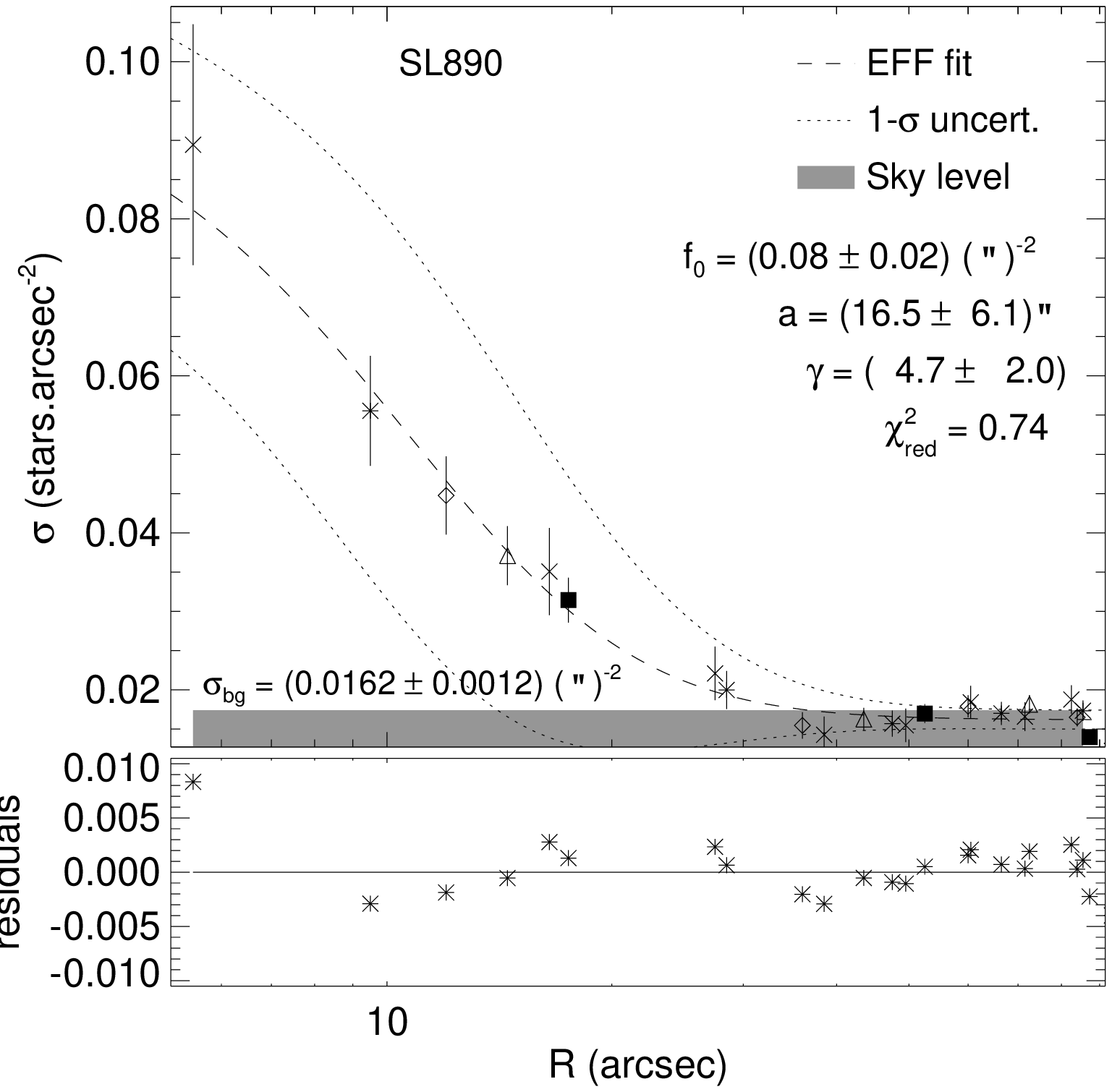}

\includegraphics[width=0.325\linewidth]{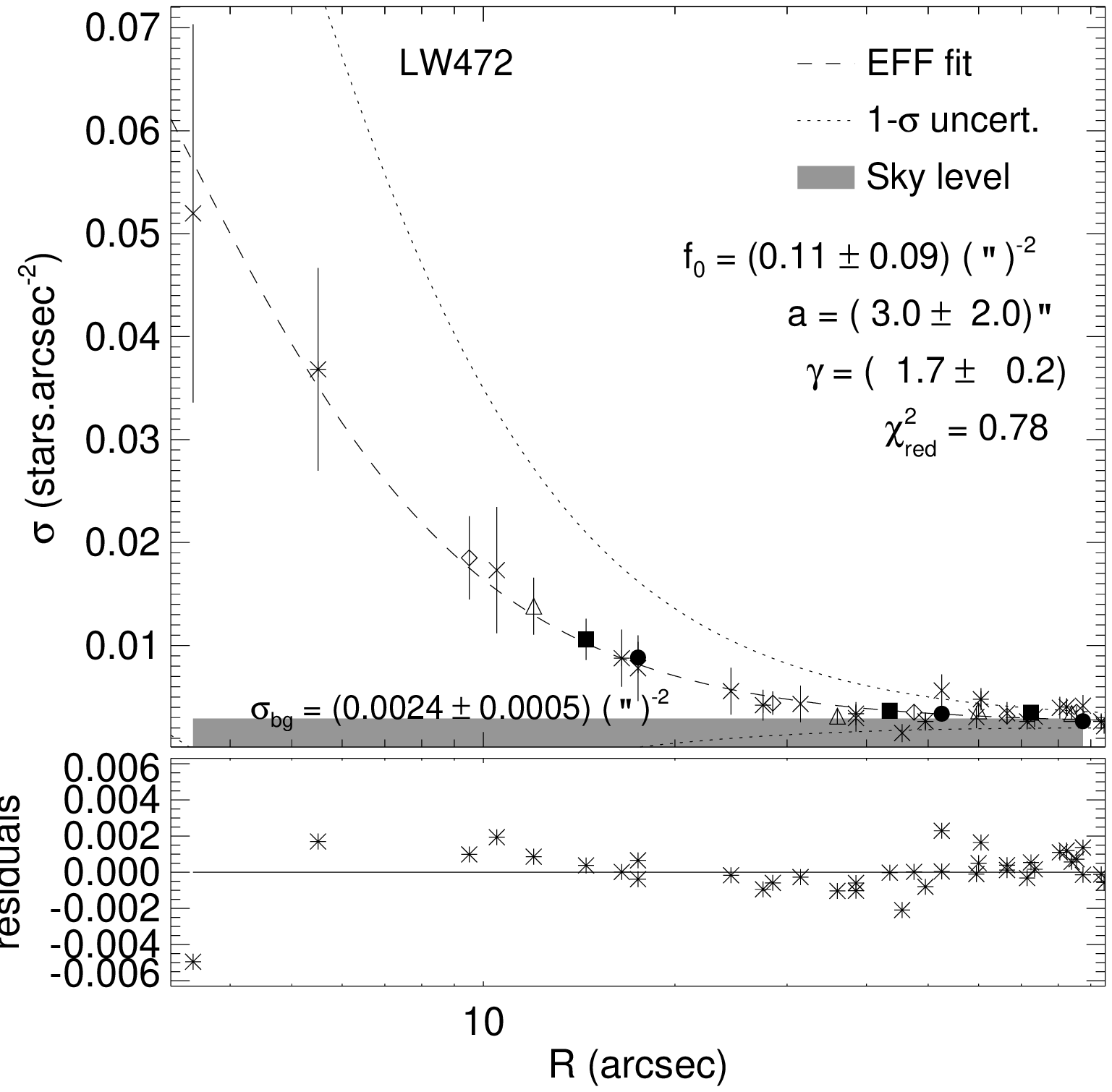}\includegraphics[width=0.325\linewidth]{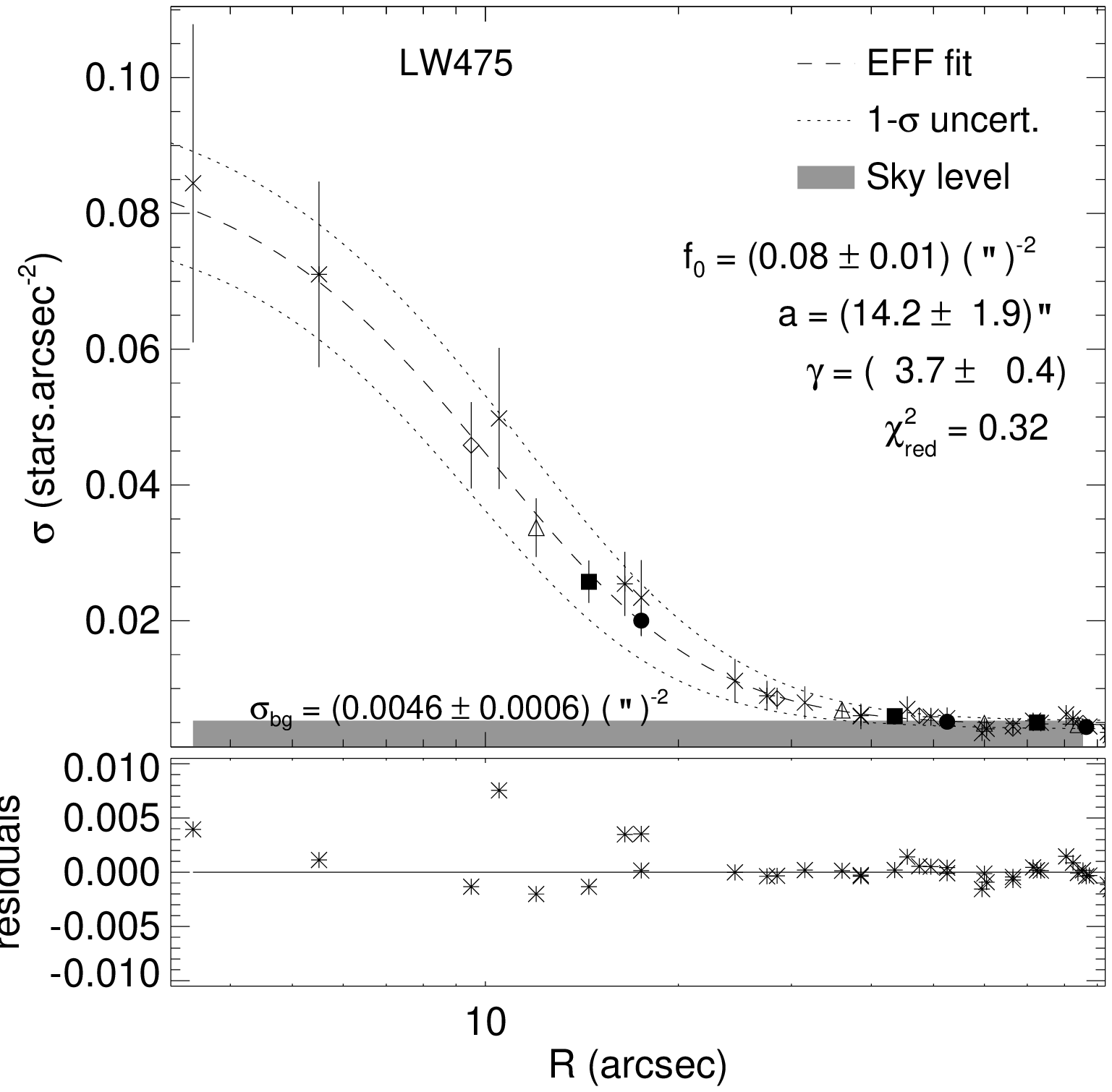}\includegraphics[width=0.325\linewidth]{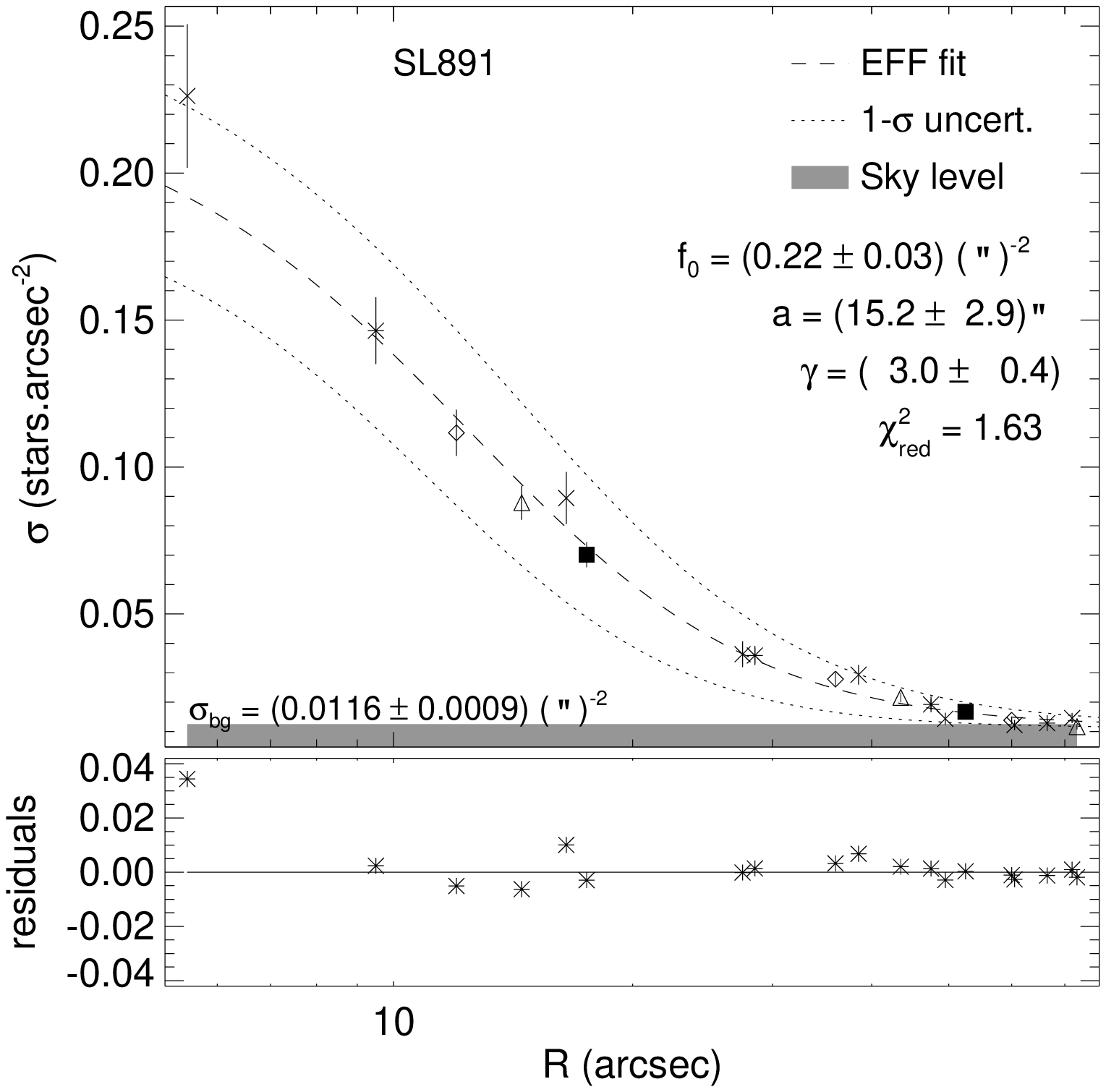}

\includegraphics[width=0.325\linewidth]{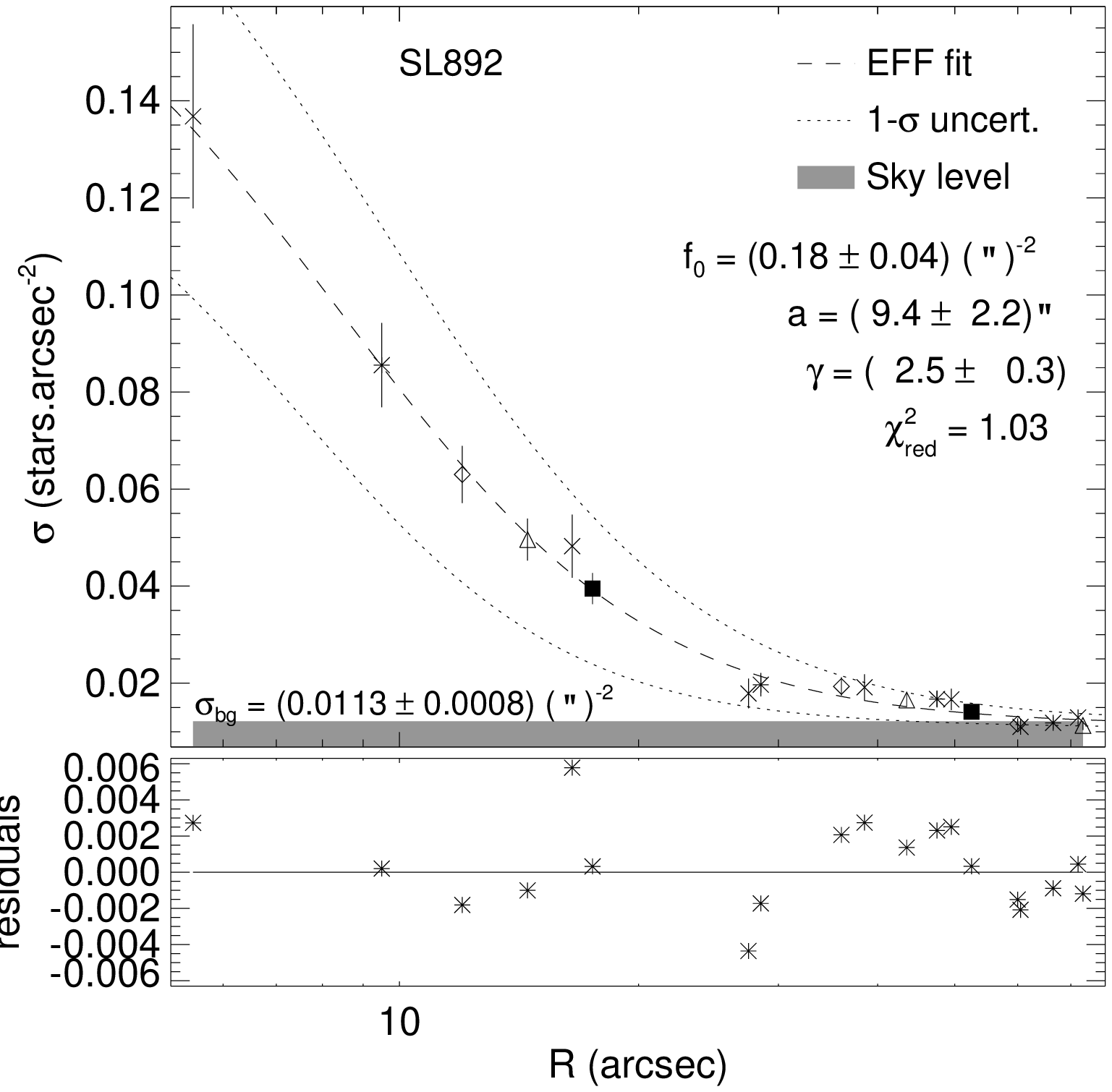}\includegraphics[width=0.325\linewidth]{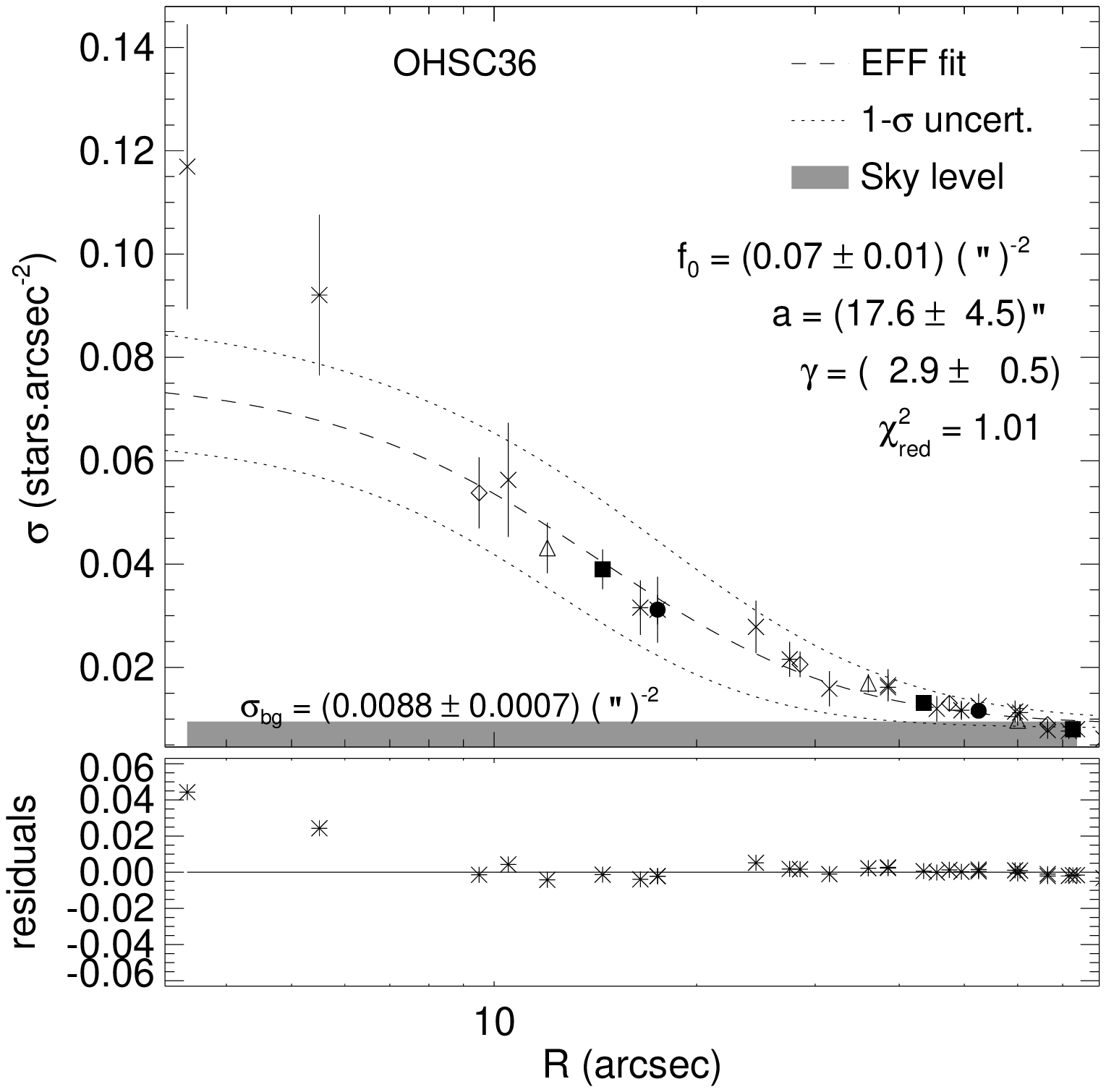}\includegraphics[width=0.325\linewidth]{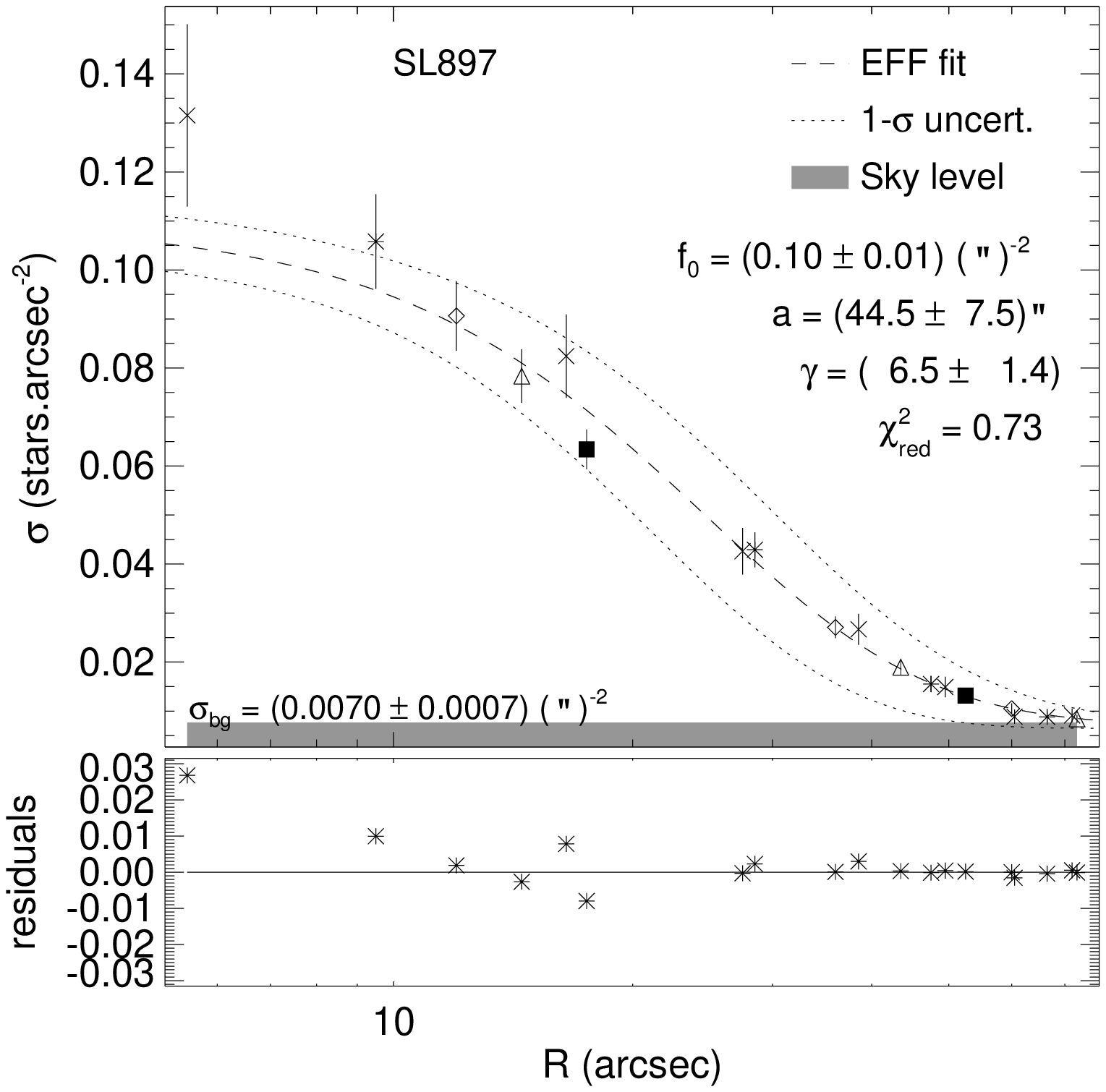}
\caption{cont.}

\end{figure*}

\begin{figure*}
\includegraphics[width=0.325\linewidth]{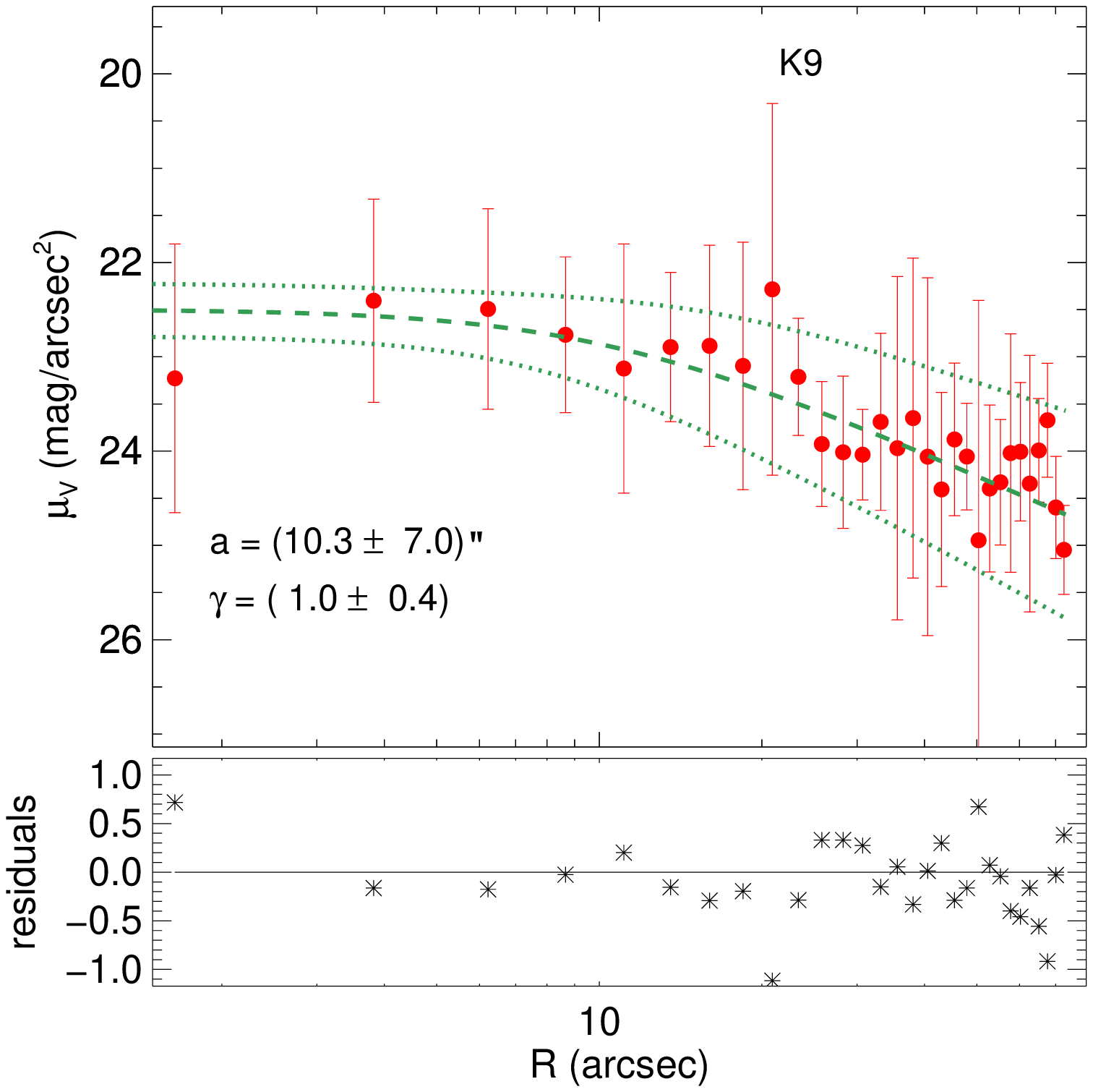}\includegraphics[width=0.325\linewidth]{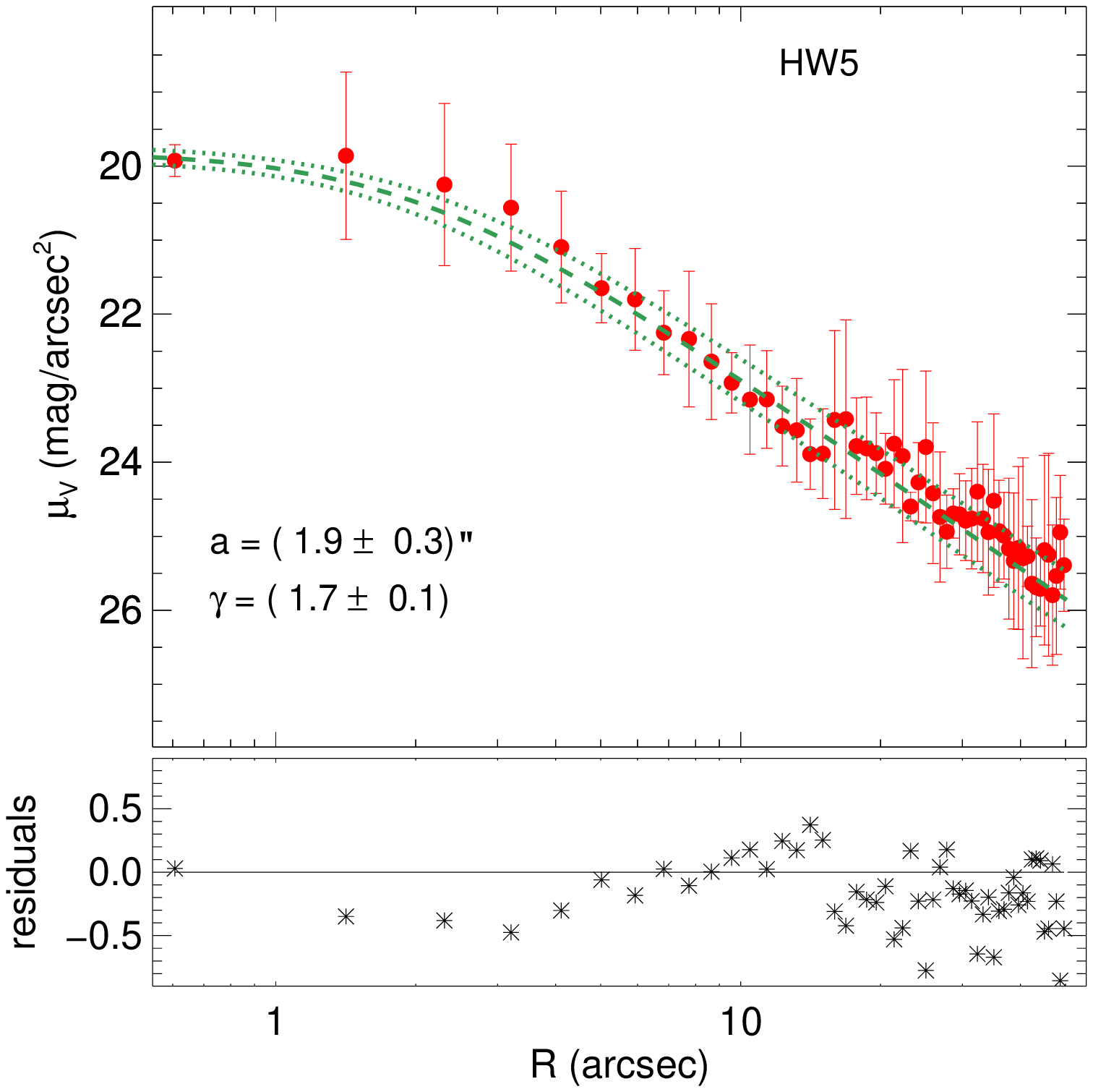}\includegraphics[width=0.325\linewidth]{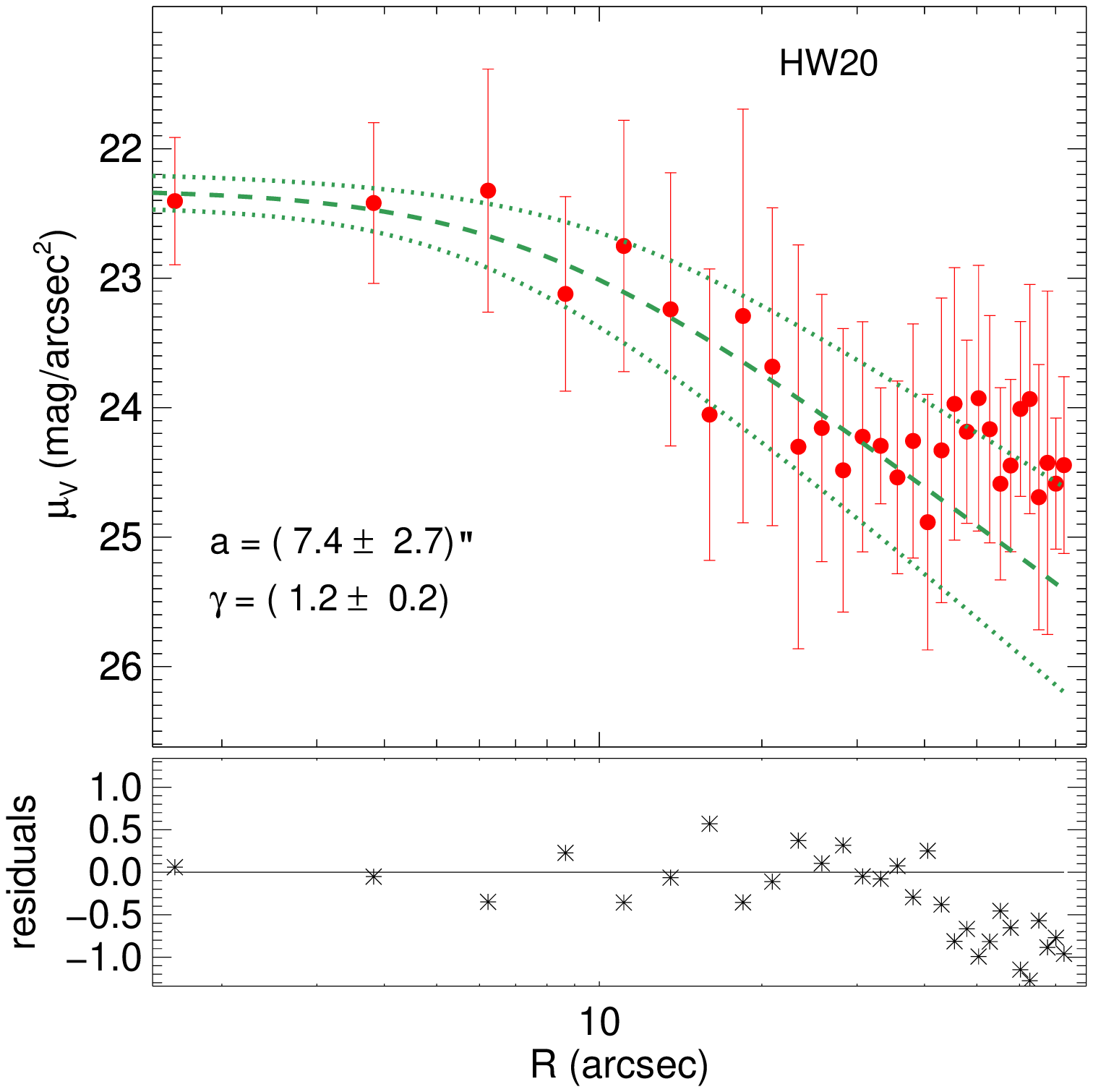}

\includegraphics[width=0.325\linewidth]{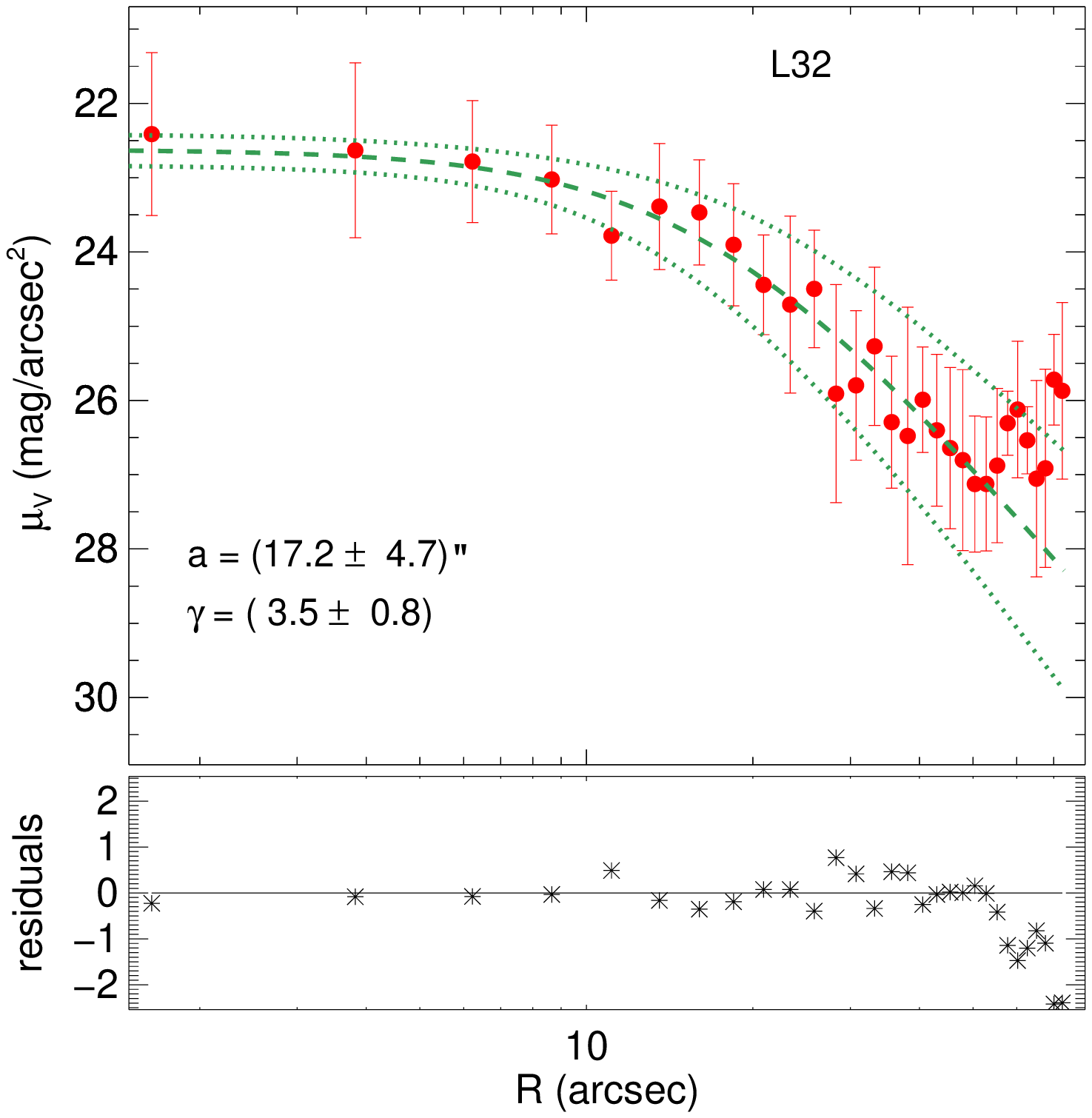}\includegraphics[width=0.325\linewidth]{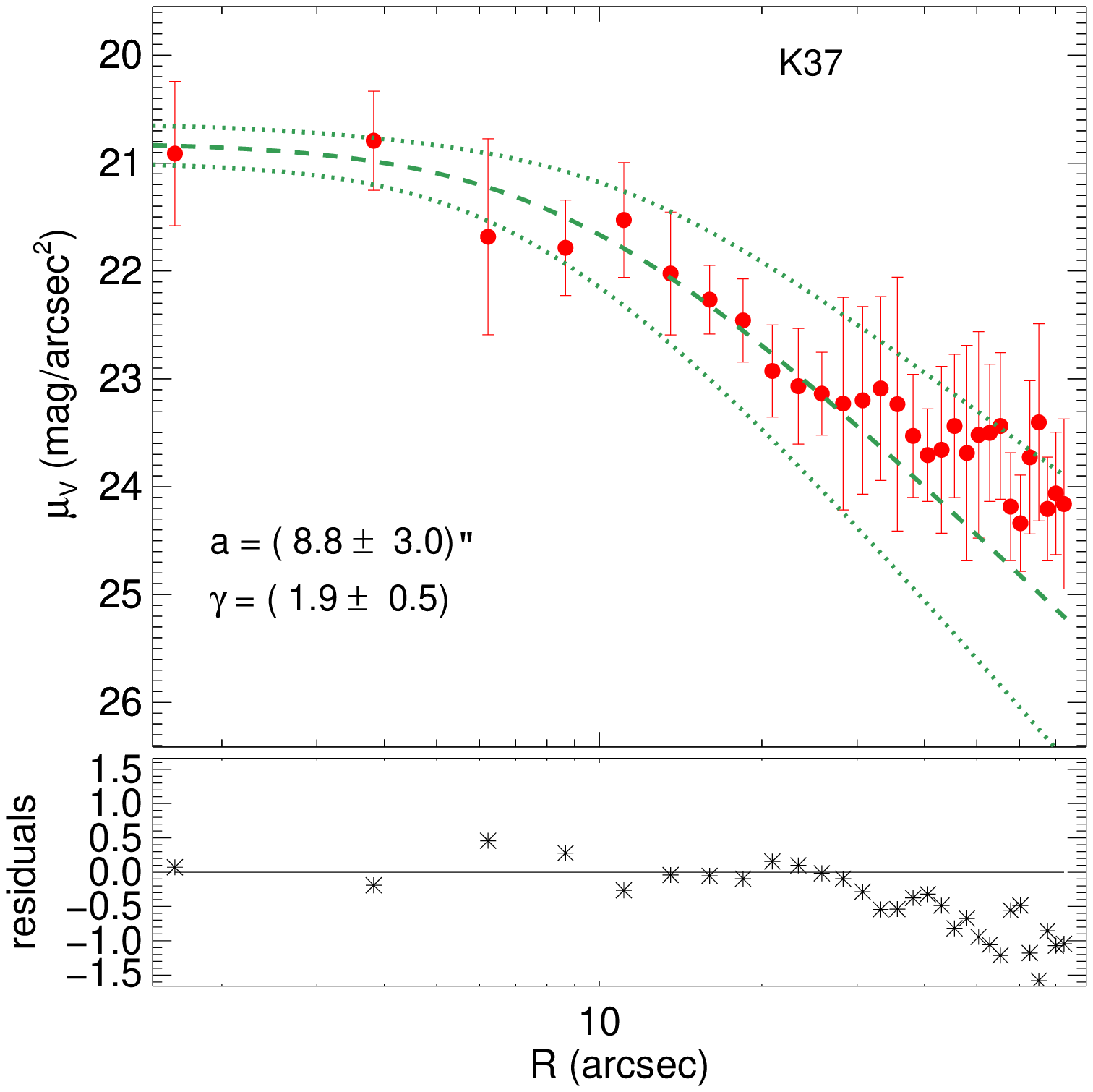}\includegraphics[width=0.325\linewidth]{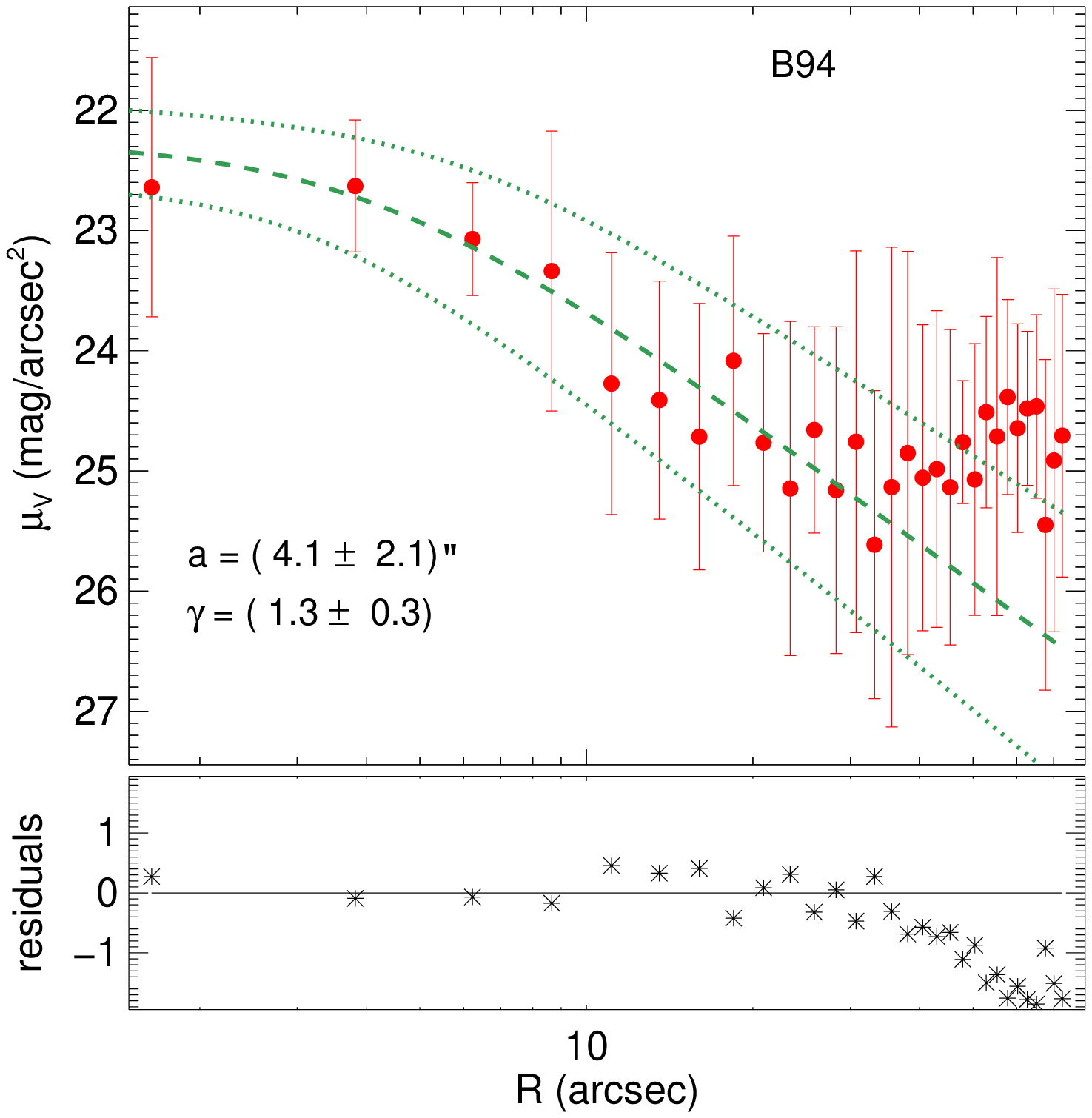}

\includegraphics[width=0.325\linewidth]{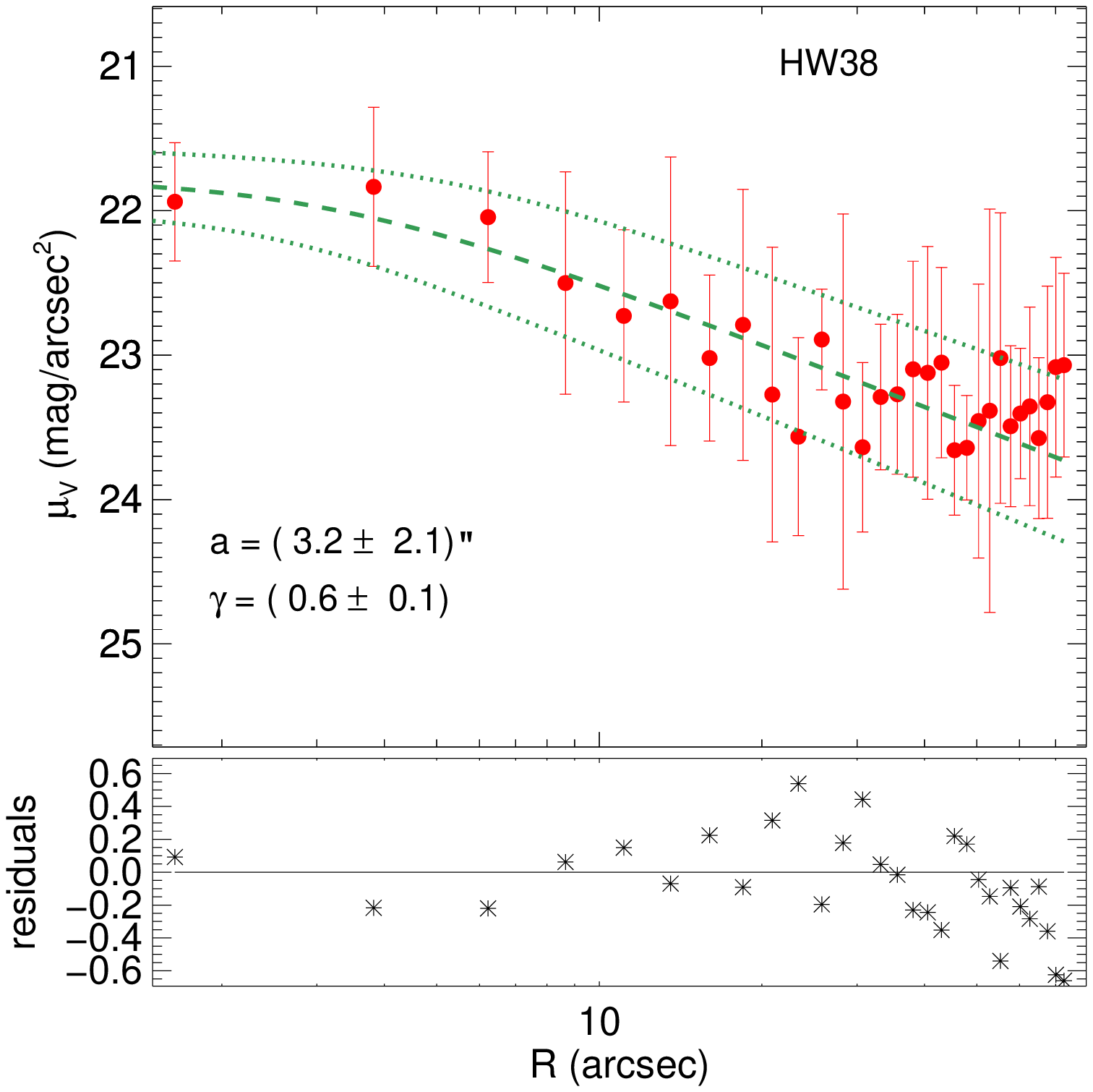}\includegraphics[width=0.325\linewidth]{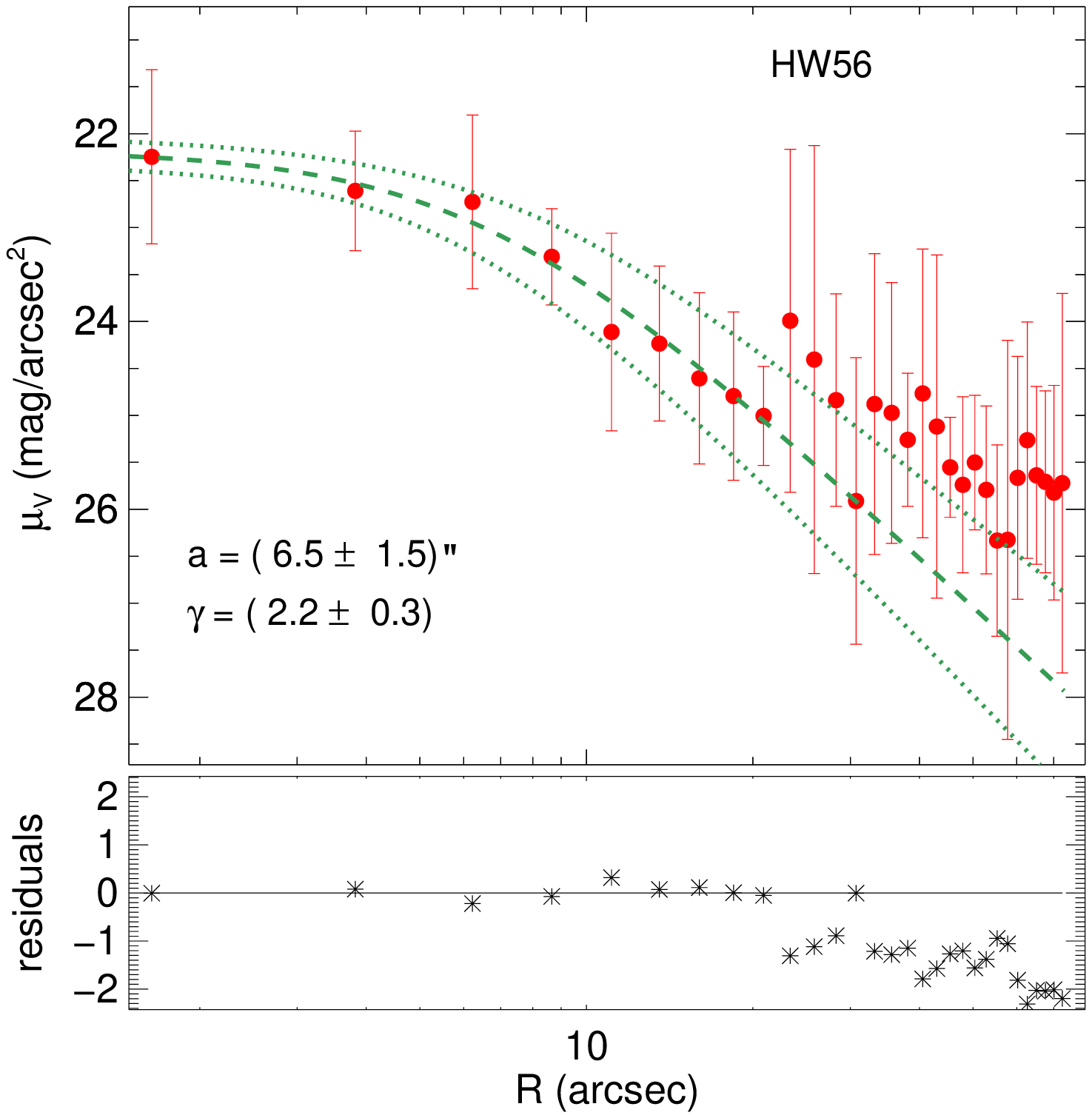}\includegraphics[width=0.325\linewidth]{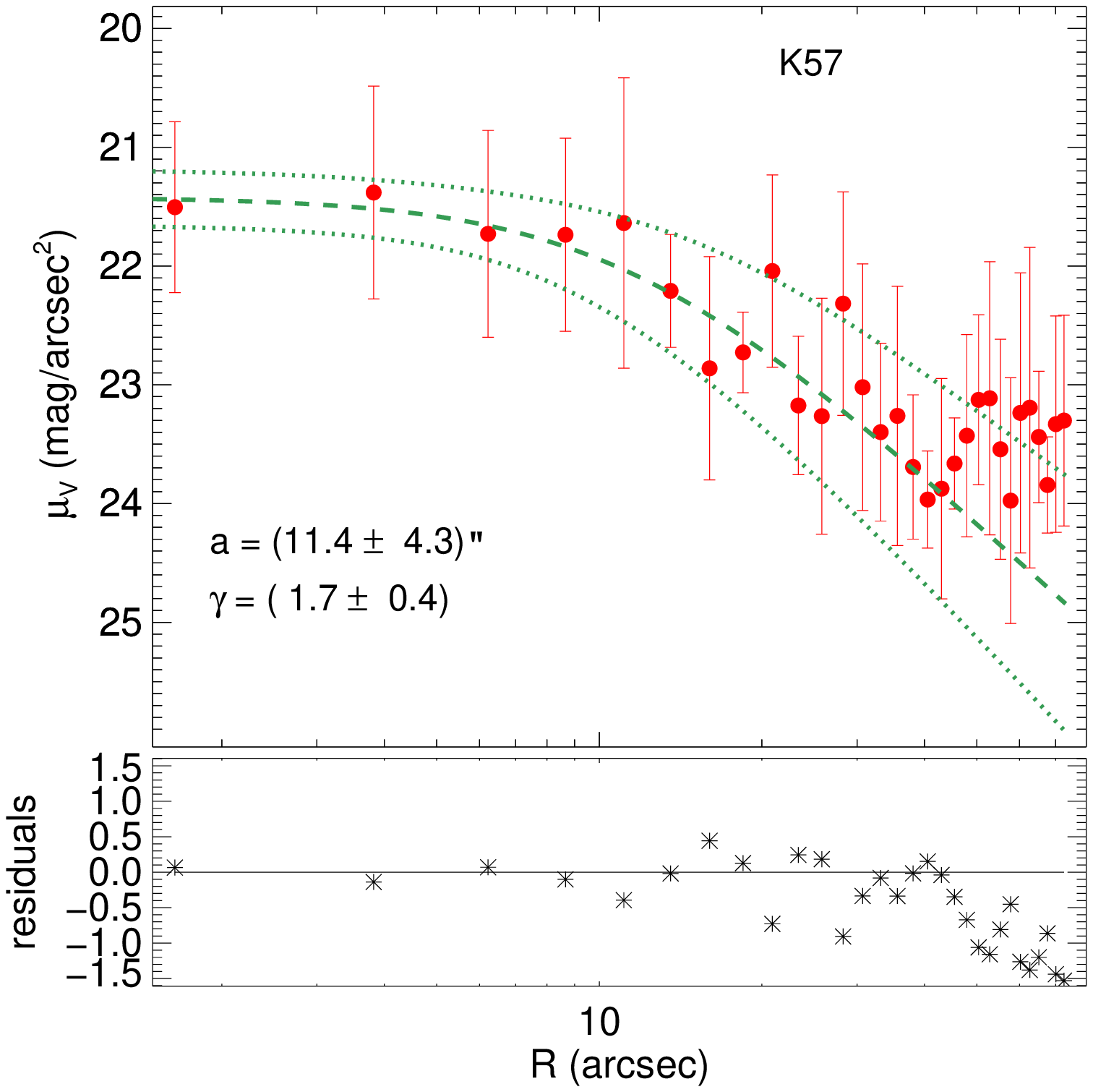}

\includegraphics[width=0.325\linewidth]{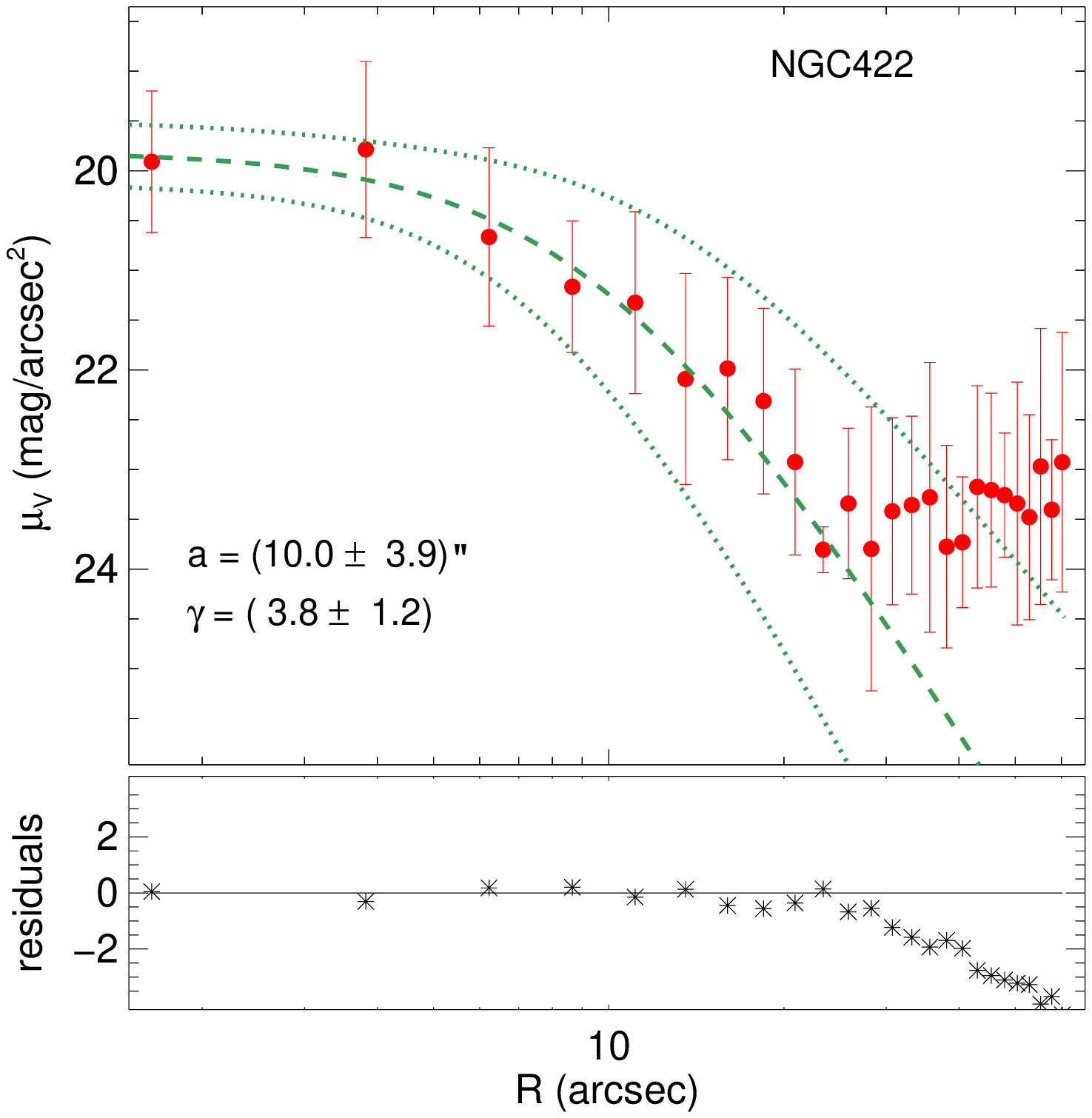}\includegraphics[width=0.325\linewidth]{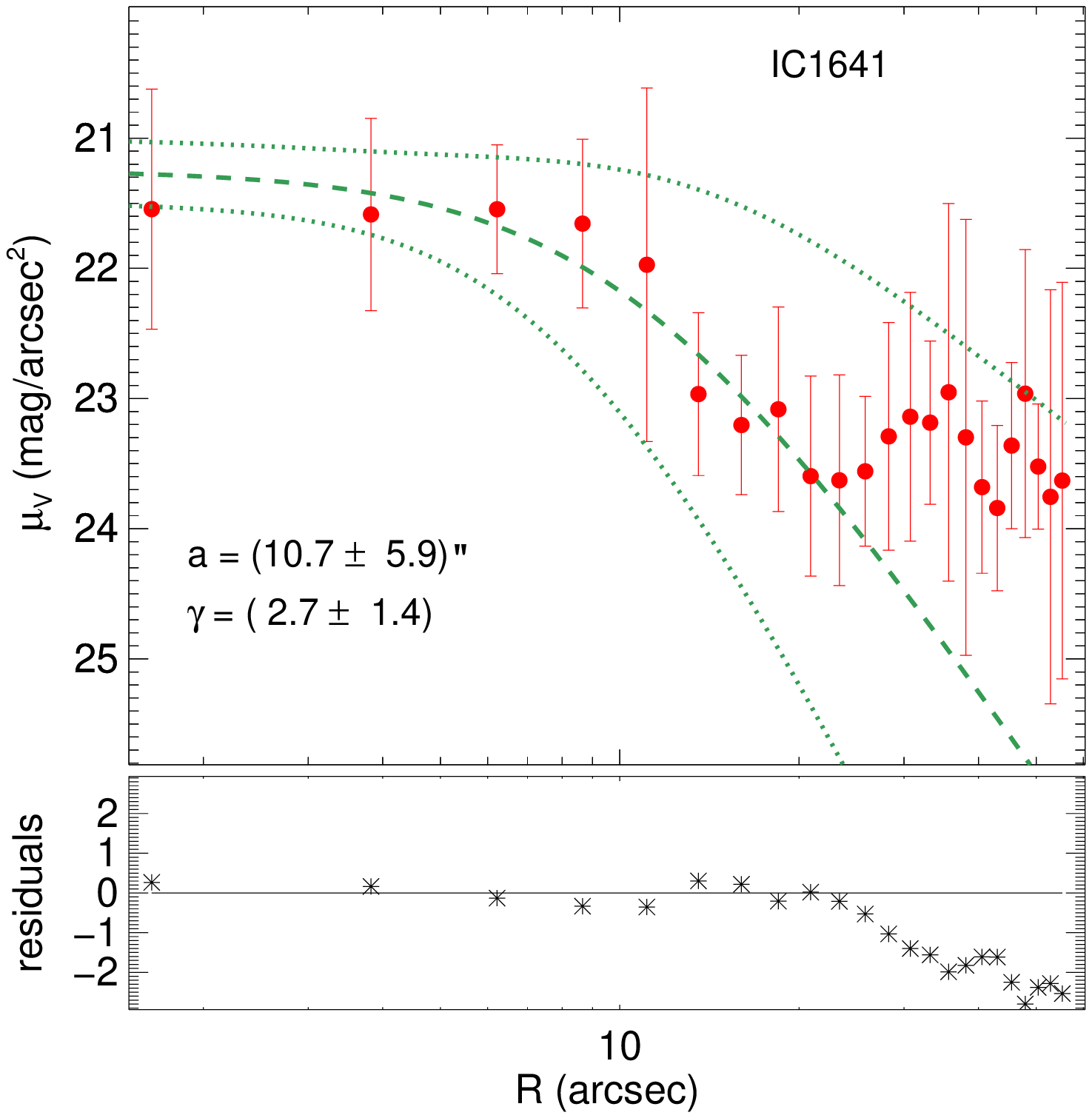}\includegraphics[width=0.325\linewidth]{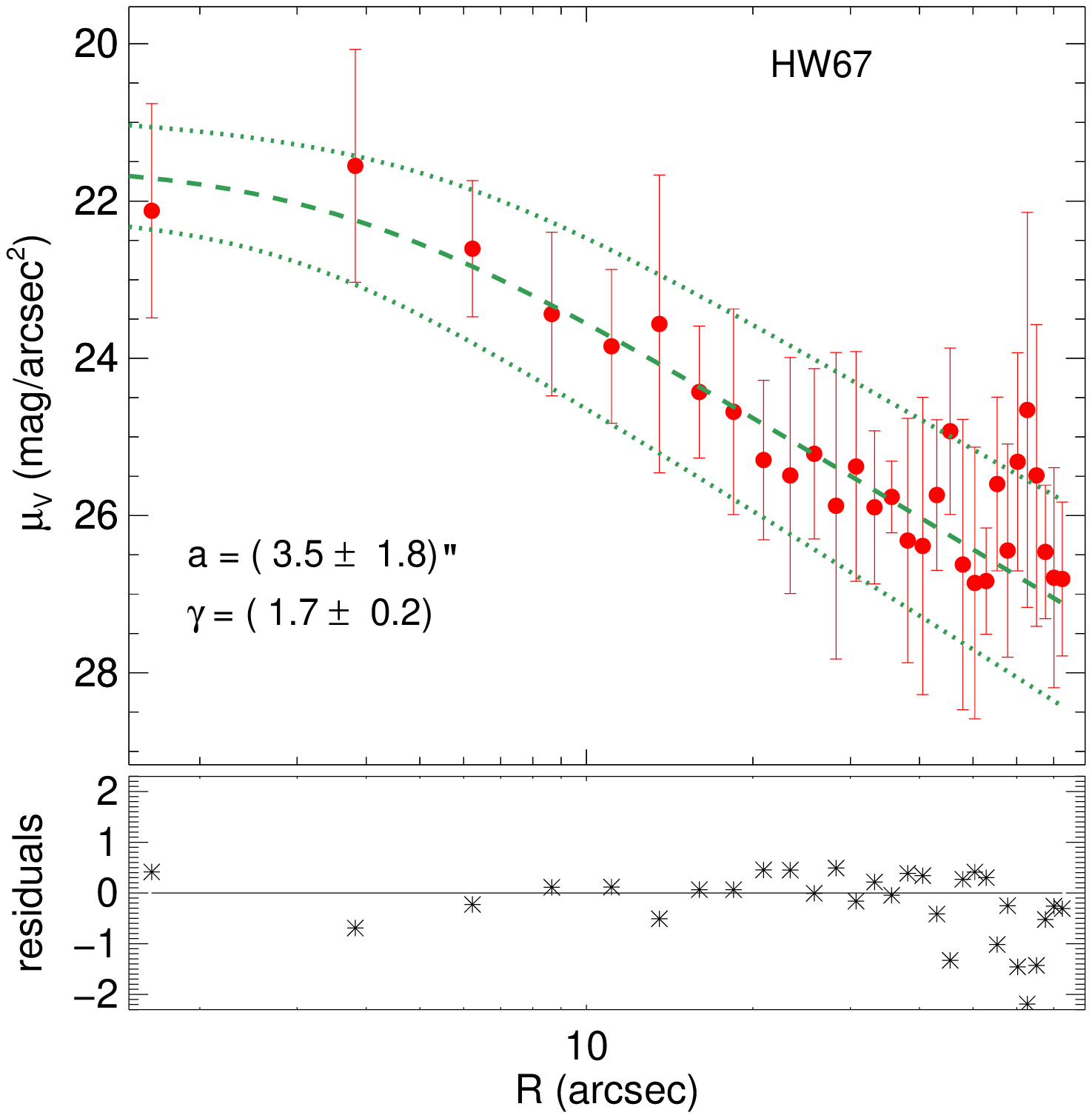}

\caption{Surface brightness profiles of additional SMC clusters complementing the sample presented in Fig.~\ref{fig:rdp_sbp}. The EFF model fits (dashed lines)  and 1\,$\sigma$ uncertainties (dotted lines) are shown. The best-fitting  parameters are indicated and the fit residuals are plotted in the lower panel.}

\end{figure*}

\setcounter{figure}{6}

\begin{figure*}

\includegraphics[width=0.325\linewidth]{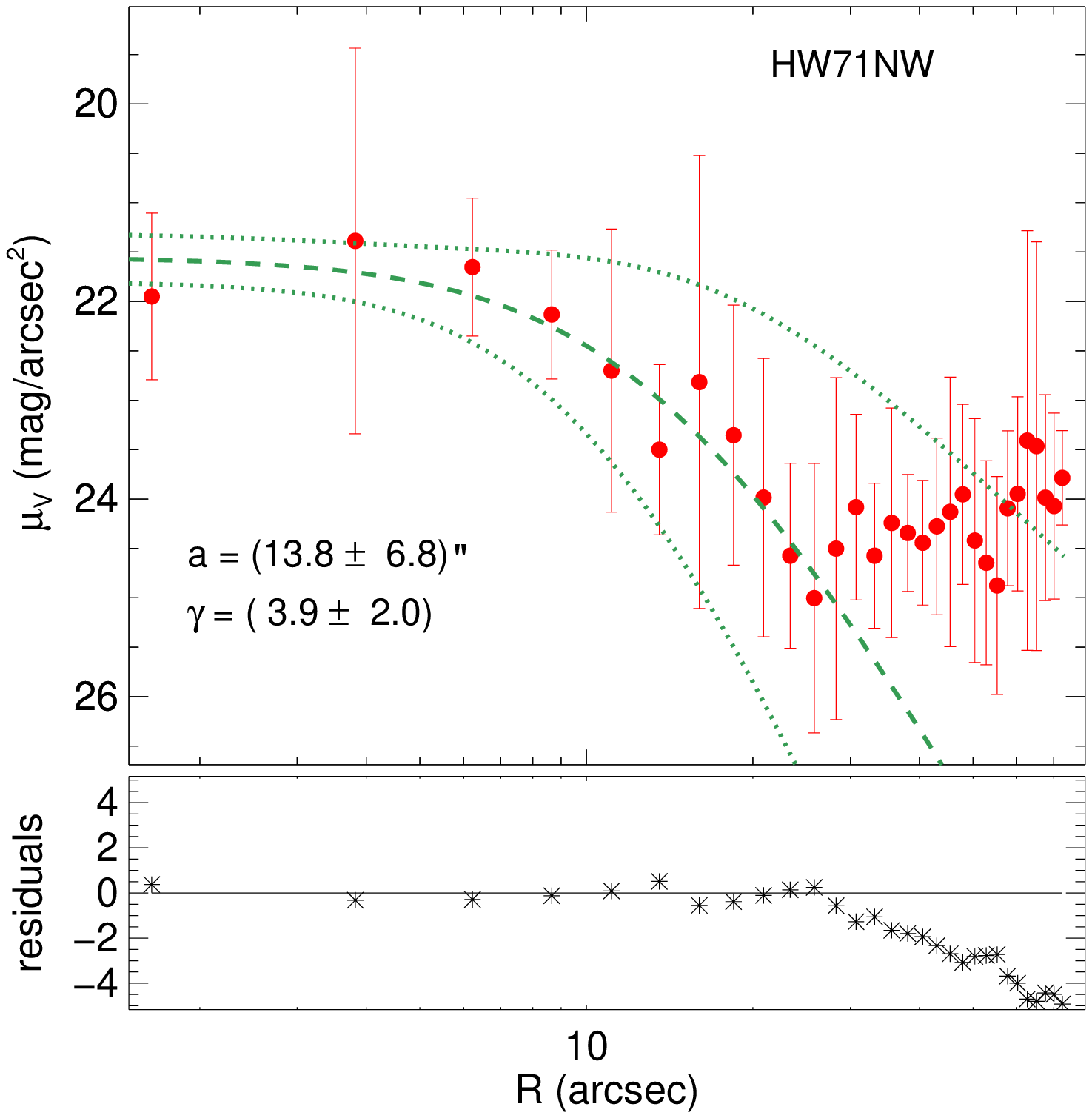}\includegraphics[width=0.325\linewidth]{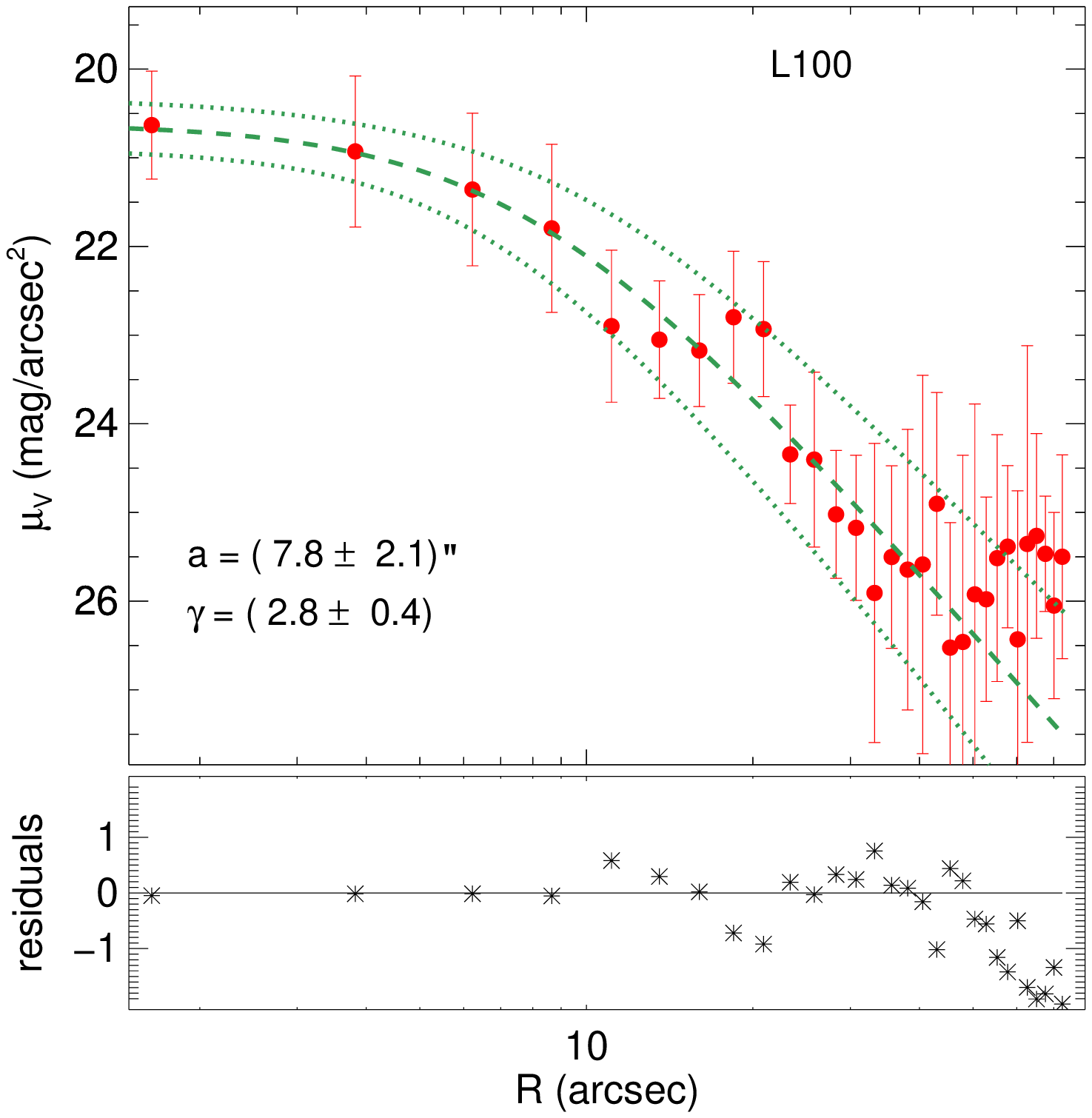}\includegraphics[width=0.325\linewidth]{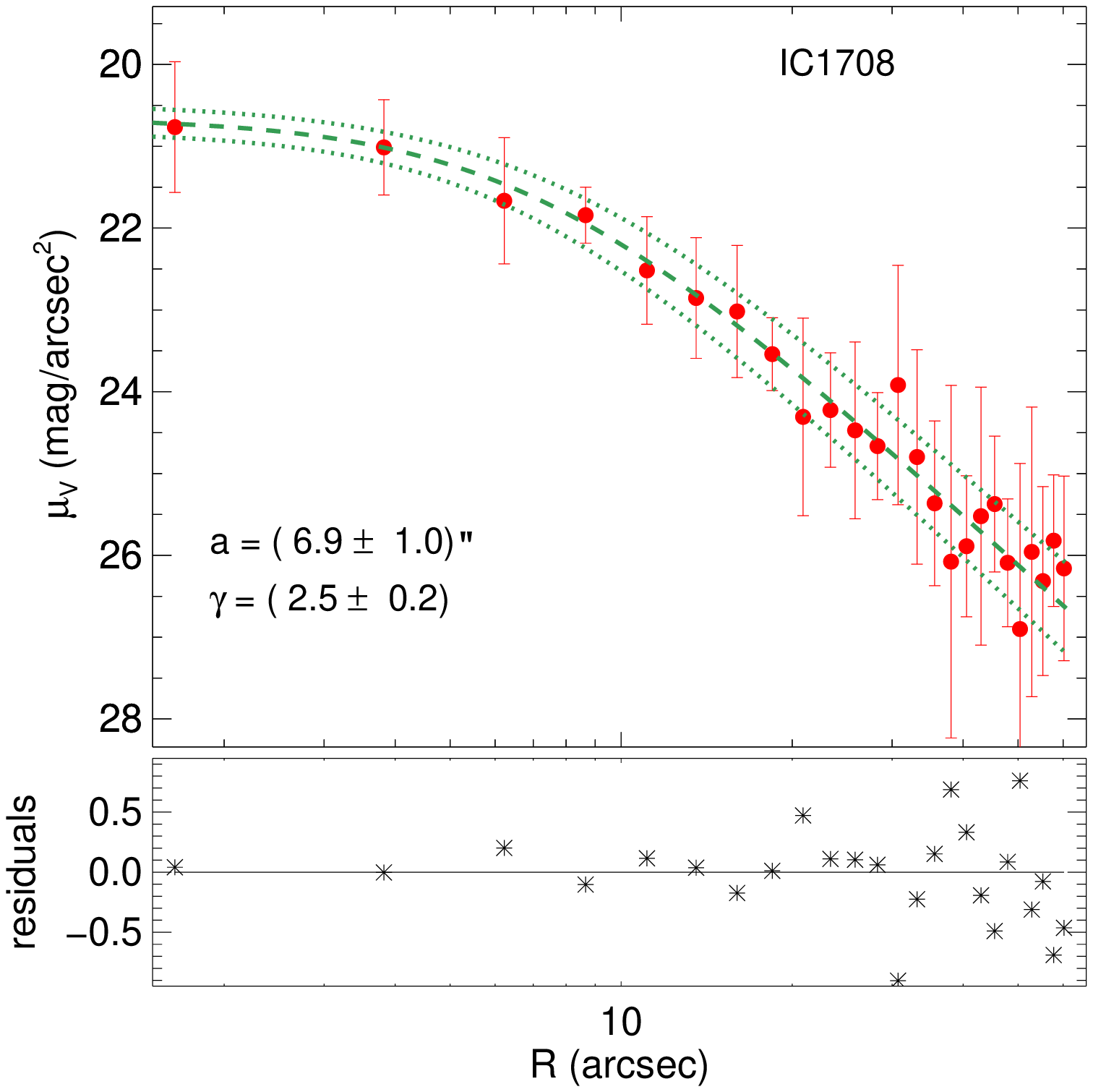}

\includegraphics[width=0.325\linewidth]{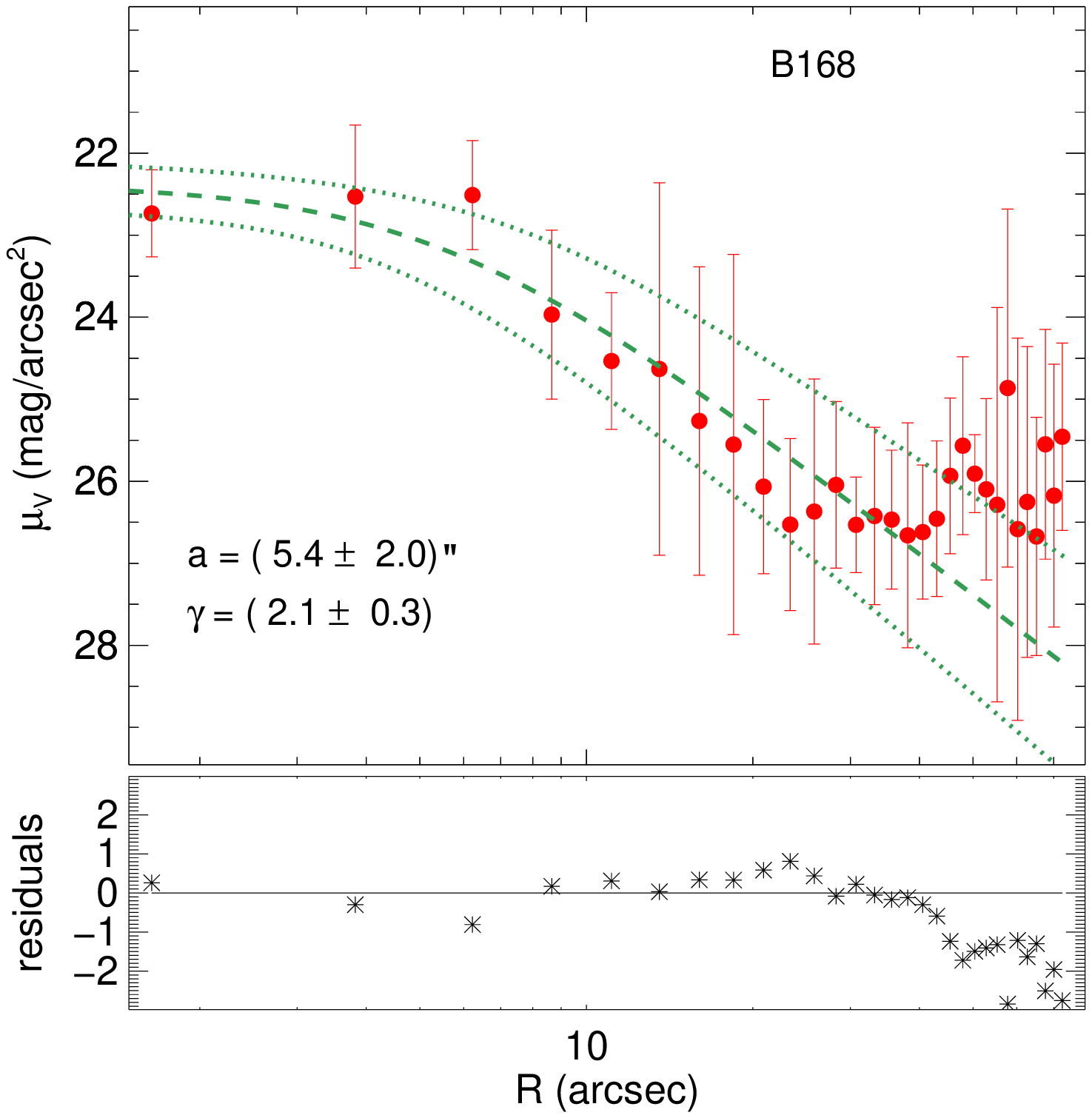}\includegraphics[width=0.325\linewidth]{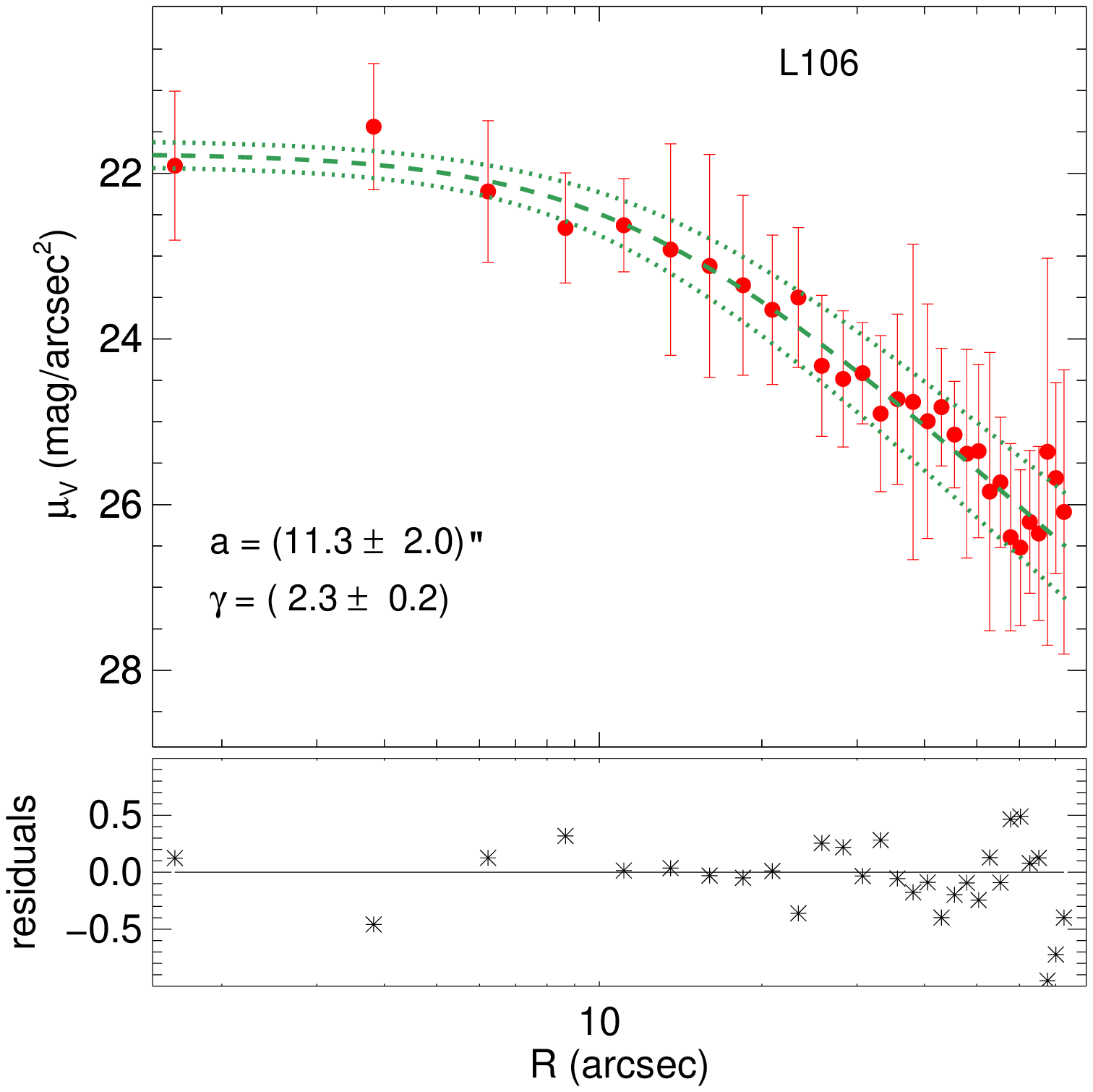}\includegraphics[width=0.325\linewidth]{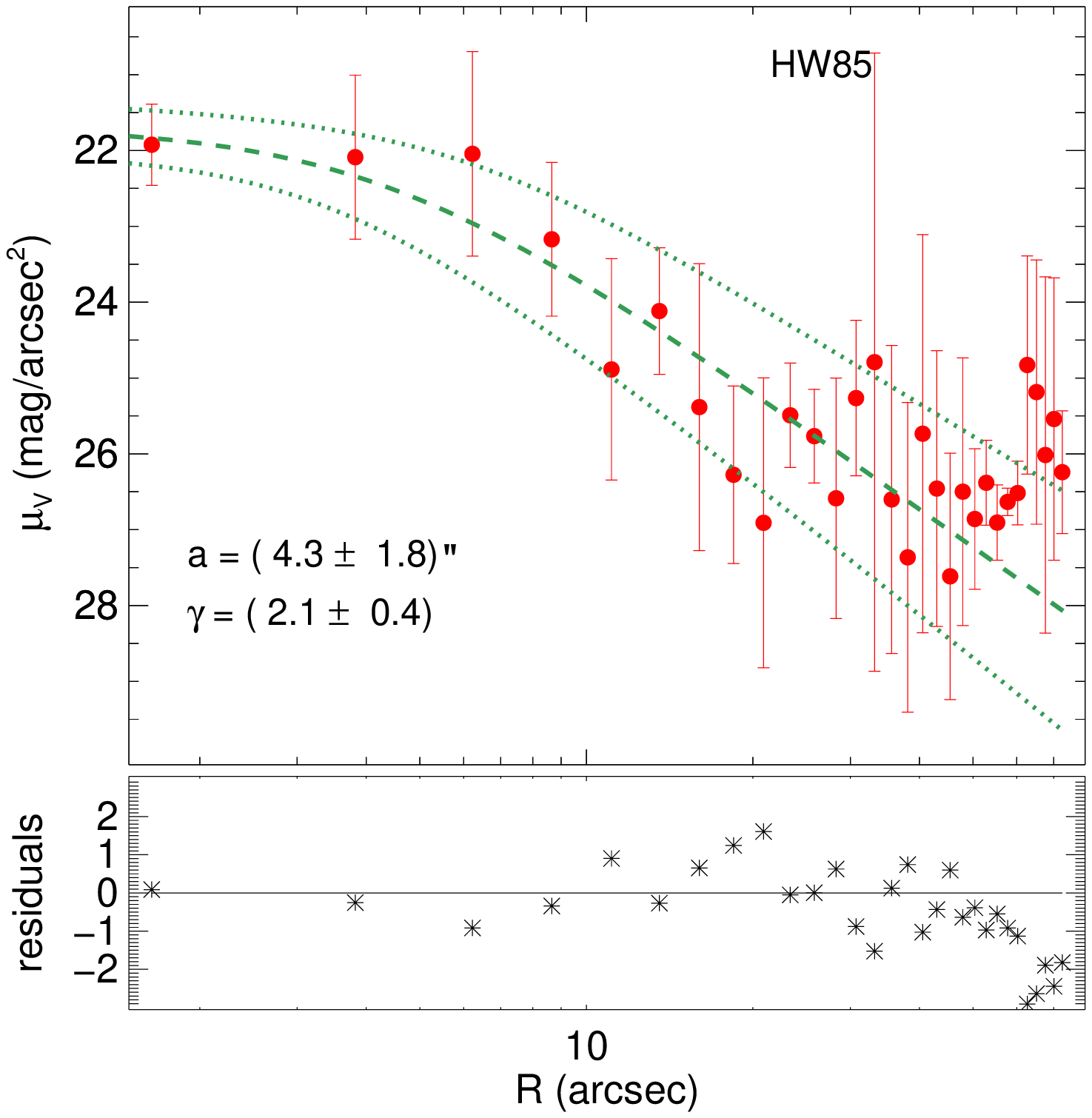}

\includegraphics[width=0.325\linewidth]{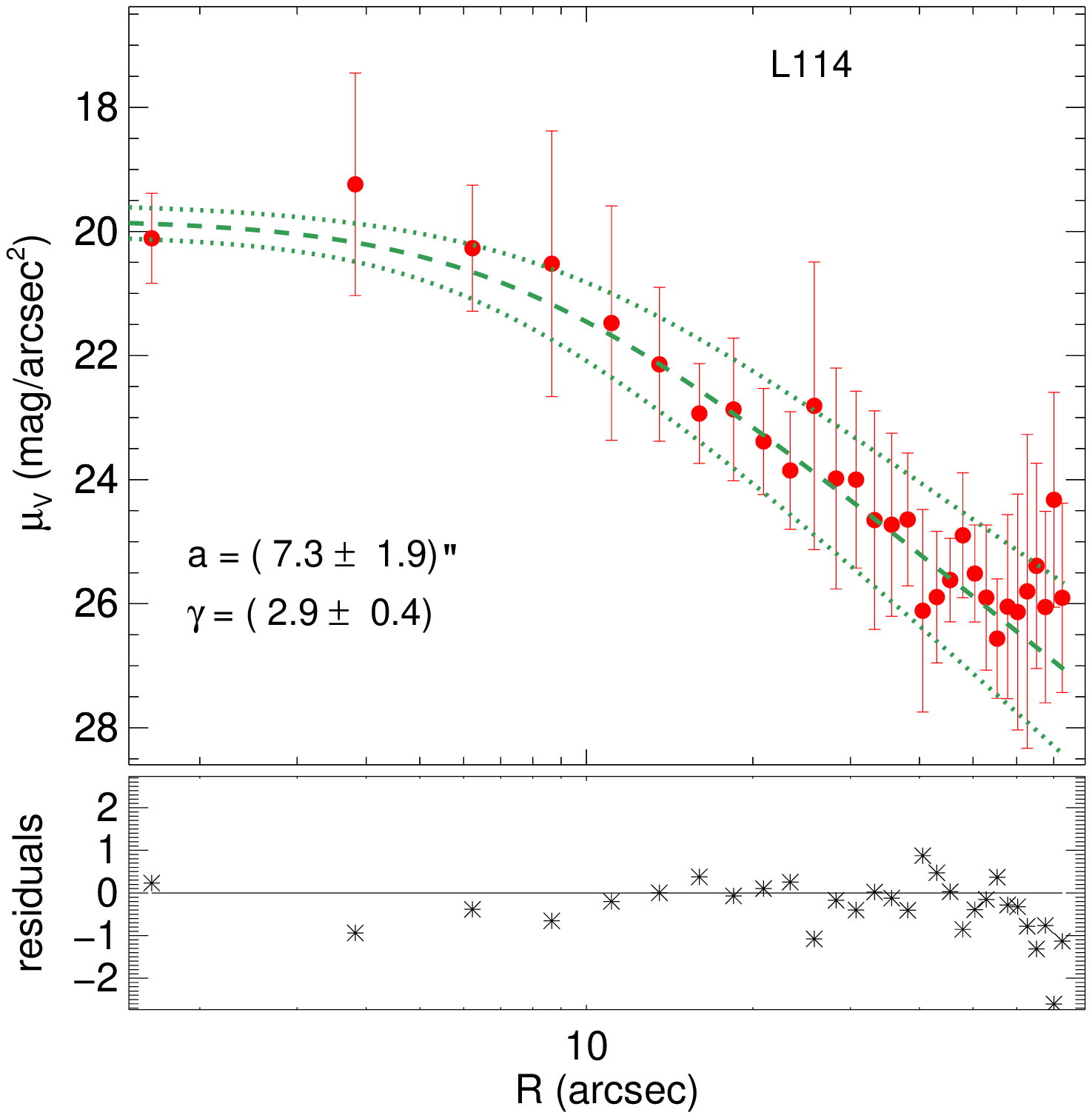}\includegraphics[width=0.325\linewidth]{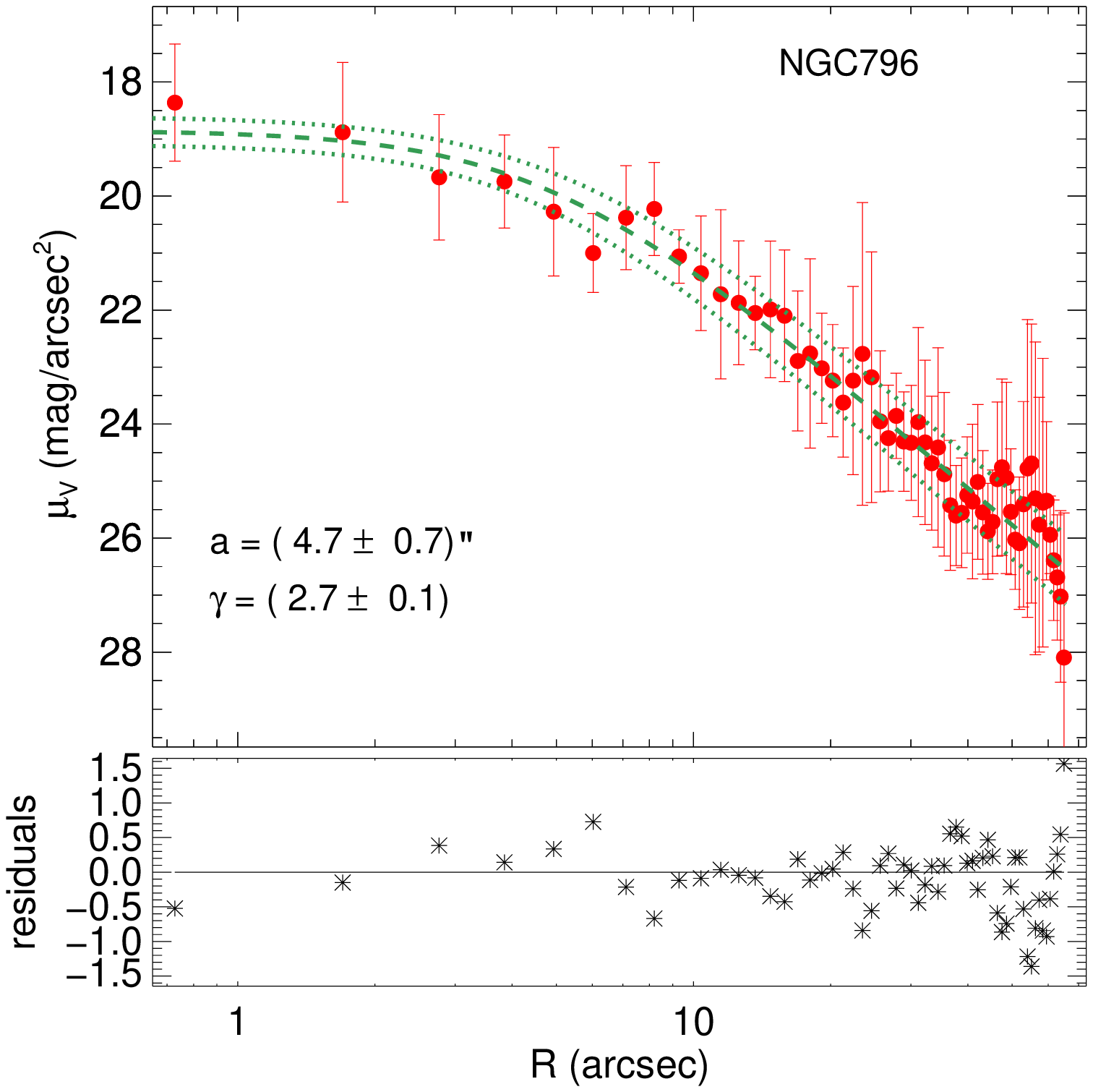}\includegraphics[width=0.325\linewidth]{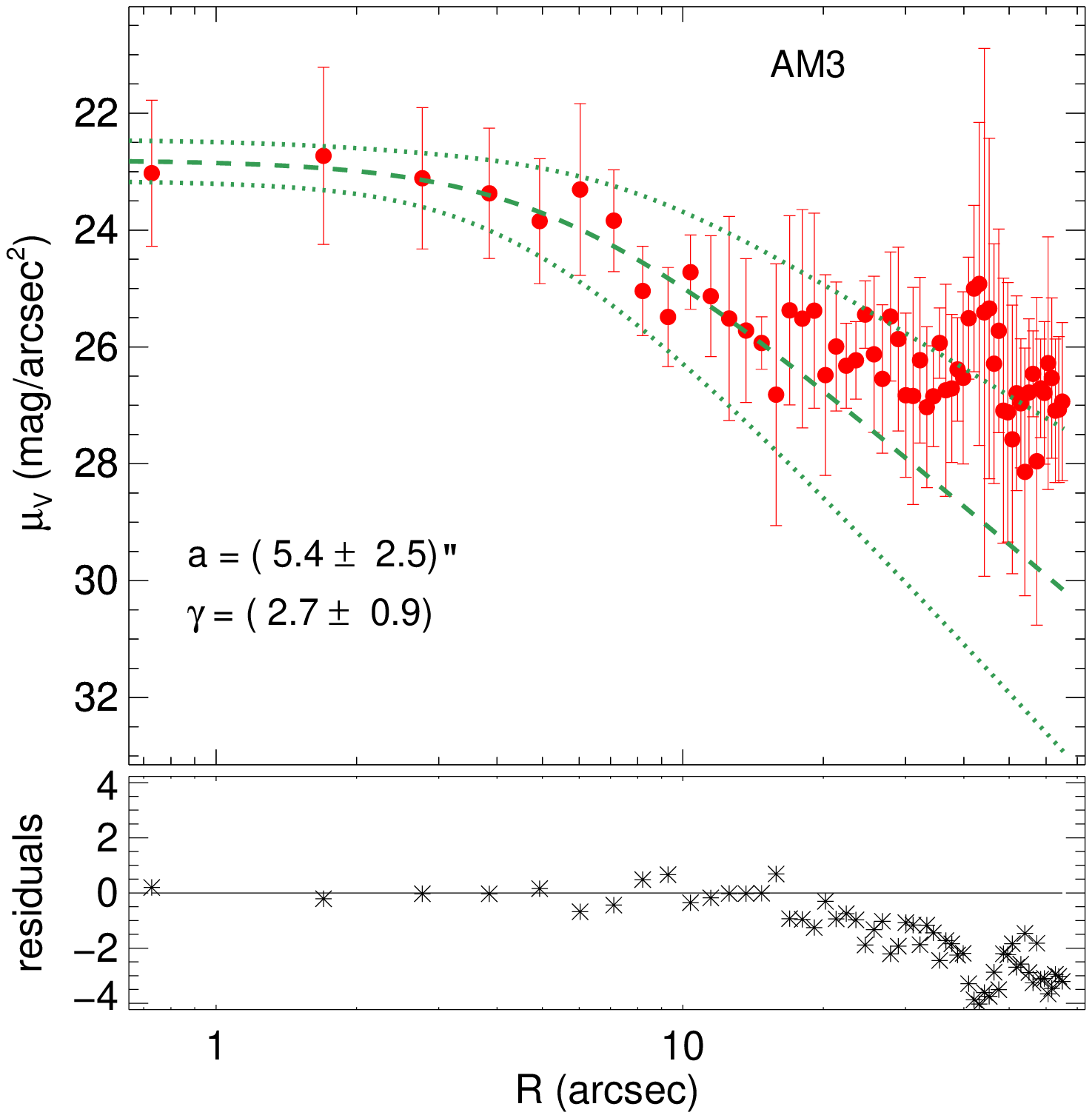}
\caption{cont.}

\end{figure*}

\begin{figure*}
\includegraphics[width=0.325\linewidth]{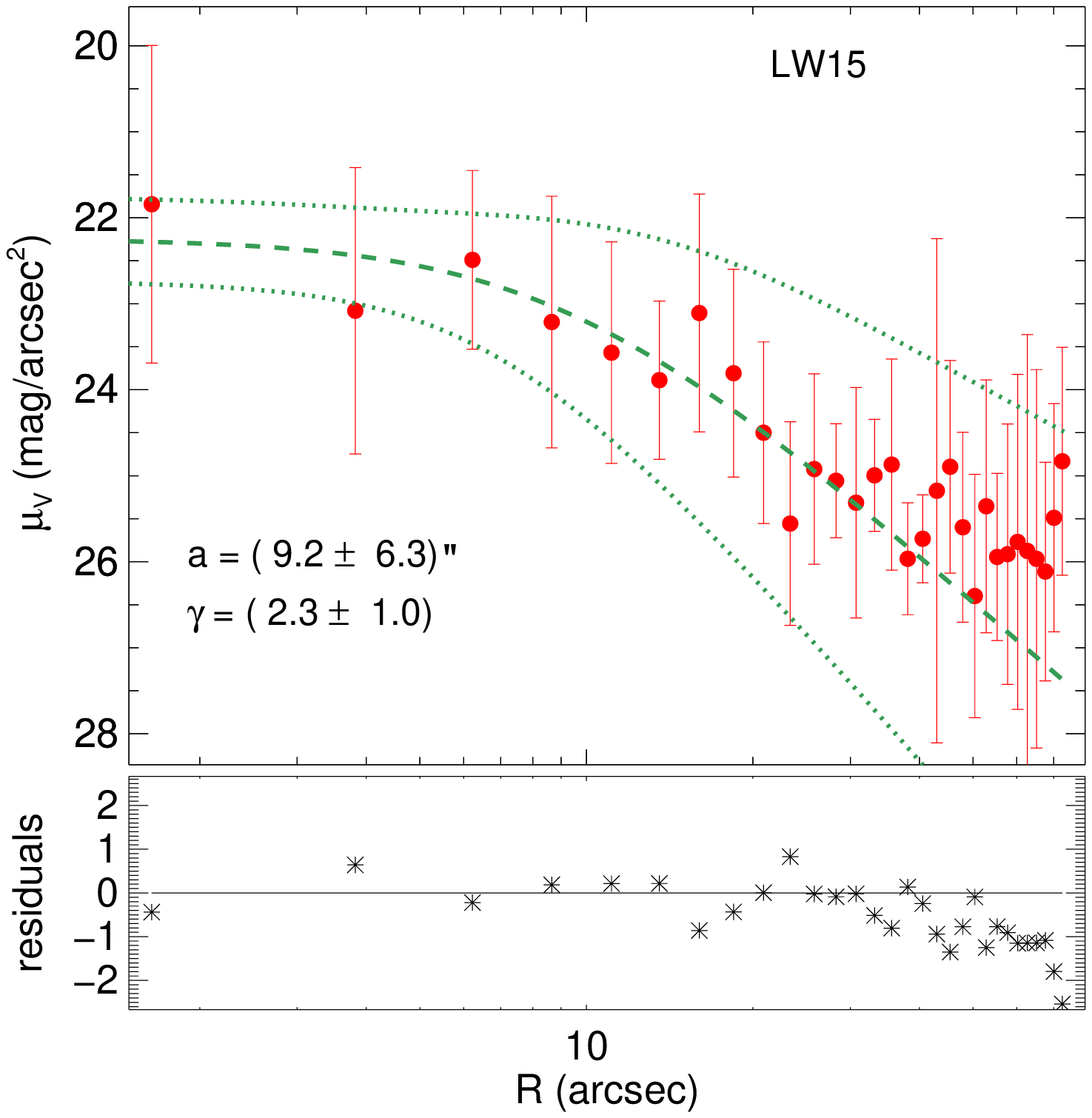}\includegraphics[width=0.325\linewidth]{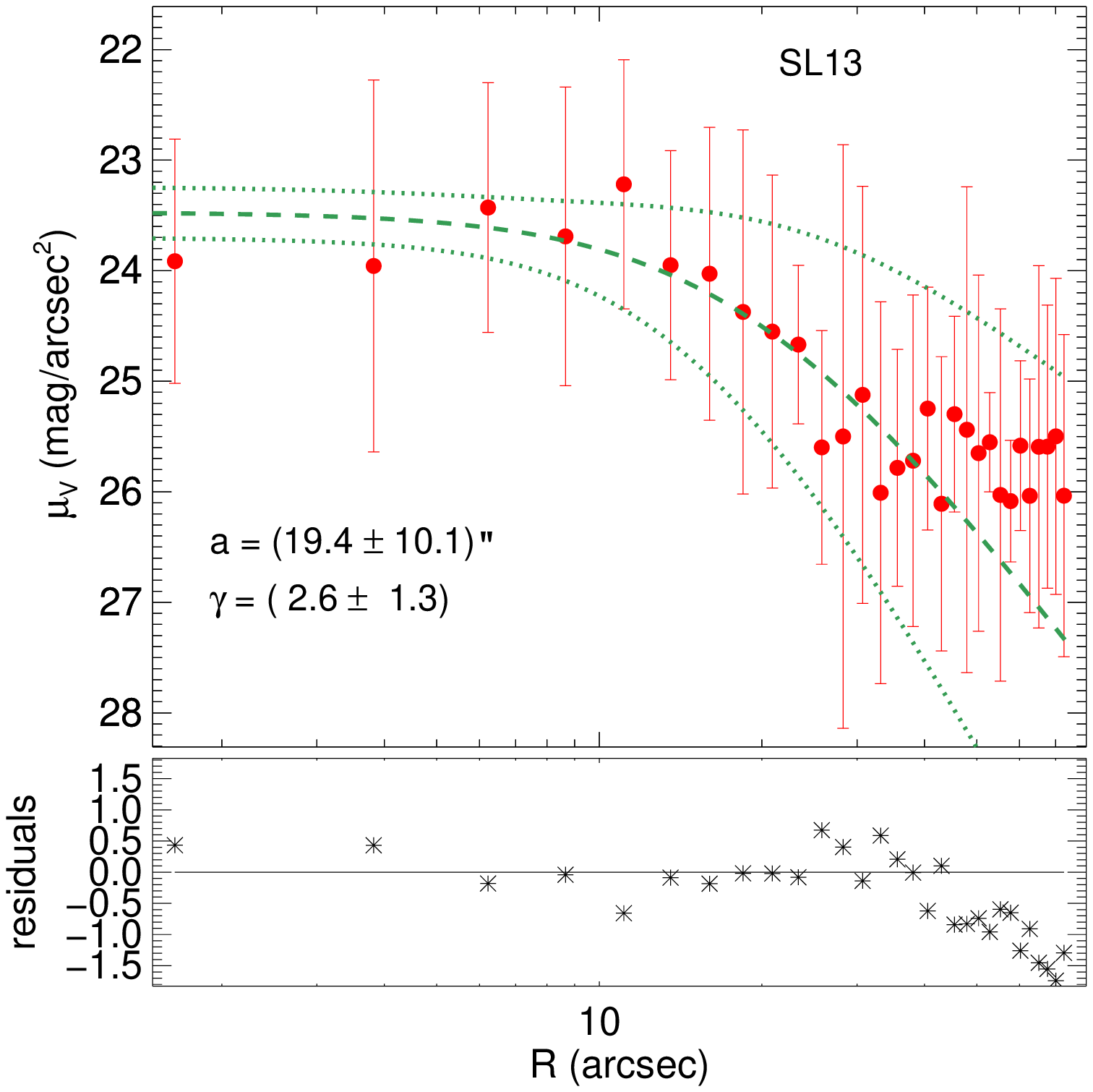}\includegraphics[width=0.325\linewidth]{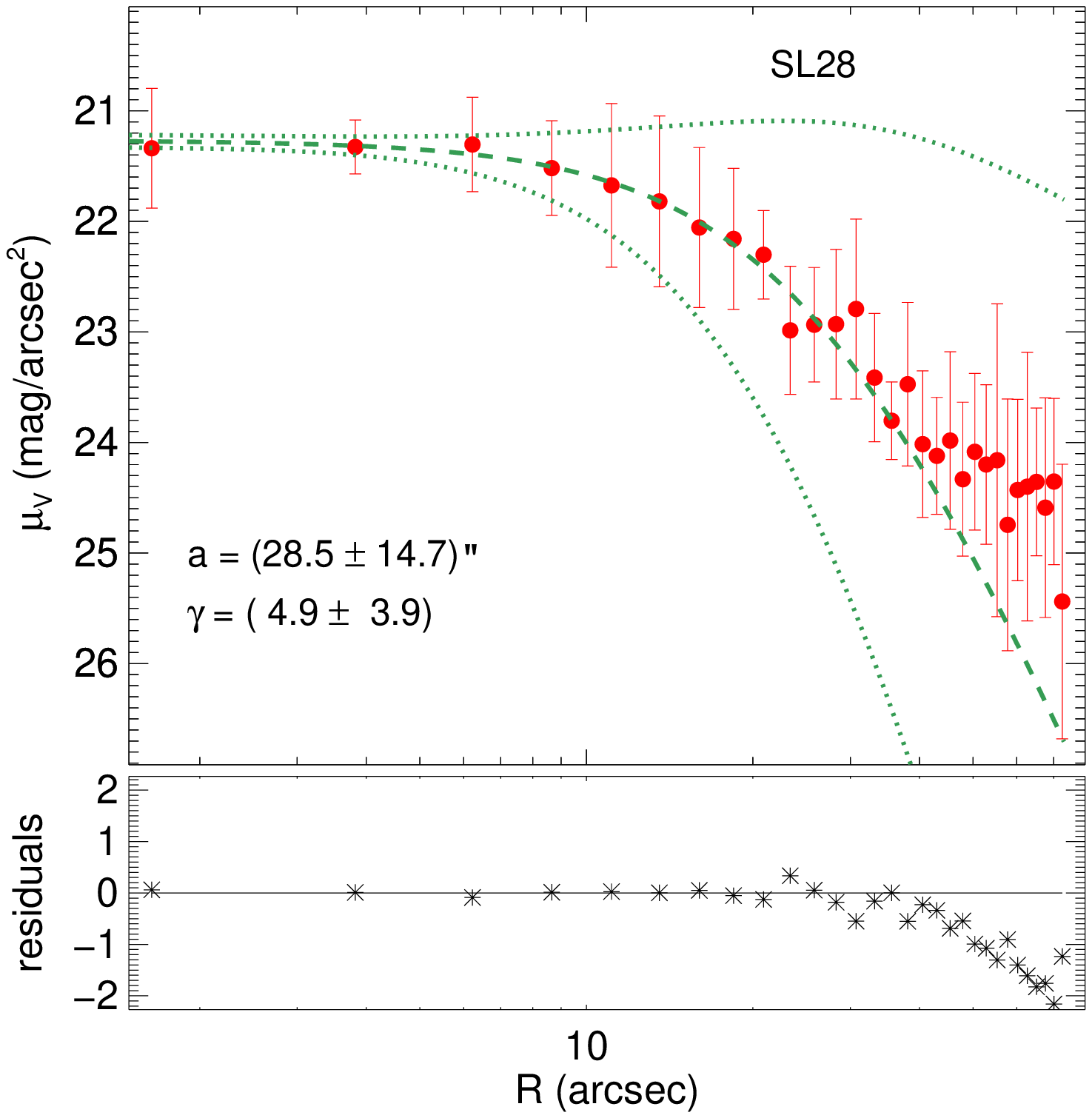}

\includegraphics[width=0.325\linewidth]{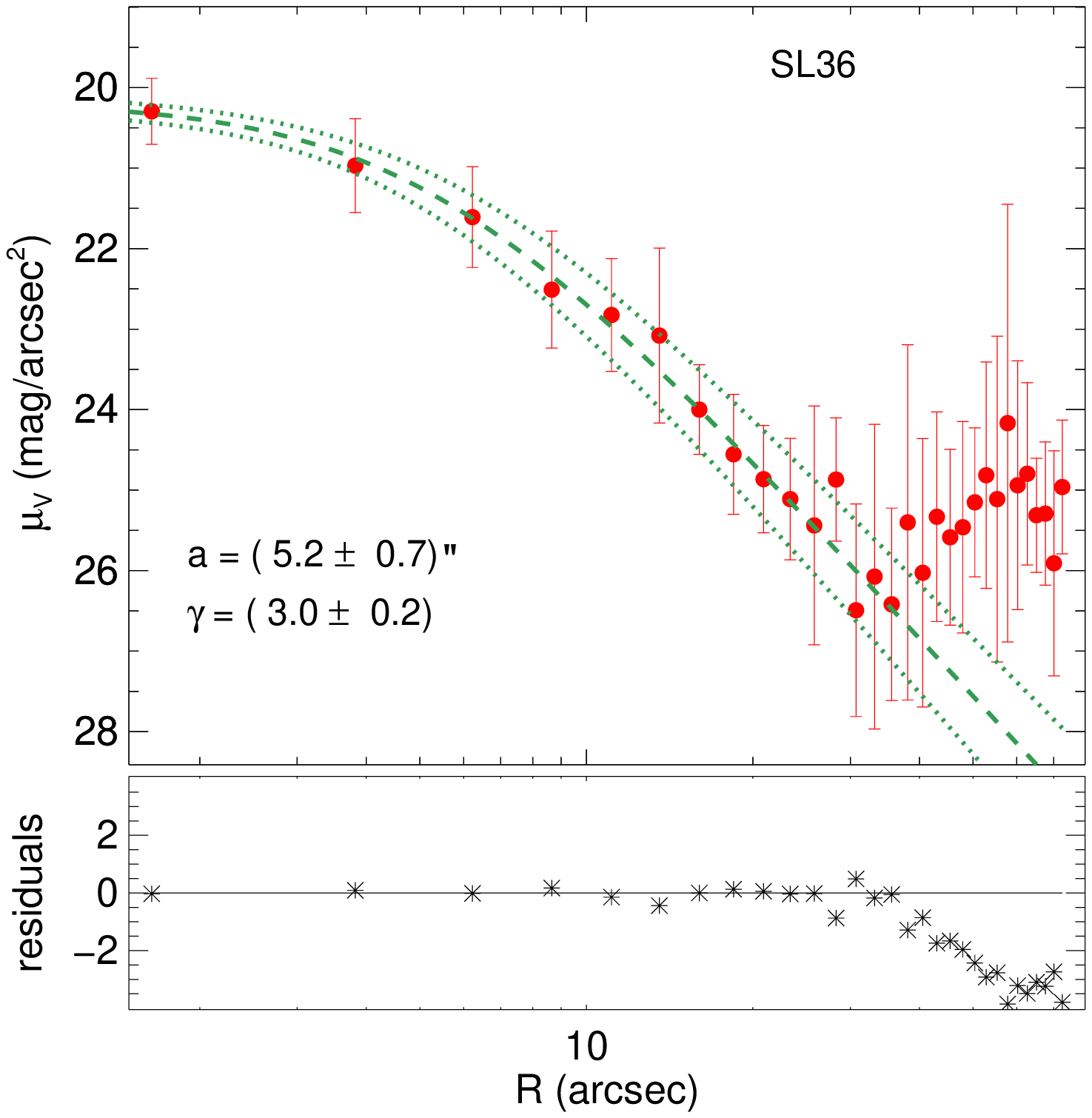}\includegraphics[width=0.325\linewidth]{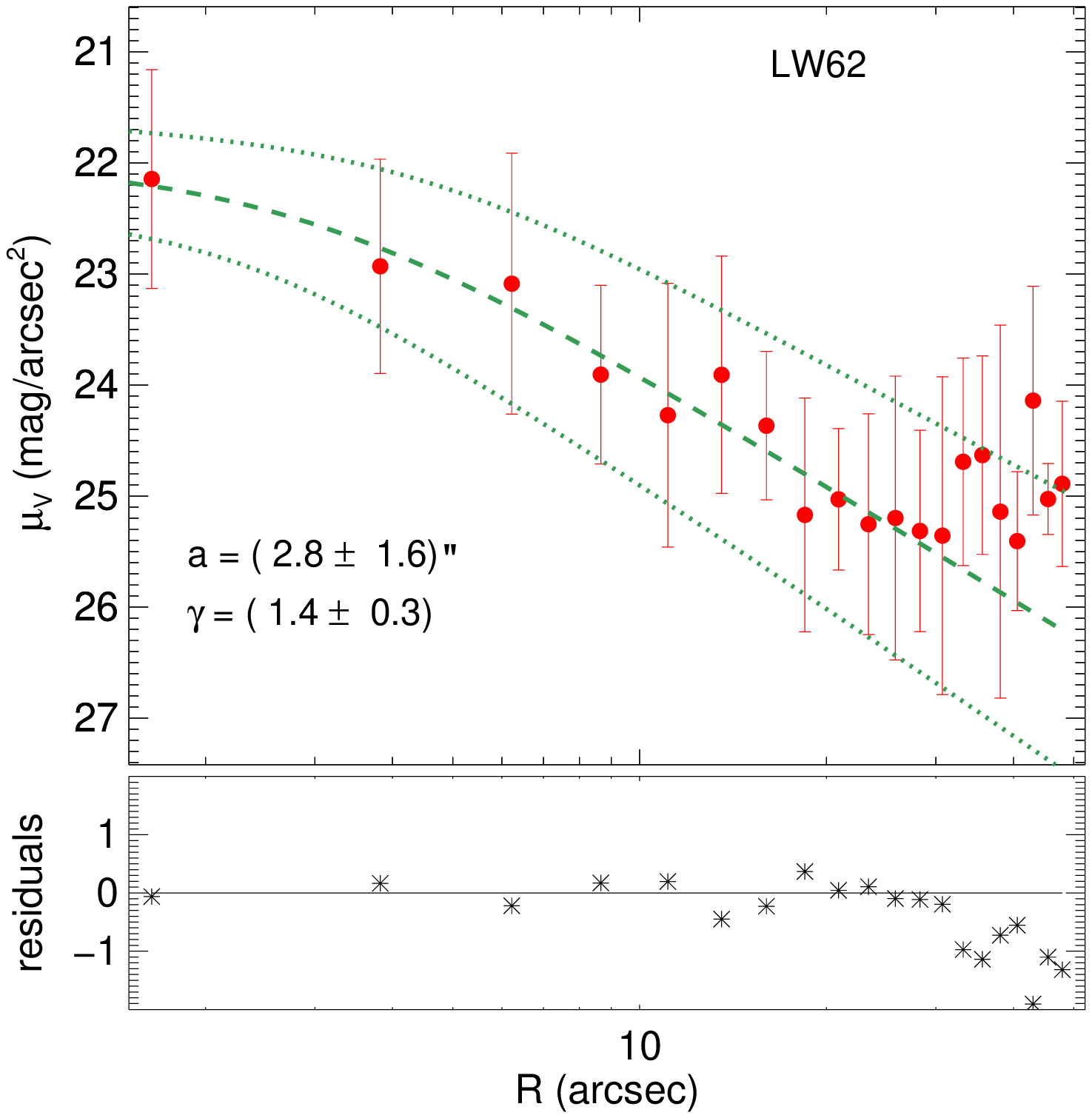}\includegraphics[width=0.325\linewidth]{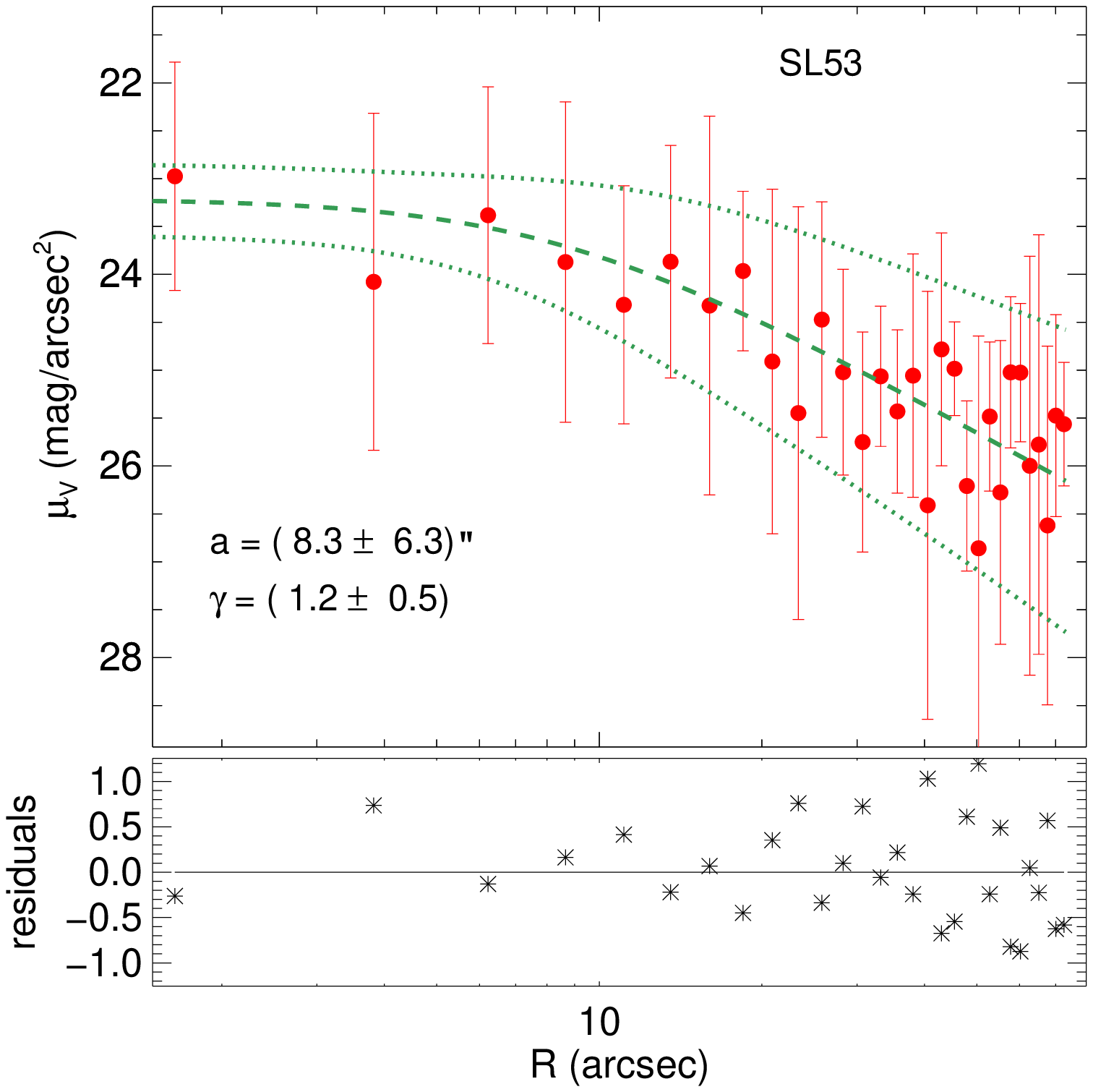}

\includegraphics[width=0.325\linewidth]{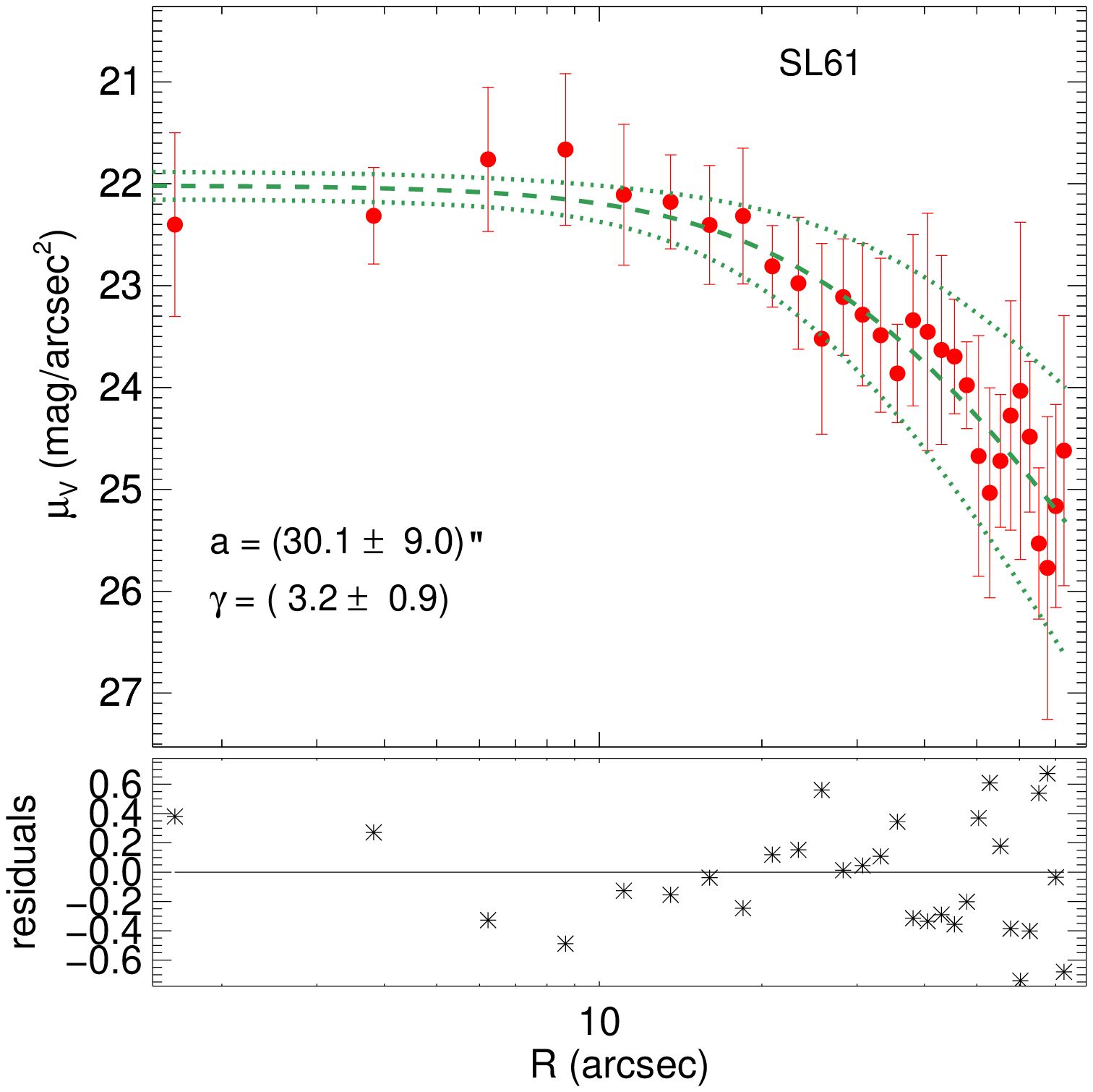}\includegraphics[width=0.325\linewidth]{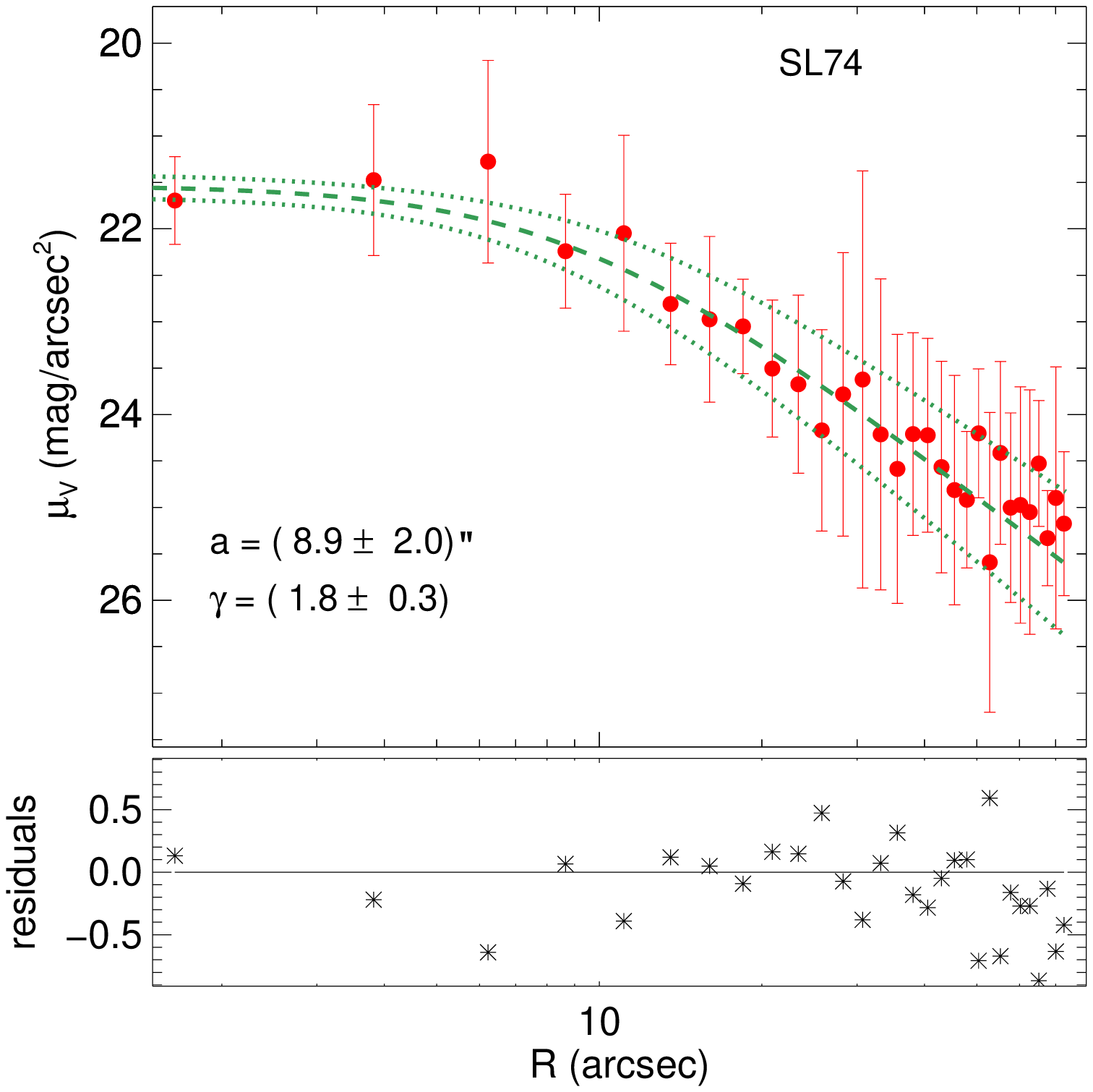}\includegraphics[width=0.325\linewidth]{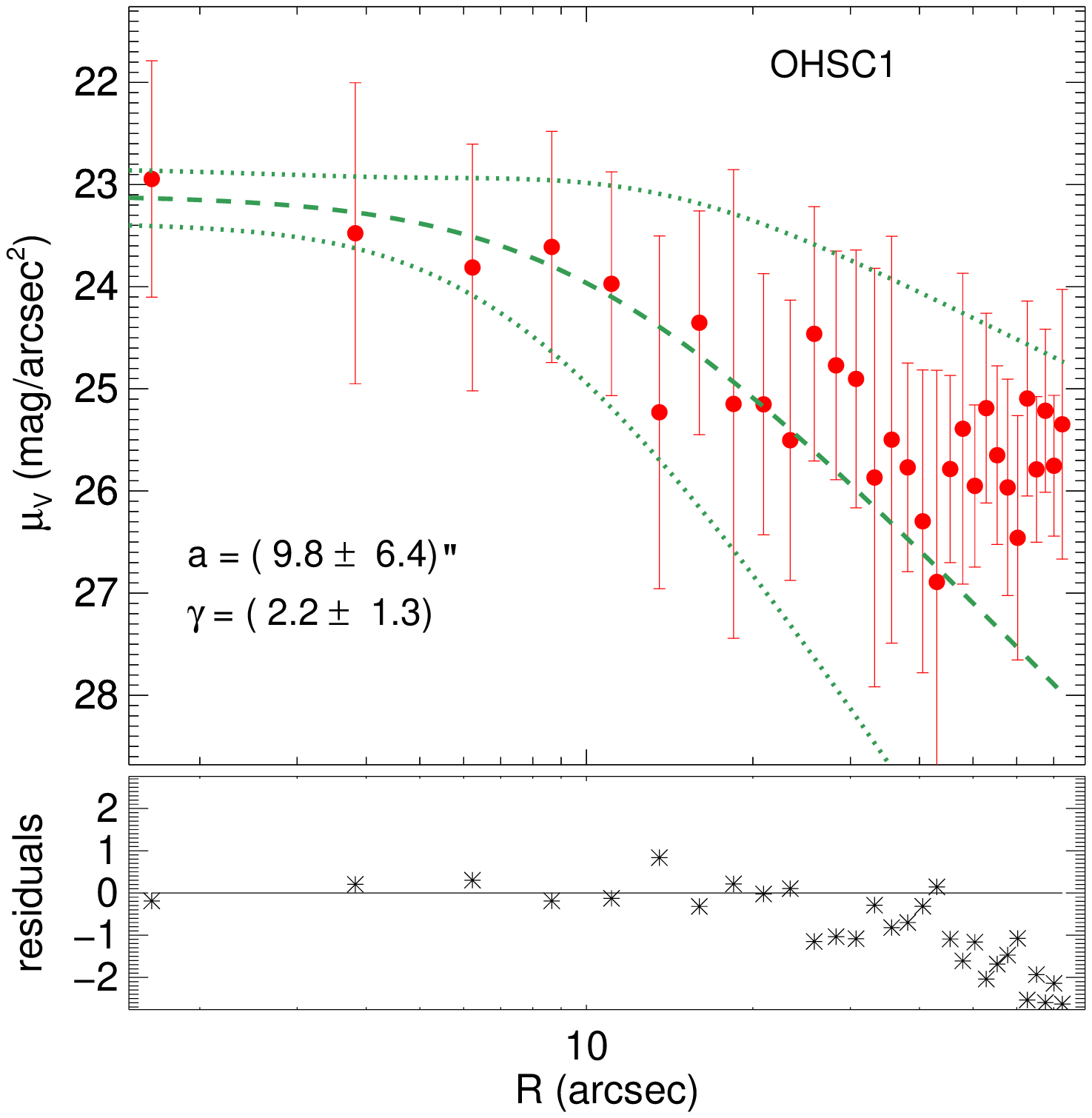}

\includegraphics[width=0.325\linewidth]{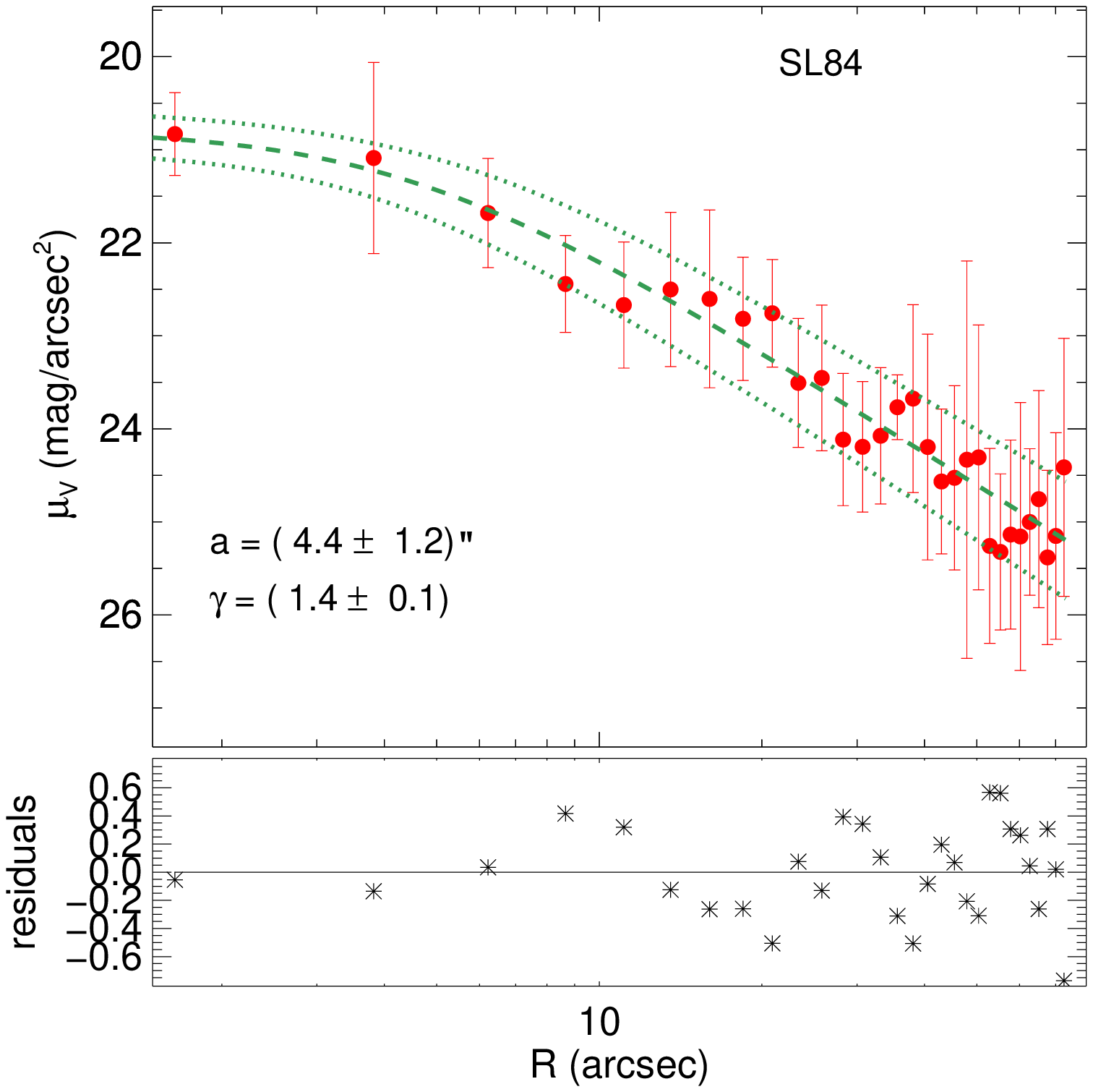}\includegraphics[width=0.325\linewidth]{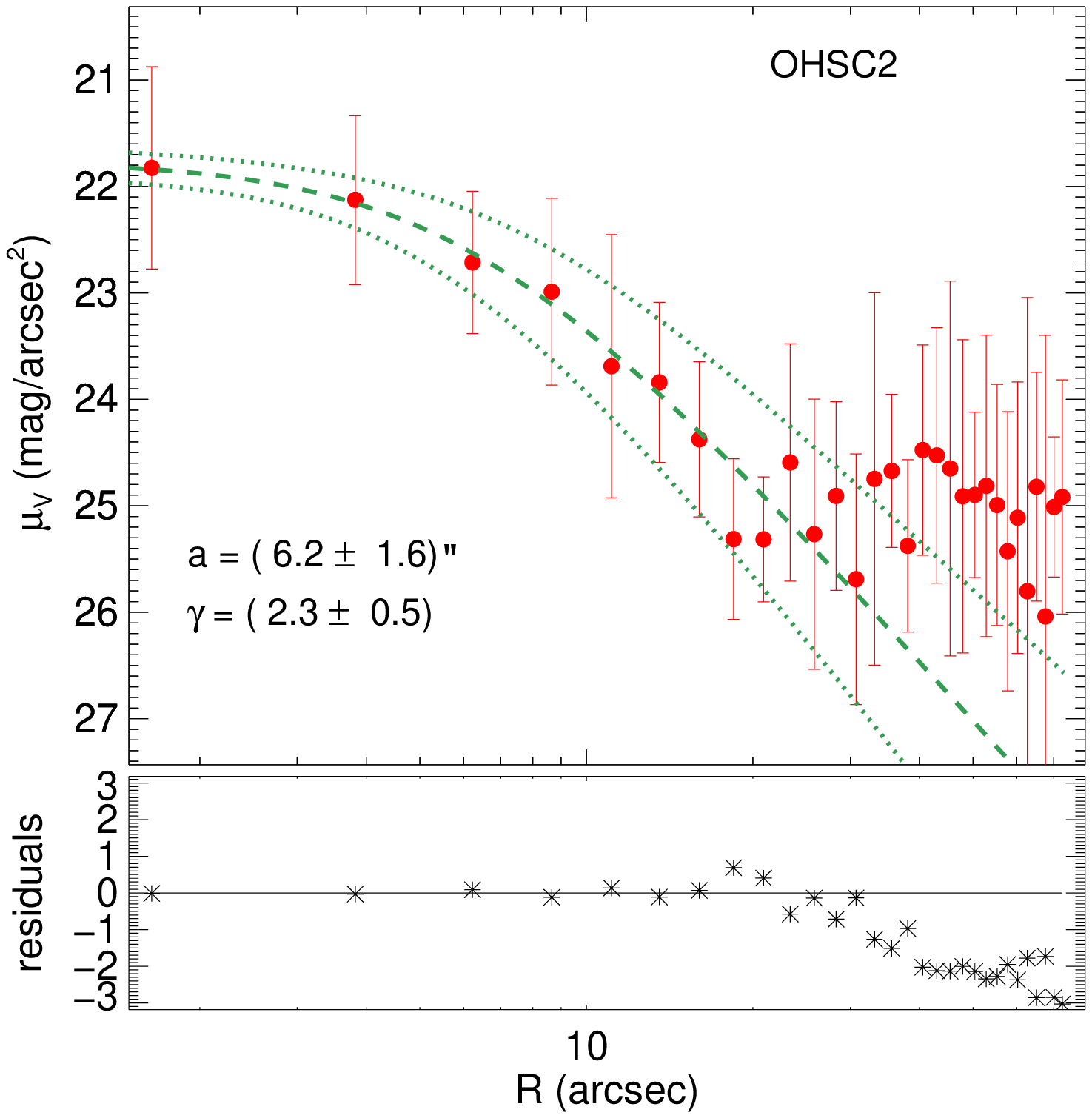}\includegraphics[width=0.325\linewidth]{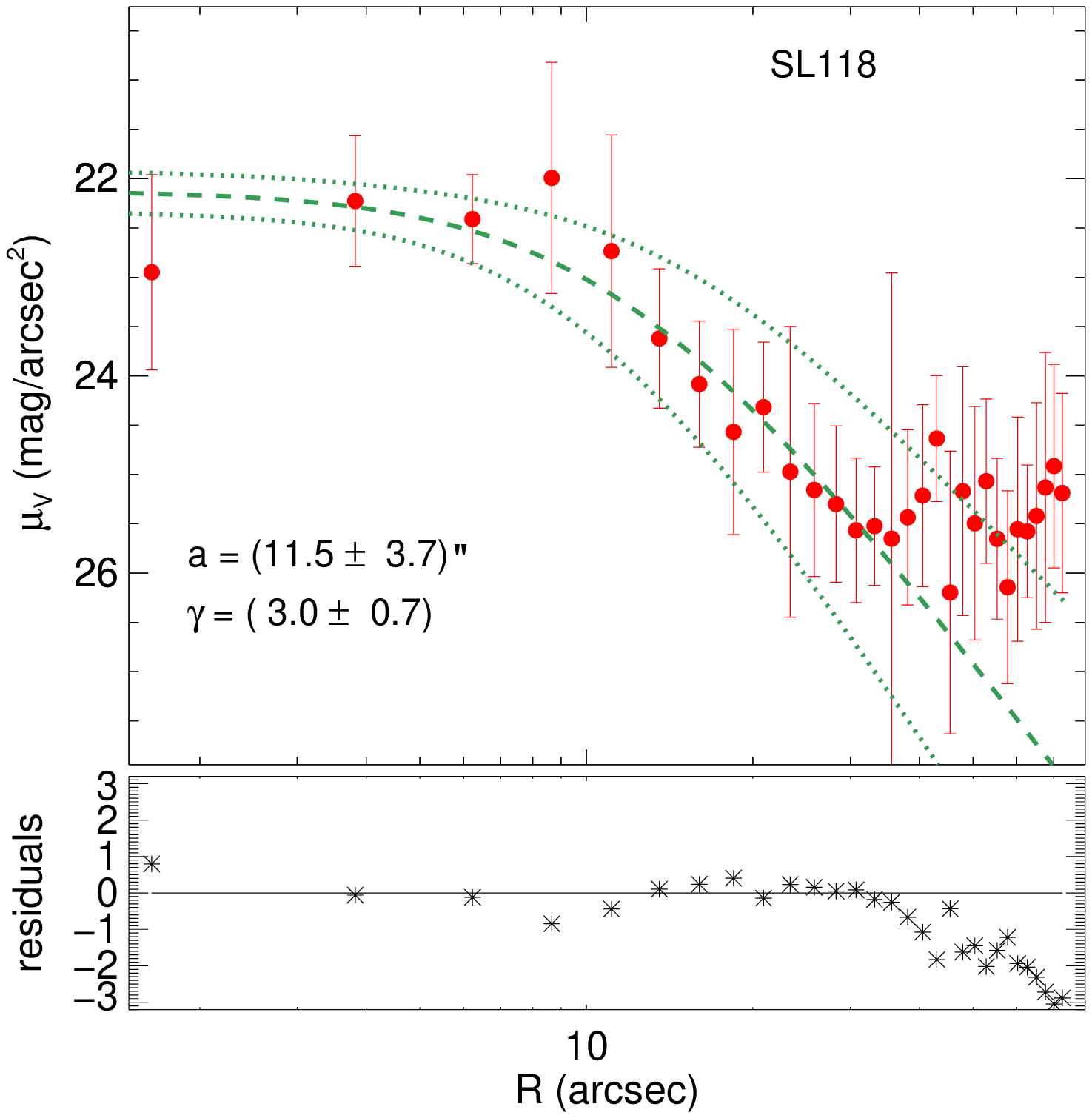}

\caption{Surface brightness profiles of additional LMC clusters complementing the sample presented in Fig.~\ref{fig:rdp_sbp}. The EFF model fits (dashed lines)  and 1\,$\sigma$ uncertainties (dotted lines) are shown. The best-fitting  parameters are indicated and the fit residuals are plotted in the lower panel.}

\end{figure*}

\setcounter{figure}{7}

\begin{figure*}

\includegraphics[width=0.325\linewidth]{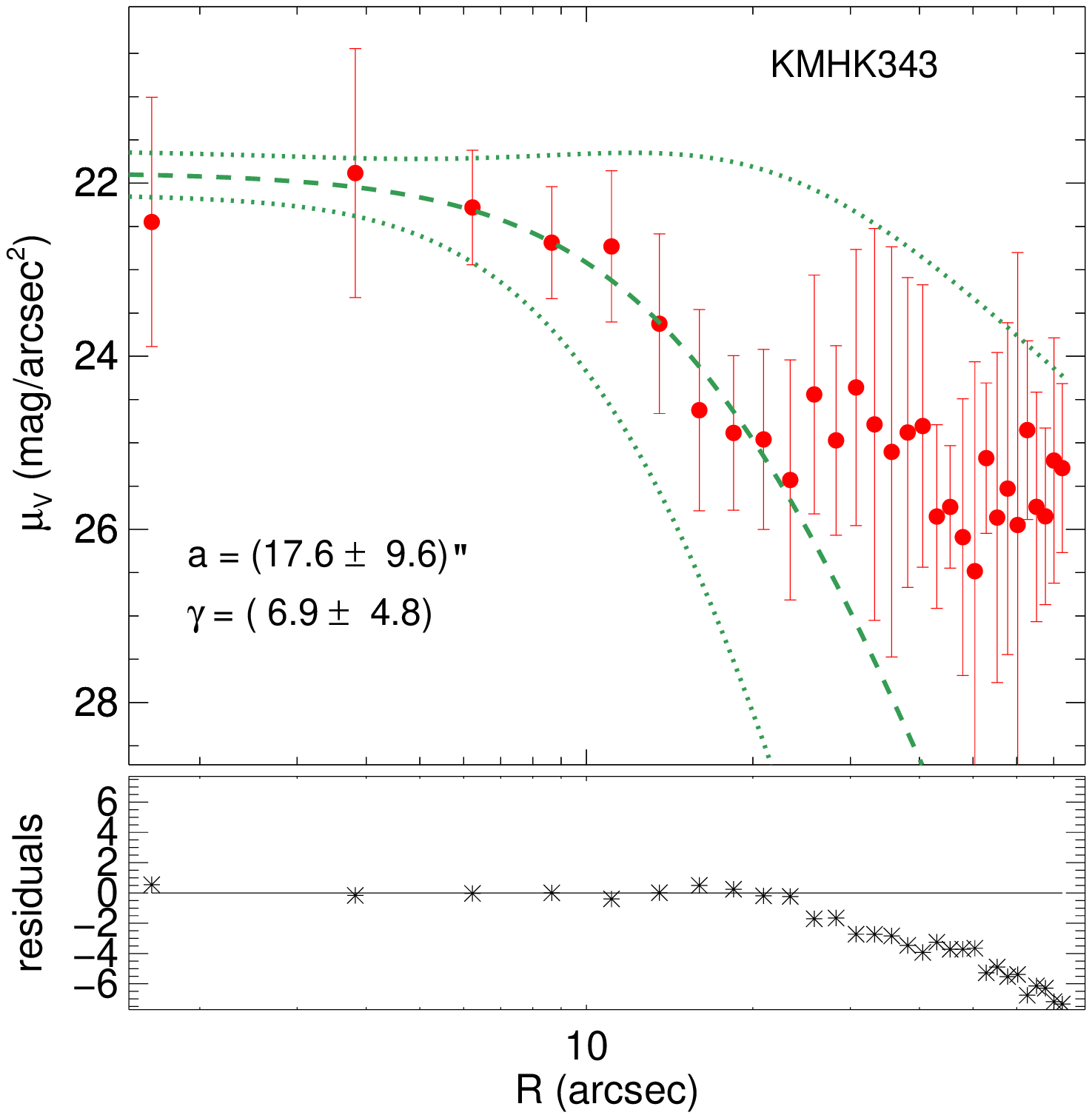}\includegraphics[width=0.325\linewidth]{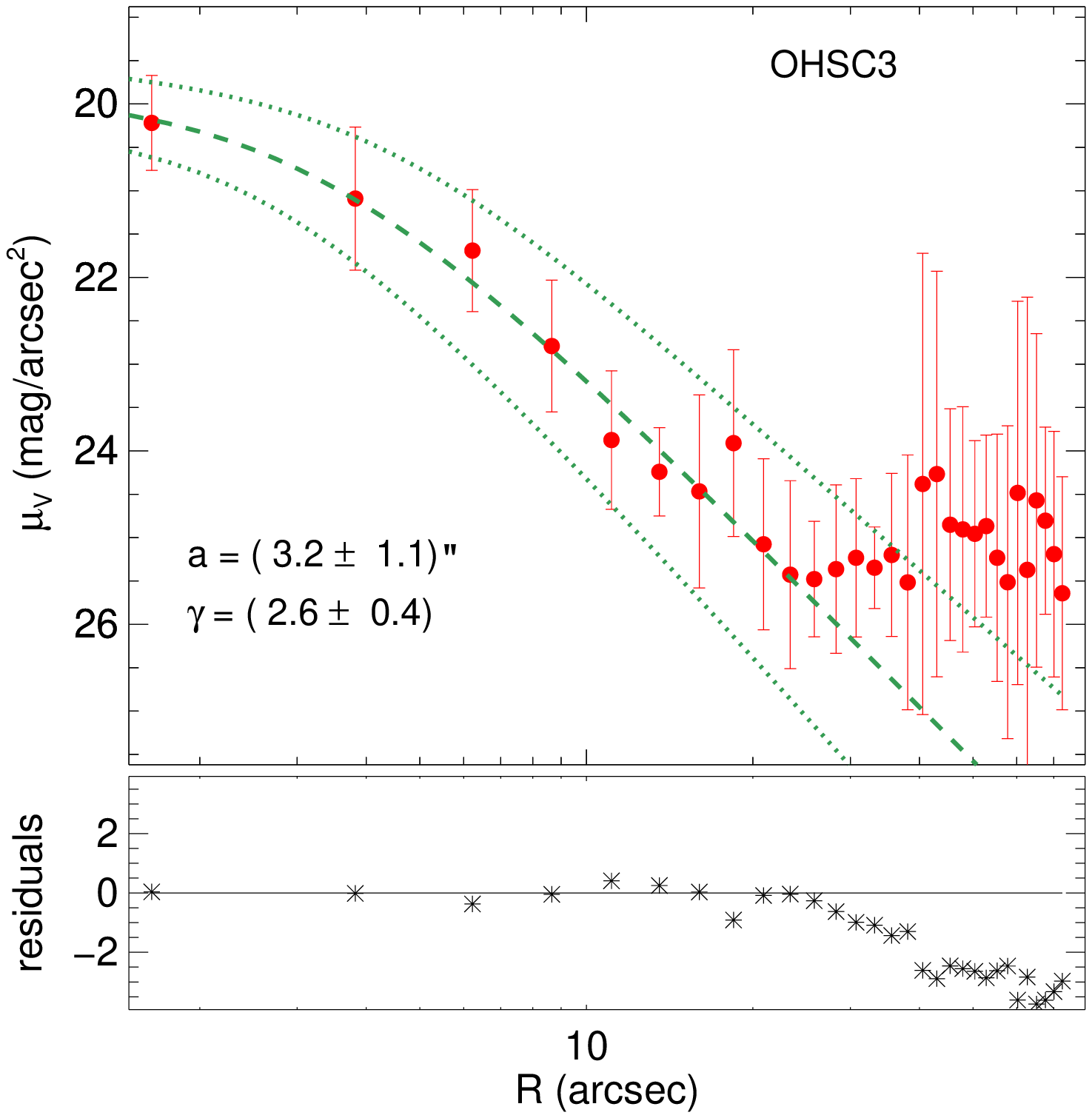}\includegraphics[width=0.325\linewidth]{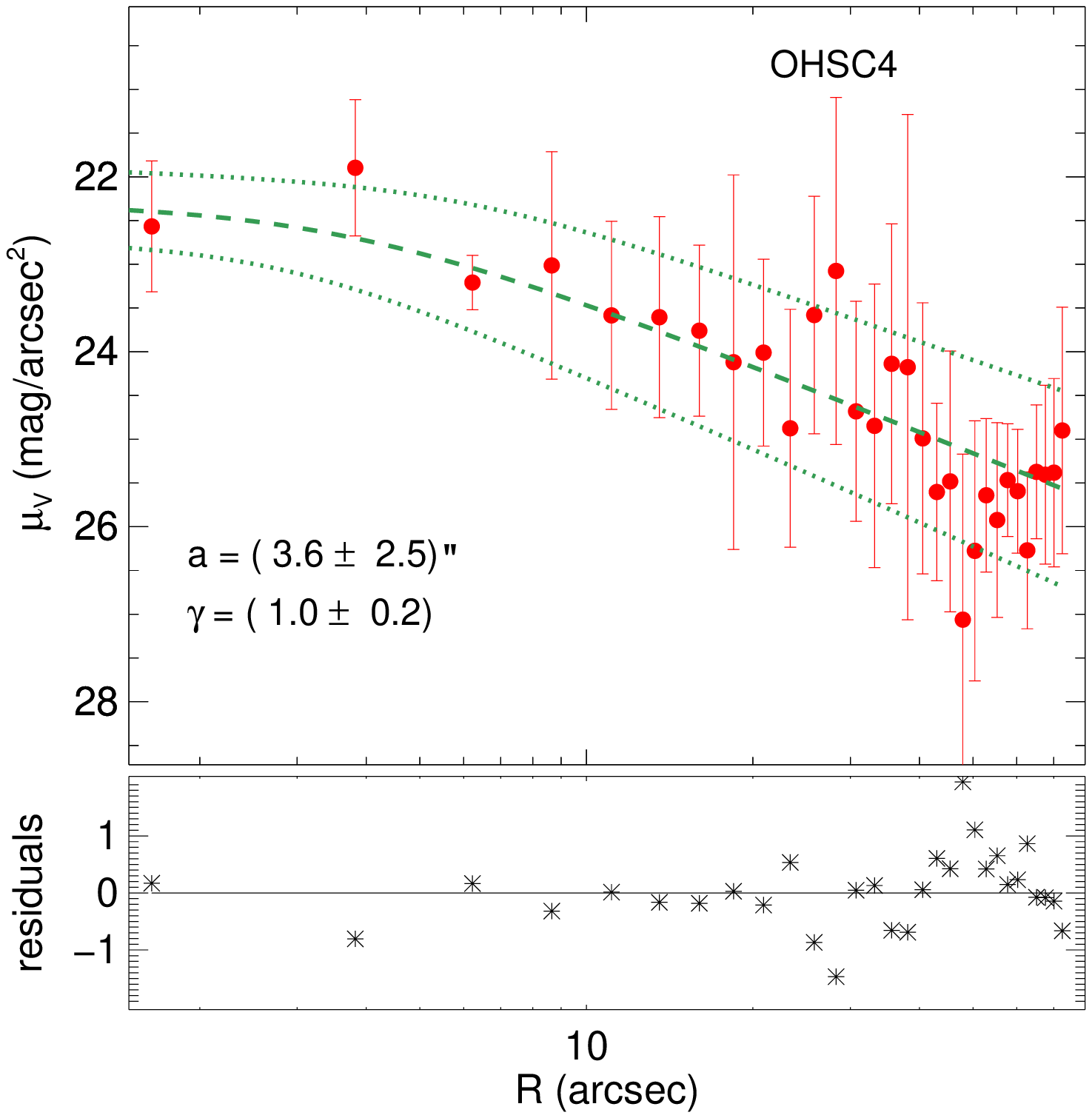}

\includegraphics[width=0.325\linewidth]{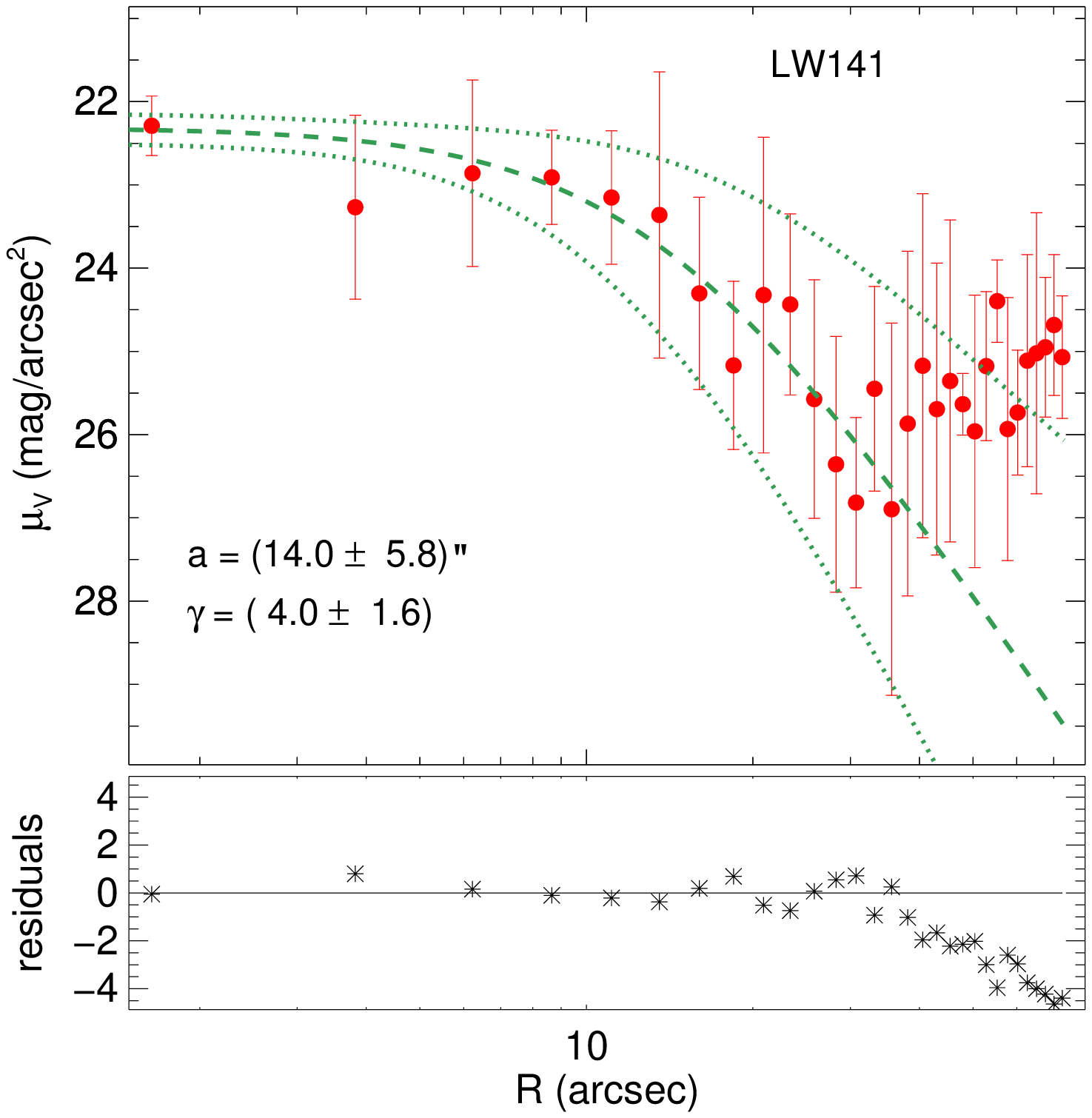}\includegraphics[width=0.325\linewidth]{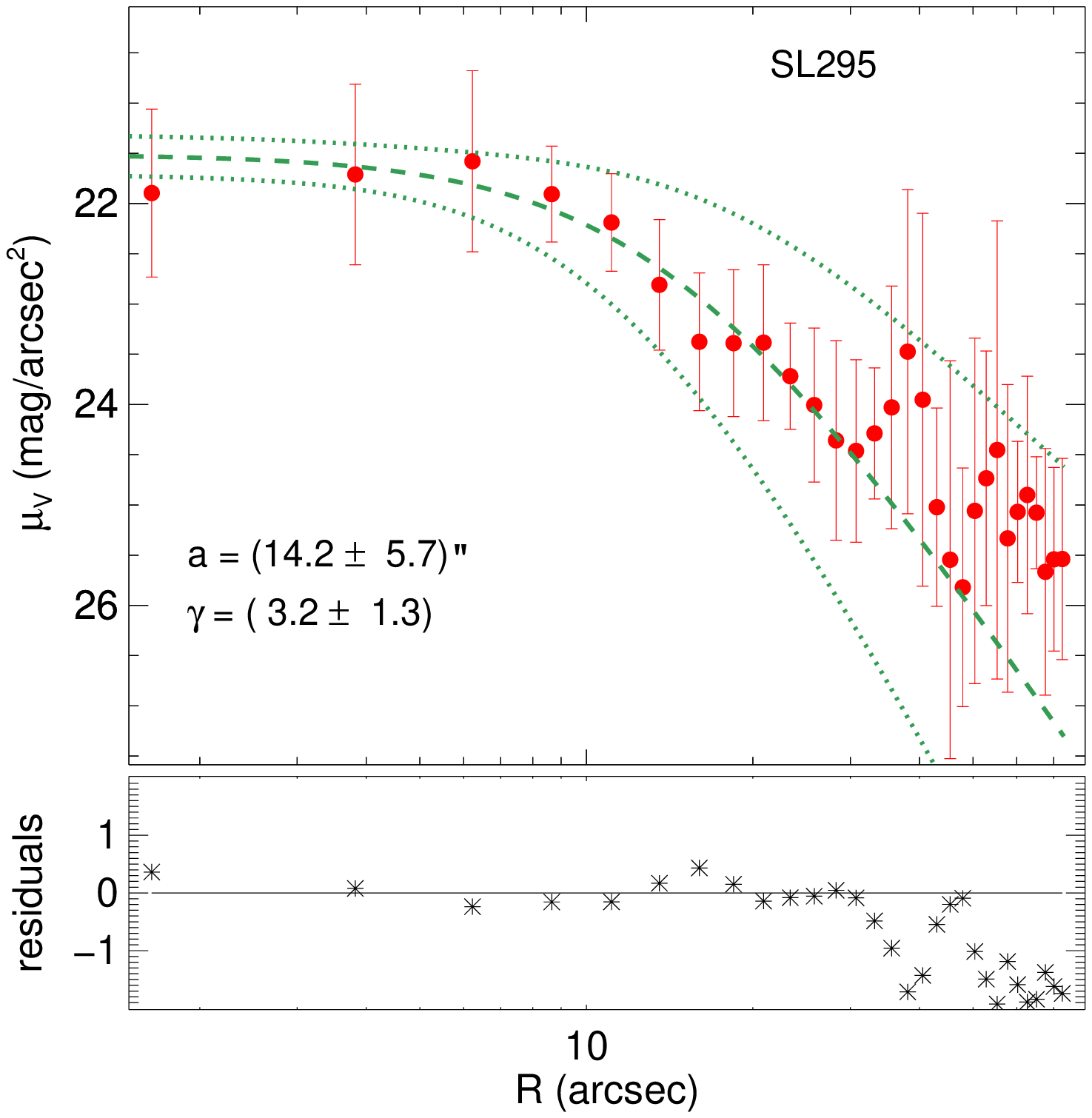}\includegraphics[width=0.325\linewidth]{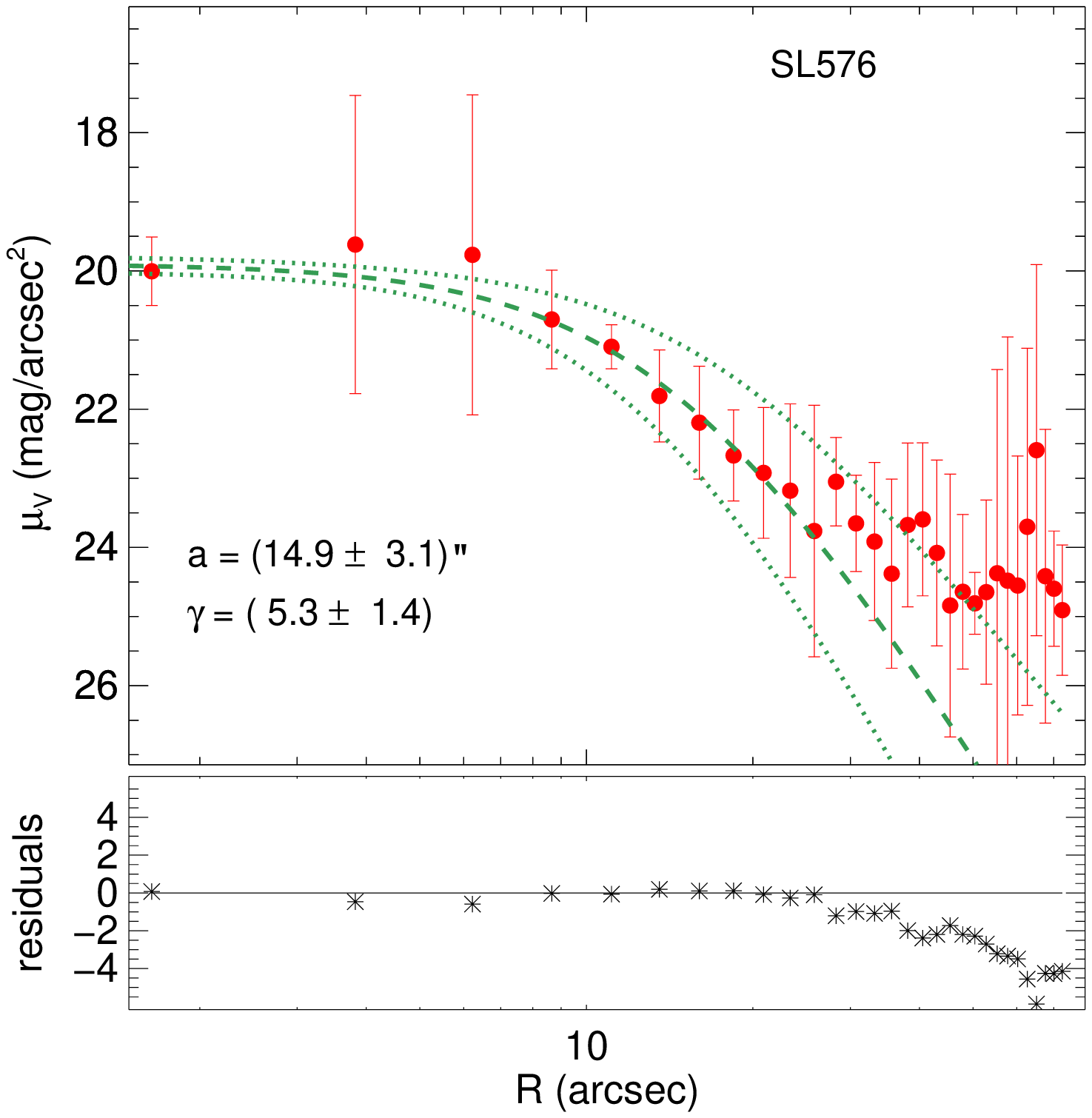}

\includegraphics[width=0.325\linewidth]{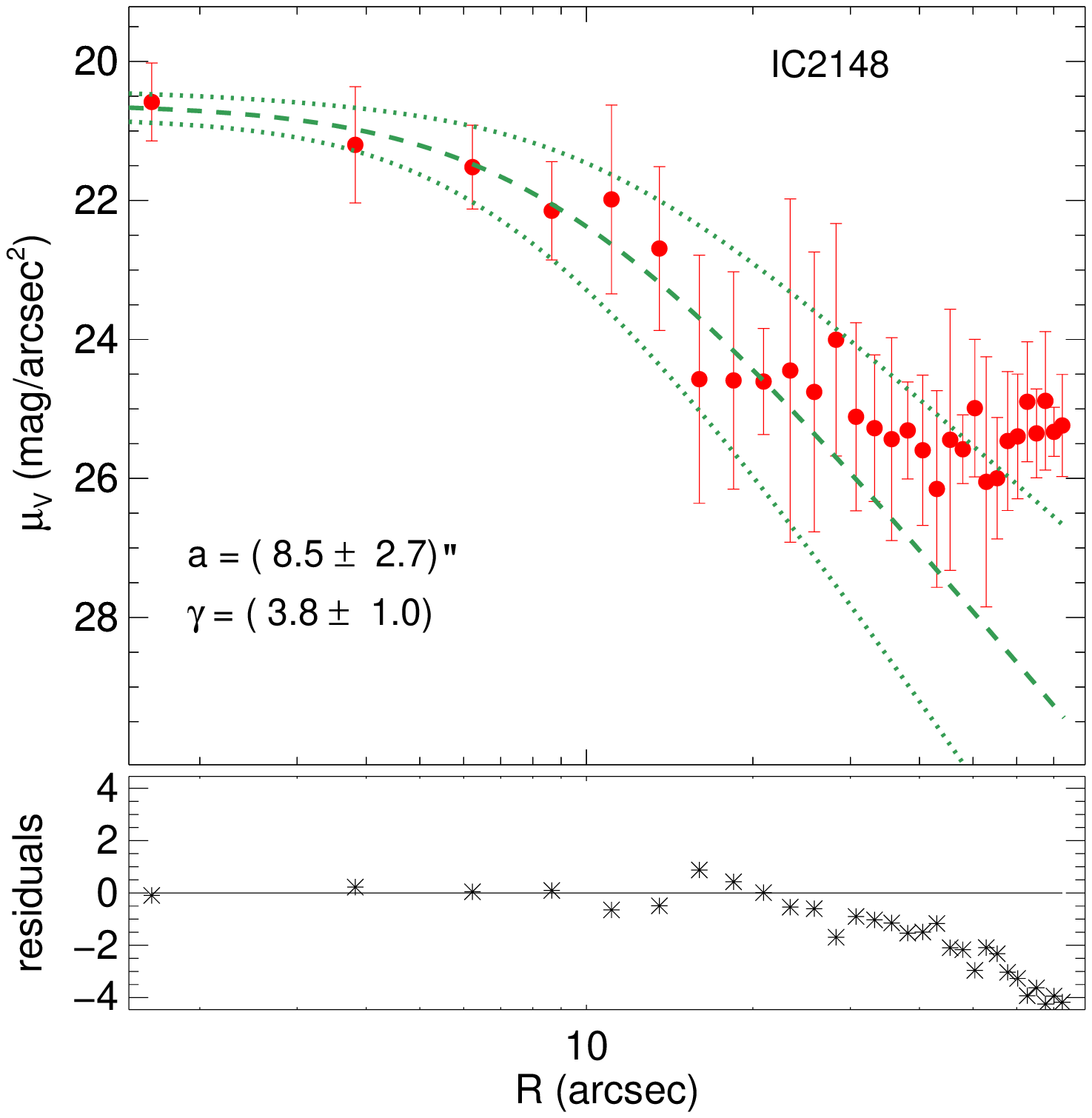}\includegraphics[width=0.325\linewidth]{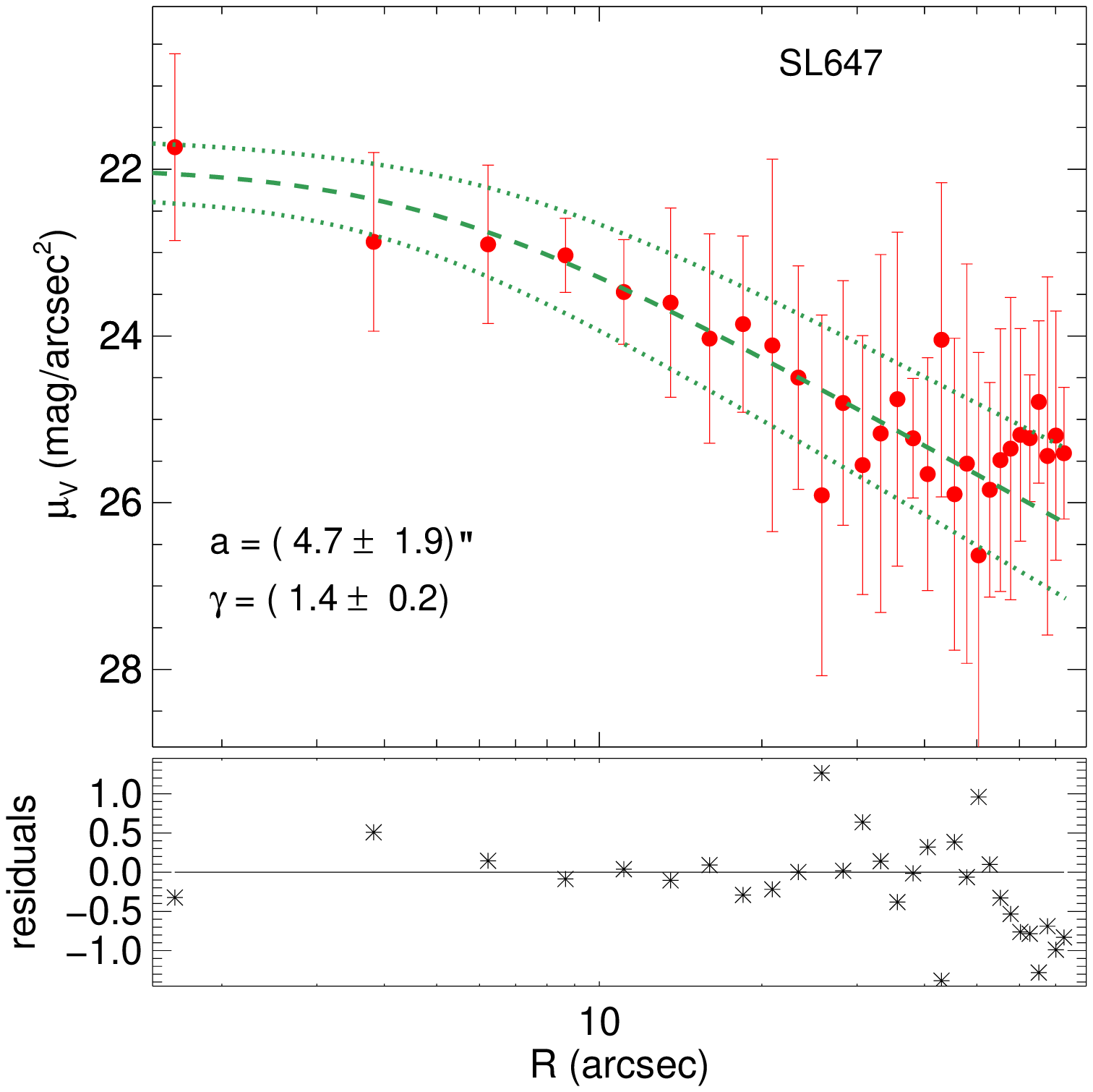}\includegraphics[width=0.325\linewidth]{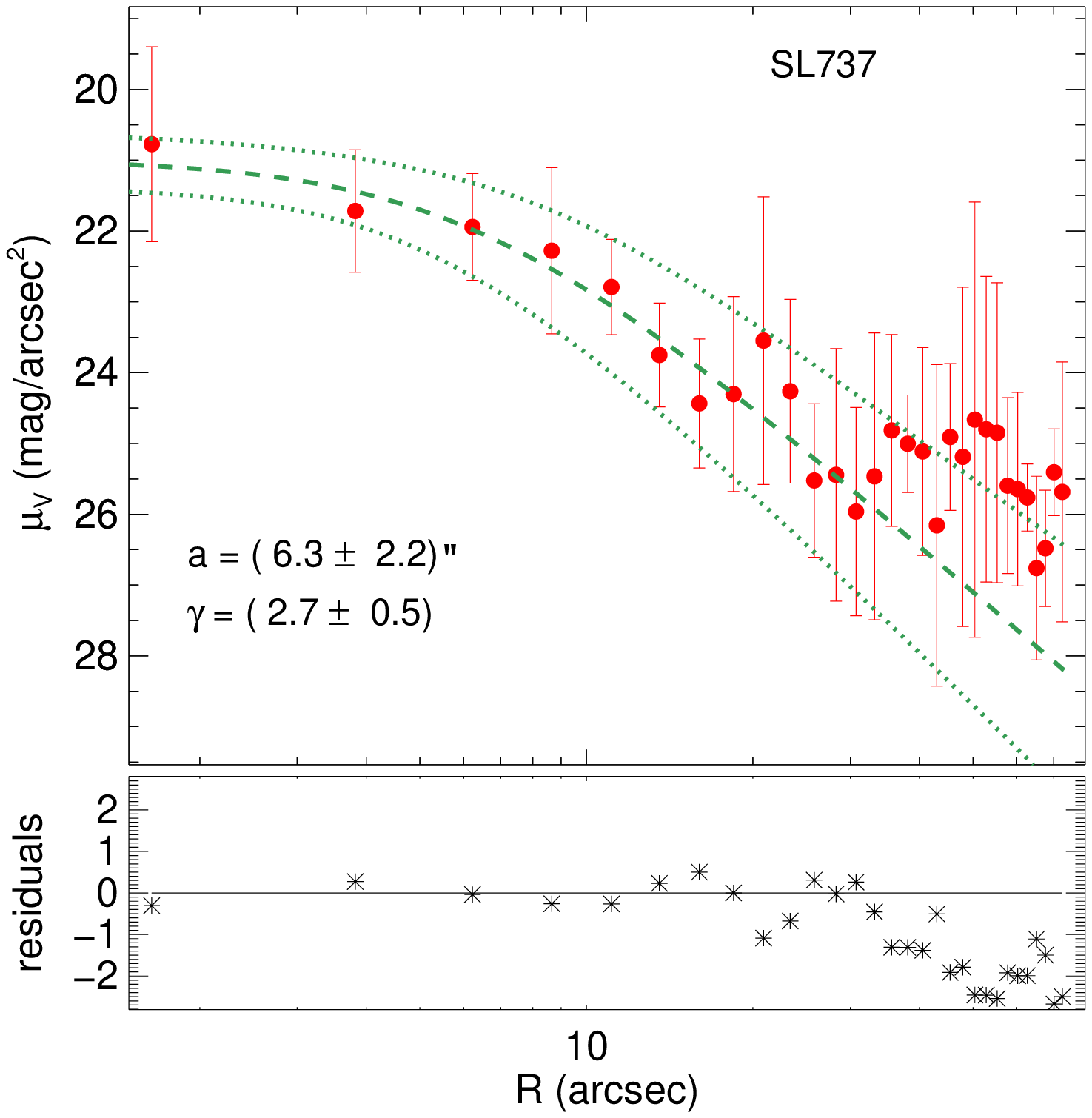}

\includegraphics[width=0.325\linewidth]{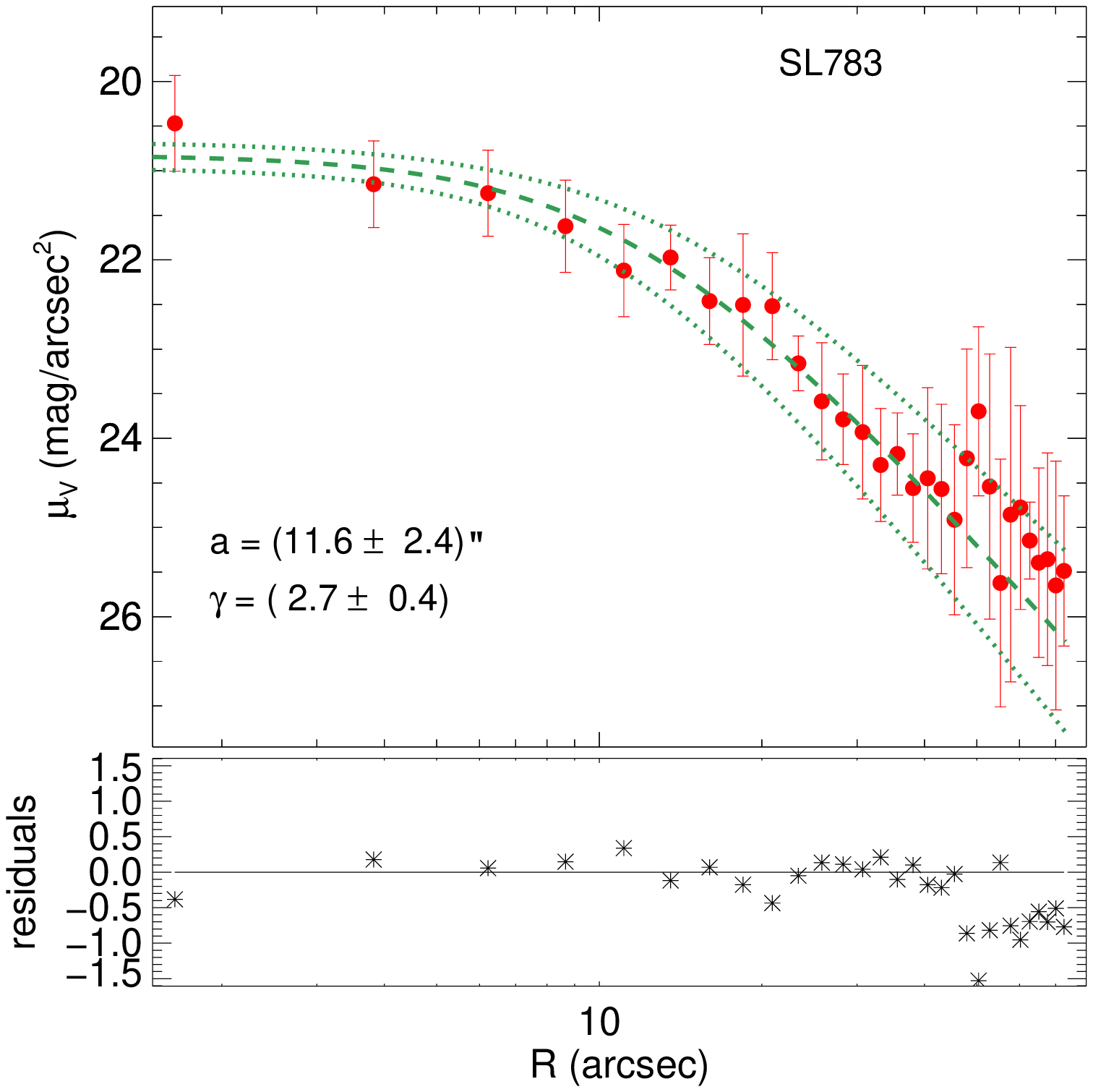}\includegraphics[width=0.325\linewidth]{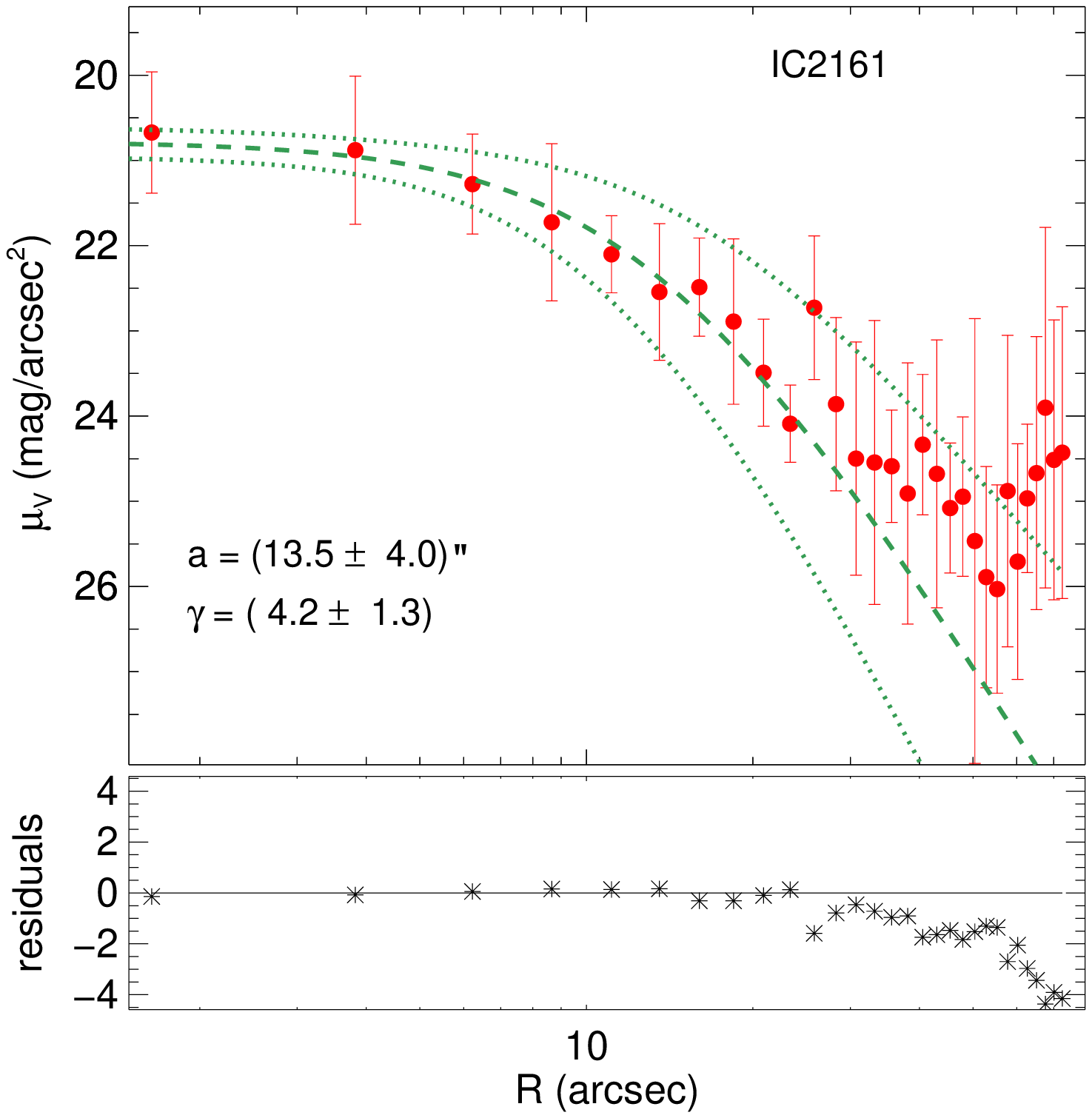}\includegraphics[width=0.325\linewidth]{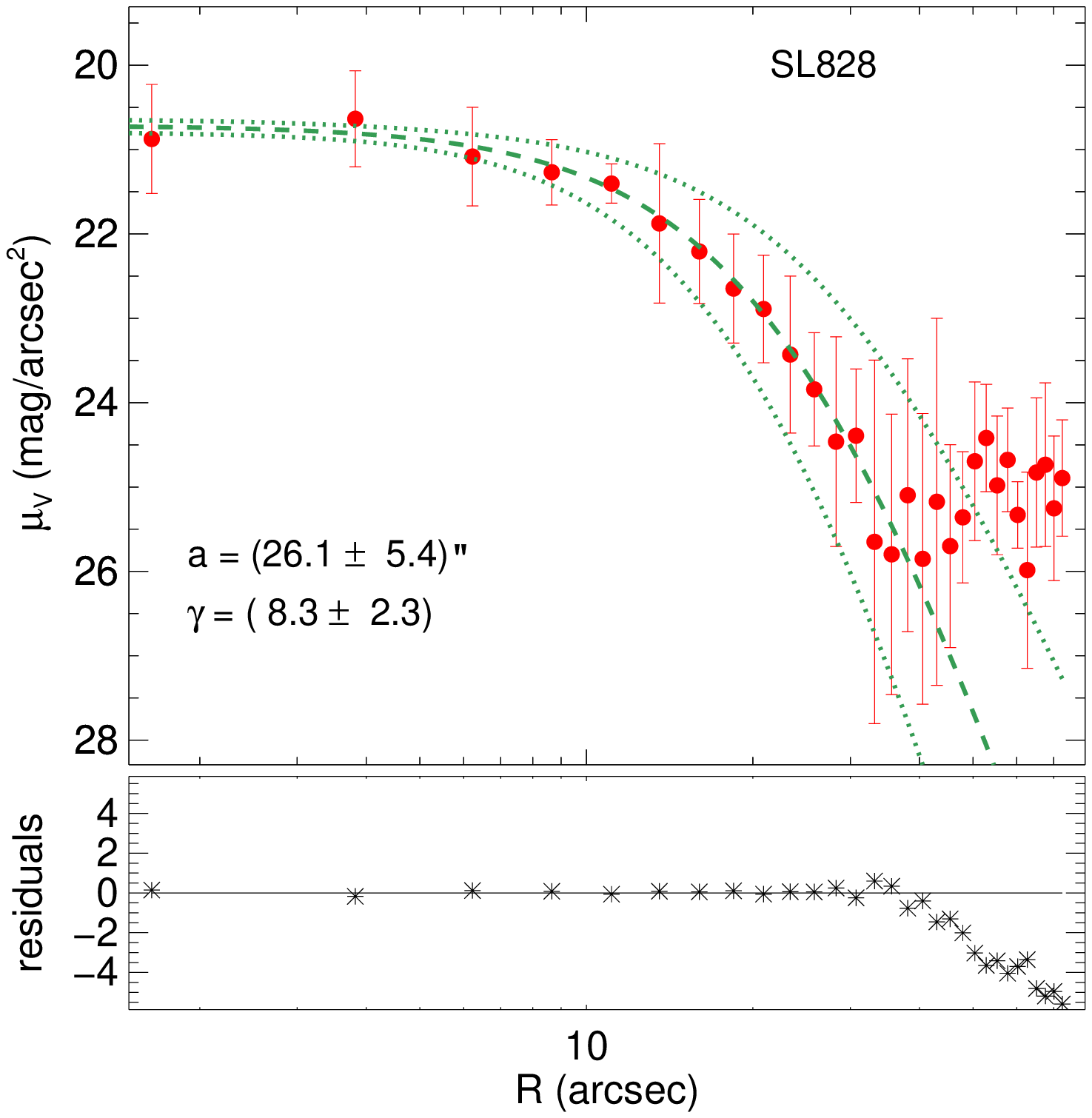}

\caption{cont.}

\end{figure*}

\setcounter{figure}{7}

\begin{figure*}

\includegraphics[width=0.325\linewidth]{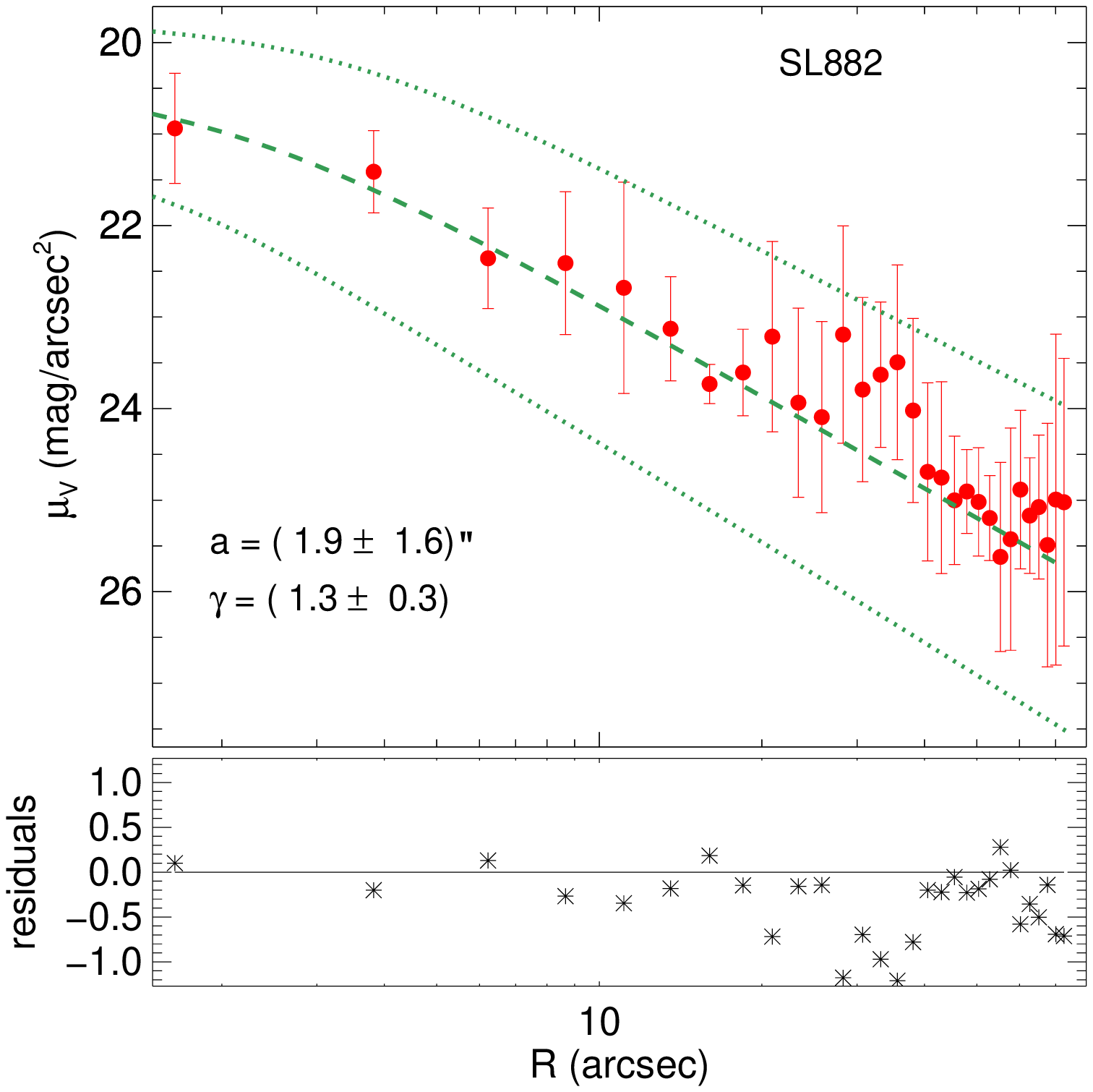}\includegraphics[width=0.325\linewidth]{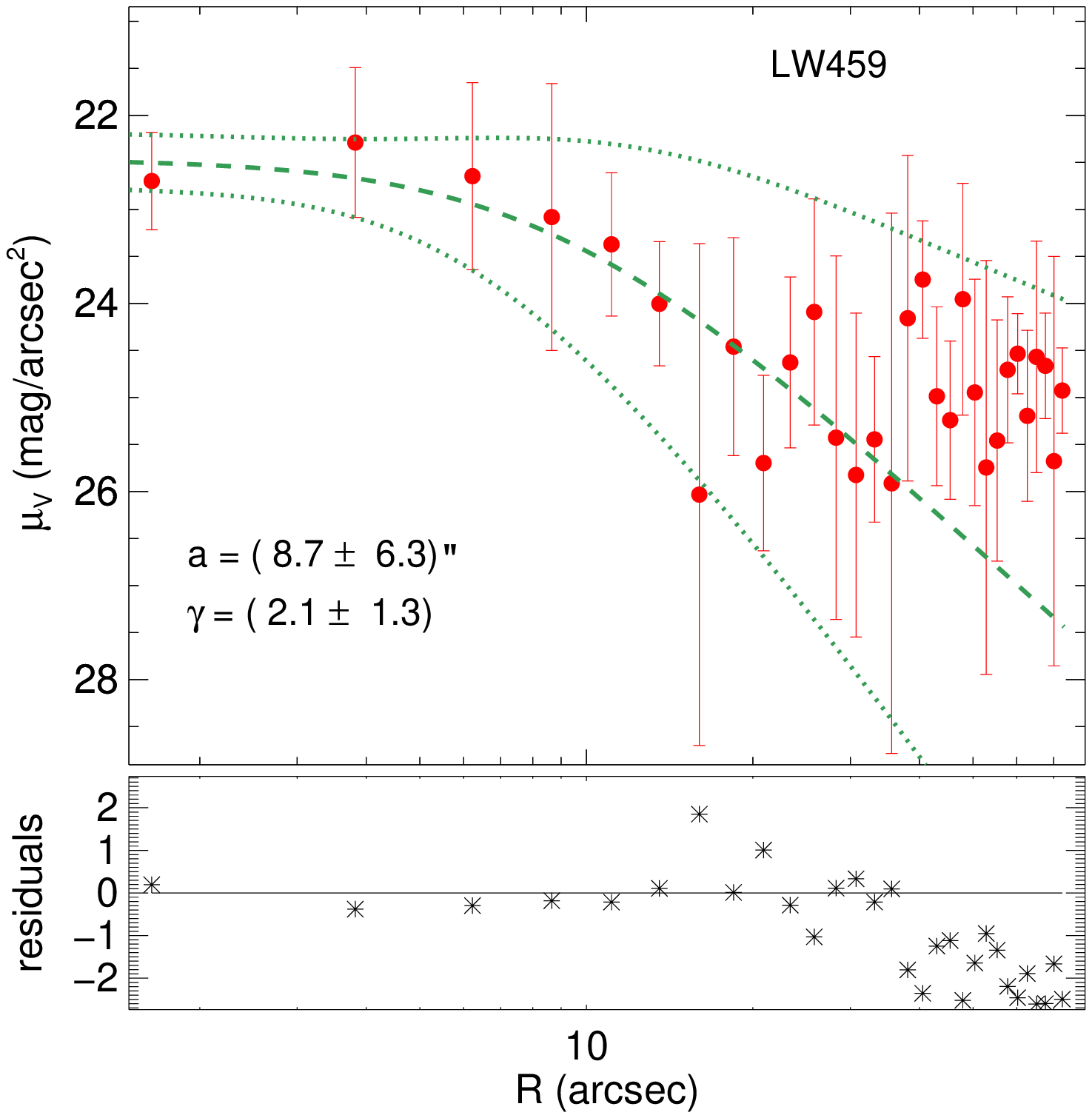}\includegraphics[width=0.325\linewidth]{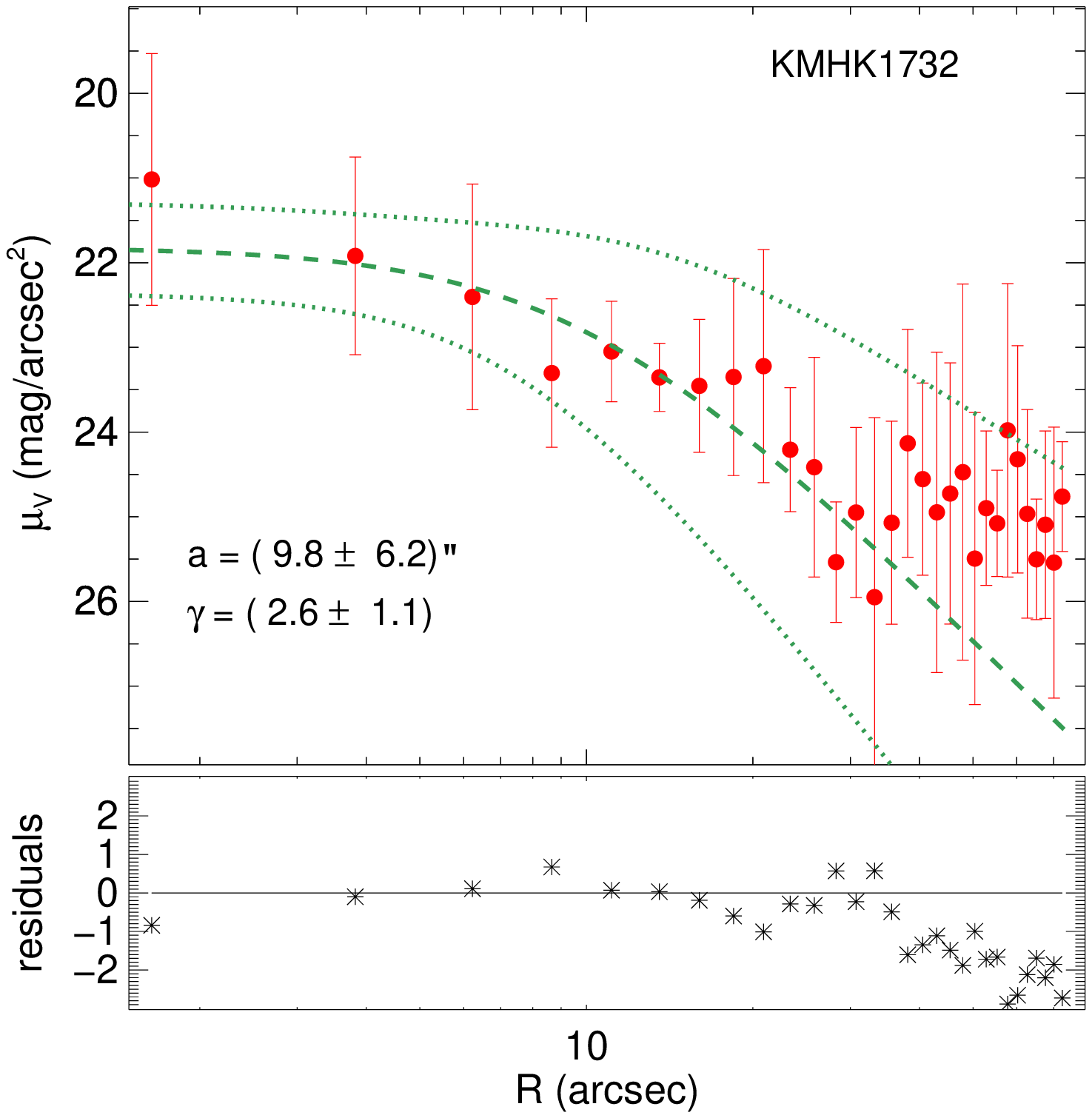}

\includegraphics[width=0.325\linewidth]{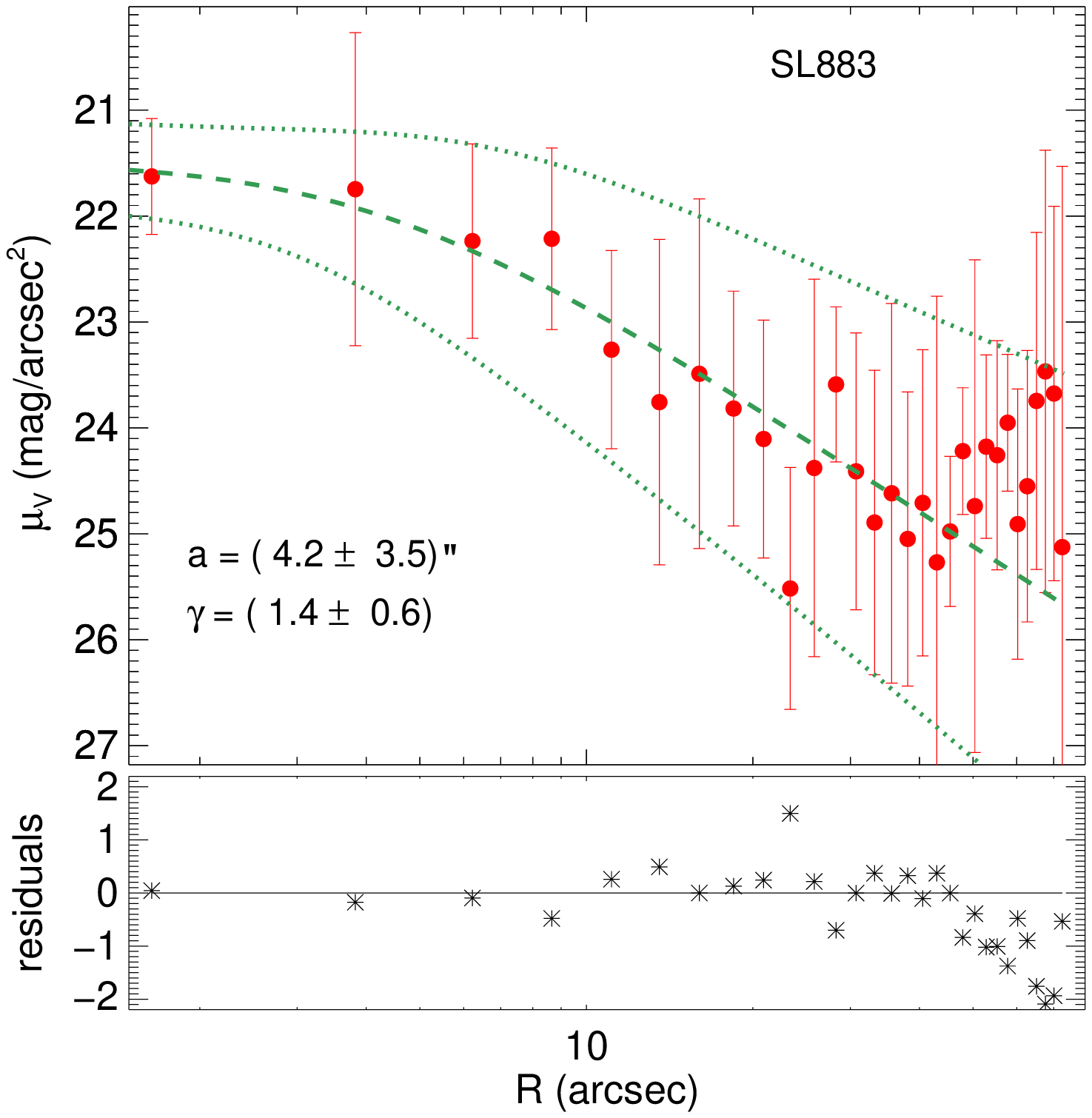}\includegraphics[width=0.325\linewidth]{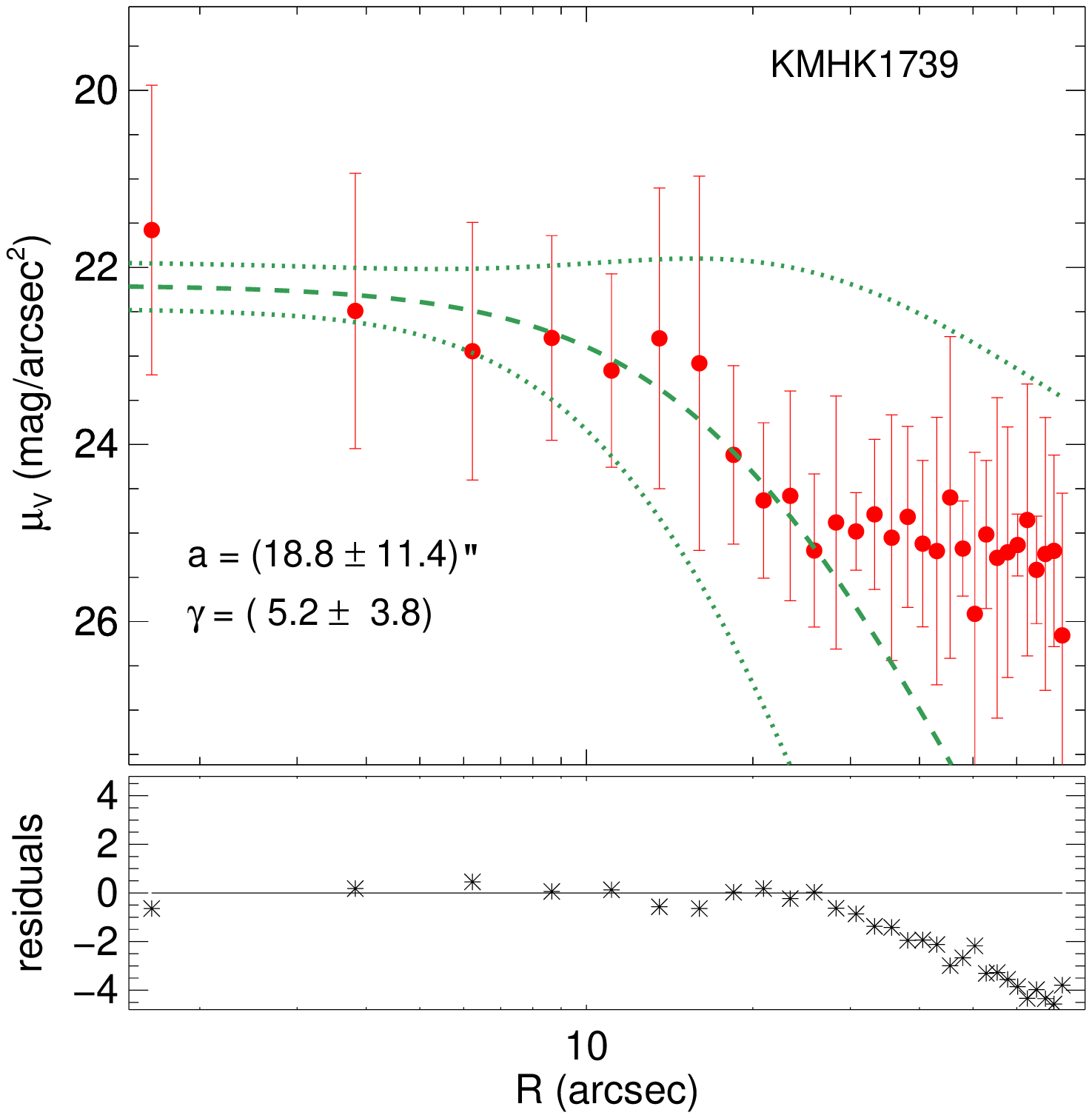}\includegraphics[width=0.325\linewidth]{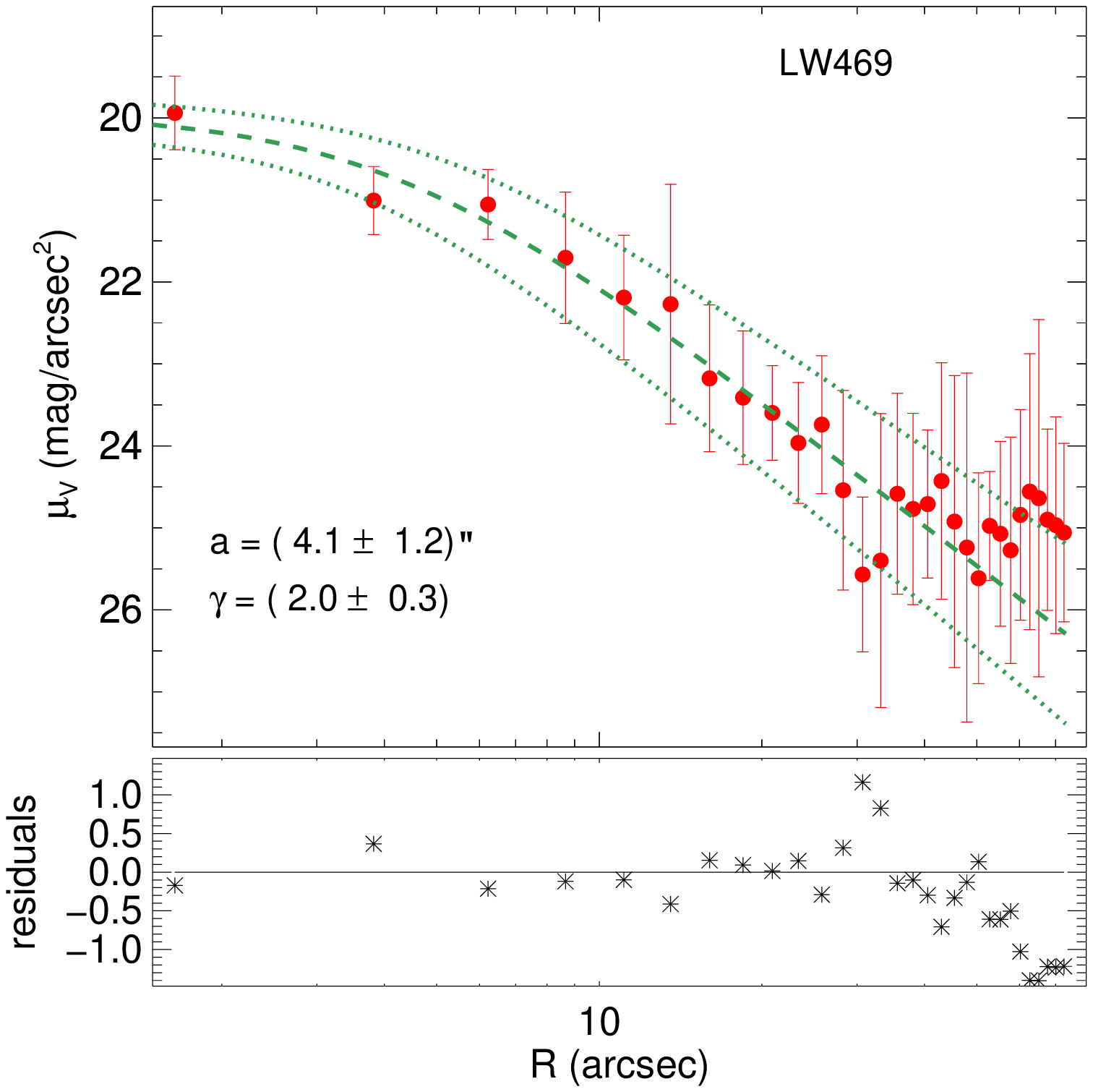}

\includegraphics[width=0.325\linewidth]{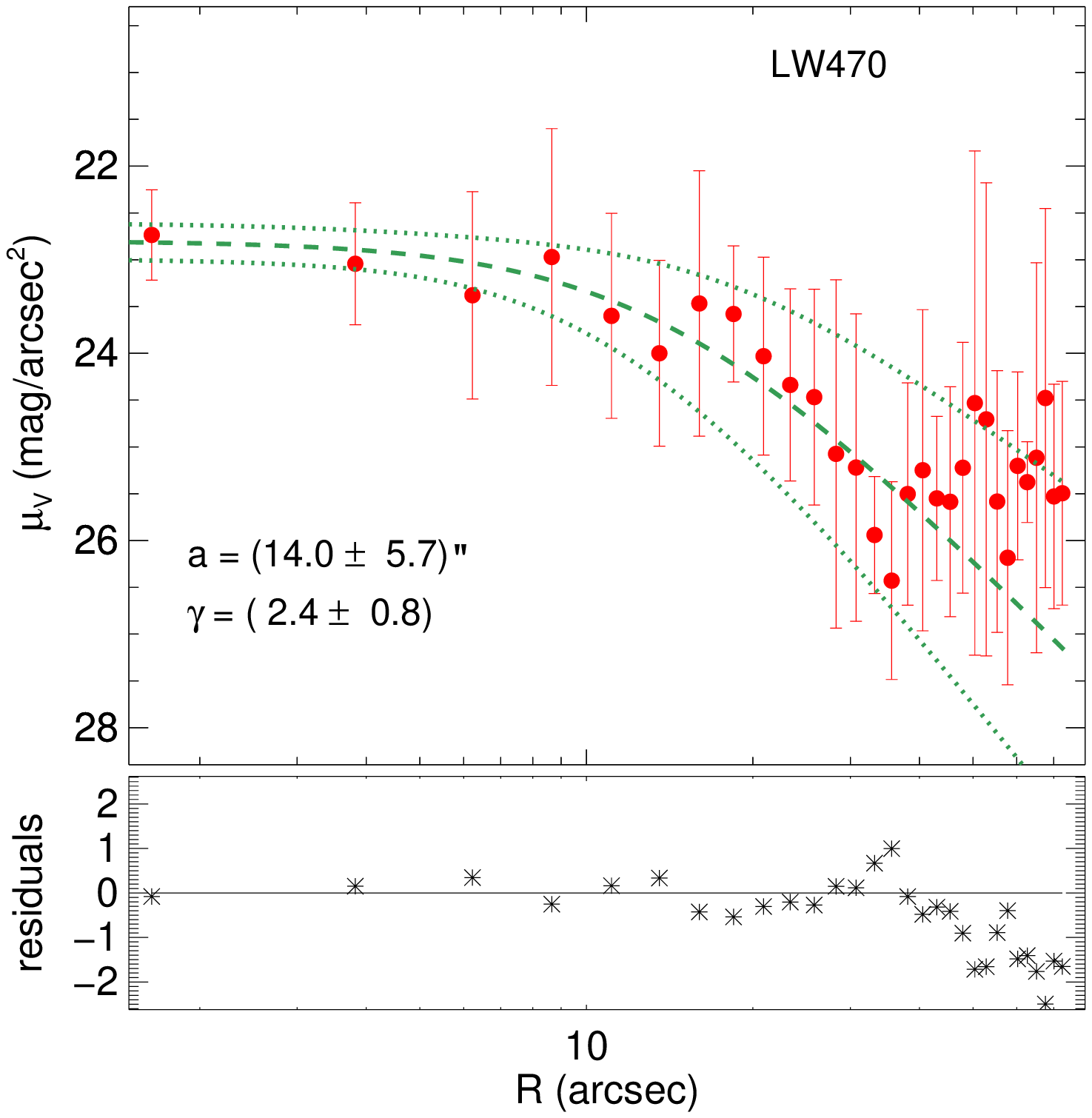}\includegraphics[width=0.325\linewidth]{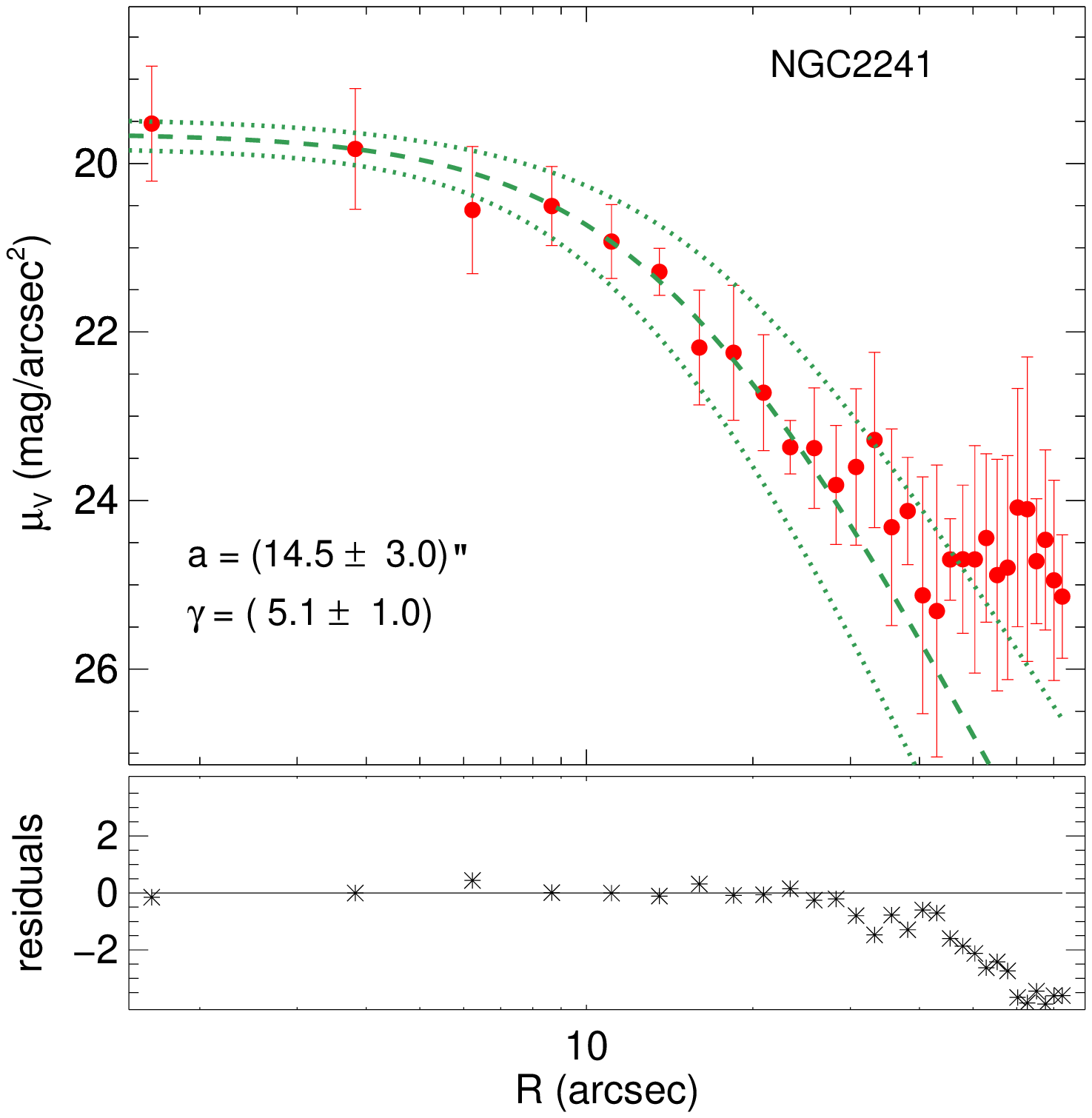}\includegraphics[width=0.325\linewidth]{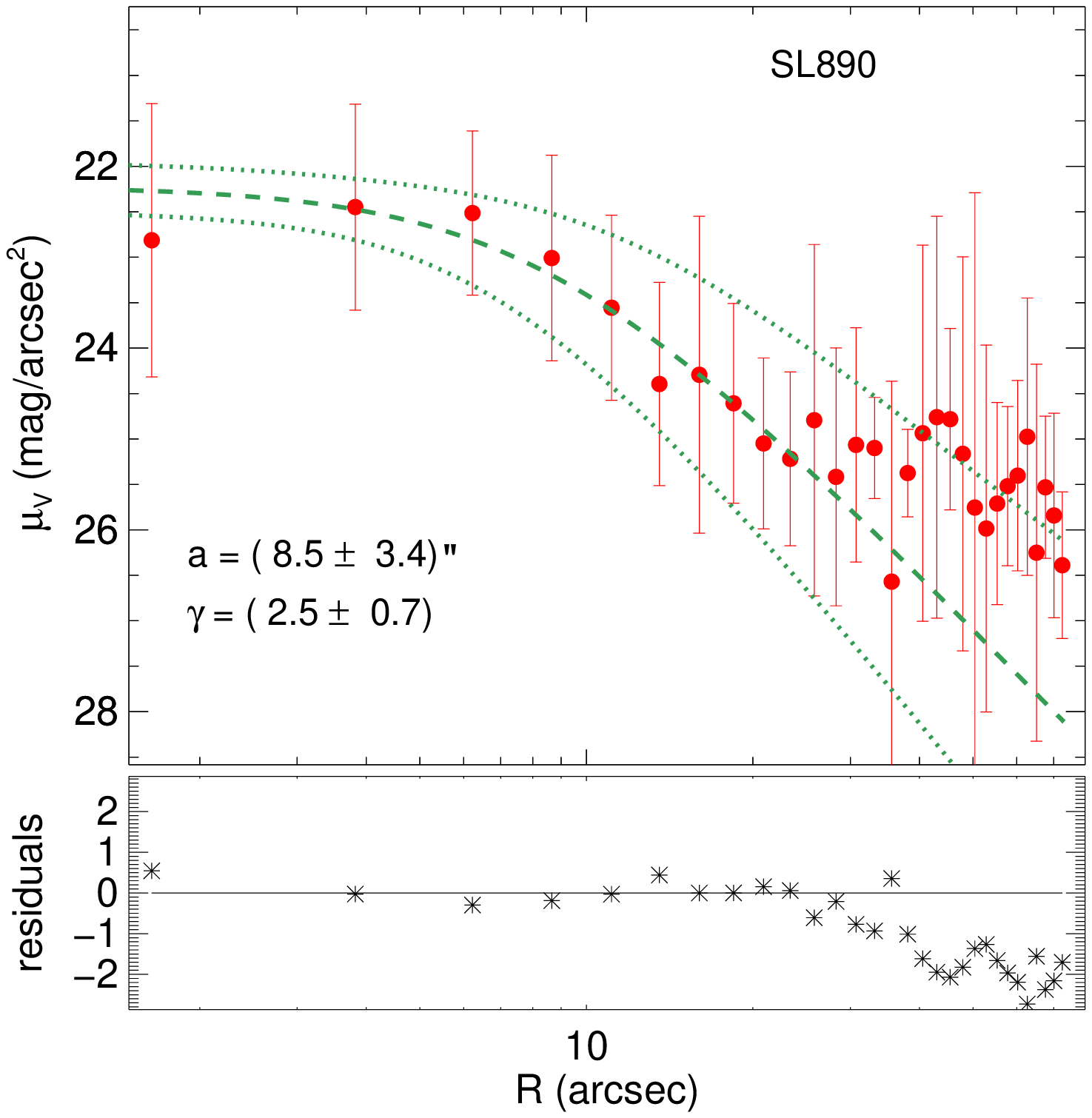}

\includegraphics[width=0.325\linewidth]{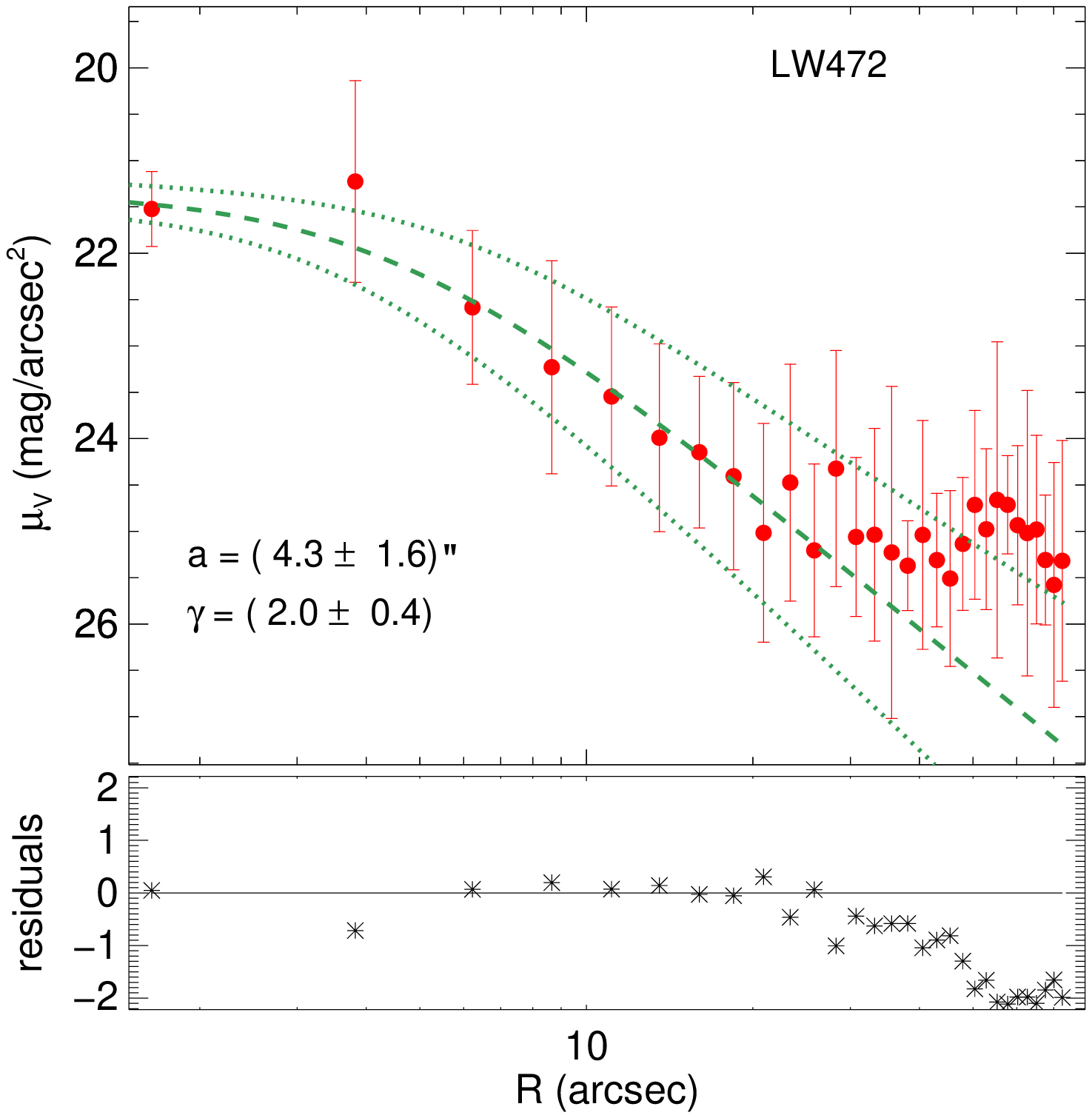}\includegraphics[width=0.325\linewidth]{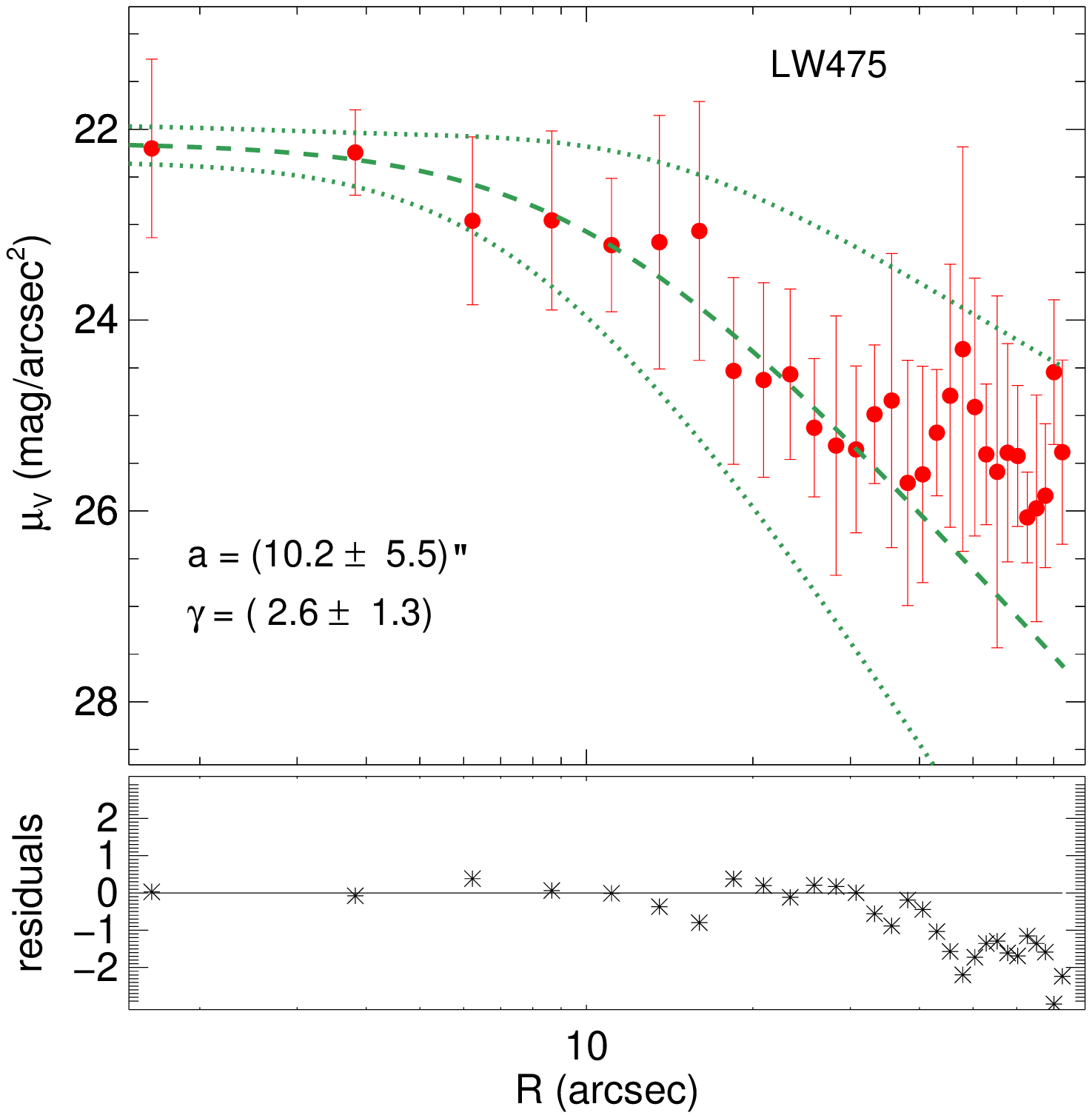}\includegraphics[width=0.325\linewidth]{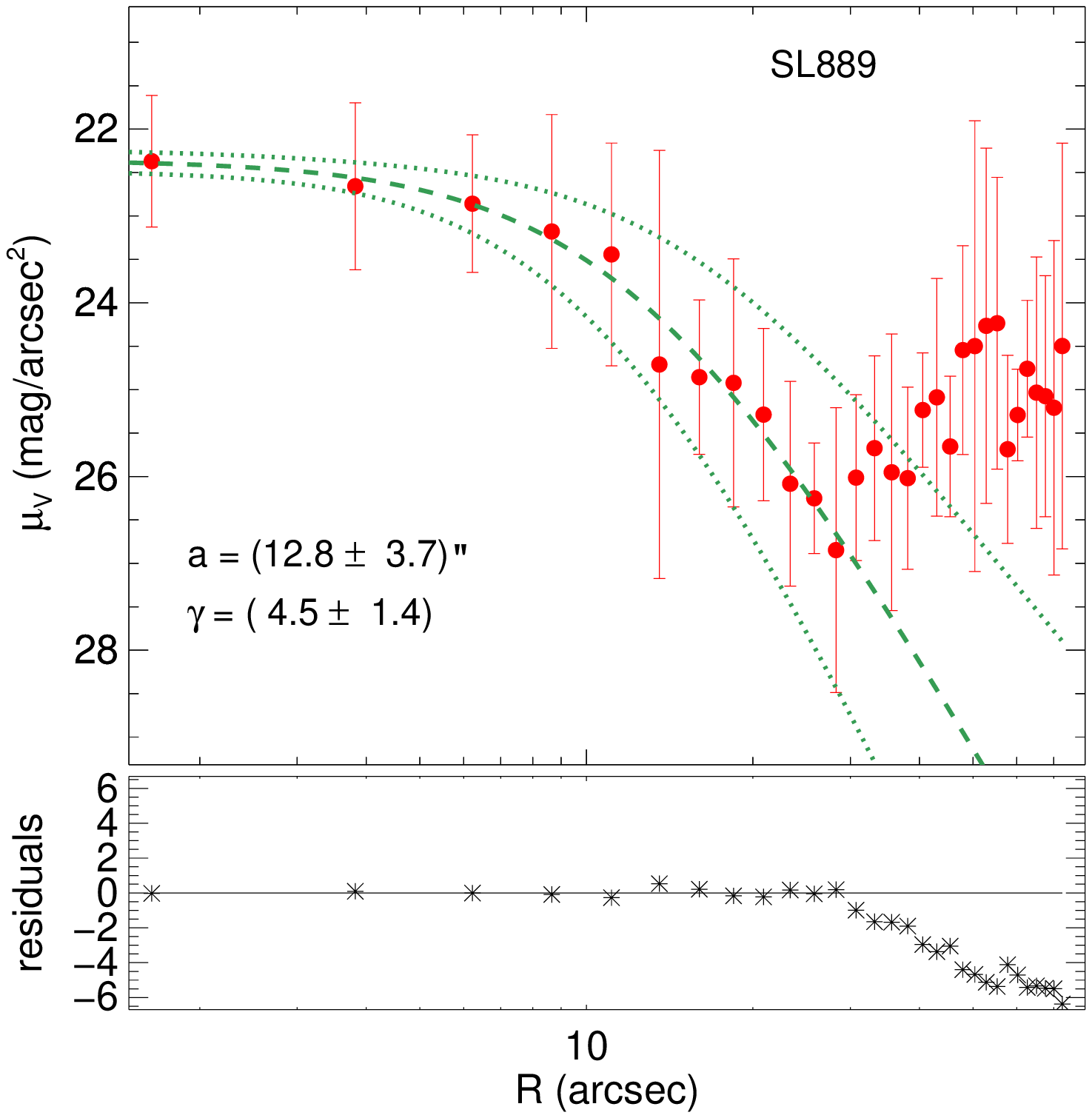}

\caption{cont.}

\end{figure*}

\setcounter{figure}{7}

\begin{figure*}

\includegraphics[width=0.325\linewidth]{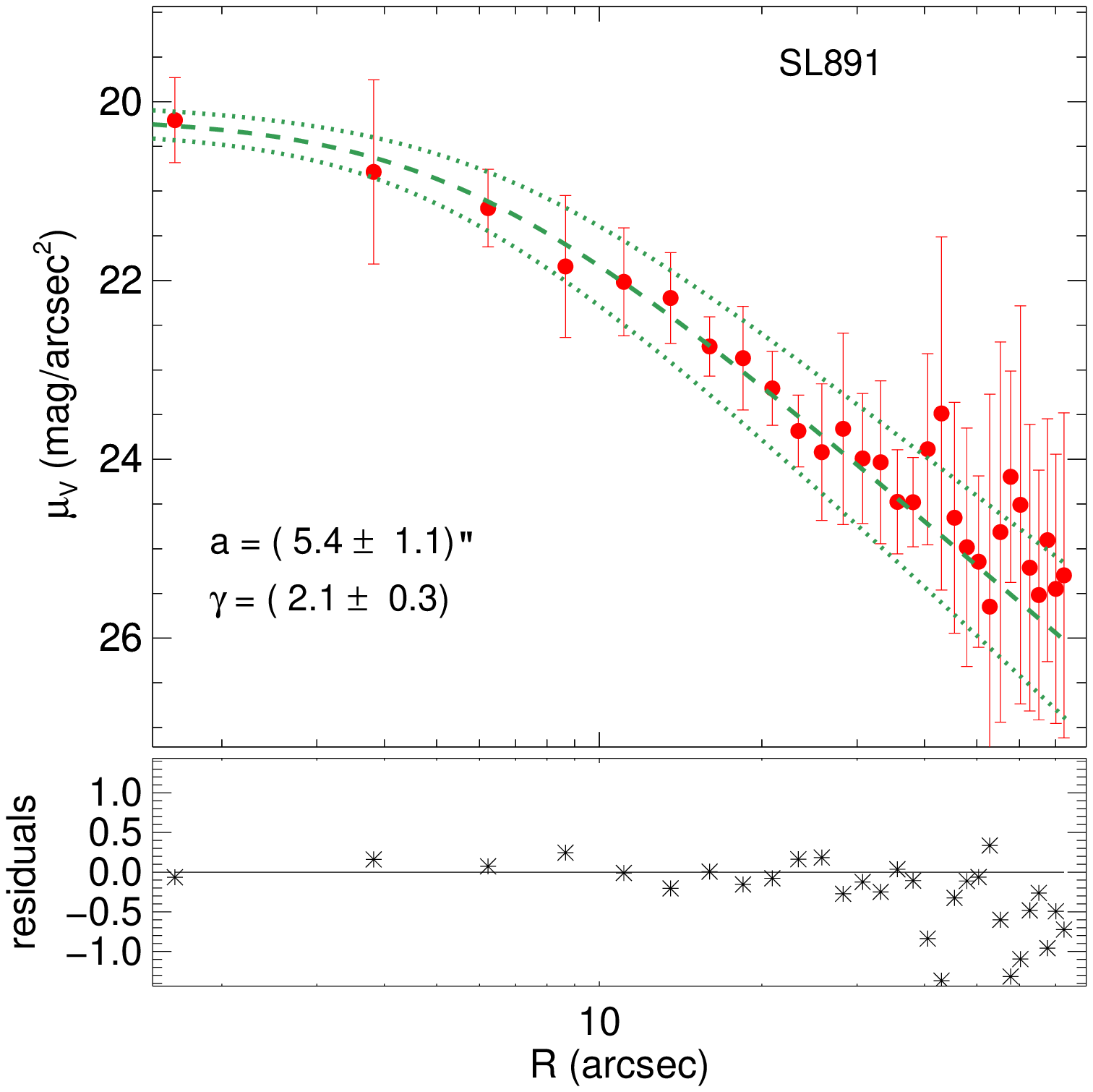}\includegraphics[width=0.325\linewidth]{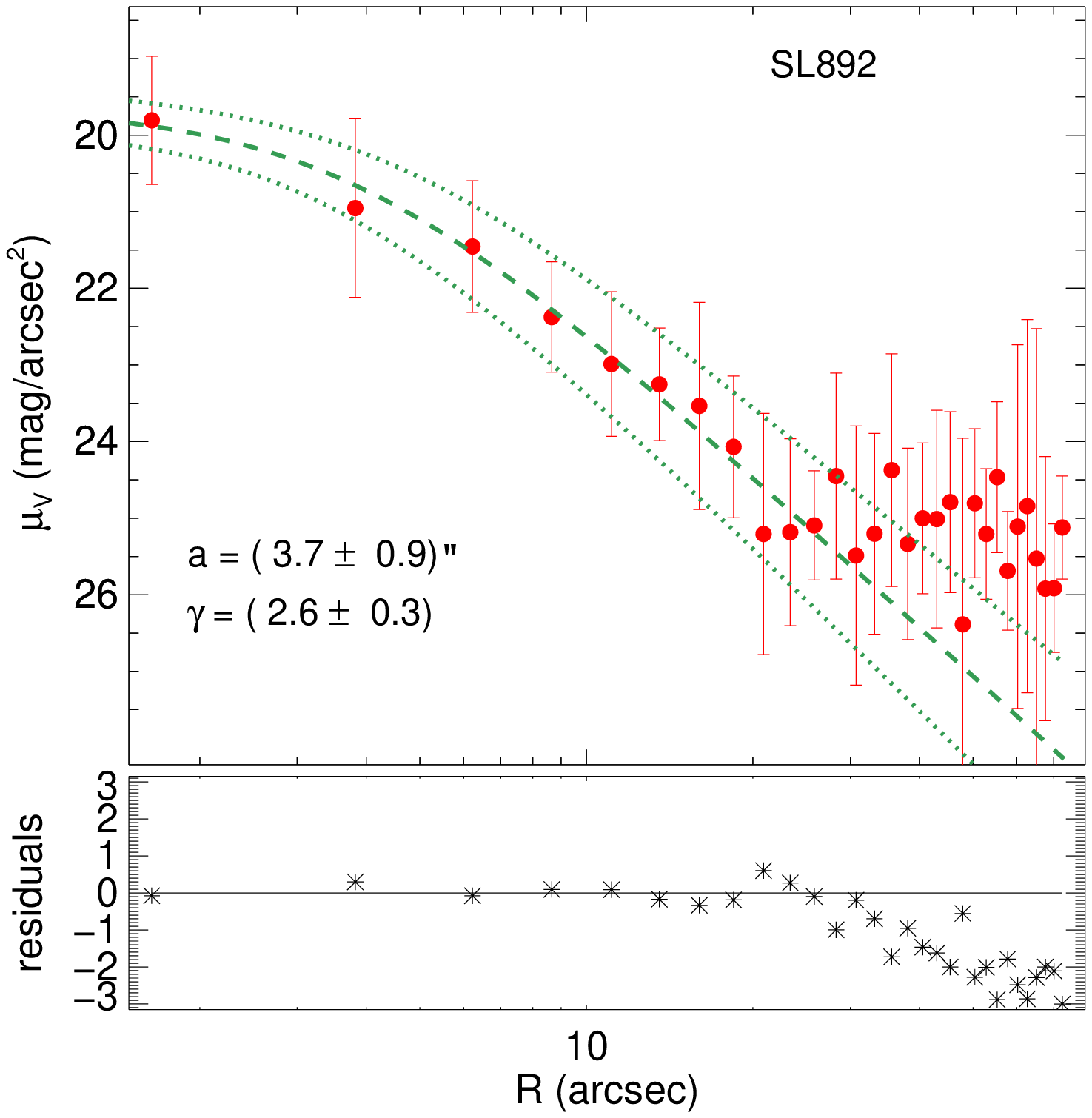}\includegraphics[width=0.325\linewidth]{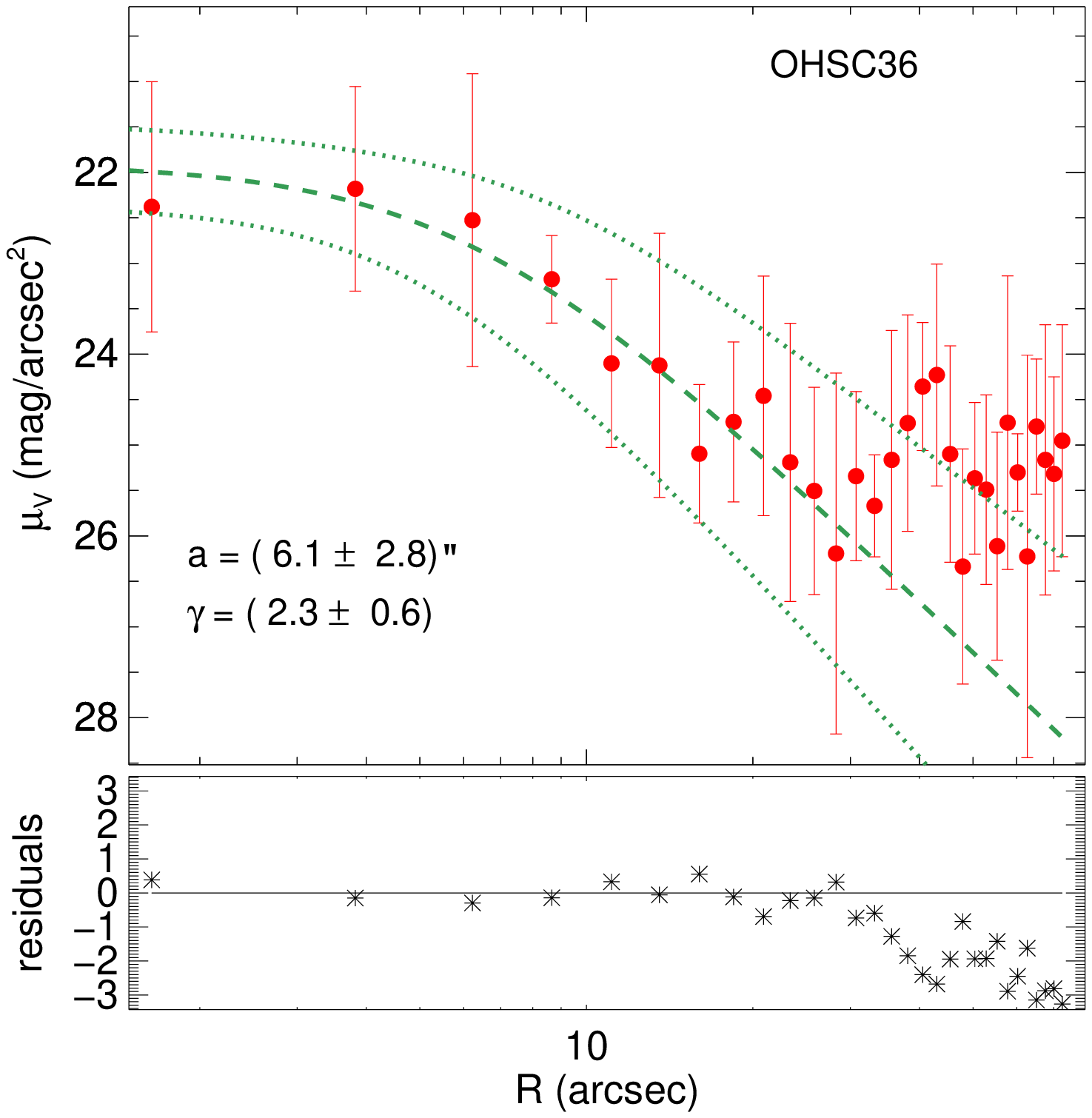}

\includegraphics[width=0.325\linewidth]{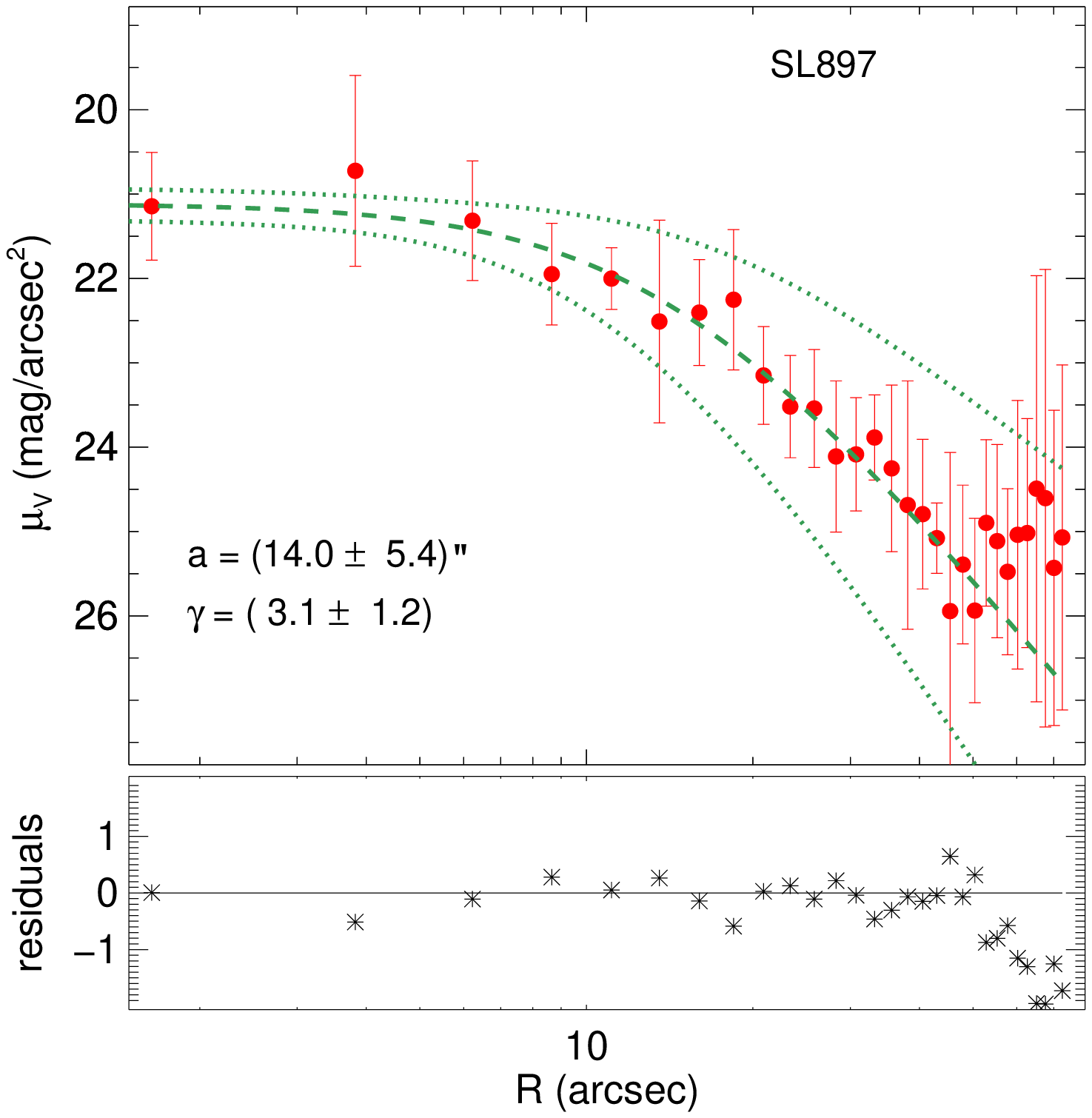}

\caption{cont.}

\end{figure*}

\bsp	

\label{lastpage}
\end{document}